\documentclass[11pt, oneside]{Thesis} 

\graphicspath{{Pictures/}} 
\pdfoutput=1
\usepackage[square, numbers, comma, sort&compress]{natbib} 
\hypersetup{urlcolor=blue, colorlinks=true} 
\title{\ttitle} 
\usepackage[T1]{fontenc}
\usepackage[english]{babel}
\usepackage{makeidx}
\usepackage{graphicx}
\usepackage{epsfig}
\usepackage{verbatim}
\usepackage{subfigure}
\usepackage{color}
\usepackage{amsmath}
\usepackage{feynmf}
\usepackage{amssymb}
\usepackage{amsmath}
\usepackage{multicol}
\usepackage{floatflt}
\usepackage{wasysym} 
\usepackage{bm}
\usepackage{amsfonts}
\usepackage{amssymb}
\usepackage{longtable}
\usepackage{eucal}
\usepackage[norules]{frontespizio}
\usepackage{hyperref}
\usepackage{afterpage}
\usepackage[bottom]{footmisc}
\usepackage{setspace}
\usepackage{multirow}
 \usepackage[table,xcdraw]{xcolor}
\usepackage{mathrsfs}
\usepackage{afterpage}

\begin{document}
\newcommand\blankpage{%
    \null
    \thispagestyle{empty}%
    \addtocounter{page}{-1}%
    \newpage}
 \newcommand{\pt}{p$_{\rm T}$ } 
 \newcommand{\pteq}{p_{\rm T} } 
 \newcommand{\sqnn}{$\sqrt{s_{\rm NN}}$ } 
 \newcommand{\sqnneq}{\sqrt{s_{\rm NN}} } 
 \newcommand{\rppb}{$R_{\rm pPb}$ } 
 \newcommand{\raa}{$R_{\rm AA}$ } 
 \newcommand{\dplus}{D$^{+} $ } 
 \newcommand{\dplusm}{D$^{+}$-meson } 
 \newcommand{\sqs}{$\sqrt{s}$ }   
 \newcommand{\fprompt}{$f_{\rm prompt}$ } 
 \newcommand{\ncoll}{N$_{\rm coll}$ } 
 \newcommand{\npart}{N$_{\rm part}$ } 
 \newcommand{\qppb}{$Q_{\rm pPb}$ } 
\frontmatter 

\setstretch{0.5} 

\fancyhead{} 
\rhead{\thepage} 
\lhead{} 

\pagestyle{fancy} 

\newcommand{\HRule}{\rule{\linewidth}{0.5mm}} 

\hypersetup{pdftitle={\ttitle}}
\hypersetup{pdfsubject=\subjectname}
\hypersetup{pdfauthor=\authornames}
\hypersetup{pdfkeywords=\keywordnames}


\begin{titlepage}
\begin{center}

\textsc{\LARGE \univname}\\[1.5cm] %
\textsc{\Large Doctoral School of Sciences and Innovative Technologies}\\[0.5cm] 
\textsc{\large Indirizzo in Fisica ed Astrofisica}\\[0.5cm] 

\HRule \\[0.4cm] 
{\huge \bfseries \ttitle}\\[0.4cm] 
\HRule \\[1.5cm] 
 
\begin{minipage}{0.4\textwidth}
\begin{flushleft} \Large
Autor:\\

Riccardo Russo
\end{flushleft}
\end{minipage}
\begin{minipage}{0.5\textwidth}
\Large
Supervisor: Stefania Beole'  \\

Co-Supervisor: Francesco Prino \\

Examiner:  Jaroslav Biel\v{c}\`ik
\end{minipage}\\[7cm]
 
\vskip 2 em 

\begin{figure}
\centering
\includegraphics[width=0.3\textwidth]{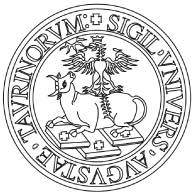}
\end{figure}

\end{center}

\end{titlepage}


%
%
%
%
%
%
\clearpage 
%
\pagestyle{empty} 
\null\vfill 
\textit{Vorrei iniziare ringraziando i miei genitori, Pino e Cristina, per avermi sostenuto in 21 lunghi anni di formazione. Loro e le mie sorelle sono  la famiglia che augurerei a chiunque di avere.\\ \\
Questo lavoro non si sarebbe mai concluso senza la preziosa e costante guida di Stefania e Francesco, che mi hanno introdotto in un ambiente di lavoro stimolante ed altamente formativo, accompagnandomi in tutte le prove che si sono presentate.\\ \\
Ringrazio Elena Bruna, collega ed amica, che e' stata una guida nella mia analisi di fisica in particolare durante il primo anno. Lavorare con lei mi ha reso piu' indipendente e mi ha permesso di proseguire piu' autonomamente l'analisi dei dati protone piombo. Inoltre ringrazio la sua disponibilita' nei giorni prima della discussione.\\
Vorrei anche ringraziare Davide Caffarri e Zaida Conesa del Valle (che se leggera', capira' benissimo anche in italiano). Loro sono stati i coordinatori del gruppo di lavoro da noi chiamato "D2H" durante il periodo in cui ho svolto  l'analisi di dottorato, e sono riusciti a coordinare ottimamente il mio lavoro e quello di altri per ottenere i risultati di fisica esposti in questa tesi. \\
Esprimo un pensiero a parte per il collega ed amico Gian Michele, che ha vissuto il periodo di scrittura della tesi un anno fa, nella scrivania accanto alla mia. E' stato di grande aiuto sia direttamente, mettendo a mia disposizione la sua maggiore esperienza, sia indirettamente mostrandomi come portare a termine un dottorato nel migliore dei modi.\\
Inoltre ringrazio:
\begin{itemize}
\item Stefania Bufalino per il grande supporto durante l'anno al CERN. Sono molto felice che si ritrasferisca da noi a Torino
\item le amiche di Torino: Sarah, Sara, Cristina ed Anastasia, la nostra new entry
\item gli amici di Torino: Maximiliano, Stefano, Niccolo', Bruno Alessandro ed il Prof. Massimo Masera 
\item Yasser, che come me si dottora il 31/03/2015 e insieme a me ha negli ultimi due anni ha subito   il furto di: quattro PC, documenti, passaporto, permesso di soggiorno, due cellulari, valigie con tutti gli indumenti e di certo sto dimenticando qualcosa.
\item gli amici di Ginevra: Andrea, Martino, Stefano, Chiara, Federica, Emilia, Raffaella, Cristina. Avete reso il mio anno a Ginevra un anno migliore
\item Grazia per il grande impegno nel prendersi cura in prima persona del Paper \rppb
\item Chiara Oppedisano per le consulenze in particolare sui Capitoli 3 e 7
\end{itemize}}
%
%
%
\clearpage 

\afterpage{\blankpage}

%
\addtotoc{Abstract} 
\abstract{\addtocontents{toc}{\vspace{1em}} 
\linespread{2.5}
This thesis describes  the measurement of \dplusm production in p--Pb  collisions at \sqnn= 5.02 TeV at the Large Hadron Collider with the ALICE detector.\\
The main goal of the experimental programs on ultra-relativistic heavy ion collisions at the LHC is the production and characterization of the Quark Gluon Plasma (QGP),  a phase of nuclear matter in which strongly interacting constituents (quarks and gluons) are deconfined.  Heavy quarks are considered effective probes of the properties of the QGP as they are created on a short time scale, with respect to that of the QGP, and subsequently interact with it. \\
Moreover, for a proper assessment of the characteristics of the matter produced in heavy-ion collisions, it is important to disentangle the final state effects due to the formation of a QGP from the initial state effects due to the fact that nuclei are present in the colliding system. Both initial and final state effects may lead to  qualitatively similar phenomena in the observables of interest.\\
The measurement of charmed meson production in proton-nucleus collisions allows to assess  initial state effects present in nuclear collisions, under the assumption that an extended deconfined medium is not created  in this kind of interactions. The nuclear modification factor of $D$ mesons in p--Pb collisions (\rppb) is essential for a complete understanding of the modification of $D$ mesons momentum distributions  observed in Pb--Pb collisions at \sqnn= 2.76 TeV, which is interpreted as due to the $c$-quark energy loss in the medium. In addition,  some of the results obtained from high-multiplicity p--Pb collisions at LHC, such as the ridge structure in the two-particle correlation function,  turned out to be unexpected, and have been interpreted   in terms of final state effects such as hydrodynamic flow. These aspects make a study of charmed meson production in p--Pb collisions as a function of the  event multiplicity of great interest.
\\ \\
In the first Chapter of this thesis, an introduction to the physics of heavy-ion collisions will be given. The second Chapter will be dedicated to a description of heavy flavour
production in proton-proton, proton-nucleus and nucleus-nucleus collisions, focusing in particular on the initial and final state effects present in nuclear interactions. In Chapter three,
the main features of the ALICE apparatus will be discussed, with a specific focus on the
detectors that are directly involved in the \dplusm  analyses. In Chapter four, the procedure used to extract the \dplusm yield from the p--Pb data sample collected in 2013 by the ALICE Collaboration is described.  Chapter five is focused on the measurement of the prompt \dplusm nuclear modification factor in p--Pb collisions at \sqnn= 5.02 TeV which was published in \cite{PaperpPb}. Chapter six describes a data-driven method for separating the contribution of prompt \dplus mesons from that of \dplus mesons coming from B-hadron decays alternative to the one based on theoretical calculations used in \cite{PaperpPb}. In the last Chapter the measurement of \dplusm production  as a function of event multiplicity will be discussed.
\\ \\
The measurements presented in this thesis were approved by the ALICE Collaboration and presented in various conferences. The few measurements that are not yet approved will report the label "This Thesis".

%

%
\addtotoc{Abstract} 
\abstract{\addtocontents{toc}{\vspace{1em}} 
\linespread{2.5}
Questa tesi espone la misura della produzione di mesoni D$^{+}$ in collisioni protone piombo a \sqnn= 5.02 TeV con l'esperimento ALICE a LHC. \\
Il principale obiettivo degli esperimenti sulle collisioni di ioni pesanti ultrarelativistiche e' la produzione e la caratterizzazione del plasma di quark e gluoni (QGP), una fase della materia nucleare in cui i costituenti elementari del nucleo (quark e gluoni) sono deconfinati. I quark pesanti sono considerati utili sonde per lo studio del QGP perche' vengono creati in tempi brevi rispetto alla durata della fase di QGP con il quale interagiscono. \\
Inoltre, per una completa caratterizzazione della materia prodotta in collisioni nucleari, e' importante distinguere gli effetti di stato finale dovuti alla formazione del QGP da quelli di stato iniziale dovuti al fatto che le particelle che collidono sono nuclei. Entrambe le categorie di effetti possono risultare in fenomeni qualitativamente simili nelle osservabili di interesse.\\
La misura della sezione d'urto di produzione di mesoni charmati in collisioni protone nucleo permette di evidenziare gli effetti di stato iniziale presenti nelle collisioni nucleari, se si assume che in queste collisioni non si crei un mezzo deconfinato esteso. Il fattore di modificazione nucleare dei mesoni $D$ in collisioni protone piombo (\rppb) e' una quantita' fondamentale per comprendere la modifica delle distribuzioni in momento dei mesoni $D$ osservata in collisioni piombo piombo ad una energia nel centro di massa \sqnn= 2.76 TeV, interpretata come un effetto della perdita di energia dei quark $c$ nel mezzo. Inoltre, alcune delle misure  ottenute in collisioni protone piombo ad alta molteplicita' a LHC hanno dato risultati inaspettati, che sono stati interpretati come effetti di stato finale e di flusso idrodinamico. Lo studio della produzione di mesoni charmati in funzione della molteplicita' in collisioni protone piombo ha quindi un suo interesse.
\\ \\
Nel primo capitolo di questa tesi verra' introdotta la fisica delle collisioni nucleari. Il secondo capitolo e' dedicato alla produzione di quark pesanti in collisioni protone protone, protone nucleo e nucleo nucleo, con un'attenzione particolare sugli effetti di stato iniziale e finale. Il terzo capitolo descrive sommariamente l'apparato dell'esperimento ALICE, approfondendone le componenti maggiormente usate nello studio dei mesoni $D$. Il quarto capitolo descrive la strategia di ricostruzione dei mesoni $D$ in collisioni protone piombo. Il quinto capitolo  e' dedicato alla misura del fattore di modificazione nucleare dei mesoni D$^{+}$  in collisioni protone piombo ad un'energia nel centro di massa di \sqnn= 5.02 TeV, risultato pubblicato in \cite{PaperpPb}. Il sesto capitolo descrive un metodo di sottrazione dei mesoni $D$ da decadimento di quark beauty basato sui dati, alternativo a quello basato su predizioni teoriche utilizzato in  \cite{PaperpPb}. L'ultimo capitolo discute la misura della produzione di mesoni D$^{+}$ in funzione della molteplicita' dell'evento protone piombo. 
\\ \\
I risultati presentati in questa tesi sono stati approvati dalla Collaborazione ALICE, e presentati a diverse conferenze internazionali. Alcuni risultati non ancora approvati sono mostrati con l'etichetta "This thesis".

\afterpage{\blankpage}
\clearpage 
%
%
%
%


\pagestyle{fancy} 

\lhead{\emph{Contents}} 
\tableofcontents 

\mainmatter 

\pagestyle{fancy} 



\chapter{Introduction: Quarks, Gluons and  Quark-Gluon Plasma} 

\label{Chapter1} 

\lhead{Chapter 1. \emph{Introduction: Quarks, Gluons and  Quark-Gluon Plasma}} 

\begin{spacing}{1.0}

\section{Quantum Chromodynamics and Confinement}
During the 60's  the strong force was by far the most poorly understood among the four fundamental interactions.
Many new particles were discovered in those years beside proton, neutron and electron in high energy experiments. 
Gell-Mann (Nobel prize, 1969) suggested that these new particles could be grouped in  multiplets (singlets, octects, decuplets) 
each one  composed of particles having  roughly  the same mass, hypothesizing the existence of an underlying new physics 
obeying an unknown internal symmetry between pointlike constituents. 
The observation of strange particles  suggested that
the symmetry group could be SU(3), and that hadrons consist of  elementary constituents named quarks,
that belong to the fundamental representation of SU(3):
the up and down quarks are the constituents of ordinary matter (protons and neutrons), while the strange
quark is present in strange particles \cite{hm}. All hadrons known at that time  were grouped in non-fundamental representations of the SU(3) group, octets and decuplets (the same way nucleons and pions were identified as representations of the isospin SU(2) symmetry).\\
Figure 1.1 shows the $\Delta$  decuplet,
which however exhibits an evident violation of the Pauli principle:
$\Delta^{++}$, $\Delta^{-}$ and $\Omega$ have 3 quarks of the same flavour in the same spin state and this forbids
their wave function 
\begin{equation}
\centering
	    \psi = \psi(x) * \psi_{flavour} * \psi_{spin}
\end{equation}
to be completely antisymmetric. \\
\begin{figure}[htbp]
\centering
\includegraphics[width=0.5\textwidth]{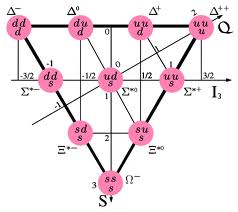}
\caption{The $\Delta$ decuplet: three axis are shown, relative to the quantum numbers that identify each particle:
isospin, electric charge and strangeness}
\end{figure}
This puzzle led to the introduction of a new degree of freedom for quarks named color, with the following properties:
\begin{itemize}
\item there are 3  colors for quarks  and 3 anticolors for antiquarks
\item all hadrons in nature show up in a colorless combination of quarks, that is 3 different colored quarks (baryons) or
a quark with one color and an antiquark with the respective anticolor (mesons); this property is necessary to explain why
this new degree of freedom never shows up in hadronic interactions
\end{itemize}
In the late 60's,
experiments carried out  to explore the electromagnetic structure of protons
by mean of electromagnetic probes (electrons and muons), showed clearly that, 
above a few GeV/$c$ of momentum transfer, scattering occurs between the accelerated electrons (or muons) 
and pointlike constituents. 
These deep inelastic scattering (DIS) experiments lead Feynman (Nobel prize, 1965) to the formulation of the parton theory,
in which protons are made of elementary constituents; partons and quarks are the same
thing, and DIS experiments were actually the first observation of quarks \cite{dis}.\\
In those years   the importance of gauge theories was becoming
evident, based on the success of Quantum ElectroDynamics (QED), a theory based 
on the U(1) symmetry group with one conserved charge. It was argued that analogously the strong interaction could be described by a gauge theory  invariant under  SU(3) group transformations (local rotations in color space) with three conserved colour charges.\\
In QED the gauge transformations correspond to changes in the phase of the wave function
\begin{equation}
\centering
\psi \rightarrow e^{i\alpha} \psi
\end{equation}
and if the transformation is local ($\alpha=\alpha(x)$) the derivative of the wavefunction varies by a non-trivial term
\begin{equation}
\centering
\partial_{\mu}\psi \rightarrow \partial_{\mu}e^{i\alpha} \psi = e^{i\alpha}(\partial_{\mu}\psi) + e^{i\alpha}(i\partial_{\mu}\alpha)\psi
\end{equation}
 The unwanted second term is cancelled by the gauge change of the electromagnetic potential $A_{\mu}$,
\begin{equation}
\centering
A_{\mu} \rightarrow A_{\mu} + \frac{1}{e} \partial_{\mu}\alpha
\label{minimalQEDcoupling}
\end{equation}
if the potential is added
to the derivative operator in minimal coupling ($\partial_{\mu}\psi +ieA_{\mu}$) .
The QED lagrangian is thus obtained as
\begin{equation}
\centering
\textit{L}_{QED} = i\bar{\psi}\gamma^{\mu}(\partial_{\mu}\psi +ieA_{\mu})\psi - m\psi\bar{\psi} -\frac{1}{4}F_{\mu\nu}F^{\mu\nu}
\end{equation}
where the electromagnetic field tensor $F_{\mu\nu} = \partial_{\mu}A_{\nu} - \partial_{\nu}A_{\mu}$ is also  invariant under the Gauge transformation of Equation
1.4.	\\
From this Lagrangian the following interaction vertex between two fermions and the QED vector boson, i.e. the photon, is obtained\\
\begin{center}
\begin{fmffile}{vertex0}

  \begin{fmfgraph*}(110,62) 
    \fmfleft{i1,i2}
\fmfright{o1}
\fmf{fermion}{i1,v1,i2}
\fmf{photon}{v1,o1}
  \end{fmfgraph*}

\end{fmffile}
\end{center}
In QCD there are three colors, so the wave function has three components in color space, $\psi = (\psi_{r}
,\psi_{b},\psi_{g})$; then a color gauge transformation is described by a 3$\times$3 unitary matrix U with det(U)=1, and $\psi \rightarrow U\psi$.
U can be written as $e^{iL}$ with L hermitian, and being det(U)=1, L must also be traceless. All traceless 3$\times$3 hermitian matrices
can be written as a linear combination of the eight Gell-Mann matrices $\lambda_{a}$
\begin{equation}
\centering
L = \sum_{i=1}^{8}\theta_{i}\lambda_{i}
\end{equation}
If the transformation operated by U is local,
\begin{equation}
\centering
U(x) = e^{i \sum_{i=1}^{8}\theta_{i}(x)\lambda_{i}}
\end{equation}
 the derivative acquires an unwanted additional term:
\begin{equation}
\centering
\partial_{\mu}(U(x)\psi) = U\partial_{\mu}\psi + (\partial_{\mu}U)\psi = U[\partial_{\mu}\psi + U^{*}(\partial_{\mu}U)\psi] 
\end{equation}
There is now an additional term $U^{*}(\partial_{\mu}U)$, where $U^{*}$ is the hermitian conjugate of U. This contribution is cancelled with the introduction of 
a color potential $A_{\mu}$ which is a 3$\times$3 matrix and can be reprensented as a linear combination of the Gell-Mann matrices
\begin{equation}
\centering
A_{\mu}(x) = \frac{1}{2}\sum_{a=1}^{8}A_{\mu}^{a}(x)\lambda_{a}
\end{equation}
where $A_{\mu}^{a}$ are eight real potentials:  one has to introduce eight color potentials instead of one like in the
electromagnetic case  (Equation \ref{minimalQEDcoupling}).
If the potential changes under local rotations as 
\begin{equation}
\centering
A_{\mu} \rightarrow U^{*}A_{\mu}U - i\frac{1}{g}U^{*}(\partial_{\mu}U)
\end{equation}
where we have introduced the coupling strength g, the minimally coupled derivative remains invariant under the gauge transformation
\begin{equation}
\centering
(\partial_{\mu} - igA_{\mu})U\psi = U(\partial_{\mu} - igA_{\mu})\psi
\end{equation}
By defining
\begin{equation}
\centering
F_{\mu\nu}^{a} = \partial_{\mu}A_{\nu}^{a} - \partial_{\nu}A_{\mu}^{a} + gf_{abc}A_{\mu}^{b}A_{\nu}^{c}
\end{equation}
where $f_{abc}$ are the structure constants of  SU(3), we obtain a gauge invariant lagrangian for QCD\footnote{the therm $m\bar{\psi}\psi$ explicitly breaks the SU(3) gauge symmetry, but when QCD is coupled to electroweak theory to complete the Standard Model the Higgs mechanism is  introduced to restore the symmetry} \cite{muller}
\begin{equation}
\centering   
L_{QCD} = -\dfrac{1}{4}F_{\mu\nu}^{(a)}(x)F^{\mu\nu}_{(a)}(x)+\bar{\psi}(x)(i\gamma_{\mu}\partial^{\mu}+g\gamma_{\mu}A_{(a)}^{\mu}(x))\psi(x)-m\bar{\psi}(x)\psi(x)
\label{QCDlag}
\end{equation}
In order to get local invariance, 8 gauge fields have been introduced: gluons are thus 
the vector bosons of the theory explaining the strong interaction. They are massles vector bosons, interacting with quarks via the following vertex present in the lagrangian 
\\
\begin{center}
\begin{fmffile}{vertex1}

  \begin{fmfgraph*}(110,62) 
    \fmfleft{i1,i2}
\fmfright{o1}
\fmf{fermion}{i1,v1,i2}
\fmf{curly}{v1,o1}
  \end{fmfgraph*}
\end{fmffile}
\end{center}
very similar to the QED vertex, but also among themselves \cite{gr}.\\
\begin{center}
\begin{fmffile}{vertex2}
\begin{fmfgraph*}(110,62)
\fmfleft{i1,i2}
\fmfright{o1,o2}
\fmf{curly}{i1,v1,i2}
\fmf{curly}{o1,v1,o2}
\end{fmfgraph*}%
\end{fmffile}
\begin{fmffile}{vertex3}
\begin{fmfgraph*}(140,56)
\fmfleft{i1,i2}
\fmfright{o1}
\fmf{curly}{i1,v1,i2}
\fmf{curly}{v1,o1}
\end{fmfgraph*}
\end{fmffile}
\\
\end{center}
The last vertices are characteristic of non-abelian theories such as QCD\footnote{QCD is non-abelian because the generators of the transformation group, the $\lambda$ matrices, do not commute among themselves}. They derive from the fact that in QCD (contrary to QED) the force mediators carry a non-zero value of the color charge.
\subsection{Confinement}
The concept of confinement was introduced by Gross, Politzer and Wilczek in 1964 \cite{Asimp} (Nobel prize in 2004). \\
The renormalization corrections in QCD include those similar to the QED ones (Figure 1.2)
\begin{figure}
\centering
\includegraphics[width=0.45\textwidth]{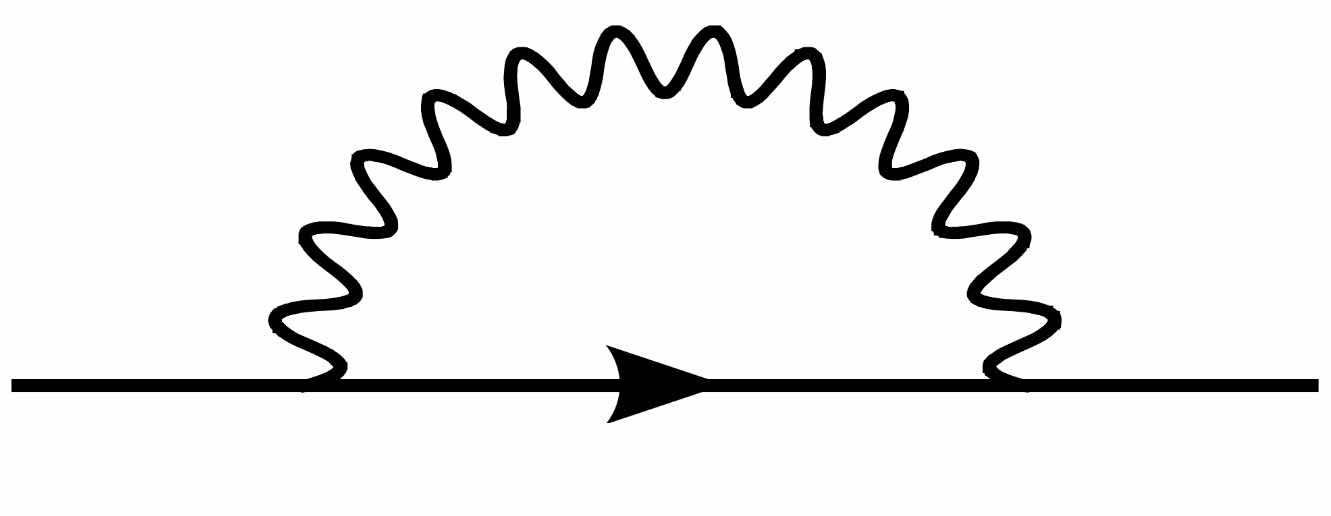}%
\quad\quad\
\includegraphics[width=0.45\textwidth]{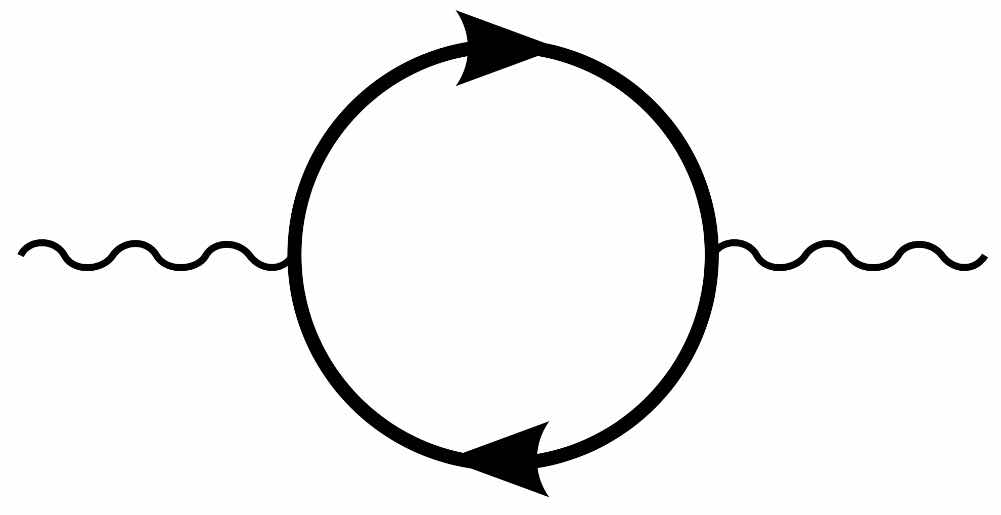}
\caption{Left: self energy of a fermion. Right: vacuum polarization diagram.}
\end{figure}
but also those characteristic of a non abelian theory (Figure 1.3)
\begin{figure}[t]
\centering
\includegraphics[width=0.5\textwidth]{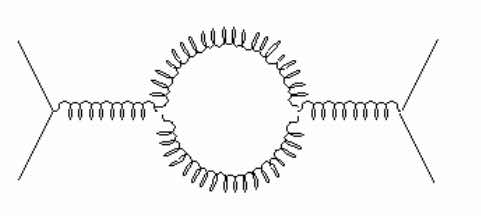}
\caption{Correction to the gluon propagator characteristics of QCD: this diagram is possible because of the self-interaction of gluons}
\end{figure}
This leads to the following scaling rule for the strong coupling constant
\begin{equation}
\centering
\alpha_{s}(|q^{2}|) = \frac{\alpha_{s}(\mu^{2})}{1+(\alpha_{s}(\mu^{2})/12\pi)(33-2n_{flavour})\ln(\frac{|q^{2}|}{\mu^{2}})} 
\end{equation}
where $\mu$ is the momentum scale considered and $n_{flavour}$ is the number of flavours considered in the theory.
Considering three flavours, the coupling constant increases with decreasing momentum scale $\mu$, and  the perturbative approach is therefore no longer valid.
The same formula may be expressed in the following way
\begin{equation}
\centering
\alpha_{s}(|q^{2}|) = \frac{12\pi}{(33-2n_{flavour})\ln(\frac{|q^{2}|}{\Lambda_{\rm QCD}})} 
\end{equation}
where $\Lambda_{\rm QCD}$ is the momentum scale at which alpha diverges. This means that as one tries to pull apart two 
colored quark constituing a colorless meson, their strong interaction gets more and more intense: the potential between them has the form 
\begin{equation}
\centering
V(r) = -\frac{\alpha}{r} + Kr
\label{eqconf}
\end{equation}
and increases linearly with distance for $r$>1 fm, so that for a certain distance it becomes energetically
convenient to create a $q\bar{q}$ pair with the right colors to create 2 new colorless mesons \cite{ba}. This simple example explains color confinement. All hadronic states we observe in nature, such as  nucleons in  heavy nuclei, are colorless states and they
interact with a force which is the residual of the color interaction among their constituents. 
\subsection{Deconfinement}
The first hint of existence for a phase transition in ordinary nuclear matter at high temperatures was given by Rolf Hagedorn's statistical model for hadronic systems \cite{hagedorn}. The model  predicted the density of a given hadron species of mass $m$ in the system
\begin{equation}
\frac{d\rho}{dm} \approx m^ae^{m/m_0}
\end{equation} 
where $a$ and $m_{0}$  are  model parameters. In 1965 Hagedorn showed that  this exponential behaviour results in  a limiting temperature T$_{c}$ for the energy density of the system not to diverge. The interpretation given at that time was that an increase of energy  at $T_{c}$ results in the creation of heavier hadrons, without increasing the average system momenta. Later Cabibbo and Parisi \cite{Cabibbo}  suggested that  Hagedorn's  exponential spectrum of hadronic states is not necessarily connected with a limiting temperature, but  is present in any system which undergoes a second order phase transition, e.g. from ordinary nuclear matter  to  a different phase  in which quarks and gluons are not confined.\\
Ten years later, shortly after the formulation of QCD, a very simple phenomenological model for hadron structure was introduced, the MIT Bag Model \cite{MITBag}. In the simplest version of this model hadrons are containers of massless quarks, free to move on a limited volume. Inside this volume quarks are allowed to move as quasi-free particles. Their wave function can be obtained from the Dirac equation and the short-distance scale characteristic of the bag allows the use of a perturbative approach to QCD. The vacuum surrounding the bag exerts an inward pressure B, which represents long-distance, non perturbative phenomena of QCD (confinement).\\
In this simplified massless model the Dirac equation
\begin{equation}
(i\gamma^{\mu}p_{\mu} - m) = 0 
\end{equation} 
becomes
\begin{equation}
(\gamma^{0}p_{0} - \vec{\gamma}\cdot\vec{p}) = 0 
\end{equation} 
In the Dirac representation of the Gamma matrices the equation can be written as
\begin{equation}
 \begin{pmatrix}
 p^0 &  \vec{\sigma}\cdot\vec{p}  \\
   \vec{\sigma}\cdot\vec{p} & -p^0  \\
 \end{pmatrix}  \begin{pmatrix}
 \varphi_+  \\
  \varphi_-  \\
 \end{pmatrix}= 0 
\end{equation}
where $\sigma$ are the 2x2 Pauli matrices. This equation can be solved analytically and its lowest energy solutions are
\begin{equation}
 \varphi_+ = Ne^{-ip^ot}j_0(p^0R)\chi_+   \  \   \   \  \   \varphi_- = Ne^{-ip^ot} \vec{\sigma}\cdot\hat{r}j_1(p^0R)\chi_-
\end{equation}
where $j_0$ and $j_1$ are the Bessel functions, $\chi_+$ ad $\chi_-$ the Weil spinors and both wave functions are normalized to the number of quarks inside the bag $N$. \\
Confinement can be imposed requiring the current flux 
\begin{equation}
J_{\mu} = \bar{\varphi}\gamma_{\mu}\varphi
\end{equation}
through the bag surface of radius R to be zero
\begin{equation}
n^{\mu}J_{\mu} = 0
\end{equation}
being $\vec{n}$ the vector normal to the bag surface. This requirements translate into
\begin{equation}
\bar{\varphi}\varphi|_{r=R} = [j_0(p^0R)]^2 -  (\vec{\sigma}\cdot\hat{r})^2[j_1(p^0R)]^2 = 0
\end{equation}
This condition is fullfilled if $p^0R$ = 2.04, which means that the  energy of the quarks inside  the bag is E = 2.04N/R. Including the potential energy of the external pressure B, the total energy becomes E = 2.04N/R +4$\pi$/3R$^3$B. The bag is in equilibrium if $\partial E$/$\partial R$ = 0, i.e.
\begin{equation}
4 \pi R^2 B - \frac{2.04N}{R^2} = 0
\end{equation}
For a proton ($N$=3, $R$ = 0.8 fm) we obtain B$^{1/4}$ = 206 MeV. We can get some quantitative estimations of the critical parameters by applying statistical thermodynamics  with some simple hypotheses.
In the high temperature limit we consider a system of non interacting, massless quarks, antiquarks and gluons, with the further hypotesis that
the net baryon number is zero. Quarks and gluons are confined inside the bag
and exert a pressure on its boundaries. The partial pressures of the different constituents inside the bag
can be obtained by adapting the Maxwell-Boltzmann equation for blackbody radiation \cite{sh};
the expression for their sum is
\begin{equation}
\centering 
      P = [g_{g} +\frac{7}{8}\times(g_{q} + g_{\bar{q}})]\frac{\pi^{2}}{90}T^{4}
\label{SBpressure}
\end{equation}
where $g_{q}$ and $g_{g}$ are the quarks and gluons number of degrees of freedom, respectively. \\
For gluons we have two polarization states, so $g_{g}=16$; for quarks we have
\begin{equation}
\centering
g_{q}=g_{\bar{q}}=N_{Colour}N_{Flavour}N_{Spin}
\end{equation}
and in the simple case of 2 flavours $g_{total}=37$.
Hence we find that the critical temperature at which the internal pressure overcomes the critical bag pressure B is T = 144 MeV.
Note that this result only holds if the quark gluon system has no boundary. \\
If  now we imagine to put a very large number of baryons into a cylinder and compress it adiabatically with a piston, keeping T = 0,
the quark wave functions will start to overlap and according to Pauli's exclusion principle the kinetic energy will rise \cite{ya}. 
The pressure of the system will thus increase until deconfinement occurs. For a bag pressure 
B$^{1/4}$ = 206 MeV, the critical value of the baryon number density in the 2 flavour hypothesis will be
\begin{equation}
\centering
n_{B}^{QGP} = 0.72/\textrm{fm}^3
\end{equation}
which is about 5 times the normal nuclear matter density \cite{wo}.\\
We discussed  two limiting cases in which deconfinement occurs via heating or compression only. In other  cases, pressure can be increased by applying compression and heating at the same time and the transition will then occur at  intermediate temperatures and baryon densities. Thus we can  draw a simple phase diagram
\begin{figure}[tbp]
\centering
\includegraphics[height=7cm]{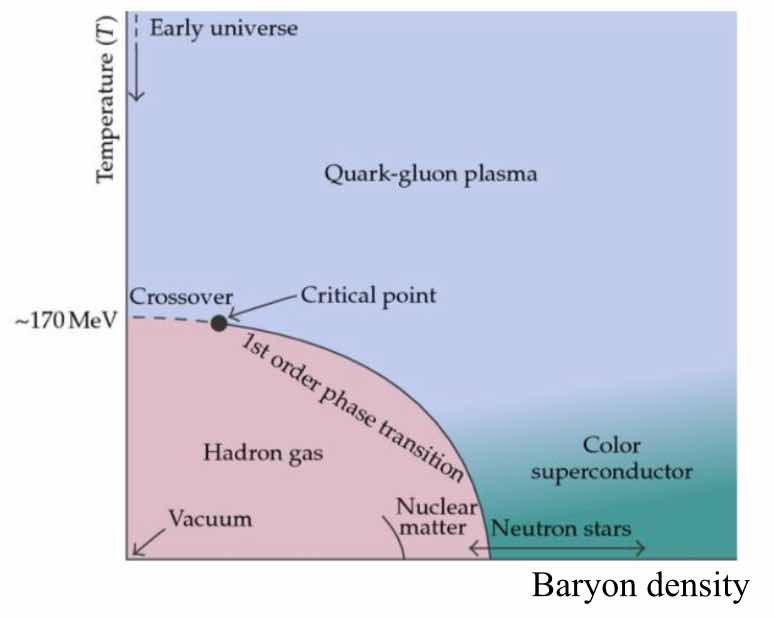}
\caption{Phase diagram of QCD}
\label{fig:phase}
\end{figure}
where  a phase transition leads to quark and gluon deconfinement. Ordinary nuclear matter lies on the
T$\sim$0 axis, with quarks and gluons having approximately a momentum of 200 MeV/c, that is 
the Fermi momentum. In the bottom left-hand  corner the behaviour of QCD can be described
in terms of a hadron gas, meaning that the dominant degrees of freedom are the hadronic ones; after deconfinement
new degrees of freedom appear (i.e. the color ones) and the physics of the system  evolves to a state called Quark-Gluon Plasma (\textbf{QGP}). This happens moving vertically towards higher temperatures.\\
Another interesting region is at low temperature and large baryon density. This is the phase for  compressed nuclear matter that might exist in the interior of neutron stars. In this region theories have been developed that predict a behaviour of quarks and gluons in nearby hadrons similar to the behaviour of electrons in superconductor, i.e. formation of Cooper pairs of quarks driven by the attractive component of the QCD interaction. This behaviour is known as Color Superconductivity \cite{csc}. \\
All  previous  results have been derived from the bag model; this model however also predicts the existence of many exotic states
that have never been observed. To have better qualitative results, one must deal with QCD, even if a 
perturbative approach to the theory is not always possible.  \\
\section{Lattice QCD}
The numerical calculations used to extract quantitative results employ the lattice approach, introduced in 1974 by  Kenneth Wilson (Nobel Prize 1982).
The fundamental idea of lattice QCD is to  discretize space and "time"\footnote{we will see later the exact meaning of time variables in this context} points, creating a ``lattice''
of points located at a certain distance from each other.  The QCD quantities of interest are computed on the lattice, 
obtaining results that depend on the distance of separation and are finally   
extrapolated  to the continuum limit to obtain physical results. By introducing a minimum distance scale a momentum cut-off is defined, making possible numerical evaluation of integrals in a non-perturbative QCD regime. \\
The approach to this method begins considering the analogy of the Feynman path integral to the partition function in quantum mechanics: 
the mechanical amplitude
for a system to be initially located at $(x_{a},t_{a})$ and to end up at $(x_{b},t_{b})$ is given by
\begin{equation}
\centering 
\textbf{Amplitude|}(x_{a},t_{a})\rightarrow(x_{b},t_{b})| = <x_{b}|e^{-iH(t_{b}-t_{a})}|x_{a}> = \sum_{all\  paths} e^{iS_{M}}
\label{partition}
\end{equation}
where $S_{M}$ is the action related to one path calculated using the Minkowsky metric tensor $g_{\mu\nu}$ with diagonal elements (1,-1,-1,-1).
In statistical quantum mechanics the partition function Z is the sum of the expectation values of the operator  $e^{-\beta_{B}H}$ ($\beta_{B}= 1/kT$), with the sum
to be carried out over all possible states of the system
\begin{equation}
\centering 
Z= \sum_{x_{a}} <x_{a}|e^{-\beta_{B}H}|x_{a}> 
\end{equation}
To reconduct ourselves to the path integral formulation of the partition function
\begin{itemize}
\item we introduce the imaginary time coordinate $\tau$ with $t=-i\tau$ and allow $\tau$ to range from $\tau_{a}=0$ to $\tau_{b}=\beta_{B}$ (so the exact meaning of time in this formulation is temperature)
\item we impose the periodic boundary condition that the configuration at $\tau_{b}$ is the same as that at $\tau_{a}$
\item we perform an additional summation over the states of the system 
\end{itemize}
The computation of the partition function is thus reduced to a path integral. A detailed explanation of these passages can be found in \cite{wo}. \\
The partition function can be evaluated by  generating quark and gluon field configurations on the discrete lattice. Each configuration will give one addendum in Equation \ref{partition}.
From the partition function many observables of the system can be evaluated, e.g. pressure
\begin{equation}
\centering 
\frac{p}{T^4}= \frac{1}{VT^3}\textrm{ln}Z
\end{equation}
However results depend on the lattice spacing used in the computation. Physics results are those extrapolated to the continuum limit.\\
The computing power needed sets a limit on the lattice spacing which can be used, tipically larger than 0.1 fm
and with $32^{3}$ lattice sites in three dimensional space and 16 points on the "time" axis \cite{ba}.
\subsection{Lattice QCD predictions}
LQCD calculations have been performed in different conditions. In the pure gauge approximation, only the gluonic degrees of freedom are considered,
while fermions and antifermions are treated as static and infinitely massive, meaning that the values of their field configurations
are not considered in the simulation process. In this framework, the potential between two static, infinitely massive quarks
can be calculated \cite{vo}.
The results of MC calculations clearly show that the potential has a linear confining part at long distances and a Coulomb attractive part at
short distances (Equation \ref{eqconf}), and has no significant dependence on the lattice spacing used in the simulation \cite{ya}.\\
The inclusion of fermionic fields in lattice calculations makes the situation more complicated. The reasons for this are:
 \begin{itemize}
\item fermion fields must be defined on the lattice sites, and their derivatives, appearing in the QCD lagrangian (Equation \ref{QCDlag}), must be approximated by differences
\item the Pauli principle requires  fermion fields to anticommute: the functional integral thus becomes a determinant running over
 all of the lattice sites,  requiring a huge computing power \cite{muller}
\end{itemize}
Many different frameworks have been developed to treat fermion fields in the proper way, usually with 2 or 3 different flavours (u,d,s) and 
different mass values. Different estimates of the transition temperature T$_{crit}$ have been obtained using different discretization schemes
in the fermion sector \cite{ppr} and are reported in Table 1.1. \\
\begin{table}[h]
\begin{center}
\begin{tabular}{|r|c|}
\hline 
&T$_{crit}$ \\ \hline
Pure Gauge & 271$\pm$2 MeV\\ \hline
2 light flavours & 173 $\pm$4 Mev\\ \hline
3 light flavours & 154 $\pm$8 MeV\\ \hline
\end{tabular} 
\label{tab:Crit}
\caption{Critical temperature values obtained with different mass and flavour hypotheses}
\end{center}
\end{table}

\vspace{-2em}
\begin{figure}[bt]
\includegraphics[width=0.45\textwidth]{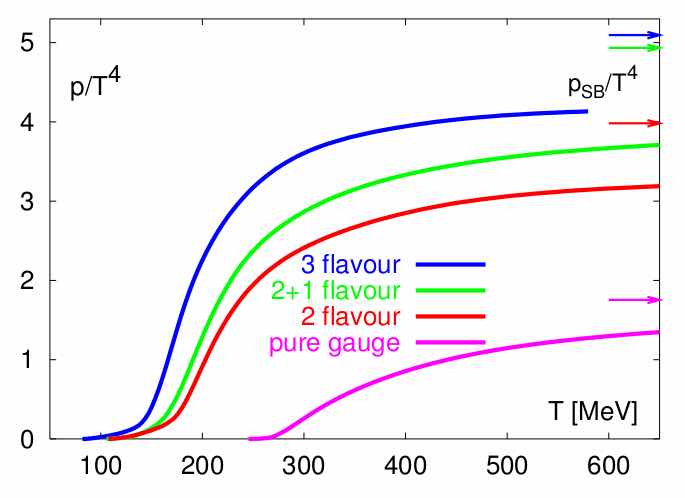}
\includegraphics[width= 0.45\textwidth]{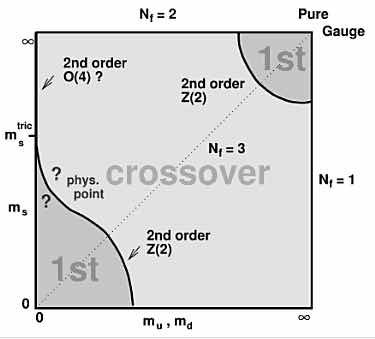}
\caption{Left: pressure vs temperature from lQCD predictions. Predictions are shown in the pure-gauge scenario, in a 2 and 3 light quarks scenario and in the (2+1) scenario, considering two lighter and one heavier quark. The arrows on the right indicate the Stephan-Boltzmann value for each scenario. Right: order of the phase transition as a function of quark masses}
\label{lQCDpred}
\end{figure}
The behaviour of  pressure is presented in Figure \ref{lQCDpred} (left).
No sharp change is seen in correspondence to the phase transition to a QGP phase, and the arrows indicate the limiting values according to the Stephan-Boltzmann Law in the different cases (Equation \ref{SBpressure}). This values 
are not reached even at the highest temperature, indicating that no ideal gas behaviour occurs \cite{sh},
and that strong interactions between  $q, \bar{q}$ and gluons are still present.\\
Lattice QCD also predicts the order of 
the phase transition: while a first order transition would lead to a discontinuity in the energy density between the plasma phase
and the hadron gas, a second order transition would lead to a rapid change in the thermodynamical variables as the critical temperature
is approached. In Figure \ref{lQCDpred} (right) we see that in the pure gauge theory the transition is of the first order. Also in the case of 2 (u,d) or 3 (u,d,s) massless (or light) 
quarks the transition is first order,  while for  intermediate values of the quark masses lQCD predicts the existence of a crossover between the 2 phases. If we look 
at the QCD phase diagram (Figure \ref{fig:phase}) the prediction is that the crossover takes place above a critical temperature, below which the transition
is of first order \cite{ba}.\\
The effective quark mass\footnote{In the low energy limit of QCD, a description of quark dynamics by means of perturbation theory is not possible. At low energy the interactions between valence quarks and sea quarks have great significance  and this is evident looking  at the mass of the proton, which is about O(100) higher than the sum of its constituents quarks "current" masses, i.e. the $m_q$ values in the QCD lagrangian. The non-perturbative behaviour of QCD  confers to quark a dynamically-generated mass, which is the "effective" mass.} \cite{Chiral}  is measured by the expectation value of the corresponding term in the Lagrangian, the chiral condensate $\langle\bar{\psi}\psi\rangle$(T). The predictions for two flavour QCD (with M$_q \sim$ 0.3 GeV/c$^2$) in Figure \ref{fig:adscft} (left)  show that  the effective quark mass has a rapid drop around T$_{crit}$, and its suceptibility defined as $ \chi_m = \partial \langle\bar{\psi}\psi\rangle(T)/\partial m_{q}$ has a peak in the same region. The meaning of this drop in the effective quark mass towards current quark mass values around T$_{crit}$ is that the QCD Lagrangian (Equation \ref{QCDlag}), which is not invariant under chiral transformation\footnote{A chiral transformation acts on a Dirac fermion as $\psi \rightarrow e^{\alpha \gamma^{5}}\psi$,. The QCD Lagrangian kinetic and gauge coupling terms are in fact invariant under this transformation. The mass term breaks this symmetry.}, recovers chiral symmetry at high temperature. In fact the QCD lagrangian chiral symmetry is broken by the mass term $m\bar{\psi}\psi$, and this is  reflected in the mass values of hadrons. If chiral symmetry was valid at low energies, all states with the same quantum numbers and opposite parity should have the same mass. This is not the case in nature, for example the $\rho$ ($J^{\rm P} = \rm 1^{-}$) and $a_1$ ($J^{\rm P} = \rm 1^{+}$) mesons have masses 770 MeV/c$^{2}$ and  1260 MeV/c$^{2}$, respectively. \\
For completeness it should be also mentioned that lattice calculations are not the only tool to acces non-perturbative phenomena in QCD. An approach based on a string theoretical construction, the Anti-de Sitter/Conformal Field Theory correspondence (AdS/CFT), has shown to be a useful tool to study non-perturbative phenomena of non-Abelian gauge theories, like those involved in QGP phenomenology \cite{adscft}.  This method can be used to compute several observable relative to the deconfined phase, e.g. the pressure as a function temperature, shown in Figure \ref{fig:adscft} (right).
\begin{figure}
\centering
\includegraphics[width= 0.35\textwidth]{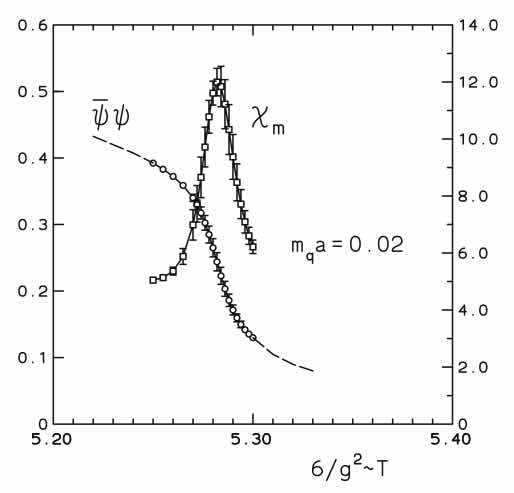}
\includegraphics[width= 0.6\textwidth]{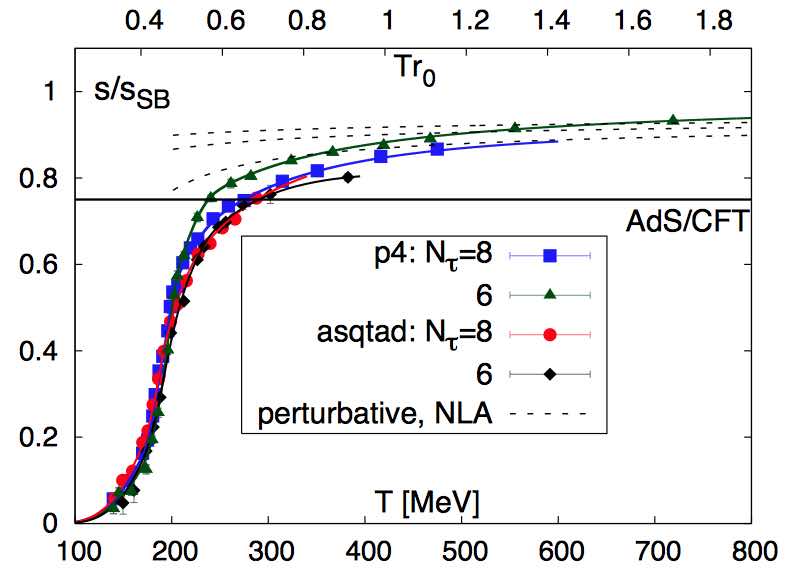}
\caption{Left: chiral condensate value vs T from lQCD predictions. Right: pressure (relative to Stephan-Boltzmann value) vs temperature diagram for different lattice QCD calculations (N$_{\tau}$ is the number of lattice sites in the $\tau$ direction). The AdS/CFT value is also included }
\label{fig:adscft}
\end{figure}
\section{Heavy Ion Collisions}
The QCD phase diagram (Figure \ref{fig:phase}) clearly shows that, to observe the deconfined state of nuclear matter described above, extreme conditions of density or temperature have to be achieved. While indirect observation of the high density region can be provided by the study of compact astrophysical objects,  the high temperature regions at low ($\approx$ 0) baryochemical potential correspond to what the universe looked like about 10$^{-6}$ s after the Big Bang. Current technologies allow access  these regions by colliding  heavy nuclei at high energies.\\
\subsection{Geometry and evolution}
\begin{figure}[bt]
\centering
\includegraphics[width=0.35\textwidth]{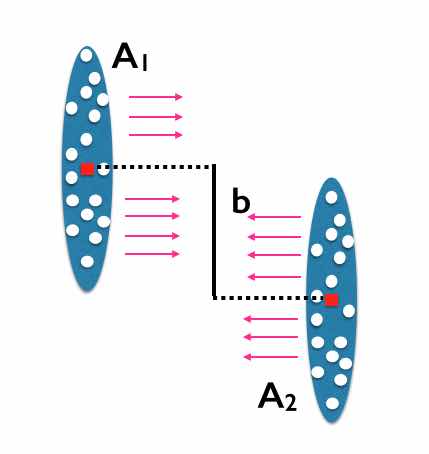}
\includegraphics[width= 0.35\textwidth]{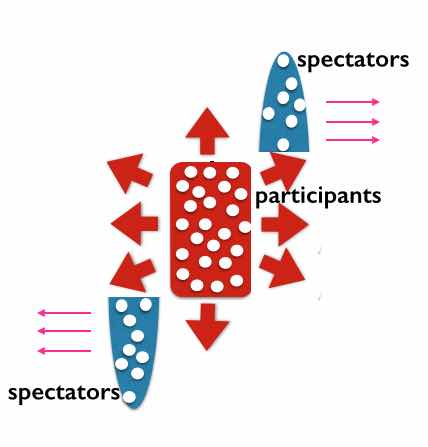}
\caption{Schematic representation of a collision between two nuclei, before (left) and after (right) the interaction}
\label{fig:coll}
\end{figure}
Figure \ref{fig:coll} schematizes the geometry of a heavy-ion collision: in the left panel, the incoming nuclei (A$_1$ and A$_2$) approach each other at ultrarelativistic speeds and are thus Lorentz contracted along the direction of their motion.  The vector that connects their two centres in the plane orthogonal to the direction of their motion is the impact parameter, \textbf{b}. In the right panel, the situation right after the collision is shown. The overlapping region of the colliding nuclei  is the volume in which multiple scattering among the nucleons occurs, while the non-overlapping regions of the nuclei move away from the interaction point. We can divide nucleons in two group: spectators nucleons do not experience any interaction during the nuclear collision, while participant nucleons experience at least one interaction. \\ 
\begin{figure}[b]
\centering
\includegraphics[width=0.55\textwidth]{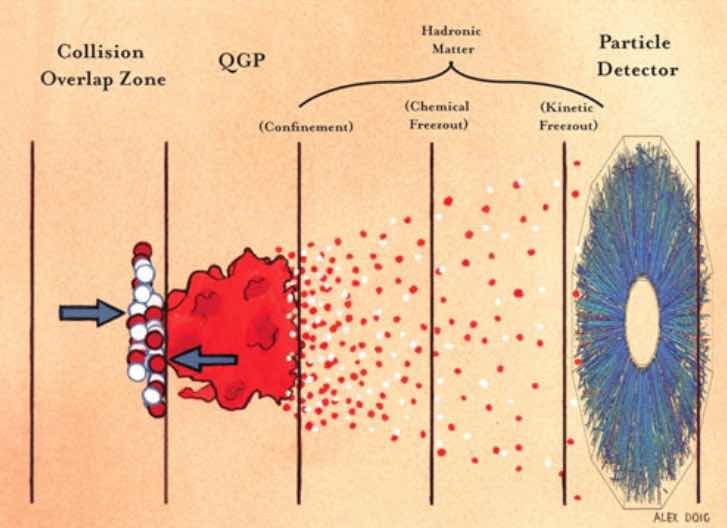}
\caption{Evolution of a heavy-ion collision}
\label{fig:collev}
\end{figure}
Now let's  concentrate on the system created in the collision, that undergoes the  evolution depicted in Figure \ref{fig:collev}:
\begin{itemize}
\item during the first stages of the collision  partons of the incoming nucleons undergo a large number of binary scatterings. The total energy is  converted into the creation of  new partons, so a high multiplicity system forms. The system is originally not in thermal equilibrium, which is expected to be reached within $\sim$ 1 fm/c;
\item the system reaches thermal equilibrium and its initial  temperature depends on the centre-of-mass energy and on the atomic mass number of the nuclei.  As  shown in Section 1.4.1, exact measurement of the initial temperature depend on models which still have large uncertaintes; anyway once equilibrium is achieved ($\tau_{\rm eq}\sim$1 fm/c) temperatures reached in heavy ion collisions  both at RHIC (\sqnn = 200  GeV) and LHC (\sqnn = 2.76 TeV) are estimated\footnote{see Section 1.4.1} to be well above the T$_{crit}$ predicted by lQCD (Table 1.1), and deconfinement is expected to occur.   The lifetime of the QGP phase depends on both centre-of-mass energy and system size, and is estimated as 1$\div$4 fm/c at RHIC energies and $\sim$10 fm/c at LHC energies 
\item this "fireball" expands cooling down until T$_{crit}$ (Table 1.1) is reached. Below this point QGP phase can no longer exist and the system hadronyzes. At this point, the hadronic gas continues its expansion and cooling
\item hadrons reinteract inelastically, so the relative hadronic abundances in the system are not fixed.  Below a certain temperature, T$_{chem}$ ($\sim$ 160 MeV/c, Section 1.4.2), inelastic interaction cease: this is the \textbf{chemical freeze-out}. The relative hadronic abundances are  fixed, however elastic interactions are still possible
\item  hadrons reinteract elastically, so the kinematic spectrum is not fixed.  Below a certain temperature, T$_{kin}$ ($\sim$ 100 MeV/c, Section 1.4.3), elastic interactions cease. The momentum spectrum of each hadron species  is  fixed: this is the \textbf{kinetic freeze-out}. Hadrons continue their path with no further interactions
\end{itemize}
The goal of experiments with heavy-ion collisions at high-energies  is to detect these particles or their decay products to investigate the overall history of the interaction. 
\subsection{Glauber Model}
We can describe  the collision represented in Figure \ref{fig:coll} at femtoscopic scales using three fundamental quantities: the impact parameter \textbf{b}, the number of participant nucleons \textbf{N$_{\rm part}$} and the total number of binary nucleon-nucleon collisions \textbf{N$_{\rm coll}$}. The Glauber model is a theoretical semi-classical tool developed to describe the geometry of the colliding nuclei and allow us to calculate N$_{\rm part}$ and N$_{\rm coll}$ as a function of the impact parameter \cite{Glauber}. \\
The model needs in principle two input quantities:
\begin{itemize}
\item the nucleon density distribution $\rho(r)$
\item the inelastic nucleon-nucleon cross section $\sigma^{NN}_{inel} $. The elastic and diffractive components of the total nucleon-nucleon cross-section are ignored  
\end{itemize}
The model views the collision of two nuclei in terms of the individual interactions of the constituents nucleons. One approach to get quantitative results with the  Glauber model is within the Optical Limit, which depends on the following assumptions 
\begin{itemize}
\item typical size of nucleon much lower that the typical size of nucleus
\item neutrons and protons are treated in the same way
\item  each binary nucleon-nucleon interaction is  independent from the presence of other nucleons in the colliding system 
\item the colliding nuclei, as well as their nucleons, are undeflected by the collision
\item each nucleon always interacts with same inelastic cross section in subsequent collisions
\end{itemize}
Let's consider Figure \ref{fig:glauberparam}: two heavy ions, A and B, collide with impact parameter $\textbf{b}$. Focusing on the area located at a coordinate  $\textbf{s}$ from the center of the nucleus A in the transverse plane, we define  $\textbf{s - b}$ the position of this area with respect to the  centre of the nucleus B. The probability per unit transverse area of a given nucleon of A to be located in the considered area is 
\begin{equation}
\hat{T}_A(\textbf{s})= \int\hat{\rho}(\textbf{s},z_A)dz_A
\end{equation}
where $\hat{\rho}(\textbf{s},z_A)$ is the normalized probability per unit volume for finding the nucleon at location  $(\textbf{s},z_A)$. A similar expression follows for a nucleon in nucleus B. The product
\begin{equation}
\hat{T}_A(\textbf{s})\hat{T}_B(\textbf{b-s})d^2s
\end{equation}
gives the  probability per unit area of nucleons being located in the respective overlapping target and projectile elementary areas $d^2s$. Integrating this quantity over the whole overlapping region  we get the thickness function 
\begin{equation}
\hat{T}_{AB}(\textbf{b}) = \int  \hat{T}_A(\textbf{s})\hat{T}_B(\textbf{b-s})d^2s
\end{equation}
which has the units of an inverse area. The probability for the two nucleons considered to interact is $\hat{T}(\textbf{b})\sigma^{\rm NN}_{\rm inel}$.
\begin{figure}
\centering
\includegraphics[width=0.75\textwidth]{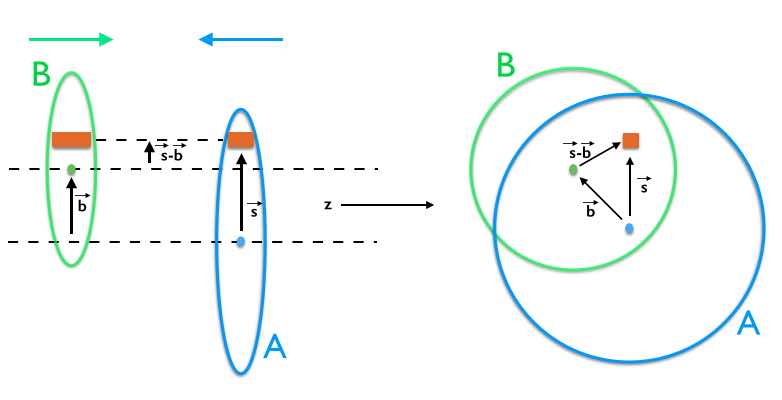}
\caption{Schematic representation of the two nuclei before the collision}
\label{fig:glauberparam}
\end{figure}
The probability of having $n$ such interactions is given by the binomial distribution
\begin{equation}
P(n, \textbf{b}) =   \left(\!  \begin{array}{c}  AB \\ n \end{array}\!\right) [\hat{T}(\textbf{b}) \ \sigma^{\rm NN}_{\rm inel}]^n[1-\hat{T}(\textbf{b}) \ \sigma^{\rm NN}_{\rm inel}]^{AB-n}
\end{equation}
where the number of combinations for finding $n$ collisions out of $AB$ possible nucleon-nucleon interactions appears.  We can get the total interaction probability between the nuclei  A and B (for a given impact parameter \textbf{b}) summing on all possible "trials"
\begin{equation}
\frac{d^2\sigma^{\rm AB}_{\rm inel}}{db^2} = \sum^{\rm AB}_{n=1}P(n,\textbf{b}) = 1 - [1-\hat{T}_{\rm AB}(\textbf{b})\sigma^{\rm NN}_{\rm inel}]^{\rm AB}
\end{equation}
The total number of nucleon-nucleon collisions is the mean value of the binomial distribution
\begin{equation}
N_{\rm coll}(b) =   \sum^{\rm AB}_{n=1}nP(n,b) = AB\hat{T}_{\rm AB}(b)\sigma^{\rm NN}_{\rm inel}
\end{equation}
Another implementation of  the Glauber model is based on Monte Carlo simulations. In this approach, the two nuclei A and B are composed of nucleons whose 3D coordinates are randomly generated according to their nuclear density distributions. As the two nuclei collide, a single nucleon-nucleon collision takes place if their distance $d$ in the plane orthogonal to the beam axis satisfies 
\begin{equation}
d < \sqrt{\sigma^{\rm NN}_{\rm inel}/\pi}
\end{equation}
If a large number of collisions with impact parameter \textbf{b} distributed according to $d\sigma/db \propto 2\pi b$ is simulated, the distribution for N$_{\rm part}$ and N$_{\rm coll}$ are obtained (Figure \ref{fig:glauberpred} - right panel). These  distributions show good agreement with the calculation in Optical Approximation. In the left panel of Figure \ref{fig:glauberpred} the total nuclear cross section vs nucleon-nucleon inelastic cross-section is shown, computed in both the Optical Approximation and in the Glauber MC  framework. Small discrepancies appear for the higher  nucleon-nucleon  cross-section values.  This is mainly due to the fact that in the Optical Approximation the incoming nucleon sees the incoming nucleus as a smooth density object and does not account for event-by-event density fluctuations. \\
\begin{figure}
\centering
\includegraphics[width=0.75\textwidth]{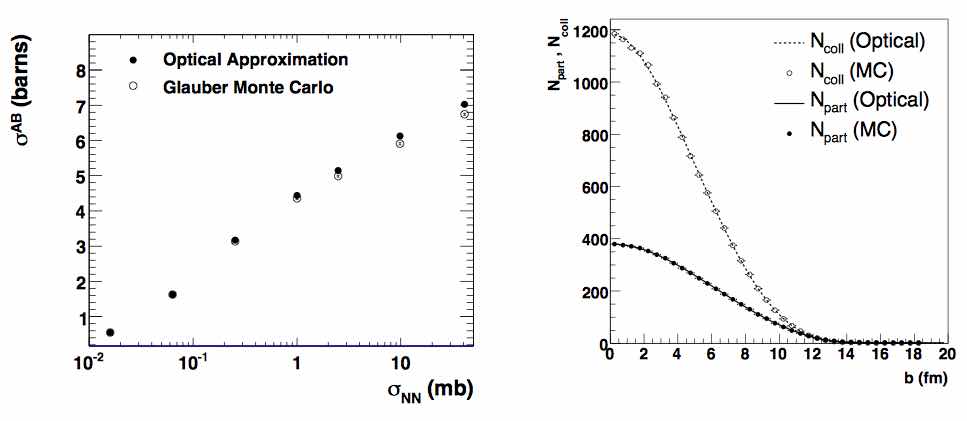}
\caption{Left: total nuclear cross section vs $\sigma^{\rm NN}_{\rm inel}$ for Au-Au collisions. Right: N$_{\rm part}$ and N$_{\rm coll}$ for Au-Au collisions.}
\label{fig:glauberpred}
\end{figure}
Of course  the impact parameter \textbf{b} is not an accesible quantity, but it can be related to experimental observables. This will be discussed in the Chapter 3. However to fully understand some of the results shown in next section it is necessary to mention that collisions with a small value of the impact parameter are called central collisions, those with large values of impact parameter are called peripheral collisions. Considering a sample of AA collisions, events are usually divided in percentiles of certains observables correlated to the impact parameter, and these percentiles are called centrality classes. For instance, the 0-10\% centrality class defined based on charged-particles multiplicity corresponds to the 10\%  events with the highest charged-particles multiplicity.
\section{Heavy-ion physics observables}
The particle multiplicities produced in ultra-relativistic heavy-ion collisions are much larger than the ones obtained in proton-proton, proton-electron and electron-electron collisions. Systems created in high-energy nuclear interactions show a higher degree of complexity, due to the structured evolution described in Section 1.2.1. \\
The study of relativistic heavy-ion collisions started in the 70s at the Lawrence Berkeley National Laboratory, where a  linear accelerator designed for heavy-ions was built. Accelerated nuclei were sent to the LBNL synchrotron (Bevatron) that could  accelerate them at about 1 $A$ GeV/c.\\
The effort continued in both Europe and US at CERN and Brookhaven, respectively.  The Alternating Gradient Synchrotron (AGS) at Brookhaven National Laboratory accelerated Si ions up to 15 $A$ GeV. The CERN heavy-ion program at the Super Proton Synchrotron started in 1986 accelerating  nuclei with equal numbers of protons and neutrons, due to the capabilities of the ion injector.  Subsequently an injector was constructed to accommodate arbitrarily heavy nuclei and the use of  Pb beams at 160 $A$ GeV started in 1995.\\
Until 2010 the highest energies in heavy ion collisions  were reached at the Relativistic Heavy Ion Collider (RHIC) in Brookhaven. The experimental activities are still ongoing, with two experiments still taking data: PHENIX and STAR. The colliding systems explored at RHIC up to now are  p-p, d--Au, Cu--Cu, Cu--Au, Au--Au and U--U, with a maximum center-of-mass energy for  Au--Au collisions \sqnn = 200 GeV.\\ Recently the Large Ion Collider at the LHC started its heavy-ion programme and the first Pb-Pb collisions at \sqnn = 2.76 TeV were recorded in 2010. Three  experiments have a heavy-ion program: ALICE, ATLAS and CMS. \\
In the next sections we will discuss a list of  observables carrying information on the system produced in ultra-relativistic heavy-ion collisions. Note that partonic energy-loss in the hot medium will not be mentioned here as it will be one of the main topics of Chapter 2. 
\subsection{Thermal photons}
Photons produced in heavy-ion collisions can be divided in three groups \cite{thermalphotontheory}:
\begin{itemize}
\item \textbf{decay photons} coming from decays of other particles (mainly $\pi^{0}\rightarrow\gamma+\gamma$). These are also present in small system collisions (p-p, e-e, etc.)
\item \textbf{prompt photons} coming  from initial partonic processes with large $Q^2$ (e.g. $g+q\rightarrow q+\gamma$); these are also present in small system collisions (p-p, e-e, etc.) and have large momenta ( $\ge$ 3 GeV/c)
\item \textbf{thermal photons} coming from thermal radiation of the QGP: these are peculiar of nuclear collisions, where the strong interacting medium is expected to form, and  have lower momenta ($\le$ 2GeV/c)
\end{itemize}
While photons belonging to the first group come from secondary vertices, prompt and thermal photons originate at the interaction point and are thus called "direct" photons. \\
For what concerns thermal photons, theory can predict their transverse momentum distributions, taking into account the full evolution of the system. The final  spectrum will be the convolution of all  photons  emitted during the temperature history of the fireball T(t). Fitting the direct photon spectra could yield the initial temperature T$_0$ of the fireball, but still theory predictions depend on a large number of parameter. These are currently  not completely understood, in particular the time it takes to the system to reach thermal equilibrium $\tau_{eq}$ and the dynamical parameters of the expansion. \\
What one can do is to consider the fireball as a static thermal source at constant temperature that emits photons  accordignly to a Maxwell-Boltzmann exponential rate $\sim e^{-p_{\rm T}/T}$. Fitting the momentum spectrum of direct photons at low \pt with this exponential will provide an  ”effective temperature”, a sort of temporal mean from the very initial stages to hadronization, which is expected to take place at T$_{c}\sim$ 170 MeV. \\
\begin{figure}[t]
\centering
\includegraphics[width=0.95\textwidth]{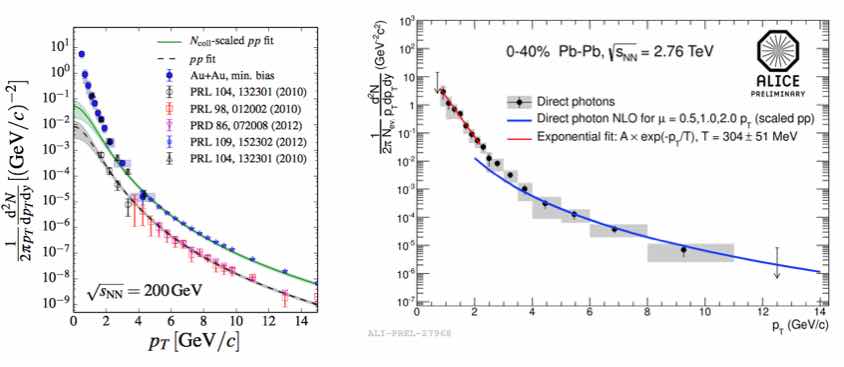}
\caption{Left: the filled blue points correspond to the \pt differential cross section of direct photons in minimum bias in Au--Au collisions at \sqnn = 200 GeV \cite{photonsPHENIX}. Right: direct photons \pt spectrum in Pb--Pb collisions at \sqnn  = 2.76 TeV \cite{PhotonsALICE}.}
\label{fig:direct}
\end{figure}
Figure \ref{fig:direct} shows direct photon measurements performed by the PHENIX (left, \cite{photonsPHENIX}) and ALICE (right, \cite{PhotonsALICE}) experiments. The exponential fit is superimposed to the low \pt points for the ALICE measurents (red curve), yielding an effective temperature T = 304 $\pm$ 51 MeV. For the PHENIX measurement,  fits have been performed dividing the minimum bias sample in four different centrality classes yielding results that are compatible within uncertaintes. For the most central events at RHIC energies one gets T = 239 $\pm$ 25$^{sys}$ $\pm$ 7$^{sys}$ MeV. Both values of effective temperature  are well above the critical temperature for deconfinement expected from lattice QCD calculations.
\subsection{Hadronic abundances}
 For central ultra-relativistic nuclear collisions  light hadrons yields  can be described very well  within the statistical hadronization model \cite{Becattini}. The chemical freeze-out temperature T$_{\rm chem}$, the baryo-chemical potential $\mu_b$\footnote{In particle physics reactions a qqq baryon is always created or destroyed pairwise with a $\bar{\rm q}\bar{\rm q}\bar{\rm q}$ anti-baryon. There is no process within QCD which can change the number of baryons N$_{\rm B}$ minus the number of anti-baryons N$_{\bar{\rm B}}$; in other words we can identify a conserved quantum number B = N$_{\rm B} $ - N$_{\bar{\rm B}}$ called baryon number. Quarks and anti-quarks carry B = $\pm$ 1/3 respectively. Now, for systems in which baryon number is allowed to vary (such as the Gran Canonical Ensamble), the most convenient thermodynamic potential to consider is the grand canonical potential $\Omega(T,V,\mu_b)=E - TS  - \mu_b B$. Thermodynamic equilibrium is reached when $\Omega$ is minimised, and for a system in equilibrium we recognise $\mu_b$ as the increase in E whenever B increases by one. } and the fireball volume V are the     parameters of the model, which assumes that the system is in chemical and thermal equilibrium.\\
The basic quantity used to compute the hadronic composition of thermal yields measured in heavy ion collisions is the partition function Z(T,V). Considering a system in equilibrium that exchanges energy and particles with the outside (Gran Canonical Enseble), the partition function for a certain hadronic species $i$ is
\begin{equation}
\ln Z_i = \frac{Vg_i}{2\pi^{2}}\int^{\infty}_{0}\pm p^2 dp \ln [1 \pm \textit{exp}(-(E_i - \mu_i)/T)]
\end{equation}
where $g_{\rm i}$ ($\mu_i$) is the spin degeneracy factor (chemical potential) for the hadron species $i$ and the + (-) is for fermions (bosons). The total number of hadrons of specie $i$ at equilibrium is given by
\begin{equation}
N_i = - T \frac{\partial \ln Z_i}{\partial \mu_{i}}
\end{equation}
This model can be used to fit experimental data of particles multiplicities and extract T and $\mu_b$ . One can  get rid of the fit parameter V fitting particles ratios. At  LHC energies the baryon/antibaryon ratio is sufficiently close to 1, so that $\mu_b$ can be fixed at 0. \\
\begin{figure}[t]
\centering
\includegraphics[width=0.99\textwidth]{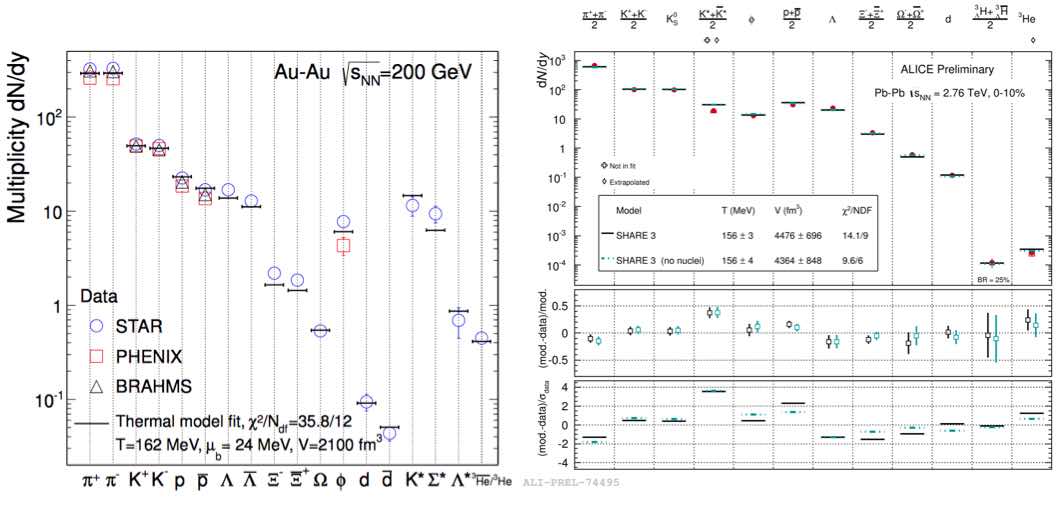}
\caption{Left: \pt integrated hadronic abundances in Au--Au at \sqnn = 200 GeV \cite{ThermalFits}. Right: \pt integrated hadronic abundances in central Pb--Pb at \sqnn = 2.76 TeV \cite{Floris}.}
\label{fig:thfit}
\end{figure}
The temperatures values extracted at low energies first increase sharply with increasing beam energy and level off near T$_{\rm chem}$ $\approx$ 162 MeV for energies \sqnn>20 GeV \cite{ThermalFits}. \\
Figure \ref{fig:thfit} shows the results for RHIC experiments \cite{ThermalFits} (left) and ALICE \cite{Floris} (right). RHIC results indicate  T$_{\rm chem}$ =162 MeV and a small, but non vanishing, chemical potential $\mu_b$ = 24 MeV. The overall trend of the data is very well reproduced by the calculations. However, the yield of protons and anti-protons from PHENIX and BRAHMS is overpredicted, while the yield of multi-strange baryons is generally underpredicted, leading to a rather poor $\chi^2$ value of the fit. ALICE results for central Pb--Pb collisions are fitted with and without considering light nuclei yields and the two methods give compatible results for T$_{\rm chem}$ =154 MeV and for the volume at chemical freeze-out  V = 4476 fm$^3$. The rather low temperature obtained from ALICE results is due to the unexpectedly low yields of protons and anti-protons. The exclusion of  protons and anti-protons from the fit leads to a very good description of all remaining data, with excellent $\chi^2$ of the fit and  T$_{\rm chem}$ =164 MeV, completely in line with expectations as pointed out in  \cite{ThermalFits}.
\subsection{Transverse momentum spectra of hadrons}
\label{sub:kfo}
The yields of most of the produced particles get fixed at chemical freeze-out. The abundances can still vary due to decays. \\
After some time, when the inter-particle distance becomes so large that the elastic interactions stop, the system  undergoes kinetic freeze-out. At this stage, the transverse momentum spectra of the produced particles are fixed. At the kinetic freeze-out the expanding system can be considered  as a static thermal source in equilibrium at a temperature T$_{\rm kin}$, so that the emission rate of hadrons can reasonably be  a Maxwell-Boltzmann exponential slope $\sim e^{-p_{\rm T}/T_{slope}}$. However, fitting the momentum spectra of different hadrons (e.g. pions and protons) one gets different $T_{slope}$ values. 
This is understood considering that the dynamic of particles at this stage is not only determined by their elastic interactions, but also by the collective expansion of the system, i.e. by the  ``radial flow'' depicted in left panel of Figure \ref{fig:spectra}.
The temperature value extracted from a fit to a given hadronic species is thus the sum of two contributions, a random thermal motion characterized by the temperature T$_{kin}$ and a collective expansion:
\begin{equation}
\centering
T_{slope} = T_{kin} + \frac{1}{2}mv_{\bot}
\end{equation}
where $v_{\bot}$ is the mean radial velocity of  particles expanding  with the fireball. \\
\begin{figure}
\centering
\includegraphics[width=0.99\textwidth]{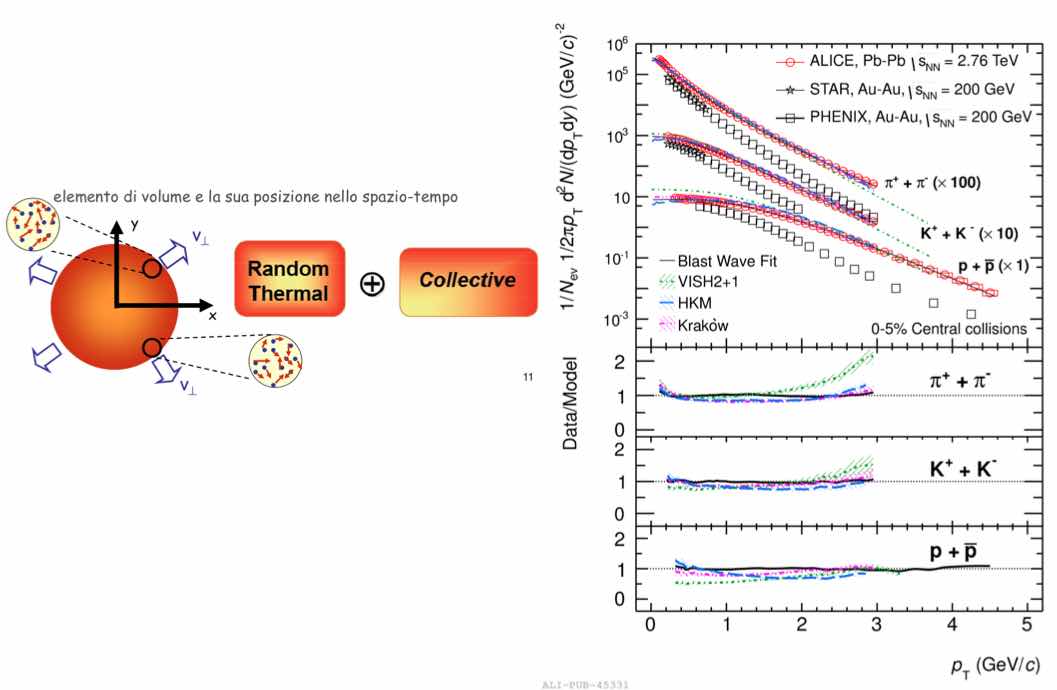}
\caption{Left: schematic representation of radial flow. Right: pion, proton and kaons \pt spectra in central AA collisions from ALICE, STAR and PHENIX  \cite{ALICESpectra}.}
\label{fig:spectra}
\end{figure}
A fit to the spectra with the so-called blast-wave function \footnote{A model that takes  into account radial flow and hadron yield from resonances decay, and assumes a thermal source at equilibrium} \cite{BlastWave} allows one to extract T$_{\rm kin}$, the temperature at which elastic interactions among particles
cease, and $\beta_{\bot}$ = $v_{\bot}/c$.\\
The right panel of Figure \ref{fig:spectra} shows the \pt distributions of pions, kaons and protons (summed to those of their respective antiparticles) from Pb--Pb collisions at \sqnn = 2.76 TeV  in the 0-5\% centrality class, measured by the ALICE Collaboration \cite{ALICESpectra}. The same measurements from PHENIX and STAR Collaborations in  Au--Au collisions at \sqnn = 200 GeV are also shown. The spectra shapes show a significant change from RHIC to LHC energies, having a distinctly harder distribution. This indicates a stronger radial flow at higher energies, and fitting ALICE spectra with the blast wave function one gets $\langle \beta_{\rm T}\rangle$ = 0.65 $\pm$ 0.02 and T$_{\rm kin}$ = 95 $\pm$ 10 MeV for central (0-5\%) Pb--Pb collisions at \sqnn = 2.76 TeV. Both values are about 10\% higher than what is extracted from RHIC data at similar centralities. \\ Calculation based on hydrodynamics also reproduce the spectra measured by the three collaborations, in agreement with the hypothesis of a collective expansion. 
\subsection{Hadronization}
\label{sub:hadronization}
\begin{figure}[t]
\centering
\includegraphics[width=0.99\textwidth]{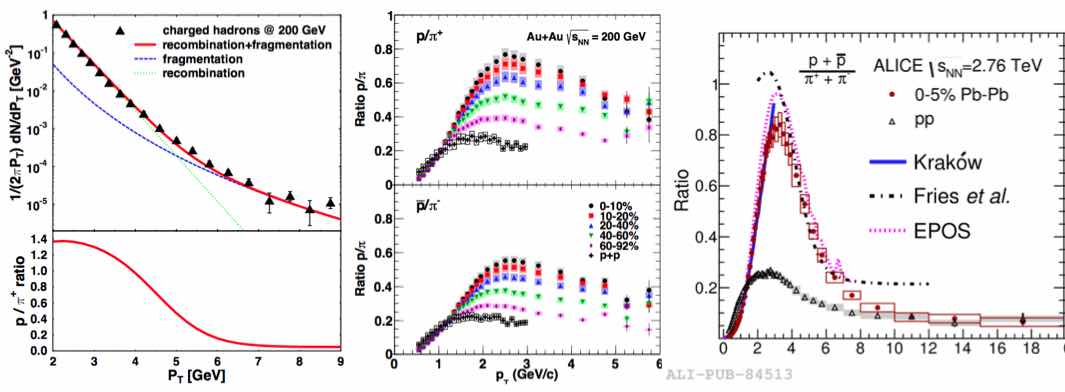}
\caption{Left panel: Fries model fit to PHENIX  charged hadron \pt spectrum in minimum bias  Au--Au collisions at \sqnn = 200 GeV  \cite{recombination1}. Central: PHENIX  p/$\pi$ and $\bar{\rm p}/\pi^-$ratios  in five centrality classes (also  ratios from pp collisions  are shown)  Au--Au collisions at \sqnn = 200 GeV \cite{poverpPHENIX}. Right panel: p/$\pi$ ratios  measured with ALICE for central Pb--Pb and pp collisions at \sqnn = 2.76 TeV  and comparison to models  \cite{poverpALICE}.}
\label{fig:pioverp}
\end{figure} 
In high-energy elementary particle collisions the transition from quarks to hadrons is a non-perturbative QCD phenomenon. The hadronisation process can be described by the fragmentation function $D^{i}_{h}(z)$, which represents the probability of parton $i$ to fragment to hadron $h$ carrying a fraction $z$ of its energy. Fragmentation functions  can be obtained from global analysis in $e^+e^-$, $e^+$p and p-p collisions. In event generators such as PYTHIA the fragmentation process  is ususally implemented via a parton shower and subsequent color string hadronization. \\
In ultrarelativistic heavy-ion collisions, however, the creation of a hot medium can affect the production of hadrons as well as their momentum distributions. Beside the influence of the radial flow already discussed, the hadronization mechanism itself can be modified \cite{recombination1}, since now a hadron can be formed not only by a fragmenting quark, but also by two or three quarks that are close to each other both in space and in momentum and recombine. This  can affect the final momentum distributions of hadrons, since while in the fragmentation processes the hadron momentum is a fraction of the parent parton one, the momentum of hadrons originating from recombination is the sum of up to three different contributions. \\ 
Recombination is expected to dominate for low \pt partons in the medium, and models have been developed to predict hadron spectra in scenarios taking into account recombination at low \pt  and fragmentation at  high-\pt, such as the Fries model in Figure \ref{fig:pioverp} (left)  \cite{recombination1}. The model is used to fit PHENIX data for the charged hadron transverse momentum spectrum and the top panel shows the recombination and fragmentation contributions to the fit function. \\
An  observable predicted by this model is the proton to pion ratio in A--A collisions , shown in  the bottom-left panel of  Figure \ref{fig:pioverp}. An enhancement of the p/$\pi$ ratio is predicted for 2<\pt<4 GeV/c, followed by a decrease.  Also other models based on recombination draw the same qualitative  conclusions  \cite{recombination2}. \\
  The central panel of  Figure \ref{fig:pioverp} shows the p/$\pi$ ratio measured as a function of \pt by PHENIX for Au--Au collisions at \sqnn = 200 GeV \cite{poverpPHENIX} for different centralities. A peak is visible for all centralities around \pt = 2.5 GeV/c, however the behaviour of the  p/$\pi$ and $\bar{\rm p}/\pi^-$ratio is substantially different from the model predictions in the low-\pt region. This is due to the fact that the model does not take into account the radial flow that differently affects  the transverse momentum distribution of pions and protons, in particular shifting heavier hadrons (in this case protons) to higher \pt values. \\
The right panel of  Figure \ref{fig:pioverp}  shows the ALICE results for central Pb--Pb collisions  at \sqnn = 2.76 TeV \cite{poverpALICE}. The peak position is around \pt = 3 GeV/c and its value is about 20\%  above the one  measured by PHENIX.  The Fries model predictions superimposed in Figure \ref{fig:pioverp} qualitatively describe the shape of the ratio down to $\sim$ 2 GeV/c. The Krakow model \cite{pioverpHydro}, which agrees well with the data below $\sim$ 2GeV/c, is based on a hydrodynamical evolution of the partonic stage followed by statistical  hadronization. 
\subsection{Elliptic flow}
\begin{figure}[b]
\centering
\includegraphics[width=0.99\textwidth]{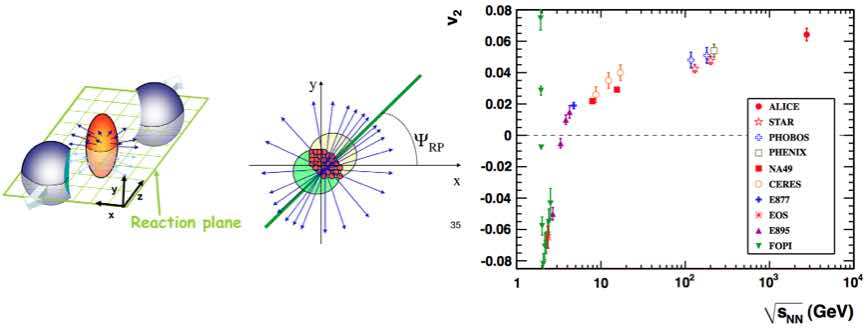}
\caption{Left: a non central heavy-ion collision showing how the reaction plane is defined. Right: \pt integrated elliptic flow values for semicentral events collisions as a function of \sqnn \cite{FlowALICE}.}
\label{figflow}
\end{figure} 
In non-central collisions, the spatial anisotropy of the overlap region of the colliding nuclei is converted into a momentum anisotropy of the final state particles due to interactions among the medium constituents. The anisotropy is quantified by the coefficients of the Fourier expansion of the distribution of the final state particle azimuthal angles relative to the reaction plane, which is defined by the impact parameter of the collision and the beam direction
\begin{equation}
\centering
\frac{dN}{d(\phi - \Psi_{RP})} = \frac{N_{0}}{2\pi}(1+2v_{1}\cos(\phi-\Psi_{RP})+2v_{2}\cos(2(\phi-\Psi_{RP}))+...)
\end{equation}
where $\Psi_{RP}$ is depicted as in Figure \ref{figflow} (center). The second
coefficient $v_{2}$ is the elliptic flow coefficient and reprensents an elliptic distribution of particles in the
transverse plane.\\
\begin{figure}[t]
\centering
\includegraphics[width=0.99\textwidth]{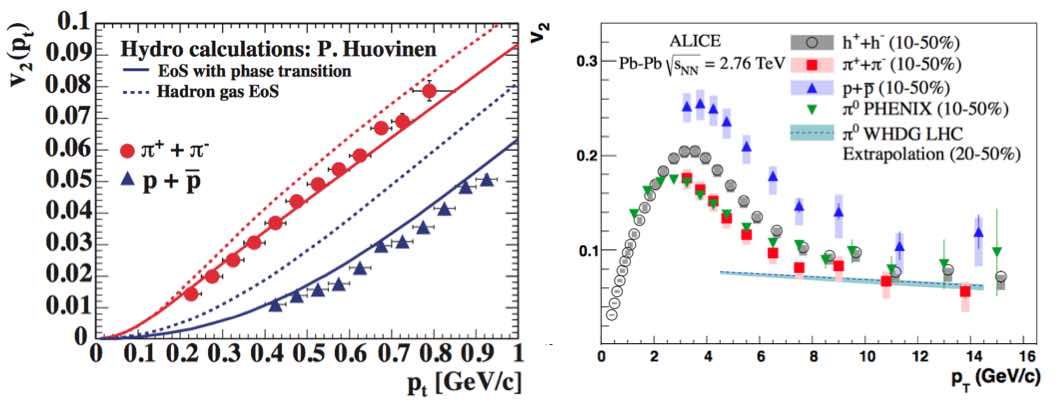}
\caption{Left: PHENIX low \pt measurements of v$_{2}$ compared to hydrodynamical calculations  \cite{snellings}. Right: ALICE results for proton and pion $v_2$ from semicentral Pb--Pb events at \sqnn = 2.76 TeV  \cite{IDFlow}}
\label{flowid}
\end{figure}  
This effect is negligible in central collisions (\textbf{b}$ \sim$0) and increases when going towards higher values of \textbf{b} (semiperipheral collisions) due to increasing eccentricity of the overlap region of the colliding nuclei. In the most peripheral collisions, however, this effect  decreases again since the particle densities achieved do not build up sufficiently strong  pressure gradients that can translate the initial spatial anisotropy into an observable momentum anisotropy. \\
The right panel of Figure \ref{figflow} shows the \pt integrated elliptic flow results from different experiments as a function of  \sqnn ranging from few GeV up to RHIC and LHC energies \cite{FlowALICE}. The ALICE results is for the 20-30 \% centrality range, other results correspond to similar centrality intervals. Going from RHIC to LHC energies,  a 30\% increase of the \pt-integrated $v_2$ values  is observed. \\
Most of the observed $v_2$ is expected to form at early times of the collision, when the  spatial anysotropy is at its maximum. However, interactions among the constituents do occur  also in the hadronic phase of the fireball's evolution, and it is reasonable to argue that part of the elliptic flow may build up  after hadronization. Figure \ref{flowid} (left panel) shows RHIC results for \pt differential $v_2$ of identified pions and protons at very low p$_{\rm T}$. Hydrodynamics predictions are superimposed for both species, one (continous line) taking into account a deconfined state that undergoes a phase transition to hadron gas, the other for a hadron gas system. The predictions including a deconfined state better agree with both data sets \cite{snellings}. The right panel of Figure \ref{flowid} shows the   \pt differential $v_2$ of identified pions and protons measured by ALICE \cite{IDFlow}. In this higher \pt region the situation is reversed, i.e. $v_2$ of protons is higher than that of pions, and could be interpreted as the effect of recombination: hadrons inherit the flow component of the recombined parent quarks, resulting in an increased elliptic flow of baryons compared to mesons \cite{flowhierachy}. However also the different radial flow push on $\pi$ and protons can produce the same qualitative effect.\\
Another important factor that needs to be taken into account when looking at elliptic flow measurements is how momentum and energy are transported inside the medium. A good way to study this theoretical aspect is to look at shear viscosity to entropy ratio $\eta/s$.   Theoretical arguments, based on the AdS/CFT conjecture, suggest a
universal lower bound of 1/4$\pi$ for the ratio of shear viscosity to entropy density \cite{etas} in the QGP phase. Viscous hydrodynamic predictions from Luzum and Romatschke reasonably reproduced RHIC elliptic flow data with shear viscosity values $\eta/s$=0.08$\div$0.16 \cite{luzum}. Figure \ref{visc} shows viscous hydrodynamic predictions for elliptic flow measurements at both RHIC and LHC energies with   $\eta/s$=0.2 \cite{Pasi}. The left panel  shows the hydrodynamically calculated differential elliptic flow for unidentified charged hadrons  compared to  STAR $v_2$ data, for four centrality classes ranging from semi-central to mid- peripheral collisions. With $\eta/s$ = 0.20, viscous hydrodynamics gives an excellent description of the STAR $v_2$ data. Extrapolating the same calculation at  LHC energies keeping the same   $\eta/s$=0.2 value  a good agreement with centrality dependent  $v_2$  measurements for pions and protons performed by the ALICE Collaboration  \cite{IDFlow} is achieved. \\
Preliminary studies of possible temperature dependent variations of $\eta/s$ are ongoing, as well as developements of  dynamical models for the pre-thermal evolution of the collision fireball to better quantify  the transport properties of the QGP and its behaviour as a non-perfect fluid with very low (but not zero) shear viscosity.\\
\begin{figure}[t]
\centering
\includegraphics[width=0.99\textwidth]{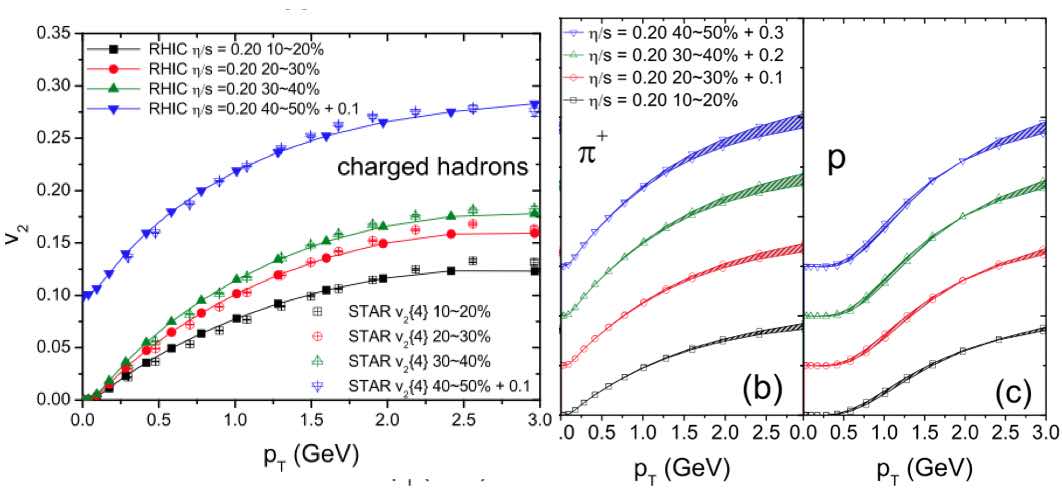}
\caption{Left: \pt differential elliptic flow  for charged hadrons from  Au+Au collisions at \sqnn= 200 GeV  of different centralities. Open symbols are experimental data from the STAR experiment for $v_2$, lines with filled symbols of the same shape are the corresponding hydrodynamic fits from \cite{Pasi}. Right: ALICE results for proton and pion \pt differential $v_2$ from  Pb--Pb events at \sqnn = 2.76 TeV at different centralities. Shaded regions correspond to hydrodynamic extrapolations at higher energies of the same calculations showed in the left panel \cite{Pasi}.}
\label{visc}
\end{figure}
\subsection{Higher order armonics}
\label{sub:hoa}
In a non-central heavy ion collision the beam axis and the impact parameter define the reaction plane $\Psi_{RP}$ as reported in previous section. Assuming a smooth matter distribution in the colliding nuclei, all odd coefficients of Equation 1.42 are zero by symmetry. However,  fluctuations in the matter distribution, including contributions from fluctuations in the positions of the  nucleons participating in the collision,  give rise to odd harmonics like $v_3$ and $v_5$. The measurement of these odd harmonics together with the fourth harmonic $v_4$ is interersting not only to study event-by-event fluctuations of the initial geometric configuration.  In fact, as shown in \cite{Triangular}, deviations from the ideal fluid behaviour governed by the $\eta/s$ shear viscosity have larger effects on higher order harmonics than on $v_2$.\\
Figure \ref{tri} shows the \pt differential    $v_2$, $v_3$, $v_4$ and $v_5$ measured by the ALICE Collaboration with a two particle correlation analysis of data from semicentral Pb--Pb collisions at \sqnn = 2.76 TeV \cite{ALICEHarmonics}. The results for $v_2$ and $v_3$ are compared to hydrodynamic predictions using Glauber initial conditions and different values of  $\eta/s$ from \cite{PredHarmonics}. At low p$_{\rm T}$, the \pt dependence of $v_2$ and $v_3$ is described well by these predictions, however the \pt dependence of $v_2$ is better described by $\eta/s$=0 while  the \pt dependence of $v_3$ is better described by $\eta/s$=0.08.\\
\begin{figure}
\centering
\includegraphics[width=0.99\textwidth]{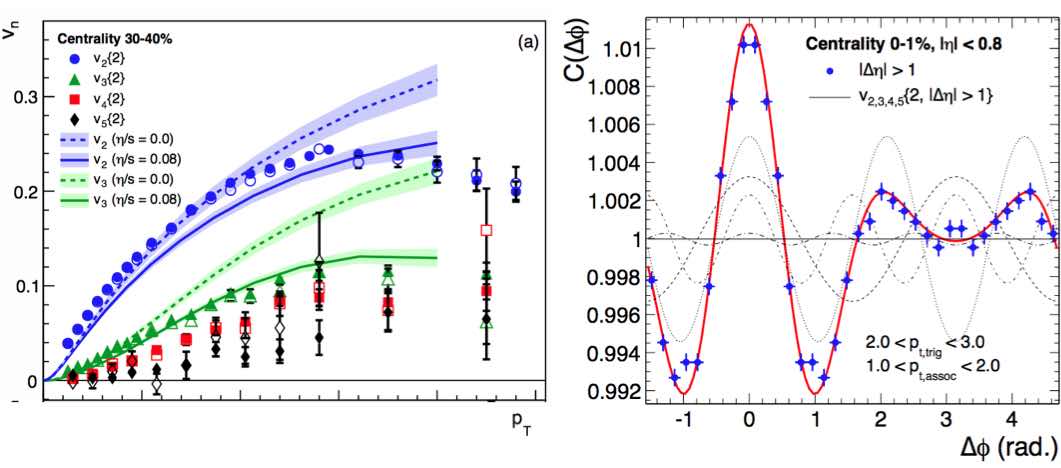}
\caption{Left: ALICE \pt differential results results for $v_2$, $v_3$, $v_4$ and $v_5$ obtained with a two particle correlation analysis of data from semicentral Pb--Pb collisions at \sqnn = 2.76 TeV \cite{ALICEHarmonics}. Right: two-particle azimuthal correlations observed in 0-1\% Pb--Pb collisions and expected azimuthal correlation shape obtained combining the Fourier harmonics weighted by the v$_n$ coefficients \cite{ALICEHarmonics}. }
\label{tri}
\end{figure} 
 The right panel of Figure \ref{tri} shows  the two-particle azimuthal correlations measured in very central Pb--Pb collisions for trigger particles in the range 2<\pt<3 GeV/c and associated particles in 1<\pt<2 GeV/c for pairs in |$\Delta\eta$|>1, measured by calculating 
\begin{equation}
C(\Delta\varphi)=\frac{N_{\rm mixed}}{N_{\rm same}}\frac{\rm{d}N_{\rm same}/\rm{d}\Delta\varphi}{\rm{d}N_{\rm mixed}/\rm{d}\Delta\varphi}
\end{equation}
where $\Delta\varphi$=$\varphi_{\rm trig}-\varphi_{\rm assoc}$, $\rm{d}N_{\rm same}/\rm{d}\Delta\varphi$ ($\rm{d}N_{\rm mixed}/\rm{d}\Delta\varphi$) is the number of associated particles as function of  $\Delta\varphi$ within the same (different) event, and N$_{\rm same}$ (N$_{\rm mixed}$) the total number of associated particles in $\rm{d}N_{same}/\rm{d}\Delta\varphi$ ($\rm{d}N_{mixed}/\rm{d}\Delta\varphi$).  A clear doubly-peaked correlation structure
centered at $\Delta\varphi$=$\pi$ with respect to the trigger particle is visible. The red line corresponds to the azimuthal correlation shape obtained combining the Fourier harmonics associated to $v_2$, $v_3$, $v_4$ and $v_5$ coefficients weighted by their values measured at the corresponding p$_{\rm T}$.  The combination of these harmonics gives a natural description of the observed  doubly-peaked correlation structure centered at $\Delta\varphi$=$\pi$, opposite to the trigger particle. This implies that the measured anisotropic flow Fourier coefficients, including the odd ones arising from initial state fluctuations in the matter distribution, give a natural description of this structure.
\subsection{Observables related to chiral symmetry restoration}
\label{sub:strangeness}
The evolution of the effective quark mass parameter  $\langle\bar{\psi}\psi\rangle$ with temperature (Figure \ref{fig:adscft}) shifts the strange quark mass from its "dressed" value of about 500 MeV/c$^2$ to the current mass value $\sim$ 150 MeV/c$^2$. This would make strange quarks creation (via gg$\rightarrow$s$\bar{\rm s}$) more frequent in a deconfined QGP medium. However, it was shown that, at the same entropy-to-baryon ratio, the plasma in equilibrium does not contain more strangeness than an equilibrated hadron gas at the same temperature. What could signal the presence of a deconfined phase followed by hadronization via recombination are multi-strange baryons. If hadronization only occurs via fragmentation (e.g. in pp collisions), multi-strange baryons formation from single s($\bar{\rm s}$) quarks  only takes place if all the other valence
quarks are created
separately \cite{strangeQGP}. In a QGP system  multi-strange baryon formation is favoured by  recombination of the more abundant s($\bar{\rm s}$) quarks.\\
ALICE measured $\Xi$ baryon \pt differential cross-section in Pb--Pb collisions at \sqnn = 2.76 TeV. Figure \ref{chiralobs} shows the strangeness enhancement factor for $\Xi$ baryons,  defined as the ratios of the $\Xi$ baryon  yield measured in Pb–Pb
collisions, normalized to the mean number of participant nucleons $\langle \rm N_{\rm part} \rangle$, to the corresponding quantity in pp (or p--Be) interactions at the same energy \cite{MultiStrange}. 
Comparing the ALICE measurements with those from the experiments NA57 at the SPS (Pb–Pb collisions at
\sqnn = 17.2 GeV) and STAR at RHIC (Au–Au collisions at \sqnn= 200 GeV), the enhancements are found to decrease with increasing centre-of-mass energy.  $\Xi$ baryon  enhancement factor is larger than unity for all  considered intervals, and increases with $\langle \rm N_{\rm part} \rangle$.\\
\begin{figure}
\centering
\includegraphics[width=0.99\textwidth]{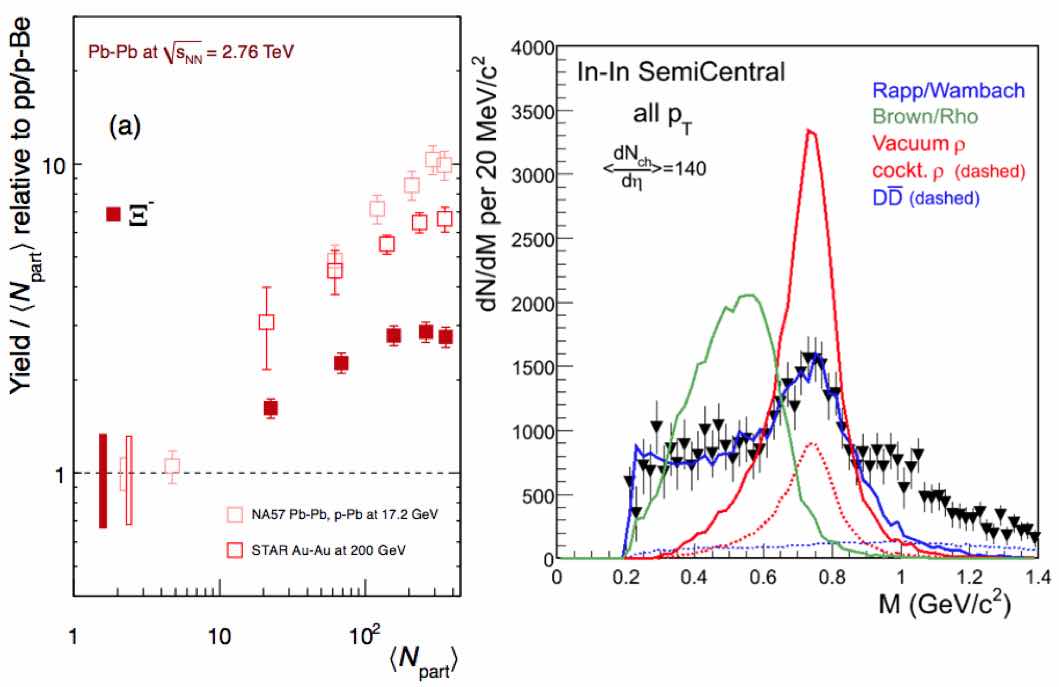}
\caption{Left: ALICE measurement of $\Xi$ meson strangeness enhancement factor as a function of $\langle \rm N_{\rm part} \rangle$. Results from NA57 and STAR are also shown \cite{MultiStrange}. Right: invariant mass spectrum of dimuons from  $\rho$ meson decays from the NA60 Collaboration \cite{RhoNA60}. }
\label{chiralobs}
\end{figure} 
The spontaneous breaking of chiral symmetry manifests itself also in the  violation of the  mass degeneracies within
chiral multiplets ($\rho$ and a$_1$, as seen in Section 1.1.3).  Its restoration should thus modify the mass and decay width of hadrons that form in the hot QCD medium, for instance theoretical models for the $\rho$ meson in-medium production predict a mass shift and a broadening of its Breit-Wigner amplitude \cite{RappRho} . The $\rho$ meson has a life-time comparable to the QGP phase duration ($\sim$ 10$^{-23}$s) and can thus decay in the medium. Its leptonic ($e^+e^-$ or $\rm \mu^+ \rm \mu^-$) decay mode provides a cleaner signal with respect to its (more probable) hadronic decays. Figure \ref{chiralobs} (right) shows NA60  precision measurements of low-mass di-muon pair invariant mass distributions in 158 $A$ GeV\footnote{fixed-target experiment} semicentral In-In collisions at the CERN SPS \cite{RhoNA60}. The vacuum $\rho$ scenario is clearly ruled out.  The in-medium broadening scenario (Rapp/Wambach \cite{RappRho}) appears more realistic. However, the nearly symmetrical broadening around the $\rho$  pole seen in the data is not reproduced by this model and no shift in mass is visible. More detailed work including precise \pt dependences is under way to consolidate these findings.
\subsection{Quarkonia}
At high temperatures,  quark-gluon plasma is expected to reduce
the range of the attractive force between heavy quarks and antiquarks, and above a given
critical temperature  it prevents the formation of bound states.\\
The $q\bar{q}$ potential present in heavy quark bound states is the one reported  in Equation 1.16 containing a Coloumbian term analogue
to the QED one and a confinement term, which takes into account non-perturbative effects of QCD.\\
QGP formation however modifies this potential; the confinement 
term disappears after QGP formation while the color attractive term between q and $\bar{\rm q}$ gets screened by the presence
of deconfined quarks and gluons. Assuming the following form 
\begin{equation}
\centering
V(r) = - \frac{\alpha}{r} e^{-\frac{r}{\lambda_{D}}}
\end{equation}
which is the expression of the Yukawa potential, where $\lambda_{D}$ is the Debye screening length, related to the maximum 
distance two quarks can have in order to form a bound state. The lenght $\lambda_{D}$ decreases as the temperature of the plasma increases. \\
\begin{figure}
\centering
\includegraphics[width=0.99\textwidth]{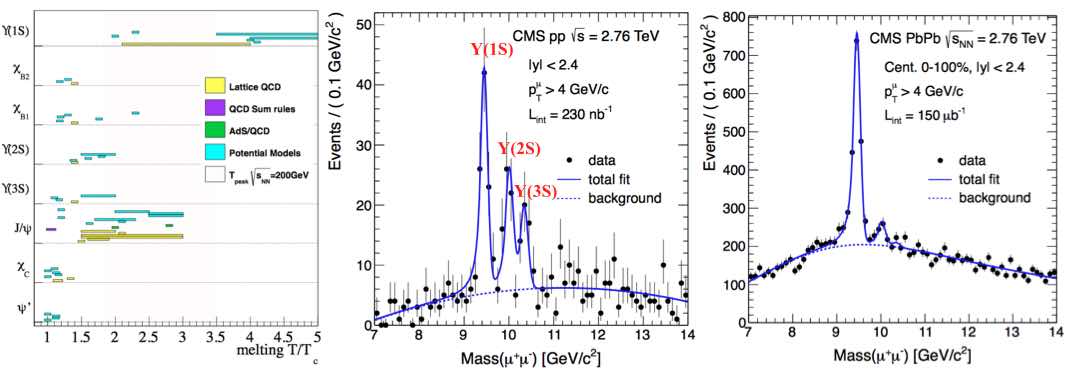}
\caption{Left panel: Compilation of dissociation temperatures for different quarkonia states obtained with different theoretical models \cite{jpsiPHENIX}. Central: CMS invariant mass distributions for di-muon in pp collisions at $\sqrt{\rm s}$ = 2.76 TeV, showing the three bottomonium peaks. Right panel: CMS invariant mass distributions for di-muon in Pb--Pb collisions at \sqnn = 2.76 TeV \cite{upsilonCMS}.  }
\label{fig:quarkonia}
\end{figure} 
Quarkonia states formed by heavy  q$\bar{\rm q}$  pairs can be used to prove this effect. Among charmonia states (c$\bar{\rm c}$ pairs) the J/$\rm \Psi$ meson is the most stable, with a binding energy $\Delta$E=0.64 GeV, while   $\Psi$' mesons are weakly bound ( $\Delta$E= $\sim$ 0.05 GeV). It follows that the $\Psi$' is expected to start dissociating at  temperatures  lower than the J/$\rm \Psi$. The same holds in the bottomonium (b$\bar{\rm b}$) family, where the most bound state is $\rm \Upsilon$(1S)( $\Delta$E= $\sim$ 1 GeV), followed by  $\rm \Upsilon$(2S)( $\Delta$E= $\sim$ 0.54 GeV) and  $\rm \Upsilon$(3S)( $\Delta$E= $\sim$ 0.20 GeV). 
Figure \ref{fig:quarkonia} (left panel)  shows a compilation of medium temperatures relative to the critical temperature (T$_{\rm crit}$) at which quarkonium states are expected to be dissociated in the quark-gluon plasma.  Each horizontal bar corresponds to one estimation and its temperature extension  represents the range where the quarkonia state undergoes a modification until it completely melts \cite{jpsiPHENIX}. The techniques used in the calculations are lattice QCD, QCD sum rules, AdS/CFT, effective field theories  and potential models.\\
Experimentally, J/$\rm \Psi$  production was studied in heavy-ion collisions at the Super Proton Synchrotron (SPS) and at the Relativistic Heavy Ion Collider (RHIC), covering a large energy range from about 20 to 200 GeV center-of-mass energy per nucleon pair. At the SPS the measured J/$\rm \Psi$ suppression is compatible with the melting of the excited states (at SPS energies about 40\% of the yield comes from $\Psi$'  and  $\chi_c$ decays) whereas the RHIC data suggest a small amount of suppression for the direct J/$\rm \Psi$ \cite{jpsiPHENIX}. Recent results from ALICE suggest that at LHC energies dissociation competes with  J/$\rm \Psi$ production via (re)combination of charm quarks in the hot medium \cite{jpsiALICE}. In fact at LHC energies the cross section for $c\bar{c}$ pair production is much higher than at SPS, making recombination effects more relevant. \\
The central and right panels of  Figure  \ref{fig:quarkonia} show the di-muon invariant mass spectrum for $\rm \Upsilon$(1S),  $\rm \Upsilon$(2S) and  $\rm \Upsilon$(3S) in pp and Pb--Pb collisions  measured by the CMS Collaboration \cite{upsilonCMS}. Their comparison clearly shows a suppression of the   $\rm \Upsilon$(2S) and  $\rm \Upsilon$(3S) peaks with respect to the $\rm \Upsilon$(1S) one in Pb--Pb collisions, in accordance with predictions of sequential melting showed in the left panel of Figure \ref{fig:quarkonia}.\\
In conclusion the study of quarkonia production can provide a direct  way to connect experimental high-energy nuclear physics to  lattice QCD predictions, such as  sequential quarkonia states suppression.

\chapter{Heavy flavours as probes of the QGP} 
\lhead{Chapter 2. \emph{Heavy flavours as probes of the QGP}} 
The production of charm quarks in  hadronic collisions is of great interest for many reasons. First, their mass value is well above the QCD confinement  energy scale, and perturbative expansion of QCD diagrams for the calculation of the cross sections should in principle be valid. Data on  particles containing $c$-quarks can thus be a benchmark for perturbative QCD calculations and provide constraints on the parameters used in the theory. Furthermore the comparison of $c$-quark related observables in $pp$ and AA collisions is sensitive to any dynamical influence of the QGP medium on the evolution of heavy quarks that   are predominantly produced in the very early stages of the nuclear interaction and traverse the entire evolution of the strongly-interacting medium.\\
Observables related to charm quark production can be divided in two cathegories, open and hidden charm:
\begin{itemize}
\item hidden charm states are all $c\bar{c}$ bound states (J/$\Psi$, $\Psi$', etc.) and are easilly reconstructed through their decay into lepton pairs ($e^+e^-$ or $\mu^+\mu^-$)
\item open charm states are hadrons in which $c$($\bar{c}$) quarks bind to other lighter quarks. They include mesons (D$^0$, D$^+$, D$^{*+}$..) and baryons ($\Lambda_c$, $\Xi_c$ ) and are experimentally detected by reconstructing their decay products:
\begin{itemize}
\item hadronic decay modes: these allow a full reconstruction of the charm hadron kinematics, i.e. the value of the open heavy-flavour hadron transverse momentum, azimuthal angle, etc. are measured
\item semileptonic decay modes:  a high-\pt lepton coming from the open heavy-flavour hadron decay is detected. To separate leptons from charm and beauty hadron decays different analysis strategy are available (impact parameter studies, $e$-$h$ correlations).  This allows an inclusive measurement of open heavy-flavour hadrons properties and theoretical models can be compared to the measured leptonic transverse momentum distributions
\end{itemize}
\end{itemize}
In this chapter we will briefly  introduce those aspects of pQCD relevant for charm quark production in hadronic collisions, and compare them to  some measurements performed in pp($\bar{\rm p }$) collisions. The influence of cold and hot nuclear matter effects on $c$-quark production and evolution in pA and AA collisions constitutes the second part of the chapter, concluded by a brief review of experimental results. 

\section{Charm quark production in pp collisions \cite{Mangano}}
\begin{figure}
\centering
\includegraphics[width=0.8\textwidth]{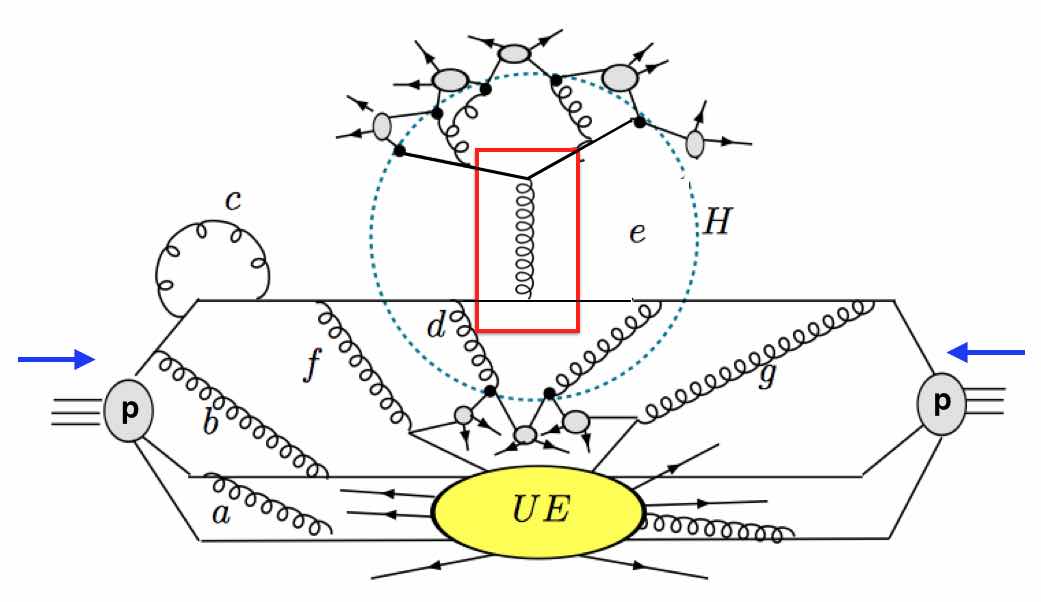}
\caption{Schematic view of a pp interaction in which a hard scattering between two valence quarks (red box) takes place}
\label{CollSketch}
\end{figure}
Let's consider the hadronic interaction depicted in Figure \ref{CollSketch}, where two protons collide inelastically at high centre-of mass energy (several hundreds GeV). The two particles interact because of the hard (large momentum transfer $Q^2$)  inelastic process among two constituent partons. In this example  two valence quark interact, but in this energy regime also sea quarks and gluons can contribute. \\
Before the interaction occurs the valence quarks of the  protons are constantly held togheter by the exchange of virtual gluons. These gluons are mostly soft, like $a$ and $b$, since they are the expression of the QCD confinement. Gluons with high virtuality\footnote{Virtual particles do not  obey the formula $m^2c^4 = E^2 - p^2c^2$ and are said to be off-shell. A virtual gluon will behave kinematically as a massive particle and is short-lived. A virtual quark will have a kinematic mass value higher than  its current mass.} like $c$ prefer to be reabsorbed by the emitting parton since their exchange is not compatible with the stability of the proton. The Heisenberg principle predicts the time scale  of these gluon exchanges, which is (in natural units) the inverse of the gluon virtuality $\tau$=1/$Q^2$ so that the hard gluon ($c$) lifetime is much shorter than the one of $a$ and $b$ which is $\sim$ 1/$m_p$.\\
The  inelastic scattering in the red box is characterized by large momentum transfers (we are considering a pp collision  with $\sqrt{s}$ at the scale of hundreds GeV) and  it also happens on a very short time scale, which means that the soft-gluon mediated interactions ($a$,$b$,$d$,$f$,$g$) of the involved parton with the rest of the proton can be neglected, since they happen on a longer time scale.  The struck quark simply does not have enough time to communicate  that it is being kicked away from the proton.\\
The proton fragments excluded from the hard scattering will continue their evolution interacting on a low momentum scale and hadronizing into colorless states, constituing the Underlying Event (UE - yellow box).\\
The hard scattering in our example generates a $q\bar{q}$ pair at high virtaulity, which could be a $c\bar{c}$ pair. These two quarks will  emit gluons, that can in turn generate quark pairs and so on (parton shower). Once the virtuality scale   of the produced particles is reduced to hadronic mass values hadronization occurs.\\
The evolution depicted above can be expressed by the factorization theorem:
\begin{equation}
\sigma_{pp \rightarrow Hx} = \sum_{a,b}PDF(x_a,Q^2) PDF(x_b,Q^2) \sigma_{ab \rightarrow q\bar{ q}} D_{q\rightarrow H}(z_q,Q^2)
 \label{factorization} 
\end{equation}
where $\sigma_{ab \rightarrow Hx}$ is the inclusive cross-section for the production of a given hadron $H$ in $pp$ collisions and:
\begin{itemize}
\item the sum runs over all the possible combinations of  parton pairs  participating to the hard scattering, including sea quarks and gluons
\item $\sigma_{ab \rightarrow q\bar{q}}$ is the partonic cross section for the production of the quark pair $q\bar{q}$. Given the large momentum transfer of the hard interaction, this cross section is calculable with perturbative QCD
\item PDF(x$_a,Q^2$) and PDF(x$_b,Q^2$) are the Parton Distribution Functions of the two partons involved in  the interaction inside the protons. They are the probability distributions of the fraction of   proton momentum carried by the interacting parton
\item $D_{q\rightarrow H}(z_q,Q^2)$ is the quark fragmentation function and represents the probability that the quark $q$ produces a hadron H carrying a fraction $z$ of its momentum
\end{itemize}
For light quarks and gluons, which have low (or zero) mass, also production via soft processes is possible.  This means that to give a theoretical description of light flavour hadron production one has  to deal with non-perturbative aspects of QCD.\\
If the  $q\bar{q}$ produced in the hard scattering are $c\bar{c}$ (or $b\bar{b}$) pairs, their mass $m_c$>>$\Lambda_{QCD}$ will set a scale for the momentum transfer of the binary partonic interaction, assuring that its cross section will be calculable by perturbative QCD. This means that the study of charmed hadrons allows to better separate the non-pertubative and perturbative factors of Equation \ref{factorization}  when observing a hadronic cross-section.\\
The members of Equation  \ref{factorization} will be now discussed.
\subsection{Parton Distribution Functions: \textbf{PDF(x$_a,Q^2$)}}
\label{sub:pdfs}
In an electron-proton scattering the proton structure is investigated by an electromagnetic  probe, i.e. an off-shell photon.  The virtuality of the photon ($Q^2$) sets the time scale of the interaction and at sufficiently high $Q^2$  it will hit a single and substantially free quark carrying a fraction $x$ of the total proton momentum. The probability $f(x)$ of finding a parton with a fraction $x$ of the total momentum depends in general on the parton type, e.g. considering valence quarks  the $u$ and $d$ quarks have different $f(x)$. However  also the $Q^2$ of the interaction plays a role, since a gluon emitted with momentum $q<Q$ by the probed parton will not have enough time to be reabsorbed before the interaction with the photon takes place, and the parton momentum fraction $x$ will appear lower. The distribution $f(x,Q^2)$ depends on the energy at which the parton is probed, and corresponds to the PDFs introduced in  Equation \ref{factorization}. The evolution with $Q^2$ of a given PDF is given by the Altarelli-Parisi differential equation
\begin{equation}
\frac{df(x,\mu)}{d\rm log \mu^2} = \frac{\alpha_s}{\pi}\int^{1}_{x}\frac{dy}{y}f(y,\mu)P_{q\bar{q}}(x/y)
\label{APEq}
\end{equation}
where $P_{q\bar{q}}(x/y)$ describes the probability that a quark with a momentum fraction $x$ emits a gluon keeping a fraction $y$ of its momentum and $\mu$ is the factorization scale at which the PDFs are evaluated. Qualitatively, the factorization scale corresponds to the resolution with which the hadron is being probed.\\
In Deep Inelastic Scattering (DIS) the virtual photon can also interact with  the sea  $q\bar{q}$ pairs generated by gluon splitting. If the virtuality of the photon is high enough, the time scale of the hard interaction is so short that gluon fluctuations in virtual $c\bar{c}$ pairs become visible. For instance, in $ep$ collisions at $x$ $\sim$ 0.01 and $Q^2$ $\sim$ 100 GeV$^2$, charm contributes approximately to 25\% of the total Neutral Current\footnote{When the $Q^2$ of the interaction overcomes the electroweak vector bosons mass threshold, the mediator of the DIS event can also be a W$^{\pm}$ or a Z boson  } (NC) DIS cross-section \cite{tesirum}.\\
PDFs are extracted from experimental data of NC and Charged Current (CC) scatterings in high-energy $ep$ collisions. The most recent PDFs sets are based on data coming from the Hadron Elektron Ring Anlage (HERA) collider, where electrons(positrons) and protons were collided at $\sqrt{s}$ up to $\sim$ 300 GeV. The ZEUS and H1 experiments have  taken data on NC and CC in DIS, measuring in particular D mesons cross-section as a function of $x$ and $Q^2$ that brings information on $c$-quark PDFs. The PDFs are extracted from a theoretical fit to the experimental data. The PDF uncertainties depend on the treatement of non-perturbative effects, the impact of missing higher-order perturbative terms and the choice of the QCD parameters ($\mu_F$, $\mu_R$\footnote{The renormalization scale $\mu_R$ is the energy scale used in the evaluation of $\alpha_s$. The factorization scale $\mu_F$ is the scale used in the evolution of the parton densities (eq. \ref{APEq} \cite{Mangano2})}, quark masses) for what concerns the theoretical inputs, and of course on the statistical and systematic uncertainties of the fitted data. As an example, in Figure \ref{pdfs} the CTEQ6M\footnote{Coordinated Theoretical-Experimental Project on QCD} PDF at  Q = 2 and 100 GeV are shown. Details on the  CTEQ6M fit can be found in \cite{cteqm6}.
\begin{figure}
\centering
\includegraphics[width=0.8\textwidth]{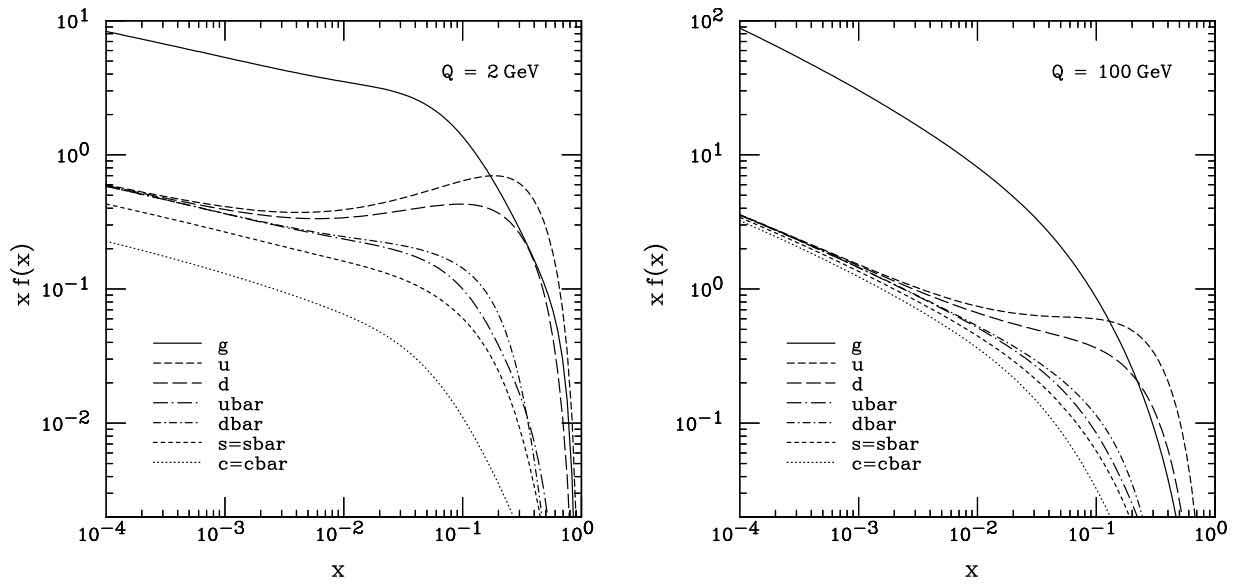}
\caption{CTEQ6M parton distribution functions at Q = 2 and 100 GeV   \cite{cteqm6}.}
\label{pdfs}
\end{figure}
\subsection{Partonic cross-section: \textbf{$ \sigma_{ab \rightarrow q\bar{ q}} $}}
Let us limit ourselves to the production of  heavy-quark pairs ($Q\bar{Q}$). In pQCD the leading order processes to be considered are:
\begin{itemize}
\item quark-antiquark annihilation $q\bar{q} \rightarrow Q\bar{Q}$
\item gluon-gluon fusion $gg \rightarrow Q\bar{Q}$
\end{itemize}
The amplitudes of both processes are finite, due the mass of the heavy quark that sets the scale both in the $s$ and $t$ production channels. At high energies  these amplitudes can be expressed as  \cite{Mangano2}:
\begin{equation}
\centering
 \lim_{\hat{s}\to\infty}\hat{\sigma}_{q\hat{q} \rightarrow Q\hat{Q}}\propto\frac{1}{\hat{s}} \ \ \ \ \ \ \ \ \lim_{\hat{s}\to\infty} \hat{\sigma}_{gg \rightarrow Q\hat{Q}}\propto\frac{1}{\hat{s}}\textsl{L}(\beta)  \ \ \ \ \ \ \ \  \textsl{L}(\beta) = \frac{1}{\beta}\log(\frac{1+\beta}{1-\beta})-2
\end{equation}
with $\beta$=$\sqrt{1-\frac{4m^2}{\hat{s}}}$ and $\hat{s}$ the partonic centre-of-mass energy. From these formulas we can draw some conclusions 
\begin{itemize}
\item At large $\hat{s}$ the  $q\bar{q} \rightarrow Q\bar{Q}$ vanishes more quickly
\item For two heavy quarks of mass $m_1$ and $m_2$ 
\begin{equation}
\lim_{\hat{s}\to\infty}\frac{\hat{\sigma}_{gg \rightarrow Q_1\hat{Q_1}}}{\hat{\sigma}_{gg \rightarrow Q_2\hat{Q_2}}}=1-\frac{\log(m_1^2/m_2^2)}{\log(s/m_2^2)} \ \ \ \ \ \ \ \ \ \lim_{\hat{s}\to\infty}\frac{\hat{\sigma}_{q\hat{q} \rightarrow Q_1\hat{Q_1}}}{\hat{\sigma}_{q\hat{q} \rightarrow Q_2\hat{Q_2}}}=1-O(m_1^4/s^2)
\end{equation}
which at $\sqrt{\hat{s}}$ = 100 GeV means
\begin{equation}
\frac{\sigma(gg \rightarrow b\hat{b})}{\sigma(gg \rightarrow c\hat{c})}\sim 0.7 \ \ \ \  \ \ \ \ \  \ \ \ \  \frac{\sigma(q\hat{q}  \rightarrow b\hat{b})}{\sigma(q\hat{q}  \rightarrow c\hat{c})}\sim 0.99
\end{equation}
\end{itemize}
\begin{figure}
\includegraphics[width=0.99\textwidth]{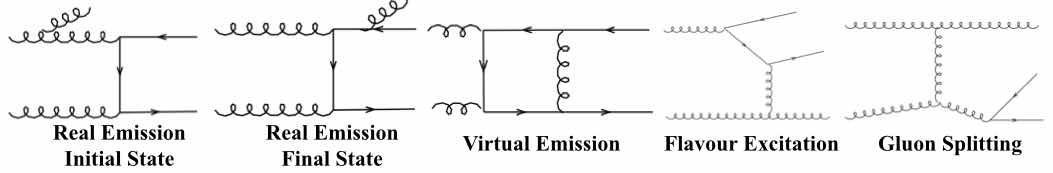}
\caption{NLO diagrams for heavy-quark pairs production}
\label{NLODiagrams}
\end{figure}
The inclusion of Next to Leading Order (NLO) diagrams in the calculation should make the pQCD picture of heavy-flavour production more realistic. In particular  uncertainties related to the factorization $\mu_F$ and  renormalization $\mu_R$ scales should be reduced.\\ NLO diagrams are shown in Figure \ref{NLODiagrams}. Ultraviolet divergences in the virtual diagrams are removed by the renormalization process. Real emission by inital state parton is reabsorbed in the PDFs as explained before.  Real  emission by final state heavy-quarks is collinear safe since the quarks mass value prevents gluon emission at small angles. However, when the momenta of the produced quarks become large with respect to the heavy-quark mass, large logarithms of the ratio $\frac{\pteq}{m}$ arise to all orders in the perturbative expansion of the cross-section, and spoil its convergence. Two examples are the Flavour Excitation and Gluon Splitting diagrams in Figure \ref{NLODiagrams},  and their inclusion in the calculation originates similar large logarithms at higher orders of  perturbative expansion. \\
Different theoretical framework have been developed to deal with these large influence of higher-order corrections in perturbative calculations:
\begin{itemize}
\item \textbf{Fixed Order Next-to-Leading Log (FONLL)} \cite{FONLL}: the \pt spectrum of heavy-flavour production is obtained at NLO accuracy in $\alpha_s$, while leading logarithmic (LL) terms of the form $\alpha_s^2(\alpha_s \log\frac{\pteq}{m})^k$ and next-to-leading logarithmic (NLL) terms of the form $\alpha_s^3(\alpha_s \log\frac{\pteq}{m})^k$ are resummed in the pQCD expansion. Uncertainties depend on the heavy-quark mass value, and the renormalization and factorization scales $\mu_R$ and $\mu_F$
\item \textbf{General-Mass Varaible-Flavour-Number Scheme (GM-VFNS)} \cite{GM-VFNS}: this framework adopts the minimal subtraction scheme\footnote{renormalization scheme used to absorb the infinities that arise in perturbative calculations beyond leading order, introduced by 't Hooft (Nobel Prize 1999) } considering a massive charm quark. The number of active quarks in the calculations depends on the factorization scale $\mu_F$: for low values of $\mu_F$ only   $g$,$u$,$d$ and $s$ are considered. Charm is included if $\mu_F > m_c$  
\end{itemize}
Both frameworks however are collinearly-factorized, which means that the transverse momenta of the incoming partons  are zero. As seen previously at LHC energies charm production is dominated by gluon fusion processes, which means that in a collinearly-factorized framework transverse momenta of produced charm quark and antiquark are  equal. Thus    collinear factorization  cannot be used for studies of correlation observables for charmed meson pairs. In the \textbf{$k_{\rm T}$-factorization} approach, off-shell leading order matrix elements for $gg\rightarrow c\bar{c}$ together with unintegrated gluon densities (UGDF) that depend on the transverse momentum of the gluon, $k_{\rm T}$, as well as the usual dependence on $x$ and on the factorization scale are used \cite{VogtFact}. Predictions for D-meson production in pp collisions have recently been published and  their comparison to ALICE measurements will be shown in Section 2.1.4.
\begin{figure}[b]
\centering
\includegraphics[width=0.8\textwidth]{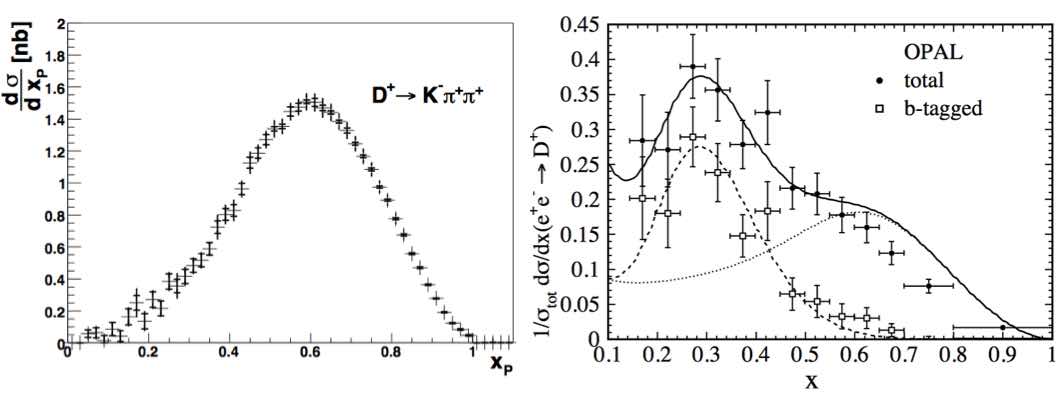}
\caption{Belle (left \cite{BelleFrag}) and OPAL (right \cite{OPAL}) results for D$^+$ meson cross-section as a function of $x_p$ }
\label{charmfrag}
\end{figure}
\subsection{Fragmentation Functions:$D_{q\rightarrow H}(z_q,Q^2)$}
The process of heavy-quarks fragmenting into hadrons is non-perturbative, but the fragmentation functions can be extracted from data.\\
In $e^+e^-$ collisions the energy of the primary heavy quark pair before the perturbative (hard gluon emission) and non perturbative (soft gluon emission, hadronization) processes is known. The measurement of hadronic cross sections in $e^+e^-$ annihilation via a virtual photon ($\gamma$) or Z boson, $e^+e^- \rightarrow \gamma(Z) \rightarrow H$ ($H$ is a generic hadron) can be fitted with electroweak theoretical calculations that take into account pQCD  (gluon radiation) and electroweak (photon radiation) effects. \\
\begin{figure}[t]
\centering
\includegraphics[width=0.8\textwidth]{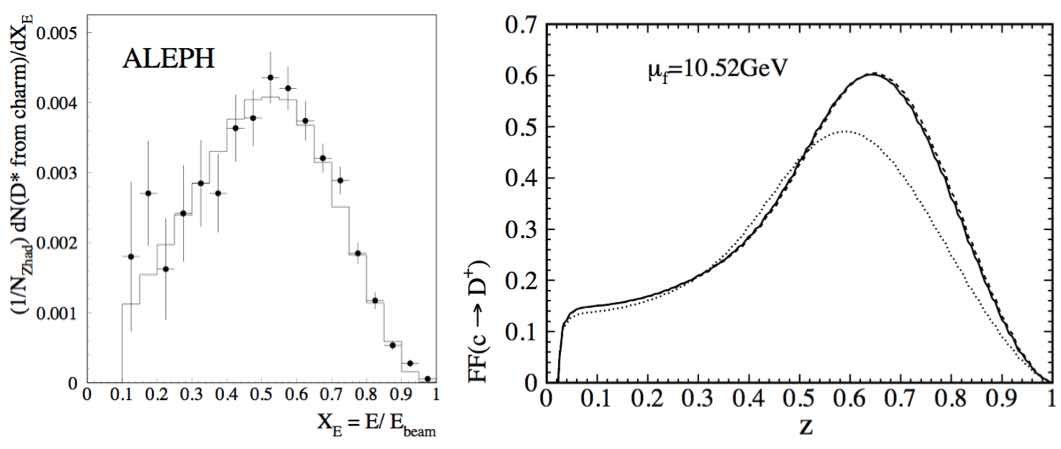}
\caption{Left: ALEPH distribution of $X_E$=E/E$^{\rm beam}$ for the D$^{*+}$ meson cross-section from the $c\bar{c}$ source  \cite{ALEPHFrag}. Right: fragmentation function of $c$-quarks into D$^+$ mesons at the reference scale $\mu_F$=10.52 GeV  from Belle and CLEO data \cite{BelleFrag}.}
\label{frag1}
\end{figure}
Figure \ref{charmfrag} (left) shows Belle results for D$^+$ meson production in $e^+e^-$ collisions at $\sqrt{s}$ = 10.6 GeV \cite{BelleFrag}. 	The D$^+$ meson  cross section is shown as a function of x$_p$ = $\overrightarrow{p}^{D^{+}}/\overrightarrow{p}^{D^{+}}_{max}$ with $\overrightarrow{p}^{D^{+}}_{max}$=$\sqrt{s/4 - m_{D^{+}}^2}$. This measurement excludes the fraction of D$^+$ mesons coming from B hadron decays, giving access to the $c$ quark fragmentation functions. \\
The right panel of Figure  \ref{charmfrag}  shows the same results from $e^+e^-$ collisions at $\sqrt{s}$ = 91.2 GeV \cite{OPAL} measured by the OPAL collaboration at LEP. Here the total D$^+$ meson normalized cross section is shown, and the cross section of $b$-quark-initiated D$^+$ is superimposed. A theoretical fit is superimposed to extract the c-quark-initiated, b-quark-initiated, and total contributions, respectively.\\
The charm fragmentation function used to extract the FONLL  predictions that will be reported in the next paragraph are obtained from fits to  D$^{*+}$ fragmentation data in   $e^+e^-\rightarrow$Z $\rightarrow c\bar{c}$ events published by the ALEPH collaboration \cite{ALEPHFrag}. The ALEPH distribution of $X_E$=E/E$^{\rm beam}$ for the D$^{*+}$ meson cross-section from the $c\bar{c}$ source (Figure \ref{frag1} - right) is fitted with a convolution of a perturbative contribution $D^{pQCD}_{N}$ (which describes the production of a c quark in the hard interaction) and a  non-perturbative fragmentation function $D^{V}_{BCFY,N}(r,z)$ describing its hadronization into the D$^{*+}$ \cite{FragCacciari}:
\begin{equation}
\sigma_N(c\rightarrow D^{*+})(r,z)=D^{pQCD}_{N}\ast D^{V}_{BCFY,N}(r,z)
\end{equation}
The  term $D^{V}_{BCFY,N}(r,z)$ depends on the non-perturbative parameter $r=\frac{M_H - M_Q }{M_H}$ ($M_H$ and $M_Q$ are the masses of the charmed hadron and of charm, respectively) and on $z$, the  longitudinal momentum fraction of the hadron relative to the fragmenting heavy quark. The functional form of $D^{V}_{BCFY,N}(r,z)$ (which describes the fragmentation of $c$-quarks into a vector meson state like D$^{*+}$) can be found in \cite{formulaccia} together with  that of $D^{P}_{BCFY,N}(r,z)$, describing the transition of a $c$-quark into a pseudoscalar meson state like D$^{+}$. The fit to ALEPH data allows to fix the $r$ parameter to 0.1, and the fragmentation functions of the different D meson states are obtained as linear combination $D^{V}_{BCFY,N}(r,z)$ and $D^{P}_{BCFY,N}(r,z)$, with the corresponding branching ratios as coefficients. As an example, for D$^{*+}$ and D$^{+}$ we get
\begin{equation}
D^{c\rightarrow D{*+}}(z)=\textrm{BR}(c\rightarrow D^{*+})D^{V}_{BCFY,N}(z)=0.233D^{V}_{BCFY,N}(z)
\end{equation}
\begin{equation}
\begin{split}
D^{c\rightarrow D{+}}(z)=\textrm{BR}(c\rightarrow D^{+})D^{P}_{BCFY,N}(z)+\textrm{BR}(c\rightarrow D^{*+})\textrm{BR}(D^{*+}\rightarrow D^{+})D^{V}_{BCFY,N}(z)=\\=0.162D^{P}_{BCFY,N}(z)+0.072D^{V}_{BCFY,N}(z)
\end{split}
\end{equation}
The GM-VFNS predictions use fragmentation functions for D$^{0}$, D$^{+}$ and D$^{*+}$  determined via a fit to  $e^+e^-$ data  at $\sqrt{s}$ = 10.52 GeV from Belle and CLEO Collaborations \cite{BelleFrag}, as exposed in \cite{FragGM}. The fitted fragmentation function of $c$-quarks into D$^+$ mesons is shown in Figure \ref{frag1} (right) at the reference scale $\mu_F$=10.52 GeV, and the Altarelli Parisi equation (Equation 2.2) predicts their evolution at other energy scales.  \\
\begin{figure}[t]
\centering
\includegraphics[width=0.99\textwidth]{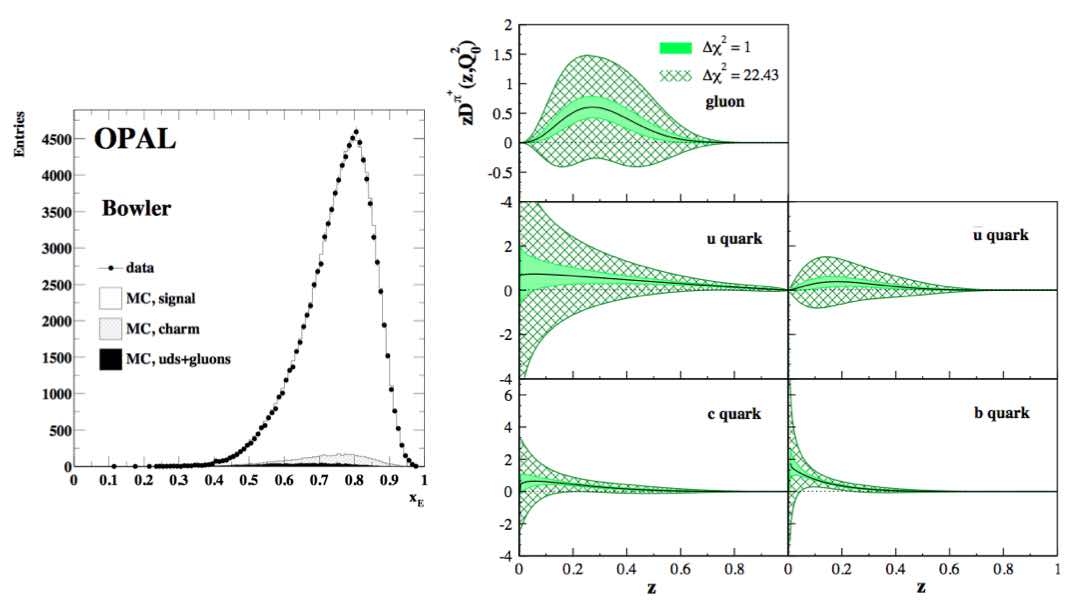}
\caption{Left:  inclusive fragmentation function of $b$ quarks into B hadrons measured by the OPAL Collaboration in  $e^+e^-$ collisions at $\sqrt{s}$ = 91.2 GeV \cite{OpalBeauty}. Right:  $\pi^+$ fragmentation functions from global fit to single-inclusive electron-positron annihilation data and   semi-inclusive deep inelastic scattering asymmetry data from HERMES and COMPASS, as detailed in \cite{pionfrag}. }
\label{frag2}
\end{figure}
The $c\rightarrow D^{+}$ fragmentation function in Figure \ref{frag1} (right) shows a peak structure at high ($\sim$0.7) values of the fractional momentum $z$. This feature is also visible in the  inclusive fragmentation function of $b$ quarks into B hadrons measured by the OPAL Collaboration in  $e^+e^-$ collisions at $\sqrt{s}$ = 91.2 GeV \cite{OpalBeauty}, shown in Figure \ref{frag2} (left) together with a MC fit based on Bowler fragmentation model \cite{BowlerFrag} highlighting the low contributions from charm, gluons and light quarks. Heavy flavored mesons  retain a large fraction of the momentum of the primordial heavy quark, in contrast to light hadrons fragmentation processes. Figure \ref{frag2} (right) shows $\pi^+$ fragmentation functions from global fit to single-inclusive electron-positron annihilation data and   semi-inclusive deep inelastic scattering asymmetry data from HERMES and COMPASS, as detailed in \cite{pionfrag}.
\subsection{Some results in pp($\bar{\rm p}$)}
\begin{figure}
\centering
\includegraphics[width=0.9\textwidth]{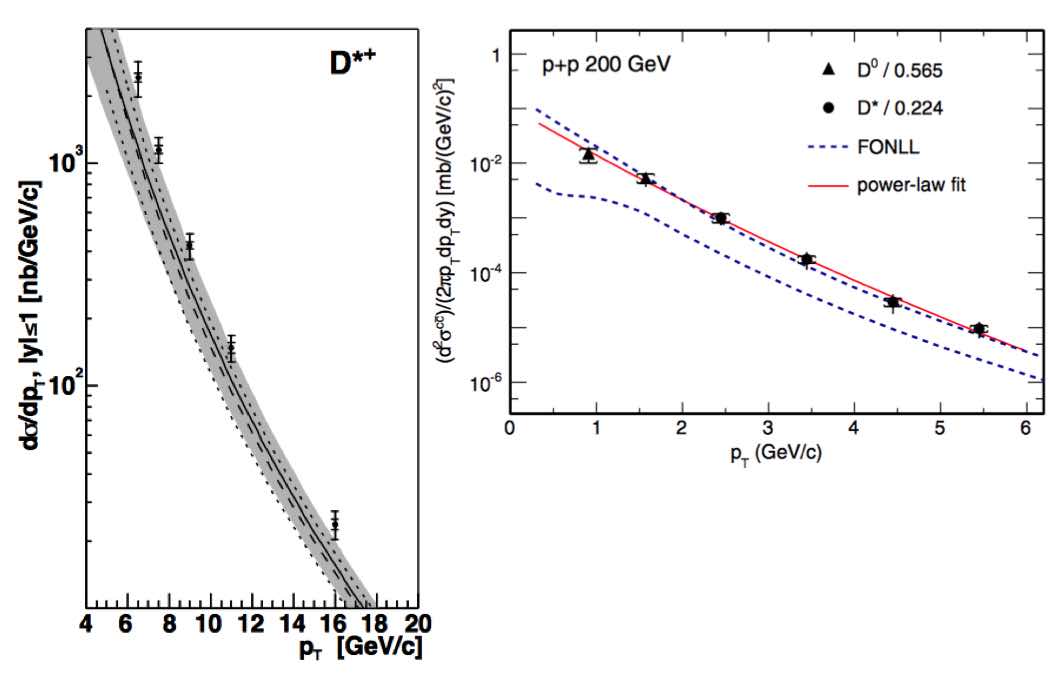}
\caption{Left: D$^{*+}$ differential cross section measurement for $|y|<1$ in $p\bar{p}$ collisions at $\sqrt{s}$ = 1.96 TeV \cite{CDFD}. The inner bars represent the statistical uncertainties; the outer bars are the quadratic sums of the statistical and systematic uncertainties. The solid curves are the theoretical predictions for FONLL, with the uncertainties indicated by the shaded bands. The dashed curve   is the theoretical prediction from GM-VFNS, the dotted lines indicate the uncertainty. Right:  \pt differential cross section of $c\bar{c}$ pairs measured by STAR in pp collisions at $\sqrt{s}$ = 200 GeV  \cite{STARD}. The dashed blue lines represent the limits of FONLL predictions. }
\label{FigPPLE}
\end{figure}
I will now discuss four measurement of open-charm production in pp or p$\bar{\rm p}$ collisions together with predictions from the pQCD calculations described previously
\begin{itemize} 
\item Figure \ref{FigPPLE} (left) shows the D$^{*+}$ \pt differential cross section in $p\bar{p}$ collisions at $\sqrt{s}$ = 1.96 TeV measured by CDF \cite{CDFD}. The cross-section is measured  only for D mesons coming from $c$-quark fragmentation (prompt D mesons) and excludes the fraction of D meson coming from beauty hadron decays (feed-down D mesons). The measurements are compared to both FONLL and GM-VFNS calculations. The uncertainties on the calculated cross sections are evaluated by varying independently the renormalization and factorization scales. The measured differential cross sections are higher than the theoretical predictions by about 100\% at low \pt and 50\% at high p$_{\rm T}$. However, they are compatible within uncertainties.
\item  Figure \ref{FigPPLE} (right)  shows the  \pt differential $c\bar{c}$ cross section in $pp$ collisions at $\sqrt{s}$ = 200 GeV measured by STAR at mid-rapidity \cite{STARD}. The $c\bar{c}$ cross-section derives from the D$^0$ and D$^{*+}$ cross-sections divided by the corresponding $c$-quark fragmentation fraction values. FONLL calculations are superimposed, the upper and lower limits derive from the variation of the factorization and renormalization scales. These results are consistent with the upper limit of the FONLL pQCD calculations. 
\end{itemize}
\begin{figure}
\centering
\includegraphics[width=0.95\textwidth]{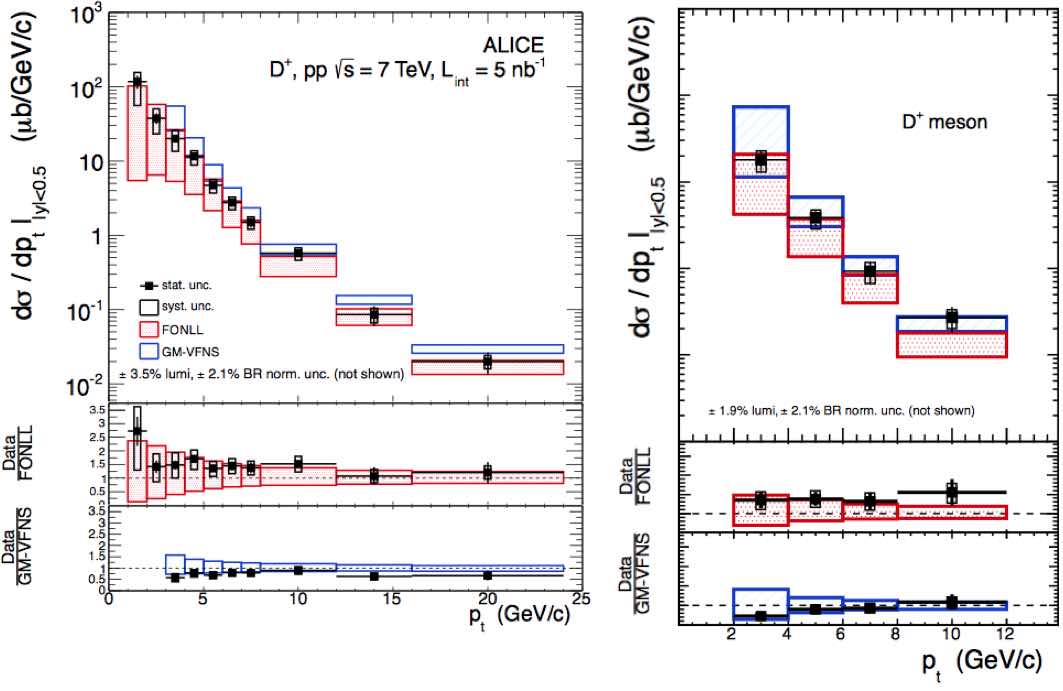}
\caption{Prompt D$^{+}$ differential cross section measurement ($|y|<0.5$) in $pp$ collisions at $\sqrt{s}$ = 7 TeV (left) and  $\sqrt{s}$ = 2.76 TeV (right). Both FONLL and GM-VFNS predictions are shown \cite{D7TeV}. }
\label{Dplus7276}
\end{figure}
\begin{figure}
\centering
\includegraphics[width=0.95\textwidth]{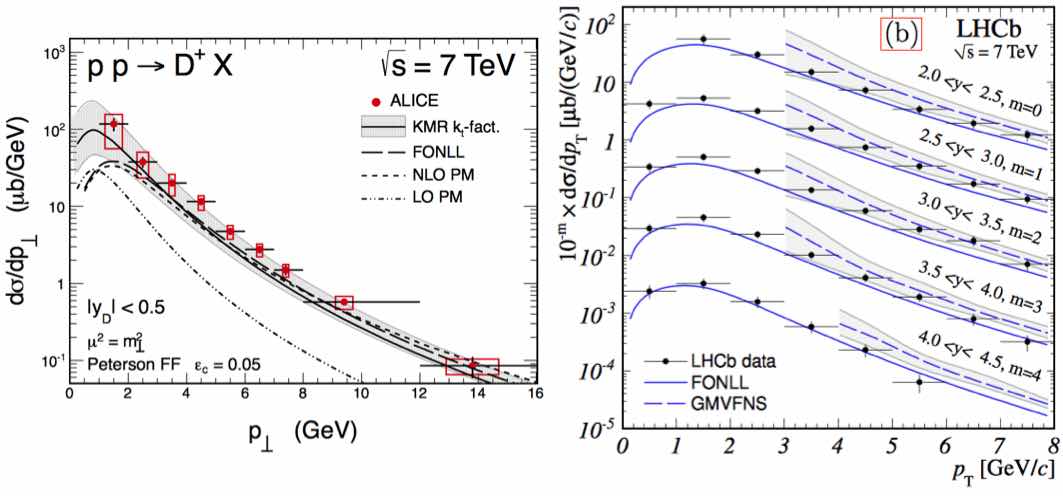}
\caption{Left: prompt D$^{+}$ differential cross section measurement ($|y|<0.5$) in $pp$ collisions at $\sqrt{s}$ = 7 TeV with FONLL, GM-VFNS and $k_T$ factorization predictions \cite{ALICEkt}. Right: prompt D$^{+}$ \pt differential cross section in $pp$ collisions at $\sqrt{s}$ = 7 TeV measured by the LHCb Collaborations in five rapidity intervals. The measurement relative to each rapidity  interval is scaled by 10$^{-m}$  where m is indicated close to each set of points  \cite{LHCb}. }
\label{ktfact}
\end{figure}
\begin{itemize}
\item Figure \ref{Dplus7276} shows the prompt D$^{+}$ \pt differential cross sections in $pp$ collisions at $\sqrt{s}$ = 7 TeV (left) and $\sqrt{s}$ = 2.76 TeV (right) measured by the ALICE Collaboration in $|y|<0.5$ compared to both FONLL and GM-VFNS models \cite{D7TeV}. Both calculations use the CTEQ6.6 parametrizations of the PDFs, and their uncertainties correspond to the variation of  $m_c$, $\mu_R$ and $\mu_F$ parameters used in the calculations. The central values of the FONLL predictions lie systematically below the data, as for the first two measurements (CDF and STAR). The GM-VFNS central points lie systematically above the data. However both models are  consistent with data within uncertainties. Figure \ref{ktfact} (left) shows the 7 TeV data (same as in the left panel of Figure \ref{Dplus7276})  together with FONLL, GM-VFNS and $k_T$ factorization predictions \cite{ALICEkt}. The  $k_T$ factorization approach reproduces the data only if the  KMR UGDFs are used. All other unintegrated gluon distributions strongly underpredict the experimental results. This may suggest that some mechanism of charm production, like double parton scattering,  is still missing.\\ The total charm production cross section was extrapolated to the full phase space using the sum of the total production cross section for D$^0$ and D$^{+}$ divided by the sum of their fragmentation ratios. The total production cross section for D$^{*+}$ divided by its fragmentation ratio using the inverse of the squared statistical uncertainties as weights. The values obtained are\\
\vskip 0.1 em 
$\sigma(c\bar{c})=8.5\pm0.5({\rm stat}) ^{+1.0}_{-2.4}({\rm syst})\pm0.1({\rm BR})\pm0.2({\rm frag})\pm0.3({\rm lum}) ^{+5.0}_{-0.4}({\rm extr})$ mb\footnote{BR =  from branching ratios uncertainties, frag = from fragmentation function uncertainties, lum = luminosity uncertainty, extr =  full phase space extrapolation uncertainty}  \\
at $\sqrt{s}$ = 7 TeV and\\
\vskip 0.1 em 
$\sigma(c\bar{c})=4.8\pm0.8({\rm stat}) ^{+1.0}_{-1.3}({\rm syst})\pm0.06({\rm BR})\pm0.1({\rm frag})\pm0.1({\rm lum}) ^{+2.6}_{-0.4}({\rm extr})$ mb \\
at $\sqrt{s}$ = 2.76 TeV.
\vskip 0.1 em 
\item Figure \ref{ktfact} (right) shows the prompt D$^{+}$ \pt differential cross section in $pp$ collisions at $\sqrt{s}$ = 7 TeV measured by the LHCb Collaboration compared to both FONLL and GM-VFNS pQCD calculations \cite{LHCb}. The cross-section is measured in five rapidity intervals in the region 2<$y$<4.5. Both models are  consistent with data within uncertainties (FONLL uncertainties are not shown) and the same behaviour (FONLL central value underestimating the data, GM-VFNS overestimating the data) observed in ALICE and CDF measurements  is found. These data together with other LHCb measurements relative to D$^{0}$ ,  D$^{*+}$  and D$^{+}_s$  mesons allow the extraction of the total charm cross-section  inside the acceptance region of the measurements, 0<\pt<8 GeV/c and 2<$y$<4.5:\\ $\sigma(c\bar{c})_{p_{\rm T}<8GeV/c,2<y<4.5}=1419\pm12({\rm stat})\pm116({\rm syst})\pm65({\rm frag})$ $\mu$b 
\end{itemize}
\section{Charm quark production in ultrarelativistc heavy-ion  collisions}
Heavy quarks  in heavy-ion collisions are produced via the same mechanism described in the previous section for pp collisions, i.e. via hard partonic scattering processes occurring on short time scales.  The formation time of a heavy-quark pair ($\tau \sim 1/Q^2$, which for charm formation at the  threshold  is $\sim$10$^{-2}$fm/c) is shorter than the time needed for the QGP to reach thermal equilibrium ($\sim$1 fm/c, see Section 1.3.1), so that heavy quark are predominantly produced in the very initial stages of the collision. In fact, even if inelastic scatterings among partons occur in the QGP fireball, the characteristic temperatures in this phase (i.e. the relevant momentum scale) are too low for thermal heavy-flavour production, whose contribution is small even at RHIC and LHC energies\footnote{Some works have shown the possibility of significant thermal contribution to charm quark production at LHC energies, but assuming  high values of initial temperature ($\sim$ 750 MeV) and bare mass of light constituents (light quarks, gluon) in the QGP \cite{cctherm} \cite{cctherm2} } \cite{cctherm}. This means that:
\begin{itemize} 
\item  heavy quarks (like charm) experience the full evolution history of the expanding fireball, and are thus good probes of hot QCD medium
\item heavy quarks production in AA collisions is proportional to the total number of binary collisions $N_{\rm coll}$ that took place in the nuclear interaction 
\end{itemize}
But how can we extract information from heavy-flavour probes in heavy-ion collisions? If we look at Equation  \ref{factorization} we have to take into account that the PDFs are in general modifed in a nuclear environment, as it will be discussed later. We can also expect that the part relative to fragmentation functions won't be valid in heavy-ion collisions, since quark fragmentation is not the only hadronization mechanism as already mentioned in Section 1.4.4. Also the full fragmentation process of hard partons could be affected by the presence of the dense QCD medium. Besides these aspects we also have to consider interactions between heavy quarks and the medium, that can modify the kinematics of  heavy quarks prior to their hadronization. We can divide the phenomena influencing heavy-flavour production in heavy-ion collisions in two cathegories:
\begin{itemize}
\item \textbf{Cold Nuclear Matter Effects} are not due to QGP formation, but to the fact that the colliding particles are nuclei: for example, as far as heavy flavour production is concerned, PDFs in nuclei differ from those in free nucleons, and influence the heavy quark production kinematic. These effects should be also visible in p--A and e--A collisions.
\item \textbf{Hot Medium  Effects} are due to QGP formation, and are expected to be present only if the energy density and temperature achieved in the collision are high enough to allow the transition to the deconfined state. They include the interaction of the heavy-quark with the medium and the way it hadronizes.
\end{itemize}
If these effects influence heavy-quark production and dynamics, we cannot consider ultrarelativistic heavy-ion collisions as a superposition of inchoerent nucleon-nucleon interactions. The \pt differential cross section for heavy-flavour  observables will not be the one measured in pp collisions multiplied by the average Glauber N$_{\rm coll}$ of the event class considered.\\
 When we measure the yield of some observable in AA collisions, we  define the \textbf{nuclear modification factor} as the ratio of the observables yield in AA  collisions and the one in pp collisions scaled by the average number of binary collisions $\langle N_{\rm coll}\rangle$ obtained from Glauber calculation for the AA centrality class considered
\begin{equation}
R_{\rm AA}  = \frac{Yield^{\rm AA}}{\langle N_{\rm coll} \rangle Yield^{\rm pp}}
\end{equation}
In the case of the modification of the \pt distributions,  $R_{\rm AA} $ can be expressed as:
\begin{equation}
R_{\rm AA} (\pteq) = \frac{dN^{\rm AA}/d\pteq}{\langle N_{\rm coll} \rangle dN^{\rm pp}/d\pteq} =  \frac{dN^{\rm AA}/d\pteq}{\langle T_{\rm AA} \rangle d\sigma^{\rm pp}/d\pteq}
\end{equation}
In case no modification (due to  either cold or hot nuclear matter effect) is present, we expect R$_{\rm AA}$ to be equal to 1, and AA collisions can be treated as a superposition of N$_{\rm coll}$ nucleon-nucleon collisions. On the contrary any modification that affects the production rate and the momenta distributions of the considered observables in  nuclear collisions leads to an R$_{\rm AA}$ different than 1. \\ The same observable can be computed in pA collisions as:
\begin{equation}
R_{\rm pA}(\pteq) = \frac{dN^{\rm pA}/d\pteq}{\langle N_{\rm coll}\rangle dN^{\rm pp}/d\pteq} =  \frac{d\sigma^{\rm pA}/d\pteq}{{\rm A} \ d\sigma^{\rm pp}/d\pteq}
\label{Eq:rppb}
\end{equation}
where A is the  mass number of the colliding nucleus. \\
In the following sections we will examine cold and hot nuclear matter effects, presenting some experimental results that will make more complete the discussion of QGP physics of Chapter 1.  
\subsection{Initial State Effects}
\label{sub:instef}
The presence of a nucleus in the initial state induces modifications in the heavy quarks production mechanism with respect to what happens in pp interactions. In this section we are going to discuss:
\begin{itemize}
\item nuclear modification of the PDFs
\item $k_{\rm T}$ broadening and Cronin enhancement
\item energy loss in cold nuclear matter
\end{itemize}
Parton distribution functions in nuclei differ from those measured in free nucleons.  This was first observed by the European Muon Collaboration (EMC) at CERN in 1983  in muon-nuclei DIS experiments. In particular in Figure \ref{NuclearPDFs} (left) we see the ratio of the DIS cross-section in nuclei (Fe, Cu) to the one in deuterium as a function of Bjorken $x$ measured by EMC and by SLAC E139 experiment. This ratio is proportional to the ratio of the structure functions F$_2^{\rm Fe,Cu}$ and F$_2^d$ in nuclei and deuterium \cite{StrucFun}. We can divide this measurement in four region of Bjorken $x$:
\begin{itemize}
\item the region 0.8$<x<$1 the ratio is above unity which is explained by the effect of the nucleon motion inside the nucleus (Fermi motion)
\item in the interval  0.3$<x<$0.8 the ratio is lower than unity and this suppression is called the EMC effect
\item for 0.1$<x<$ 0.3, the ratio is above unity; the effect is small (of the order of a few percent) and does not reveal a dependence on the atomic mass number. This is the  nuclear anti-shadowing region.
\item for small values of Bjorken $x$,  $x<$ 0.05 - 0.1, the ratio is noticeably suppressed. The suppression increases with  increasing  atomic number A and  decreasing  $x$. This is the  nuclear shadowing region.
\end{itemize}
\begin{figure}
\centering
\includegraphics[width=0.9\textwidth]{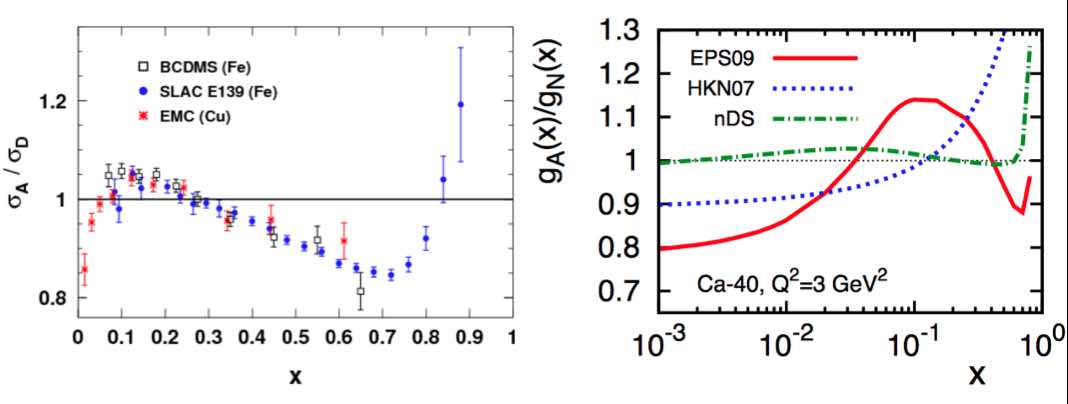}
\caption{Left: measurement of nuclear PDFs modification in Fe and Cu from several experiments. Right: The EPS09 fit to the nuclear PDFs data, together with other models}
\label{NuclearPDFs}
\end{figure}
This modifications influence charm quark production, as it can be deduced when looking at Equation \ref{factorization}. For what concerns the theoretical explanation of  the \textbf{EMC} effect, several phenomenological hypotheses have been proposed \cite{EMCStatus}:
\begin{itemize} 
\item \textbf{$x$ rescaling:}  values of $x$  of partons in nuclei are shifted to lower values with respect to the free nucleon case. This  may  either arise from conventional nuclear binding  leading to a reduced effective mass of partons (and thus lower  effective $x$ values) or from the presence of $\pi$ fields in the intense nuclear potential that enhance the number of degrees of freedom among which the total nuclear momentum is shared
\item \textbf{$Q^2$ rescaling:} the effective  $Q^2$ value with which parton in nuclei are probed in DIS experiments is larger then the measured one, if we assume that quarks are confined into larger volumes when bound into nuclei with respect to the free nucleon case. We can expect confinement scale to be larger if we consider the overlap of nucleon wave functions or the formation of nucleon clusters
\end{itemize}
In the \textbf{shadowing region}, two main explanations have been proposed:
\begin{itemize} 
\item a modification of the interaction of the virtual photon with the atomic nucleus due to fluctuations of the virtual photon into quark-antiquark pairs (i.e. vector mesons having the same quantum numbers as the photon). Such a pair then interacts with the nucleus via the strong interaction. Since the strength of the latter is much larger than the electromagnetic one, the interaction does no longer happen incoherently with all the nucleons in the nucleus but preferentially with those at the front surface. These photonic fluctuations are visible when small Bjorken $x$ values are being proved.  The nucleons being in the ‘shadow’ of the nucleons at the front surface then do not or much less contribute to the interaction, resulting in a reduction of the cross section
\item in the laboratory rest-frame ultrarelativistic heavy-ions are Lorentz-contracted. This enhances the partonic densities and in particular the gluon density is expected to saturate at very low-$x$, leading to  gluon fusion process. The BKFL equations are evolution equations (like DGLAP) that predict the evolution of PDFs going to lower $x$-values. They don't take into account gluon fusion processes and lead to an uncontrolled increase of  gluon densities at low-$x$. The Color Glass Condensate (CGC) is a phenomenological  model for shadowing that assumes the existence of a saturation scale $Q^2_s$ below which the BKFL also take into account recombination effects \cite{CGC}.
\end{itemize}
\begin{figure}
\centering
\includegraphics[width=0.9\textwidth]{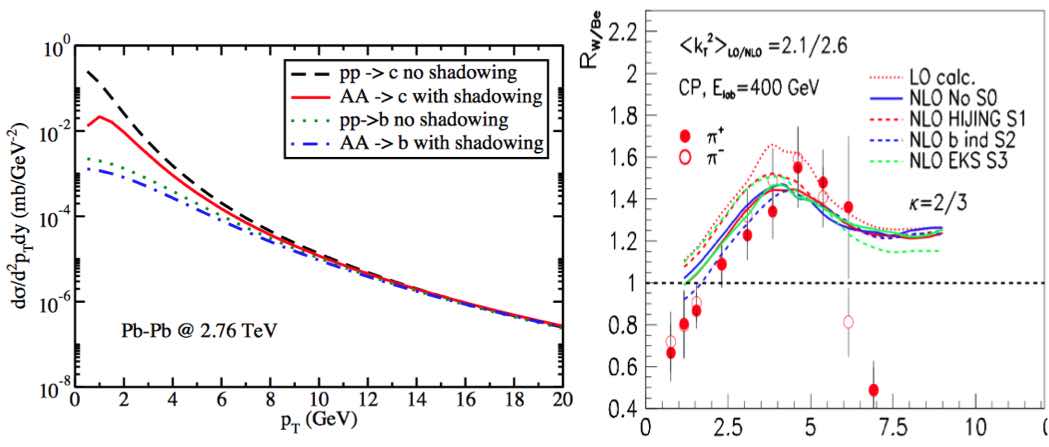}
\caption{Left: NLO pQCD predictions for $c\bar{c}$ and $b\bar{b}$ production in Pb--Pb collisions at \sqnn = 2.76 TeV with and without the EPS09 parametrization  of nuclear PDFs. Right: the ratio of the pion nuclear modification factors obtained colliding 400 GeV/$c$ protons on W and Be target, respectively}
\label{CroninFig}
\end{figure}
QCD global analyses on experimental data allow to parametrize the nuclear PDFs as well as their dependence on $Q^2$ values and atomic mass number.  The \textbf{EPS09}\footnote{Eskola, Paukkunen, Salgado} parametrization of PDFs  is extracted from a NLO pQCD analysis with three different  experimental inputs, deep inelastic $l$--A scattering, Drell-Yan dilepton production in p--A collisions, and inclusive pion production in d--Au and p-p collisions at RHIC \cite{EPS09}. The fit to experimental data is shown in Figure \ref{NuclearPDFs} (right) together with  parametrizations from other analyses (HKN07, more details in  \cite{HKN07}).\\
The influence of these modifications of the PDFs on the trasverse momentum spectrum of heavy-flavour hadrons can be visualized if we consider the collision of two nucleons  at a certain $\sqrt{s}$. Two partons undergoing a hard scattering process will have (neglecting their intrinsic transverse momenta) the four-momenta ($x_1,0,0,x_1$)$\sqrt{s}$ and ($x_2,0,0,-x_2$)$\sqrt{s}$, respectively.  If a heavy-quark pair is created, the square of their invariant mass M$_{Q\bar{Q}}$ is given by
\begin{equation}
M_{Q\bar{Q}} = x_1x_2s
\end{equation}
and their rapidity $y_{Q\bar{Q}}$
\begin{equation}
y_{Q\bar{Q}} = \frac{1}{2}\ln{\frac{E+p_z}{E-p_z}} = \frac{1}{2}\ln{x_1/x_2}
\end{equation}
From these two relations we can derive
\begin{equation}
x_1 = \frac{M_{Q\bar{Q}}}{s_{\rm NN}}e^{+y_{Q\bar{Q}}} \ \ \ \ \ \  x_2 = \frac{M_{Q\bar{Q}}}{s_{\rm NN}}e^{-y_{Q\bar{Q}}} 
\end{equation}
that can be easily \cite{TesiDainese} generalized in case we collide two nuclei of atomic and mass number (A$_1$,Z$_1$) and (A$_2$,Z$_2$) with the same accelerating device. In this case the centre-of-mass energy per nucleon pair is \sqnn = $\sqrt{\frac{Z_1Z_2}{A_1A_2}}\sqrt{s}$ and the following relation holds
\begin{equation}
x_1 = \frac{Z_1}{A_1}\frac{M_{Q\bar{Q}}}{s_{\rm NN}}e^{+y_{Q\bar{Q}}} \ \ \ \ \ \  x_2 = \frac{Z_2}{A_2}\frac{M_{Q\bar{Q}}}{s_{\rm NN}}e^{-y_{Q\bar{Q}}} 
\label{xvalues}
\end{equation}
From these relations we deduce that the $x$ values probed at mid-rapidity ($y$=0) are  of the same order of magnitude for the two quarks taking part to the hard interaction. The value of $x$ decreases with increasing centre-of-mass energy. At LHC energies,  low (compared to the quark mass) transverse momentum charm-quarks probe $x$ values of the order of 10$^{-4}$,  which are in the shadowing region \cite{ProcDainese}.\\
Figure \ref{CroninFig} shows the cross section for  $c\bar{c}$ and $b\bar{b}$ production at mid-rapidity in pp and Pb--Pb collisions at \sqnn=2.76 TeV, respectively. The predictions for Pb--Pb are obtained from NLO pQCD (as the pp ones) and the EPS09 parametrization of PDFs. The effect is more evident for $c\bar{c}$, where at low \pt the cross-section (rescaled by the number of binary collisions) is lower than in pp collisions. This can be understood if we look at Equation \ref{xvalues}:  at low \pt we are probing the shadowing region of Bjorken $x$ , where  nuclear PDFs are suppressed, leading to a reduction of the $c\bar{c}$ production cross-section according to Equation \ref{factorization}. \\
The so-called \textbf{Cronin enhancement} was first observed in p--W fixed-target experiments at Fermilab \cite{CroninEffect}. This effect consists in an enhancement above unity of the nuclear modification factor of produced particles at intermediate \pt (2$\div$6 GeV/$c$). It can be interpreted  in terms of multiple elastic interactions of the projectile parton in the nucleus prior to the hard scattering. These elastic interactions confer to the projectile an extra quantity of transverse momentum before the hard scattering occurs, leading to a broadening of  the momentum distributions for the produced particles.  The results of the measurement of the  $\pi$ nuclear modification factor in p--W collisions relative to the one in p--Be collisions is shown in Figure \ref{CroninFig}(right). The models describing the effect may be classified according to the physical object which is undergoing rescatterings (the projectile hadron or its partons), and to the hardness of the rescattering processes taken into account \cite{CroninTheory}.\\
Another effect that could be present is the energy loss of partons traversing cold nuclear matter.  In the BDMPS model \cite{CNMEL} the  energy loss 
\begin{equation}
\langle\Delta E\rangle \propto \alpha_s C_R \hat{q} L^2
\end{equation}
is proportional to the square of the traversed distance $L^2$,  $C_R$ is the colour charge of the parton projectile (4/3 for quarks, 3 for gluons) and $\hat{q}$ is a parameter depending on the characteristics of the medium. The BDMPS model, which is also valid in case the traversed medium is the QGP, will be discussed  in next Section. Here the previous formula will be used to have some  estimates on the energy loss in cold  medium: for $\hat{q}$= 0.01 GeV$^{2}$/fm, $C_R$=4/3 and L=10 fm, the expected energy loss is $\langle\Delta E\rangle \approx$ 2 GeV ($L$/10fm) \cite{CNMEL}.\\
The study of cold nuclear matter effects  is not straightforward when one looks at A--A results, since any observed modification of transverse momentum spectra with respect to pp collisions could be also due to the creation of the QGP affecting particle production. However cold nuclear matter effects can be studied in p--A collisions, under the assumption that energy densities and temperature achieved in this kind of collisions are not high enough for an extended QGP phase to form. 
\subsection{Hot Nuclear Matter Effects}
As mentioned at the beginning of this section, $c\bar{c}$  pairs are produced in the very early stages of a AA collision and are thus good probes of the high-density medium created. Their interactions with the medium constituents  modifies their dynamical properties, and this modifications can be  studyied by measuring  charmed hadron production as introduced at the beginning of this chapter. In particular in this section we will discuss heavy quarks energy loss in the hot medium, to which both collisional and radiative processes contribute. \\
\subsubsection{Collisional Energy Loss}
High energy partons propagating through a QGP suffer  energy loss via elastic scattering off  particles (quarks and gluons)  in the plasma. This mechanism is very similar in structure  to ionization energy loss of  charged particles in ordinary matter via electromagnetic interactions. The first quantitative calculations of this effect were performed by Bjorken \cite{BjorkenColl} and a similar formalism was carried out by Peigne' in \cite{PeigneEnLoss} for heavy and light quarks with initial energy E >> T,$m_q$, where $m_q$ is the mass of the light constituents of the medium. In the case the incoming particle is a heavy-flavour quark Q of mass M >> T and energy E >> M , we can neglect $s$-channel diagrams with quarks or antiquarks of the same flavour in the plasma (because their aboundance is negligible), and the scattering amplitude  is given by the diagrams shown in Figure \ref{colldiagr}, where elastic scattering occurs with gluons and $u$,$d$ and $s$ quarks. Summing these amplitudes at the leading order gives \cite{PeigneEnLoss}
\begin{equation}
\frac{dE(Q)}{dx} = \frac{4\pi \alpha_s T^2}{3} \bigg[ (1+n_f/6)\ln(\frac{ET}{m_D})+\frac{2}{9}\ln(\frac{ET}{M})+O(1) \bigg]
\end{equation}
where $n_f$ is the number of flavours considered in the scattering diagrams of Figure \ref{colldiagr} and $m_D$ is the Debye screening mass of the plasma $m_D$=4$\pi \alpha_s T^2(1+n_f/6)$.
The analogous calculation for a light quark $q$ (in the limit E >> T) also has to take into account $s$-diagrams for elastic scatterings and gives
\begin{equation}
\frac{dE(q)}{dx} = \frac{4\pi \alpha_s T^2}{3} \bigg[ (1+n_f/6)\ln(\frac{ET}{m_D})+ O(1) \bigg]
\end{equation}
Due to the additional logarithm in Equation 2.12, when E >> M/T the collisional loss is  larger for a heavy than for a light  quark. Taking E $\rightarrow \infty$ and $n_f$ =3,
\begin{equation}
\frac{\frac{dE(Q)}{dx}}{\frac{dE(q)}{dx}} = 1.15
\end{equation}
\begin{figure}
\centering
\includegraphics[width=0.9\textwidth]{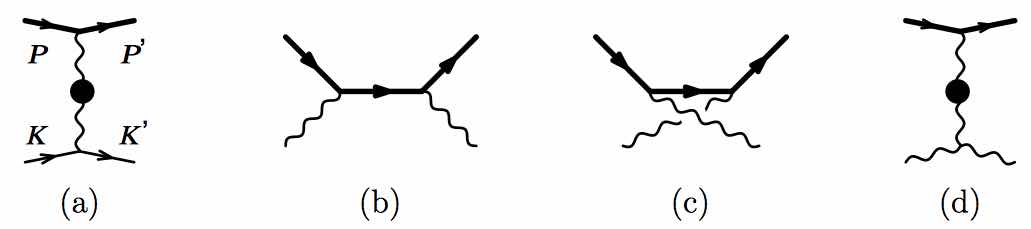}
\caption{Amplitudes for heavy quark elastic scattering in a QGP. A curly line denotes a gluon.}
\label{colldiagr}
\end{figure} 
\subsubsection{Radiative Energy Loss}
Another source of energy loss is gluon radiation. This process is analogous to the QED brehmstrahlung of high-energy electrons traversing ordinary matter,  and is often named "gluonstrahlung". The gluon emission off high energy partons happens via interactions with the medium constituents,  and the radiated gluon of energy $\omega$ undergoes a  number of rescatterings in the medium. This number is  given by  its formation time $t_{form}=\omega/k^{2}_{\bot}$ and its mean free path $\lambda_g$ in the medium. The gluon will thus accumulate a transverse momentum $k^{2}_{\bot}\sim \mu^2 t_{form}/\lambda$ and the total number of scattering centres participating coherently to the emission of the gluon is $N_{coh}=t_{form}/\lambda=\sqrt{\omega/\mu^2\lambda}$. As shown in \cite{DokKha} and \cite{BaierEnLoss} the differential radiated power is 
\begin{equation}
\frac{dW}{dzd\omega}=\frac{\alpha_sC_R}{\pi \omega \lambda} \sqrt{\frac{\mu^2 \lambda}{\omega}}==\frac{\alpha_sC_R}{\pi \omega} \sqrt{\frac{\hat{q}}{\omega}}
\end{equation}
where $C_R$ is the colour charge of the parton projectile (4/3 for quarks, 3 for gluons). The parameter $\hat{q}=\mu^2/\lambda$ is the transport coefficient and is proportional to the density of the scattering centres in medium. Typical estimated values are $\hat{q}\sim$0.01 GeV$^2$/fm for cold nuclear matter and  $\hat{q}\sim$0.2 GeV$^2$/fm for hot nuclear matter. \\
The medium induced radiation vanishes for $\omega$<$\hat{q}L^2$, where  $L$ is the distance traversed in the medium, as the formation time of such gluons starts to exceed the length of the medium.\\
 Integrating over $\omega$ and multipling by $L$  as explained in \cite{BDMPS}, one obtains the average energy loss\footnote{in the approximation where the medium consists of static scattering centers and the the projectile and the outgoing gluon undergo many soft scatterings with the medium (multiple-soft scattering approximation)}
\begin{equation}
\langle\Delta E\rangle \propto  \frac{1}{4}\alpha_s C_R \hat{q} L^2
\end{equation}
For hot QCD matter having temperature T=250 MeV the above formula  for L=10 fm gives  $\langle\Delta E\rangle = 20 \alpha_s $ GeV. If $\alpha_s = \frac{1}{3} \div \frac{1}{2}$ we get  $\langle\Delta E\rangle \sim$ 10 GeV/fm. \\
\begin{figure}[b]
\centering
\includegraphics[width=0.9\textwidth]{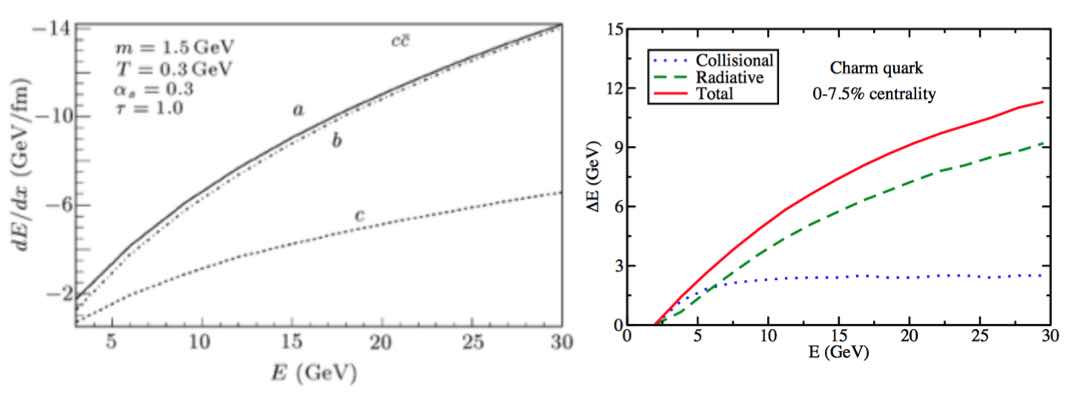}
\caption{Left: dE/dx predictions for $c$ quarks as a function of the initial $c$-quark energy. The solid line is the calculation performed for light quarks, the pin-dotted line is for $c$ quarks without dead-cone effect and the dotted line is for $c$ quarks taking into account dead-cone effect \cite{Cinesi}. Right: energy loss of charm quarks as a function of their initial energies \cite{Cao}.}
\label{collvsrad}
\end{figure} 
 No dependence on the projectile parton mass is present. However as shown in \cite{DokKha}, when we look at gluon emission off heavy quarks the angular emission  probability  is
\begin{equation}
dP = \frac{\alpha_s C_F}{\pi}\frac{d\omega}{\omega}\frac{k^2_{\bot}dk^2_{\bot}}{(k^2_{\bot}+\omega^2\theta^2_0)^2}
\end{equation}
and we see that the emission is suppressed for angles $\theta<\theta_0=M/E$. This effect is known as the dead cone phenomenon. Reference \cite{DokKha} also demonstrates that this effect is relevant up to an energy of the heavy quark E$_{HQ}=M\sqrt{\hat{q}L^3}\sim$ 100 GeV at  LHC energies.   Figure \ref{collvsrad}(left) shows a quantitative estimation of the effect calculated as in \cite{Cinesi}. The parameter values used in the calculation are also shown in the figure, $m$ is the $c$ quark mass value, and the three curves represent the energy loss per unit length  as a function of the initial partonic energy for light quarks and  charm quarks with and without including the dead cone effect. The inclusion of the dead cone effect clearly reduces the energy loss of heavy quarks, and the suppression of gluon emission for  $\theta<\theta_0=M/E$ implies the following relation:
\begin{equation}
\Delta E_{g}>\Delta E_{u,d,s}>\Delta E_{c}>\Delta E_{b}
\end{equation}
and one can expect:
\begin{equation}
R_{\rm AA}(b)>R_{\rm AA}(c)>R_{\rm AA}(g, \ u, \ d, \ s)
\end{equation}
However this relation has to be interpreted  keeping in mind that the inital transverse momentum spectrum of  quarks produced in the collisions plays a role in the final R$_{\rm AA}$ of hadrons that one observes. In fact for a given value of $\Delta E$, the steeper the trasverse momentum distribution of the quark, the lower the resulting nuclear modification factor will be.\\
The right panel of Figure  \ref{collvsrad} shows an estimate of the average energy loss of charm quarks as a function of their initial energy separating the collisional and radiative contributions. These predictions are obtained describing the heavy quark evolution in the QGP medium with a modified version of the  Langevin equation \cite{Cao} that also takes into account gluon radiation:
\begin{equation}
\frac{d\overrightarrow{p}}{dt}=-\nu_D(p)\overrightarrow{p} + \overrightarrow{\xi} +\overrightarrow{f_g}
\label{langevin}
\end{equation}
where the first two terms on the right-hand side are the drag force and the thermal random force that describe the collisional contribution, and the third term describes the recoil force exerted on heavy quarks due to gluon radiation. In this model, collisional energy loss dominates at low energies, while medium-induced gluon radiation dominates at high energies where the dead cone effect gets suppressed. \\
Another effect affecting $c$ quark related observables is the hadronization mechanism in the QGP, which was described in Section 1.4.4.\\
\subsection{Charm quark thermalization}
The elliptic flow,  already introduced in the previous section, can be compared to models that take into account the hydrodynamic evolution of a thermalized medium. The thermalization of light components of the QGP is reached via multiple scatterings and is expected within 1 fm/c from the nuclear interaction (Section 1.3.1). Due to their mass values, heavy quarks thermalization requires a longer time interval to be established and it is even possible that it is not reached at all in the heavy-ion collisions studied so far.\\
In the previous sections the energy loss mechanisms of heavy quarks in the medium was introduced, whose effect is the sum of multiple elastic and inelastic interactions. Since on average the momentum transfer of each interaction is small compared to the heavy quark energy, the Langevin approach of Equation 2.23 can be used. This approach was applied for charm quark in an expanding  finite-size QGP medium in \cite{CaoTherm}. The initial conditions of the hydrodynamic calculation are tuned to describe the hadronic data in the soft sector, such as hadron yields, spectra, rapidity-distributions as well as radial and elliptic flow from RHIC data. The QCD medium experiences a sudden thermalization  at an initial time $\tau_0$ =0.6 fm/c, at which the hydrodynamic evolution begins. The charm quark thermalization criterion is based on a comparison of the  charm quark energy distribution to the medium temperature.  Once the energy and momentum distributions of the charm quark ensemble yield thermal distributions with a temperature corresponding to the medium  temperature,  the selected charm quark ensemble has thermalized in the medium at that given temperature and selected time step. The result of this sudy is that charm quark thermalization  does not occur within the lifetime of the QGP phase, and a quasi-equilibrium state is achieved after about 8 fm/c, exceeding the expected lifetime of the QGP phase at RHIC  (1$\div$4 fm/c, Section 1.3.1). \\
In case of complete thermalization, then the pressure gradients acting on heavy quarks are the same acting on light quarks, and due to the higher mass the  $v_2$ of charm quark should be smaller than the one of light quarks. In case no thermalization is reached, charm quark $v_2$ should in principle be zero at low \pt,   but at high \pt energy loss plays a role, since heavy quarks directed out of plane have to traverse a longer path inside the medium, loose more energy and this reflects in a non-zero $v_2$. \\
\begin{figure}[t]
\centering
\includegraphics[width=0.5\textwidth]{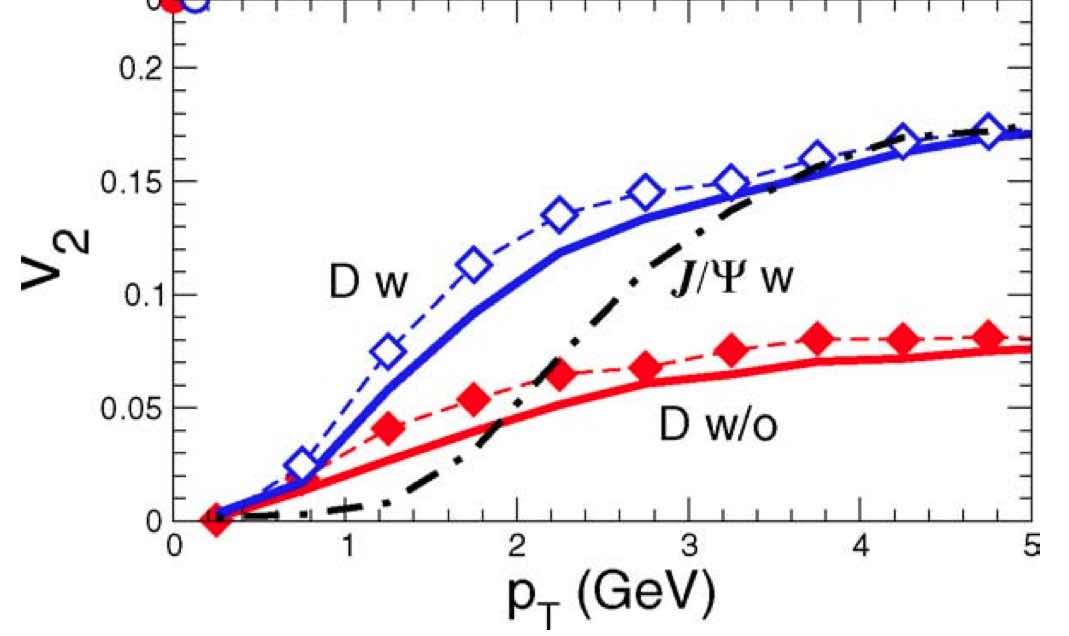}
\caption{Predictions for D meson and heavy-flavour decay electron $v_2$ obtained from a coalescence model with and without considering full thermalization of charm quarks in the medium  \cite{GKR}.}
\label{Greco}
\end{figure} 
The elliptic flow of charmed hadrons is a useful tool to extract information on charm quark thermalization in the hot medium. Greco, Ko and Rapp  \cite{GKR} used a coalescence model to extract predictions on D mesons elliptic flow for Au--Au collisions at \sqnn = 200 GeV in two extreme scenarios:
\begin{itemize}
\item no charm quark rescattering in the medium occurs (spectra are taken from PHYTHIA)
\item complete thermalization including transverse expansion of the bulk matter
\end{itemize}
The results are shown in Figure \ref{Greco}. The red line is obtained in the no thermalization scenario (solid line: D-meson, diamonds: heavy quarks decay electron), the blue one is for thermalization and transverse expansion. The D-meson elliptic flow in the case of complete thermalization deviates from the no-thermalization scenario by about a factor 2 for \pt>1.5 GeV/c. However also if thermalization is not considered, D meson elliptic flow is non-zero even at low \pt,  since D mesons are also composed of light quarks that carry a fraction of $v_2$, indipendently from charm quarks degree of  thermalization. \\
\subsection{Some results from pA collisions}
\label{sub:pAresults}
We will now show some experimental results obtained in pA collisions and compare them to models taking into account the cold nuclear matter effects described so far. The measurements relative to open charm production in p--Pb collisions at LHC energies will not be presented in this chapter, since they are the main topic of this thesis. They will be discussed after the analysis procedure is described in detail. \\
\begin{figure}[t]
\centering
\includegraphics[width=0.9\textwidth]{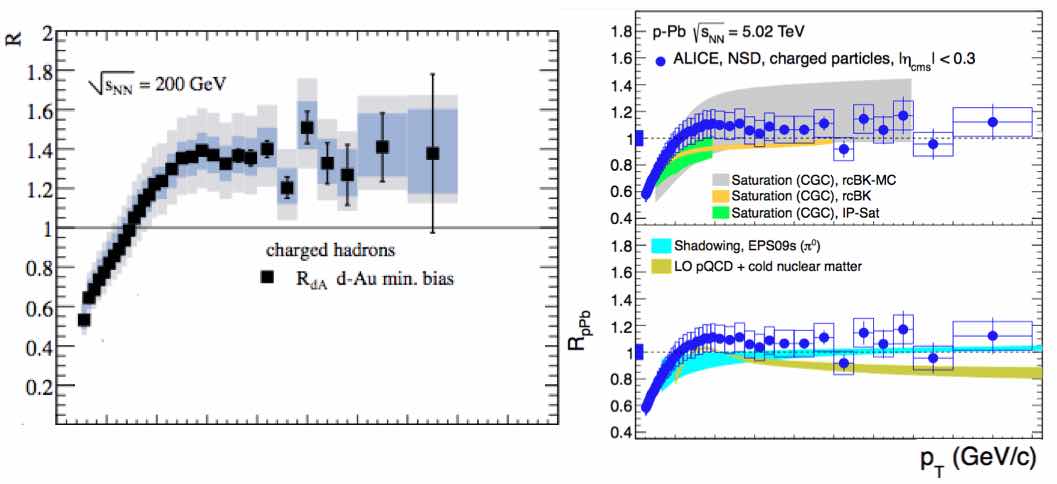}
\caption{Left: charged hadron R$_{\rm dAu}$ measured by the PHENIX Collaboration  in minimum bias d-Au collisions at \sqnn = 200 GeV \cite{PHENIXRdAu}. Right: R$_{\rm pPb}$ of charged hadrons measured by the ALICE Collaboration compared to models in p--Pb collisions at \sqnn = 5.02 TeV \cite{ALICERpPbLight}.}
\label{lhrpa}
\end{figure} 
\begin{figure}[b]
\centering
\includegraphics[width=0.9\textwidth]{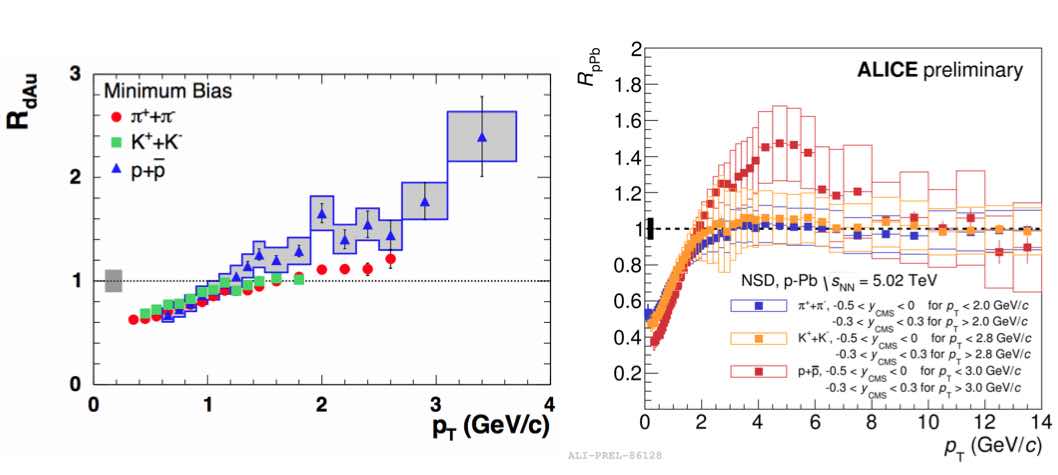}
\caption{Left: nuclear modification factor of identified pions, kaons and protons in minimum bias d--Au collisions at \sqnn=200 GeV measured by the PHENIX Collaboration \cite{poverpPHENIX}. Right: nuclear modification factor of identified pions, kaons and protons in minimum bias p--Pb collisions at \sqnn=5.02 TeV measured by the ALICE Collaboration }
\label{CroninSpecies}
\end{figure} 
We first show some results not directly related to charm quark production. Figure \ref{lhrpa} (left) shows  the nuclear modification factor of charged hadrons in minimum bias d-Au collisions at \sqnn = 200 GeV measured by the PHENIX collaboration  \cite{PHENIXRdAu}. The measurement is performed at mid-rapidity and shows an enhancement in inclusive charged particle production for \pt > 2 GeV/c that can be attributed to Cronin effect. The right panel of the same figure shows the same measurement performed by the ALICE Collaboration at mid-rapidity in p--Pb collisions at \sqnn = 5.02 TeV \cite{ALICERpPbLight}, where a smaller magnitude of the Cronin effect is seen at higher energies.  The top panel of this figure shows a comparison with  CGC models (Section 2.2.1), that are consistent with the data. The bottom panel shows the comparison to two different predictions. NLO pQCD calculations with EPS09 parametrization of the nuclear PDFs  (Section 2.2.1) show good agreement with data.\\
Figure \ref{CroninSpecies} shows the nuclear modification factor of identified pions, kaons and protons in minimum bias d--Au collisions at \sqnn=200 GeV measured by the PHENIX Collaboration \cite{poverpPHENIX}  
(left) and in minimum bias p--Pb collisions at \sqnn=5.02 TeV measured by the ALICE Collaboration (left\footnote{ALICE Physics Preliminary, not yet published in a paper}). PHENIX results show a species dependence in the Cronin effect. The Cronin effect for charged pions is small, while the nuclear enhancement for protons and antiprotons is considerably larger. The kaon measurement has a more limited kinematic range, but the R$_{\rm dA}$ is in agreement with that of the pions at comparable \pt. ALICE results show no enhancement for pions and kaons and a pronounced peak for protons. Models including initial state multiple scattering as well as geometrical shadowing do not predict a species dependent Cronin effect, as initial state parton scattering precedes fragmentation into the different hadronic species. This might point to relevant  final state effects in p--A collisions.\\ 
\begin{figure}[t]
\centering
\includegraphics[width=0.9\textwidth]{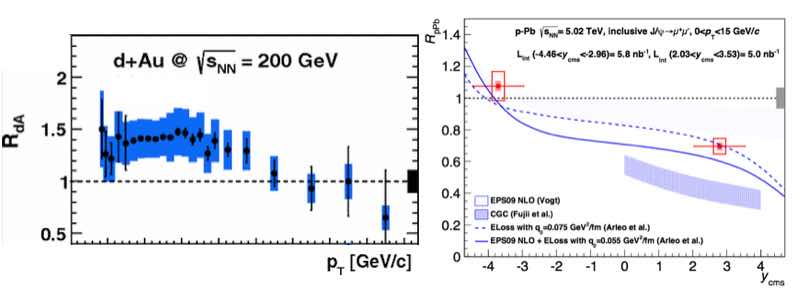}
\caption{Left:  heavy-flavour decay electrons R$_{\rm dAu}$ measured at mid-rapidity by the PHENIX Collaboration  in minimum-bias  d-Au collisions at \sqnn = 200 GeV \cite{PHENIXEl}. Right: $J/\Psi$ R$_{pPb}$ at forward and backward rapidity measured by the ALICE Collaboration in p--Pb collisions at \sqnn = 5.02 TeV \cite{jpsipPb}}
\label{charmpA}
\end{figure} 
Figure \ref{charmpA} (left) shows the nuclear modification factor of electrons from open heavy-flavour decays measured by the PHENIX collaboration in minimum-bias  d-Au collisions at \sqnn = 200 GeV \cite{PHENIXEl}. The R$_{dAu}$ shows an enhancement above unity up to \pt = 5 GeV/c while at higher \pt results are compatible with unity within uncertainties. \\
The right panel of Figure \ref{charmpA} shows the nuclear modification factor of inclusive  $J/\Psi$ in p--Pb collisions at \sqnn = 5.02 TeV measured by the ALICE collaboration \cite{jpsipPb}. The measurement is performed in two different regions of rapidity measured in the centre-of-mass system: positive rapidity (2.03<$y_{\rm cms}$< 3.53) corresponds to the p-going direction, negative rapidity (-4.46<$y_{\rm cms}$< -2.96) corresponds to the Pb-going direction. From Equation \ref{xvalues} we obtain that the $x$ value of the  partons in the Pb nucleus involved in the $c\bar{c}$ pair production is $\sim$ 10$^{-5}$  for the positive $y_{\rm cms}$ interval and $\sim$ 10$^{-2}$ for the negative interval. At forward rapidity the inclusive $J/\Psi$ production is suppressed with respect to the one in binary-scaled pp collisions, whereas it is unchanged at backward rapidity. Four different models are superimposed to data. A CGC-based model overpredicts the observed suppression. A model based on NLO pQCD calculations with EPS09 nuclear PDFs parametrization qualitatively reproduces the results. The last two models also include a partonic energy loss  in cold nuclear matter. The $\hat{q}$ values used are shown in the legend, and both models are able to describe the data. 
\subsection{Some results from AA collisions}
\begin{figure}
\centering
\includegraphics[width=0.9\textwidth]{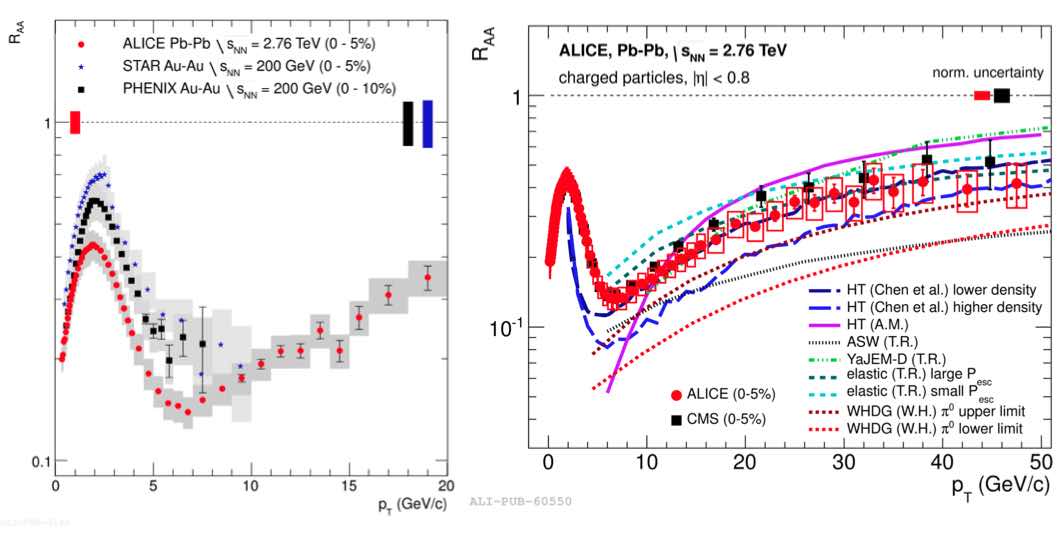}
\caption{Left: charged-hadron nuclear modification factor measured in central Pb--Pb collisions at \sqnn=2.76 TeV by the ALICE Collaboration \cite{ALICEChargedRaa}.  The results of charged-hadron nuclear modification factor measured in central Au--Au collisions at  \sqnn=200 GeV by the PHENIX and STAR \cite{STARCharged} Collaborations at RHIC are  superimposed.  Right: charged-hadron nuclear modification factor measured in central Pb--Pb collisions at \sqnn=2.76 TeV by the ALICE Collaboration \cite{ALICEChargedRaa} and CMS Collaboration \cite{CMSCharged}. Predictions of different theoretical calculations based on in-medium energy loss are superimposed.}
\label{ChargedRaa}
\end{figure} 
The nuclear modification factor defined in Equations 2.9 and 2.10 is  sensitive to possible energy loss mechanisms that could play a major role in A--A collisions. Before discussing results relative to heavy-flavour, results on charged hadron R$_{AA}$ will be discussed.\\
The left panel of Figure \ref{ChargedRaa} shows charged-hadron nuclear modification factor measured in central Pb--Pb collisions at \sqnn=2.76 TeV by the ALICE Collaboration \cite{ALICEChargedRaa}. The results of charged-hadron nuclear modification factor measured in central Au--Au collisions at  \sqnn=200 GeV by the PHENIX and STAR \cite{STARCharged} Collaborations at RHIC are  superimposed. The peak structure centred at \pt$ \approx$ 2 GeV/c has been interpreted in terms of different hadronization mechanisms (quark coalescence) in AA collisions with respect to pp collisions, or as a consequence of radial flow. At higher \pt, in which the effects of in-medium energy loss start to be dominant, we can summarize the results as follows:
\begin{itemize}
\item STAR and PHENIX results show that hadron production in 6<\pt<10 GeV/c is suppressed by a factor 4$\div$5 with respect to pp collisions. This suppression cannot be reproduced by models taking into account Cronin enhancement and shadowing alone, but it is reproduced if partonic energy loss in dense matter is included \cite{STARCharged}.
\item ALICE results show that R$_{AA}$ reaches a minimum of R$_{AA}\approx$ 0.14 at \pt = 6-7 GeV/c. A significant rise of R$_{\rm AA}$ by about a factor of two is observed for  7<\pt<20 GeV/c
\end{itemize}
At the larger LHC energy the density of the medium is expected to be higher than at RHIC, leading to a larger energy loss of high \pt partons. On the other hand, the less steeply falling spectrum at the higher energy would lead to a smaller suppression in the \pt spectrum of charged particles, for a given magnitude of partonic energy loss. The measurement of Figure \ref{ChargedRaa} show that the nuclear modification factor measured by ALICE in 6<\pt<7 GeV/c is smaller than at RHIC. This suggests an enhanced energy loss at LHC and therefore a denser medium. It can also indicate a more relevant fraction of high-\pt hadrons which originate from fragmentation of gluons, that according to Equation 2.21 loose more energy in the medium.\\
Figure \ref{ChargedRaa} (right) shows the same ALICE results of the left panel together with the corresponding CMS results \cite{CMSCharged}. CMS result is fully in agreement with the ALICE measurement within the uncertainties. Predictions of different theoretical calculations based on in-medium energy loss are superimposed. All selected models use RHIC data to calibrate the medium density. A variety of energy loss formalisms is used. An increase of R$_{\rm AA}$ due to a decrease of the relative energy loss with increasing \pt is seen for all the models.\\
We now switch to the heavy-flavour sector. 
\begin{figure}
\centering
\includegraphics[width=0.9\textwidth]{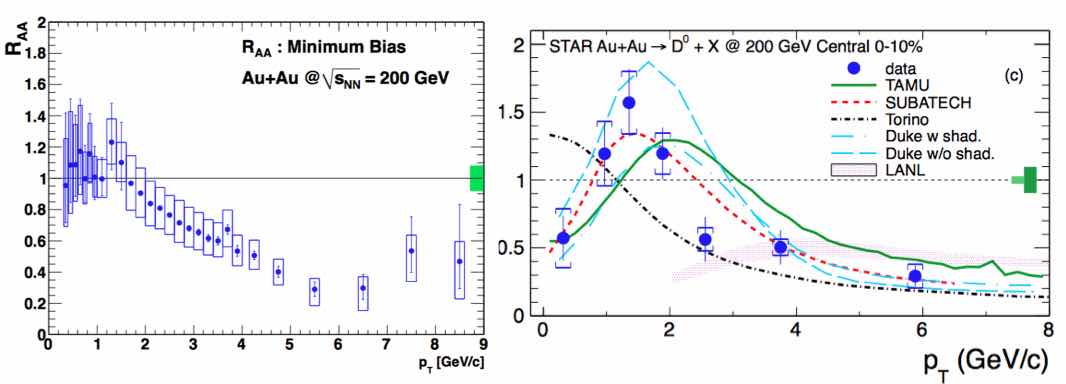}
\caption{Left: heavy-flavour decay electron R$_{\rm AA}$  measured by PHENIX from  Au--Au collisions at \sqnn = 200 GeV in the  0-10\% centrality class \cite{PHENIXAuAu}. Right: D$^0$  R$_{\rm AA}$ measured by STAR from  Au--Au collisions at \sqnn = 200 GeV in the 0-10\% centrality class  \cite{STARD0}. The SUBATECH curve correspond to the MC@sHQ+EPOS model. The TORINO curve corresponds to POWLANG. The DUKE curve corresponds to Cao,Qinn,Bass. The LANL curve corresponds to  Vitev.   }
\label{RHICAU}
\end{figure} 
Figure \ref{RHICAU} shows two measurements performed at RHIC in Au--Au collisions at \sqnn = 200 GeV. The left panel shows heavy-flavour decay electron R$_{\rm AA}$  measured by PHENIX Collaboration in the  0-10\% centrality class \cite{PHENIXAuAu}. The measurement is inclusive (i.e. both charm and bottom hadrons decay electrons are included) and shows that heavy-flavour electron production is compatible to binary scaling of pp cross-section up to \pt$\sim$ 2 GeV/c, while above this \pt value a suppression is clearly seen, reaching a value $\sim$ 4 at \pt = 6 GeV/c.  \\
The right panel of Figure \ref{RHICAU} shows the \pt differential nuclear modification factor of D$^0$ mesons measured by STAR in the 0-10\% centrality class \cite{STARD0}.  A suppression of D$^0$ meson yield is observed for \pt > 2.5 GeV/c, and different predictions are superimposed\footnote{they are the same theoretical models of the ALICE measurement in Figure \ref{RaaDmeson}, but they are tagged in a different way in the figure - as explained in the caption}. POWLANG calculation misses the intermediate-\pt enhancement structure  for \pt < 3 GeV/c, indicating that a coalescence type hadronization mechanism  is important in modeling charm-quark hadronization at low and intermediate \pt.\\
\begin{figure}
\includegraphics[width=0.9\textwidth]{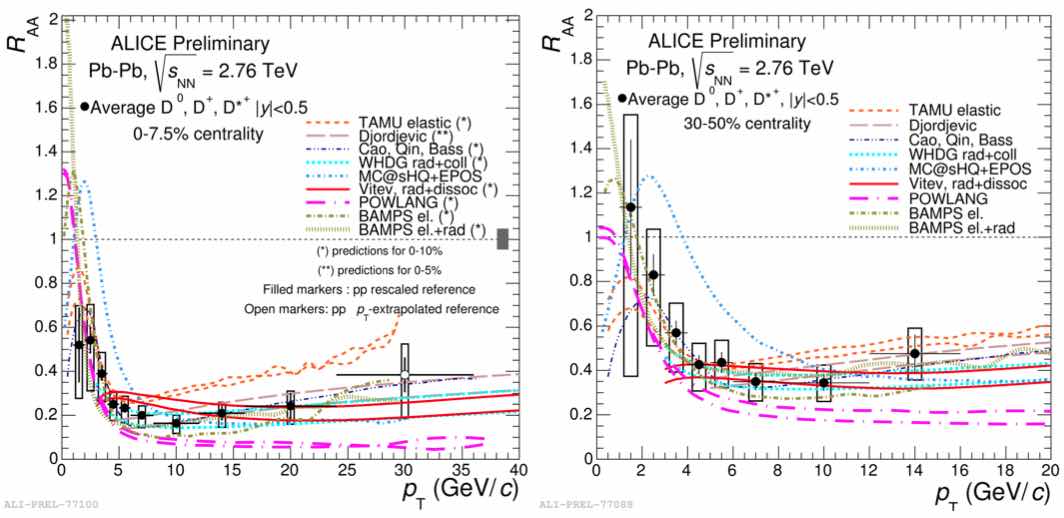}
\caption{D-meson R$_{\rm AA}$ (average of D$^{0}$, D$^{+}$ and D$^{*+}$) measured by the ALICE Collaboration in Pb--Pb collisions  at \sqnn = 2.76 TeV in the 0-7.5\% (left) and 30-50\% (right) centrality classes. }
\label{RaaDmeson}
\end{figure} 
 Figure \ref{RaaDmeson} (left) shows the \pt differential prompt D-meson nuclear modification factor measured by the ALICE collaboration in the 0-7.5\% centrality class in Pb--Pb collisions at \sqnn = 2.76 TeV. This measurement is obtained from the weighted average of D$^0$, D$^+$ and D$^{*+}$ results. It shows a strong (factor 4$\div$5) suppression of D mesons trasverse momentum spectra with respect to pp collisions  for \pt > 5 GeV/c, comparable to the one seen for charged particles at the same energy in Figure  \ref{ChargedRaa}. This big suppression indicates a large energy loss of charm quarks, similar to the one of light quarks and gluon and in contrast to what would be expected considering the radiative energy loss discussed  in Section 2.2.2.2.\\
Several theoretical model are superimposed in the plot, briefly:
\begin{itemize}
\item \textbf{TAMU}  \cite{TAMU} \cite{TAMU2} predictions take into account elastic scattering with medium constituents. In the hypothesis that the typical momentum transfer in a single elastic interaction with a light partons is small compared to the heavy quark energy,  the diffusion process can be simulated using relativistic Fokker-Planck dynamics for elastic scattering in a hydrodynamic background
\begin{equation}
d\textrm{x}=\frac{\textbf{p}}{E}dt  \  \ \ \ \ d\textbf{p}=-\Gamma(p)\textbf{p}dt+\sqrt{2\Gamma(\textbf{p+dp})E(\textbf{p+dp})Tdt}\rho
\end{equation}
where \textbf{x}, \textbf{p} and $E$ are the position, momentum and energy of the heavy quark, $\Gamma$ is the drag coefficient, and $\rho$ a noise term. The evolution is constrained by light-flavour hadron spectra and elliptic flow data.  The heavy-quark transport coefficients in the medium are obtained from non-perturbative $T$-matrix interactions which build up resonant correlations close to the transition temperature. The latter also form the basis for hadronization of heavy quarks into heavy-flavor mesons via recombination with light quarks from the medium.  Hadronization  via  coalescence (fragmentation) is found to dominate at low  (high) \pt.
\item  \textbf{Djordjevic}  predictions  \cite{Djordjevic} compute an energy loss probability  which includes both radiative and collisional processes in a realistic finite size dynamical QCD medium.  The quenched spectra of partons and hadrons are calculated from the following convolution
\begin{equation}
\frac{E_fd^3\sigma(H_Q)}{dp^3_f}=\frac{E_id^3\sigma(Q)}{dp^3_i}\otimes P(E_i\rightarrow E_f)\otimes D(Q\rightarrow H_Q)
\end{equation}
where $Q$ denotes quarks and gluons, $H_Q$ heavy-flavour hadrons, $P(E_i\rightarrow E_f)$ is the energy loss probability and $D(Q\rightarrow H_Q)$ is the fragmentation function. The initial heavy quark spectrum $\frac{E_id^3\sigma(Q)}{dp^3_i}$ is computed at next-to-leading order with FONLL, using CTEQ5M PDFs with collinear factorization. The $P(E_i\rightarrow E_f)$ term takes into account path-length fluctuations, multigluon fluctuations (i.e. the fact that energy loss is a distribution), magnetic mass of QGP constituents and running coupling in the assumption of large $E_f$, so that the Eikonal approximation can be employed. This model does not simulate the hydrodynamic evolution of heavy-quark in the medium.
\item \textbf{Cao,Qinn,Bass} is based on a modified version of the Langevin equation that also takes into account radiative energy loss,  and was discussed in Section 2.2.2.2 (Equation \ref{langevin}).
\item \textbf{WHDG}  approach \cite{WHDG} is based on Equation 2.27, with the initial heavy quark spectrum computed as in Djordjevic predictions. A realistic collision geometry based on the Glauber
model is used, without hydrodynamical expansion, and anisotropy results from path-length dependent energy loss.   The $P(E_i\rightarrow E_f)$ term  takes into account path length and energy loss fluctuations. The collisional contribution to energy loss for a parton with Casimir factor $C_R$ in an ideal ultrarelativistic QGP with $n_f$ active flavours and temperature T is
\begin{equation}
\frac{dE^{el}}{dx}=C_R\pi \alpha^2_sT^2(1+n_f/6)\log{B_c}
\end{equation} 
where the  parameter $B_c$ takes into account minimum and maximum momentum transfers.The radiative energy loss is given by the DGLV formula that can be found in \cite{DGLV}. 
\item \textbf{MC@sHQ+EPOS}  \cite{MCat} is based on Monte-Carlo propagation of heavy quarks in the medium. The medium fluid dynamical expansion is based on the EPOS model, including initial conditions obtained from a flux tube approach \cite{FluxTube} and according to the equations of ideal hydrodynamics within a 3 + 1 D fluid dynamically expanding plasma, namely the local energy-momentum conservation and the conservation of  baryon number, strangeness, and electric charge.  The elastic cross sections are obtained from pQCD matrix elements in Born approximation including a running coupling $\alpha_s$.   The incoherent emission of bremsstrahlung gluons is included via matrix elements from scalar QCD, while the coherent emission of gluons, i.e., the QCD generalization of the Landau-Pomeranchuk-Migdal (LPM\footnote{A high energy particle undergoing multiple soft scatterings from a medium will experience interference effects between adjacent scattering sites. As the longitudinal momentum transfer gets small the particles wavelength will increase, and if the wavelength becomes longer than the mean free path in the medium (the average distance between scattering sites) then the scatterings can no longer be treated as independent events. This is the LPM effect. The Bethe–Heitler spectrum for multiple scattering induced radiation (Equation 2.20) assumes that the scatterings are independent, the quantum interference between successive scatterings caused by the LPM effect leads to suppression of the radiation spectrum relative to that predicted by Bethe–Heitler.}) effect is included via an effective reduction of the spectrum. The hadronization of the heavy quarks  takes place at the transition temperature  via coalescence (fragmentation) for low (high) \pt heavy quarks.
\item \textbf{Vitev} \cite{Vitev} is a perturbative QCD description of heavy flavor dynamics in the thermal medium that combines D and B meson in-medium formation and dissociation with parton-level charm and beauty quark radiative energy loss. Heavy-quarks initial distributions are taken from FONLL predictions and the model also takes into account bound state formation above the deconfinement temperature T$_c$ 
\item \textbf{POWLANG} \cite{POWLANG} uses the relativistic  Langevin equation,  
\begin{equation}
\frac{d\overrightarrow{p}}{dt}=-\nu_D(p)\overrightarrow{p} + \overrightarrow{\xi} 
\end{equation}
Only collisional energy loss is considered.
A sample of c and b quarks is generated using POWHEG \cite{POWHEG}, a code which implements pQCD at NLO accuracy, and CTEQ6M PDFs with EPS09 nuclear corrections. Heavy quarks are initially distributed in the transverse plane according to the nuclear overlap function corresponding to the selected impact parameter \textbf{b}. At a given equilibrium time $\tau_0$ an iterative procedure is started to follow the stochastic evolution of the heavy quarks in the plasma until hadronization: the Langevin transport coefficients are evaluated at each step according to the local 4-velocity and temperature $T(x)$ of the expanding background medium. Hadronization only happens via fragmentation. 
\item \textbf{BAMPS}   \cite{BAMPS} is a 3+1 dimensional partonic transport model that solves the Boltzmann equation
\begin{equation}
\bigg( \frac{\partial}{\partial t}+ \frac{\textbf{p$_{i}$}}{E_i} \frac{\partial}{\partial r} \bigg) f_i(\textbf{r}, \textbf{p$_{i}$},t)= C_i^{2\rightarrow 2}+C_i^{2\rightarrow 3}+...
\end{equation}
 for on-shell partons and allows elastic interactions among  gluons, light quarks and heavy quarks. $C_i$ are the relevant collision integrals, and  $f_i(\textbf{r}, \textbf{p$_{i}$},t)$ the one-particle distribution function of partons.  Heavy quarks, in particular, interact with the rest of the medium via binary scatterings with a running coupling and a Debye screening length. The lack of radiative processes in the heavy flavor sector is accounted for by scaling the binary cross section with a phenomenological factor K = 3.5, which describes well the elliptic flow $v_2$ and nuclear modification factor $R_{\rm AA}$ at RHIC.
\end{itemize}
The right panel of Figure \ref{RaaDmeson} shows the same measurement in the 30-50\% centrality class. The modification of D meson trasverse momentum spectra with respect to pp collisions is still present but of smaller magnitude, as expected from the smaller size of the system created.\\
\begin{figure}
\centering
\includegraphics[width=0.9\textwidth]{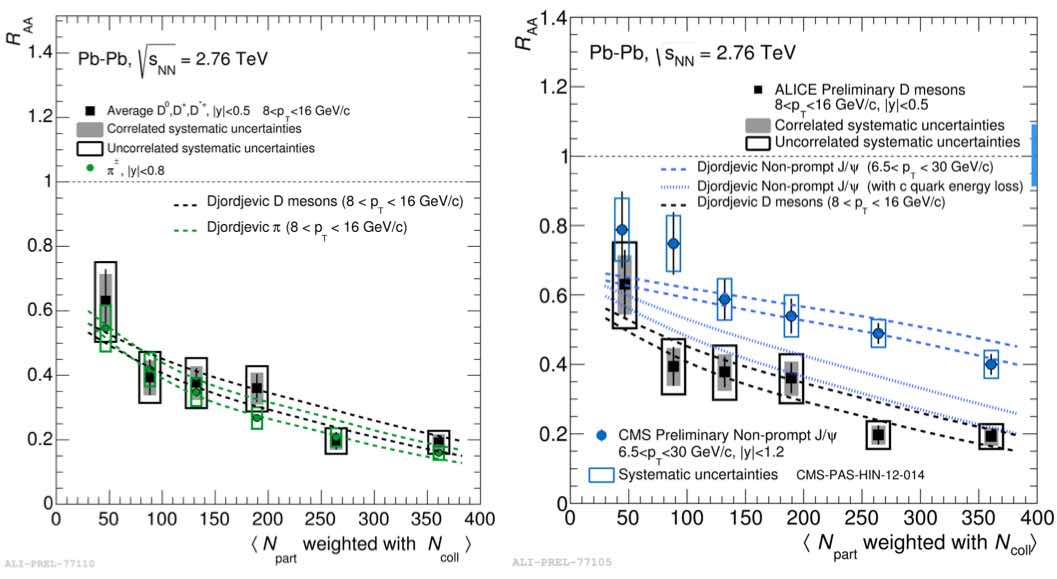}
\caption{ Comparison of D-meson R$_{\rm AA}$ as a function of average N$_{\rm part}$ to the one of  $\pi$ (left) and  non-prompt $J/\Psi$ (right) measured by the ALICE and CMS Collaborations. Predictions from \cite{Djordjevic} are superimposed, in particular in the right plot the non-prompt $J/\Psi$ R$_{\rm AA}$ is computed considering $c$ (blue dotted line) and $b$ (blue dashed line) energy loss. }
\label{RAAComp}
\end{figure} 
Figure \ref{RAAComp} (left) shows a comparison of D-meson and charged $\pi$ R$_{\rm AA}$ as a function of centrality expressed in terms of number of participant nucleons.  D-meson and  charged $\pi$ R$_{\rm AA}$ are measured in the 8<\pt<16 GeV/c range. The two measurements are compatible within uncertainties and with Djordjevic \cite{Djordjevic} model that takes into account energy loss hierarchy due to color charge and quark mass.  The calculation closely describes the similarity of the D meson and charged $\pi$ R$_{\rm AA}$ over the entire centrality range. In this calculation the colour-charge dependence of energy loss introduces a sizeable difference in the suppression of the gluon and c quark production. The softer fragmentation and \pt spectrum of gluons with respect to those of c quarks, together with the increase of the parton-level R$_{\rm AA}$ with increasing \pt, lead to a compensation effect that results in very similar R$_{\rm AA}$  for D mesons and pions.  \\
Figure \ref{RAAComp} (right)  shows a comparison of D-meson and non-prompt $J/\Psi$ (measured by CMS \cite{nonpromptCMS}) R$_{\rm AA}$ as a function of centrality expressed in terms of number of participant nucleons. D-meson  R$_{\rm AA}$ is measured in the 8<\pt<16 GeV/c range, in the 6.5<\pt<30 GeV/c range for non-prompt $J/\Psi$, which means that the \pt of the B meson  originating the non-prompt $J/\Psi$ and that  of the D mesons are comparable. Djordjevic model  \cite{Djordjevic} predicts a difference between the D-meson and non-prompt $J/\Psi$ R$_{\rm AA}$ similar to that observed in the data. In order to study the origin of this large difference in the calculation, the result for a test case with the energy loss of b quarks calculated using the c quark mass was considered. In this case the R$_{\rm AA}$ of non-prompt $J/\Psi$ was found to be quite close to that of D mesons, indicating that the difference between D-meson and non-prompt $J/\Psi$ R$_{\rm AA}$ observed in Djordjevic model   is mainly due to the quark mass dependence of parton energy loss. \\
\vskip 0.5 em 
To conclude this section we will discuss some measurement on the elliptic flow of heavy-flavour decay electrons and D mesons, which are sensitive to the degree of thermalization of charm quarks in the medium: 
\begin{itemize}
\item PHENIX Collaboration measured the heavy-flavour decay electrons \pt differential $v_2$ in minimum bias Au--Au collisions at \sqnn = 200 GeV \cite{PHENIXAuAu}. The resulting $v_2$ is large ($v_2 \sim$ 0.05$\div$0.1, not compatible with zero within uncertainties) in the range 0.5<\pt<3 GeV/c. This result is shown in the left panel of Figure \ref{ALICEDv}.
\item  the ALICE Collaboration measured the  D-meson \pt differential  $v_2$ in Pb--Pb collisions at \sqnn = 2.76 TeV for the 0-10\%, 10-30\% and 30-50\% centrality class \cite{ALICEv2}. A positive $v_2$ is observed in mid-central collisions (30-50\% centrality class) with a mean value of 0.204$^{+0.099}_{-0.036}$(tot. unc.) in the interval 2<\pt<6 GeV/c, which decreases towards more central collisions. Figure \ref{ALICEDv} (right) shows the \pt differential $v_2$  in the 30-50\% centrality class. This measurement is obtained from the weighted average of D$^0$, D$^+$ and D$^{*+}$ results.  The measured $v_2$ of D mesons is comparable in magnitude to that of light-flavour hadrons shown in previous chapter. \\
\end{itemize}
The right panel of Figure  \ref{ALICEDv} also shows the theoretical predictions of $v_2$ based on the same models used in Figure  \ref{RaaDmeson} for  $R_{\rm AA}$. 
\begin{figure}
\centering
\includegraphics[width=0.9\textwidth]{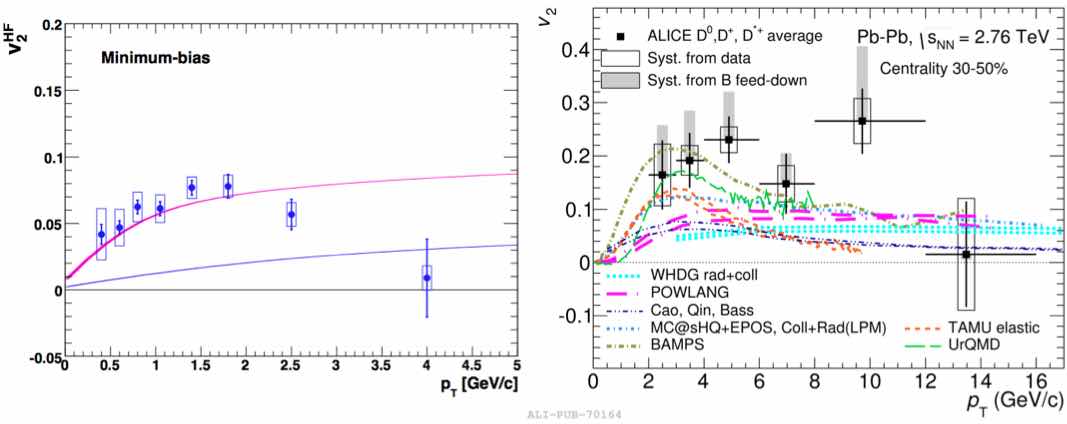}
\caption{ Left: \pt differential $v_2$ in minimum bias Au--Au collisions at \sqnn = 200 GeV \cite{PHENIXAuAu}. Right: D-meson \pt differential  $v_2$ in Pb--Pb collisions at \sqnn = 2.76 TeV for the 0-10\%, 10-30\% and 30-50\% centrality class \cite{ALICEv2}. }
\label{ALICEDv}
\end{figure} 
Models including both charm quark energy loss in a geometrically anisotropic medium and mechanisms that transfer to charm quarks the elliptic flow induced during the system expansion  qualitatively describe the observed anisotropy.  Models that do not include a collective expansion of the medium or lack a contribution to the hadronization of charm quarks from recombination with light quarks from the medium predict in general a smaller anisotropy than observed in the data. \\The comparison for R$_{\rm AA}$ and $v_2$ shows that it is challenging to simultaneously describe the large suppression of D mesons in central collisions and their anisotropy in non-central collisions. In general, the models that are best in describing R$_{\rm AA}$ tend to underestimate  $v_2$ and the models that describe $v_2$ tend to underestimate the measured R$_{\rm AA}$  at high \pt. In conclusion, it is challenging for models to describe simultaneously the large suppression of D mesons in central collisions and their anisotropy in non-central collisions. The higher statistic expected for LHC Run 2 will help restricting the experimental  uncertainties allowing a better discrimination between models. 
 
\chapter{p--Pb collisions with the ALICE detector} 
\lhead{Chapter 3. \emph{p--Pb collisions with the ALICE detector}} 
\begin{figure}[b]
\centering
 \includegraphics[width=0.99\textwidth]{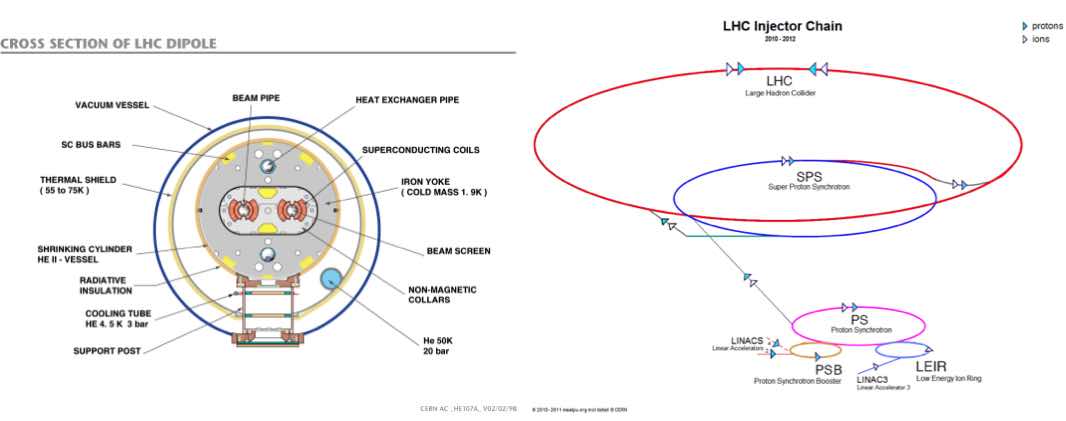}
\caption{Left: Section of the LHC two-in-one accelerating dipole. Right: LHC injection scheme}
\label{magnet}
\end{figure}
Even if  it was not mentioned in the inital LHC Design Report (2004), the ALICE Collaboration asked for a p--Pb data taking period in 2011, and after some feasability tests (2011) and pilot runs (2012), p--Pb runs took place at the beginning of 2013.\\
Since the LHC is mainly conceived as a particle-particle collider, the two beams can not be accelerated in a single ring. The super-conducting magnet design  adopted  is  a "two in one" dipole as in Figure \ref{magnet} (left). The identical bending field in
both apertures of the dipole  fixes the relation between the momenta of the beams in the two rings
\begin{equation}
{\Large p}_{\rm Pb} = Z p_{\rm p}
\end{equation}
As seen in the previous chapter, circulating particles with charges $Z_1$ and $Z_2$ in the rings with the  magnetic field set to accelerate protons with momentum $p_{\rm p}$ will result in:
\begin{equation}
\sqnneq = \sqrt{s_p}\sqrt{\frac{Z_1Z_2}{A_1A_2}} \ \ \ \  y_{\rm NN} = \frac{1}{2}\log\frac{Z_1A_2}{Z_2A_1}
\end{equation}
For the proton beam energy of 4 TeV as in the 2012 pp runs one gets  \sqnn = 5.02 TeV. The centre-of-mass of the p--Pb system in the laboratory frame has a rapidity $y_{\rm cms}$ = 0.465 in the proton direction.\\ The acceleration chain (Figure \ref{magnet} - right) at the  LHC begins with a linear accelerator (LINAC2) accelerating proton beams to 50 MeV, followed by the Proton Synchrotron Booster (PSB) that further accelerates them to 1.4 GeV. Proton beams are then extracted from PSB and sent to the  Proton Synchrotron (PS) reaching an energy of 25
GeV and finally to the Super Proton Synchrotron (SPS) that accelerates them to 450 GeV. At this point protons are ready to be injected to the LHC where they are further accelerated and finally collided. \\
The accelerating chain differs for Pb ions in the  first two steps, since Pb ions are first accelerated in the LINAC3 at 4.5 MeV per nucleon and then in the Low Energy Ion Ring (LEAR) which delivers them  to the PS. The complete ionization of Pb atoms is more complex than $^1$H ionization. The process begins before the injection in the  LINAC3, where an electron-cyclotron resonance source ionize them to Pb$^{28+}$. Full ionization is then achieved sending the beam through two different copper stripper foils, one after LINAC3 and one after the PS.
\begin{figure}[b]
\centering
 \includegraphics[width=0.99\textwidth]{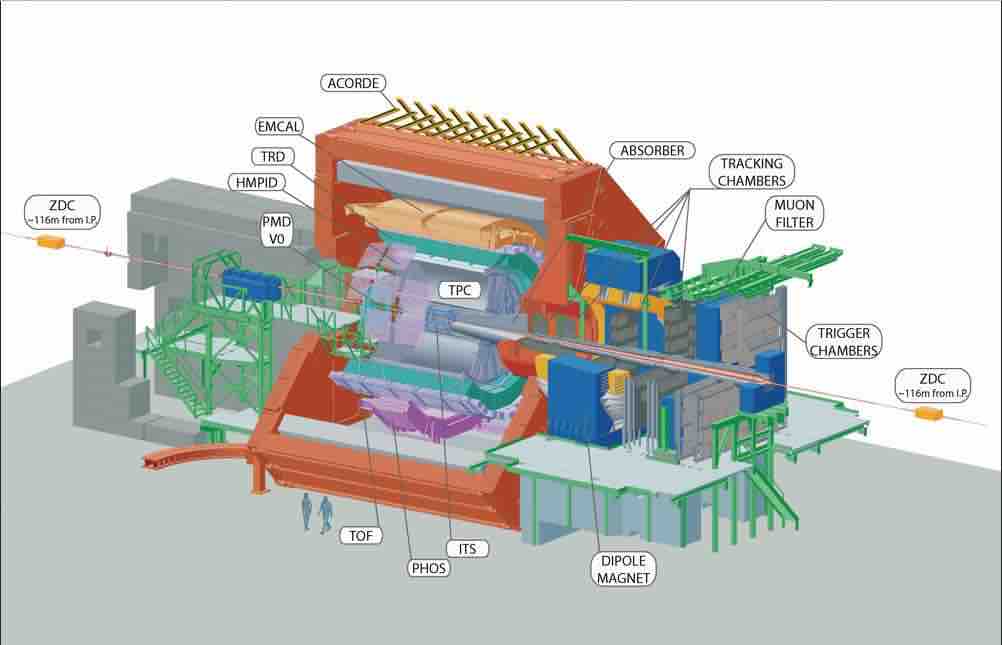}
\caption{ALICE Detector}
\label{ALICEOverview}
\end{figure} 
Once p and Pb beams reach the LHC, they are further accelerated and finally collided in the four interaction points.  \\
The ALICE detector is located at Interaction Point 2 of the LHC and an overview of its components is shown in Figure \ref{ALICEOverview}. ALICE consists of a central barrel, a forward muon arm and several detectors for trigger and event characterization. \\
The direction of the p and Pb beams where reversed during the 2013 data taking to allow the muon arm to study p--Pb collisions in two different rapidity regions, allowing measurements like the one shown in Figure \ref{charmpA} of Chapter2\footnote{During the first part of 2013 data taking, the proton beam was directed towards the muon spectrometer: this will be referred as the p--Pb data taking. Then the beam configuration was reversed with the Pb beam directed towards the muon spectrometer: this is the Pb--p configuration.}.  The muon arm  consists of an absorber, a large dipole magnet and tracking stations; a second muon filter at the end of the spectrometer and four planes of RPC which are are used for muon identification and triggering. This part of the ALICE detector is not used for the analysis presented in this thesis and won't be discussed further.\\
In this chapter I will describe the central barrel detectors used in the reconstruction of D-meson hadronic decays, concentrating in particular on the vertexing and  tracking  capabilites of to the ALICE detector. The determination of centrality in both Pb--Pb and p--Pb collisions will also be discussed. \\ 
 \section{ALICE Detector}
The ALICE detector has been designed as a high-energy nuclear experiment able to analyze the complex systems emerging from Pb--Pb collisions at several TeV of  centre-of-mass energy. This requires good performance in:
\begin{itemize}
 \item measuring and identifying hadrons, electrons and photons in the central rapidity region
 \item tracking  particles in broad range of transverse momentum (0.1-100 GeV/c)
 \item reconstructing primary and secondary  vertices with a good spatial resolution 
 \item operate in a high-multiplicity environment 
\end{itemize}
The  multiplicity  in the 5\% most central  Pb-Pb events  was measured during  the first Pb-Pb runs in November 2010 \cite{mult}
\begin{equation}
 \frac{dN_{ch}}{d\eta} = 1584 \pm 4(stat) \pm 76(syst)
\end{equation}
 from which we can extract an average multiplicity per participant pair of 8.3, a factor of 2.2 larger than that found at 
highest RHIC energies (200 GeV per nucleon pair).\\
Having a good tracking performance in such a high multiplicity environment requires the use of high granularity detectors which can only operate at a  reduced luminosity
in pp interactions.  Luminosity reduction is obtained by beam defocusing (the bunches are broadened near the interaction point and refocused after it)
and beam displacing (the two colliding bunches do not collide in their central regions but in their tails). \\After three years of operation at the LHC, experience has shown that the maximum pp interaction rate at which all ALICE detectors can be safely operated is around 700 kHz, corresponding to  a luminosity of  $\mathscr{L}$ = 10$^{29}$ s$^{-1}$cm$^{-2}$ for minimum bias triggers \cite{ALICEPerfPaper}.  In Pb--Pb collisions the highest luminosity was achieved in 2011 ($\mathscr{L}$ = 10$^{26}$ s$^{-1}$cm$^{-2}$) while the maximum manageable interaction rate for p–Pb collisions was 200 kHz, roughly corresponding to a luminosity  $\mathscr{L}$ = 10$^{29}$ s$^{-1}$cm$^{-2}$.\\
 The value of the magnetic field $B$ (0.5 T) has been choosen 
in order to find a compromise between two requirements: tracking of  low \pt
hadrons and  momentum resolution at high \pt. In fact a 200 MeV/c pion has a curvature radius $R\approx\frac{70}{B}$cm, expressing the magnetic field in Tesla, which for CMS magnetic field (4 T) would result in $R\approx$ 18 cm. The lower magnetic field value of 0.5 T allows  the same pion to reach the ALICE Time Projection Chamber (Section 3.1.2), which is located at 90 cm from the beam pipe. However the lower the magnetic field, the smaller the curvature of high \pt tracks, which results in a worse \pt resolution. \\
The next paragraphs are dedicated to the detectors used in the analysis presented in this thesis. For information on the other detectors more details can be found in \cite{ALICEJinst}  can be consulted.
\subsection{Inner Tracking System (ITS)}
\label{sub:its}
The ITS surrounds the beam pipe and consists of six cylindrical 
layers of silicon detectors, located
at radial distances from the beam line  between 4 and 43 cm. The whole detector covers a pseudorapidity range of |$\eta$|<0.9 for all vertices located within 5.3 cm along the beam axis from the detector center, but the  innermost layers have a wider rapidity range that can contribute to multiplicity 
measurements.\\
The main tasks of this detector are:
 \begin{itemize}
  \item  localize primary and secondary vertices 
 \item  track and identify particles with momentum below 200 MeV/c 
\item  contribute to the Minimum Bias (MB) trigger
 \item  improve momentum and angle resolution for particles reconstructed in the TPC  (main tracking device in ALICE)
 \item  reconstruct particles passing through dead regions of the TPC
 \end{itemize}
To satisfy these requirements in the high-multiplicity environment of heavy-ion collisions at the LHC the detector has been designed as follows: 
\begin{itemize}
\item the two innermost  layers are based on silicon pixel technology  (SPD)
\item the  two intermediate layers are  Silicon Drift Detectors (SDD) equipped with analogue read-out of charge deposition
\item the two outer layers have double-sided Silicon Strip Detectors (SSD) with analogue read-out of charge deposition
\end{itemize}
\begin{figure}
\centering
 \includegraphics[width=0.99\textwidth]{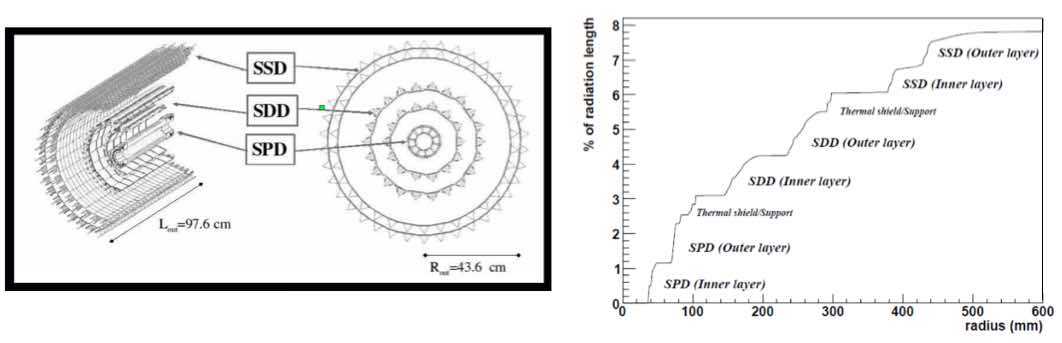}
\caption{Left: transverse and side view of the ITS. Right: \% of radiation lenght traversed by a particle crossing ITS as a function of the radial distance from the beam axis.}
\label{is1}
\end{figure}
The four outer layers (SPD and SDD) can provide dE/dx measurement for Particle Identification (PID) in the non-relativistic (dE/dx$\propto 1/\beta^2$) region of the Bethe-Block curve.\\
Since track momentum and impact parameter resolution are dominated by multiple  scattering effects at low momentum, the material budget
of ITS has been kept to a minimum using thin silicon layers ($\sim 300\mu$m) and
taking care of the total material budget of support, cables and cooling system. The support consists of lightweight carbon-fibre
structures. For particles crossing ITS in the radial direction the material budget is about 8\% of a radiation length (Figure \ref{is1}). This value slightly depends on 
the azimutal angle.
The design parameters of the different layers are summarized in table \ref{TabellaITS}. \\
\begin{table}[b]
 \begin{center}
  \begin{tabular}{|c|c|c|c|}
   \hline Parameter & Silicon Pixel & Silicon Drift & Silicon Strip \\
  \hline Spatial precision $r\varphi$ ($\mu$m) &12&35&20\\
\hline Spatial precision z ($\mu$m) & 100&25&830\\
\hline Two track resolution $r\varphi$ ($\mu$m)& 100 & 200 & 300\\
\hline Two track resolution z ($\mu$m) & 850 & 600 & 2400\\
\hline Active area per module (mm$^{2}$) & 12.8$\times$69.6 & 72.5$\times$75.3& 73$\times$40\\
\hline Readout channels per module & 40960 & 2 $\times$ 256 & 2$\times$ 768\\
\hline Total number of modules &240&260&1698\\
\hline Total number of readout channels (k) & 9835 & 133 & 2608\\
\hline Total number of cells (M) & 9.84 & 23 & 2.6\\
\hline
\end{tabular}
\end{center}
\caption{Main characteristics of the ALICE ITS; maximum occupancy is referred to central Pb-Pb collisions}
\label{TabellaITS}
\end{table}
The Silicon Pixel Detector (SPD) is based on a two dimensional silicon matrix of reverse biased silicon detectors diodes bump-bonded to readout chips. 
The sensor matrix includes 256 x 160 cells measuring 50 $ \mu$m ($r \varphi$\footnote{The z axis is parallel to the mean beam direction at ALICE interaction point and points along the LHC anticlockwise direction. The $r\varphi$ direction is tangent to a circumference perpendicular to the beam and centred on the interaction point }) by 425 $\mu$m (z) with a thickness of 200 $\mu$m. \\
The inner SPD layer is located at an average distance of 3.9 cm from the beam axis, while the outer one at an average distance of 7.6 cm. SPD are used to reconstruct  \textbf{tracklets},  obtained aligning clusters in the two SPD layers with the interaction vertex.  \\
The third and fourth layers of the ITS (Silicon Drift Detector - SDD) provide 2D position measurement and dE/dx 
of incident particles via   measurements of drift time (for $r\varphi$ coordinate)  and charge deposition. The electrons produced by ionizing particles crossing the detectors  (Figure \ref{its2} - left) drift in an electric field produced  by an array of  cathode strips connected by resistors to form a detector voltage divider kept at -1.8 kV and are collected  by 512 anodes divided in two  rows located at the two opposite edges of the sensor. The inner SDD layer is located at an average radial distance of 15 cm from the beam axis, while the outer one at an average radius of 23.9 cm.\\
The two outer layers which are fundamental for the matching of tracks from the TPC to the ITS  consist of double-sided Silicon Strip Detectors
(SSD) mounted on carbon-fibre support structures identical to the ones which support the SDD.
 Each module (Figure \ref{its2} - right) is a double sided silicon sensor  with 768 p-type strips on one side and  768 n-type strips on the other side. The spacing between adjacent strips is 95 $\mu$m and the strips are
oriented with an angle of 7.5 mrad with respect to the short edge of the sensor on
the p-side and with an angle of 27.5 mrad on the n-side. This configuration allows  2D position measurement on the surface of the module. \\
The spatial resolutions in both $z$ and $r\varphi$ direction depend on the layer and are summarized in Table 3.1. The dE/dx resolution achieved in Pb--Pb collisions combining SDD and SSD signals  is $\sigma(dE/dx) \approx$ 10-15\%.\\
\begin{figure}
\centering
 \includegraphics[width=0.99\textwidth]{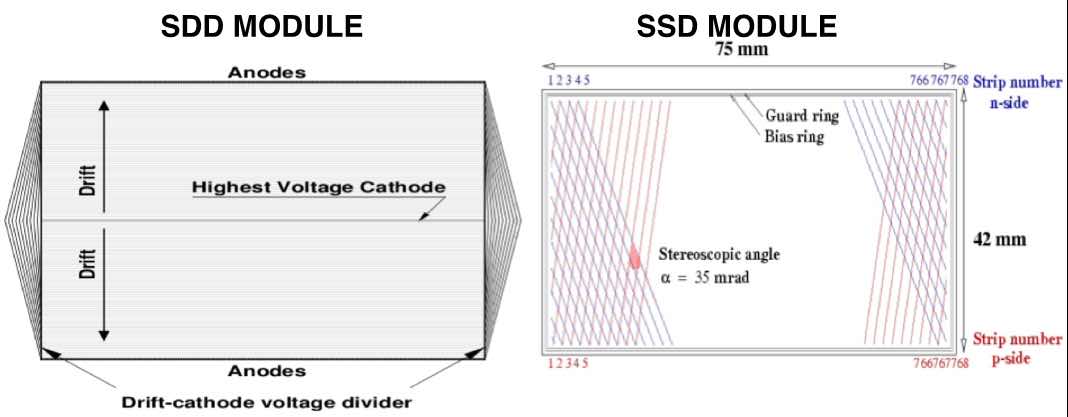}
\caption{Left: SDD sensor. Right: SSD sensor}
\label{its2}
\end{figure}
\subsection{Time Projection Chamber (TPC)}
The TPC (Figure \ref{TPCTOF} - left) is the main tracking detector of the ALICE central barrel. The phase space covered by this detector is:
\begin{itemize}
\item pseudo-rapidity $|\eta|$<0.9 for tracks that also match ITS, TOF and TRD; for reduced track lenght an acceptance up to about $|\eta|$<1.5 is accesible
\item full azimuthal coverage, except for dead zones between the sectors
\item \pt range covered from about 0.2 GeV/c to 100 GeV/c with a \pt resolution\\ $\sigma(\pteq)/\pteq\sim$1\% at 1 GeV/c, $\sigma(\pteq)/\pteq\sim$10\% at 50 GeV/c
\end{itemize}
The TPC consists of a cylindrical volume with an inner radius of 85 cm, an outer radius of 250 cm and an overall length along the beam direction of 5 m filled with 90 m$^3$ of Ne/CO2/N2 gas mixture. A central high voltage (100 kV)  electrode  creates a highly uniform electrostatic field ($\sim$ 400 V/cm) in which electrons and ions produced by the particles traversing the gas drift along the beam direction. The maximum drift time in the volume is $\sim$ 92 $\mu$s. The readout system is located at the two edges of the cylinder and  is based on multi-wire proportional chambers with catode pad read-out.  There are about 560000 readout pads of size ranging from 4 $\times$ 7.5 mm$^2$ to 6 $\times$ 15 mm$^2$, since the inner region has a higher occupancy and finer segmentation is needed. The read-out pads provide space as well as dE/dx and drift time measurents,  from which energy loss and 3D track reconstruction can be extracted. The relative dE/dx resolution was measured to be about 6\% for tracks that cross the entire detector, the spatial resolution is 1 mm in both $z$ and $r\varphi$ directions. \\
\begin{figure}
\centering
 \includegraphics[width=0.99\textwidth]{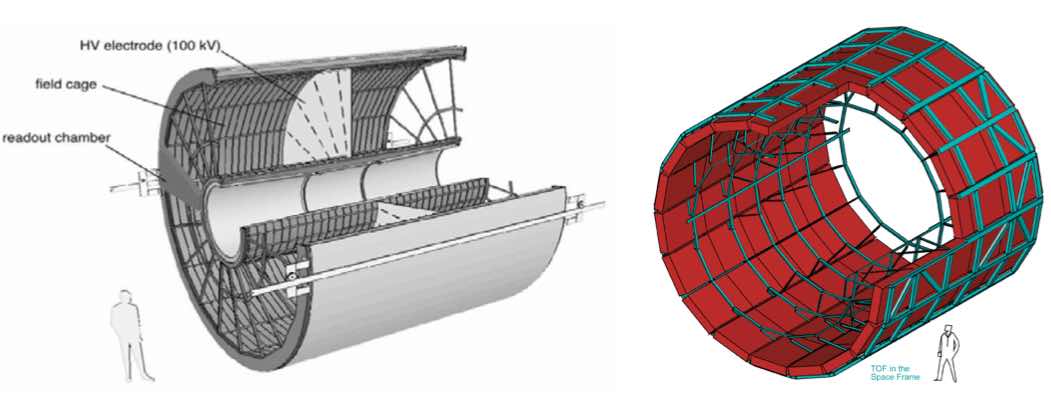}
\caption{Left:  structure of the ALICE TPC. Right: schematic view of the ALICE TOF}
\label{TPCTOF}
\end{figure}
\subsection{Time of Flight Detector (TOF)}
\label{sub:TOF}
The TOF (Figure \ref{TPCTOF} - right) is located at a radial distance of about 4 m from the beam pipe and it  is separated from the TPC by the Transition Radiation Detector (TRD). It is a large area array of Multi-gap Resistive Plate Chambers (MRPCs) covering the full azimuth in the central rapidity region ($|\eta|$<0.9). Unlike  standard RPCs, in  MRPCs the gas gap between the electrodes is divided by means of internal plates which are physical barriers stopping the avalanche from growing too much; in this way it is possible to apply a very intense electric field that provokes the avalanche process immediately after primary ionization occurs in the gas volume. The MRPCs  are designed as double-stack strips each having  5 gas gaps of 250 $\mu$m width. The resistive plates between the gaps are made of high-resistivity glass of 0.4 mm thickness. The two stacks are placed on each side of a central anode.  On the outer surface of the external plates a very intense electric field ($\sim$ 100 kV/cm) is applied.\\  A time-of-flight measurement is based on the quantity \textbf{t}$_{hit}$ - \textbf{t}$_0$, where \textbf{t}$_{hit}$ is the time measured by the TOF detector while \textbf{t}$_0$ is the time of the interaction that can be measured  with the ALICE T0 subdetector. However, due to the detector acceptance, expecially in pp collisions, no signals are observed by the T0 for a fraction of the events and an alternative method was developed where the TOF information itself is used to determine the event time when at least three tracks have an associated TOF signal.  This is done by means of a combinatorial algorithm which compares the measured TOF times to the expected times of the tracks, assuming a common event time  \textbf{t}$_0$. This latter quantity is obtained from a $\chi^2$ minimization procedure.  The overall time resolution of the TOF detector is given by  $\sigma^{2}_{\rm tot}$=$\sigma^{2}_{\rm TOF}$+$\sigma^{2}_{\rm t_0}$, where $\sigma^{2}_{\rm TOF}$ is the intrinsic TOF time resolution and $\sigma^{2}_{\rm t_0}$ is the resolution on the event time. Both $\sigma^{2}_{\rm TOF}$  and $\sigma^{2}_{\rm t_0}$
depend on the number  of tracks matched to TOF: for $n_{\rm tracks}$= 30, $\sigma^{2}_{\rm TOF}\approx$80 ps and  $\sigma^{2}_{\rm t_0}\approx$25 ps \cite{TOFBologna}.
\subsection{V0 detector}
The V0 detector (Figure \ref{V0ZDC} - left) consists of two arrays of scintillator counters, called V0A and V0C, which are installed on either side of the ALICE interaction point. The V0A detector is located 340 cm from the detector centre on the side opposite to the muon  spectrometer whereas V0C is located on the muon spectrometer  side at   90 cm distance from the detector centre. They cover the pseudo-rapidity ranges 2.8<$\eta$<5.1 (V0A) and -3.7<$\eta$<-1.7 (V0C) and each one  is segmented into 32 individual counters  distributed in four rings.  The scintillation light deposited by particles in each counter is collected by optical fibers and then guided to photo-multipliers. This detector has several functions:
\begin{itemize}
\item triggering
\item background rejection 
\item luminosity measurements
\item centrality and multiplicity measurements
\end{itemize}
\begin{figure}
\centering
 \includegraphics[width=0.99\textwidth]{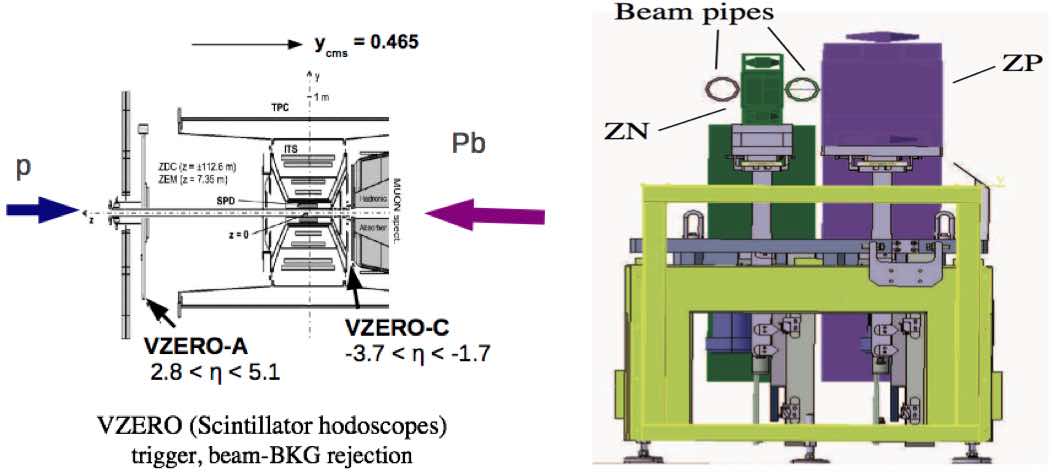}
\caption{Left: location of the two V0 scintillators along the beam line. Right: location of the ZDC detectors, transverse view}
\label{V0ZDC}
\end{figure}
\subsection{Zero Degree Calorimeter (ZDC)}
Two sets of hadronic calorimeters are located at 112.5 m (114 m until 2011) from the detector centre on either side of the Interaction Point (IP) to detect spectator nucleons. Since spectator protons and nuclear fragments are deflected  by the magnetic field of LHC while neutrons are undeflected, each ZDC set is composed of two elements: one for spectator neutrons (ZN) placed between the two beam pipes at 0 degrees relative to the LHC axis, and one for spectator protons (ZP) placed outside the beam pipe on the side where positive particles are deflected (Figure \ref{V0ZDC} - right). Nuclear fragments with Z/A$\approx$1/2 (like Pb nuclei) are not detected since they remain inside the beam pipe. \\The ZN dimensions are 7.04$\times$7.04$\times$100 cm$^3$, the ZP dimensions are 12$\times$22.4$\times$150 cm$^3$. They are hadronic sampling  calorimeters using  W-alloy (ZN) and brass (ZP) as passive material. The shower generated by incident particles produces Cherenkov radiation in quartz fibers interspersed in the passive material. The fibre spacing is smaller than the radiation length of the passive material in order to avoid electron absorption. The Cherenkov light is then sent to photomultipliers. \\
As will be discussed in next sections, ZDC is used for event selection (Section 3.2.2) and centrality measurements (Section 3.3).
\section{ALICE Performance}
\subsection{Trigger}
\label{sub:Trigger}
During the data taking  the trigger decision is generated by the Central Trigger Processor (CTP)  based on detector signals and information about the LHC bunch filling scheme. \\
 The CTP evaluates trigger inputs from the trigger detectors every LHC machine clock cycle ($\sim$ 25 ns). The Level 0 trigger decision (L0) is made $\sim$ 0.9 $\mu$s after the collision and is  delivered to the detectors with a latency of about 300 ns. The events accepted at L0 are further evaluated by the Level 1 (L1) trigger algorithm in the CTP. The L1 trigger decision is made $\sim$6.5 $\mu$s after L0. This delay is caused by the computation time (TRD and EMCal) and propagation times (ZDC, located at 112.5 m from IP2). The L0 and L1 decisions, trigger the buffering of the event data in the detector front-end electronics.
The trigger decisions is complemented by the information as to whether there are bunches coming from both A-side and C-side, or one of them, or neither, with a resolution of 25 ns.\\
ALICE operates with minimum-bias (MB) triggers, mainly based on V0 and SPD, and with rare triggers that are optimized to select particular classes of events such as events containing jets or muons or high-multiplicity events. In Pb--Pb collisions the V0 information has been used for triggering central and semicentral events. The decision in this case depended on the summed-up amplitude of the signals coming from both scintillators.\\
Figure \ref{trigger} (right) shows the integrated luminosity collected with the different triggers used in the 2013 p--Pb runs as a function of time. In this thesis events selected with   the \textbf{MBAND} trigger are used. This minimum bias trigger requires the presence of signal from both V0A and V0C detectors. From Figure \ref{trigger} (left) we see that an integrated luminosity of more than 7 nb$^{-1}$ has been collected using this trigger. 
\subsection{Background rejection}
\label{sub:backrej}
\begin{figure}[t]
\centering
 \includegraphics[width=0.99\textwidth]{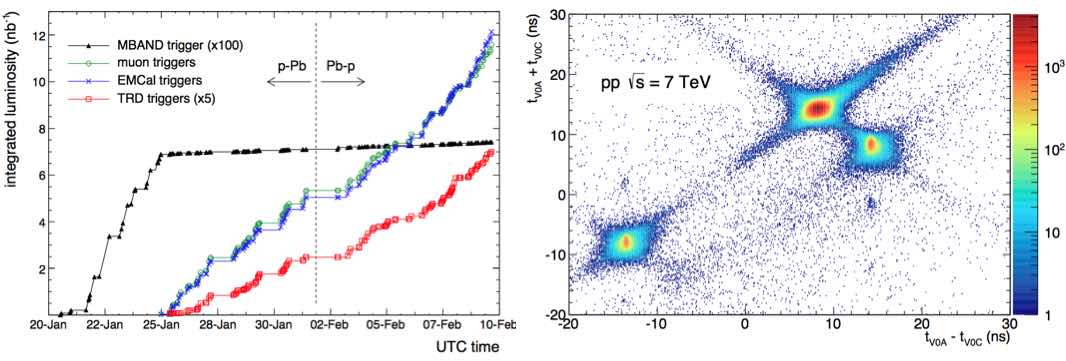}
\caption{Left: integrated luminosity per trigger type in the 2013 p-Pb run. Right: Correlation between the sum and difference of signal times in V0A and V0C.
}
\label{trigger}
\end{figure}
During the operation of the LHC a significant amount of background events arises from interactions of the beams with the residual gas present in the beam vacuum tubes. The most relevant component of beam background in ALICE is produced close to the experimental region by inelastic beam–gas (BG) interactions occurring within the first 40 m  on either side of the detector centre. 
Given the requirement of a reduced luminosity, in pp running the background rate in ALICE can be of the same order of magnitude as the interaction rate.\\
 ALICE  background rejection consists of an online and an offline procedure. During the data taking the signal from either sides of the V0 detector is used. Since the background caused by beam-gas interactions is produced upstream of the V0 in the side from which the beam arrives, it  produces an “early” signal in the corresponding V0 scintillator when compared with the time corresponding to a collision in the nominal interaction point. The difference between the expected beam and
background signals is on average 22.6 ns in the A side and 6 ns in the C side. As shown in Figure \ref{trigger} (right), background events accumulate mainly in two peaks in the time sum-difference plane, well separated from the main  peak that corresponds to events occuring in the interaction point.  Background from Beam 1 concentrates at (-14.3 ns, -8.3 ns), background from Beam 2 concentrates at (14.3 ns, 8.3 ns). Events are not rejected only if their timing signal from  both V0A and V0C detectors is within a fiducial window.  \\
The offline background rejection is based on a cut on the correlation between clusters and tracklets reconstructed in the SPD (Figure  \ref{BGBack} - left). Background particles usually cross the pixel layers in a direction parallel to the beam axis.
Therefore, only random combinations of BG hits can build a reconstructed track pointing to a fake vertex. In the cluster-tracklet plane of Figure  \ref{BGBack} (left) background events which have many clusters and only few tracklets  populate the region above the green dashed line, which is the cut used offline to exclude BG events from data analysis.  Only a very small fraction of background events (0.02\%) survive the
above-mentioned cuts in Pb–Pb collisions while in pp collisions  the amount of background surviving the cuts is strongly dependent on the running conditions and on the specific trigger configuration under study, reaching a value of 17\% in some of the 2012 runs. In p--Pb the fraction of background events surviving the physics selection cuts is $\sim$ 0.11\%.\\
\begin{figure}[b]
\centering
 \includegraphics[width=0.99\textwidth]{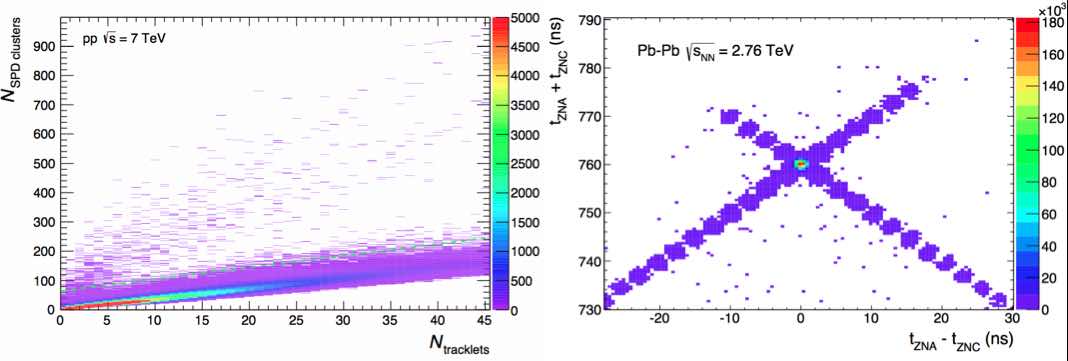}
\caption{ Left: correlation between reconstructed SPD clusters and tracklets. Two bands corresponding to physics collisions
and background are visible. The dashed cyan line represents the cut used in the offline selection: events lying in the region above the line are tagged as BG and rejected. Left: correlation between the sum and the difference of times recorded by the neutron ZDCs on either side (ZNA and ZNC) in Pb-Pb collisions.}
\label{BGBack}
\end{figure}
Collisions of main bunches and satellite bunches\footnote{The RF operations involved in the delivery of proton and lead ion bunches to the LHC can result in satellite bunches of varying intensity occupying the nominally empty RF buckets} located at short distance from the main bunch are also a source of background. These satellite events are rejected using the correlation between the sum and the difference of times measured in the ZDC, as shown in Figure \ref{BGBack} (right) for Pb--Pb collisions. The large cluster in the middle corresponds to collisions between ions in the nominal RF bucket on both sides, while the small clusters along the diagonals (spaced by 2.5 ns in  time difference) correspond to collisions in which one of the ions is displaced from the main bucket by one or more RF buckets.
\subsection{Track and vertex reconstruction}
\label{sub:tvrec}
\begin{figure}[t]
\centering
 \includegraphics[width=0.99\textwidth]{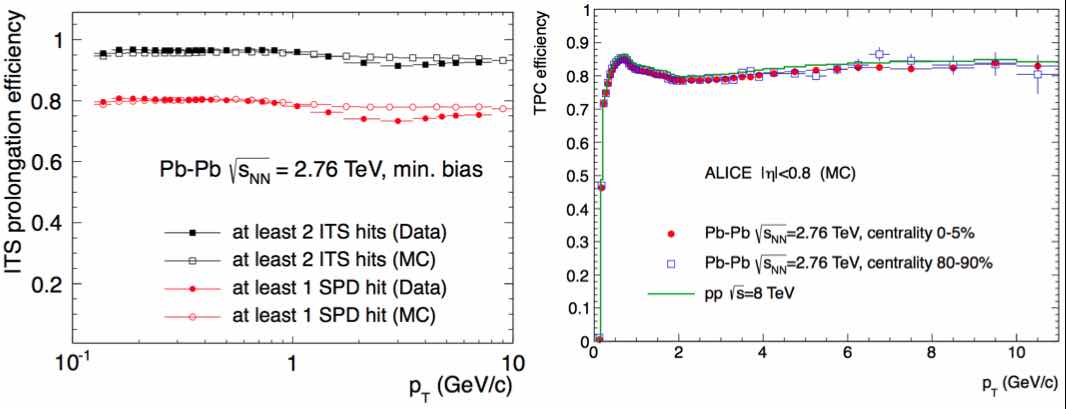}
\caption{Left: ITS-To-TPC matching efficiency as a function of \pt  for data and MC in minimum bias Pb--Pb collisions. Tracks used in the evaluation of the matching efficiency are required to have at least 2 ITS hits or at least 1 SPD hit. Right: TPC track finding efficiency from MC simulations for pp collisions at $\sqrt{s}$=8 TeV, central and peripheral Pb--Pb collisions at \sqnn= 2.76 TeV.}
\label{itstpcmatch}
\end{figure}
\begin{figure}[b]
\centering
 \includegraphics[width=0.99\textwidth]{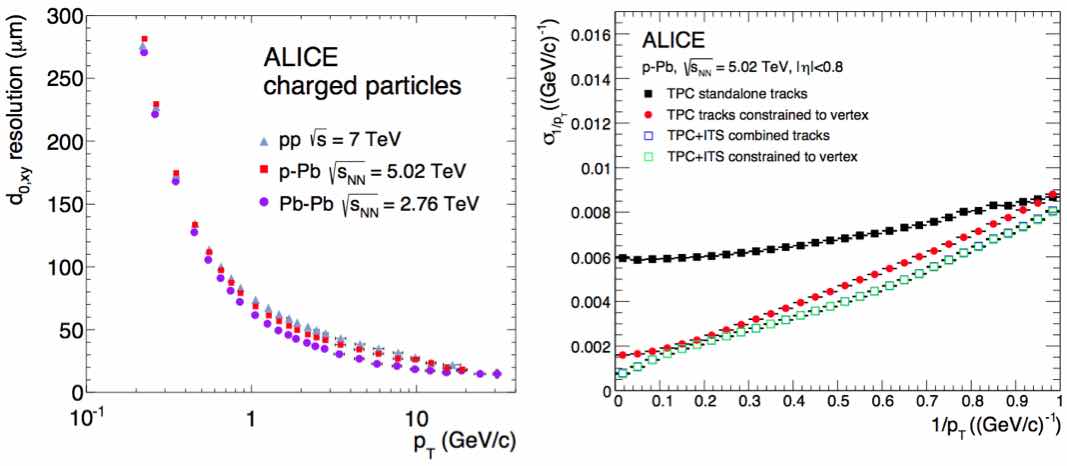}
\caption{Left: resolution on the projection of the  impact parameter in the transverse plane for charged particles in pp, Pb--Pb and p--Pb collisions as a function of \pt.  Right: inverse-\pt resolution $\sigma_{1/\pteq}$ as a function of \pt  in p--Pb collisions, for TPC and ITS-TPC tracks without and with  vertex constrain}
\label{ipptres}
\end{figure}
Tracking in ALICE is performed offline and consists of the following steps:
\begin{itemize}
\item the clusters in the two layers of SPD are used to reconstruct the primary vertex, defined as the space point to which a maximum number of tracklets (Section \ref{sub:its}) converge.  When  a 3D reconstruction of the primary vertex is not feasable (essentially in very low-multiplicity events) the algorithm  performs a minimization using information from the distance of closest approach (DCA) of tracklets to the  beam axis. This method only gives access to the $z$ coordinate of the vertex
\item tracking starts at the TPC outer radius. Track seeds are built first with two TPC clusters and the vertex point, then with three clusters and without the vertex constraint. The seeds are propagated inward and, at each step, updated with the nearest cluster found by the algorithm (Kalman filter \cite{Kalman}) until the inner radius of the TPC is reached. 
\item  tracks are then propagated to the outermost ITS layer and become the seeds for track finding in the ITS. Each TPC track produces a tree of track hypotheses in the ITS and the candidates are selected  according to their reduced $\chi^2$. The matching efficiency of this step, defined as the ratio of the fraction of TPC tracks that have a prolongation in the ITS,  is shown in Figure \ref{itstpcmatch} (left) for data and simulations and different requirements on the ITS hits attached to the track
\item the ITS clusters not used in the TPC-ITS reconstruction are used to perform an ITS stand-alone reconstruction. In fact the TPC track finding  efficiency drops at low \pt (Fig \ref{itstpcmatch} - right) whereas the ITS still has a good performance in this \pt region. This ITS stand-alone reconstruction also allows to reconstruct track traversing dead regions of the TPC or decaying before the inner TPC radius
\item once the reconstruction in the ITS is complete, all tracks are extrapolated to their point of closest approach to the  interaction vertex, and the outward propagation starts.  Once the track reaches the TRD  an attempt is made to match it with a TRD tracklet (track segment within a TRD layer) in each of the six TRD layers. Tracks reaching the TOF detector are matched to TOF clusters.  The tracks are then propagated further for matching with signals in EMCal, PHOS, and HMPID.
\item at the final stage of the track reconstruction, all tracks are propagated inwards starting from the outer radius of the TPC. In each detector (TPC and ITS), the tracks are refitted with the previously found clusters. The track position, direction, inverse curvature, and its associated covariance matrix are determined. The vertex position is recomputed using these tracks.
\end{itemize}
\begin{figure}[b]
\centering
 \includegraphics[width=0.99\textwidth]{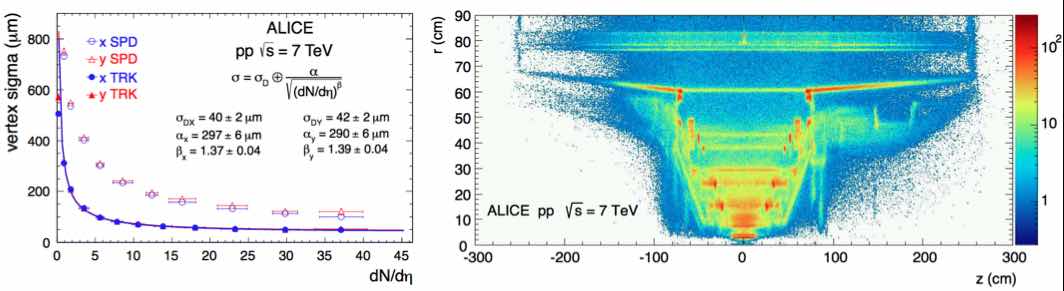}
\caption{Left:  transverse resolution of the preliminary interaction vertices found with SPD and  with global tracks. Right: Distribution of secondary vertices from hadronic interactions in the ALICE material.}
\label{seco}
\end{figure}
The left panel of Figure \ref{ipptres} shows the resolution on the transverse distance of closest approach of the track to the primary vertex (defined as impact parameter) for all charged particle tracks in pp, Pb--Pb and p--Pb collisions. One can notice an improvement of the resolution in p--Pb and Pb--Pb collisions thanks to the more precise determination of the primary   vertex for higher multiplicities.\\
The absolute  resolution on 1/\pt for TPC standalone tracks and ITS–TPC combined tracks for p--Pb collisions, is shown in Figure  \ref{ipptres} (right). The effect of constraining the tracks to the primary vertex is shown as well. The  absolute inverse-\pt resolution $\sigma_{1/\pteq}$, plotted in this figure, is connected to the relative transverse momentum resolution via $\sigma_{\pteq}/\pteq = \pteq \sigma_{1/\pteq}$.\\
The  resolution  on the position in the transverse plane of the  interaction vertex  reconstructed with SPD and  with global tracks are shown as a function of multiplicity in Figure \ref{seco} for the case of pp collisions at $\sqrt{s}$=7 TeV.  Both resolutions scale with the square root of the number of tracklets/tracks used in the vertex determination.\\
Once the tracks and the interaction vertex have been found and reconstructed, a search for photon conversions and secondary vertices from strange particle decays is performed. Tracks with a distance of closest approach to the interaction vertex exceeding a certain minimum value (0.5 mm in pp and 1 mm in Pb–Pb) are selected and their point of closest approach is calculated. Figure \ref{seco} shows the 2D distribution of such vertices that mostly come from photon conversion in the material. The ITS layers, the inner TPC containment vessel, and the inner TPC field cage are visible.
\subsection{Particle Identification}
\label{sub:PID}
Many of the ALICE subdetectors have particle identification capabalities:
\begin{itemize}
\item ITS: The outer  four layers  of the Inner Tracking System have an analog readout to measure the deposited charge, thereby providing a dE/dx measurement. For each track, the dE/dx is calculated using a truncated mean: the average of the lowest two points if four points are measured, or a weighted sum of the lowest (weight 1) and the second-lowest points (weight 1/2), if only three points are measured. The dE/dx resolution achieved in Pb--Pb collisions combining SDD and SSD signals  is $\sigma$(dE/dx)$\approx$ 10-15\%.
\item TPC  measures the charge deposited on up to 159 pad rows. A truncated mean dE/dx (40\% highest-charge clusters discarded) is calculated and used for a wide range of momenta. The relative dE/dx resolution was measured to be about 6\% for tracks that cross the entire detector.
\item TOF  measures the arrival time of particles with a resolution of $\sim$ 100 ps
\item  the High Momentum Particle Identification Detector (HMPID) is a ring-imaging Cherenkov detector that covers |$\eta$|<0.6 in pseudorapidity and 57.6 degrees in azimuth, corresponding to 5\% acceptance of the central barrel, and provides proton/kaon separation up to \pt = 5 GeV/c.
\item The Transition Radiation Detector (TRD) identifies electrons based on their specific energy loss and transition radiation (TR) and covers the full azimuth for |$\eta$|<0.9.
\item  The Electromagnetic Calorimeter (EMCal) identifies electrons by measuring their energy deposition and comparing it to the measured track momentum  (E/p).
\item The Photon Spectrometer (PHOS) is a high-granularity electromagnetic calorimeter that can also identify electrons using the E/p method.
\end{itemize}
In the analysis that will be exposed in this thesis, particle identification is exploited to  identify the pions and kaons coming from D$^+$ mesons decay. To do this, PID information from TPC and TOF is used. \\
\begin{figure}
\centering
 \includegraphics[width=0.99\textwidth]{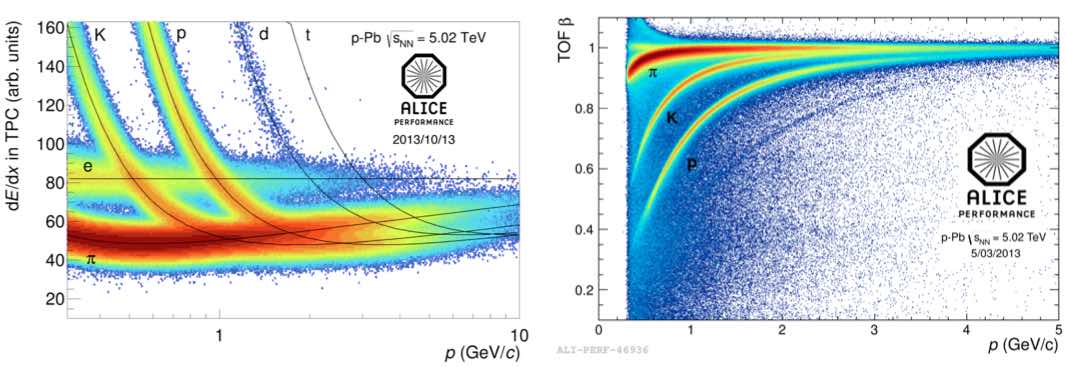}
\caption{Left: Specific energy loss (dE/dx) in the TPC vs. particle momentum in p--Pb collisions. The lines show the ALEPH parametrizations of the expected mean energy loss. Right: TOF measured velocity $\beta$ distribution as a function of momentum.}
\label{PIDTPCTOF}
\end{figure}
In the TPC particle identification is performed by simultaneously measuring the specific energy loss (dE/dx), the  charge, and the  momentum of each particle traversing the detector gas. The energy loss, described by the Bethe-Bloch formula, is parametrized by a function originally proposed by the ALEPH collaboration  
\begin{equation}
f(\beta \gamma) = \frac{P_1}{\beta^{P_4}}(P_2 - \beta^{P_4} - \ln(P_3+\frac{1}{(\beta \gamma)^{P_5}}))
\end{equation}
where $\beta$ is the particle velocity, $\gamma$ is the Lorentz factor, and P$_{1-5}$ are fit parameters. The energy loss is also reproduced by parametrizations developed by the ALICE Collaboration  that are periodically tuned on the data. Figure \ref{PIDTPCTOF} (left) shows the measured dE/dx vs. particle momentum in the TPC, demonstrating the clear separation between the different particle species. The lines correspond to the ALEPH parametrization. While at low momenta (p<1 GeV/c) particles can be identified on a track-by-track basis, at higher momenta particles can still be separated on a statistical basis via multi-Gaussian fits. Indeed, for long tracks ($\sim$ 130 dE/dx  samples) and with the truncated-mean method the resulting dE/dx ditribution is Gaussian down to at least 3 orders of magnitude from the peak.\\
The TOF particle identification performance depends on its intrinsic resolution and on the start time measurement.The start time for the TOF measurement is provided by the T0 detector, which consists of two arrays of Cherenkov counters T0C and T0A, positioned at opposite sides of the interaction point (IP) at -3.28 < $\eta$< -2.97 and  -4.61 < $\eta$ < -4.92, respectively. However, due to the detector acceptance, expecially in pp collisions, no signals are observed by the T0 for a fraction of the events and an alternative method was developed where the TOF information itself is used to determine the event time when at least three tracks have an associated TOF signal.  This is done by means of a combinatorial algorithm which compares the measured TOF times to the expected times of the tracks, assuming a common event time  \textbf{t}$_0$. This latter quantity is obtained from a $\chi^2$ minimization procedure. Figure \ref{PIDTPCTOF} (right) illustrates the performance of the TOF detector by showing the measured velocity $\beta$ distribution as a function of track  momentum (measured by the TPC and ITS). The bands relative to electrons, muon, pions, kaons, protons and deuterons are visible. The background is due to tracks that are incorrectly matched to TOF hits in high-multiplicity p–Pb collisions.\\
The separation of hadron species can be  improved by combining information from TOF and TPC, thus allowing a further extension of the momentum range for identified particle measurements. In Figure \ref{combinedPID} the difference between the measured and expected PID signals for TPC and TOF are represented in the pion mass hypothesis. It is evident that cuts or fits using a combination of the variables provide a better separation than just considering their projections. This can improve the  PID performance   in the identification of  kaon and pions from D$^+$ meson decays.
\begin{figure}
\centering
 \includegraphics[width=0.75\textwidth]{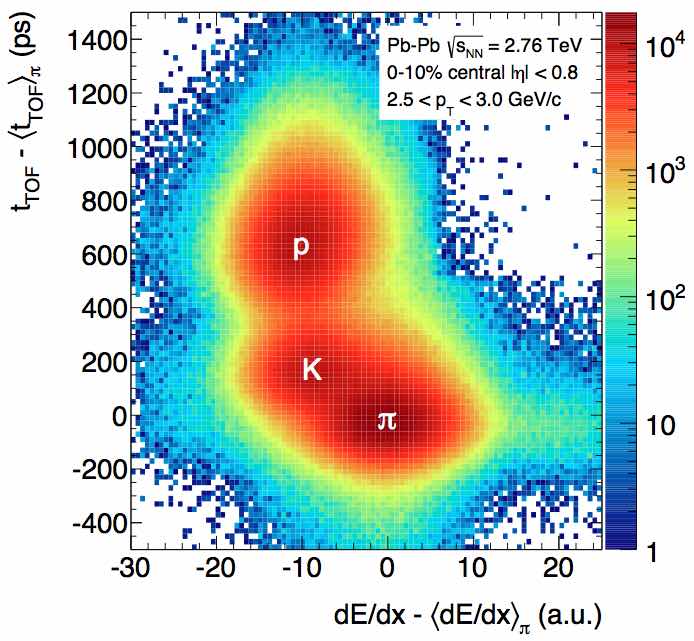}
\caption{Combined pion, kaon and proton identification with TOF and  dE/dx in the TPC for central Pb--Pb collisions}
\label{combinedPID}
\end{figure}
\section{Centrality determination with ALICE}
For spherical nuclei, the geometry of heavy-ion collisions is characterized by the impact parameter, which is the  vector \textbf{$\overrightarrow{b}$} connecting the centers of the two colliding nuclei in the plane transverse to the beam direction. It is usual to express the centrality of nuclear collisions not in terms of the impact parameter b but via a percentage of the total hadronic interaction cross section $\sigma_{\rm AA}$. The centrality percentile $c$ of an AA collision with impact parameter 0<\textbf{b}<\textbf{b$_{max}$} is defined as
\begin{equation}
c(b)=\frac{\int^{b'}_{0}\frac{d\sigma}{db'}db'}{\int^{\infty}_{0}\frac{d\sigma}{db'}db'}=\frac{1}{\sigma_{\rm AA}}\int^{b_{max}}_{0}\frac{d\sigma}{db'}db'
\end{equation}
Experimentally, the centrality is defined as the fraction of cross section with the largest detected charged-particle multiplicity N$_{\rm ch}$ or the smallest zero-degree energy $E_{\rm ZDC}$
\begin{equation}
c \approx \frac{1}{\sigma_{\rm AA}}\int^{\infty}_{N_{ch}}\frac{d\sigma}{dN_{ch}'}dN_{ch}' \approx   \frac{1}{\sigma_{\rm AA}}\int^{E_{ZDC}}_{0}\frac{d\sigma}{dE_{ZDC}'}dE_{ZDC}'
\end{equation}
The cross section may be replaced with the number of observed events $n$ (corrected for the trigger efficiency and for the non-hadronic interaction background)
\begin{equation}
c \approx \frac{1}{N_{ev}}\int^{\infty}_{N_{ch}}\frac{dn}{dN_{ch}'}dN_{ch}' \approx  \frac{1}{N_{ev}}\int^{E_{ZDC}}_{0}\frac{dn}{dE_{ZDC}'}dE_{ZDC}'
\end{equation}
We are assuming  that, on average, 
\begin{itemize}
\item the particle multiplicity at midrapidity increases monotonically  with centrality.
\item the  zero-degree energy decreases monotonically with centrality.
\end{itemize}
For the zero-degree energy measurement, this assumption holds only for central collisions because nuclear fragments emitted in peripheral collisions may be deflected out of the acceptance of the zero-degree calorimeter, leading to low signals indistinguishable from those seen in central collisions.
\subsection{Centrality determination in Pb--Pb collisions}
The Glauber MC introduced in Section 1.3.2 is used  to simulate Pb--Pb collisions. The impact parameter b is randomly selected from the geometrical distribution dP/db $\propto$ b up to a maximum b$_{max} \approx$ 20fm > 2R$_{\rm Pb}$. The maximum value of the impact parameter b$_{max}$  is chosen large enough to simulate collisions until the interaction probability becomes zero.\\
The number of collisions N$_{\rm coll}$ and the number of participants N$_{\rm part}$ are determined by counting, respectively, the binary nucleon collisions and the nucleons that experience at least one
collision. The geometric nuclear overlap function T$_{\rm AA}$ is then
calculated as T$_{\rm AA}$= N$_{\rm coll}$/$\sigma_{\rm inel}$ where $\sigma_{\rm inel}$ is the nucleon-nucleon inelastic cross-section. For nuclear collisions at \sqnn = 2.76 TeV, $\sigma_{\rm inel}$ = (64 $\pm$ 5) mb is used, estimated by interpolation of pp data at different center-of-mass energies and from cosmic rays,  and subtracting the elastic scattering cross section from the total cross section. The impact parameter and N$_{\rm part }$ distributions are shown in Figure \ref{CentrPbPb2}. The centrality classes are defined by sharp cuts on the the impact parameter distribution, the corresponding N$_{\rm part}$ distributions are shown in Figure \ref{CentrPbPb2} (right). \\
\begin{figure}
\centering
 \includegraphics[width=0.95\textwidth]{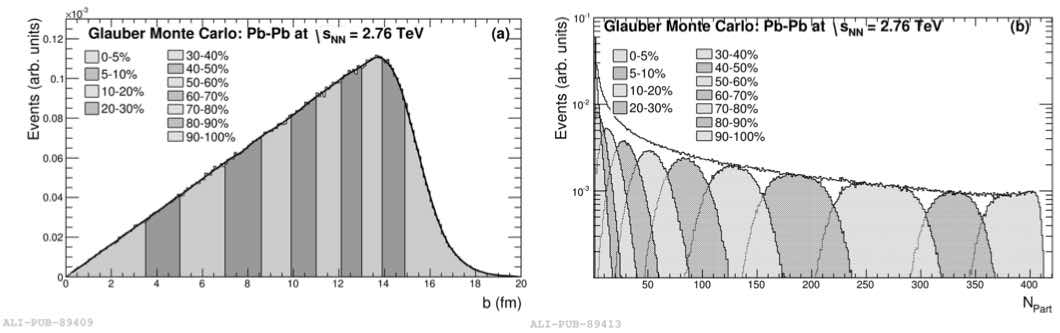}
\caption{Left: impact parameter distribution obtained from Glauber MC simulations for Pb--Pb collisions at \sqnn = 2.76 TeV. Right: the corresponding N$_{\rm part}$ distributions for different intervals of impact parameter values.}
\label{CentrPbPb2}
\end{figure}
To divide the events collected by ALICE in centrality classes, a connection between the  impact parameter and an experimental observable has to be established. This can be done using the Glauber MC to generate N$_{\rm part}$ and N$_{\rm coll}$ and parametrizing the particle multiplicity per nucleon-nucleon collision with a Negative Binomial Distribution with parameters $\mu$ and $k$:
\begin{equation}
P_{\mu,k}(n)=\frac{\Gamma(n+k)}{\Gamma(n+1)\Gamma(k)}\cdot\frac{(\mu/k)^n}{(\mu/k+1)^{n+k}}
\end{equation}
 To apply this model to any collision with a given  N$_{\rm part}$ and N$_{\rm coll}$ one has to  introduce the concept of “ancestors”, i.e. independently emitting sources of particles. We assume that the number of ancestors N$_{\rm ancestors}$ can be parameterized by \\  N$_{\rm ancestors}$ = f$\cdot$N$_{\rm part}$+(1-f )$\cdot$N$_{\rm coll}$. This is inspired by two-component models which decompose nucleus-nucleus collisions into soft and hard interactions, where the soft interactions produce particles with an average multiplicity proportional to N$_{\rm part}$, and the probability for hard interactions to occur is proportional to N$_{\rm coll}$. Experimentally this framework is used to fit different observables:
\begin{itemize}
\item  the sum of the signal amplitudes of the  V0A and V0C scintillators, which is proportional to the number of particles hitting them
\item the number of clusters in the second layer of the SPD
\item the number of tracks reconstructed in the TPC
\end{itemize}
Figure \ref{CentrPbPb} (left) shows the distribution of the sum of amplitudes in the V0 scintillators. The distribution is fitted with  a parametrization based on the Glauber MC and the NDB distribution (NBD-Glauber fit in the following) and the parameters extracted are shown: f is the parameter that appears in the definition of N$_{\rm anchestors}$, while $\mu$ and k are the parameters of the NBD distribution. \\
\begin{figure}[b]
\centering
 \includegraphics[width=0.95\textwidth]{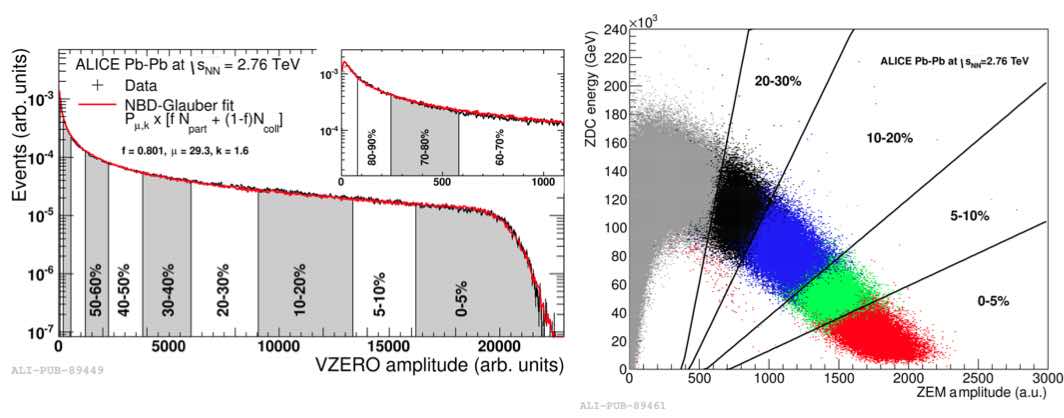}
\caption{Left:  sum of amplitudes in the V0 scintillators for Pb--Pb collisions at \sqnn = 2.76 TeV. The distribution is fitted with the NBD-Glauber distribution. Right: two dimensional distribution for ZEM amplitude and ZDC energy.}
\label{CentrPbPb}
\end{figure}
Another way to determine the centrality is to measure the energy deposited by the spectator nucleons in the ZDC. In theory this would easily relate to N$_{\rm part}$ via
\begin{equation}
N_{\rm part} = 2A - E_{ZDC}/E_A
\end{equation}
where $E_{ZDC}$ is the energy measured by the ZDC and $E_{A}$ is the beam energy per nucleon. In practice however some nuclear fragments have charge over mass ratio similar to the lead ions in the beam and remain inside the beam pipe. A quantity that has a monotonic increasing behaviour with N$_{\rm part}$  is the energy deposited in the   ZEM detectors, a set of two electromagnetic calorimeters placed close to the beam pipe at 7.5 m from the interaction point covering a region 4.8<$|\eta|$<5.7. Figure  \ref{CentrPbPb} (right) shows the two dimensional distribution of ZEM amplitude and ZDC energy. The centrality here is defined by cutting the plane into regions defined by straight lines. As one can see from the figure, the slope of these straight lines increases going to peripheral collisions, where the correletion between ZEM amplitude and ZDC energy is partially lost, making this measurement trustable in 0-30\% centrality region only. \\
The accuracy of the experimental determination of centrality with the different methods was evaluated in \cite{CentPbPb} and turned out to depend in the rapidity coverage of the detector used. The best centrality resolution is achieved when combining V0A and V0C detectors  (Figure \ref{CentrPbPb}-left), due to 4.3 units of rapidity coverage, and ranges from 0.5\% resolution for central collisions to 2\% for peripheral collisions. \\
The standard method typically used in ALICE physics analyses to extract N$_{\rm part}$ and N$_{\rm coll}$ per centrality class in Pb--Pb collisions is the NBD-Glauber fit to the sum of V0A and V0C amplitudes. The other methods described are used to asses the  systematic uncertainty on the centrality determination.
\subsection{Centrality determination in p--Pb collisions}
\label{sub:centpPb}
\begin{figure}
\centering
 \includegraphics[width=0.99\textwidth]{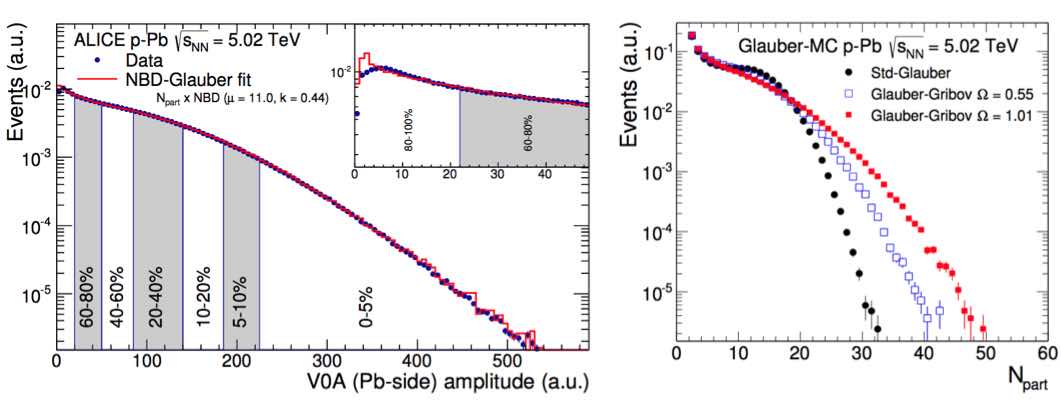}
\caption{Left: distribution of amplitudes in the V0A scintillators during the p--Pb runs. Right: N$_{\rm part}$ distributions obtained from standard Glauber MC and from  Glauber MC coupled to Glauber-Gribov fluctuations.}
\label{CentrpPb1}
\end{figure}
In Figure \ref{CentrpPb1} (left)  the distribution of amplitudes in the V0A scintillator array  during the p--Pb runs (the V0A is thus in the Pb-going direction) is shown together with  the NBD-Glauber fit and with the centrality classes indicated by the vertical lines. The  NBD-Glauber approach used in p--Pb collisions is similar to the one described in the previous section, but here N$_{\rm ancestors}$ is assumed to be equal to N$_{\rm part}$ and the final multiplicity distributions scales with a NDB$\times$N$_{\rm part}$ law. Furthermore in hadron-nucleus collisions one has to take into account the particular quark-gluon configuration of the hadron, which is frozen during the collision leading to a different interaction  strength  compared to $\sigma_{\rm NN}^{\rm inel}$. This leads to fluctuations of the number of collisions which are significantly larger than in the standard Glauber MC approach \cite{FlucGlauber}. In the Glauber MC shown in Figure \ref{CentrpPb1} (left) these fluctuations are  implemented by an effective scattering cross-section.\\ Another approach taking into account these fluctuations is the Glauber-Gribov model,  which includes event-by-event variations of $\sigma_{\rm NN}^{\rm inel}$. The width of the variations in  $\sigma_{\rm NN}^{\rm inel}$ is controlled by the  parameter $\Omega$. The N$_{\rm part}$ distributions obtained from standard Glauber MC and from  Glauber MC coupled to Glauber-Gribov fluctuations is shown in  Figure \ref{CentrpPb1} (right), where the Glauber-Gribov computations have been performed for  $\Omega$=0.55 and $\Omega$=1.01.  One can see that the N$_{\rm part}$ distribution becomes broader when considering  Glauber-Gribov fluctuations. I recall that in p--Pb collisions N$_{\rm part}$=N$_{\rm coll}$+1. \\
In Figure \ref{CentrpPb2} (left)  the same V0A distribution of Figure \ref{CentrpPb1} (left) is shown together with the standard NBD-Glauber fit and with a   Glauber-Gribov ($\Omega$=0.55) fit.  Both methods have been repeated with different assumptions on the multiplicity scaling law: NDB$\times$N$_{\rm part}$ and NDB$\times$N$_{\rm coll}$. The standard NBD-Glauber fits yield satisfactory results using either the N$_{\rm part}$ or the N$_{\rm coll}$ scaling, which result in similar average number of collisions $\langle N_{\rm coll}\rangle$ evaluated for each of the centrality intervals as shown in Table 3.2. The Glauber-Gribov fits with $\Omega$= 0.55 provide an equally good description of the measured V0A distribution as the standard Glauber, indicating that the fits cannot discriminate between the models. As expected, the Glauber-Gribov values of $\langle N_{\rm coll}\rangle$ also shown in Table 3.2 are larger (smaller) for central (peripheral) collisions than those obtained from the standard Glauber, as a consequence of the different shapes of the N$_{\rm part}$ distributions in these models.\\
\begin{figure}
\centering
 \includegraphics[width=0.99\textwidth]{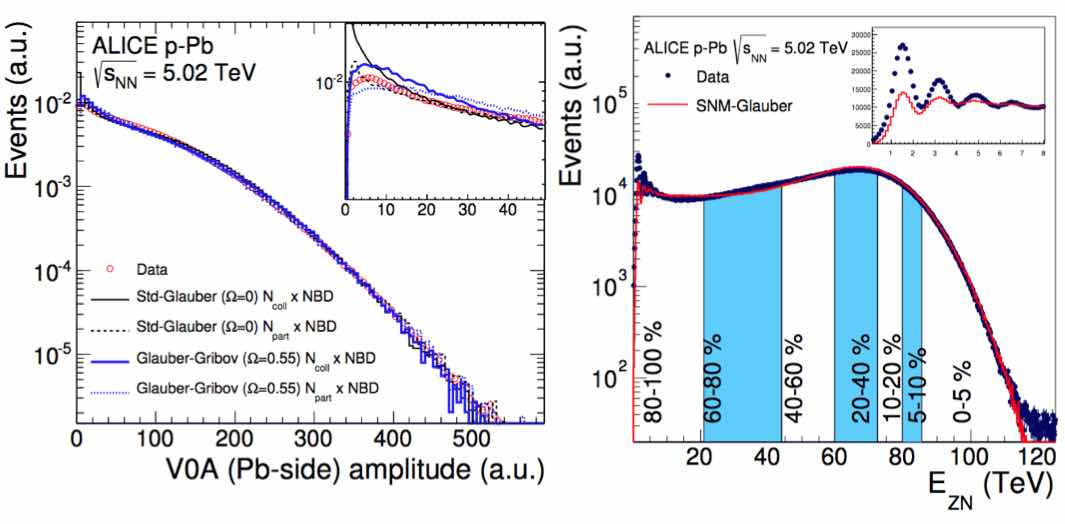}
\caption{Left: distribution of amplitudes in the V0A hodoscopes during the p--Pb runs with standard Glauber and Glauber-Gribov fits. Right:  distribution of the neutron energy spectrum measured in the ALICE neutron calorimeter (ZN) on the A-side.}
\label{CentrpPb2}
\end{figure}
Another estimator used for centrality determination in p--Pb is the so-called \textbf{CL1}, which is given by the number of clusters in the outer layer of the SPD in the range |$\eta$|<1.4. A NBD-Glauber fit on the CL1 signal similar to the one performed on V0A amplitute, performed assuming \npart scaling, gives the CL1 \ncoll values that are also shown in Table \ref{tablencollppb}.\\
The ZDC detects the slow nucleons emitted during the proton-nucleus interaction that can give information on the centrality  of the collision. Emitted nucleons are classified as “black” or “grey”. This terminology originates from emulsion experiments:
\begin{itemize}
\item black particles, typically defined to have velocity $\beta \lesssim$0.25 in the nucleus rest frame, are produced by nuclear evaporation processes
\item gray particles, 0.25$\lesssim \beta \lesssim$ 0.7, are mainly nucleons knocked out from the nucleus
\end{itemize}
These slow nucleons have been studied at lower energies.  The similarities in their emission found at different energies suggested the idea that their production  is dictated by nuclear geometry only \cite{SlowNucleons}. Therefore one can expect that slow nucleon emission at LHC can be described by the same models used at lower energy after an appropriate scaling of the parameters. \\
\begin{table}[t]
\tiny
\centering
\begin{tabular}{|c|c|c|c|c|c|c|cl}
\hline
\multicolumn{1}{|l|}{\multirow{2}{*}{Centrality (\%)}} & \multicolumn{2}{c|}{V0A NBD-Glauber}                                                              & \multicolumn{2}{c|}{V0A Glauber-Gribov}                                                           & \multirow{2}{*}{ZNA SNM-Glauber}  & \multirow{2}{*}{CL1  NBD-Glauber}  \\ \cline{2-5}
\multicolumn{1}{|l|}{}                                 & \multicolumn{1}{l|}{N$_{\rm part} \times$NBD} & \multicolumn{1}{l|}{N$_{\rm coll} \times$NBD} & \multicolumn{1}{l|}{N$_{\rm part} \times$NBD} & \multicolumn{1}{l|}{N$_{\rm coll} \times$NBD} &          &                    \\ \hline
0-5                                                    & 14.8                                          & 15.3                                          & 17.8                                          & 19.2                                          & 15.7             & 15.6           \\ \hline
5-10                                                   & 13.0                                          & 13.4                                          & 14.4                                          & 15.2                                          & 13.9            & 13.6             \\ \hline
10-20                                                  & 11.7                                          & 12.0                                          & 12.0                                          & 12.5                                          & 12.4           & 12.0              \\ \hline
20-40                                                  & 9.36                                          & 9.62                                          & 8.82                                          & 9.04                                          & 9.99           & 9.49              \\ \hline
40-60                                                  & 6.42                                          & 6.40                                          & 5.68                                          & 5.56                                          & 6.53           & 6.18              \\ \hline
60-80                                                  & 3.81                                          & 3.42                                          & 3.33                                          & 2.89                                          & 3.04           & 3.40              \\ \hline
80-100                                                 & 1.94                                          & 1.85                                          & 1.72                                          & 1.43                                          & 1.24          & 1.76               \\ \hline
\end{tabular}
\caption{N$_{\rm coll}$ values obtained from various fits to V0A, CL1 and ZNA distributions.}
\label{tablencollppb}
\end{table}
In Figure \ref{CentrpPb2} (right)  the distribution of the neutron energy spectrum measured in the ALICE neutron calorimeter (ZN) on the A-side (Pb-going direction) is shown. In the following this variable will be called ZNA energy. The distribution is fitted using the Glauber MC coupled to a slow nucleon emission model (Slow Nucleon Model - SNM) obtained based on the parametrization of experimental results at lower energies (E910 at BNL, Cooler Synchrotron COSY). The average $\langle N_{\rm coll} \rangle$ values obtained with this method are also shown in Table 3.2. \\
\subsection{Biases on centrality  in p--Pb collisions}
\label{sub:biaspPb}
I will now discuss three different sources of bias that have been observed in p--Pb collisions.
\subsubsection{Multiplicity bias}
\begin{figure}[t]
\centering
 \includegraphics[width=0.999\textwidth]{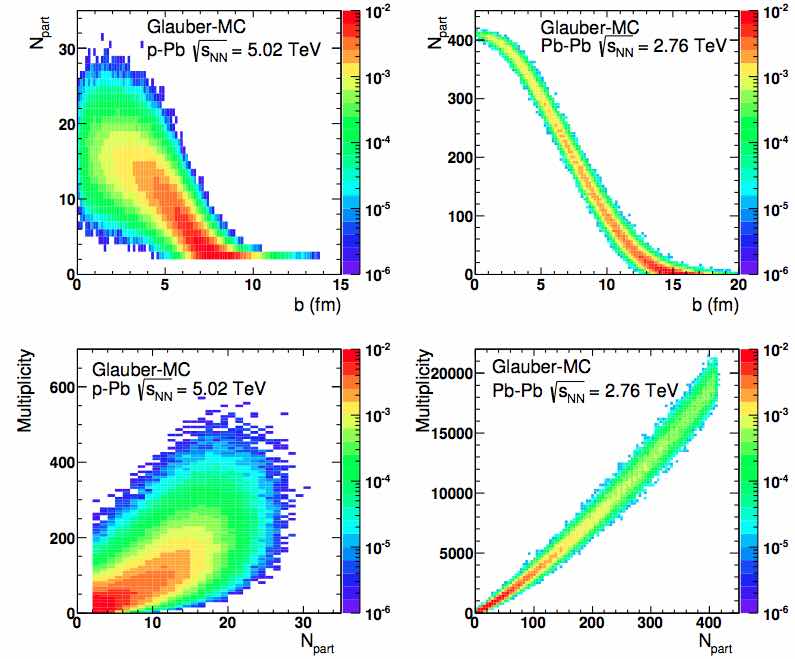}
\caption{Top: Scatter plot of number of participating nucleons versus impact parameter. Bottom: Scatter plot of multiplicity versus the number of participating nucleons from the Glauber fit to V0A amplitudes. The quantities are calculated with a Glauber Monte Carlo of p--Pb (left) and Pb--Pb (right) collisions.}
\label{multfluc}
\end{figure}
The left panels of Figure \ref{multfluc} show the scatter plot of N$_{\rm part}$ as a function of the impact parameter \textbf{b} of the collision (top) and  of the particle multiplicity as a function of N$_{\rm part}$ (bottom) in the case of a Glauber MC simulation of p--Pb collisions at \sqnn= 5.02 TeV. The right panels are the same plots obtained with the Glauber MC for Pb--Pb collisions at \sqnn= 2.76 TeV. 
Looking at the top left plot, it can be observed that the range of N$_{\rm part}$ values generated by the Glauber MC for a given value of \textbf{b} is large and it is comparable comparable in size to the full range of \npart values that appear in the plot, except for  \textbf{b} values exceeding 7$\div$8 fm. This is not the case in the top right plot, corresponding to Pb--Pb collisions.\\
Also the multiplicity fluctuations present in p--Pb collisions for a given \npart are large. For instance let's consider the bottom right plot for Pb--Pb   and concentrate on N$_{\rm part}$= 100 and  N$_{\rm part}$= 300: the ranges of multiplicity values generated by the Glauber-NBD simulation are about 2000-400 and 10000-13000, respectively, and do not overlap. Considering the same scatter plot for p--Pb collisions (bottom left) at  N$_{\rm part}$= 10 and N$_{\rm part}$= 20,   the ranges of multiplicity values generated largely overlap. \\
These fluctuations lead to the fact that in p--Pb collisions a centrality classification of  events based on multiplicity may select a sample with an average number on N--N collisions which is biased compared to a sample defined by cuts in the impact parameter \textbf{b}.\\
\begin{figure}[t]
\centering
 \includegraphics[width=0.999\textwidth]{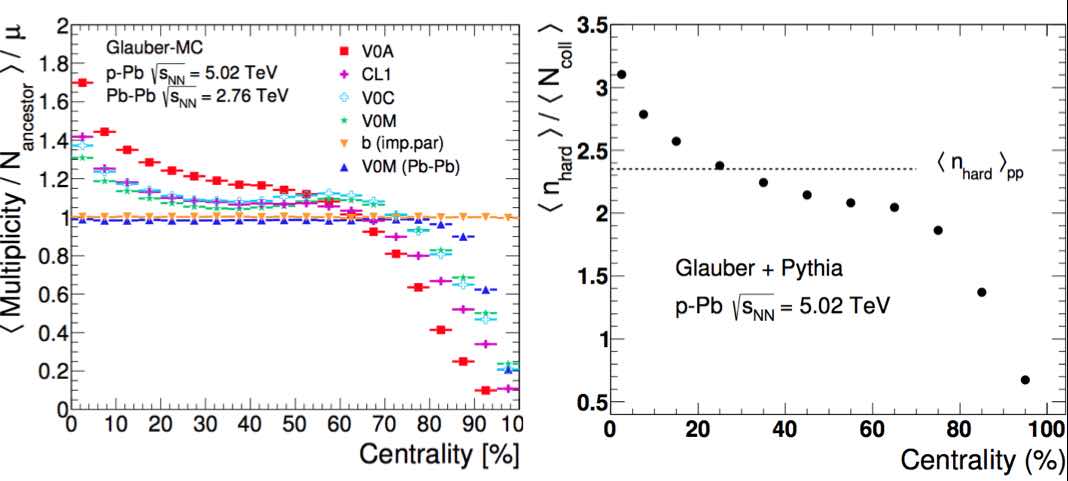}
\caption{Top: ratio between the average multiplicity per ancestor and the average multiplicity of the NBD mean multiplicity $\mu$ as a function of centrality for different estimators based on multiplicity in p--Pb collisions at \sqnn= 5.02 TeV and for V0M estimator in Pb--Pb collisions at \sqnn= 2.76 TeV. Bottom: $\langle$N$_{\rm hard}\rangle$/$\langle$N$_{\rm coll}\rangle$ in the different centrality classes for a  p--Pb data sample at \sqnn= 5.02 TeV generated with Glauber MC + PHYTHIA.}
\label{bias1}
\end{figure}
\begin{figure}[b]
\centering
 \includegraphics[width=0.999\textwidth]{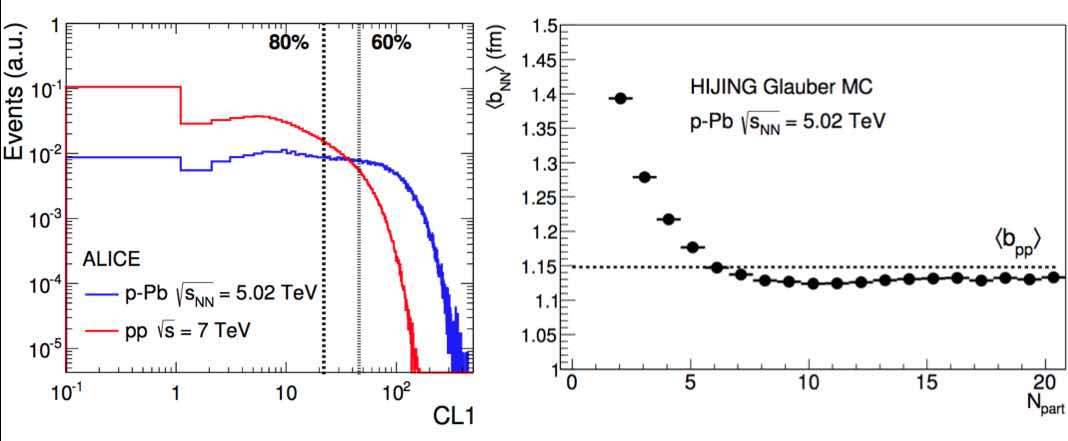}
\caption{Top: CL1  multiplicity distribution  used as centrality estimator in p--Pb collision at \sqnn= 5.02 TeV, compared to the distribution in pp collisions at \sqs= 7 TeV. The dashed lines mark the 80\% and the 60\% percentile of the p--Pb cross-section respectively. Bottom: average impact parameter between the proton and each wounded nucleon of the Pb-nucleus (\textbf{$\langle$b$_{\rm NN}\rangle$}), calculated from a  Glauber MC  simulation implemented in HIJING as a function of N$_{\rm part}$.}
\label{bias2}
\end{figure} 
The left panel of Figure \ref{bias1} shows the ratio between the average multiplicity per ancestor and the average multiplicity of the NBD mean multiplicity $\mu$ as a function of centrality for different estimators based on multiplicity (V0M is the sum of the V0A and V0C amplitudes). In Pb--Pb collisions, where the width of the plateau of the \npart distribution is large with respect to multiplicity fluctuations, the ratio (estimated for the V0M centrality estimator) deviates from unity only for the most peripheral collisions. As expected, in p--Pb collisions the ratio differs from unity for all centralities and all estimators with large deviations for the most central  and most peripheral collisions; the most central (peripheral) collisions have on average much higher (lower) multiplicity per participant. When selecting event classes using impact parameter \textbf{b} intervals, there is no deviation from unity, as expected. \\
The bias on multiplicity also corresponds to a bias on  the average number of hard scatterings N$_{\rm hard}$ with respect to the number of binary nucleon-nucleon collisions \ncoll estimated wtih Glauber MC. Figure  \ref{bias1} (right) shows the results obtained using Glauber MC to generate a sample of p--Pb events at \sqnn= 5.02 TeV, each having \ncoll  binary collisions. For each  generated event, PYTHIA is used \ncoll times  to generate \ncoll independent pp collisions at \sqs= 5.02 TeV. The average number of hard scattering per event in pp collisions at this energy in PYTHIA is $\langle$N$_{\rm hard}^{pp}\rangle\approx$2.4. Dividing the p--Pb data sample generated with Glauber MC + PYTHIA in centrality classes based on multiplicity, and calculating $\langle$N$_{\rm hard}\rangle$/$\langle$N$_{\rm coll}\rangle$ in the different centrality classes, a bias is observed as shown in  Figure \ref{bias1}  (right). Central collisions tend to have on average a higher number of hard scatterings per binary collision.\\
\subsubsection{Jet-veto bias}
High \pt particles contribute to the overall event multiplicity rising with parton energy and can thus introduce a  correlation between the centrality estimator and the presence of a high \pt particle in the event. Figure \ref{bias2} (left) shows the  distribution of the number of clusters in the outer layer of SPD (CL1) for p–Pb collisions, compared to the same distribution in pp collisions at \sqs= 7 TeV. The dashed lines mark the 80\% and the 60\%  percentile of the p–Pb cross-section. The fraction of the  cross collisions having a multiplicity in the range defined by the  80- 100\% (60-100\%) CL1  multiplicity class in p--Pb collisions is 0.8 (0.97). This means that the  80-10\% p--Pb multiplicity class includes events with multiplicity  values lower than those reached in  the 20\% of  pp events with the highest multiplicies. This represent an effective veto on hard processes, leading to a \rppb < 1 for low multiplicity p--Pb collisions.\\
\subsubsection{Geometrical bias}
Another source of bias arises from the geometric configuration of p--Pb interactions:  Figure  \ref{bias2} (right) shows the average impact parameter between the projectile proton and each wounded nucleon of the Pb-nucleus (\textbf{$\langle$b$_{\rm NN}\rangle$}), calculated from a  Glauber MC simulation implemented in HIJING as a function of N$_{\rm part}$. \textbf{$\langle$b$_{\rm NN}\rangle$} is almost constant for central collisions, but rises significantly for N$_{\rm part}$<6. This reduces the average number of interactions for most peripheral events. \\
\vskip 2 em
All the elements discussed above introduce a bias in the determination of \ncoll for a given centrality class. Similarly to \rppb, we define Q$_{\rm pPb}$ as
\begin{equation}
Q_{\textrm{pPb}}^{\textrm{est.}}(\pteq, \textrm{centr})=\frac{dN^{\rm pPb}_{\rm cent}/d\pteq}{\langle N_{\rm coll}^{\rm est.}\rangle dN^{\rm pp}/d\pteq}=\frac{dN^{\rm pPb}_{\rm cent}/d\pteq}{\langle T_{\rm pA}^{\rm est.}\rangle d\sigma^{\rm pp}/d\pteq}
\end{equation}
for a given centrality percentile according to a particular centrality estimator \textbf{"est."} (in our case \ncoll is estimated using V0A or CL1 signals with Glauber-NBD fit or ZNA signal with SNM fit). The \qppb is different from \rppb because it is influenced by potential biases due to the centrality estimator which are not related to nuclear effects. Hence \qppb can differ from
unity even in the absence of nuclear effects. Figure \ref{qpch1} shows the \qppb of charged particles in |$\eta$|<3 measured by the ALICE collaboration in p--Pb collisions at \sqnn= 5.02 TeV as a function of \pt \cite{CentpPb}. Data are divided in seven centrality classes by mean of Glauber-NDB fit applied to V0A amplitude (Q$_{\rm pPb}^{\rm V0A}$ - left) or CL1 distribution (Q$_{\rm pPb}^{\rm CL1}$ - right). The $N_{\rm coll}$ values used are reported in Table \ref{tablencollppb}. \qppb strongly deviates from unity at high \pt in all centrality classes, with values well above unity for central collisions and below unity for peripheral collisions. These results reflect all biases discussed above. However, the spread of \qppb values between centrality classes reduces with increasing rapidity gap between the tracking region and the centrality estimation region, i.e. going from CL1 to V0A results. The classes selected by the ZNA (Figure  \ref{qpch2} - left) present Q$_{\rm pPb}^{\rm ZNA}$ values much more similar to each other and close to unity than those of CL1 and V0A estimators, except for peripheral collisions (60-80\% and 80-100\%). In fact the ZNA does not measure multiplicity and should not be sensitive to the bias sources described above except for the geometrical bias on \textbf{$\langle$b$_{\rm NN}\rangle$}. The behaviour of Q$_{\rm pPb}^{\rm ZNA}$  in 60-100\%  is rather due to a limit in the Slow Nucleon Model   to correctly describe the  most peripheral events. \\
\begin{figure}
\centering
 \includegraphics[width=0.999\textwidth]{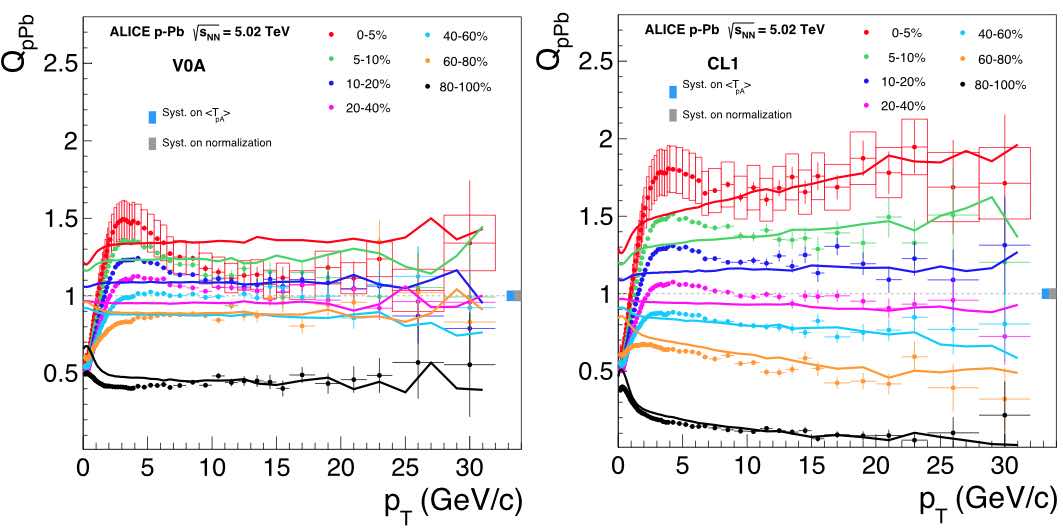}
\caption{Left: Q$_{\rm pPb}^{\rm V0A}$ measured by the ALICE Collaboration in p--Pb collisions at \sqnn = 5.02 TeV \cite{CentpPb}. Right: Q$_{\rm pPb}^{\rm CL1}$ measured by the ALICE Collaboration in p--Pb collisions at \sqnn = 5.02 TeV \cite{CentpPb}.}
\label{qpch1}
\end{figure}
The \qppb results show that the bias in the determination of \ncoll depends on the rapidity gap between the regions where centrality is estimated and the regions in which the charged particles production is measured. We will now discuss this rapidity dependence in more detail.\\ 
To provide an unbiased centrality estimator, the ALICE Collaboration has developed a method (\textbf{hybrid} method in the following) that relies on two main assumptions: 
\begin{itemize}
\item a centrality event selection based on ZN does not introduce any bias on the bulk of particle production at mid-rapidity and on high-\pt particle production
\item charged-particle multiplicity measured at mid-rapidity scales with \npart
\end{itemize}
The first step is to study the scaling law of particle production in different pseudorapidity regions. To do this we divide the events in seven centrality classes using the energy deposition in the Zero-Degree Neutron Calorimeter in the A side (ZNA), as shown in Figure \ref{CentrpPb2}. For each centrality class $i$, the average value of the following signals $\langle s_i\rangle$ is considered
\begin{itemize}
\item  charged-particle density d$N_{\rm ch}$/d$\eta$ in    -1<|$\eta$|<0, measured with  SPD  tracklets (Section 3.1.1)
\item  charged-particle density d$N_{\rm ch}$/d$\eta$ in    -2<|$\eta$|<-1.5, measured with  SPD  tracklets (Section 3.1.1)
\item  charged-particle density d$N_{\rm ch}$/d$\eta$ in    1.5<|$\eta$|<2, measured with  SPD  tracklets (Section 3.1.1)
\item raw signal in the inner ring of the V0A detector
\item raw signal in the inner ring of the V0C detector
\end{itemize}
The average values $\langle s_i\rangle$ are then compared to the average signal in minimum bias collisions to get  the normalized signals $\langle s_i\rangle$/$\langle s_{\rm MB}\rangle$. Figure \ref{qpch2} (right) shows the correlation  of these ratios for the seven ZNA centrality classes considered (centrality decreasing from left to right).  One can note that the correlation exhibits a clear dependence on the pseudorapidity region in which the normalized signal has been measured. Using the assumption that charged-particle multiplicity measured at mid-rapidity scales with N$_{\rm part}$, d$N_{\rm ch}$/d$\eta |_{-1<|\eta|<0}\propto$ \npart, we look for a  relation between the other observables in Figure \ref{qpch2} (right) and \npart, assuming a  linear  dependence.\\
\begin{figure}
\centering
 \includegraphics[width=0.999\textwidth]{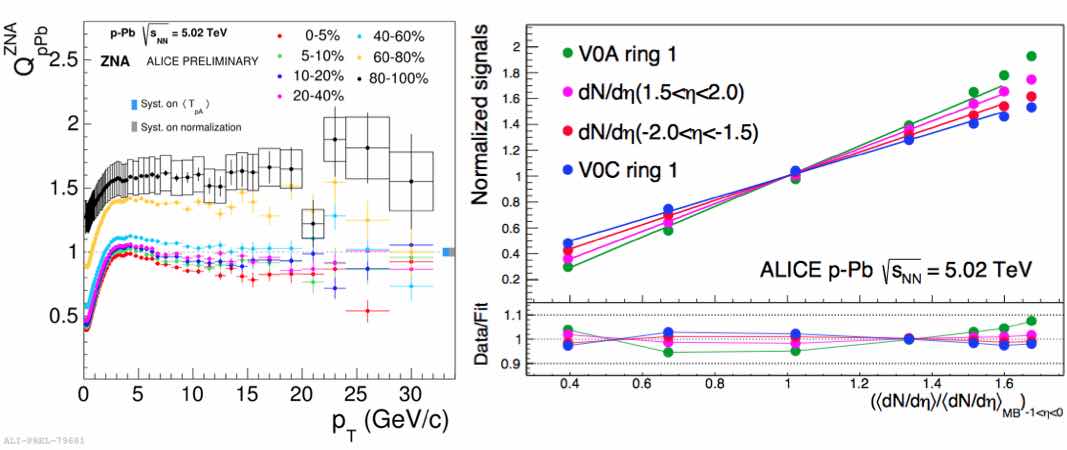}
\caption{Left: Q$_{\rm pPb}^{\rm ZNA}$ measured by the ALICE Collaboration in p--Pb collisions at \sqnn = 5.02 TeV \cite{CentpPb}. Right: normalized signals $\langle s_i\rangle$/$\langle s_{\rm MB}\rangle$ for V0A and V0C (inner ring) and charged particle density in -2<|$\eta$|<-1.5 and 1.5<|$\eta$|<2 measured with SPD, vs normalized signals for charged particle density in -1<|$\eta$|<0 measured with SPD. Each set of data points is relative to one of the centrality classes measured with ZNA defined in Figure \ref{CentrpPb2}. Centrality increases from left to right.}
\label{qpch2}
\end{figure}
The linear dependence can be parameterized as  $\langle s_i\rangle\propto$ \npart - $\alpha$, so that $\alpha$=0 is equivalent to $\langle s_i\rangle\propto$ \npart and $\alpha$=1 is equivalent to $\langle s_i\rangle\propto$ N$_{\rm coll}$. The normalized signals can then be expressed as 
\begin{multline}
\langle s_i\rangle/\langle s_{\rm MB}\rangle=\frac{(\langle N_{\rm part}\rangle_i - \alpha)}{(\langle N_{\rm part}\rangle_{\rm MB} - \alpha)} = \\ = \frac{\langle N_{\rm part}\rangle_{\rm MB} }{(\langle N_{\rm part}\rangle_{\rm MB} - \alpha)}\cdot  \bigg (\frac{\langle dN_{\rm ch}/d\eta \rangle_i}{\langle dN_{\rm ch}/d\eta  \rangle_{\rm MB}}\bigg )_{-1<\eta<0} - \frac{\alpha}{(\langle N_{\rm part}\rangle_{\rm MB} - \alpha)}
\end{multline}
where $\langle N_{\rm part}\rangle_{\rm MB}$ = 7.9 is the average number of participating nucleons in minimum bias collisions. The relation is used to find $\alpha$ for each of the normalized signals by a fit to the data. The ratio of the data and the fit results are shown in the lower panel of Figure \ref{CentrpPb2} (left).\\
Figure \ref{qpch3} shows the results for $\alpha$  as a function of the pseudorapidity in the centre-of-mass frame, $\eta_{\rm cms}$. To increase the pseudorapidity coverage, data from both p--Pb and Pb--p collisions have been used. Negative pseudorapidity values in the c.m.s. correspond to the Pb going direction in both configurations. At mid-rapidity, the fit is also performed for high-\pt particles (red cross in Figure \ref{qpch3}). \\ The result of the fits indicates a smooth and continuous change of the scaling behaviour for charged particle production with pseudorapidity. High \pt particle production at midrapidity scales  almost proportionally to  \ncoll ($\alpha\approx$1), as already observed when discussing the Q$_{\rm pPb}^{\rm ZNA}$ in 0-60\% centrality.   At large negative pseudorapidity (Pb-going direction) the value of the parameter $\alpha$  obtained  from fits to V0 normalized signals gets close to unity, indicating that the scaling behaviour approaches \ncoll scaling. \\
\begin{figure}[t]
\centering
 \includegraphics[width=0.999\textwidth]{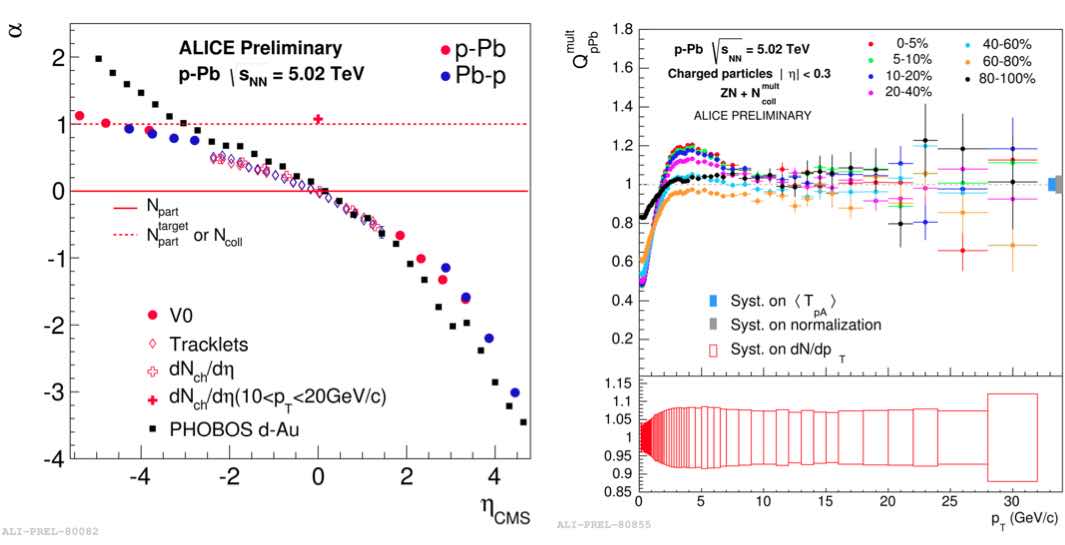}
\caption{Left:  results for $\alpha$  as a function of the pseudorapidity in the centre-of-mass frame, $\eta_{\rm cms}$. Data from both p--Pb and p--Pb collisions are used. Negative pseudorapidity values in the c.m.s. correspond to the Pb going direction in both configurations. At mid-rapidity, the fit is also performed for high-\pt particles (red cross). Right:  Q$_{\rm pPb}^{\rm mult}$ measured by the ALICE Collaboration in p--Pb collisions at \sqnn = 5.02 TeV \cite{CentpPb}. Hybrid method is used to estimate \ncoll.}
\label{qpch3}
\end{figure}
The evolution of the fit parameter $\alpha$ reported in Figure \ref{qpch3} (left) gives an explanation of the bigger bias observed for CL1: under the assumption that the ZNA signal is unbiased,  Figure \ref{qpch3} demonstrates that the signals of V0A and CL1 centrality estimators do not scale with \ncoll, and that the signal of CL1  deviates more from the \ncoll scaling law than that of V0A.\\
We can then easily extract the $\langle N_{\rm coll}\rangle_i$ relative to each ZNA centrality class $i$ in the hybrid approach as
\begin{equation}
\langle N_{\rm coll}\rangle_i^{\rm mult} =  \langle N_{\rm part}\rangle_i - 1 =  \langle N_{\rm part}\rangle_{\rm MB} \cdot  \bigg (\frac{\langle dN_{\rm ch}/d\eta \rangle_i}{\langle dN_{\rm ch}/d\eta  \rangle_{\rm MB}}\bigg )_{-1<\eta<0} - 1
\end{equation}
The $\langle N_{\rm coll}\rangle_i^{\rm mult}$ obtained with this method are shown in Table \ref{hybrid}. These values agree within 15\% with those obtained with the SNM approach applied to ZNA signal, except for the most peripheral regions where  the Slow Neuclon Model is inaccurate.  We can now use these  $\langle N_{\rm coll}^{\rm mult}\rangle_i$  values to compute $Q_{\rm pPb}^{\rm mult}$ in ZNA centrality classes. \\ The results of Q$_{\rm pPb}^{\rm mult}$ in the hybrid approach are shown in  Figure \ref{qpch3} (right) and remain consistent with unity at high-\pt for all centrality classes. This confirms the absence of initial state effects, already observed for minimum bias collisions. The Cronin enhancement, which has already been noted in minimum bias collisions, is observed to be stronger in central collisions and nearly absent in peripheral collisions. The geometry bias is still present and uncorrected, even with this method. Its effect is limited to  peripheral classes only, resulting in $Q_{\rm pPb}^{\rm mult}$ < 1 for 80-100\%.\\
In summary, we have seen how the determination of centrality classes in p--Pb collisions is affected by several sources of bias on the extraction of $\langle N_{\rm coll}\rangle$. This  forces us to question the usage of the word "centrality", therefore we will use instead  the term "event activity" or "multiplicity" in the next chapters.  
\begin{table}[h]
\centering
\begin{tabular}{|c|c|}
\hline
Centrality (\%) & $N_{\rm coll}^{\rm mult}$ \\ \hline
0-5             & 12.2                      \\ \hline
5-10            & 11.6                      \\ \hline
10-20           & 11.0                      \\ \hline
20-40           & 9.56                      \\ \hline
40-60           & 7.08                      \\ \hline
60-80           & 4.30                      \\ \hline
80-100          & 2.11                      \\ \hline
\end{tabular}
\caption{$\langle N_{\rm coll}\rangle_i^{\rm mult}$ values obtained in the hybrid approach.}
\label{hybrid}
\end{table}
 
\chapter{Reconstruction of prompt D$^+$ mesons in p--Pb collisions} 
\lhead{Chapter 4. \emph{Reconstruction of prompt D$^+$ mesons in p--Pb collisions}} 
\begin{figure}[b]
\centering
 \includegraphics[width=0.60\textwidth]{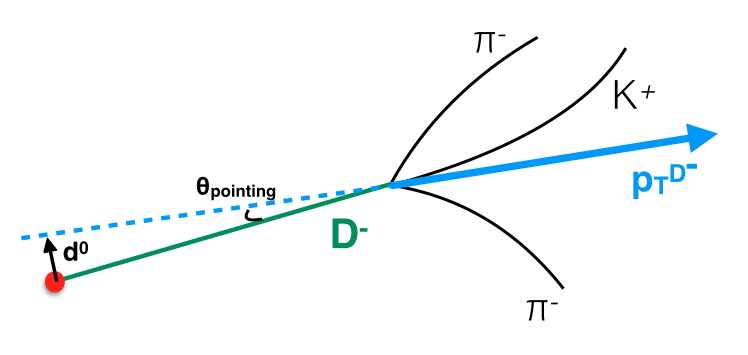}
\caption{Sketch of D$^+$ to $K^{-}\pi^{+}\pi^{+}$ hadronic decay.}
\label{decaysketch}
\end{figure}
In this chapter I describe the general strategy used to extract the prompt D$^{+}$ meson signal from  the data sample of p--Pb collisions at \sqnn=5.02 TeV shown in Chapter 3.\\
In the analysis I consider the hadronic decay of D$^+$ into $K^{-}\pi^{+}\pi^{+}$ and its charge conjugate (D$^-$ into $K^{+}\pi^{-}\pi^{-}$). The relevant quantities that characterize this  meson  are shown in Table \ref{tab:decays}\footnote{PDG average \cite{PDG}}  . The same informations are  given  for the other D-meson species reconstructed up to now with the ALICE detector (D$^0$, D$^{*+}$, D$^{+}_{s}$). The  branching ratio\footnote{About 20\% of this branching ratio is given by resonance decays, namely D$^{+}\rightarrow \bar{K}^{*}_{0}(1430)\pi^{+} \rightarrow K^{-}\pi^{+}\pi^{+}$(BR=1.21\%), D$^{+}\rightarrow \bar{K}^{*}(892)\pi^{+} \rightarrow K^{-}\pi^{+}\pi^{+}$ (BR=1.01\%), D$^{+}\rightarrow \bar{K}^{*}(1680)\pi^{+} \rightarrow K^{-}\pi^{+}\pi^{+}$ (BR=2.1 10$^{-4}$\%) and D$^{+}\rightarrow \bar{K}^{*}_{2}(1430)\pi^{+} \rightarrow K^{-}\pi^{+}\pi^{+}$ (BR=2.2 10$^{-4}$\%) } is relatively high, compared to the other decays shown in Table \ref{tab:decays},  and the mean life assures an average decay length of few  hundreds $\mu$m. \\
\begin{table}[t]
\begin{tabular}{|l|l|l|l|l|}
\hline
Particle                       & Mass (MeV/c$^2$)                   & Decay Channel                                 & Branching Ratio (\%)                      & Mean Life (s)                                \\ \hline
D$^{0}$                        & 1864.84$\pm$0.07                      & $K^{-}\pi^{+}$                                & ( 3.88 $\pm$ 0.05 )                       & (410.1$\pm$1.5) 10$^{-15}$                     \\ \hline
{\color[HTML]{FE0000} D$^{+}$} & {\color[HTML]{FE0000} 1869.5$\pm$0.4} & {\color[HTML]{FE0000} $K^{-}\pi^{+}\pi^{+}$}  & {\color[HTML]{FE0000} ( 9.13 $\pm$ 0.19)} & {\color[HTML]{FE0000} (1040$\pm$7) 10$^{-15}$} \\ \hline
D$^{*+}$                       & 2010.26$\pm$0.07                      & D$^{0}\pi^{+}\rightarrow K^{-}\pi^{+}\pi^{+}$ & (67.7 $\pm$ 0.5)                          & strong decay                                 \\ \hline
D$^{+}_{s}$                    & 1969.0$\pm$1.4                      & $\Phi\pi^{+}\rightarrow K^{+}k^{-}\pi^{+}$    & (2.32 $\pm$ 0.14)                         & (500$\pm$7) 10$^{-15}$                         \\ \hline
\end{tabular}
\caption{Characteristics of D mesons hadronic decays reconstructed with the ALICE detector}
\label{tab:decays}
\end{table}
Looking at the D$^{-}$ decay sketch in Figure \ref{decaysketch} we define few quantities that will be used in the following:
\begin{itemize}
\item \textbf{Pointing angle $\theta_{\rm pointing}$}: the angle between the D-meson flight line (joining the primary and the secondary vertices) and the direction of the reconstructed D-meson momentum. Usually the cosine of this quantity is used 
\item \textbf{d$_0$}: the impact parameter of the D-meson
\item \textbf{decay length}: the distance between the primary and the decay vertex
\end{itemize}
It is also interesting to estimate  the momentum of the three daughters: considering the 3-body decay in the rest frame of the mother meson, kinematic calculations give
\begin{equation}
\begin{aligned}
p_{K,max}=\frac{1}{2M_{D^+}}\sqrt{[M_{D^+}^2 - (m_{K}+2m_{\pi})^2][M_{D^{+}}^2 - (2m_{\pi}-m_{K})^2]}\\
p_{\pi,max}=\frac{1}{2M_{D^+}}\sqrt{[M_{D^+}^2 - (2m_{\pi}+m_{K})^2][M_{D^{+}}^2 - (m_{\pi}+m_{K}-m_{\pi})^2]}
\end{aligned}
\end{equation}
where $M$ is the mass of the mother particle and $m_{\pi /K}$ the mass of the daughters. 
Figure \ref{ptDvsptDaughters} shows the scatter plot of the D$^{+}$ mesons \pt vs the \pt of its daughters and the \pt of the daughter with the lowest momentum. These simulations are obtained with PYTHIA 6. \\  
\begin{figure}[b]
\centering
 \includegraphics[width=0.60\textwidth]{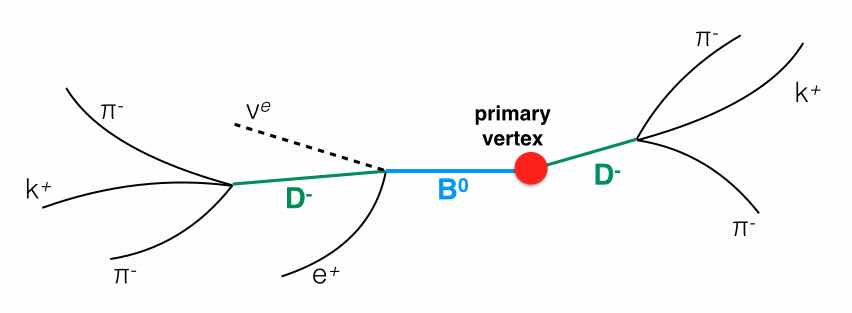}
\caption{Sketch of prompt and feed-down D$^+$ to $K^{-}\pi^{+}\pi^{+}$ hadronic decay}
\label{decaysketchpfd}
\end{figure}
Also important is the word "prompt" appearing in the title of this chapter. Prompt \dplus mesons are those coming from the hadronization of a $c$-quark, or from the decay of an excited D-meson state such as D$^{*+}$(2010). Figure \ref{decaysketchpfd} shows the decay sketch for a prompt D$^{-}$ meson (right) and a feed-down  D$^{-}$ meson. The latter comes from the decay of a B hadron, in this case a semi-leptonic decay of   B$^0$-meson\footnote{all B meson states have significant inclusive branching fraction into D-meson} and is on average more displaced from the primary vertex, given the typical  mean lives of B mesons ($\sim$10$^{-12}$ s).\\
In  this chapter we will show the strategy used to extract the D meson signal from the 2013 p--Pb data sample.
\section{D$^{+}$ meson signal in p--Pb collisions}
\begin{figure}[t]
\centering
 \includegraphics[width=0.97\textwidth]{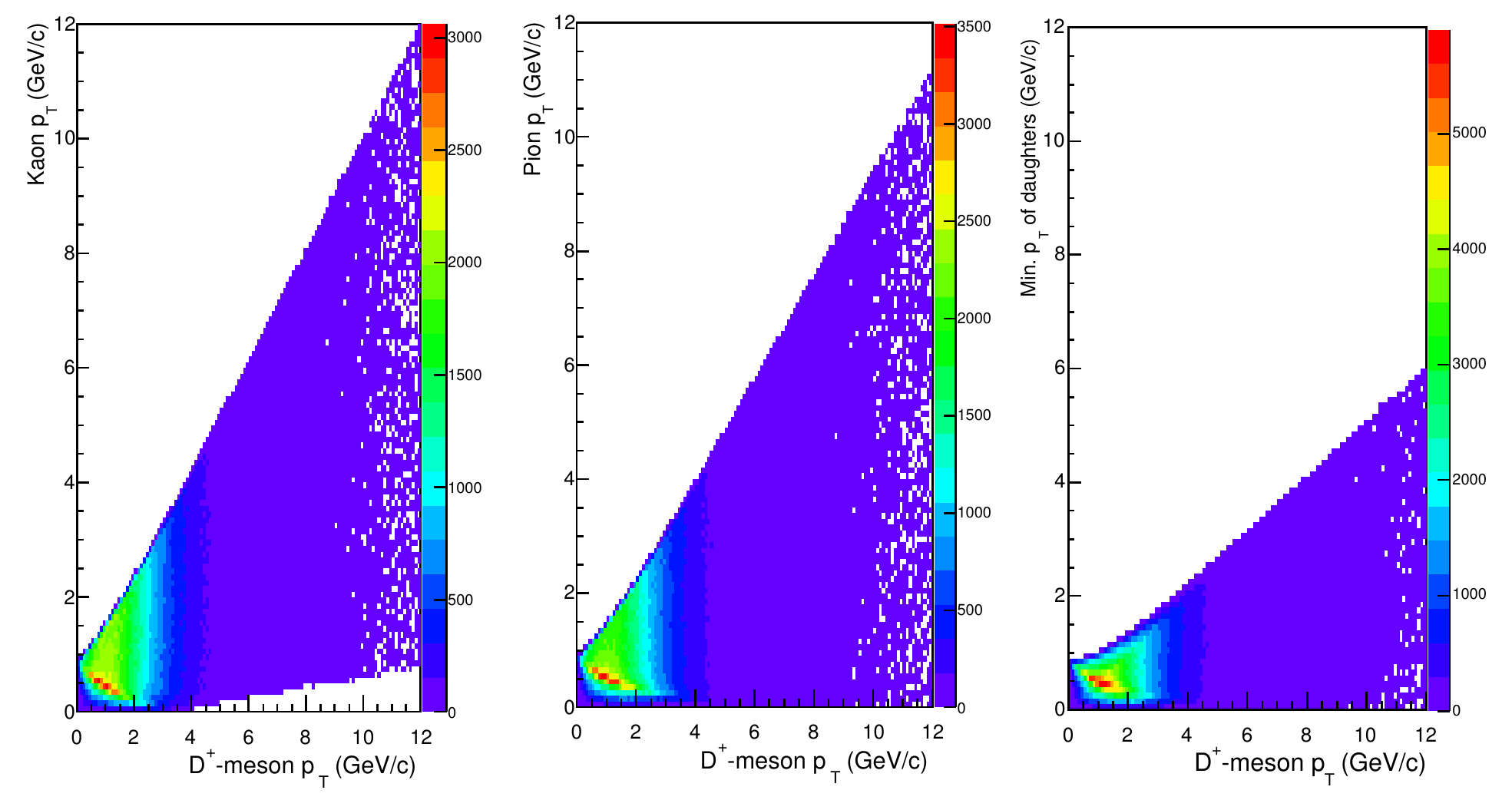}
\caption{Left: scatter plot of  D$^{+}$ mesons \pt vs the \pt of the  decay kaons. Centre: scatter plot of  D$^{+}$ mesons \pt vs the \pt of the  decay pion with the highest momentum. Right: scatter plot of  D$^{+}$ mesons \pt vs the \pt of the  decay particle with the lowest momentum.}
\label{ptDvsptDaughters}
\end{figure}
\begin{figure}[b]
\centering
 \includegraphics[width=0.55\textwidth]{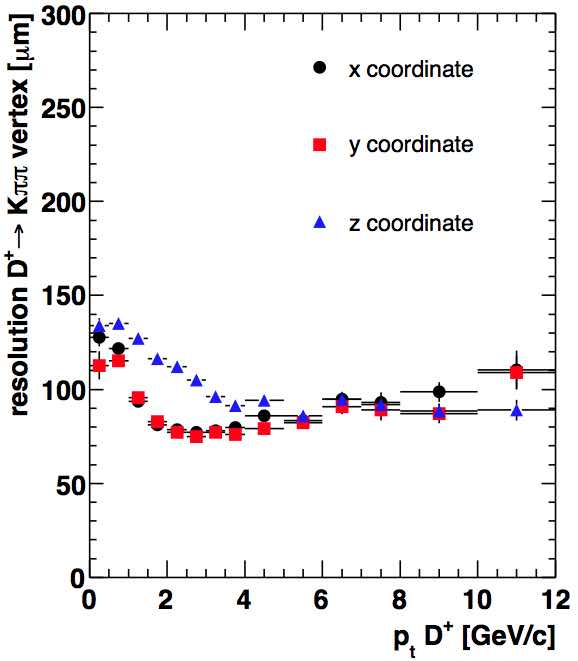}
\caption{$x$, $y$ and $z$ resolution on secondary vertex postition as a function of \dplus \pt \cite{secver}.}
\label{SecVerRes}
\end{figure}
The signal coming from D$^{+}$ mesons is obtained via an invariant mass analysis of $K^{-}\pi^{+}\pi^{+}$ decay topologies. \\
The first step of this analysis is the identification, for each event, of secondary vertices from  triplets of tracks with the proper sign combination: two positive tracks and one negative in the case of D$^{+}$ candidates and the opposite to obtain D$^{-}$ candidates. Tracks are approximated as straight lines in the vicinity of the primary vertex and the coordinates of the primary vertex are given by finding the point of minimum distance among the tracks. The resolution on the secondary vertex position was estimated in \cite{secver} for pp collisions and is shown in Figure \ref{SecVerRes}. For low \pt  D$^{+}$, the daughter particles have low momentum and suffer multiple scattering, resulting in a poorer  resolution on the secondary vertex position. At high p$_{\rm T}$, an improvement of the resolution would be expected since  tracks have large momentum and are less affected by multiple scattering, however the decay particles are more collinear with the direction of the parent D$^{+}$ momentum and  this makes the determination of the vertex position less precise. \\
 For each vertex the invariant mass is obtained as
\begin{equation}
M_{\rm inv}=\sqrt{(E_K+E_{\pi_1}+E_{\pi_2})^2-(\vec{p_K}+\vec{p_{\pi_1}}+\vec{p_{\pi_2}})^2}
\label{eqinvmass}
\end{equation}
where the energy values $E_K,E_{\pi_1},E_{\pi_2}$ are the energies of the kaon and of the pions, respectively. The D$^{+}$ candidates  \pt is obtained as
\begin{equation}
\pteq^{D^+}= \pteq^K + \pteq^{\pi_1} + \pteq^{\pi_2}
\end{equation}
 Very important in the D$^{+}$ case is the fact that the decay configuration allows only one possible value for the invariant mass, since the kaon mass in Equation \ref{eqinvmass} is always assigned to the track with opposite charge sign with respect to the other two. This is not the case e.g. in the D$^{0} \rightarrow K^{-}\pi^{+}$, where each candidate vertex has two values of the invariant mass. \\
\subsection{Event Selection}
\label{sub:EvSel}
The analysis is performed on  $\sim$ 110 10$^{6}$  p--Pb collisions collected with a minimum bias trigger requiring signal in both V0 scintillators (Section \ref{sub:Trigger} - Figure \ref{trigger}).
Only events  that are not rejected by the physics selection cuts described in  Section \ref{sub:backrej} are selected for the analysis.\\
A further selection is performed on the basis of the reconstructed vertex. From this point of view events can be divided  in four groups:
\begin{itemize}
\item events where no vertex has been  reconstructed
\item  events where the vertex has been reconstructed with global (ITS+TPC and TPC only) tracks (Section \ref{sub:tvrec})
\item  events where the vertex has been reconstructed only with SPD tracklets and the 3D ($x,y,z$) coordinates have been determined (Section \ref{sub:tvrec})
\item  events where the vertex has been reconstructed only with SPD tracklets and  only the $z$ coordinate has been determined (this happens in case the 3D SPD vertexing algorithm fails, e.g. at low multiplicity) (Section \ref{sub:tvrec})
\end{itemize} 
The reconstruction of the vertex with SPD has higher efficiency due to the wider $\eta$ coverage of the SPD and of the less stringent request applied to tracklets w.r.t. tracks in the vertex calculations. The efficiency of vertex reconstruction $\epsilon^{vertex}$ was estimated in minimum bias pp collisions at \sqs= 7 TeV. For the SPD vertex reconstruction (3D+$z$)  $\epsilon^{vertex}_{\rm SPD}\approx$ 96\%, while the reconstruction based on tracks has $\epsilon^{vertex}_{\rm TRK}\approx$ 81\%.\\
\begin{figure}[t]
\centering
 \includegraphics[width=0.97\textwidth]{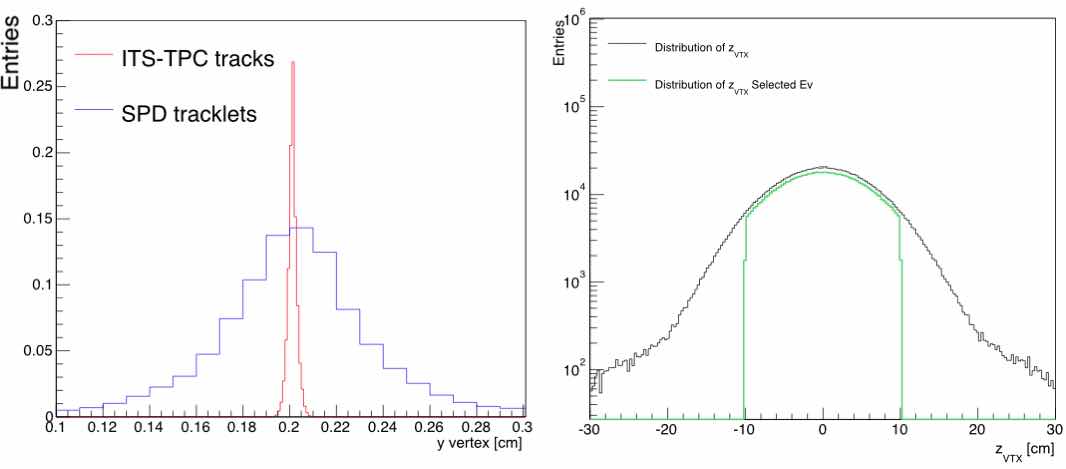}
\caption{Left:  distribution of the $y$ coordinate of vertices reconstructed with SPD(blue) and global tracks (red). Right: distribution  of the $z$ coordinate of  vertices reconstructed with global tracks before (black) and after (green) the selction used in this analyis.}
\label{VertexSel}
\end{figure}
In the analysis we only keep  events with vertex reconstructed using global tracks. The reason for this choice is that this class of vertices has a better resolution in the transverse plane compared to vertices reconstructed with the SPD tracklets, as it can be seen in   Figure \ref{VertexSel} (left) that shows  the distribution of the $y$ position of the reconstructed vertices with SPD and global tracks. This better resolution is crucial in the reconstruction of \dplusm, since as we will see in the next section, the topological selections applied rely on a precise measurement of the primary and secondary vertex position.\\
Only vertices in a $|z|$<10 cm window are used  in the analysis, this to keep  a uniform acceptance on the  tracks reconstructed by ITS and TPC in |$\eta$|<0.8. \\ Figure \ref{VertexSel}  (right) shows the distribution of the $z$ coordinates for vertices reconstructed with global tracks before and after the event selection.  The sharp cut at 10 cm is visible in the distribution of selected events.\\
 The inclusion of the \dplusm decay tracks introduces a bias in the calculation of the primary vertex position and covariance matrix. For this reason, in this analysis for each \dplusm candidate the primary vertex was recomputed excluding its decay tracks.\\
In the p--Pb minimum bias data sample used in this analysis the amount of pile-up events still present after the background rejection described in Section \ref{sub:backrej}  is low, below the per cent level. Events are identified as pile-up and rejected if the SPD vertexer has reconstructed two vertices at a minimum distance of 0.8 cm, the lower multiplicity  vertex having at least 5 contributors.\\
\begin{figure}
\centering
 \includegraphics[width=0.97\textwidth]{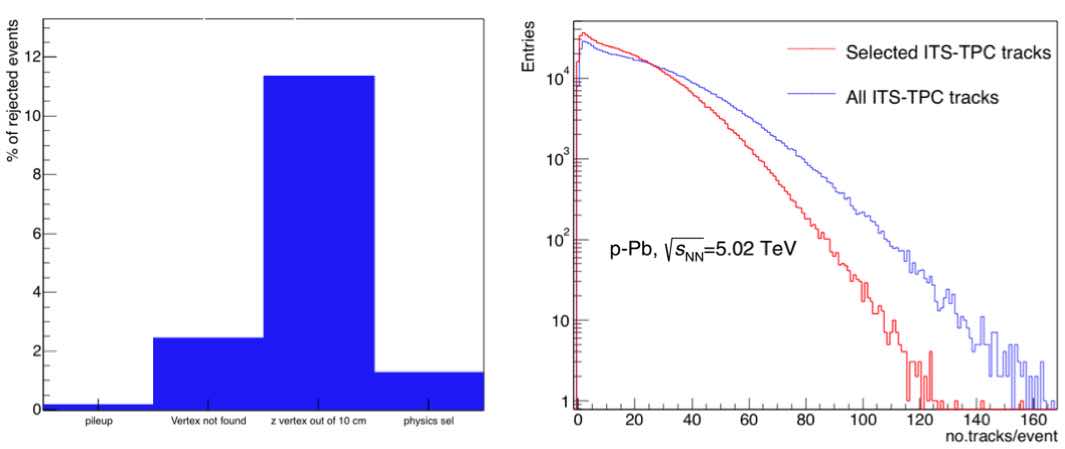}
\caption{Left: fractions of events rejected during vertex and physics selection. Right: distribution of tracks per event before (blue) and after(red) the single track selections, for events passing the selections described in Section \ref{sub:EvSel}.}
\label{VertexRej}
\end{figure}
Figure \ref{VertexRej} (left) shows the percentage of events rejected at the different selection steps mentioned above. The most important contribution comes from the vertex position cut, rejecting $\sim$ 12\% of the events. 
\subsection{Track Selection}
\label{sub:TrackSel}
The secondary vertices of D meson candidates are reconstructed using ITS-TPC tracks with $|\eta|$<0.8, \pt>0.3 GeV/c. Selected tracks are required to have been successfully fitted by the Kalman Filter algorithm during the final inward propagation (from outer detectors to inner ones - see Section \ref{sub:tvrec}) both in the ITS and TPC; these conditions in ALICE software framework are called kITSrefit and kTPCrefit. \\
Furthermore, selected tracks have at least 70 (out of a maximum of 159) associated clusters in the TPC, and a $\chi^2$/$ndf$<2 for the track momentum fit in the TPC. \\
Tracks are also required to have at least one associated cluster in the  SPD layers. This condition improves the impact parameter resolution of the daughter tracks and of the secondary vertex, since it ensures a spatial measurement as close as possible to the primary and secondary decay vertices.\\
Figure \ref{VertexRej} (right) shows the distribution of number of the ITS-TPC tracks per selected event before and after the track selection cuts. After track selections on average $\sim$18 tracks per selected event are kept for  further analysis. \\
The black histogram in Figure \ref{InvmassNoSel}  shows the invariant mass distribution for  D$^{+}$ and D$^{-}$ candidates in the \pt bin 2<$\rm p_{T}^{D^{+}}$<3 GeV/c obtained after applying the event and track selection cuts discussed so far.   No peak is visible in the D$^{+}$ mass region. In fact, a number of selections has to be performed on \dplusm candidates to sufficiently reduce the background and make the signal peak emerge, as  in the red histogram of the same figure, also corresponding to D$^{+}$ and D$^{-}$ candidates in the \pt bin 2<$\rm p_{T}^{D^{+}}$<3 GeV/c.  These selections will be discussed in the following paragraphs.
\begin{figure}[b]
\centering
 \includegraphics[width=0.6\textwidth]{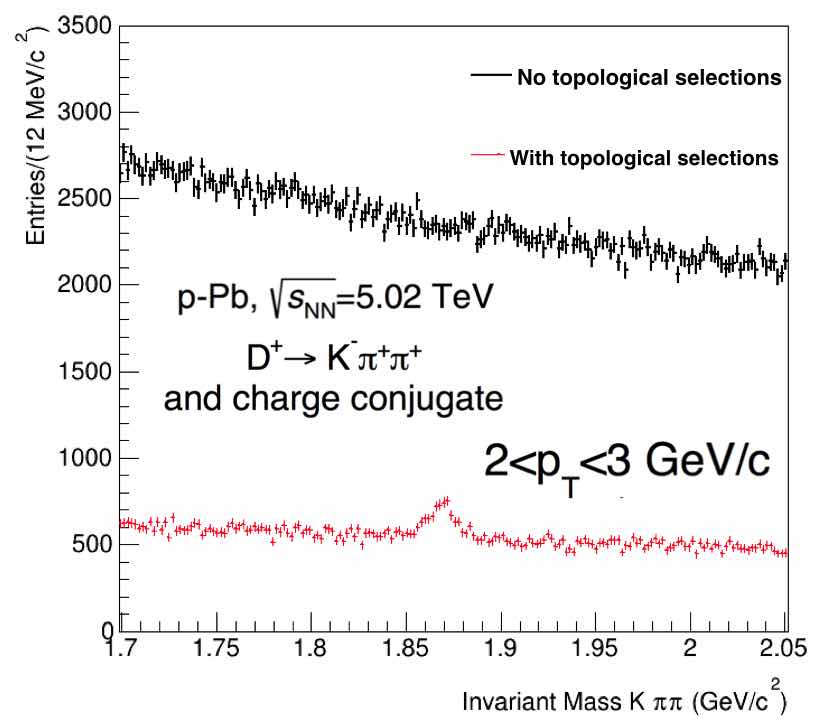}
\caption{Invariant mass spectrum of D$^{+}$ candidates before and after applying topological selections.}
\label{InvmassNoSel}
\end{figure}
\subsection{Kinematical and topological selection}
\label{sub:TopSel}
\begin{figure}[t]
\centering
 \includegraphics[width=0.99\textwidth]{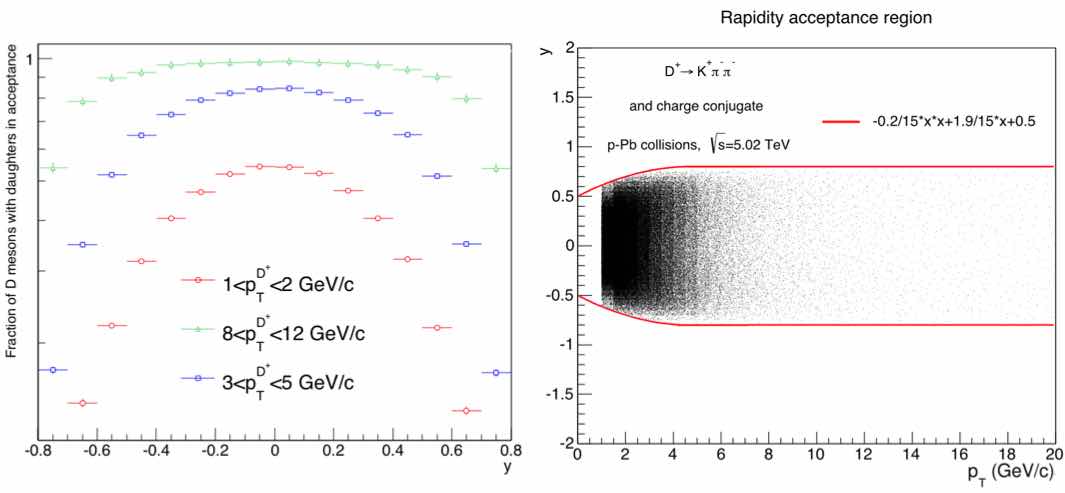}
\caption{Left: fraction of D$^{+}$ mesons with daughters in  |$\eta$|<0.8 and \pt> 0.1 acceptance region vs  D$^{+}$ meson rapidity. Right: \pt dependent y fiducial cut.}
\label{DYAcc}
\end{figure}
\begin{figure}[b]
\centering
 \includegraphics[width=0.99\textwidth]{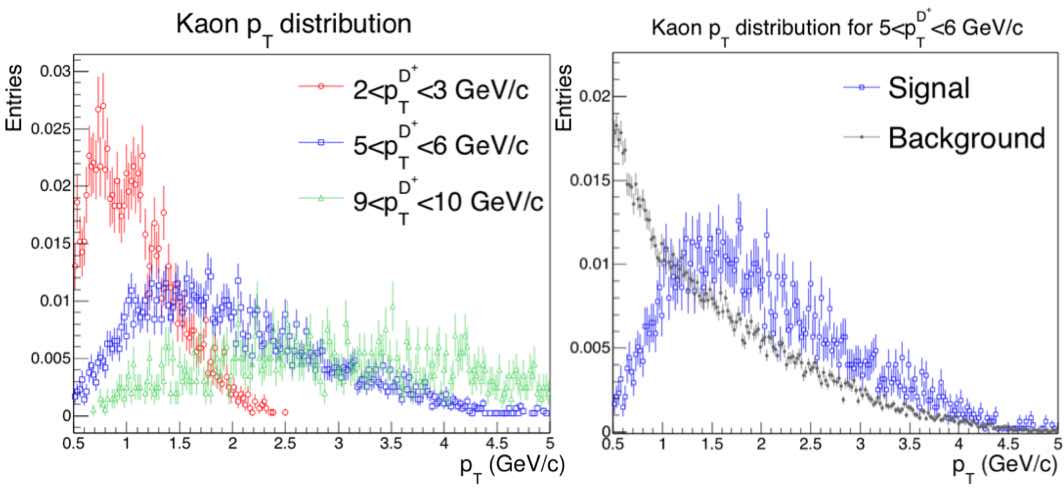}
\caption{Distributions of the decay kaon transverse momentum}
\label{kaon}
\end{figure}
In this section we will describe all selections that are perfomed at the level of D$^{+}$ candidate, based on the kinematical and topological characteristics of the reconstructed decay. This selections exploit the different  distribution of topological and kinematic variables for signal and background  candidates. \\ 
Figure \ref{DYAcc}  (left) shows the $y$-differential fraction of D$^{+}$ that have all daughters in  |$\eta$|<0.8, obtained from the same MC simulations that will be described in detail in Section 4.2. One can see that low \pt  D$^{+}$ mesons have a lower acceptance, since decay tracks have a larger dispersion around the D$^{+}$ meson flight direction. The acceptance further decreases towards higher absolute values of D$^{+}$ meson rapidity (|y|>0.5), where the D$^{+}$ meson flight line approaches the borders of the tracking region. D$^{+}$-meson candidates kept for the analysis have a \pt dependent fiducial cut on their rapidity values. The cut values ranges from |y|<0.5 at low \pt (1<\pt<2) to |y|<0.8 above 4 GeV/c and is visualized by the red lines in the right panel of Figure  \ref{DYAcc}. \\
Figure \ref{kaon} (left) shows the   distributions (normalized to their area) of transverse momentum of kaons coming from D$^{+}$ meson decays from MC simulations for three different intervals of \dplus mesons \pt. One observes that at higher \pt  of the parent D$^{+}$  the distribution becomes harder and peaks at higher values of kaon \pt. The right panel of Figure \ref{kaon}  shows the same \pt distributions (normalized to their area) for signal and background candidates for \dplus with   5<\pt<6 GeV/c from MC simulations. The background distribution is softer, so that a lower cut on the kaon transverse momentum can help excluding a significant fraction of background keeping most of the signal candidates.\\
 The situation for the pion transverse momentum distributions is similar if one considers the pion emitted with the highest momentum, which distribution is shown in Figure \ref{pion} (left). When looking at the pion emitted with  lowest momentum, the distribution is softer and more similar to the background distribution (Figure \ref{pion} - left).\\
\begin{figure}[t]
\centering
 \includegraphics[width=0.99\textwidth]{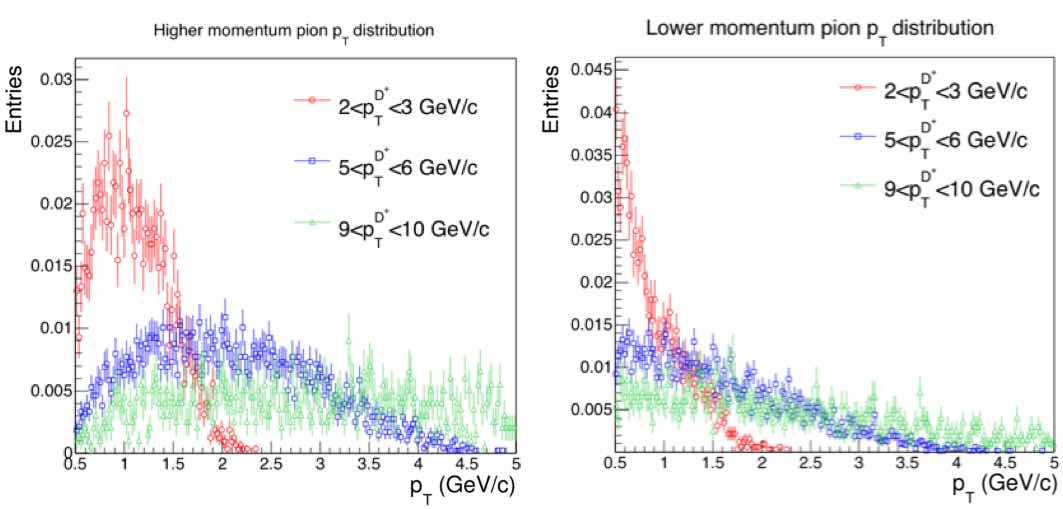}
\caption{Distributions  of the decay pion transverse momentum}
\label{pion}
\end{figure}
Figure \ref{declplots} (left) shows the  distributions (normalized to their area) of the decay length, measured as the 3D distance of the primary and secondary vertices,  for three different prompt D$^+$ meson \pt intervals.     The decay length in the different \pt intervals is governed by the mean proper decay length and by the Lorentz boost: $\gamma * c\tau$. In the case shown here, $\gamma \sim 1.4$ for 1<\pt<2, $\gamma \sim 3$ for 5<\pt<6, $\gamma \sim 6$ for 8<\pt<12. The comparison with the decay length  distributions of background D$^+$ meson candidates, shown in the right panel of Figure \ref{declplots}, shows that a cut on the decay length can be useful to reject a significant amount of background. However this cut has to be handled carefully if the purpose is to select prompt D$^+$ mesons, since the distribution of feed-down candidates extends to even larger decay lengths, as seen in the left panel of Figure \ref{NormDecL}. In fact in the feed-down case the total distance between the primary and secondary vertices of the D$^+$ mesons  includes also the B-meson decay length, which is tipically of the order of several hundreds of $\mu$m ( B$^+$ $c\tau$=491 $\mu$m,  B$^0$ $c\tau$=457 $\mu$m).\\
\begin{figure}[t]
\centering
 \includegraphics[width=0.99\textwidth]{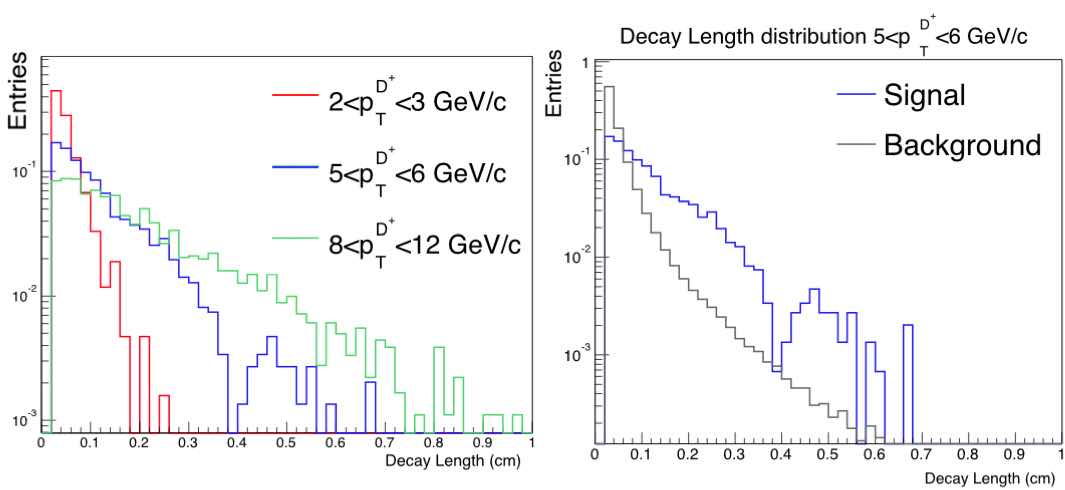}
\caption{Distributions of the decay length of D$^+$ mesons.}
\label{declplots}
\end{figure}
\begin{figure}[b]
\centering
 \includegraphics[width=0.99\textwidth]{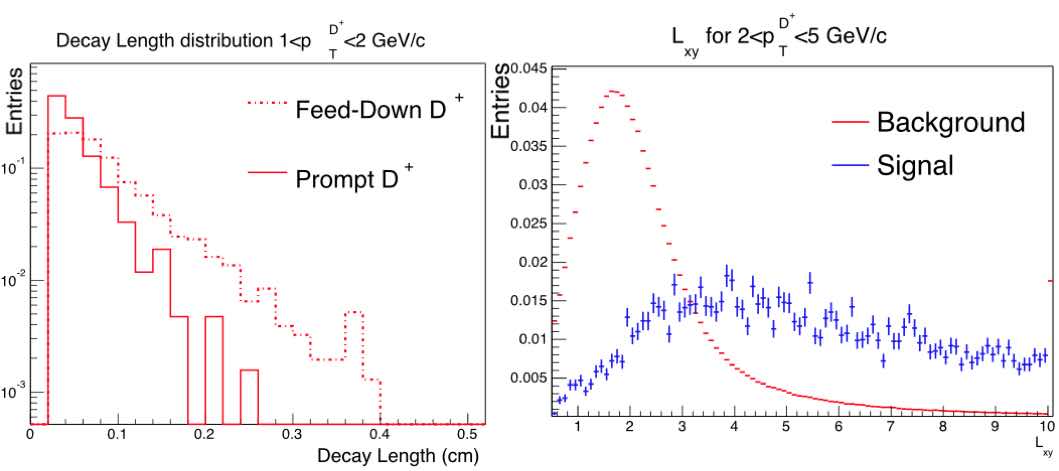}
\caption{Left: Distributions of the decay length of prompt and feed-down D$^+$ mesons. Right: Distributions of the L$_{xy}$ variable.}
\label{NormDecL}
\end{figure}
As seen in Table \ref{TabellaITS}, the  spatial resolution on the ITS points is higher in the $r\varphi$ direction than in the z direction. This results in a better resolution on the vertex position  in the transverse plane. Thus a better resolution is achieved for the decay length projection on $xy$ plane with respect to the decay length in 3D. The decay length in the transverse plane divided by its uncertainty, obtained from the secondary vertex covariance matrix, is therefore used in the selection of \dplus signal  candidates
\begin{equation}
L_{xy} = \frac{Decay Length_{xy}}{\Delta_{Decay Length_{xy}}}
\end{equation}
The distribution of this variable for prompt  D$^+$ mesons and background is shown in Figure \ref{NormDecL} (right) for candidates in  2 <\pt<5 GeV/c. The background distribution shows a peak at $L_{xy}$=1$\div$2, and a negligible part of the distribution is found at L$_{xy}$ values higher than 5. On the contrary signal candidates distribution are characterized by   higher values of L$_{xy}$. This is due to the fact that signal candidates truly coming from a secondary vertex have on average a smaller  error on vertex position, which combined with the different decay length distribution of prompt and background D$^+$ mesons shown in Figure \ref{declplots}  results in these distributions.\\
A selection on the quality of the D$^+$ decay vertex was also applied looking at the daughter track dispersion around the  secondary decay vertex. The selection variable  is defined as
\begin{equation}
\sigma_{\rm vertex} = \sqrt{\rm d_1^2 + d_2^2+  d_3^2}
\end{equation}
where d$_i$ is the distance of minimal approach between the decay track $i$ and the position of the secondary vertex. For signal candidates, all  daughters are coming from the same vertex and this dispersion should be close to 0 with a distribution whose width   is determined by the resolution on the track parameters. On the contrary the background, as visible in Figure \ref{SigmaVertex}, has a broader distribution peaked at larger values. An upper cut on the value of $\sigma_{\rm vertex} $ is therefore applied.\\
Finally the cosine of the pointing angle defined at the beginning of this chapter can be used as a selection variable. In fact if the reconstructed secondary vertex does not correspond to a real D$^+$ meson, the directions of the reconstructed \dplus momentum and of its flight line will be less correlated and the cosine of the pointing angle will deviate from unity. The  distributions for signal and background D$^+$-meson candidates (normalized to their area) are shown in Figure \ref{SigmaVertex} (right). The lower panel shows the distributions of the cosine of the pointing angle in the transverse plane for signal and background. Thanks to the better resolution on the vertex coordinates in the transverse plane, the signal peak is narrower (the entries are more concentrated close to $\cos{\theta_{\rm pointing}^{xy}}$=1). For the $xy$ component, however, the background distribution is also more peaked at 1. For this reason a lower cut is  applied on both $\cos{\theta_{\rm pointing}}$ and $\cos{\theta_{\rm pointing}^{xy}}$ with the aim of optimizing the amount of background candidates rejected and the amount of signal kept. \\
\begin{figure}[t]
\centering
 \includegraphics[width=0.99\textwidth]{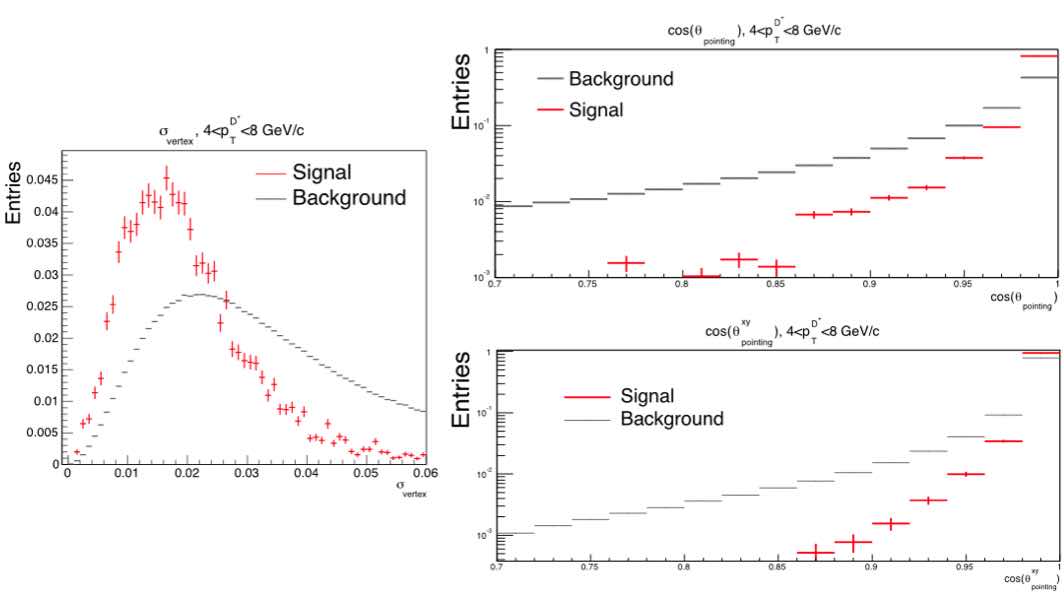}
\caption{Left: Distributions of $\sigma_{\rm vertex}$. Right: Distributions relative to the cosine of the pointing angle (top) and its projection on the transverse plane (bottom).}
\label{SigmaVertex}
\end{figure}
Convenient values of cut variables result in invariant mass spectra that clearly reveal the   D$^+$ meson  peak around its PDG mass value of 1.869 GeV/c$^2$. To extract the convenient values of the cut variables a cut optimization strategy has been adopted. This strategy, performed separately for different \pt regions, can be better explained with the help of Figure \ref{cutoptexplained}  which represent a 3 dimensional array of cut values (cells), each dimension corresponding to a different variable, in this case the kaon p$_{\rm T}$, the cosine of the pointing angle and the decay length. Each element of this array is characterized by three cut values, one for each quantity, and has a corresponding invariant mass histogram filled with   the candidates passing the selection cuts of  the considered cell.\\
\begin{figure}[t]
\centering
 \includegraphics[width=0.7\textwidth]{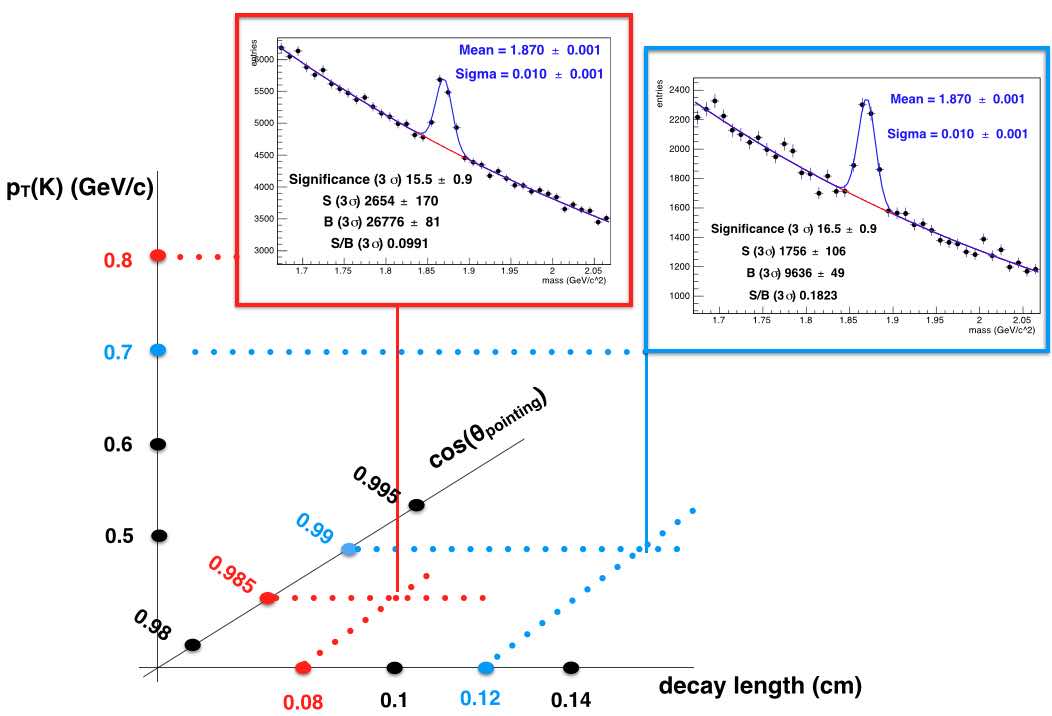}
\caption{Schematization of the cut optimization procedure}
\label{cutoptexplained}
\end{figure}
In the case of Figure \ref{cutoptexplained}, we focus on two particular cells, one with looser selections (red) and one with tighter selections (blue). Once all candidates in the data sample have been analyzed, a fit  is performed on each cell: 
\begin{itemize}
\item an exponential function is used to fit the invariant mass region on the two sides of the peak. The output of this fit is the red function in Figure \ref{InvMassFit}
\item the total distribution is then fitted with the sum of an exponenetial and a gaussian. The exponential is initialized using the parameters obtained in the previous step and the gaussian is initially centred on the  D$^+$ mass value from PDG
\item the D$^+$ meson signal is then extracted as the integral of the fitted   gaussian 
\end{itemize}
The signal  yields are shown for each of the two invariant mass distributions of Figure \ref{cutoptexplained} together with their statistical error. The significance, defined as $\frac{\rm S}{\sqrt{\rm S + B}}$ is also shown. A significance higher than 5 indicates that the probability for the peak being a fluctuation of the exponential background is lower than 2.87$\times 10^{-7}$\%.\\
As an example Figure \ref{griglia} shows the values of significance extracted from the fit described above as a function of two cut variables, namely the decay length and the cosine of the pointing angle, for 8<p$_{\rm T}^{D^{+}}$<12 GeV/c. All other cut variables are fixed in this example. The significance shows maximum values for intermediate   decay length cut values, while the highest significance values in the plot are found for the tightest cut value on pointing angle. However the choice of the central cut values has to be performed considering 
\begin{itemize}
\item the stability of the fit parameters (mean, sigma of the gaussian)
\item the amount of signal and its statistical error 
\item the fraction of prompt \dplus mesons that is being selected. As an example, the tighter the cuts on decay length, the higher the fraction of \dplus mesons from B-hadron decays that is being selected 
\end{itemize}
\begin{figure}[t]
\centering
 \includegraphics[width=0.7\textwidth]{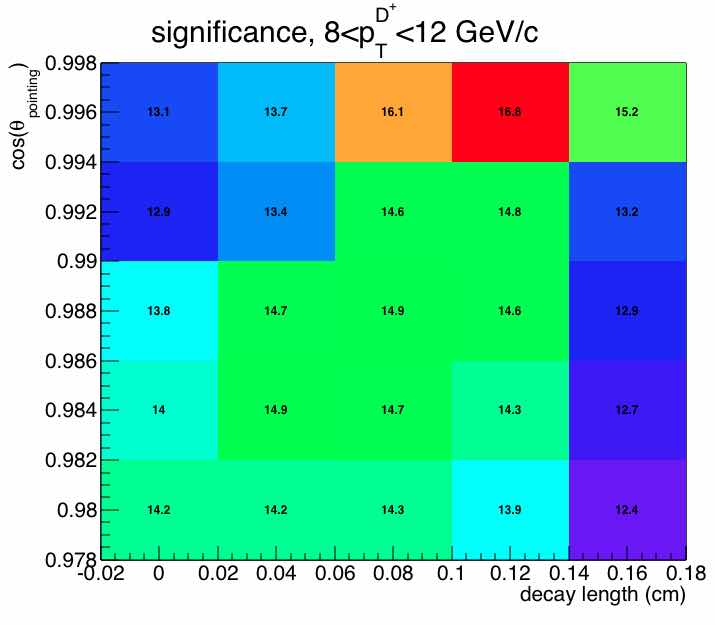}
\caption{Significance extracted from the fit described above as a function of  decay length and the cosine of the pointing angle, for 8<p$_{\rm T}^{D^{+}}$<12 GeV/c. All other cut variables are fixed in this example.}
\label{griglia}
\end{figure}
The procedure described is a semplification of what is done, since the cut variables used are 7 and so a 7 dimensional array is used. Note that using e.g. 5 intervals for each variables results in 80k cells, which is not an easily manageable number in terms of computing costs. It is thus more convenient to start with a limited number of variables, identify a convenient working point for their cut values, and further optimize the rest of them. \\
\begin{table}[b]
\footnotesize
\centering
\begin{tabular}{|c|c|c|c|c|c|c|c|c|c|}
\hline
 \pt (GeV/c) & [1,2] & [2,8] & [8,9] & [9,10] & [10,11]& [11,12] & [12,14] & [14,16] & [16,24] \\
\hline 
 $|\Delta M_{D^{+}}|$ (GeV/$c^{2}$) & 0.2& 0.2&0.2&0.2&0.2&0.2&0.2&0.2&0.2  \\
\hline
$\sigma_{vertex}$ (cm) & 0.03& 0.03 & 0.035 &0.035 & 0.035 & 0.07 & 0.07 & 0.09 & 0.03 \\
\hline
$p_{\rm T}^K$ (GeV/c) & 0.2 & 0.2 & 0.2 & 0.2 & 0.2 &0.2 &0.2 & 0.2 & 0.2  \\
\hline
$p_{\rm T}^\pi$ (GeV/c) & 0.2& 0.35 & 0.35 & 0.35 & 0.35 & 0.35 &0.35&0.35  &0.35  \\
\hline
Decay Length  (cm) & 0.02 & 0.04 & 0.04 & 0.04 & 0.04 & 0.04 & 0.1 & 0.1 & 0.15 \\
\hline 
L$_{xy}$ & 9 & 8 & 8 & 8& 8 &6 & 6 & 9 & 5 \\
\hline 
$\cos(\theta_{pointing})$ & 0.99 & 0.99 & 0.99 & 0.99 & 0.99 & 0.99 & 0.99 & 0.99 & 0.99 \\
\hline 
$\cos(\theta_{pointing, xy})$ & 0.995 & 0.99 & 0.99 & 0.99 & 0.99 & 0.99 & 0.99 & 0.99 & 0.99 \\
\hline
\end{tabular}
\caption{Summary table of the D$^+$ analysis cuts optimized for the minimum bias p--Pb data sample.}
  \label{TableCutsDplus2}
\end{table}
The results of the cut optimization performed on the minimum bias data p--Pb sample are shown in Table  4.2, and will be used as topological cut values in next section. They allow to extract the invariant mass spectrum of \ref{InvMassFit}. The fit described above is applied to extract signal, significance, signal over background ratio. The mean and width of the fitted gaussian are also shown. The mean value is compatible with the D$^+$ meson PDG mass value. The width of the peak is a detector resolution effect and depends on p$_{\rm T}$. In fact, looking at Equation 4.2, we see that the invariant mass of the D$^+$ meson candidates depends on the measured momenta of the three daughters. Looking at Figure \ref{ipptres}, we deduce that the higher the D$^+$ meson p$_{\rm T}$, the harder are its daughter, and the worse the resolution on momenta will be, yielding a constant broadening  of the gaussian sigma going at high p$_{\rm T}$. The gaussian sigma from the fits applied in different \pt bins are  shown in the right panel of Figure \ref{InvMassFit} as a function of p$_{\rm T}$. The gaussian sigma values expected from MC simulations are superimposed. Some discrepancies are present, and they will be taken into account in the evaluation of the systematic error as will be discussed in next Chapter. \\
\begin{figure}[t]
\centering
 \includegraphics[width=0.99\textwidth]{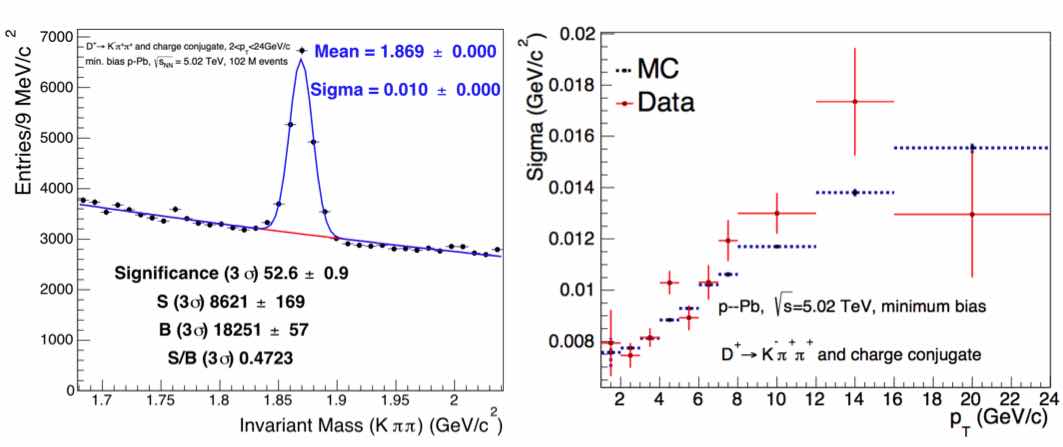}
\caption{Left: invariant mass distribution  for $D^{+}$ mesons candidates in the \pt range 1<\pt<24 GeV/c extracted from minimum-bias p--Pb collisions. The fit function is superimposed. Right: width of the gaussian of the $D^{+}$ peak as a function of \pt for minimum-bias p--Pb collisions and from the MC data sample that is described in next Section.}
\label{InvMassFit}
\end{figure}
 \subsection{Particle Identification}
\label{sub:pid}
Further background rejection is achieved using PID on the decay tracks.  The rejection is based on the identification of triplets of tracks in which:
\begin{itemize}
\item D$^+\rightarrow K^{-} \pi{+} \pi^{+}$: the negative track is compatible with the kaon hypothesis, the two positive tracks are compatible with the pion hypothesis
\item D$^-\rightarrow K^{+} \pi^{-} \pi^{-}$: the positive track  is compatible with the kaon hypothesis, the two negative tracks are compatible with the pion hypothesis
\end{itemize} 
As explained in Sec. \ref{sub:PID}, this PID selection is performed using the time of fligth of the particles and their energy loss in the TPC.\\ 
Each particle has its own resolution $\sigma$ on both time of fligth and energy loss measurements, which depends on the track momentum. To each track in the triplet, four values n$\sigma^{\rm TOF}_{\pi}$, n$\sigma^{\rm TOF}_K$ , n$\sigma^{\rm TPC}_{\pi}$ , n$\sigma^{\rm TPC}_K$  are computed
\begin{itemize}
\item n$\sigma^{\rm TOF}_{\pi}$: the number of sigmas the track time of flight deviates from the pion expected time of flight
\item n$\sigma^{\rm TOF}_{K}$: the number of sigmas the track time of flight deviates from the kaon expected time of flight
\item n$\sigma^{\rm TPC}_{\pi}$: the number of sigmas the track dE/dx deviates from the pion expected value 
\item n$\sigma^{\rm TPC}_{K}$: the number of sigmas the track dE/dx deviates from the kaon expected value 
\end{itemize}
Figure \ref{pidcomp} shows the distribution of n$\sigma^{\rm TOF}_{\pi}$ as a function of n$\sigma^{\rm TPC}_{\pi}$ (left) and of n$\sigma^{\rm TOF}_{K}$ as a function of n$\sigma^{\rm TPC}_{K}$ (right). In the left (right) panel, the dense region around (0,0) contains tracks whose signal is  close to the one expected for pions (kaons) in both TPC and TOF, and are thus very likely to be real pions (kaons). Other dense regions are visible in the distribution, and they are populated by  kaons (pions), protons and electrons.\\
From this plot it is clear that a convenient cut on the number of $\sigma$ for the selected tracks can help in rejecting the background. \\
\begin{figure}[t]
\centering
 \includegraphics[width=0.99\textwidth]{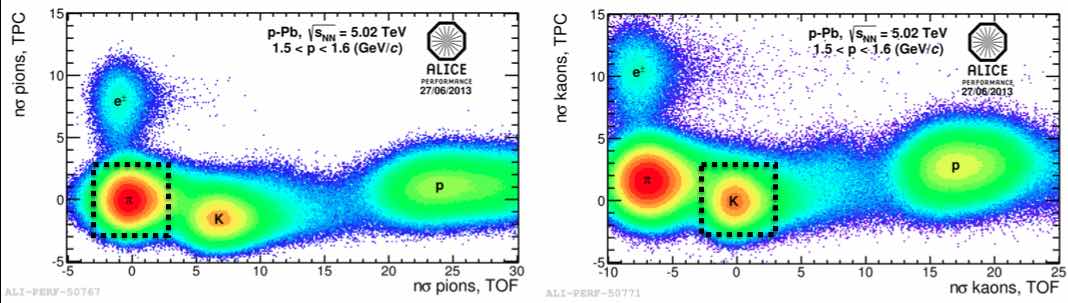}
\caption{Left:  n$\sigma^{\rm TOF}_{\pi}$ vs n$\sigma^{\rm TPC}_{\pi}$. Right: n$\sigma^{\rm TOF}_{K}$ vs n$\sigma^{\rm TPC}_{K}$}
\label{pidcomp}
\end{figure}
\begin{table}[b]
\begin{tabular}{l|c|l|l|}
\cline{2-4}
TPC                                     & \multicolumn{1}{l|}{$p_{\rm T}$<0.6 GeV/c} & 0.6<$p_{\rm T}$<0.8 GeV/c & $p_{\rm T}$>0.8 GeV/c   \\ \hline
\multicolumn{1}{|l|}{$n\sigma$<1}         & \multicolumn{2}{c|}{IDENTIFIED as $\pi$/K}                                      & COMPATIBLE with $\pi$/K \\ \hline
\multicolumn{1}{|l|}{1<$n\sigma_{\rm TPC}$<2} & IDENTIFIED as $\pi$/K                                 & \multicolumn{2}{c|}{COMPATIBLE with $\pi$/K}      \\ \hline
\multicolumn{1}{|l|}{2<$n\sigma_{\rm TPC}$<3} & \multicolumn{3}{c|}{COMPATIBLE with $\pi$/K}                                                 \\ \hline
\multicolumn{1}{|l|}{$n\sigma$>3}         & \multicolumn{3}{c|}{REJECTED}                                                              \\ \hline
\end{tabular}
\end{table}
\begin{table}[b]
\begin{tabular}{l|c|c|}
\cline{2-3}
TOF                                   & $p_{\rm T}$<1.5 GeV/c & $p_{\rm T}$>1.5 GeV/c   \\ \hline
\multicolumn{1}{|l|}{$n\sigma_{TOF}$<3} & IDENTIFIED as $\pi$/K            & COMPATIBLE with $\pi$/K \\ \hline
\multicolumn{1}{|l|}{$n\sigma_{TOF}$>3} & \multicolumn{2}{c|}{REJECTED}               \\ \hline
\end{tabular}
\caption{PID selection regions}
\label{tab:TPCTOFSigma}
\end{table}
The dE/dx and time of flight   of each track are compared to their expected values in the $\pi$ and $K$ hypothesis giving a PID response.  Three different cases  relative to different n$\sigma$ intervals in both TPC and TOF are defined in  Table \ref{tab:TPCTOFSigma}
\begin{itemize}
\item IDENTIFIED as $\pi$, $K$ in case of positive identification of the tracks 
\item COMPATIBLE with the  $\pi$, $K$ hypothesis 
\item REJECTED, in this case the tracks is very likely not to be a $\pi$ or a $K$
\end{itemize}  For instance all tracks outside the black dashed square in the left (right) panel of Figure \ref{pidcomp} are rejected as pions (kaons), while all the tracks inside the box are compatible or positively identified as pions (kaons). \\
The information from the two detectors is collected for all three daughters in the triplet and combined to finally decide if the candidate has to be kept or rejected. In the analyses that will be presented in the next chapters, two slightly different approaches have been used.  The STANDARD PID, used for candidates with \pt>2GeV/c
\begin{itemize}
\item no track in the triplet has to be positively identified as proton while being rejected as k/$\pi$  by either TOF or TPC
\item the number of tracks positively identified as kaon by either TPC or TOF and rejected as pion by either TPC or TOF does not have to exceed one
\item the number of tracks rejected as kaon  by either TPC or TOF does not have to exceed two
 \item the two tracks having the same charge sign as the D$^{+/-}$ mother have to be at least compatible with the pion hypothesis in both TPC and TOF
 \item the  track having the opposite charge sign with respect to  the D$^{+/-}$ mother has to be at least compatible with the kaon hypothesis in both TPC and TOF
\end{itemize}
For \pt<2GeV/c   stronger requirements are adopted (STRONG PID)
\begin{itemize}
\item the first three points of the STANDARD PID are unmodified
 \item the two tracks having the same charge sign as the D$^{+/-}$ have to be positively identified as  pions by either TPC or TOF
 \item the  track having the oppposite charge sign with respect to  the D$^{+/-}$ mother has to be  identified as kaon by either TPC or TOF
\end{itemize}
\begin{figure}[t]
\centering
 \includegraphics[width=0.99\textwidth]{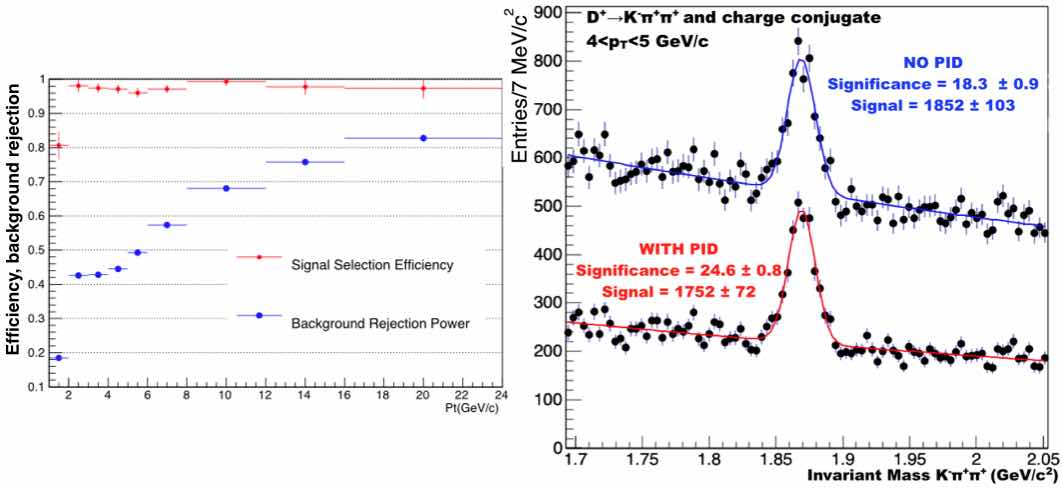}
\caption{Left: PID signal efficiency and background rejection power as a function of \pt. STRONG PID is applied  in the bin 1<\pt<2 GeV/c, STANDARD PID for \pt>2GeV/c. Right: D$^+$ mesons invariant mass distributions with and without the use of STANDARD PID.}
\label{pideff}
\end{figure}
The signal selection efficiency and background rejection power of these PID techniques have been evaluated via MC simulations and are shown in Figure \ref{pideff} (left). The STRONG PID, which is applied in the bin 1<\pt<2 GeV/c,  has a lower signal selection efficiency compared  STANDARD PID which is applied elsewhere.  However it turns out to have a  higher ($\sim$ 20\%) background rejection power\footnote{Defined as the number of background candidates that pass the PID selection}. \\
The right panel of Figure \ref{pideff} shows the fit to the D$^{+}$ invariant mass distribution obtained applying topological cuts only (top) and topological cuts plus STANDARD PID (bottom) in the \pt bin 4<\pt<5 GeV/c. An increase of significance with a minor signal loss is achieved with the  PID selections. 
\section{Monte Carlo simulations of D$^{+/-}$ mesons in ALICE}
Reliable MC simulations are needed to extract the efficiency and acceptance values used to correct the raw yields  measured in the various data samples collected with the ALICE detector. In particular it is crucial to have a precise description of the detector conditions and of track parameter resolution, as well as of the PID performance, since they affect several parameters used in the analysis, e.g. the width of the D$^{+/-}$ peak  and the distributions of the selection variables.
\subsection{Monte Carlo data sample}
\label{sub:MCsample}
The HIJING (Heavy Ion Jet INteraction Generator) Monte Carlo generator \cite{HIJING} has achieved a very good quantitative understanding of high-energy heavy ion interactions. However, in order to have acceptable values of  the statistical uncertainties on \dplusm efficiencies keeping low computation costs, a charm enriched MC data sample was generated.\\ 
The MC  sample  was constructed starting from $\sim$ 90 M pp  events  generated with PYTHIA v6.4.21, each of which had to satisfy one of the following conditions with the same probability (25 $\%$ each):
\begin{itemize}
\item an event which contains a $c\bar{c}$ pair with at least one of the quarks having $|y| <$1.5. D mesons from the hadronization of $c$-quarks are forced to decay in the  hadronic  channels of Table 4.1
\item an event which contains a $b\bar{b}$  pair with at least one of the quarks having $|y|<$1.5. In this case, D mesons that are produced by the decay of a B meson are forced to decay in one of the   hadronic  channels of Table 4.1
\item an event with at least one $c\bar{c}$  and one heavy-flavour decay electron with $y<$1.2;
\item an event with at least one $b\bar{b}$  pair and one heavy-flavour decay electron with $y<$1.2.
\end{itemize}
For each  event, a number of binary nucleon-nucleon collisions $N_{\rm coll}$  is extracted from a  distribution obtained starting from a Glauber MC simulation of p--Pb collisions. If the extracted number of binary collisions is larger than 1, a HIJING p--Pb event is added as  underlying event to the PYTHIA pp collision. The underlying event makes the multiplicity distribution in the MC more realistic, which is crucial since the resolution on vertex position depends on multiplicity as shown in Figure \ref{seco}.
The simulations used the GEANT3 particle transport package together with a detailed description of the geometry of the apparatus and of the detector response to reconstruct the generated particles. The simulation was configured to reproduce the conditions of the luminous region and of all of the ALICE subsystems, in terms of active electronic channels, calibration level, and their time evolution within the 2013 p--Pb data taking period.
\begin{figure}[b]
\centering
 \includegraphics[width=0.99\textwidth]{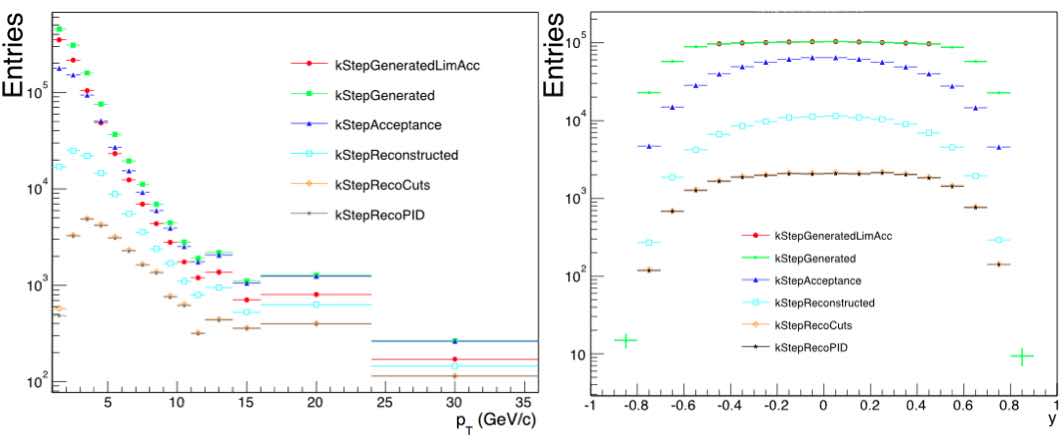}
\caption{Left: \pt distribution of prompt D$^{+/-}$ mesons at the different step of selections defined in the CF Container. Right: rapidity distribution of prompt D$^{+/-}$ mesons at the different step of selections defined in the CF Container.}
\label{cfcont}
\end{figure}
\subsection{Acceptance and efficiency computation}
\label{sub:acceff}
The calculation of the acceptance and efficiency correction is based on the Correction Framework included in the ALICE analysis software. The Correction Framework includes classes that allow one to store in a container the number of candidates at specific stages (steps) of the generation, reconstruction or selection procedure on an N-dimensional grid (called “CF container” from now on). The containers can be filled in for prompt D$^{+/-}$ mesons only, feed-down D$^{+/-}$ only or both of them, and stores, for each \dplusm, the \pt, $y$, $\varphi$ values as well as the vertex position and the multiplicity of the event. The steps used to store the candidates are:
\begin{itemize}
\item  \textit{kStepGenerated}: at this stage the Correction Framework container is filled with the generated D$^{+/-}$ particles, for collisions with |$z_{\rm vertex}$|<10 cm.
\item  \textit{kStepGeneratedLimAcc}: at this stage the D$^{+/-}$ mesons generated in the rapidity range |y| < 0.5 are counted and stored in the CF container, for collisions with |$z_{\rm vertex}$|<10 cm.
\item  \textit{kStepAcceptance}: at this step the CF container is filled for the D$^{+/-}$ mesons for which the 3 daughters in the final state fulfill the acceptance conditions |$\eta$| < 0.9 and \pt > 0.1 GeV/c of the ALICE central barrel, for collisions with |$z_{\rm vertex}$|<10 cm.
\item  \textit{kStepReconstructed}: the CF container is filled with reconstructed D$^{+/-}$ mesons that pass minimal cuts  (called filtering cuts) on decay track quality
\item  \textit{kStepRecoCuts}: at this stage the D$^{+/-}$ mesons that pass the kinematical and topological cuts used in the analysis are stored in the CF container.
\item  \textit{kStepRecoPID}: in this final step, also the Particle Identification selection is applied and the D$^{+/-}$ passing also these cuts, on top of the topological ones applied in the previous RecCuts step are used to fill the CF container.
\end{itemize}
Figure \ref{cfcont} shows the \pt and rapidity distributions of prompt D$^{+/-}$ mesons at the steps listed above. It should be noted that in the plot as a function of rapidity   the  \textit{kStepGenerated} and the \textit{kStepGeneratedLimAcc} distribution coincide for |y|<0.5. \\
 The efficiencies can then be computed as a function of the  variables stored in the grid for each D$^{+/-}$ and at the various selection steps by performing the ratios of the contents of the CF container. The CF container structure also allows the possibility to reweight the efficiencies as a function of different variables, e.g. of \pt or multiplicity. The weighting is used to get more realistic distributions in the MC, and will be used in next Chapter. \\  
\begin{figure}[t]
\centering
 \includegraphics[width=0.99\textwidth]{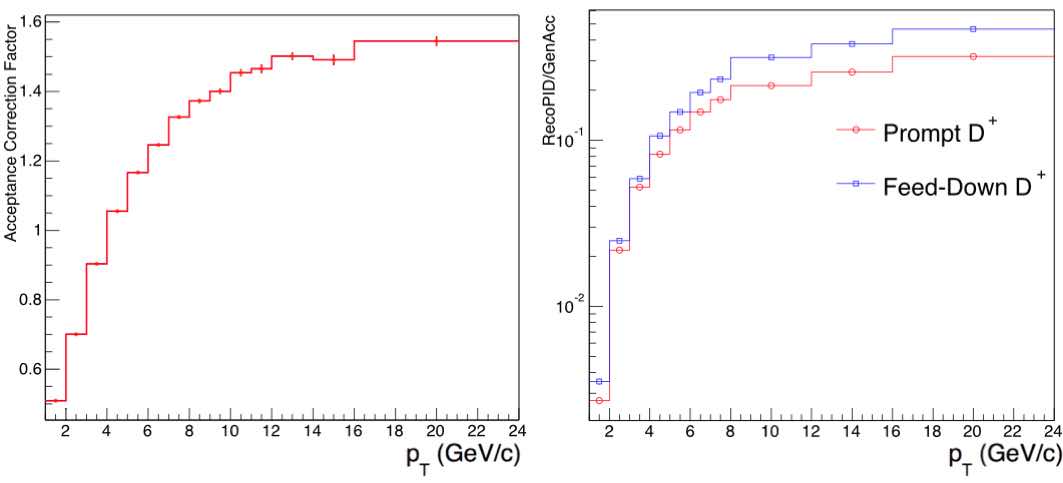}
\caption{Left: Acceptance correction factor for prompt D$^+$ mesons. Right: efficiencies for prompt D$^+$ mesons and for D$^+$ coming from beauty decays for the selection cuts used for the p--Pb analysis.}
\label{acceff}
\end{figure}
The acceptance correction factor is obtained as the ratio of the \textit{kStepAcceptance} and \textit{kStepGeneratedLimAcc} containers, and it is shown as a function of \pt  in Figure \ref{acceff} (left).\\
The efficiency correction factor is obtained as the ratio of the \textit{kStepRecoPID} and \textit{kStepAcceptance} containers. The right panel of Figure  \ref{acceff} shows the efficiencies for prompt D$^+$ mesons and for D$^+$ coming from beauty decays for the selection cuts used for the p--Pb analysis. The observed higher efficiency for the secondary D mesons is due to the fact that the D mesons coming from B feed-down are more displaced from the primary vertex and they are therefore preferentially selected by the cuts based on track/vertex displacement.

\chapter{Prompt D$^{+}$-meson production in  p--Pb collisions } 
\lhead{Chapter 5. \emph{Prompt D$^{+}$-meson production in  p--Pb collisions}} 
As shown in Figure \ref{RaaDmeson} , the D-meson nuclear modification factor \raa measured by the ALICE Collaboration in central Pb--Pb collisions at \sqnn= 2.76 TeV indicates a strong modification of the D-meson spectrum for \pt>2 GeV/c with respect to the spectrum in pp collisions at the same centre-of-mass energy scaled by the average number of nucleon-nucleon collisions $\langle$\ncoll$\rangle$. The suppression is interpreted as   $c$-quark in-medium energy loss due to the presence of QGP in the final state. In semi-central Pb--Pb collisions the ALICE Collaboration also observed a positive $v_2$ of D mesons with 2<\pt<6 GeV/c as shown in Figure \ref{ALICEDv}, which can be interpreted as due to the interaction of $c$-quarks with the medium constituents, which transfer to the $c$-quark the anysotropy of the expanding medium and may lead to  partial thermalization of $c$-quark. \\
However  a  complete understanding of the D-meson results in Pb--Pb collisions requires an understanding of cold nuclear matter effects in the initial and final state, which can be accessed by studying p–Pb collisions assuming that the QGP is not formed in these interactions. \\
As discussed in Section \ref{sub:instef}, in the initial state of the collisions several effects can influence heavy flavour production:
\begin{itemize}
\item  the nuclear environment affects the quark and gluon PDFs, which are modified in bound nucleons with respect to free nucleons. The modification depends  on the parton fractional momentum $x$ and on the atomic mass number A. Models including parton saturation as modelled in the Color  Glass Condensate theory \cite{CGC} or EPS09 parametrization of the nuclear PDFs \cite{EPS09} are able to reproduce ALICE data of charged hadron \rppb, as shown in Figure \ref{lhrpa}
\item partons experience transverse momentum broadening due to multiple soft collisions prior to the hard scattering. This leads to an \rppb higher than unity at intermediate \pt. A species dependent Cronin enhancement has been reported in both d--Au (Figure \ref{CroninSpecies} - left) and p--Pb (Figure \ref{CroninSpecies} - right) collisions
\item partons can also lose energy in the initial stages of the collision via  radiation in cold nuclear matter. In particular this modifies the effective  centre-of-mass energy of the partonic system undergoing the hard scattering. Models including this effect reasonably reproduce J/$\Psi$ production in p--Pb collisions at \sqnn=5.02 TeV, as shown  in Figure \ref{charmpA}
\end{itemize}  
 Nevertheless the presence of final-state effects in small collision systems is suggested by recent studies in both d--Au and p--Pb collisions. ALICE Collaboration measured the $\Psi$' nuclear modification factor in p--Pb collisions at forward and backward rapidities \cite{psiprimoLHC}, measuring a larger suppression with respect to  J/$\Psi$ which is not reproduced by models including only initial state effects such as cold nuclear matter energy loss or nuclear PDFs modification. Similar results were also obtained at RHIC \cite{PsiPrimoRHIC}. \begin{figure}[b]
\centering
 \includegraphics[width=0.99\textwidth]{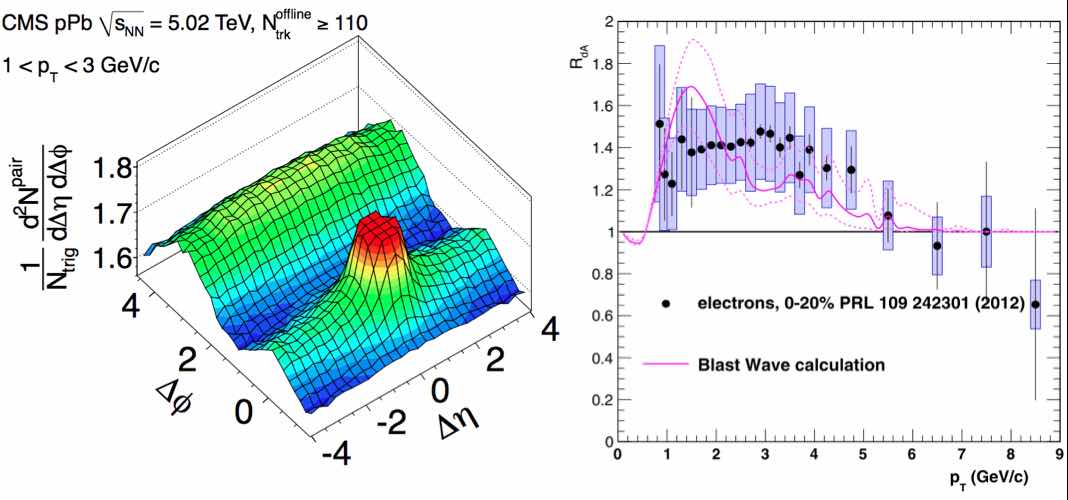}
\caption{Left: two-particle correlation function for 5.02 TeV p--Pb collisions with charged particle multiplicity N$_{\rm tracks}$> 110 \cite{RidgepPbCMS}. Right: heavy-flavour decay electron $R_{\rm dAu}$  already reported in Figure \ref{charmpA} (left) together with blast-wave calculations (Section 1.4.3) implementing collective radial flow in d-Au collisions  \cite{Sickles}.}
\label{hneppb}
\end{figure}
Also measurements of  long-range correlations of charged hadrons  in p–Pb collisions have shown unexpected results: as an example, Figure \ref{hneppb} (left) shows the two-particle ($\Delta\eta,\Delta\Phi$) correlation function\footnote{The analysis is very similar to the one described in Section \ref{sub:hoa} (Equation 1.43, Figure \ref{tri}), except for the fact that trigger and associated particles are selected in the same \pt interval  1<\pt< 3 GeV/c and that the correlation is not only studied as a function of $\Delta \Phi$, but also of $\Delta \eta$.} for  p--Pb collisions at \sqnn=5.02 TeV with charged particle multiplicity N$_{\rm tracks}$> 110 \cite{RidgepPbCMS}. The dominant features of the correlation plot are  the correlation peak near ($\Delta \eta$,$\Delta \Phi$)=(0,0) for pairs of particles originating from the same jet and the elongated structure at $\Delta \Phi\approx\pi$ for pairs of particles from back-to-back jets. These two structures have also been observed in pp collisions for a wide range of centre-of-mass energies \cite{RidgeppCMS}, however in Figure \ref{hneppb} a pronounced “ridge”-like structure emerges at $\Delta \Phi\approx$0 extending up to  |$\Delta \eta$|=4. This additional structure is similar to the one observed by the CMS Collaboration in Pb--Pb collisions at \sqnn=2.76 TeV \cite{RidgePbPbCMS}. The presence of this structure in high multiplicity p--Pb collisions was confirmed by the  ALICE \cite{RidgeALICE} and ATLAS \cite{RidgeATLAS}  Collaborations and it is  quantitatively predicted in models assuming a collective hydrodynamic expansion of the system \cite{modella}. Hydrodynamic calculations also give a natural intepretation to the  species dependent Cronin enhancement observed  by ALICE and PHENIX Collaborations.\\
Moreover  a description in terms of hydrodynamic flow in small collision systems has been recently proposed for the results  on   heavy-flavour decay electrons $R_{\rm dAu}$ . Figure \ref{hneppb} (right) shows the heavy-flavour decay electron $R_{\rm dAu}$  already reported in Figure \ref{charmpA} (left) together with blast-wave calculations \cite{Sickles}  (Section \ref{sub:kfo}) implementing collective radial flow in d-Au collisions. The magnitude of  $R_{\rm dAu}$ enhancement expected from the blast-wave calculation is in good agreement with the data.\\
All elements cited above suggest that the measurement of D$^{+}$ mesons production cross-section and nuclear modification factor \rppb in p--Pb collision can shed light on  nuclear matter effects influencing  nuclear interactions  and, in particular, it can clarify whether the results from Pb--Pb collisions shown in  Figure \ref{RaaDmeson} are the consequence  of genuine hot QCD matter effects.
\section{Signal Extraction}
\label{sec:SE}
\begin{table}[b]
\footnotesize
\centering
\begin{tabular}{|c|c|c|c|c|c|c|c|c|c|}
\hline
 \pt (GeV/c) & [1,2] & [2,8] & [8,9] & [9,10] & [10,11]& [11,12] & [12,14] & [14,16] & [16,24] \\
\hline 
 $|\Delta M_{D^{+}}|$ (GeV/$c^{2}$) & 0.2& 0.2&0.2&0.2&0.2&0.2&0.2&0.2&0.2  \\
\hline
$\sigma_{vertex}$ (cm) & 0.03& 0.03 & 0.035 &0.035 & 0.035 & 0.07 & 0.07 & 0.09 & 0.03 \\
\hline
$p_{\rm T}^K$ (GeV/c) & 0.2 & 0.2 & 0.2 & 0.2 & 0.2 &0.2 &0.2 & 0.2 & 0.2  \\
\hline
$p_{\rm T}^\pi$ (GeV/c) & 0.2& 0.35 & 0.35 & 0.35 & 0.35 & 0.35 &0.35&0.35  &0.35  \\
\hline
Decay Length  (cm) & 0 & 0.04 & 0.04 & 0.04 & 0.04 & 0.04 & 0.1 & 0.1 & 0.15 \\
\hline 
L$_{xy}$ & 9 & 8 & 8 & 8& 8 &6 & 6 & 9 & 5 \\
\hline 
$\cos(\theta_{pointing})$ & 0.99 & 0.99 & 0.99 & 0.99 & 0.99 & 0.99 & 0.99 & 0.99 & 0.99 \\
\hline 
$\cos(\theta_{pointing, xy})$ & 0.995 & 0.99 & 0.99 & 0.99 & 0.99 & 0.99 & 0.99 & 0.99 & 0.99 \\
\hline
\end{tabular}
\caption{Summary table of the D$^+$ analysis cuts.}
  \label{TableCutsDplus}
\end{table}
\begin{figure}[b]
\centering
 \includegraphics[width=0.99\textwidth]{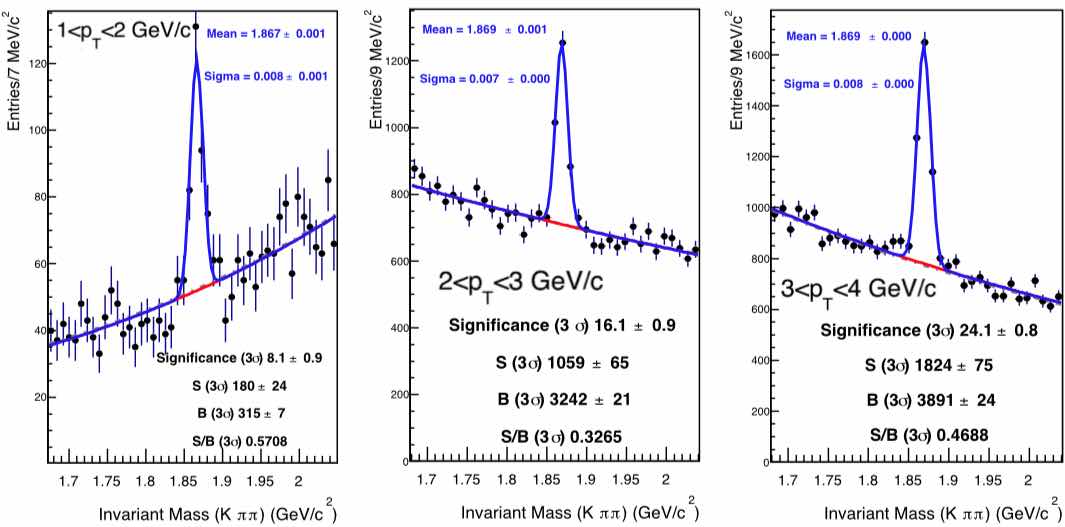}
\caption{\dplus candidates (and  charge conjugates) invariant mass distributions from minimum bias p--Pb collisions at  \sqnn= 5.02 TeV in the \pt bins [1,2] GeV/c (left), [2,3] GeV/c (centre), [3,4] GeV/c (right). }
\label{IM1}
\end{figure}
The analysis was performed using the p--Pb data sample collected in 2013 with a minimum-bias trigger that required the arrival of bunches from both directions and coincident signals in both scintillator arrays of the V0 detector (V0AND - Section \ref{sub:Trigger}). The integrated luminosity collected during the p--Pb beam configuration is shown in Figure \ref{trigger}. \\ Events were selected as described in Section \ref{sub:EvSel}.  About 110 10$^6$ events, corresponding to an integrated luminosity  $L_{\rm int}$=(48.6$\pm$1.6) $\mu$b$^{-1}$\footnote{Computed as N$_{\rm p--Pb,MB}$/$\sigma_{\rm p--Pb,MB}$ where N$_{\rm p--Pb,MB}$ is the number of p--Pb collisions passing the minimum-bias trigger condition and $\sigma_{\rm p--Pb,MB}$=2.09 b$\pm$3.5\% (syst)  is the cross section of the V0AND trigger.}, passed the selection criteria.\\  The secondary vertices of D$^+$ meson candidates are reconstructed using ITS-TPC tracks selected as described in Section \ref{sub:TrackSel}. The  selection on \dplusm candidates rapidity,  ranging from |$y_{\rm lab}$|<0.5 at low \pt (1<\pt<2) to |$y_{\rm lab}$|<0.8 above 4 GeV/c, is applied.\\
\begin{figure}[t]
\centering
 \includegraphics[width=0.99\textwidth]{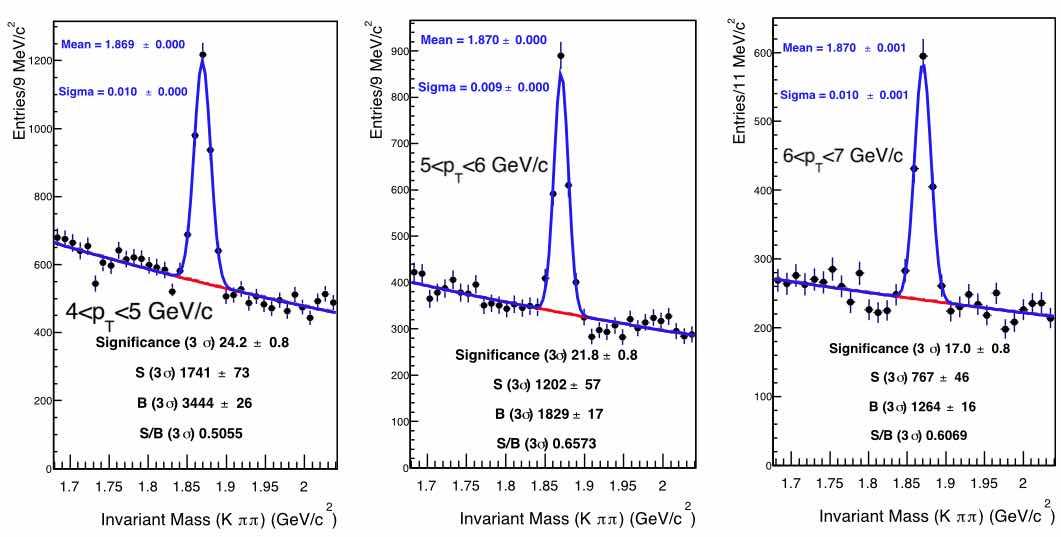}
\caption{\dplus candidates (and  charge conjugates) invariant mass distributions from minimum bias p--Pb collisions at  \sqnn= 5.02 TeV in the \pt bins [4,5] GeV/c (left), [5,6] GeV/c (centre), [6,7] GeV/c (right).}
\label{IM2}
\end{figure}
\begin{figure}[b]
\centering
 \includegraphics[width=0.99\textwidth]{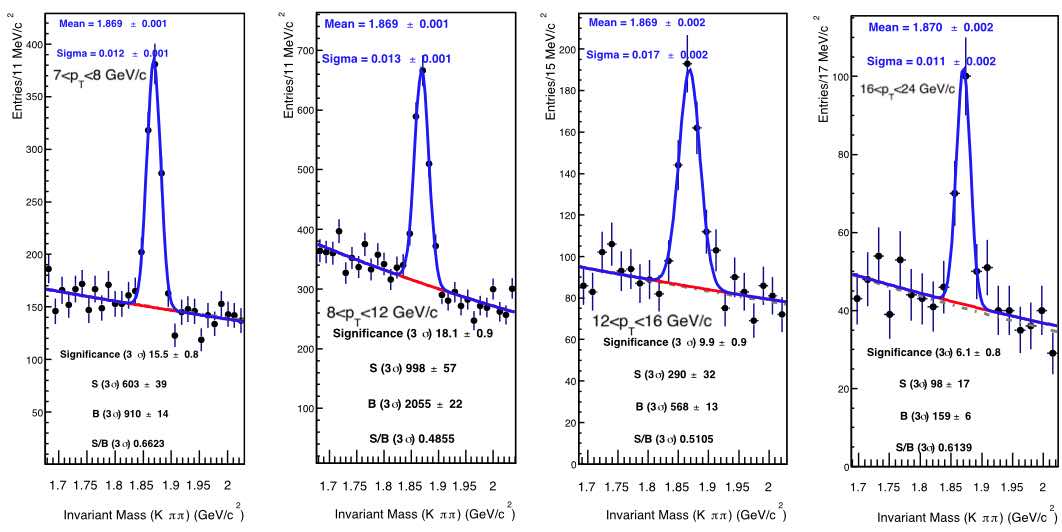}
\caption{\dplus candidates (and  charge conjugates) invariant mass distributions from minimum bias p--Pb collisions at  \sqnn= 5.02 TeV in the \pt bins [7,8], [8,12], [12,16] and [16,24] GeV/c. }
\label{IM3}
\end{figure}
\begin{table}[t]
\centering
\begin{tabular}{|c|c|c|l|}
\hline
p$_{\rm T}$ & \multicolumn{3}{c|}{D$^{+}$}                                                     \\ \hline
(GeV/c)     & Significance (3 $\sigma$) & Signal/Background (3 $\sigma$) & Signal \\ \hline
{[}1,2{]}   & 8.1 $\pm$ 0.9             & 0.57                           & 180 $\pm$ 24        \\ \hline
{[}2,3{]}   & 16.1 $\pm$ 0.9            & 0.33                           & 1059 $\pm$ 65       \\ \hline
{[}3,4{]}   & 24.8 $\pm$ 0.8            & 0.47                           & 1824 $\pm$ 65       \\ \hline
{[}4,5{]}   & 24.2 $\pm$ 0.8            & 0.51                           & 1741 $\pm$ 73       \\ \hline
{[}5,6{]}   & 21.8 $\pm$ 0.8            & 0.66                           & 1202 $\pm$ 57       \\ \hline
{[}6,7{]}   & 17.0 $\pm$ 0.8            & 0.62                           & 767 $\pm$ 46        \\ \hline
{[}7,8{]}   & 15.5 $\pm$ 0.8            & 0.66                           & 603 $\pm$ 39        \\ \hline
{[}8,12{]}  & 18.1 $\pm$ 0.9            & 0.49                           & 998 $\pm$ 57        \\ \hline
{[}12,16{]} & 9.9 $\pm$ 0.9             & 0.51                           & 290 $\pm$ 32        \\ \hline
{[}16,24{]} & 6.1 $\pm$ 0.8             & 0.61                           &      98 $\pm$ 17                \\ \hline
\end{tabular}
\caption{Significance, signal-to-background ratios and raw yields for \dplus mesons.}
 \label{TableSignal}
\end{table}
The topological cut values, optimized as described in Section  \ref{sub:TopSel}, are shown in Table \ref{TableCutsDplus}, where also the \pt bins used in the analysis are defined. The Particle Identification strategy discussed in Section \ref{sub:pid} was applied to further reduce the background: the STANDARD PID selection was used in the \pt range  2<\pt<24 GeV/c, the STRONG PID  selection was used in the \pt range 1<\pt<2 GeV/c.\\
Figures \ref{IM1}, \ref{IM2} and \ref{IM3} show the fit to the invariant mass distributions of \dplus meson candidates (and their charge conjugates) obtained after applying the selections described above in the ten \pt intervals used in the analysis. The fitting function is composed of an exponential function for the background and a Gaussian function for the signal. The peak position and width for \dplus mesons as obtained from the fit to data and to the  MC data sample described in Section \ref{sub:MCsample} are shown in Figure \ref{SigmaDataMC}. The
 peak position (right panel of Figure \ref{SigmaDataMC}) extracted from the data in the \pt range 1-24 GeV/c is compatible within two sigmas with the MC value in al \pt bins. It is also compatible within two sigmas with the PDG values of \dplusm mass in all \pt bins. The
 peak position (right panel of Figure \ref{SigmaDataMC}) extracted from the data in the \pt range 1-24 GeV/c is compatible within two sigmas with the MC value in most \pt bins.The peak width (left panel of Figure \ref{SigmaDataMC}) increases with \pt due to the worse resolution on track momentum at high-\pt.  A discrepancy of more than 2 sigmas in the peak width between data and MC is present in the \pt bin 4<\pt<5 GeV/c and  will be taken into account in the evaluation of  systematic errors. \\
The \dplus mesons raw yields are extracted by integrating the Gaussian fit functions over the whole invariant mass range. Table  \ref{TableSignal} shows the raw yields, signal-to-background ratios and significance values extracted from the fits in Figures \ref{IM1}, \ref{IM2} and \ref{IM3}.\\
\begin{figure}[t]
\centering
 \includegraphics[width=0.99\textwidth]{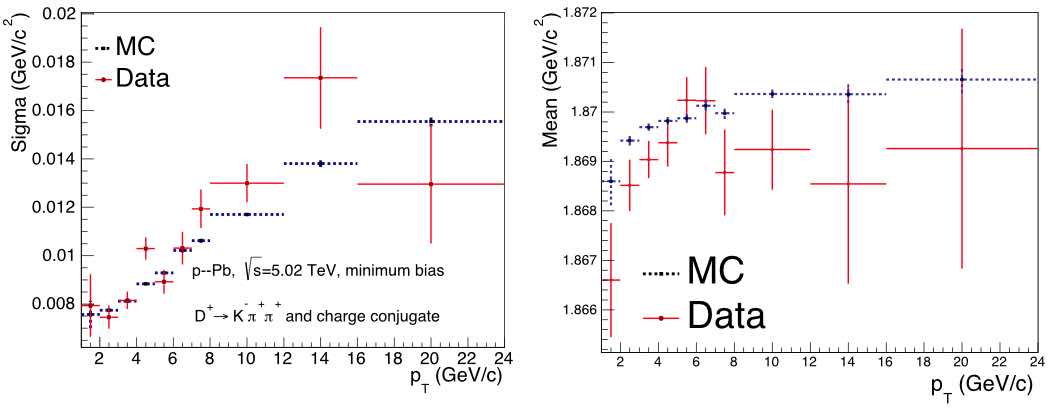}
\caption{\pt differential peak width (left) and position (right) for \dplus mesons as obtained from the fit to data and to the  MC data sample described in Section \ref{sub:MCsample}}
\label{SigmaDataMC}
\end{figure}
\begin{figure}[t]
\centering
 \includegraphics[width=0.9\textwidth]{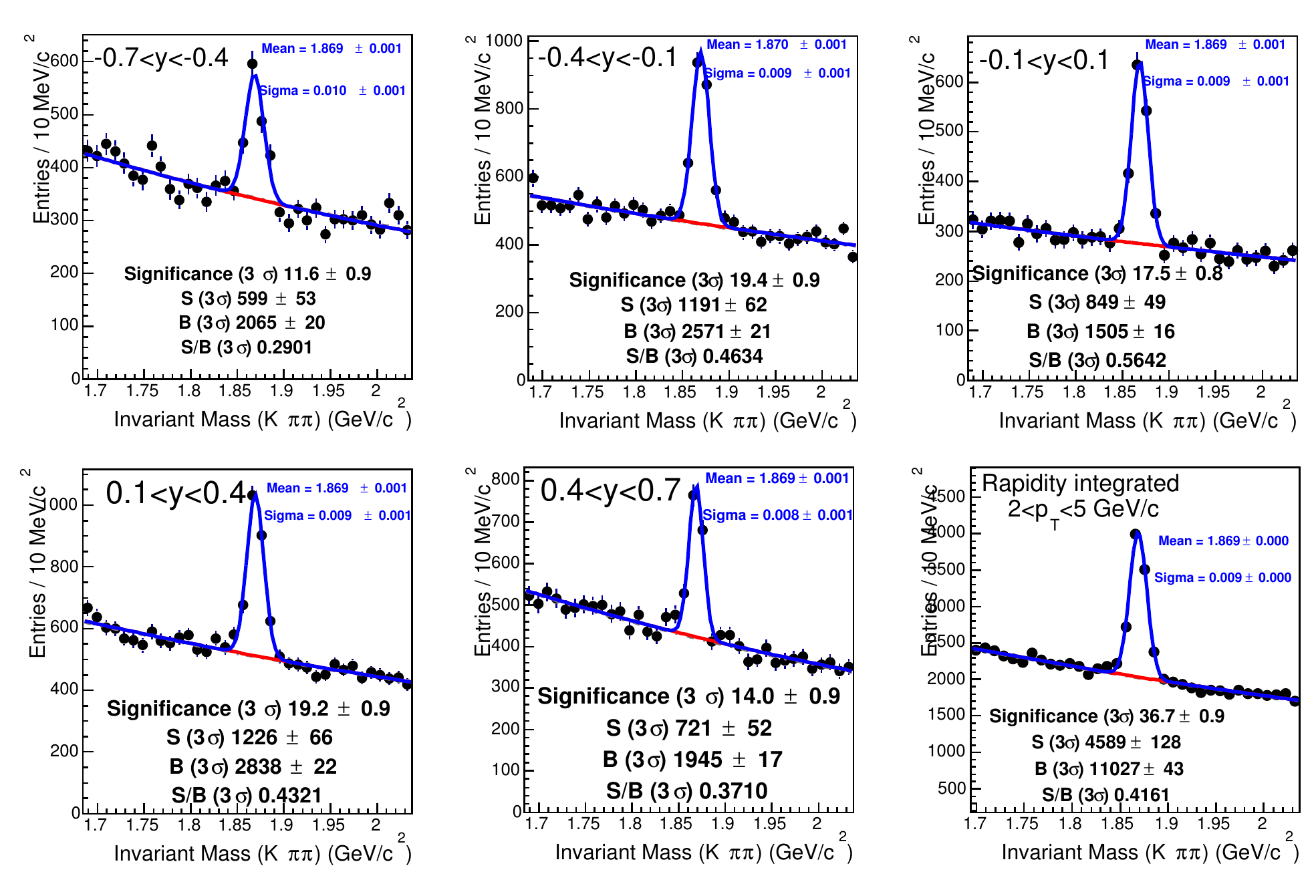}
 \includegraphics[width=0.9\textwidth]{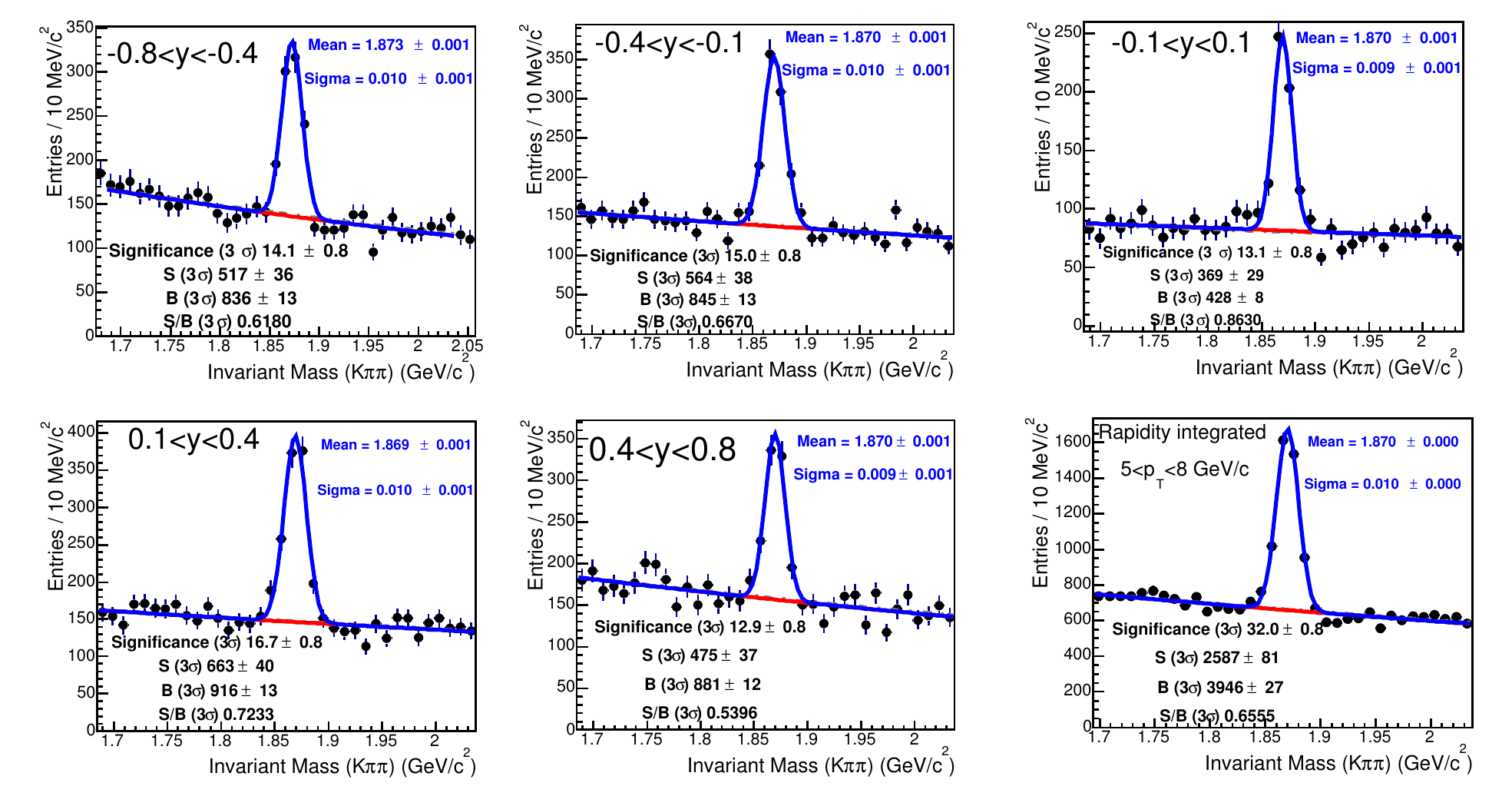}
\caption{Two top rows: \dplus candidates (and charge conjugate) invariant mass distributions from minimum bias p--Pb collisions at \sqnn= 5.02 TeV  for 2<\pt<5 GeV/c, in the five rapidity bins considered in the analysis and in the full rapidity range -0.7<$y_{\rm lab}$<0.7. Two bottom rows: \dplus candidates (and charge conjugate) invariant mass distributions from minimum bias p--Pb collisions at \sqnn= 5.02 TeV for 5<\pt<8 GeV/c, in the five rapidity bins considered in the analysis and in the full rapidity range -0.8<$y_{\rm lab}$<0.8.}
\label{DvsRap25}
\end{figure}
Apllying the same selections described above, the \dplus meson signal yield has also been extracted in five rapidity sub-intervals, namely -0.8<$y_{\rm lab}$<-0.4, -0.4<$y_{\rm lab}$<-0.1, -0.1<$y_{\rm lab}$<0.1, 0.1<$y_{\rm lab}$<0.4 and 0.4<$y_{\rm lab}$<0.8. The signal for each rapidity interval has been extracted in the \pt ranges 2<\pt<5  GeV/c, 5<\pt<8  GeV/c and 8<\pt<16 GeV/c. For the \pt interval 2<\pt<5 GeV/c the first and last rapidity intervals are  -0.7<$y_{\rm lab}$<-0.4 and 0.4<$y_{\rm lab}$<0.7 due to the fiducial acceptance cut. Figures \ref{DvsRap25} and  \ref{DvsRap816} show the $y_{\rm lab}$  and \pt differential mass plots. The raw yields extracted from these fits are shown in Table  5.3. \\ 
\begin{table}[h]
\begin{center}
\begin{tabular}{|c|c|c|c|c|c|c|}
\hline
\multirow{2}{*}{\begin{tabular}[c]{@{}c@{}}p$_{\rm T}$\\ (GeV/c)\end{tabular}} & \multicolumn{6}{c|}{Rapidity $y_{\rm lab}$}                                                                                                                                                                                                      \\ \cline{2-7} 
                                                                               & integrated   & [-0.7/-0.8, -0.4] &  [-0.4, -0.1]  &  [-0.1, 0.1] &  [0.1, 0.4] & [0.4, 0.7/0.8] \\ \hline
2-5                                                                            & 4589$\pm$128 & 599$\pm$53                                   & 1191$\pm$62                             & 849$\pm$49                             & 1226$\pm$66                           & 721$\pm$52                                \\ \hline
5-8                                                                            & 2587$\pm$81  & 517$\pm$36                                   & 564$\pm$38                              & 369$\pm$29                             & 663$\pm$40                            & 475$\pm$37                                \\ \hline
8-16                                                                           & 1264$\pm$64  & 247$\pm$28                                   & 303$\pm$34                              & 127$\pm$22                             & 319$\pm$30                            & 289$\pm$30                                \\ \hline
\end{tabular}
\caption{Summary table of the signal in 3 $\sigma$ per \pt and rapidity bin for  \dplus.}
\end{center}
\label{YieldsRap}
\end{table}
\begin{figure}[t]
\centering
 \includegraphics[width=0.99\textwidth]{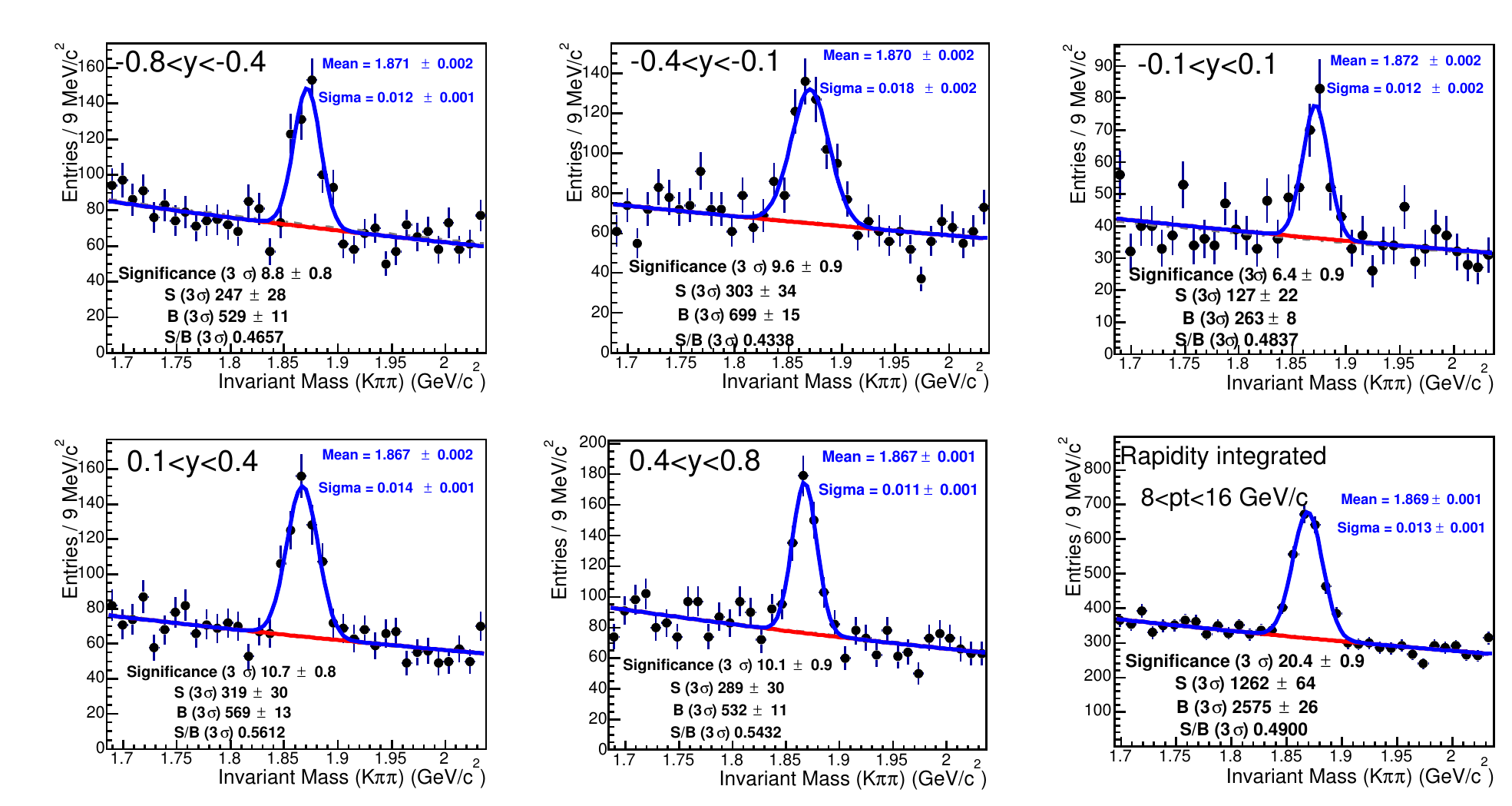}
\caption{\dplus candidates (and charge conjugate) invariant mass distributions from minimum bias p--Pb collisions at \sqnn= 5.02 TeV for 8<\pt<16 GeV/c, in the five rapidity bins considered in the analysis and in the full rapidity range -0.8<$y_{\rm lab}$<0.8. }
\label{DvsRap816}
\end{figure}
\begin{figure}[b]
\centering
 \includegraphics[width=0.99\textwidth]{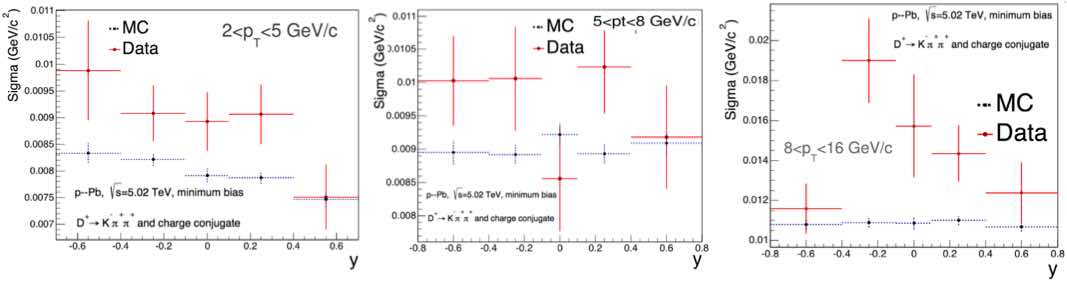}
\caption{\dplus mesons peak width  as a function of rapidity   as obtained from the fit to data and to the  MC data sample described in Section \ref{sub:MCsample} in the \pt bins [2,5] GeV/c (left), [5,8] GeV/c (centre), [8,16] GeV/c (right). }
\label{DataMCDvsY}
\end{figure}
 A discrepancy in the peak width obtained from the fit to data (Figure \ref{DataMCDvsY}) and MC is present in most of the \pt and $y$ bins. This discrepancy is within 10-15\% for \pt<8 GeV/c,within 20$\div$30\% in the \pt interval 8<\pt<16 GeV/c where it reaches 80\% in the rapidity interval -0.4<$y_{lab}$<-0.1. Similar deviations have been observed in the invariant mass distributions of the D$^0$ meson candidates. These discrepancies are accounted for in the evaluation of the systematic error. \\
\section{Cross section calculation}
The corrected per-event yields of prompt \dplus mesons in p--Pb collisions were obtained starting
from the raw yields $N^{D^{+/-}}_{\textrm{raw}}|_{|y|<y_{fid}}$ as follows:
\begin{equation}
\frac{dN^{D^{+}}}{d\pteq}=\frac{1}{2}\frac{1}{\Delta y\Delta\pteq}\frac{f_{\textrm{prompt}}\cdot N^{D^{+/-}}_{\textrm{raw}}|_{|y|<y_{fid}}}{(\textrm{Acc}\times \epsilon)^{\textrm{prompt}}\cdot BR\cdot N_{\textrm{ev}}}
\label{eq:corrsp}
\end{equation}
and the corrected invariant cross section as:
\begin{equation}
\frac{d\sigma^{D^{+}}}{d\pteq}=\frac{1}{2}\frac{1}{\Delta y\Delta\pteq}\frac{f_{\textrm{prompt}}\cdot N^{D^{+/-}}_{\textrm{raw}}|_{|y|<y_{fid}}}{(\textrm{Acc}\times \epsilon)^{\textrm{prompt}}\cdot BR\cdot L_{\textrm{int}}}
\label{eq:corrcs}
\end{equation}
where  $\Delta y$ and $\Delta\pteq$ are the rapidity and \pt interval width, respectively. The rapidity coverage of this cross section in the laboratory frame  is equivalent to a rapidity coverage  -0.96<$y_{\rm cms}$<0.04 due to the fact  that the centre-of-mass frame moves with a rapidity |$\Delta y$|=0.465 in the proton direction. \fprompt is the fraction of prompt \dplus mesons, $(\textrm{Acc}\times \epsilon)^{\textrm{prompt}}$ is the acceptance and efficiency correction for prompt \dplus mesons, BR is the branching ratio of the D$^{+} \rightarrow K^{-}\pi^{+}\pi^{+}$  decay channel, the factor 1/2 accounts for the fact that we are measuring the raw yields for the sum of  \dplus and D$^{-}$ and $N_{\rm ev}$ is the number of analyzed events.\\
It is necessary to stress  that the last two equations represent a visible cross section for \dplus mesons  counted in events triggered by the V0AND that pass the  physics selection,  the pile-up rejection and with a  primary vertex reconstructed from ITS TPC tracks and |$z_{\rm RecoVert}$|<10 cm. Since the \textit{kStepGenerated} step of the Correction Framework (\ref{sub:acceff})  requires  |$z_{\rm Generated Vert}$|<10 cm, the efficiency factor also corrects for those events in which  |$z_{\rm Generated Vert}$|<10 cm, but no vertex has been reconstructed. For this reason $N_{\rm ev}$ is computed as
\begin{equation}
\begin{split}
N_{\rm ev} = N^{\rm Reco Vert}(|z_{\rm Reco Vert}|<10 \textrm{cm})+N^{\rm No Vert}(|z_{\rm Reco Vert}|<10 \textrm{cm})=  \\ = N^{\rm Reco Vert}(|z_{\rm Reco Vert}|<10 \textrm{cm})+N^{\rm No Vert}_{\rm tot} - N^{\rm No Vert}(|z_{\rm Reco Vert}|>10 \textrm{cm})= \\ = N^{\rm Reco Vert}(|z_{\rm Reco Vert}|<10 \textrm{cm})+N^{\rm No Vert}_{\rm tot} - N^{\rm No Vert}_{\rm tot}\frac{N^{\rm Reco Vert}(|z_{\rm Reco Vert}|>10 \textrm{cm})}{N^{\rm Reco Vert}_{\rm tot}}
\end{split}
\label{eq:norm}
\end{equation}
This allows  to normalize the raw yield to the number of minimum-bias interactions (V0AND triggers) with the $z$ position of the vertex within 10 cm from the detector centre, independently from the fact that the vertex was reconstructed.
The nuclear modification factor is  computed as
\begin{equation}
R_{\rm pA}(\pteq) =   \frac{d\sigma_{\rm pA}^{D^{+}}/d\pteq}{{\rm A} \ d\sigma_{\rm pp}^{D^{+}}/d\pteq}
\end{equation}
More details on this formula can be found in Appendix \ref{AppendixA}. In the next sections the  ($\textrm{Acc}\times \epsilon$) and \fprompt correction factors will be discussed in more details.
\section{Acceptance and efficiency corrections}
\label{AccEffRppB}
\begin{figure}[t]
\centering
 \includegraphics[width=0.99\textwidth]{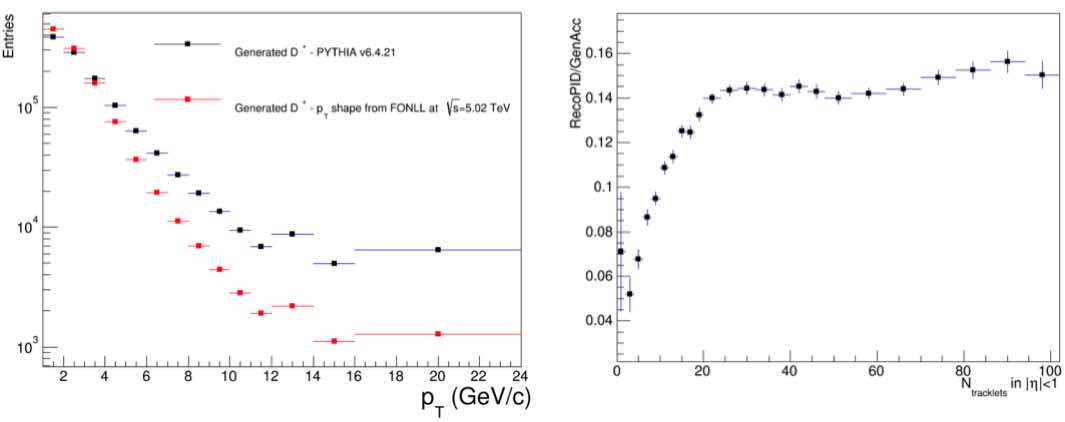}
\caption{Left: \pt distribution of generated \dplus mesons obtained from   PYTHIA v6.4.21 and from  FONLL. Right: efficiency for prompt \dplus mesons selection as a function of the multiplicity, quantified by the number of SPD
 tracklets in |$\eta$|<1}
\label{eff1}
\end{figure}
\begin{figure}[b]
\centering
 \includegraphics[width=0.6\textwidth]{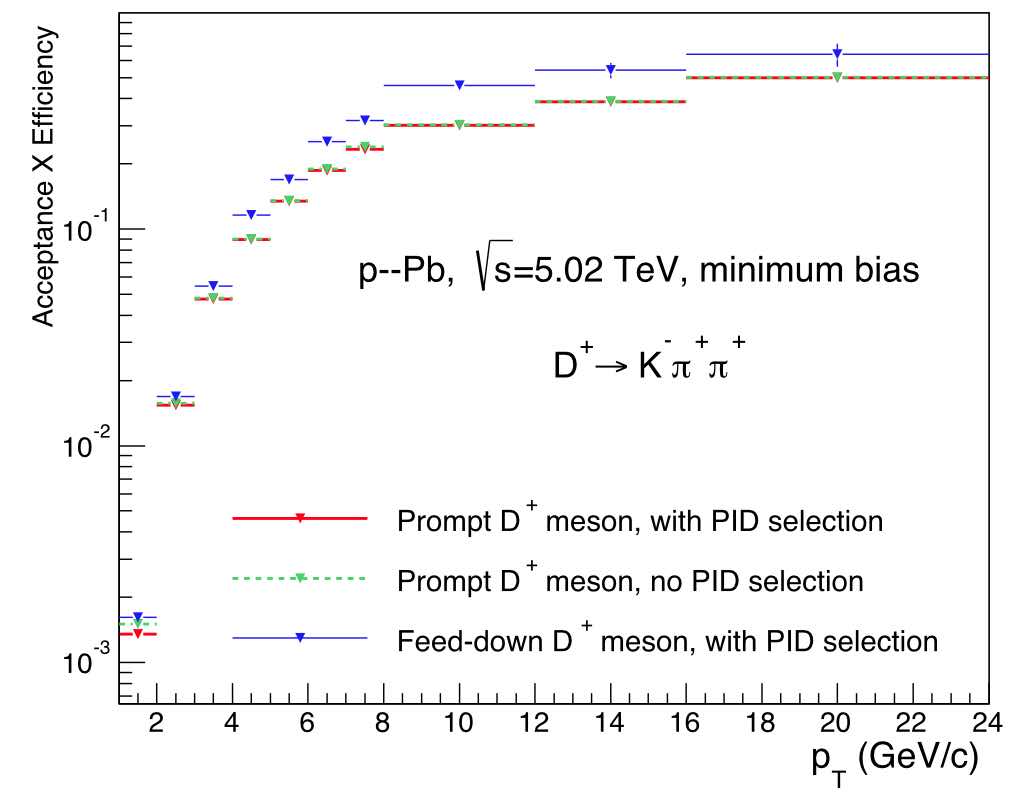}
\caption{Left: Acceptance times efficiency correction factor for prompt (with and without PID selection) and feed-down \dplus mesons as a function  of \pt, obtained after applying multiplicity and \pt shape corrections, for minimum bias p--Pb collisions at \sqnn= 5.02 TeV. }
\label{eff3}
\end{figure}
The efficiency and acceptance correction (Acc$\times \epsilon$) is obtained from the MC data sample generated using HIJING for simulating the underlying p--Pb event and PYTHIA v6.4.21 to injiect a pp collision with a $c\bar{c}$ pair as described in Section \ref{sub:MCsample}. As discussed therein,  the acceptance correction factor is obtained as the ratio of the \textit{kStepAcceptance} and \textit{kStepGeneratedLimAcc} containers (the definition of container can be found in Section \ref{sub:acceff}) and the efficiency correction factor is obtained as the ratio of the \textit{kStepRecoPID} and  \textit{kStepAcceptance} containers\footnote{Due to the definition of the \textit{kStepGeneratedLimAcc} and \textit{kStepAcceptance} in Section  \ref{sub:acceff}, the Acc$\times \epsilon$ discussed here actually also includes the $\Delta y$ factor present in Equations \ref{eq:corrsp} and \ref{eq:corrcs}}. In both cases the statistical uncertainties on the ratio are computed as binomial errors, since in both cases the numerator is a sub-sample of the denominator. We have already seen how the  efficiency of D$^{+}$ meson reconstruction and selection shown in Figure \ref{acceff} (right) varies from 0.5-1\% for \pt < 2 GeV/c to 20-30\% for \pt > 12 GeV/c because of the larger displacement of the decay vertex of high-\pt candidates due to the Lorentz boost. The efficiency of feed-down \dplus meson selection is also higher than the one for prompt \dplus meson because at the same p$_{\rm T}$, feed-down \dplus are on on average more displaced from the primary vertex. The efficiencies of Figure \ref{acceff} (right) are directly  obtained from the MC data sample without any further correction. However we will now discuss two corrections that  one has to take into account to get the appropriate values of the efficiencies: 
\begin{itemize}
\item corrections for the \pt shape of the generated \dplus mesons 
\item  corrections for the different event multiplicity distributions in data and simulations
\end{itemize}
\begin{figure}[t]
\centering
 \includegraphics[width=0.9\textwidth]{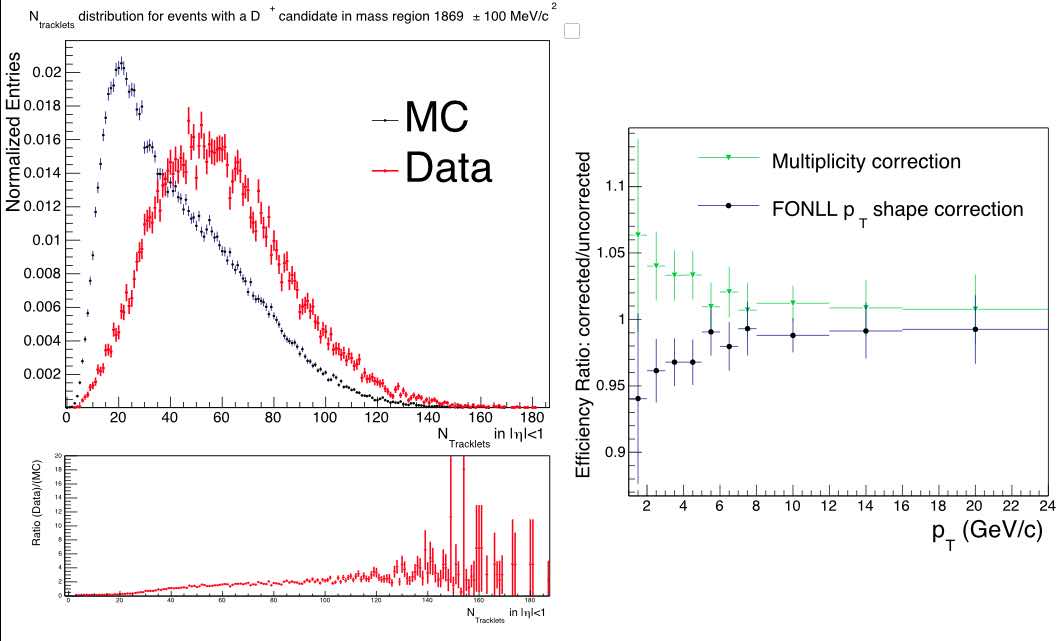}
\caption{Left: N$_{\rm tracklets}$ distributions for events that contain at least one \dplus candidate in the  invariant mass region 1869$\pm$100 MeV/c$^{2}$ for data (red) and HIJING+PHYTHIA simulation (black). In the bottom panel the ratio of the two distributions is shown. Right: ratio of the efficiency values obtained applying the two corrections separately and those obtained without any correction.}
\label{eff2}
\end{figure}
In order to avoid a bias in the efficiency values due to a difference in the generated \pt shape between data and MC, the  \dplus meson spectrum used to calculate the efficiencies was re-weighted to reproduce the shape given by FONLL  calculations at \sqs= 5.02 TeV. The difference in the \pt spectrum of generated \dplus meson obtained from   PYTHIA v6.4.21 and from  FONLL  is shown in Figure \ref{eff1}.\\
The efficiency also depends  on the multiplicity of charged particles produced in the collision since the primary vertex resolution, and consequently the resolution of the topological selection variables, improves with increasing multiplicity. As an example, the selection efficiency for \dplus mesons as a function of the multiplicity, quantified by the number of SPD
 tracklets in |$\eta$|<1, is shown in the right panel of Figure \ref{eff1}. Due to its multiplicity dependence, the efficiency  has to be
 estimated using a Monte Carlo sample that reproduces the multiplicity distribution observed in data. The
 N$_{\rm tracklets}$ distributions for events that contain at least one \dplus candidate in the  invariant mass region 1869$\pm$100 MeV/c$^{2}$ are shown in
 the top panel of Figure \ref{eff2} (left) for data and simulated events. The distribution for data  is plotted in red, that from the HIJING+PHYTHIA simulation  in black. The ratio of the two distributions is shown in
 the bottom panel of Figure \ref{eff2} (left). This ratio can be used to re-weight the multiplicity distribution of generated and reconstructed obtained from the MC data sample,
 in order to reproduce the multiplicity distribution of the data. To do this, the output of the correction framework was projected, for each \pt interval, on
 the N$_{\rm tracklets}$ variable for the steps  relevant for the efficiency calculation, namely   \textit{kStepRecoPID} and  \textit{kStepAcceptance}. The two N$_{\rm tracklets}$  distributions were then multiplied, bin by bin,  by the weight function, and the efficiencies were evaluated as the ratio of the \textit{kStepRecoPID} and  \textit{kStepAcceptance} containers after the re-weighting.\\
\begin{figure}[t]
\centering
 \includegraphics[width=0.6\textwidth]{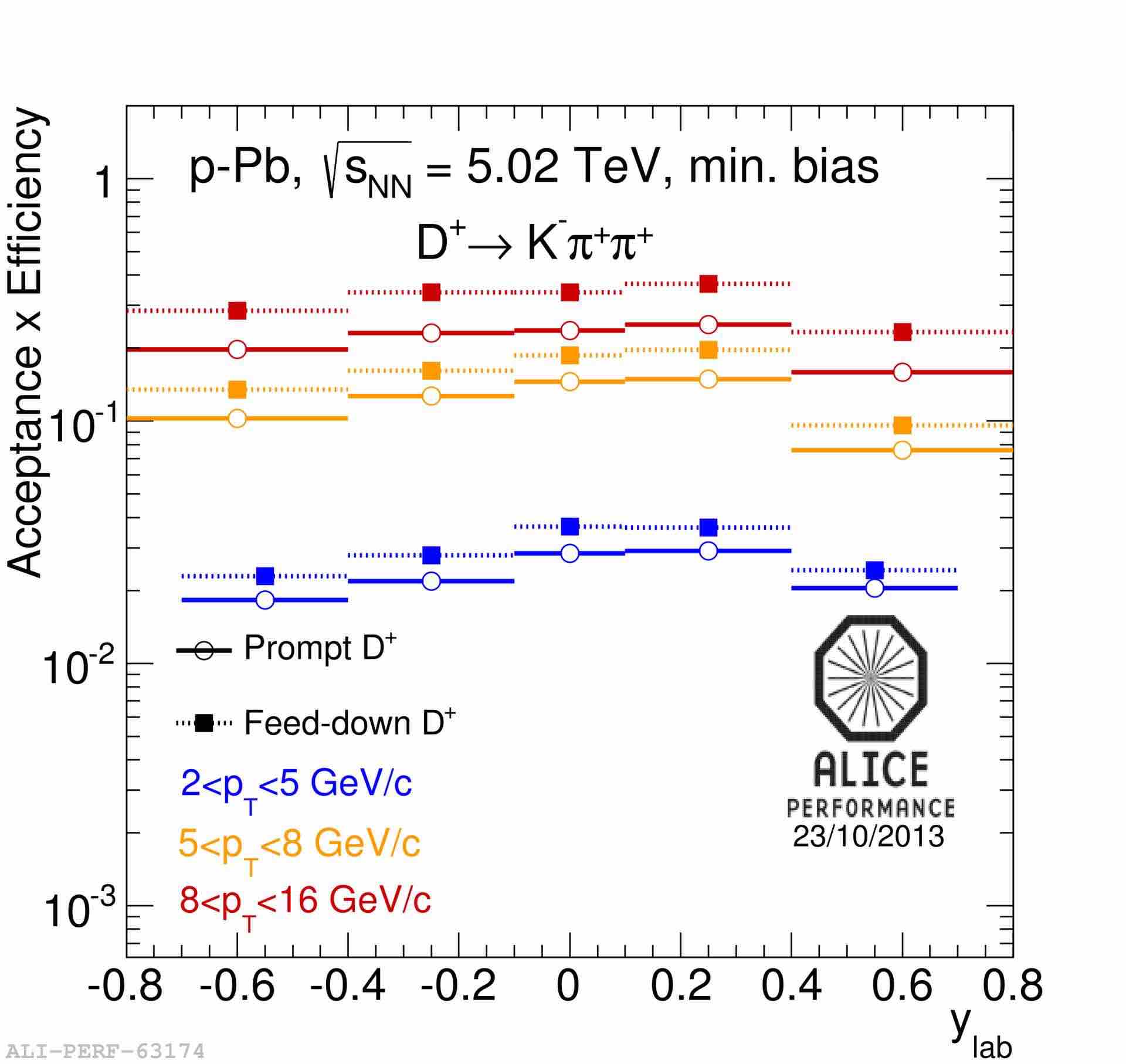}
\caption{Left:  Acceptance times efficiency correction factor for prompt and feed-down \dplus mesons as a function  of rapidity measured in the laboratory frame $y_{lab}$, obtained in three \pt bins after applying multiplicity and \pt shape corrections, for minimum bias p--Pb collisions at \sqnn= 5.02 TeV. }
\label{effy}
\end{figure}
To quantify the effect of  the two corrections described above, Figure \ref{eff2} (right) shows the ratio of the efficiency values obtained applying the two corrections separately (namely the MC \pt shape correction and the multiplicity correction) and those obtained without any correction. The vertical bars represent the stastical errors. The effect of the multiplicity weights is at the level of +4\% at low \pt  and about +2\% at high \pt. The effect of the \pt shape correction   is at the level of -4\% at low \pt  and about  -2\% at high \pt.\\
The final values of the acceptance times efficiency correction are shown in Figure \ref{eff3} as a function of \pt for prompt and feed-down \dplus mesons. The  acceptance times efficiency correction factor for prompt \dplus mesons without PID selections is also shown. For  \pt > 2 GeV/c, the efficiency is almost identical ($\sim$2\% higher in the NO PID case). For \pt< 2 GeV/c the efficiency is 15\%  lower if PID is applied, because in this \pt range the STRONG PID strategy (Section \ref{sub:pid}) has to be  applied in order to get a satisfactory significance.\\
Figure \ref{effy} shows the acceptance times efficiency correction as a function of rapidity for the three \pt intervals considered in Table 5.3. There is  a slight difference in the extraction of this correction factor with respect to the rapidity integrated one, since as shown in Figure \ref{cfcont} the \textit{kStepGeneratedLimAcc} is by definition emtpy for |$y$|>0.5 (\ref{sub:acceff}) and the acceptance correction factor here is obtained  as \textit{kStepAcceptance} over \textit{kStepGenerated}. The same multiplicity an \pt shape corrections described above are applied to get the efficiency correction factor. 
\section{Beauty Feed-down subtraction}
\label{sec:bfd}
The \dplus meson raw yields $N^{D^{+/-}}_{\textrm{raw}}$ contain the contribution of both prompt ($c\rightarrow$\dplus, $c\rightarrow D^{*+}(2010)\rightarrow$\dplus and other resonant decays cited in Chapter 4) and feed-down ($b\rightarrow$B$\rightarrow$\dplus)
 yields. As seen in the previous section, prompt and feed-down contributions to the total raw yield have different reconstruction and selection efficiencies. Since the aim of this chapter is the extraction of the nuclear modification factor of prompt \dplus mesons, the feed-down contribution has to be subtracted. \\
FONLL calculations are used to estimate the expected B-hadron production cross section, varying the theoretical parameters within the following ranges
\begin{itemize}
\item 0.5<$\mu_{F}/\mu_{0}$<2 (central value: 1, 0.5<$\mu_{R}/\mu_{0}$<2 (central value: 1), with the constraint 0.5<$\mu_{F}/\mu_{R}$<2 
\item  4.5<$m_b$<5.0 GeV/c$^2$ (central value: 4.75 GeV/c$^2$) 
\end{itemize}
where $\mu_0$=$\sqrt{m^2_{b}+p^2_{\rm T,Q}}$, with  $m_{b}$ is the mass of the heavy quark. \\
\begin{figure}[t]
\centering
 \includegraphics[width=0.95\textwidth]{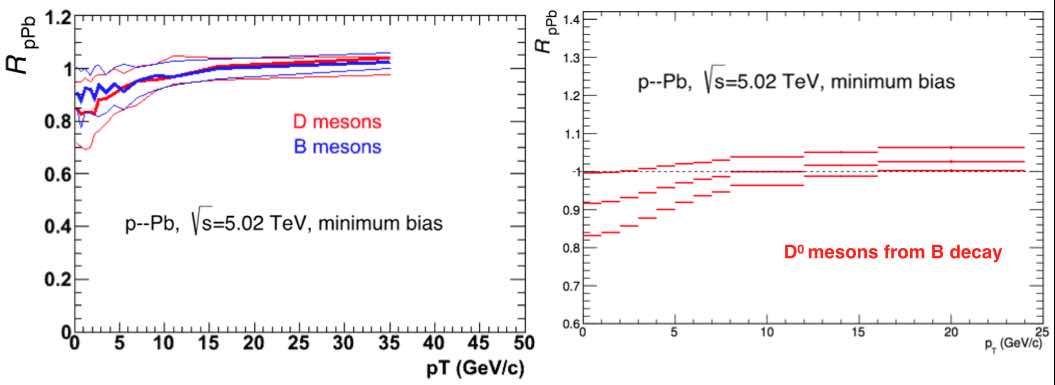}
\caption{Left: R$_{\rm pA}$ of prompt D and B mesons from NLO pQCD calculations including EPS09 shadowing modification of nuclear PDFs. Right: \rppb of D$^{0}$ mesons from B decays.}
\label{nbfc}
\end{figure}
 The B$\rightarrow$\dplus+ $X$ decay kinematic is simulated using the EvtGen\footnote{The EvtGen package is a framework for the implementation of physics processes relevant to decays of B mesons and other resonances} \cite{evtgen} generator to get the  cross-section of D$^{+}$ mesons from B hadron decays  $(\frac{d\sigma}{d\pteq dy})^{\textrm{FONLL}}_{\textrm{feed-down}}$. The production cross section of \dplus mesons from B-hadron decays were corrected by their  acceptances and efficiencies $(\textrm{Acc}\times \epsilon)^{\textrm{feed-down}}$ in each \pt bin,
 in order to evaluate the expected raw yields of \dplus mesons originating from B hadron decays. The fraction of prompt \dplus mesons $f_{\rm prompt}$ is then calculated according to 
\begin{equation}
\begin{split}
f_{\textrm{prompt}} = 1 - \langle T_{\textrm{pA}}\rangle\cdot\bigg(\frac{d\sigma}{d\pteq dy} \bigg)^{\textrm{FONLL}}_{\textrm{feed-down}}\cdot R_{\textrm{pA}}^{\textrm{feed-down}}(\pteq)\cdot \\ \cdot \frac{(\textrm{Acc}\times \epsilon)^{\textrm{feed-down}}\cdot \Delta y\Delta\pteq\cdot BR\cdot N_{\textrm{events}}}{N^{D^{+}}_{\textrm{raw}}/2}
\end{split}
\label{eq:Nb}
\end{equation}
that in the following will be called $N_b$ subtraction method.  The  $\langle T_{\textrm{pA}}\rangle$=0.098$\pm$0.007 mb$^{-1}$ is the minimum bias average nuclear overlap function and is calculated as  $\langle T_{\textrm{pA}}\rangle$= $\langle$N$_{\textrm{coll}}\rangle$/$\sigma_{inel}^{NN}$. The nuclear modification factor of feed-down \dplus mesons $R_{\textrm{pA}}^{\textrm{feed-down}}(\pteq)$ is unknown, since it has never been measured, so an assumption on its value has to be made.
 For the computation of the central value of  \fprompt we will assume $R_{\textrm{pA}}^{\textrm{feed-down}}(\pteq)$=$R_{\textrm{pA}}^{\textrm{prompt D}}(\pteq)$. 
The latter assumption is equivalent to  $R_{\textrm{pA}}^{\textrm{feed-down}}(\pteq)$/$R_{\textrm{pA}}^{\textrm{prompt D}}(\pteq)$=1, which is used in an alternative method to compute \fprompt, according to the following equation
\begin{equation}
f_{\textrm{prompt}} = \bigg( 1 + \frac{\frac{d\sigma}{d\pteq dy}^{\textrm{FONLL}}_{B\rightarrow D}\cdot (\textrm{Acc}\times \epsilon)^{\textrm{feed-down}}\cdot R_{\textrm{pA}}^{\textrm{feed-down}}(\pteq)  }{\frac{d\sigma}{d\pteq dy}^{\textrm{FONLL}}_\textrm{prompt D}\cdot (\textrm{Acc}\times \epsilon)^{\textrm{prompt D}}\cdot R_{\textrm{pA}}^{\textrm{prompt D}}(\pteq)}  \bigg)^{-1}
\label{eq:fc}
\end{equation}
that in the following will be called $f_c$ subtraction method and relies on  the ratio between the prompt and feed-down FONLL predictions for \dplus meson cross-section, instead that on the value of $(\frac{d\sigma}{d\pteq dy})^{\textrm{FONLL}}_{\textrm{feed-down}}$ like in the $N_b$ method\footnote{One difference between the $N_b$ and  $f_c$ method is that in the latter the central value for the $R_{\textrm{pA}}^{\textrm{prompt D}}(\pteq)$ follows directly from the assumption  $R_{\textrm{pA}}^{\textrm{feed-down}}(\pteq)$/$R_{\textrm{pA}}^{\textrm{prompt D}}(\pteq)$=1. In the $N_b$ method the central value of  $R_{\textrm{pA}}^{\textrm{prompt D}}(\pteq)$ is obtained generating \fprompt for a range of  $R_{\textrm{pA}}^{\textrm{feed-down}}(\pteq)$ values, and calculating  $R_{\textrm{pA}}^{\textrm{prompt D}}(\pteq)$.}.\\
The choice  $R_{\textrm{pA}}^{\textrm{feed-down}}(\pteq)$=$R_{\textrm{pA}}^{\textrm{prompt D}}(\pteq)$ has been done considering an estimate on the influence of   shadowing effects  on the production of  prompt and feed-down D mesons,  performed using NLO pQCD calculations, which included EPS09  modification of nuclear PDFs. The resulting \rppb of prompt D and B mesons can be seen in Figure \ref{nbfc}  (left). The EvtGen  decayer is used in order to extract the expected \rppb of D$^{0}$ mesons from B decays (Figure \ref{nbfc}  (right)) from that of B hadrons, which results to be similar to that of prompt D mesons. \\The hypothesis  is  varied in the range  0.9<$R_{\textrm{pA}}^{\textrm{feed-down}}(\pteq)$/$R_{\textrm{pA}}^{\textrm{prompt D}}(\pteq)$<1.3 to evaluate the systematic uncertainty, as will be discussed in  Section \ref{sec:sysrppb}.\\
\section{The pp reference at 5.02 TeV }
A pp reference at $\sqrt{s}$= 5.02 TeV is required in order to compare heavy flavour production in p--Pb and pp collisions via the nuclear modification factor \rppb. However a pp data sample at $\sqrt{s}$= 5.02 TeV does not exist yet, and the pp reference is extracted from pQCD calculations that allow us to extrapolate to lower energies the \pt-differential cross sections of D$^+$ meson measured at 7 TeV and shown in Chapter 2 (Figure \ref{Dplus7276} - left).\\
This procedure was already used in the evaluation of the D-meson nuclear modification factor in Pb--Pb collisions at \sqnn= 2.76 TeV, shown in Chapter 2 (Figure \ref{RaaDmeson}), and is described in the following.\\
The results of prompt \dplus meson \pt-differential production cross section in pp collisions at \sqs= 7 TeV are shown in Figure \ref{Dplus7276} together with FONLL and GM-VFNS predictions. The uncertainty band of the FONLL predictions is obtained considering a parameter variation as follows:
\begin{itemize}
\item 0.5<$\mu_{F}/\mu_{0}$<2 (central value: 1)
\item 0.5<$\mu_{R}/\mu_{0}$<2 (central value: 1)
\item with the constraint 0.5<$\mu_{F}/\mu_{R}$<2 
\item 1.3<$m_c$<1.7 GeV/c$^2$ (central value: 1.5 GeV/c$^2$) and   4.5<$m_b$<5.0 GeV/c$^2$ (central value: 4.75 GeV/c$^2$) 
\end{itemize}
where $\mu_0$=$\sqrt{m^2_{\rm Q}+p^2_{\rm T,Q}}$, with  $m_{\rm Q}$ is the mass of the heavy quark. \\
\begin{figure}[t]
\centering
 \includegraphics[width=0.99\textwidth]{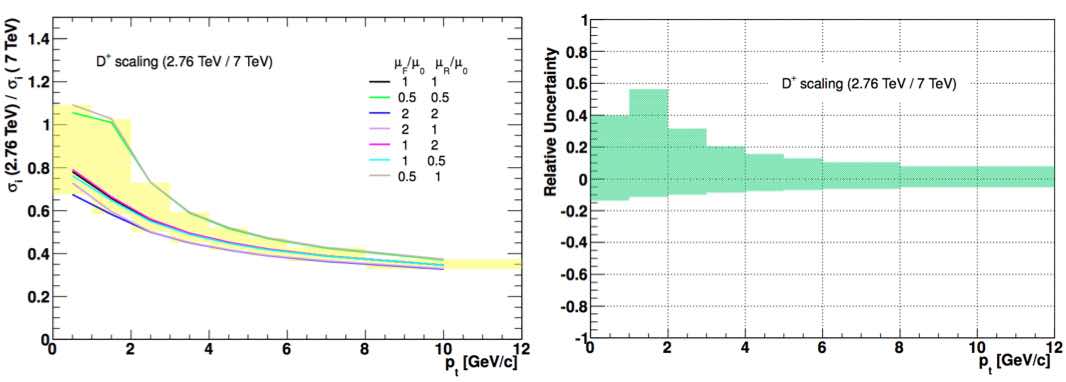}
\caption{Left: FONLL scaling factor for \dplus mesons from \sqs=7 TeV to \sqs= 2.76 TeV. Right: relative uncertainty on the scaling factor.}
\label{Ref1}
\end{figure}
In order to compute the scaling factor from 7 TeV to a given energy $\alpha$ TeV:
\begin{itemize}
\item the FONLL predictions at 7 TeV $\sigma(7)$ and at $\alpha$ TeV $\sigma(\alpha)$ are rebinned according to the results of prompt \dplus meson production in pp collisions at \sqs= 7 TeV  shown in Figure \ref{Dplus7276}
\item the ratio $\sigma(\alpha)/\sigma(7)$ is calculated considering that
\begin{itemize}
\item the central value is the ratio of the central predictions at both energies
\item its uncertainty is defined by the envelope of the ratio of the calculations for the different sets of parameter. The theoretical calculation parameters are correlated (equal) at different energies
\end{itemize}
\item multiply the  cross-section measured at 7 TeV by the FONLL  $\sigma(\alpha)/\sigma(7)$ binned ratio
\item propagate the uncertaintes:
\begin{itemize}
\item on the FONLL ratios
\item on the uncertaintes of the 7 TeV measurement
\item combine these uncertaintes summing them in quadrature
\end{itemize}
\end{itemize}
\subsection{pp reference at \sqs= 2.76 TeV}
 I will first show some results on the scaling performed to get the pp reference at \sqs= 2.76 TeV, in order to prove the stability of the method.
The FONLL scaling factor for \dplus from \sqs= 7 TeV to \sqs= 2.76 TeV  is shown in Figure \ref{Ref1} (left) as a function of $p_{\rm T}$. The scaling factor obtained with the different sets of scales are drawn with solid lines, while the resulting global scaling envelope is depicted by a yellow filled band. The central value (black dashed line) of the scaling is obtained with $\mu_{F}/\mu_{0}$ =$\mu_{R}/\mu_{0}$ = 1 and $m_{c}$=1.5 GeV/c$^{2}$. The values of the scales for the other sets are reported in the legend ($\mu_{F}/\mu_{0}$,$\mu_{R}/\mu_{0}$). The scaling factor depends mainly on the value of the factorization scale $\mu_{F}$, since for the same charm quark \pt, different Bjorken $x$ ranges are probed at 2.76 TeV and at 7 TeV, and changing the factorization scale affects the $x$ dependence of the parton distribution functions (PDFs). The right panel of Figure \ref{Ref1} shows the relative uncertainty of the scaling factor, obtained from the envelope of the different computations in the left panel. The relative uncertainty is larger (40$\div$50\%) at low p$_{\rm T}$, and decreases to about 8\% for \pt> 8 GeV/c. The scaling factor does not show a significant dependence on the value used for the charm quark mass in the calculation as shown in \cite{ppref}.\\
\begin{figure}[t]
\centering
 \includegraphics[width=0.99\textwidth]{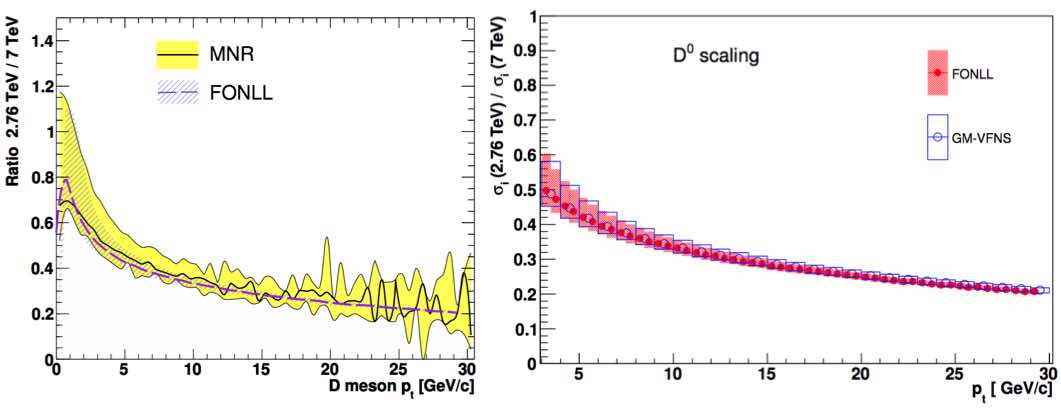}
\caption{Left: Scaling factor from \sqs=7 TeV to \sqs= 2.76 TeV for D$^{0}$ mesons obtained from FONLL and POWHEG. Right:Scaling factor from \sqs= 7 TeV to \sqs= 2.76 TeV for D$^{0}$ mesons obtained from FONLL and GM-VFNS. }
\label{Ref2}
\end{figure}
\begin{figure}[b]
\centering
 \includegraphics[width=0.99\textwidth]{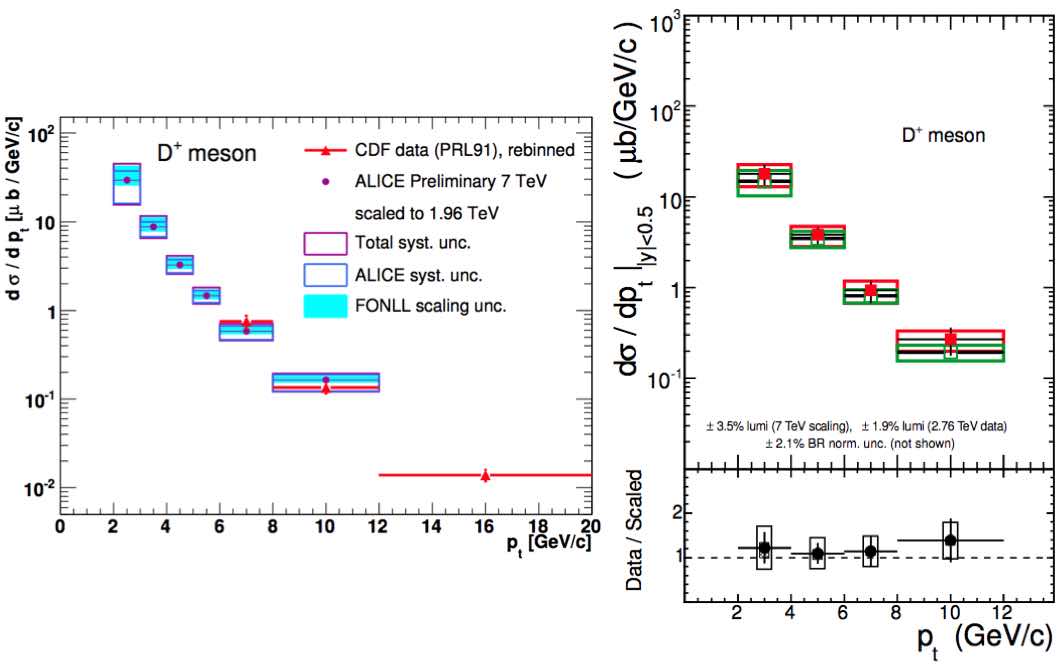}
\caption{Left: ALICE pp data at \sqs=7 TeV scaled with  FONLL calculations  to \sqs= 1.96 TeV and  D$^{+}$ meson cross-section measured by the CDF Collaboration in p$\bar{\rm p}$ collisions at \sqs= 1.96 TeV \cite{CDFD}. Right:  ALICE pp data at \sqs=7 TeV scaled with  FONLL calculations  to \sqs= 1.96 TeV and  D$^{+}$ meson cross-section measured by the ALICE Collaboration in pp collisions at \sqs= 2.76 TeV \cite{CDFD}.}
\label{Ref3}
\end{figure}
Figure \ref{Ref2} shows  the scaling factor from \sqs= 7 TeV to  \sqs= 2.76 TeV  obtained for D$^0$ with FONLL calculations and its uncertainties together with:
\begin{itemize}
\item the scaling factor obtained with MNR calculations (\cite{POWHEG} - Section 2.2.5) cross-sections and their uncertainties (left).The scaling factors agree with each other, however the uncertainties are larger for the MNR case
\item the scaling factor obtained with GM-VFNS cross-sections (\cite{GM-VFNS} - Section 2.1.2) and its uncertainties (right).  In the GM-VFNS case the parameters that are varied within the calculations to obtain the uncertaintes are the renormalization scale, the factorization scale for initial state singularities and the factorization scale for final state singularities. In this case the scaling central values and their uncertainties agree very well with those given by FONLL
\end{itemize}
Figure \ref{Ref2} (left) shows the ALICE pp data at \sqs=7 TeV scaled with FONLL calculations  to \sqs= 1.96 TeV compared to  the D$^{+}$ meson cross-section measured by the CDF Collaboration in p$\bar{\rm p}$ collisions at \sqs= 1.96 TeV \cite{CDFD}. The CDF measurements and the ALICE rescaled data agree within uncertaintes. \\
Figure \ref{Ref2} (left) compares the ALICE pp data at \sqs=7 TeV scaled with FONLL calculations  to \sqs=2.76 TeV with the D$^{+}$ meson cross-section measured by the ALICE Collaboration in pp collisions at \sqs= 2.76 TeV \cite{CDFD}. The integrated luminosity used to compute the  D$^{+}$ meson cross-section at \sqs=2.76 TeV is 1.1 nb$^{-1}$. The  measurements at \sqs = 7 TeV were rebinned to match the \pt binning used at \sqs=2.76 TeV.   The ALICE data scaled from \sqs= 7 TeV  agree with the ones measured at \sqs= 2.76 TeV within uncertainties in all \pt bins. The results are compatible within statistical uncertainties only, and their central values coincide within 5–10\% in almost all \pt bins, confirming the stability and appropriateness of the energy scaling procedure. \\
\subsection{pp reference at \sqs= 5.02 TeV}
\label{sub:sc502}
\begin{figure}[t]
\centering
 \includegraphics[width=0.7\textwidth]{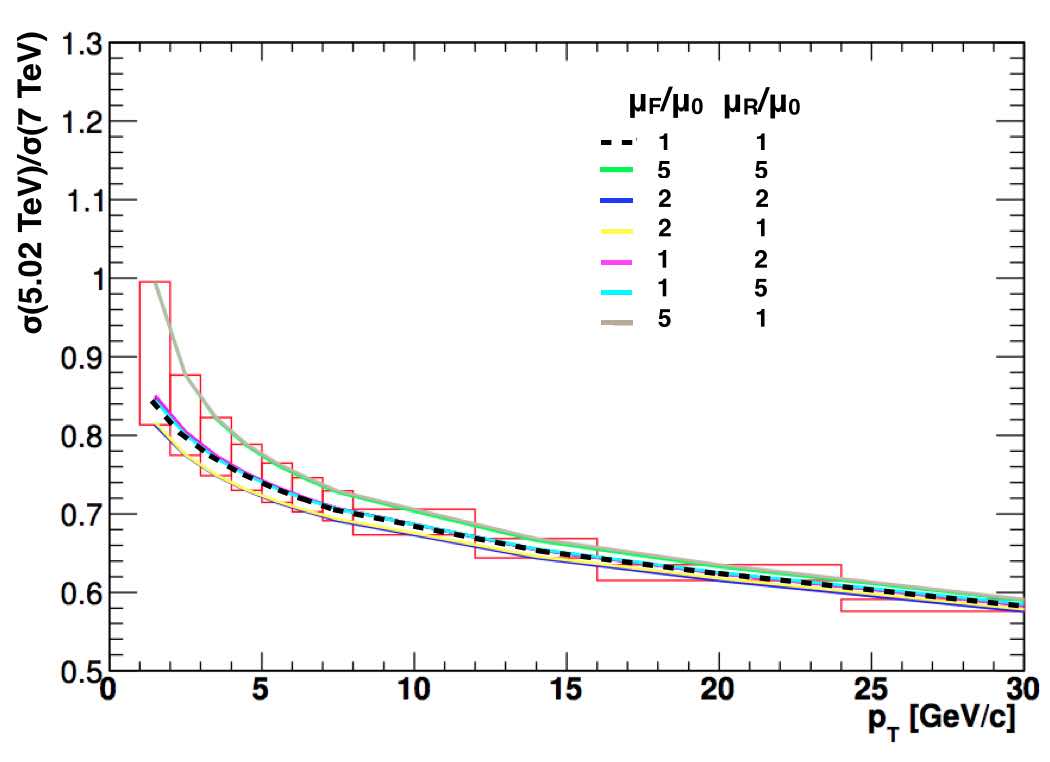}
\caption{FONLL scaling factor for \dplus mesons from \sqs= 7 TeV to \sqs= 5.02 TeV}
\label{Ref4}
\end{figure}
\begin{figure}[t]
\centering
 \includegraphics[width=0.7\textwidth]{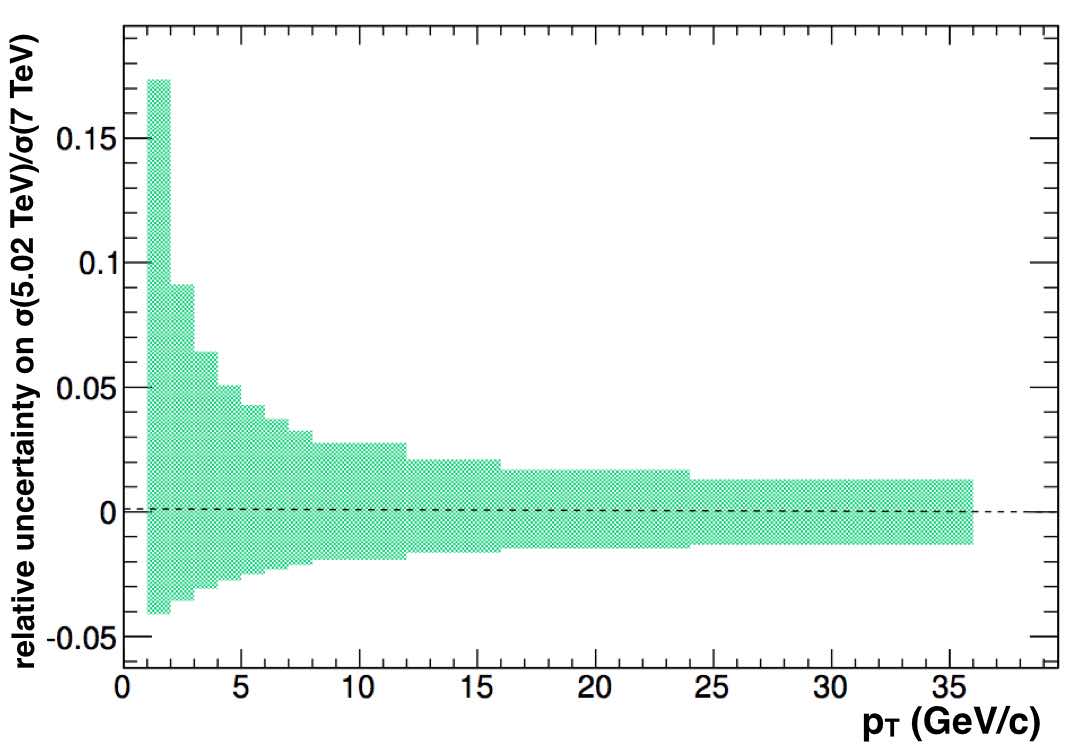}
\caption{Relative uncertainties on the FONLL scaling factor for \dplus mesons from \sqs= 7 TeV to \sqs= 2.76 TeV.}
\label{Ref5}
\end{figure}
As seen in the previous paragraph, the FONLL scaling factors agree very well with those obtained with other pQCD calculations, so only the approach based on FONLL will be shown for the case of scaling to \sqs= 5.02 TeV. The scaling factor obtained with the FONLL approach is the one that will be used to compute the \rppb that will be shown in the end of this chapter.\\
The FONLL calculations were performed using the CTEQ 6.6 PDFs parametrization (Section \ref{sub:pdfs}), and varying the factorisation and renormalisation scales within the same ranges quoted in the previous paragraph. The resulting scaling factor is shown in Figure \ref{Ref4}. As in the case of the scaling to \sqs=2.76 TeV, it depends mainly on the value of the factorization scale due to its effect on the $x$ dependence of PDFs. The  uncertainties  on the scaling factor  are shown in Figure \ref{Ref5}. The relative uncertainty is larger (15\%) at low \pt, and decreases to about 2\% for \pt> 8 GeV/c. The uncertinties are smaller with respect to the scaling from \sqs= 7 TeV to \sqs= 2.76 TeV, as expected due to the smaller energy gap.
\section{Systematic uncertaintes }
\label{sec:sysrppb}
The sources of systematic uncertainty considered in this analysis are as follows: 
\begin{enumerate}
\item systematic uncertainty on the extraction of the raw yields from the invariant mass distributions;
\item systematic uncertainty on the efficiency determination due to the imperfect description of the cut variables in the MC;
\item systematic uncertainty on the PID selection  efficiency;
\item systematic uncertainty on the efficiency due to the \pt shape of the generated D$^{+}$;
\item systematic uncertainty on the tracking efficiency. 
\item systematic uncertainty on the beauty feed-down subtraction
\item systematic uncertainty on the pp reference at \sqs= 5.02 TeV
\item systematic uncertainty on the integrated  luminosity the and branching ratio
\end{enumerate} 
 \subsection{Systematic uncertainty due to the yield extraction}
\label{sub:systYield}
The systematic error on the yield extraction was estimated by comparing the   extracted raw yields when the invariant mass fit parameters and constraints were varied. In particular, the following  tests were made:
\begin{itemize}
\item extending and reducing the  invariant  mass range used in the fit;
\item using  a different bin width for the invariant mass distribution (twice the bin width and half of the bin width);
\item fixing the peak width to that extracted from MC;
\item using a parabolic  background fit function instead of the  exponential one;
\item using a bin counting method based on counting the entries in the invariant mass histogram within $3\sigma$ from the peak centre after subtracting the background  estimated via the fit;
\end{itemize}
\begin{figure}[t]
\centering
 \includegraphics[width=0.9\textwidth]{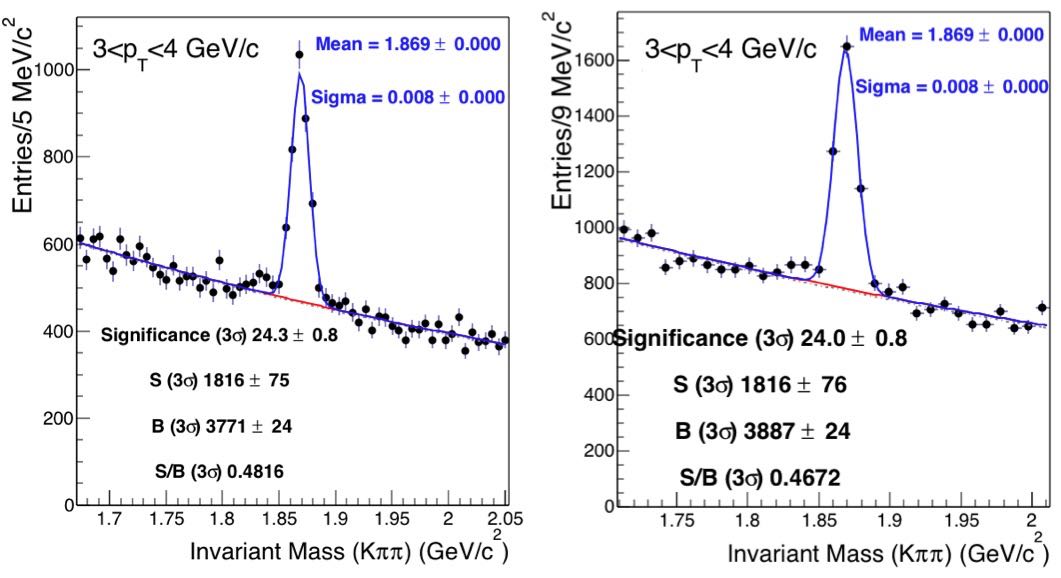}
\caption{Invariant mass fit in 3<\pt<4 GeV/c for the multiplicity integrated case  using  half  the bin width (left) and restricting the fit range (right).}
\label{exampleye}
\end{figure}
As an example, Figure \ref{exampleye} shows the fit to the  invariant distribution  in 3 < \pt<4 GeV/c for the  cases:
\begin{itemize}
\item using  half  the bin width (left)
\item restricting the fit range (right)
\end{itemize}
The ratio between the  raw yields obtained with these variations  and the ones of the default configuration  was evaluated to estimate the systematics. The ratios  are reported in Figure
 \ref{YEPlot} in the rapidity integrated case (left) and as a function of rapidity in 5<\pt<8 GeV/c. The statistical errors on the extracted raw yields are treated as uncorrelated in the ratio. However some degree of correlation is present among the raw yields extracted with different fit techniques, so the errors of Figure  \ref{YEPlot} are slightly overestimated.The systematic uncertainties were assigned so as to obtain  the best estimate of an RMS uncertainty by  removing the "pathological" cases in which  the values of mean and sigma of the invariant mass fit deviate from the expected values, and by trying not to include effects of statistical fluctuations in the estimated systematic uncertainty. The resulting values of systematic uncertainties   are summarized in Table \ref{tab:YESys} for the rapidity integrated case and for the five rapidity intervals.\\
The systematic uncertainties increase at low \pt is due to the lower significance. At high \pt the higher systematic uncertainties are due to the discrepancy in the Gaussian $\sigma$ observed between data and MC.
\begin{figure}[t]
\centering
 \includegraphics[width=0.9\textwidth]{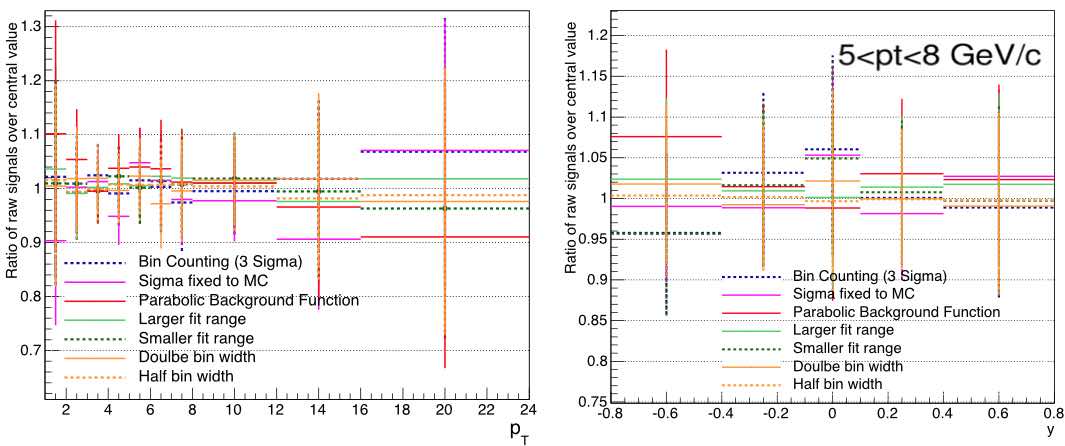}
\caption{Left: relative variation of the raw yield obtained with the different fit strategies with respect to the central values of Figures \ref{IM1},  \ref{IM2} and  \ref{IM3} as a function of transverse momentum. Right: the same in the \pt bin [5,8] GeV/c as a function of rapidity.}
\label{YEPlot}
\end{figure}
\begin{table}[h]
\begin{tabular}{|c|c|c|c|c|c|c|c|c|c|c|}
\hline
p$_{\rm T}$ (GeV/c)                & 1-2  & 2-3    & 3-4    & 4-5    & 5-6    & 6-7    & 7-8    & 8-12        & 12-16       & 16-24 \\ \hline
rapidity integrated                & 10\% & 8\%    & 5\%    & 5\%    & 5\%    & 5\%    & 5\%    & 5\%         & 8\%         & 8\%   \\ \hline
-0.8/-0.7\textless$y_{\rm lab}$\textless-0.4 & x    & \multicolumn{3}{c|}{7\%} & \multicolumn{3}{c|}{5\%} & \multicolumn{2}{c|}{10\%} & x     \\ \hline
-0.4\textless$y_{\rm lab}$\textless-0.1      & x    & \multicolumn{3}{c|}{5\%} & \multicolumn{3}{c|}{5\%} & \multicolumn{2}{c|}{10\%} & x     \\ \hline
-0.1\textless$y_{\rm lab}$\textless0.1       & x    & \multicolumn{3}{c|}{5\%} & \multicolumn{3}{c|}{5\%} & \multicolumn{2}{c|}{5\%}  & x     \\ \hline
0.1\textless$y_{\rm lab}$\textless0.4        & x    & \multicolumn{3}{c|}{5\%} & \multicolumn{3}{c|}{5\%} & \multicolumn{2}{c|}{5\%}  & x     \\ \hline
0.4\textless$y_{\rm lab}$\textless0.7/0.8    & x    & \multicolumn{3}{c|}{7\%} & \multicolumn{3}{c|}{5\%} & \multicolumn{2}{c|}{7\%}  & x     \\ \hline
\end{tabular}
\caption{Systematic uncertainty on the extraction of raw yields from the \dplus meson invariant mass distributions}
\label{tab:YESys}
\end{table}
\subsection{Systematic uncertainty due to topological cut efficiency}
\label{sub:systCut}
The systematic uncertainty  due to the imperfect description in the Monte Carlo of the variable used in  the  topological selection  was evaluated by repeating the analysis with four different sets of cuts and looking at the relative variations of the corrected yield in each \pt and rapidity interval. Two of the set of cuts adopted are looser (i.e. have higher selection efficiencies) with respect to those reported in Table \ref{TableCutsDplus}, while the other two are tighter (i.e. have lower selection efficiencies). In the plots of Figure \ref{CutVar} they are labeled according to the cut value used on the cosine of the pointing angle (central value = 0.99). The variation of the corrected spectra with the modified sets of cuts relative to the one obtained with the cuts of  Table \ref{TableCutsDplus}, shown in Figure  \ref{CutVar}, is used to assign the systematic uncertainty as a best estimate of an RMS uncertainty. This was done, for example,  checking that the values of mean and sigma of the invariant mass fit are  stable
for each set of cuts, and that the extracted significance is reasonably high (>3 in all cases). In addition, at high \pt the full spread of the ratios shown in Figure  \ref{CutVar} has to be disentangled from the high  statistical uncertainty arising from the limited statistics.  The statistical errors on the computed  corrected spectra are treated as correlated in the ratio.  Some degree of correlation is present among the raw yields extracted with different fit techniques, so the errors of Figure  \ref{YEPlot} are slightly overestimated.\\
The assigned values of  relative systematic uncertainties due to topological cut efficiencies  are reported in Table \ref{tab:CVSys}.
\begin{table}[h]
\centering
\begin{tabular}{|c|c|c|c|c|c|c|c|c|c|c|}
\hline
p$_{\rm T}$ (GeV/c)                & 1-2  & 2-3    & 3-4    & 4-5    & 5-6    & 6-7    & 7-8    & 8-12        & 12-16       & 16-24 \\ \hline
rapidity integrated                & 10\% & 10\%    & 6\%    & 6\%    & 6\%    & 5\%    & 5\%    & 5\%         & 5\%         & 5\%   \\ \hline
-0.8/-0.7\textless$y$\textless-0.4 & x    & \multicolumn{3}{c|}{7\%} & \multicolumn{3}{c|}{7\%} & \multicolumn{2}{c|}{8\%} & x     \\ \hline
-0.4\textless$y$\textless-0.1      & x    & \multicolumn{3}{c|}{7\%} & \multicolumn{3}{c|}{7\%} & \multicolumn{2}{c|}{8\%} & x     \\ \hline
-0.1\textless$y$\textless0.1       & x    & \multicolumn{3}{c|}{7\%} & \multicolumn{3}{c|}{7\%} & \multicolumn{2}{c|}{8\%}  & x     \\ \hline
0.1\textless$y$\textless0.4        & x    & \multicolumn{3}{c|}{7\%} & \multicolumn{3}{c|}{7\%} & \multicolumn{2}{c|}{8\%}  & x     \\ \hline
0.4\textless$y$\textless0.7/0.8    & x    & \multicolumn{3}{c|}{7\%} & \multicolumn{3}{c|}{7\%} & \multicolumn{2}{c|}{8\%}  & x     \\ \hline
\end{tabular}
\caption{Systematic uncertainty due to topological cut efficiency}
\label{tab:CVSys}
\end{table}
\begin{figure}[t]
\centering
 \includegraphics[width=0.9\textwidth]{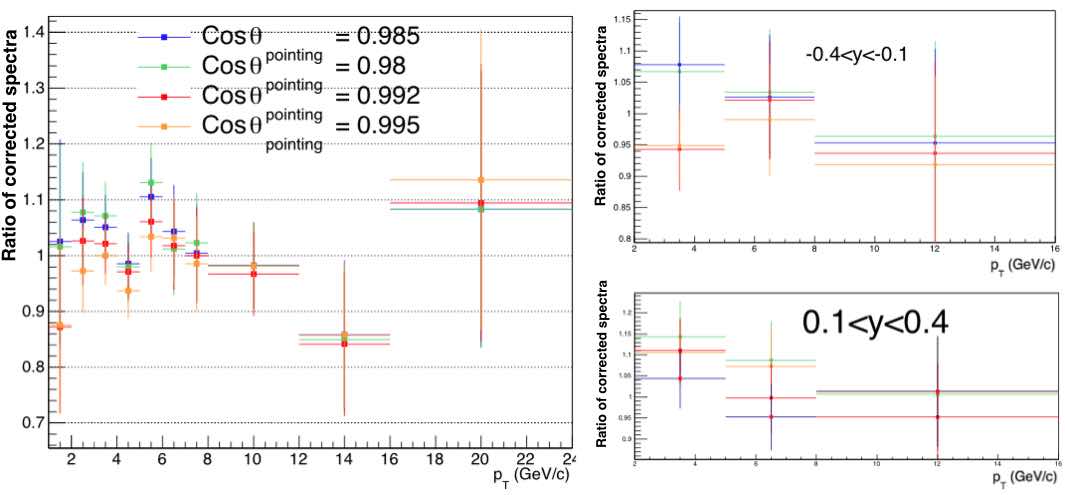}
\caption{Left: relative variation of the corrected spectra obtained with  different sets of cuts as a function of \pt. Right: the same in the \pt bin [5,8] GeV/c as a function of rapidity.}
\label{CutVar}
\end{figure}
\subsection{Systematic uncertainty due to PID}
\label{sub:systPID}
\begin{figure}[t]
\centering
 \includegraphics[width=0.6\textwidth]{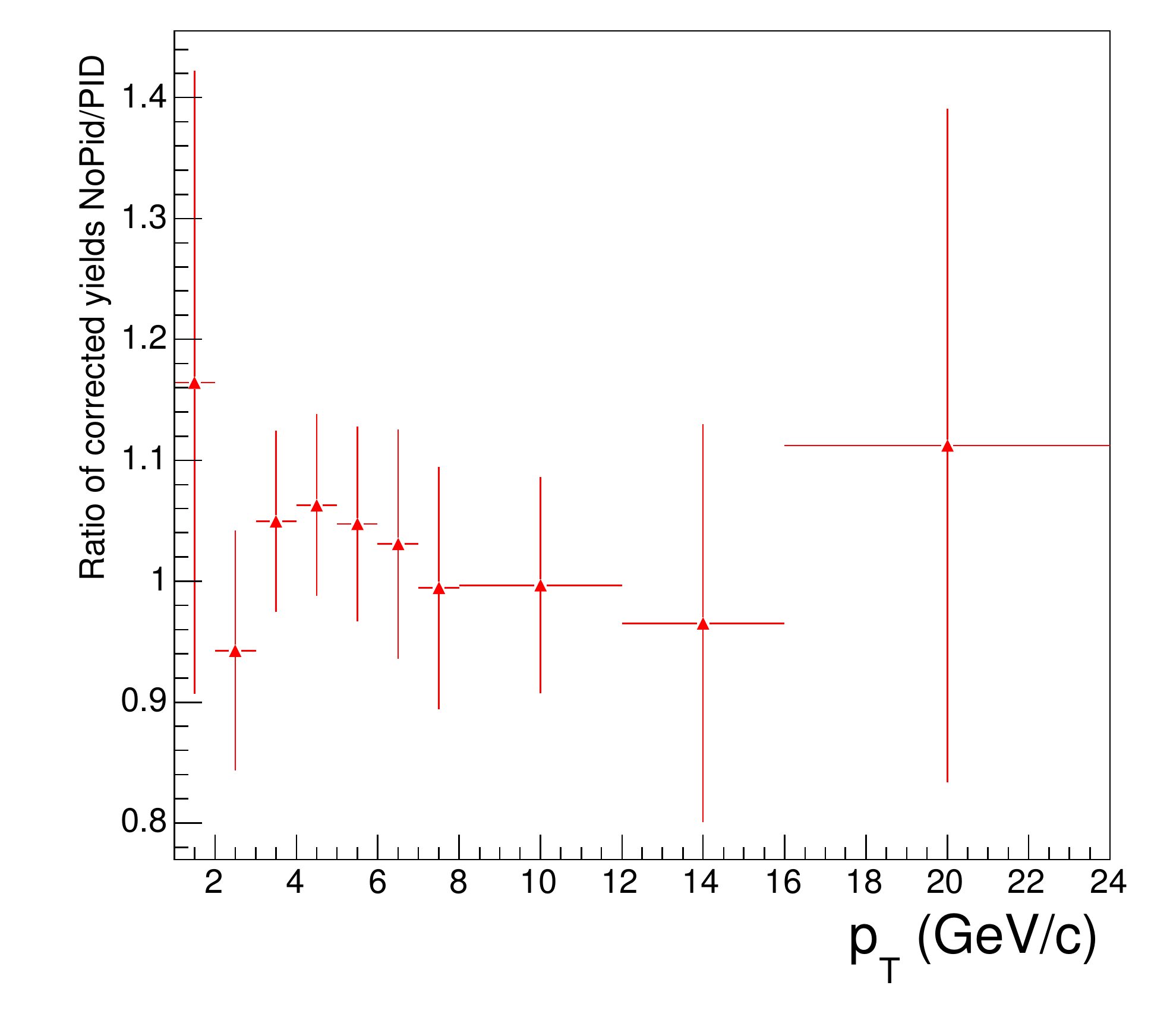}
\caption{Relative variation of the corrected \dplus meson yields obtained with and without applying PID.}
\label{PidNoPid}
\end{figure}
The possible systematic effects  due to  the PID selection were studied by comparing  the corrected \dplus meson yields  (Equation \ref{eq:corrsp}) obtained with and without applying the  PID selections to the decay tracks. The ratios of the corrected yields with and without PID as a function of \pt for the rapidity integrated analysis are reported in Figure \ref{PidNoPid}. The statistical errors on the extracted corrected spectra are treated as correlated in the ratio. The ratios are compatible with unity, and the highest discrepancy is found in the \pt interval 1<\pt<2 GeV/c, where the STRONG PID is applied. No PID systematic uncertainty is assigned in the range 2<\pt<24. For \pt <2  GeV/c  a 10\% uncertainty is estimated from Figure \ref{PidNoPid}.  Since no rapidity dependence of the PID response is expected, and after checking that the $y_{\rm lab}$ differential results are
 compatible with the $y_{\rm lab}$ integrated ones, no PID uncertainty is assigned to the $y_{\rm lab}$ differential analysis.
\subsection{Systematic due to the  \pt shape of generated \dplus mesons}
\label{sub:systMCPtShape}
The \pt distribution of the generated \dplus mesons  within the \pt intervals affects the efficiency, since the efficiency depends on \pt (see Figures \ref{eff3} and \ref{effy}) and is estimated from the relative variation of the  efficiencies obtained using \pt shapes from FONLL and those of the MC data sample described in Section  \ref{sub:MCsample}.  This relative variation has been shown in Figure  \ref{eff2}.  The rapidity integrated uncertainty values are shown in Table \ref{tab:ptSys}.\\
The efficiency variations due  to the generated  \pt shape have also  been evaluated in the $y_{\rm lab}$ differential analysis. The uncertaintes are reported in Table \ref{tab:ptSys}. They are larger  than in the rapidity integrated analysis, because here the \pt intervals are larger.\\
\begin{table}[h]
\centering
\begin{tabular}{|c|c|c|c|c|c|c|c|c|c|c|}
\hline
p$_{\rm T}$ (GeV/c)                & 1-2  & 2-3    & 3-4    & 4-5    & 5-6    & 6-7    & 7-8    & 8-12        & 12-16       & 16-24 \\ \hline
rapidity integrated                & 3\% & 0\%    & 0\%    & 0\%    & 0\%    & 0\%    & 0\%    & 4\%         & 4\%         & 4\%   \\ \hline
-0.8/-0.7\textless$y$\textless-0.4 & x    & \multicolumn{3}{c|}{10\%} & \multicolumn{3}{c|}{5\%} & \multicolumn{2}{c|}{5\%} & x     \\ \hline
-0.4\textless$y$\textless-0.1      & x    & \multicolumn{3}{c|}{10\%} & \multicolumn{3}{c|}{5\%} & \multicolumn{2}{c|}{5\%} & x     \\ \hline
-0.1\textless$y$\textless0.1       & x    & \multicolumn{3}{c|}{10\%} & \multicolumn{3}{c|}{5\%} & \multicolumn{2}{c|}{5\%}  & x     \\ \hline
0.1\textless$y$\textless0.4        & x    & \multicolumn{3}{c|}{10\%} & \multicolumn{3}{c|}{5\%} & \multicolumn{2}{c|}{5\%}  & x     \\ \hline
0.4\textless$y$\textless0.7/0.8    & x    & \multicolumn{3}{c|}{10\%} & \multicolumn{3}{c|}{5\%} & \multicolumn{2}{c|}{5\%}  & x     \\ \hline
\end{tabular}
\caption{Systematic uncertainty due to the p$_{\rm T}$ shape of generated \dplus mesons.}
\label{tab:ptSys}
\end{table}
\subsection{Systematic uncertainty due to  track reconstruction efficiency}
The systematic uncertainty related to the tracking efficiency was estimated by varying the track quality selection criteria, such as the minimum number of TPC clusters for analyzed tracks (default value = 70, systematic checks with 50 TPC clusters). 
The resulting uncertainty is estimated to be 3\% per track, independent of \pt and $y$. Therefore, the uncertainty from tracking efficiency is  9\% for the 3-body decay of \dplus considered in this analysis in all \pt and rapidity intervals.
\subsection{Systematic uncertainty due to beauty feed-down subtraction}
\label{sub:fds}
\begin{figure}[b]
\centering
 \includegraphics[width=0.99\textwidth]{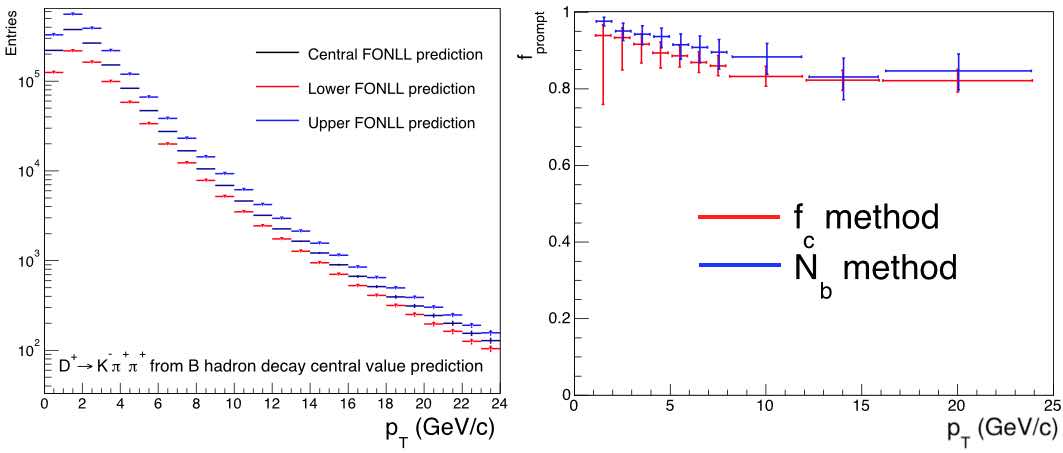}
\caption{Left: FONLL predictions for  \dplus  from B decays cross section at \sqs= 5.02 TeV. Central, upper and  lower values are shown. Right: \fprompt values obtained with the $f_c$ and $N_b$ methods taking into account the FONLL prediction uncertainties.}
\label{FDSys}
\end{figure}
The beauty feed-down subtraction procedure described in Section \ref{sec:bfd} contains two sources of systematics.\\
The first source of systematic uncertainty is given by the uncertainties  on the FONLL predictions used in  Equations \ref{eq:Nb} and \ref{eq:fc}. In fact, the FONLL predictions used in this analysis for prompt and feed-down \dplus meson cross section at \sqs= 5.02 TeV are composed of a central, an upper and a lower value, shown for \dplus mesons from B decays in Figure \ref{FDSys} (left). This means that \fprompt is evaluated for both methods $f_c$ and $N_b$ three times, yielding a central, a maximum and a minimum value for \fprompt. Note that in the $f_c$ case, where the ratio of FONLL predictions for  feed-down and prompt \dplus mesons are needed, a "conservative approach" is used, which means that  the to extreme values of \fprompt are obtained from Equation  \ref{eq:fc} using the ratio of the upper (lower) FONLL predictions for feed-down \dplusm to the upper (lower) FONLL predictions for prompt \dplusm: the extreme case obtained from the ratio of  the upper  FONLL predictions for feed-down \dplusm to the  lower FONLL predictions for prompt \dplusm is not considered. The two histograms in Figure \ref{FDSys}  (right) show the \fprompt values obtained with the $f_c$ and $N_b$ methods: the central values are those obtained with the central FONLL predictions, the asymmetric  error bars represent the \fprompt variation obtained using the upper/lower values of the FONLL predictions.\\
\begin{figure}[t]
\centering
 \includegraphics[width=0.6\textwidth]{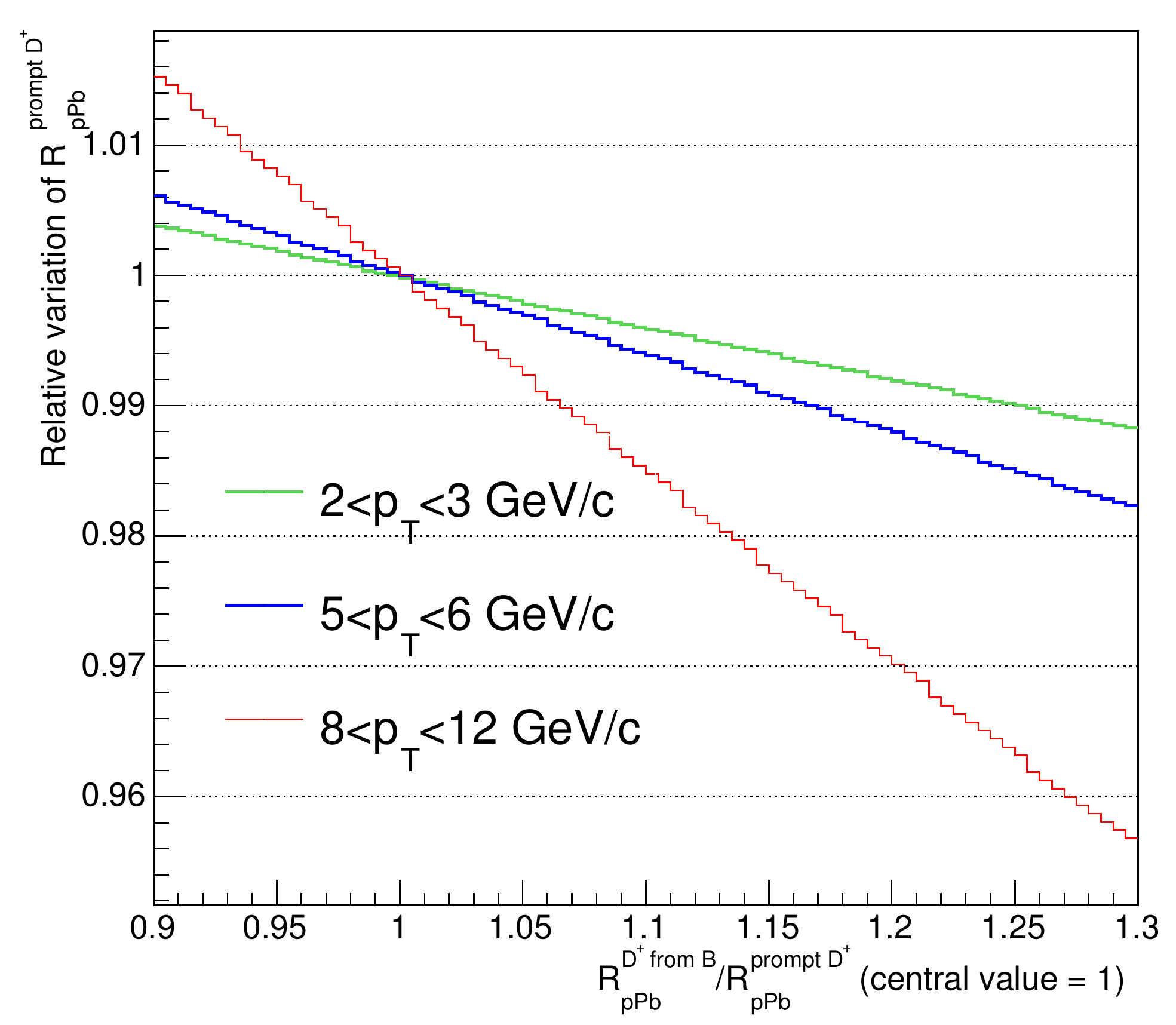}
\caption{Relative variation of prompt \dplusm \rppb as a function of the hypothesis on $R_{\textrm{pA}}^{\textrm{feed-down}}(\pteq)$/$R_{\textrm{pA}}^{\textrm{prompt D}}(\pteq)$ for three \pt intervals.}
\label{EnLoss}
\end{figure}
The second source of systematic uncertainty is the hypothesis on the nuclear modification factor of feed-down \dplus mesons $R_{\rm pPb}^{\rm feed-down}$. This contribution is evaluated varying the ratio of feed-down and prompt \dplus mesons \rppb in the range \\ 0.9<$R_{\textrm{pA}}^{\textrm{feed-down}}(\pteq)$/$R_{\textrm{pA}}^{\textrm{prompt D}}(\pteq)$<1.3, and evaluating the corresponding variation of prompt \dplus $R_{\rm pPb}$. This range is chosen for the reasons discussed in  Section \ref{sec:bfd}. Figure \ref{EnLoss} shows the relative variation of prompt \dplus mesons \rppb as a function of the hypothesis   on $R_{\textrm{pA}}^{\textrm{feed-down}}(\pteq)$/$R_{\textrm{pA}}^{\textrm{prompt D}}(\pteq)$.\\
For both methods $f_c$ and $N_b$, the two sources of systematic are summed in quadrature. The total systematic uncertainty due to beauty feed-down is defined as the full envelope of the uncertainties of the two methods $N_b$ and $f_c$. The central points for both prompt \dplus \rppb and cross section are taken as those obtained with the $N_b$ method using the central FONLL predictions. \\	
Table \ref{tab:SysFD} summarizes the total systematic uncertainties due to feed-down subtraction. For what concerns the rapidity dependence of this sources of  systematic uncertainty, it was considered that FONLL calculations for  prompt and feed-down 	\dplus meson production predicts the cross section as a function of $y$ to be flat within 4\% in the rapidity range considered, and that the ratio $R_{\textrm{pA}}^{\textrm{feed-down}}(\pteq)$/$R_{\textrm{pA}}^{\textrm{prompt D}}(\pteq)$  to vary within 3-4\% in |$y_{\rm lab}$|<1. Therefore the systematic uncertainty due to beauty feed-down in the rapidity differential analysis is evaluated as  in the rapidity integrated case. 
\begin{table}[h]
\centering
\begin{tabular}{|c|c|c|c|c|c|c|c|c|c|c|}
\hline
p$_{\rm T}$ (GeV/c)                & 1-2  & 2-3    & 3-4    & 4-5    & 5-6    & 6-7    & 7-8    & 8-12        & 12-16       & 16-24 \\ \hline
rapidity integrated               & $_{-31\%}^{+2\%}$&$  _{-11\%}^{+3\%}$&$ _{-7\%}^{-3\%}$&$_{-5\%}^{+3\%}$&$_{-4\%}^{+3\%}$&$_{-4\%}^{+3\%}$&$^{+3\%}_{-3\%}$&$_{-3\%}^{+4\%}$&$_{-3\%}^{+5\%}$&$_{-3\%}^{+4\%} $    \\ \hline
-0.8/-0.7\textless$y$\textless-0.4 & x    & \multicolumn{3}{c|}{$_{-10\%}^{+3\%}$}& \multicolumn{3}{c|}{$^{+3\%}_{-8\%}$} & \multicolumn{2}{c|}{$^{+4\%}_{-10\%}$} & x     \\ \hline
-0.4\textless$y$\textless-0.1      & x    & \multicolumn{3}{c|}{$_{-10\%}^{+3\%}$} & \multicolumn{3}{c|}{$^{+3\%}_{-8\%}$} & \multicolumn{2}{c|}{$^{+4\%}_{-10\%}$} & x     \\ \hline
-0.1\textless$y$\textless0.1       & x    & \multicolumn{3}{c|}{$_{-10\%}^{+3\%}$} & \multicolumn{3}{c|}{$^{+3\%}_{-8\%}$} & \multicolumn{2}{c|}{$^{+4\%}_{-10\%}$}  & x     \\ \hline
0.1\textless$y$\textless0.4        & x    & \multicolumn{3}{c|}{$_{-10\%}^{+3\%}$} & \multicolumn{3}{c|}{$^{+3\%}_{-8\%}$} & \multicolumn{2}{c|}{$^{+4\%}_{-10\%}$}  & x     \\ \hline
0.4\textless$y$\textless0.7/0.8    & x    & \multicolumn{3}{c|}{$_{-10\%}^{+3\%}$} & \multicolumn{3}{c|}{$^{+3\%}_{-8\%}$} & \multicolumn{2}{c|}{$^{+4\%}_{-10\%}$}  & x     \\ \hline
\end{tabular}
\caption{Systematic uncertainty due to beauty feed-down subtraction.}
\label{tab:SysFD}
\end{table}
\subsection{Systematic uncertainty on pp reference (only for  measurement)}
The pp reference in the denominator of the \rppb formula is also affected by  systematic uncertainties. In particular:
\begin{itemize}
\item the measured \dplus meson cross section at \sqs= 7 TeV  is affected by the same  sources of systematic uncertainties discussed for the p--Pb data sample: yield extraction, cut efficiency, PID, tracking efficiency and \pt shape in the MC. 
\item the uncertainty on the scaling factor from \sqs= 7 TeV to   \sqs= 5.02 TeV, shown in Figure  \ref{Ref5}
\item the uncertainty on the beauty feed-down subtraction, which in pp collisions does not include the contribution of beauty energy loss hypothesis
\end{itemize}
In the calculation of \rppb these contributions are summed in quadrature to the  systematic uncertainties on  the p--Pb cross section, with two exceptions: the systematic  errors relative to tracking  efficiency and those relative to the FONLL scale on beauty feed-down subtraction are considered as completely correlated. The systematic uncertainty values due to the pp reference are reported in Table \ref{SysPP}.
\begin{table}[h]
\footnotesize
\centering
\begin{tabular}{|c|c|c|c|c|c|c|c|c|c|c|c|}
\hline
\pt (GeV/c) & [1,2] & [2,3] & [3,4] & [4,5] & [5,6]& [6,7] & [7,8] & [8,12] & [12,16] & [16,24] \\
\hline 
Uncertainty (\%) & $^{+33}_{-37}$&  $^{+33}_{-34}$& $^{+30}_{-30}$&$^{+19}_{-19}$&$^{+19}_{-19}$&$^{+17}_{-17}$&$^{+17}_{-17}$&$^{+17}_{-17}$&$^{+21}_{-21}$&$^{+19}_{-19}$ \\
\hline
\end{tabular}
\caption{Summary table of systematics on pp reference (data+scaling).}
  \label{SysPP}
\end{table}
\subsection{Luminosity and branching ratio}
The integrated luminosity was evaluated as the ratio between the number of events passing the V0AND (\ref{sub:Trigger}) trigger selection with |$z_{\rm vertex}$|<10 cmand the corresponding trigger cross section in p--Pb collisions at \sqnn=5.02 TeV, $\sigma_{\rm V0AND}$. The V0A trigger is 100\% efficient for \dplus mesons with \pt  > 1 GeV/c and |$y_{\rm lab}$| < 0.5. The value of the V0AND trigger cross section was estimated with van der Meer scans  as $\sigma_{\rm V0AND}$=2.09$\pm$3.5\%(syst) b with negligible statistical uncertainty \cite{V0AND}. The luminosity is measured via the V0  trigger counts, corrected for pileup and for background contamination, divided by the corresponding cross sections. Since  fluctuations of the order of 1\% were observed in the  run-by-run integrated luminosity, an additional 1\% systematic is assigned, yielding a total systematic error on $L_{\rm int}$ of 3.7\%.  \\	
The systematic uncertainty on the branching ratio is 2.1\%, as shown in table \ref{tab:decays}.
\section{Results}
\subsection{D$^{+}$-meson cross section}
\begin{figure}[b]
\centering
 \includegraphics[width=0.99\textwidth]{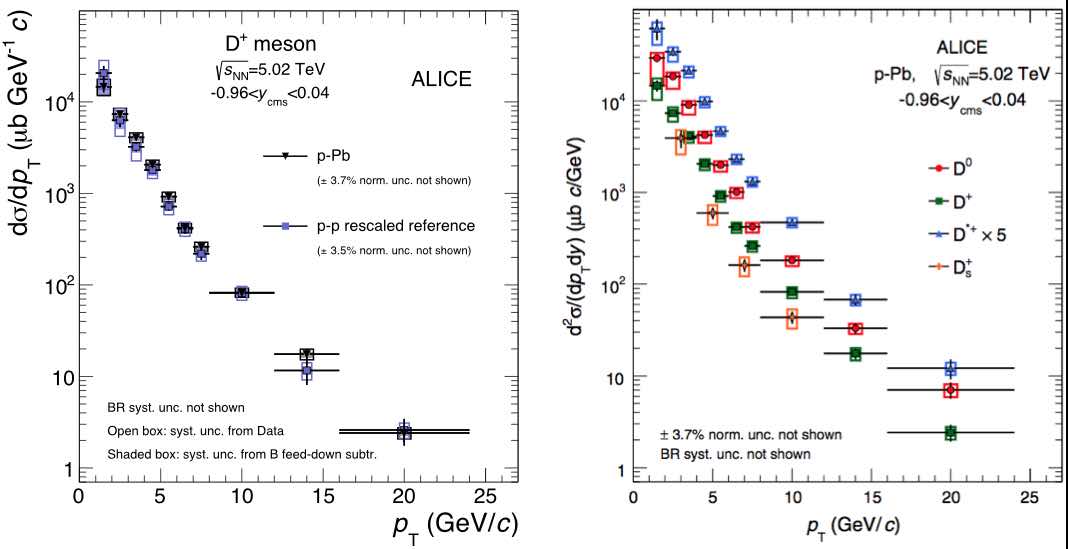}
\caption{Left:  \pt differential production cross section of prompt \dplus mesons  together with the pp  reference at \sqs= 5.02 TeV multiplied by the  mass number (A=208) of Pb nuclei. Right:  \pt differential production cross section of prompt \dplus mesons together with those of D$^{0}$, D$^{*+}$ and D$^{+}_{s}$ mesons \cite{PaperpPb}.}
\label{Res1}
\end{figure}
\begin{figure}[t]
\centering
 \includegraphics[width=0.99\textwidth]{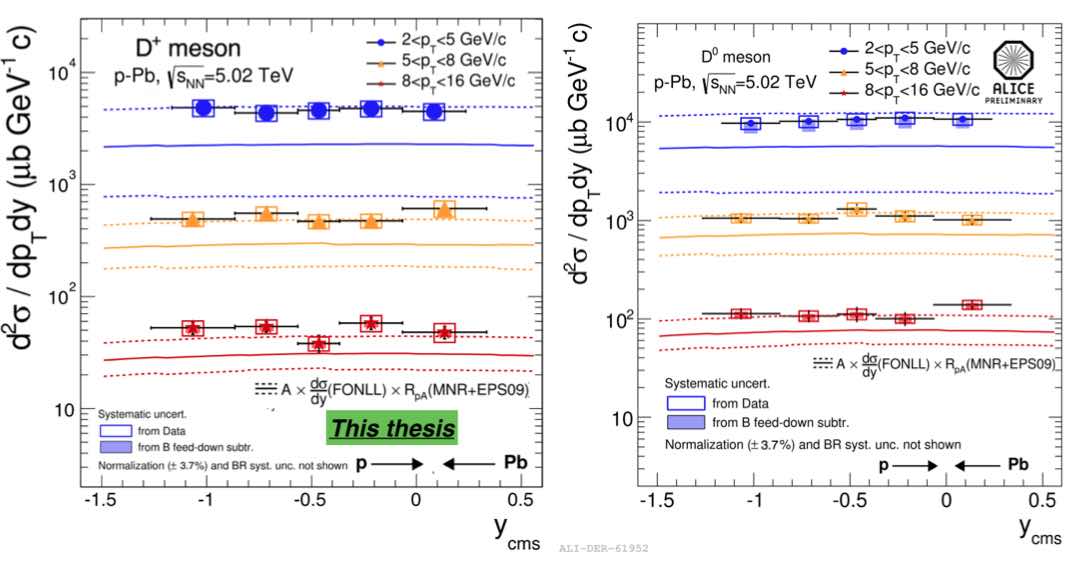}
\caption{Left: \dplusm cross section as a function of $y_{\rm cms}$ in the three \pt intervals considered in this analysis compared to FONLL predictions taking into account EPS09 parametrization of PDFs. Right: D$^{0}$-meson cross section as a function of $y_{\rm cms}$ in the three \pt intervals considered for this analysis. }
\label{Res2}
\end{figure}
The \pt differential production cross section of prompt \dplus mesons in p--Pb collisions at \sqnn= 5.02 TeV is shown in Figure \ref{Res1} (left) together with the pp  reference at \sqs= 5.02 TeV multiplied by the  mass number (A=208) of Pb nuclei. The \pt range covered is 1 < \pt< 24 GeV/c. The  vertical bars  represent the statistical error. The empty boxes represent the systematic uncertainties on data, i.e. yield extraction, cut efficiency, tracking efficiency, PID, \pt shape of generated \dplus mesons and in the pp case those arising from the scaling factor. The filled boxes represent the systematic uncertainties on beauty feed-down.\\
The right panel of Figure \ref{Res1} shows the \pt differential production cross sections of prompt \dplus mesons together with those of D$^{0}$, D$^{*+}$ and D$^{+}_{s}$ mesons measured in the hadronic decay channels listed in Table \ref{tab:decays} \cite{PaperpPb}. Due to the limited statistics, the D$^{+}_{s}$  measurement has a different \pt binning. The relative abundances of D mesons in p--Pb collisions are compatible within uncertainties with those measured in pp, $ep$, and $e^{+}e^{-}$ collisions at different energies \cite{citare}.\\
Figure \ref{Res2} (left) shows the \dplusm cross section as a function of $y_{\rm cms}$ in the three \pt intervals considered for this analysis. The empty boxes represent the systematic uncertainties on data, the filled boxes represent the systematic uncertainties on beauty feed-down. Systematic errors on branching ratio and normalization are not shown. The sketch on the p and Pb beams configuration present in the bottom right corner of Figure \ref{Res2} shows that the positive rapidity values are those in the direction of the p beam.  No significant rapidity dependence of the \dplus meson cross section in the measured $y_{\rm cms}$ interval is present. The results are compared to the predictions obtained with FONLL pQCD calculations with EPS09 parametrization of the PDFs, and are compatible within statistical and systematics uncertainties  with the upper values of the predictions. This behaviour is also observed for the $y_{\rm cms}$ differential cross section of D$^{0}$ mesons, shown in the right panel.
\subsection{D$^{+}$-meson nuclear modification factor}
\begin{figure}[h]
\centering
 \includegraphics[width=0.5\textwidth]{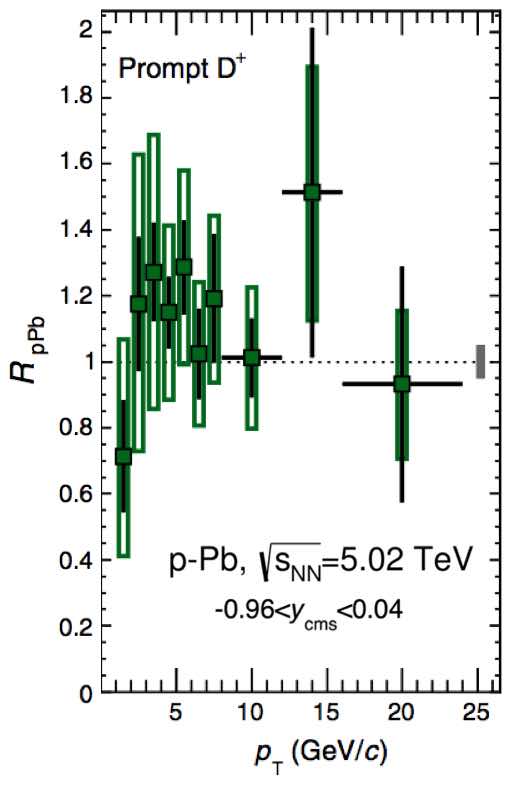}
\caption{ \rppb of \dplus mesons in p--Pb collisions at \sqnn=5.02 TeV as a function of \pt. }
\label{Res3}
\end{figure}
The \rppb of \dplus mesons in p--Pb collisions at \sqnn=5.02 TeV is shown in in Figure \ref{Res3}. It is  compatible with unity within the statistical and systematic uncertainties in the measured \pt range, indicating that the  \dplusm  production in p-Pb collisions is consistent  with 
the one in pp collisions scaled by the average number of binary collisions.\\ 
The \dplusm \rppb is also consistent with those of D$^{0}$, D$^{*+}$ and D$^{+}_{s}$ mesons measured in the hadronic decay channels listed in Table \ref{tab:decays}, as shown in Figure \ref{Res4}. In particular, the  D$^{+}_{s}$ \rppb is not only interesting to study possible modification of $c$-quark dynamic in p--Pb collisions, but also because the presence of a strange quark makes it more sensible to the hadronization mechanisms and to a possible enhancement of strangeness production in p--Pb collisions. Both aspects were discussed together with some results in Sections \ref{sub:strangeness} and \ref{sub:hadronization}.\\
\begin{figure}[h]
\centering
 \includegraphics[width=0.99\textwidth]{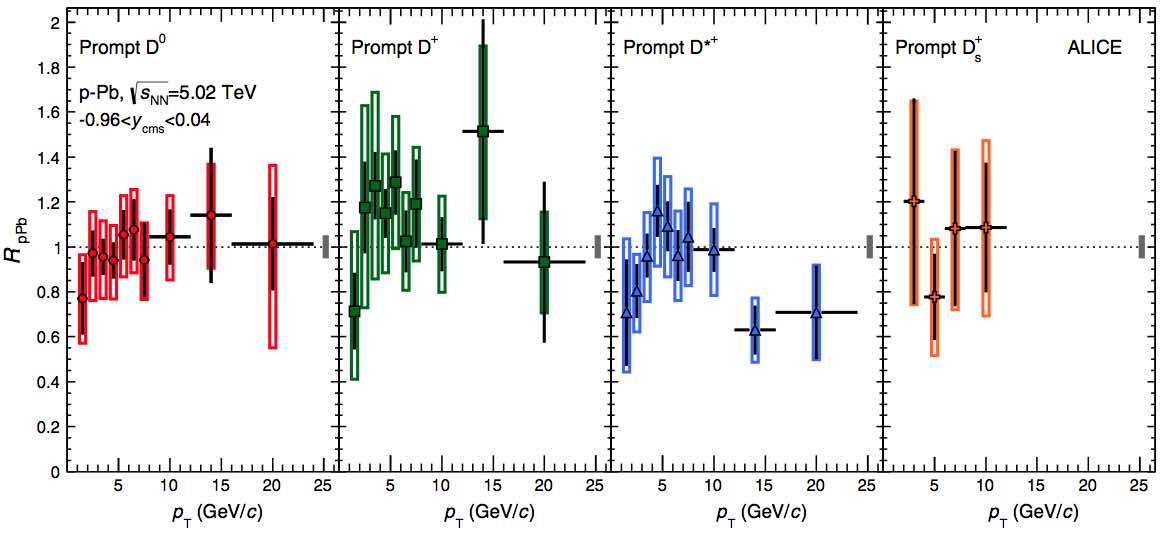}
\caption{\rppb  of D mesons in  p--Pb collisions at \sqnn=5.02 TeVas a function of \pt: from left to right D$^{0}$, \dplus, D$^{*+}$ and D$_{s}^{+}$ mesons. }
\label{Res4}
\end{figure}
\subsection{Comparison to model calculations}
The average  \rppb of D$^{0}$, D$^{*+}$ and D$^{+}$ mesons in the \pt range 1<\pt<24 GeV/c was calculated by using the relative statistical uncertainties as weights. The systematic error on the average was calculated by propagating the uncertainties through the weighted average, where the contributions from tracking efficiency, B feed-down correction, and scaling of the pp reference were taken as fully correlated among the
three species. Figure \ref{Res5} shows the average \rppb compared to theoretical calculations:
\begin{itemize}
\item next-to-leading order  pQCD calculations (Mangano, Nason, and Ridolfi (MNR) [50]) of D-meson production, including the EPS09 \cite{EPS09} (Section \ref{sub:instef}) nuclear modification of the CTEQ6M PDFs \cite{cteqm6} (Section \ref{sub:pdfs})
\item calculations based on the color glass condensate \cite{CGC} (Section \ref{sub:instef}). In this calculation gluon merging dynamic is encoded in an effective unintegrated gluon distribution (uGD) for Pb nuclei and is more prominent in the momentum region lower than the saturation scale $Q_s^2$ \cite{CGCpPn}
\item  calculations taking into account the $k_{\rm T}$ broadening of incoming partons that arises from initial-state scattering, nuclear shadowing and radiative energy loss in cold nuclear matter \cite{VitevpPb}
\end{itemize}
All these models only take into account initial state effects. Data are well described by all of them within statistical and systematic uncertainties, however the current uncertainties on both data and predictions do not allow to discriminate among models. \\
\begin{figure}[b]
\centering
 \includegraphics[width=0.99\textwidth]{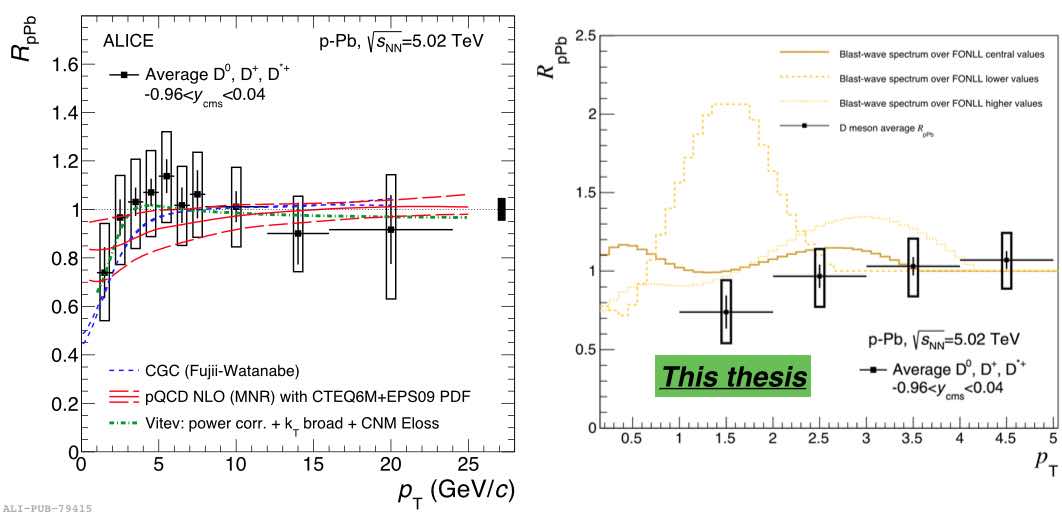}
\caption{Left: average  \rppb of D$^{0}$, D$^{*+}$ and D$^{+}$ mesons \cite{PaperpPb} compared to models including initial state effects. Right:  average  \rppb of D$^{0}$, D$^{*+}$ and D$^{+}$ mesons \cite{PaperpPb} compared to a model implementing radial expansion of the system created in the p--Pb collision \cite{Sickles}. }
\label{Res5}
\end{figure}
The possible effects due to the formation of a hydrodynamically
expanding medium as calculated in \cite{Sickles} was shown in Figure \ref{hneppb} compared to $R_{\rm dAu}$ of heavy-flavour decay electrons. The same  results from a similar calculation  for D-meson \rppb in p--Pb collisions at \sqnn= 5.02 TeV are shown in Figure \ref{Res5} (right). The  blast-wave  spectra for D mesons in p--Pb collisions at \sqs=5.02 TeV are  divided by  the \pt distributions of D mesons obtained from FONLL calculations. The blast wave curves  are obtained using initital parameters ($\beta_{\bot}$ and T$_{\rm kin}$, Section \ref{sub:kfo}) obtained from $\pi$, $K$ and $p$ spectra in minimum bias p--Pb collisions.  With respect to the predictions for RHIC energy, here the \rppb modulation expected from radial flow  is smaller and confined to lower values of \pt.\\
With the current uncertainties on both data and blast-wave predictions it is hard to draw conclusion on possible hydrodynamic effects in p--Pb collisions.
\subsection{Comparison to results with heavy-flavour decay electrons and muons}
\begin{figure}[t]
\centering
 \includegraphics[width=0.99\textwidth]{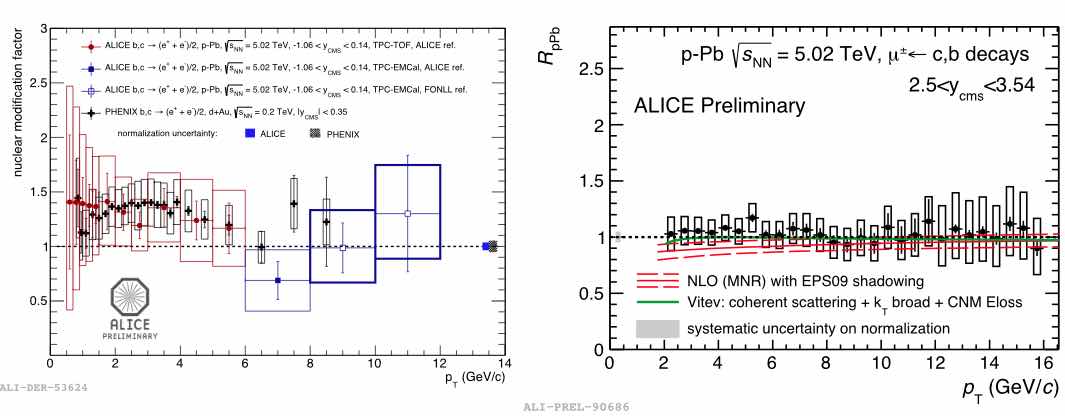}
\caption{Left:  \rppb of electrons from heavy-flavour (charm + bottom) decays in p--Pb collisions at \sqnn= 5.02 TeV measured by the ALICE Collaboration and in d--Au collisions at \sqnn= 200 GeV measured by the PHENIX Collaboration. Right:  \rppb of muons from heavy-flavour (charm + bottom) decays in p--Pb collisions at \sqnn= 5.02 TeV measured by the ALICE Collaboration. }
\label{Res6}
\end{figure}
The preliminary \rppb of electrons from heavy-flavour (charm + bottom) hadron decays in p--Pb collisions at \sqnn= 5.02 TeV measured by the ALICE Collaboration is shown in Figure \ref{Res6} (right). The measurement is performed in -1.06<$y_{\rm lab}$<0.14 in the \pt range 1 to 12 GeV/c. The pp reference is obtained following a strategy similar to the one described in Section \ref{sub:sc502} except for \pt> 8 GeV/c, where  pp data at 7 TeV are not available and the reference is obtained from FONLL predictions only. The \rppb of heavy-flavour decay electrons is compatible with unity within (large) uncertainties.  The central values are systematically above unity of a factor $\sim$ 0.3 up to \pt= 8 GeV/c.\\
The \rppb of electrons from heavy-flavour (charm + bottom) hadron decays in d--Au collisions at \sqnn= 200 GeV measured by the PHENIX Collaboration\footnote{These results are slightly different than those in Figure \ref{hneppb}, since these ones are for minimum-bias d--Au collisions.} is also  shown in Figure \ref{Res6} (left) and shows agreeement within uncertaintites with the ALICE results. However in the \pt range 1.5 to 6 GeV/c PHENIX data, which have smaller uncertainties, seem to indicate an enhancement above unity of the $R_{\rm dAu}$.\\
The preliminary  \rppb of muons from heavy-flavour (charm + bottom) hadron decays in p--Pb collisions at \sqnn= 5.02 TeV measured by the ALICE Collaboration is shown in Figure \ref{Res6}. The measurement is performed in 2.5<$y_{\rm lab}$<3.54 in the \pt range 2 to 16 GeV/c. The \rppb of heavy-flavour decay muon is compatible with unity within uncertainties.  The measurement is compared with two of the models already compared to the D-meson \rppb, namely next-to-leading order  pQCD calculations   including the EPS09  nuclear modification of the CTEQ6M PDFs  and the Vitev calculations \cite{VitevpPb}. As in the case of D mesons,  data and models show good agreement.\\
\subsection{Comparison to some results in the beauty sector}
The preliminary \rppb of electrons from beauty  in p--Pb collisions at \sqnn= 5.02 TeV measured by the ALICE Collaboration is shown in Figure \ref{Res7}.  Electrons from charm and beauty  are separated via a selection on their impact parameter exploiting the larger lifetime of B mesons with respect to D mesons. The \rppb measurement is performed in -1.06<$y_{\rm lab}$<0.14 in the \pt range 1.2 to 7 GeV/c and is consistent with unity within uncertainties.\\
In Figure \ref{Res7} (right)  the \rppb of B$^{+}$ hadrons measured by the CMS Collaboration in p--Pb collisions at \sqnn= 5.02 TeV is shown \cite{gian}.  B$^{+}$ mesons are reconstructed in their hadronic decay into  J/$\Psi$+$K^{+}$ and their signal is extracted via an invariant mass analysis. The \rppb is computed using  FONLL calculations as pp reference, since no data are currently available for B$^{+}$ mesons \pt differential cross-section in pp collisions at high energies. The resulting \rppb of B$^{+}$ mesons, measured in 10<\pt<60 GeV/c, is compatible with unity within uncertainties. Similar results are obtained for the \rppb of B$^0$  mesons. \\
\begin{figure}[t]
\centering
 \includegraphics[width=0.99\textwidth]{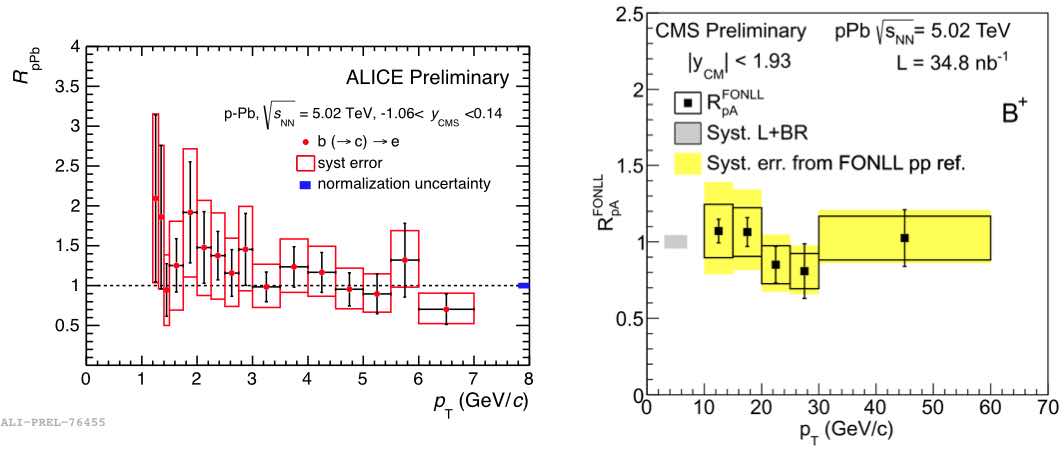}
\caption{Left:  \rppb of electrons from beauty  in p--Pb collisions at \sqnn= 5.02 TeV measured by the ALICE Collaboration. Right: \rppb of B$^{+}$ hadrons measured by CMS in p--Pb collisions at \sqnn= 5.02 TeV \cite{gian}. }
\label{Res7}
\end{figure}
\subsection{Conclusions}
The results on D meson \rppb  discussed in this section show that within the current uncertainties D-meson production in p--Pb collisions is consistent   with the binary  scaling of
 pp collisions. Similar conclusions  come   from measurements of electrons and muons from heavy-flavour hadron decays at different energies, as well as by measurements performed in the beauty sector. \\ 
We can now go back to what stated in the introduction of this Chapter about the role of p--Pb measurements in the interpretation of Pb--Pb results of D mesons nuclear modification factor, and try to draw some conclusions from Figure \ref{Res9}. The average D-meson \rppb is shown together with the average D-meson $R_{\rm AA}$ measured in central (0-20\%) and peripheral (40-80\%) Pb--Pb collisions at 	\sqnn= 2.76 TeV \cite{RAAvecchio}. \\
With the caveat that Pb--Pb and p--Pb measurement are performed at different centre-of-mass energies and for different values of $y_{cms}$, we observe that
\begin{itemize}
\item \rppb values are not consistent within uncertainties with $R_{\rm AA}$ values for central Pb--Pb collisions in the whole \pt range covered by the Pb--Pb measurement
\item \rppb values are  consistent within uncertainties with models taking into account initial state effects only
\end{itemize}
\begin{figure}[t]
\centering
 \includegraphics[width=0.6\textwidth]{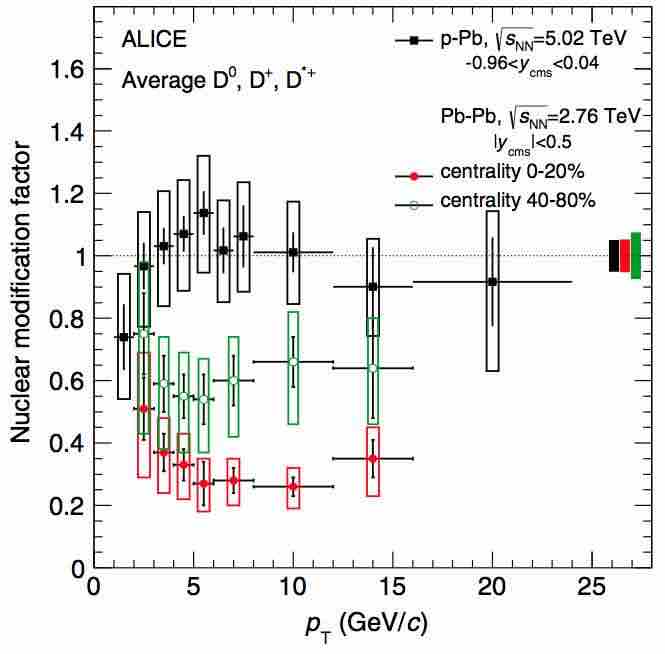}
\caption{ average D-meson \rppb  together with average D-meson $R_{\rm AA}$ measured in central (0-20\%) and semiperipheral (20-40\%) Pb--Pb collisions at 	\sqnn= 2.76 TeV \cite{RAAvecchio}. }
\label{Res9}
\end{figure}
The fact that the measured $R_{\rm AA}$ values can not be explained in terms of initial state effects only was  already observed by comparing the $R_{\rm AA}$ measured in central (0-20\%)  Pb--Pb collisions at 	\sqnn= 2.76 TeV with calculations based on next-to-leading order  pQCD calculations   including  EPS09 modification of nuclear PDFs\footnote{ This calculations are similar to the ones   shown in Figure \ref{Res5} for \rppb. The effect of shadowing is essentially at low \pt and is larger for Pb--Pb collisions were it affects both colliding nuclei}, as shown in Figure \ref{Res8}.
 This observation is now confermed by p--Pb data on D mesons production, thus  reinforcing the statement that  the suppression of D-meson production observed in central Pb-Pb collisions for \pt > 2 GeV/c is predominantly induced by final-state effects, e.g., the charm quark energy loss in the hot and dense medium.\\
\begin{figure}[t]
\centering
 \includegraphics[width=0.6\textwidth]{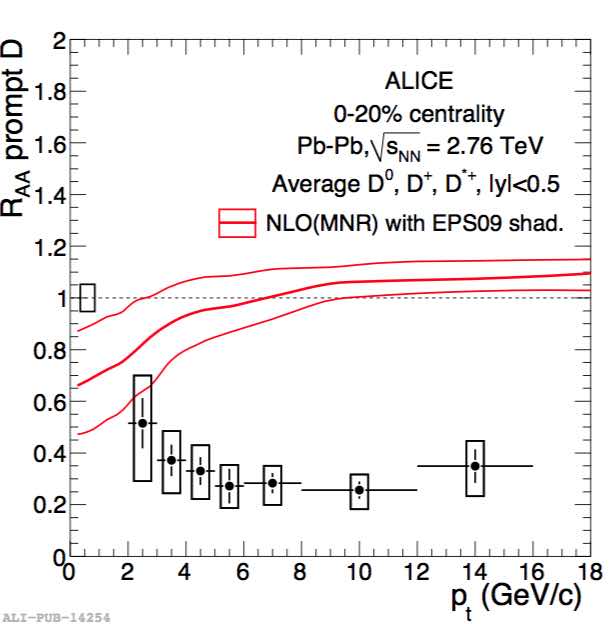}
\caption{ Average D-meson $R_{\rm AA}$ measured in central (0-20\%)  Pb--Pb collisions at 	\sqnn= 2.76 TeVcompared to next-to-leading order  pQCD calculations   including  EPS09 parametrizations of nuclear PDFs. }
\label{Res8}
\end{figure}
\chapter{Data-driven beauty feed-down subtraction}
\lhead{Chapter 6. \emph{Data-driven beauty feed-down subtraction}} 
The results on prompt \dplusm cross-section and nuclear modification factor described  in the previous chapter  are obtained with  the beauty feed-down subtraction technique described in Section \ref{sec:bfd}.
This feed-down subtraction method  relies on two theoretical hypotheses: the validity of FONLL predictions on B-hadron cross sections and  the energy loss of b quarks (\rppb of \dplusm from B hadrons decay).\\
A data-driven approach to beauty feed-down subtraction would have the advantage of not  depending  on theoretical calculations and on ad-hoc assumptions on the \rppb of feed-down D mesons. Moreover we have seen in  Section \ref{sub:fds}  that the systematic uncertainty  on  theory driven feed-down subtraction is large at low \pt (30\% in 1<\pt<2 GeV/c).\\
In this chapter I will present an alternative, data-driven, feed-down subtraction method based on the analysis of the impact parameter distributions of D$^{+}$ mesons passing the selection cuts described in the previous chapters. Figure \ref{PrincipleFD} shows the sketch of the decay of a prompt \dplusm and of a feed-down \dplusm from the decay of a B$^{0}$ hadron originating from the primary vertex.  I recall that the impact parameter is defined as the distance of closest approach of the \dplusm flight line (defined by its momentum direction) and the primary vertex.\\
\begin{figure}[b]
\centering
 \includegraphics[width=0.7\textwidth]{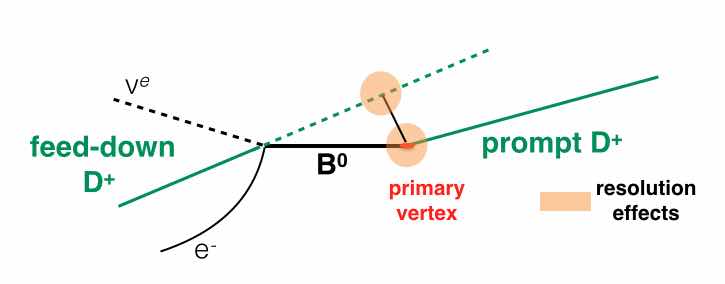} 
\caption{Sketch representing the effect of detector resolution on impact parameter measurements}
\label{PrincipleFD}
\end{figure}
The true impact parameter of prompt \dplus mesons is zero by definition. However the finite resolution  on impact parameter of the ALICE detector shown in Figure \ref{itstpcmatch} smears the measured values around zero. The effect of resolution on impact parameter is represented in Figure \ref{PrincipleFD} by an orange shadowed circle. \\
The true impact parameter of feed-down \dplusm is different form zero and depends on the B-hadron decay length and on the angle between the B-hadron and D-meson flight line, as will be discussed in next sections. Also in this case resolution effects are  present.\\
The different shapes of the  impact parameter distributions of prompt and feed-down \dplus mesons can thus be used as a tool to estimate the feed-down contribution to the total \dplusm raw yield. The prompt  fraction \fprompt of \dplusm yield can be extracted from a fit to the  impact parameter distributions as discussed in next sections.\\
It is nonetheless worth to anticipate that this beauty feed-down subtraction method was already used by the CDF Collaboration to extract the cross sections  shown in Figure \ref{FigPPLE}. The CDF analysis, performed in the \pt range  6<\pt<20 GeV/c, can count on a statistics  much higher than the one  presented in this thesis: the integrated luminosity $L_{\rm int}^{CDF}\approx$ 60 pb$^{-1}$ results in 3 10$^{11}$ events, to be compared to the 10$^{8}$ events used in the minimum bias analysis presented in the previous Chapter: more than 3 orders of magnitude of difference. However when one compares the reconstructed \dplusm yields in  the two analyses this difference in available statistics is strongly reduced, due to several factors: the increase of the \dplusm cross-section going from 1.96 TeV to 5.02 TeV of centre-of-mass energy, the \ncoll scaling of \dplusm production in p--Pb collisions and the difference in selection and reconstruction efficiencies, which in the CDF analysis are about a factor 100 lower than the ones of Figure \ref{acceff} for the minimum-bias p--Pb analysis of ALICE.  As an example, in the \pt range 6<\pt<7 GeV/c,  $N^{D^{+/-}}_{CDF}\approx$ 5700 while $N^{D^{+/-}}_{ALICE}\approx$ 600, as shown by the different number of entries in the impact parameter distributions of the right and left panels of Figure \ref{CompCDFALICE}. In 8<\pt<12 GeV/c $N^{D^{+/-}}_{CDF}\approx$ 13400 while $N^{D^{+/-}}_{ALICE}\approx$ 1000.\\
Figure \ref{CompCDFALICE}, showing the difference in available statistics between CDF and ALICE, also includes the fit function used to extract \fprompt that will be described in the following sections. In order to properly fit the data  a minimum amount of entries in required. This minimum amount of candidates in this analysis was set to 200, so that, with respect to the analysis presented in Chapter 5, the first and the last \pt bins (1<\pt<2 GeV/c and 16<\pt<24 GeV/c) will be excluded here, because the number of \dplusm candidates passing the selection cuts is too small.\\
\begin{figure}[t]
 \includegraphics[width=0.999\textwidth]{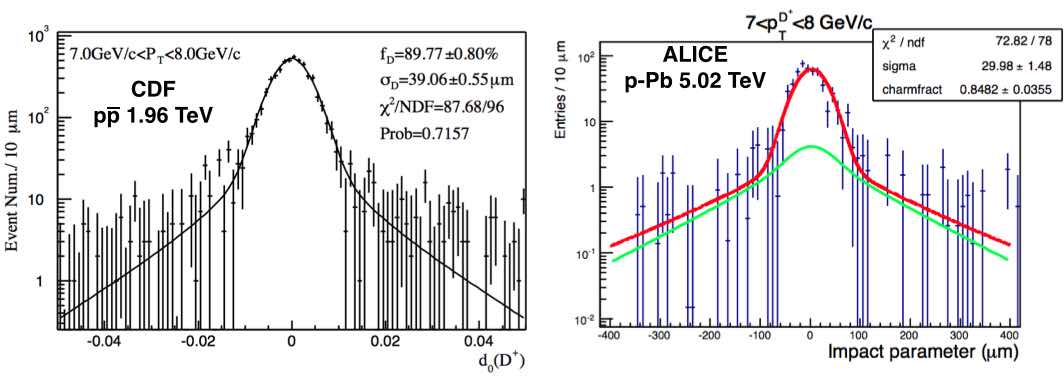} 
\caption{Impact parameter distributions of \dplus mesons used to extract \fprompt from the $p\bar{p}$ CDF data sample (left) and from the p--Pb ALICE data sample (right).}
\label{CompCDFALICE}
\end{figure}
\section{Impact parameter distributions}
\label{sec:distr}
In this section I will show the impact parameter distributions in the transverse plane ($d_0^{xy}$)  expected for \dplus mesons obtained from the MC data sample described in Section \ref{sub:MCsample}. The impact parameter distributions obtained from the  minimum-bias p--Pb data sample collected in 2013 will be also shown, together with the background subtraction method used in the analysis. 
\subsection{Prompt D$^{+}$ mesons in MC simulations}
The impact parameter ($d_0^{xy}$) distributions in the transverse plane of prompt \dplusm are shown in Figure \ref{PlotPromptImpParGaus} for eight \pt intervals in the \pt range 2<\pt<16 GeV/c. These distributions are obtained from the MC  sample described in Section  \ref{sub:MCsample} after applying the same event, track quality, topological and PID selections used to extract the \dplusm cross-section and nuclear modification factor shown in  Chapter 5. \\
For prompt \dplus mesons the impact parameter, defined as the distance of closest approach of the \dplusm trajectory to the primary vertex, is zero by definition. However resolution effects on track and primary vertex reconstruction make the distributions  of Figure \ref{PlotPromptImpParGaus} symmetrically smeared around zero. \\
Fitting the impact parameter distributions of Figure \ref{PlotPromptImpParGaus} with a gaussian function gives reasonable results at low \pt, however at high \pt the $\chi^2$ values of the fit indicate that a simple gaussian is not describing properly the distribution.  In Figure \ref{PlotPromptImpParGE} the same distributions are fitted with a gaussian with exponential tails, expressed by 
\begin{equation}
F^{\rm prompt}(d_0^{xy})=A\bigg[(1-f_g)\frac{1}{2\lambda_{\rm prompt}}e^{-\frac{|d_0^{xy}-\mu_{\rm prompt}|}{\lambda_{\rm prompt}}}+f_g\frac{1}{\sqrt{2\pi}\sigma_{\rm prompt}}e^{-\frac{(d_0^{xy}-\mu_{\rm prompt})^2}{2\sigma_{\rm prompt}^2}} \bigg]
\label{eq:gausexp}
\end{equation}
where $A$ is a normalization factor, $\lambda_{\rm prompt}$ is the slope of the exponential contribution, $\mu_{\rm prompt}$ is the  common mean of the gaussian and exponential contributions and $f_g$ is the fraction of the   integral  contained in the Gaussian function with width $\sigma_{\rm prompt}$.  The $\chi^2$ values obtained with this functional form at high \pt are smaller than those obtained with a simple gaussian. For \pt<5 GeV/c the situation does not vary significantly. Indeed the $f_g$ parameters obtained at low \pt  indicate that the exponential tails contribution to the total fitted function is less than 2\%.\\
Figure \ref{SigmaPromptImpPar} shows the  $\sigma_{\rm prompt}$ values obtained with both the simple gaussian fit and the fit of Equation \ref{eq:gausexp}  as a function of \pt. In both cases the MC predicts a decreasing trend of $\sigma_{\rm prompt}$ values with increasing p$_{\rm T}$, due to the better resolution on both impact parameter and \pt (Figure \ref{ipptres}).
\begin{figure}
 \includegraphics[width=0.999\textwidth]{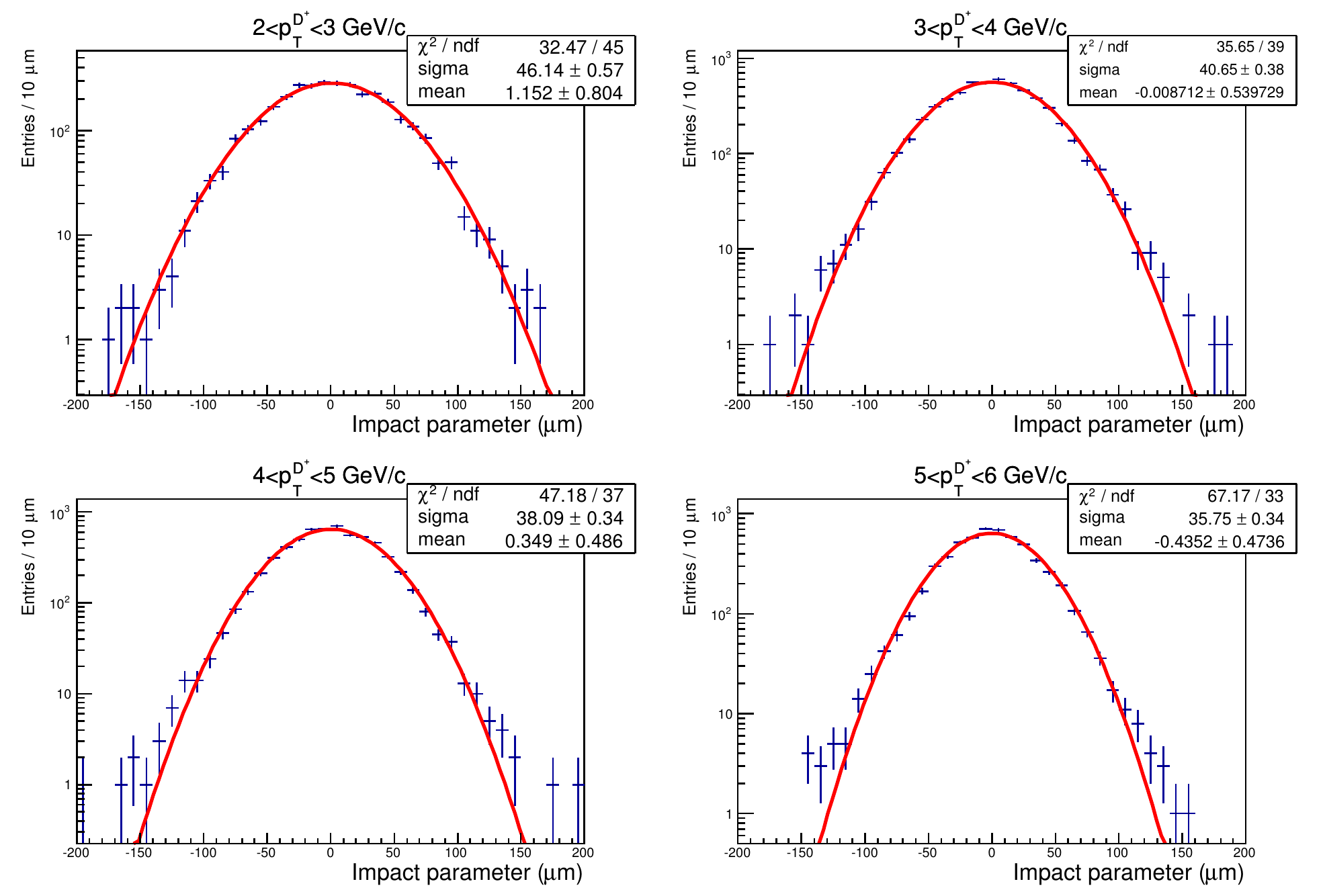} 
\includegraphics[width=0.999\textwidth]{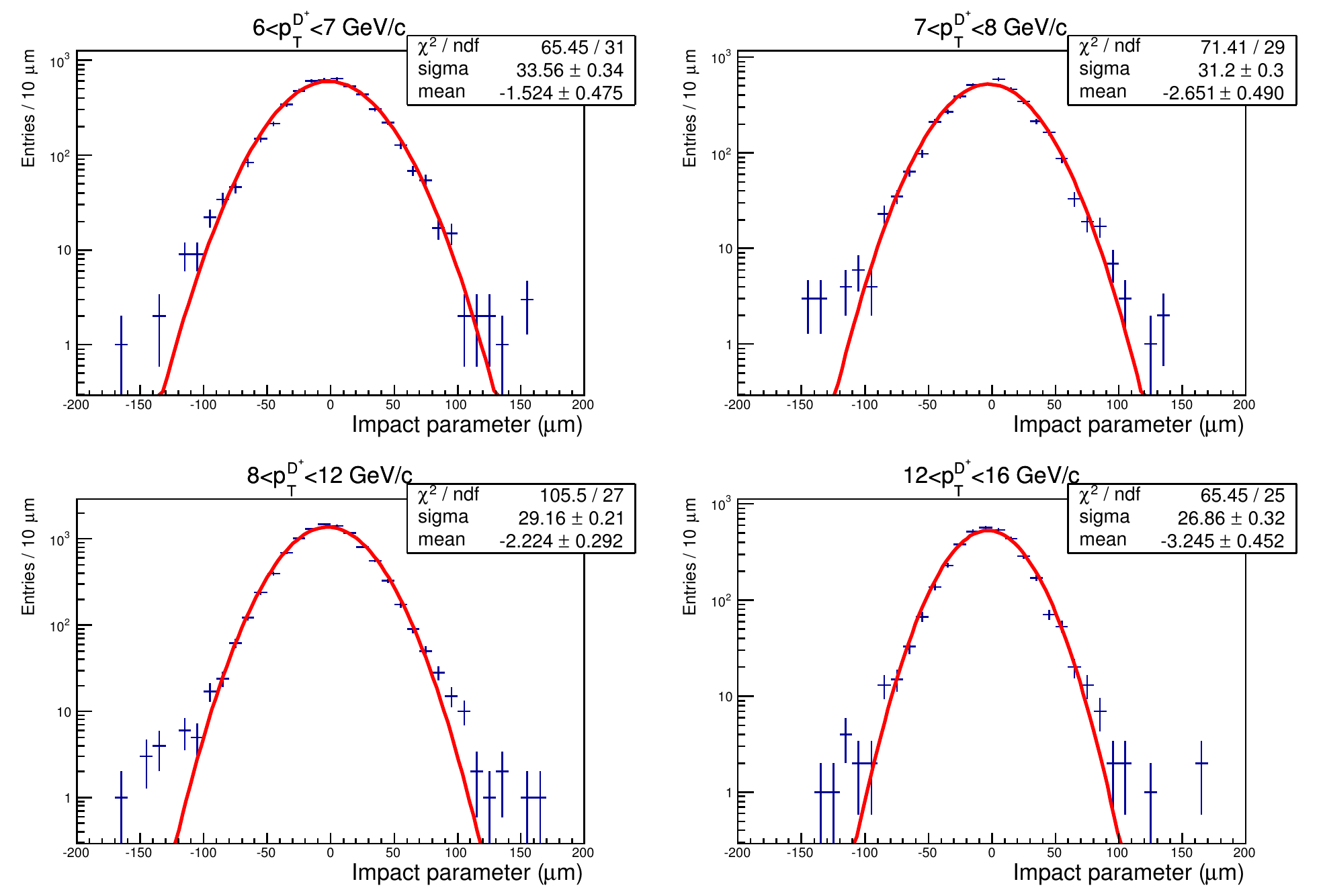}
\caption{Impact parameter ($d_0^{xy}$) distribution of prompt \dplus mesons in eight \pt bins from MC simulations. The gaussian fit is also shown.}
\label{PlotPromptImpParGaus}
\end{figure}
\begin{figure}
 \includegraphics[width=0.999\textwidth]{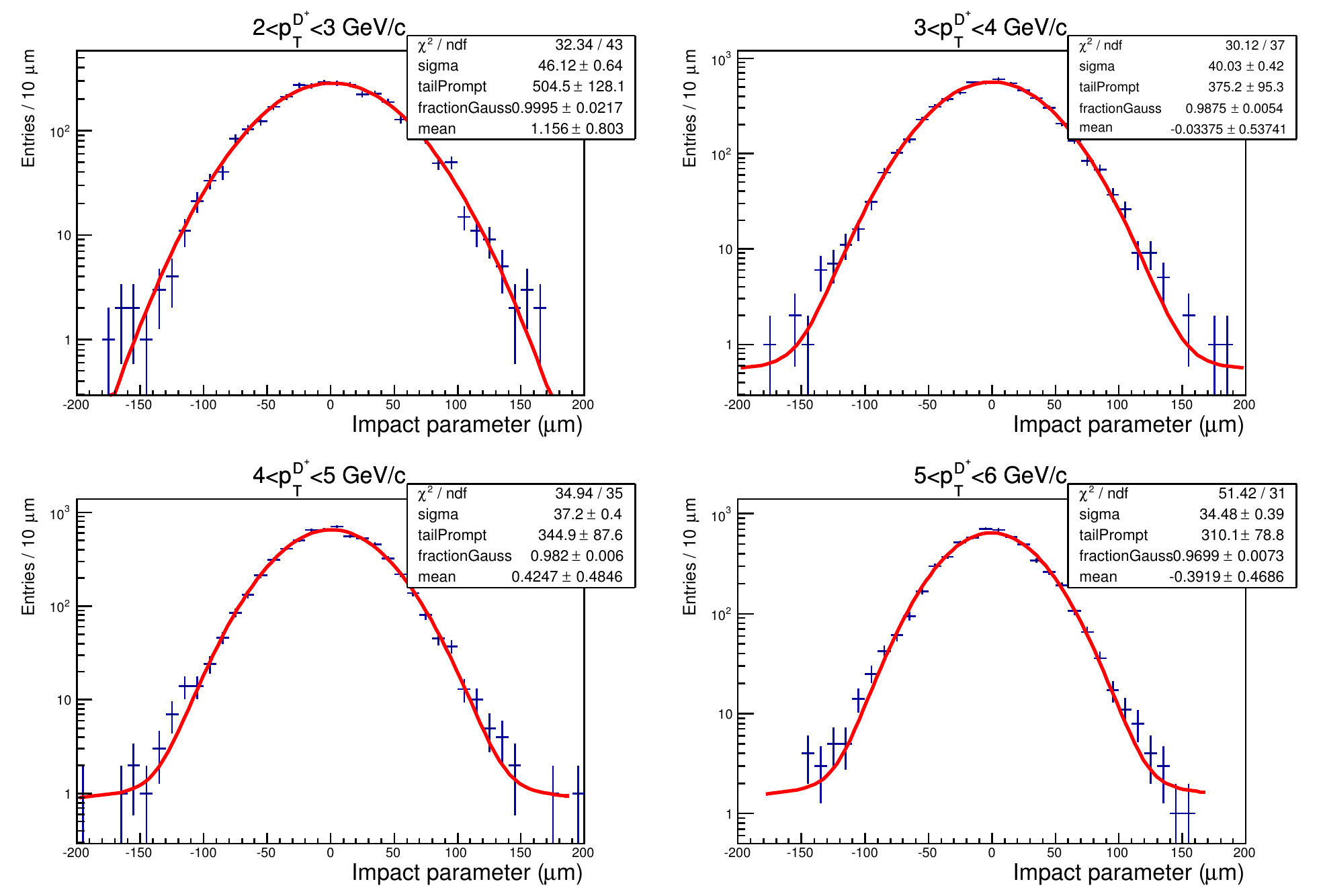} 
\includegraphics[width=0.999\textwidth]{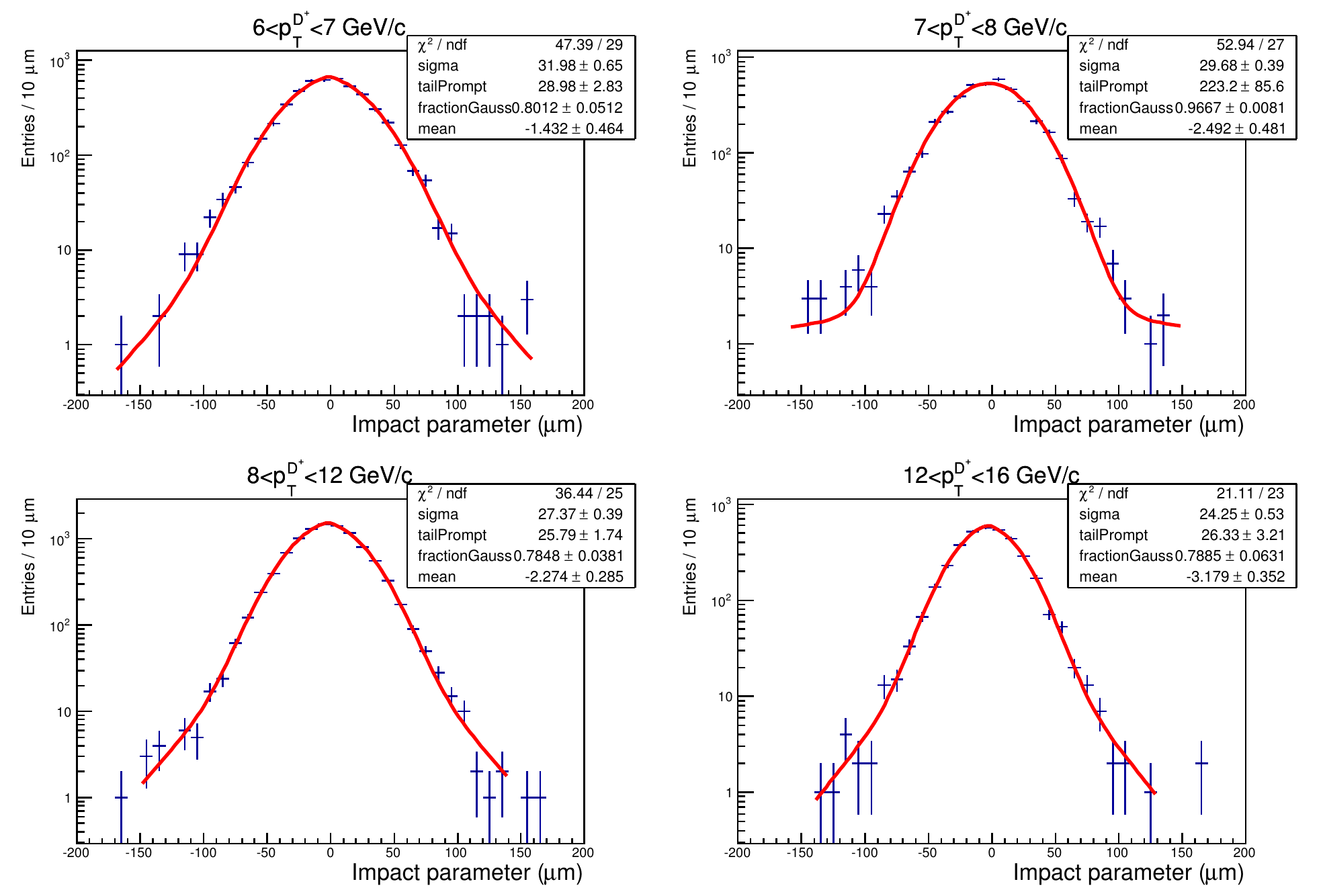}
\caption{Impact parameter ($d_0^{xy}$) distribution for prompt \dplus mesons in eight \pt bins from MC simulations. The gaussian-exponential fit of Equation \ref{eq:gausexp}  is also shown.}
\label{PlotPromptImpParGE}
\end{figure}
\begin{figure}
\centering
 \includegraphics[width=0.6\textwidth]{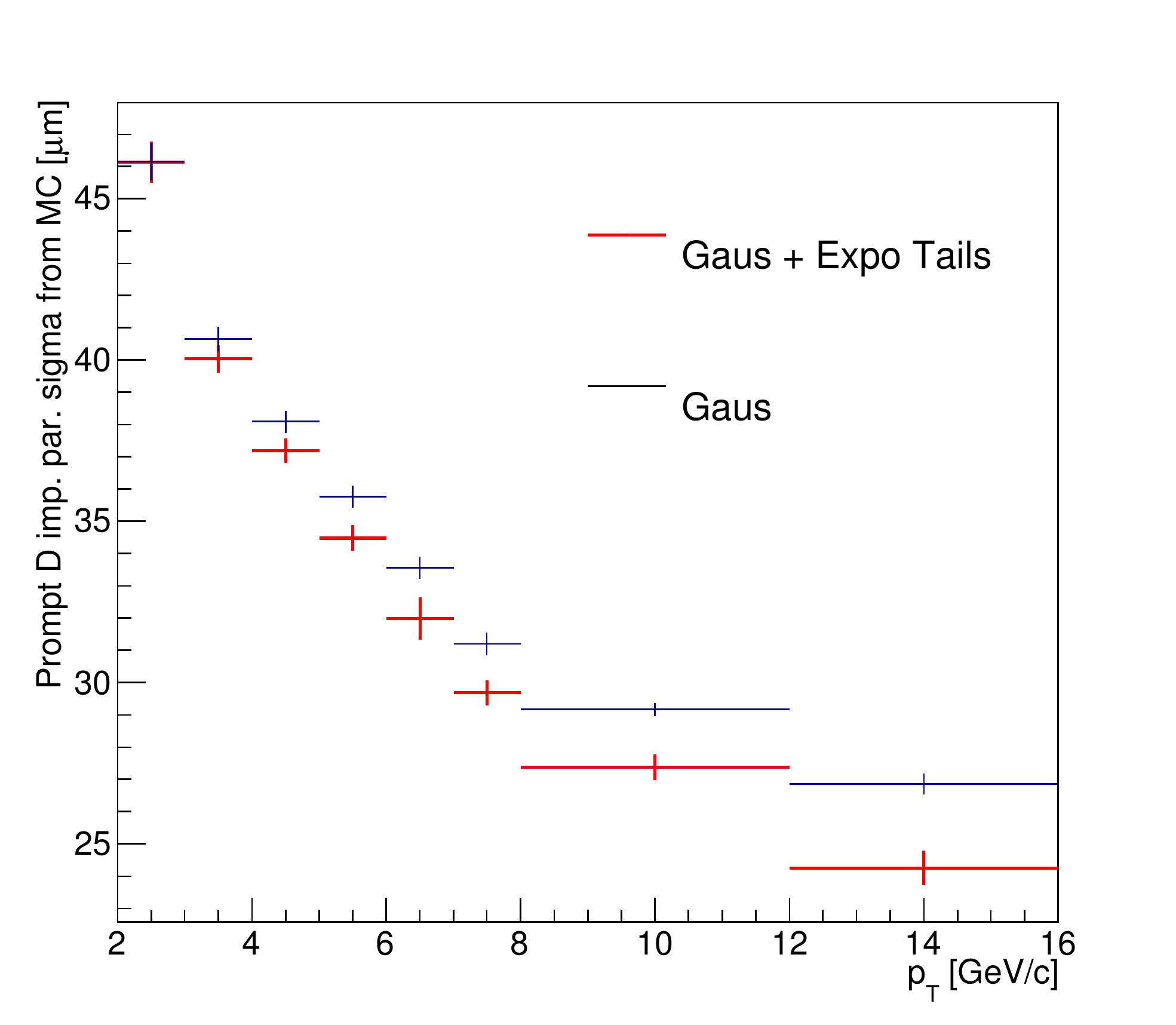} 
\caption{Gaussian width $\sigma_{\rm prompt}$ values obtained with both the simple gaussian fit and the fit of Equation \ref{eq:gausexp}  as a function of \pt.}
\label{SigmaPromptImpPar}
\end{figure}
\subsection{Feed-down \dplusm in MC simulations}
Before showing the impact parameter distributions of feed-down \dplus mesons extracted from the MC data sample of Section  \ref{sub:MCsample}, a few considerations are worth. \\
For feed-down \dplus mesons we expect an impact parameter different from zero. This is  due to the fact that feed-down \dplus meson  momenta  won't in general be parallel to those of the parent B hadrons. However the shape of the distributions in the various \pt intervals is more difficult to model with respect to the ones for prompt \dplus mesons, and this is due to several reasons.  \\
First of all, as visible in Figure \ref{sketchBdeclen} (left), for the same angle between the B-hadron and \dplusm flight lines $\Delta\varphi$ (we limit ourselves to the plane transverse to the beam axis\footnote{For the definition of the ALICE Coordinate System see Appendix \ref{AppendixB}}), the resulting \dplusm impact parameter depends on the distance travelled by the B-hadron $L_{\rm B}$: $d_0^{xy}$=$L_{\rm B}\sin{\Delta\varphi}$. The distribution of $L_{\rm B}$ in turn  depends on the \pt of the beauty hadron due to the Lorentz boost.\\
\begin{figure}[b]
 \includegraphics[width=0.99\textwidth]{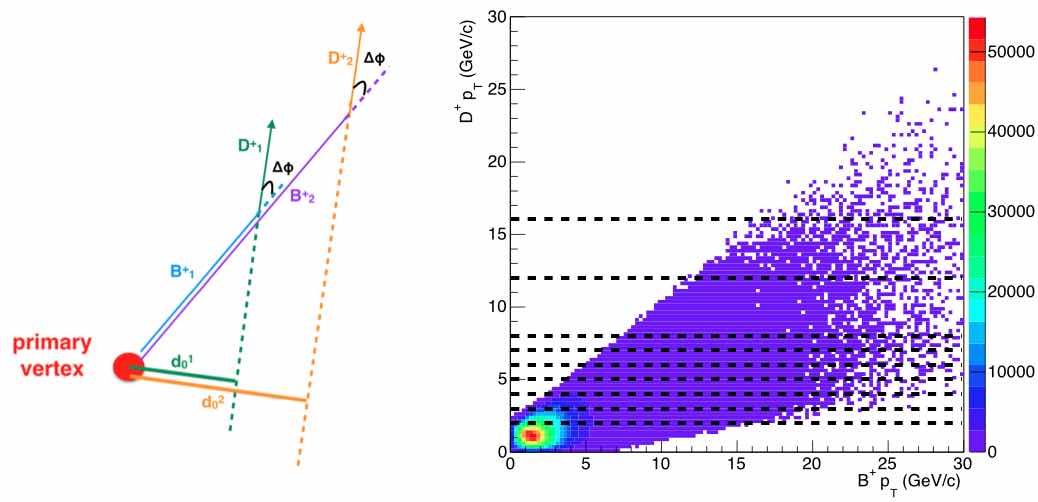} 
\caption{Left: Projection on the plane transverse to the beam axis of the decay of two B$^{+}$ mesons with different decay lengths into D$^{+}$. Both decays have the same $\Delta\varphi$ between parents and daughters tracks. Right: scatter plot of parent B$^{+}$ \pt vs daughter \dplus \pt  from PYTHIA simulations.}
\label{sketchBdeclen}
\end{figure}
Figure \ref{sketchBdeclen} (right) is obtained generating B$^+$ mesons with PYTHIA 6 and making them decay in channels containing a \dplus meson, and corresponds to the scatter plot of  daughter \dplus \pt as a function of parent B$^{+}$ p$_{\rm T}$.  The \pt spectrum of the generated B$^+$ mesons is taken from FONLL. The \dplusm \pt bins used in this analysis are shown as dashed lines.  As one can see, for a given \dplus \pt interval the range of values for the parent B$^+$ meson \pt  is broad (10$\div$20 GeV/c), which results in a wide \pt distribution of B$^+$-hadron decay length. We thus expect the distribution of \dplusm impact parameter in a given \dplusm   transverse momentum bin to receive contributions from  B$^+$ mesons with a wide range of decay lengths, given by the proper decay length $c\tau$ and by the $\gamma$ factor of the B$^+$ meson. \\
\begin{figure}
 \includegraphics[width=0.99\textwidth]{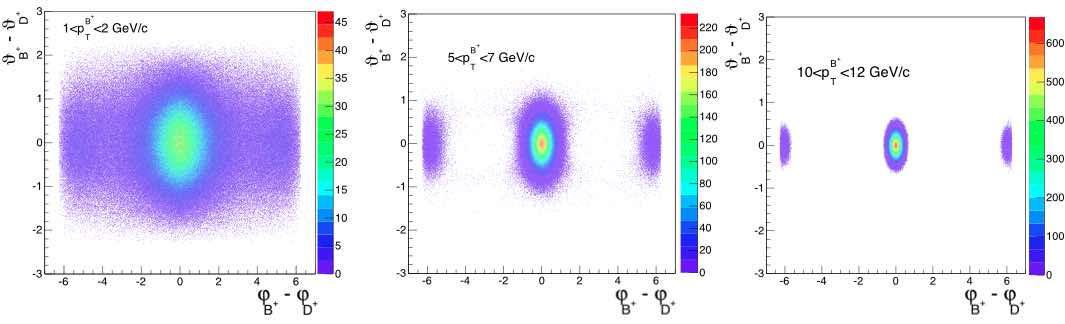} 
\caption{Scatter plot of   the difference in radial angle $\Delta\varphi$ and azimuthal angle $\Delta\theta$ between the generated parent B$^{+}$ meson and the daughter D$^{+}$ meson as obtained with  PYTHIA 6 in 1<p$_{\rm T}^{B^{+}}$<2 GeV/c, 5<p$_{\rm T}^{B^{+}}$<7 GeV/c, 10<p$_{\rm T}^{B^{+}}$<12 GeV/c.}
\label{angles}
\end{figure}
However another aspect has to be taken into account. Figure \ref{angles} shows the scatter plot of   the difference in polar angle\footnote{See Appendix \ref{AppendixB}} $\Delta\theta$ and azimuthal angle $\Delta\varphi$ between the generated B$^{+}$ meson and the D$^{+}$ meson from its decay as obtained with  PYTHIA 6 for three B$^{+}$ \pt bins.  Due to the Lorentz boost  at high \pt  the \dplus mesons flight line tends to be parallel to that of the parent B$^{+}$ hadron. The  effect of the Lorentz boost results on average in a decrease of the feed-down \dplusm impact parameter. \\
\begin{figure}
 \includegraphics[width=0.999\textwidth]{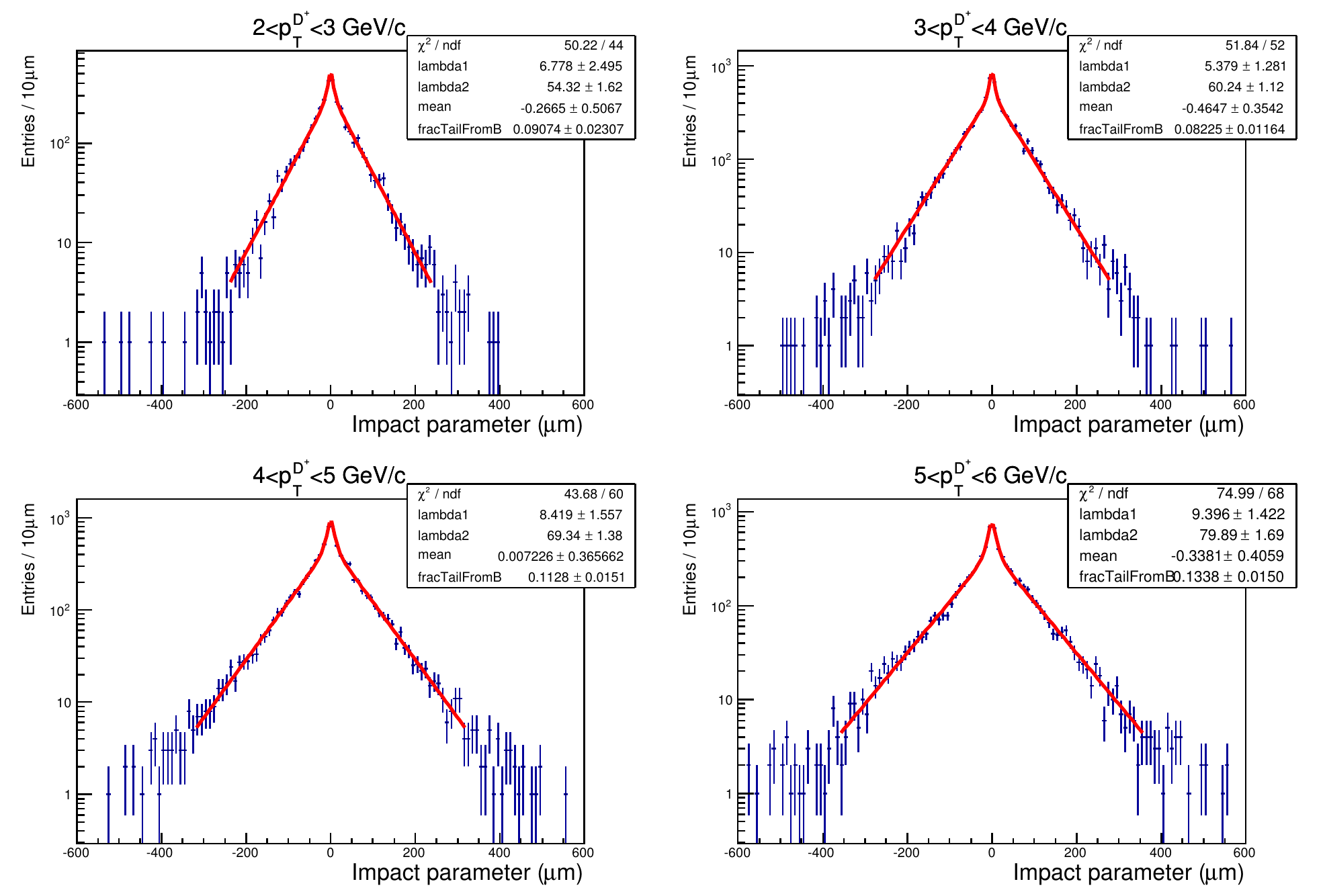} 
\includegraphics[width=0.999\textwidth]{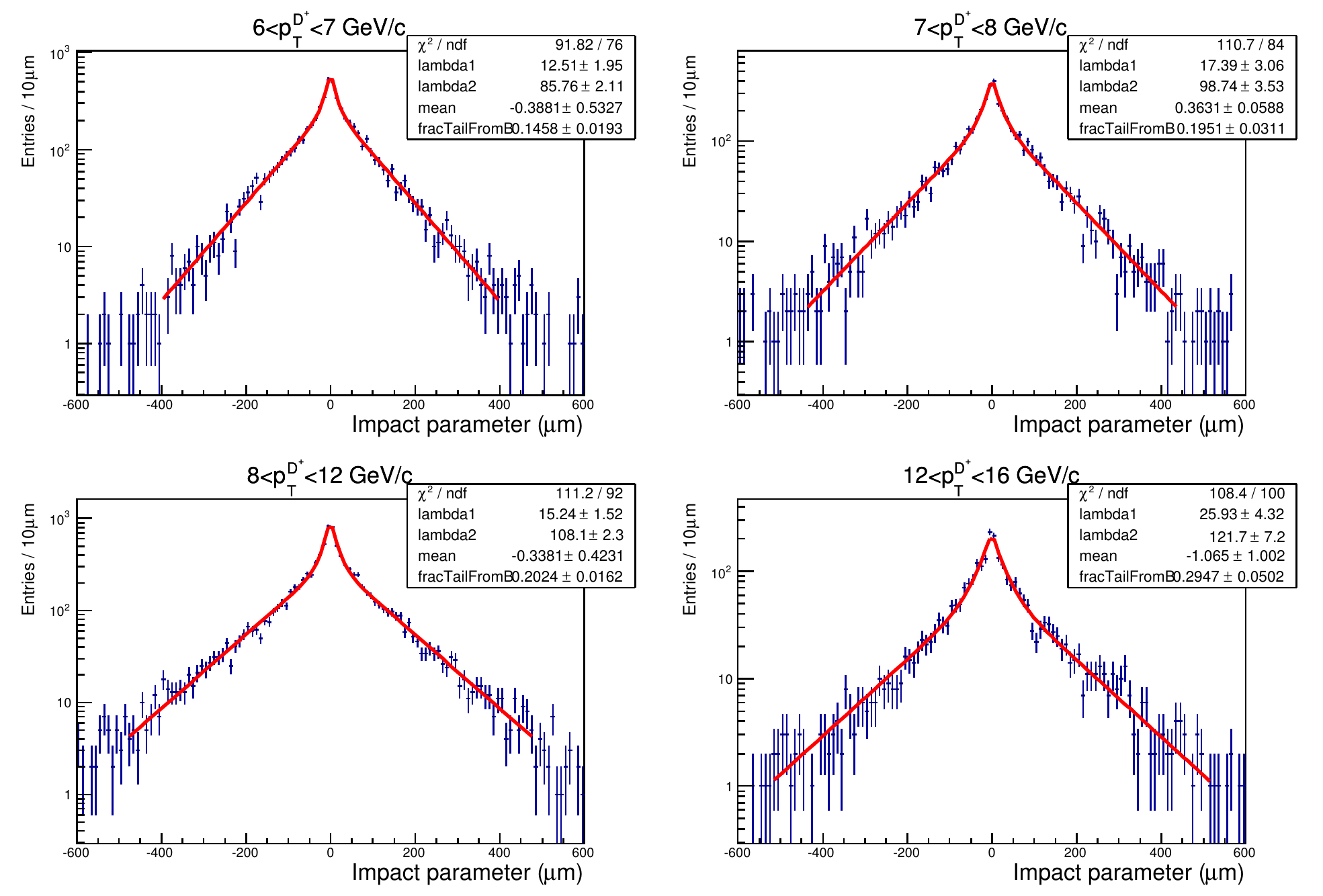}
\caption{Impact parameter ($d_0^{xy}$) distributions of  feed-down \dplusm in eight \pt bins from MC simulations, using the true value of \dplus impact parameter. The double-exponential fit of Equation \ref{eq:fdfucn}  is also shown.}
\label{PlotFDTrue}
\end{figure}
The true  impact parameter ($d_0^{xy}$) distributions of feed-down \dplusm obtained from the MC data sample described in Section \ref{sub:MCsample} are shown in Figure \ref{PlotFDTrue} in the eight \pt bins considered in this analysis. Only the feed-down \dplus mesons passing the selection criteria described in Section \ref{sec:SE} are included. The distributions are fitted with the sum of two   exponentials
\begin{equation}
f^{\rm feed-down}_{\rm true}(d_0^{xy})=A\bigg[(1-f_{\lambda_2^{\rm FD}})\frac{1}{2\lambda_1^{\rm FD}}e^{-\frac{|d_0^{xy}-\mu|}{\lambda_1^{\rm FD}}}+f_{\lambda_2^{\rm FD}}\frac{1}{2\lambda_2^{\rm FD}}e^{-\frac{|d_0^{xy}-\mu|}{\lambda_2^{\rm FD}}}\bigg]
\label{eq:fdfucn}
\end{equation}
Each  exponential is governed by a slope parameter, $\lambda_1^{\rm FD}$ and $\lambda_2^{\rm FD}$, respectively,  and they have common mean value $\mu$. The fit range considered increases with increasing \pt to take into account the broadening of the distributions with transverse momentum. The $\chi^2$ values show that this functional form gives an acceptable description of the distributions from MC simulations. \\
\begin{figure}
 \includegraphics[width=0.999\textwidth]{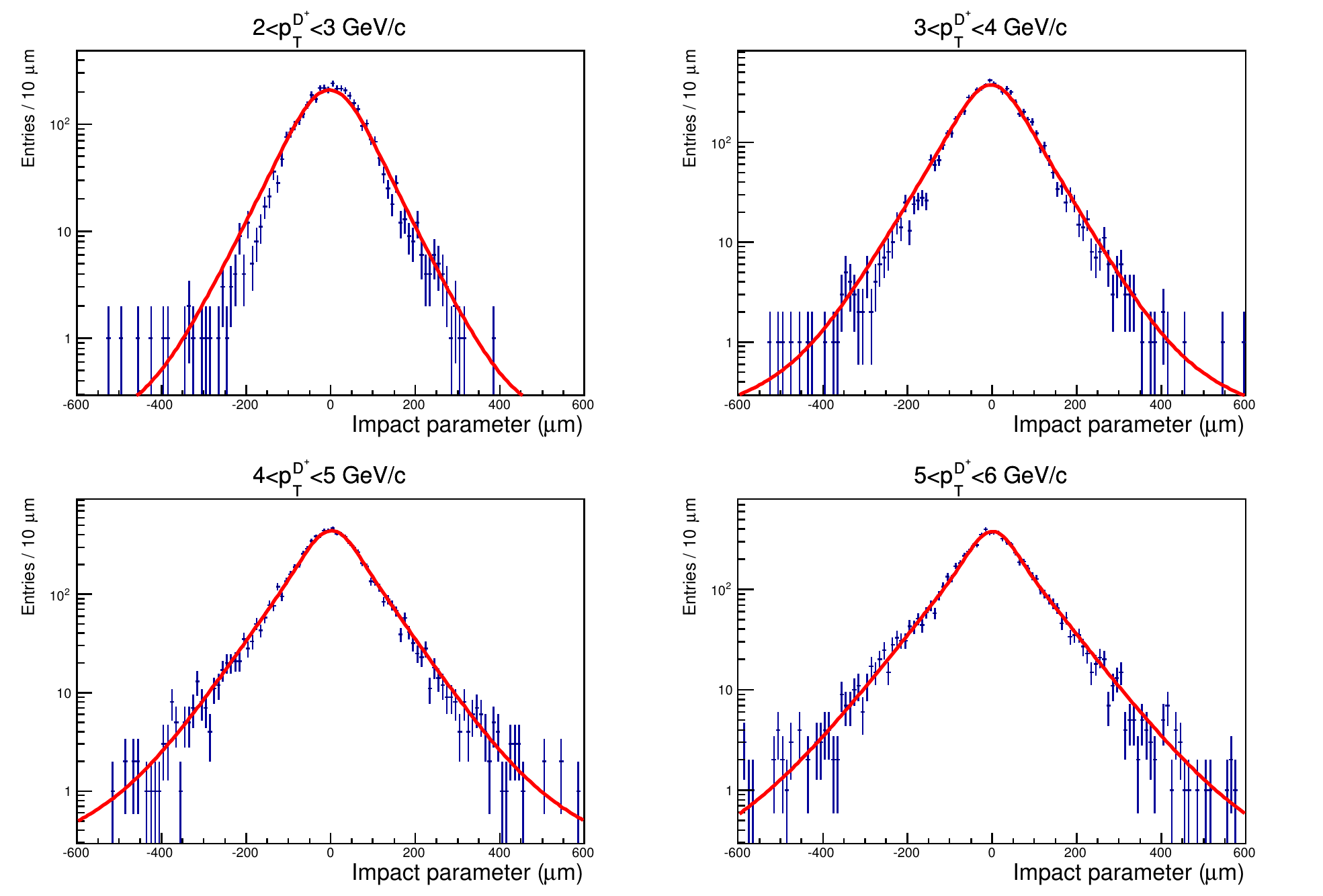} 
\includegraphics[width=0.999\textwidth]{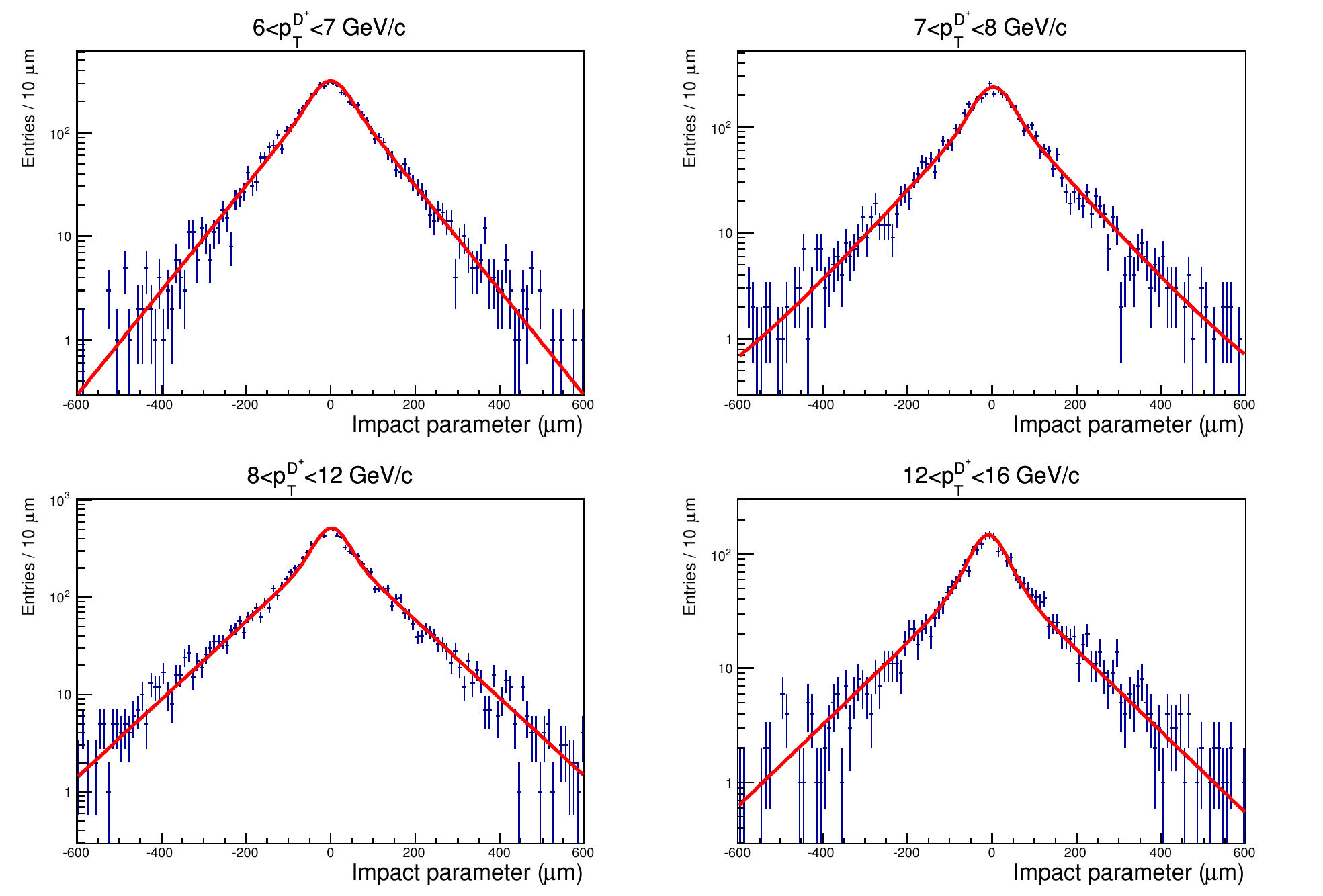}
\caption{Impact parameter ($d_0^{xy}$) distributions of  feed-down \dplusm in eight \pt bins from MC simulations, using the reconstructed value of \dplus impact parameter. The convolution of Equation \ref{convolution}  is also shown.}
\label{PlotFDReco}
\end{figure}
The impact parameter distributions for \dplus mesons after the reconstruction are shown in Figure \ref{PlotFDReco}.  Only the feed-down \dplus mesons passing the selection criteria described in Section \ref{sec:SE} are included. The functions superimposed in  Figure \ref{PlotFDReco} are not a fit to the distributions. They are the convolution of $f^{\rm feed-down}_{\rm true}(d_0)$ obtained from the fit to the \dplusm true impact parameter distributions shown in Figure \ref{PlotFDTrue} and a Gaussian which has the same width $\sigma_{\rm prompt}$ of the prompt \dplusm impact parameter distribution in the corresponding \pt bin
\begin{equation}
F^{\rm feed-down}(d_0^{xy})=A \int^{d_0^{\rm min}}_{d_0^{\rm max}}\bigg[f^{\rm feed-down}_{\rm true}(d_0';\lambda_1^{\rm FD},\lambda_2^{\rm FD})\bigg]\bigg[ \frac{1}{\sqrt{2\pi}\sigma_{\rm prompt}}e^{-\frac{(d_0^{xy}-d_0')^2}{2\sigma_{\rm prompt}^2}} \bigg]\textrm{d}d_{0'}
\label{convolution}
\end{equation}
where A is the integral of the impact parameter distributions for \dplus mesons after the reconstruction  in the corresponding \pt bin, $d_{0}^{\rm min}$=-600 $\mu$m and $d_{0}^{\rm max}$=600 $\mu$m.
The functions show good agreement with the distributions except for \pt< 4 GeV/c. In these \pt bins a fit performed with the same functional form of Equation \ref{eq:gausexp} gives a better description of the MC distribution as shown in Figure \ref{PlotFDReco2}. \\ 
\begin{figure}
\centering
 \includegraphics[width=0.85\textwidth]{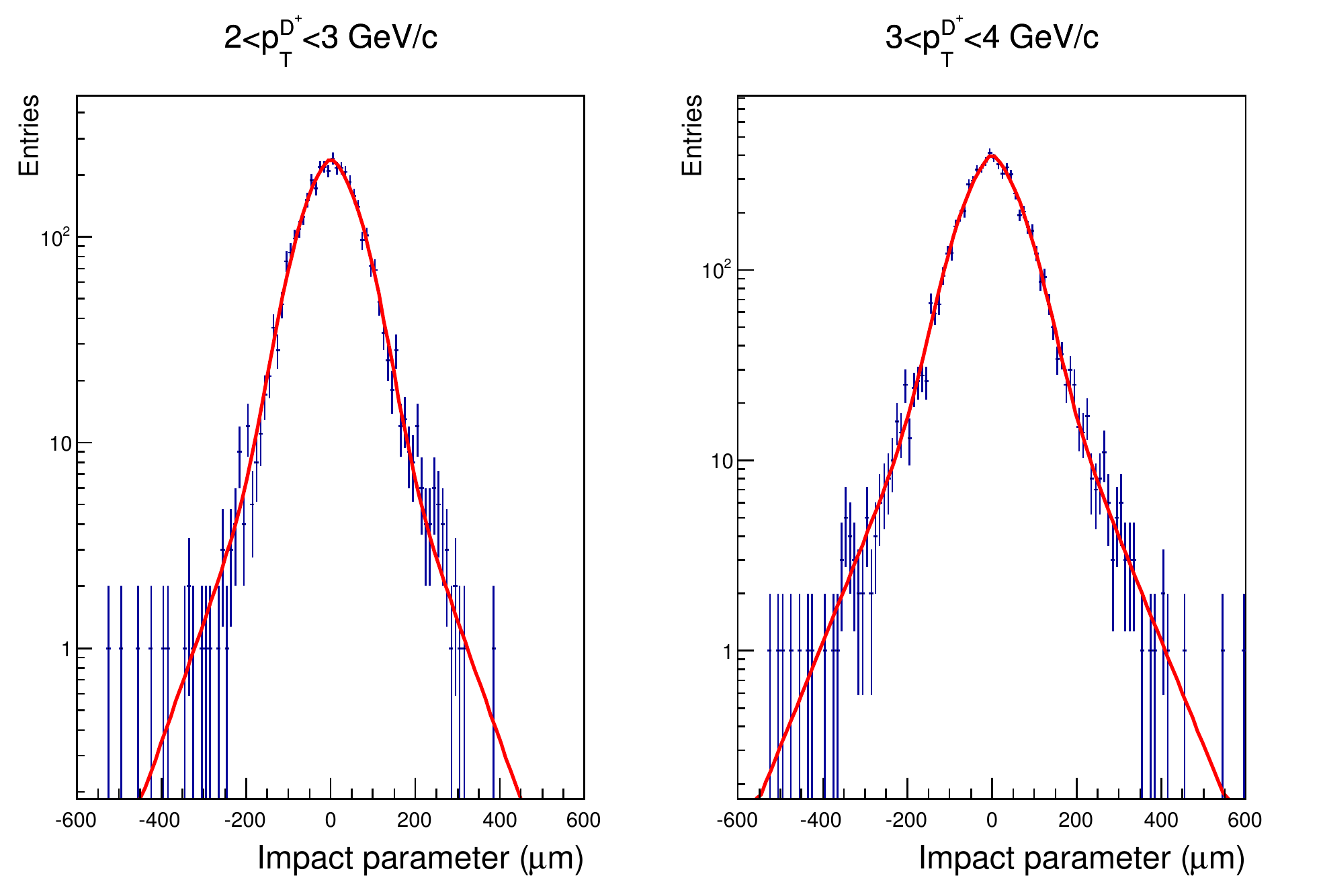} 
\caption{Impact parameter ($d_0^{xy}$) distributions of  feed-down \dplusm in 2<\pt<3 GeV/c (left) and 3<\pt<4 GeV/c from MC simulations, using the reconstructed value of \dplus impact parameter. The distributions are fitted with the functional form of Equation  \ref{eq:gausexp}.}
\label{PlotFDReco2}
\end{figure}
\subsection{Impact Parameter distributions in data}
\begin{figure}[b]
 \includegraphics[width=0.99\textwidth]{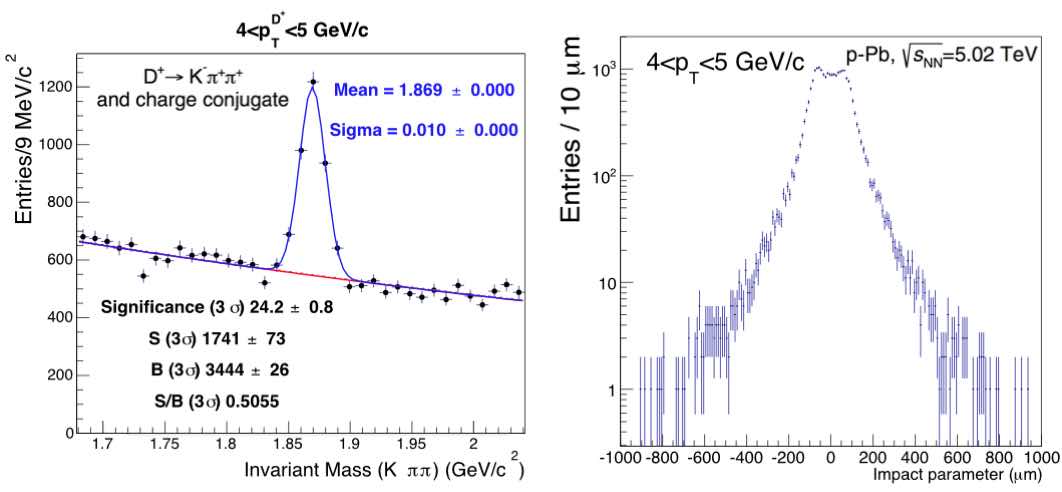} 
\caption{Left: invariant mass ($M$) distribution of the selected \dplusm candidates in the \pt interval 4<\pt<5 GeV/c after applying the same event, track quality, topological and PID cuts used  in  Chapter 5. Right:  impact parameter ($d_0^{xy}$) distribution of the same candidates shown in the left panel.}
\label{SBSub1}
\end{figure}
The impact parameter distributions  from the p--Pb data sample collected with the minimum bias trigger described in Section  \ref{sub:MCsample} are obtained as described in the following. I will show the procedure in the \pt bin 4<\pt<5 GeV/c.\\
After applying the same event, track quality, topological and PID selections used to extract the \dplusm cross-section and nuclear modification factor shown in  Chapter 5, we obtain  the invariant mass ($M$)  distribution of the selected \dplusm candidates as shown in  Figure \ref{SBSub1} (left).  The impact parameter distribution for the same selected candidates in the \pt bin 4<\pt<5 GeV/c is shown in Figure \ref{SBSub1} (right). \\
The left panel of  Figure \ref{SBSub1} (left) also shows the  fit to the invariant mass distribution, where the signal is described by a Gaussian with width $\sigma^{\rm peak}$=10 MeV/c and mean $M^{\rm peak}$=1.869 GeV/c. We define as \textbf{sideband} region the invariant mass region for which |$M$-$M^{\rm peak}$|>3$\sigma^{\rm peak}$.\\
\begin{figure}[t]
 \includegraphics[width=0.99\textwidth]{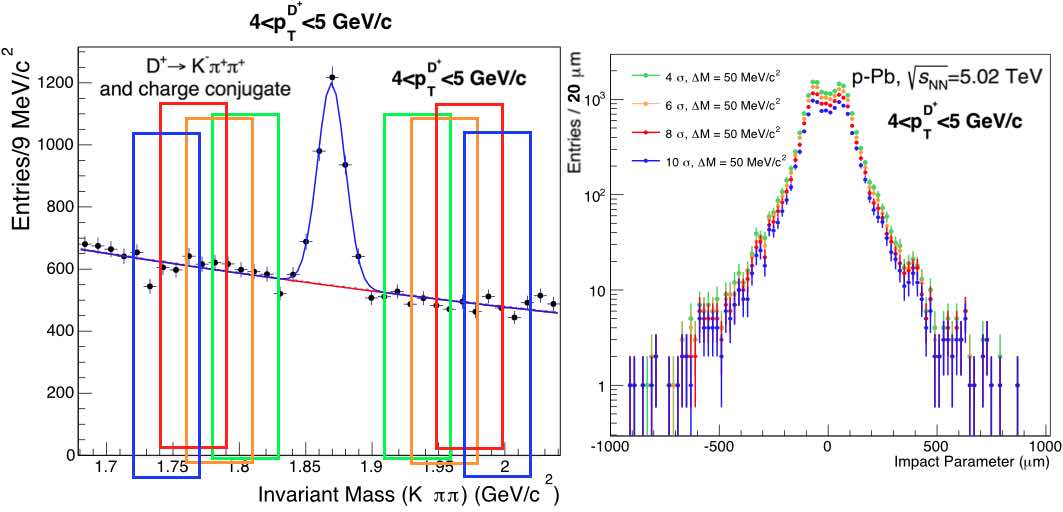} 
\caption{Left: same invariant mass distribution shown in Figure \ref{SBSub1}. Right: impact parameter distributions of the candidates with invariant mass lying within the four coloured boxes in the left panel.}
\label{SBSub2}
\end{figure}
\begin{figure}[b]
 \includegraphics[width=0.99\textwidth]{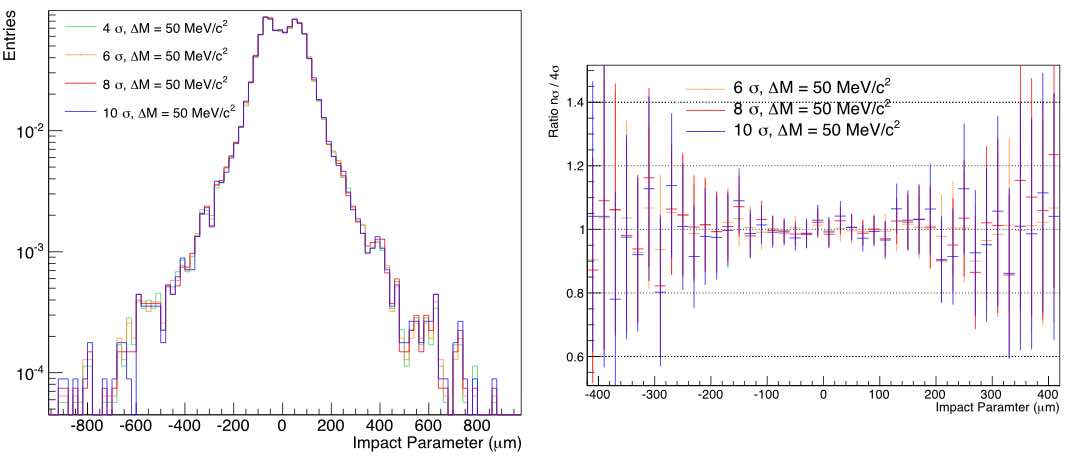} 
\caption{Left: same impact parameter  distributions shown in Figure \ref{SBSub2} normalized to their area. Right: ratio of the impact parameter distributions obtained at 6, 8 and 10 $\sigma^{peak}$ from the invariant mass peak and that obtained at 4 $\sigma^{\rm peak}$.}
\label{SBSub3}
\end{figure}
The method of the sideband subtraction aims at extracting the impact parameter distributions for signal candidates which are truly \dplus mesons. To achieve this goal it is necessary to eliminate the \dplusm background candidates  under the \dplusm mass peak.\\ The starting assumption is that the impact parameter distribution of  background candidates under the \dplusm peak    are  similar to the average of the same distributions of candidates in the sideband regions.  \\
\begin{figure}[t]
 \includegraphics[width=0.99\textwidth]{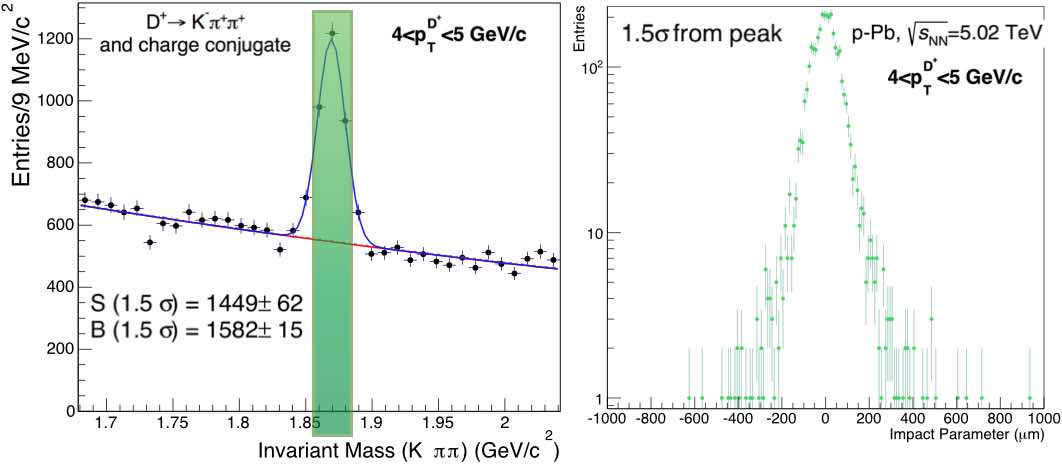} 
\caption{Left: same invariant mass distribution shown in Figure \ref{SBSub1}. Right: impact parameter distributions of the candidates within 1.5 $\sigma^{peak}$ from the   \dplusm peak.}
\label{SBSub4}
\end{figure}
\begin{figure}[b]
 \includegraphics[width=0.99\textwidth]{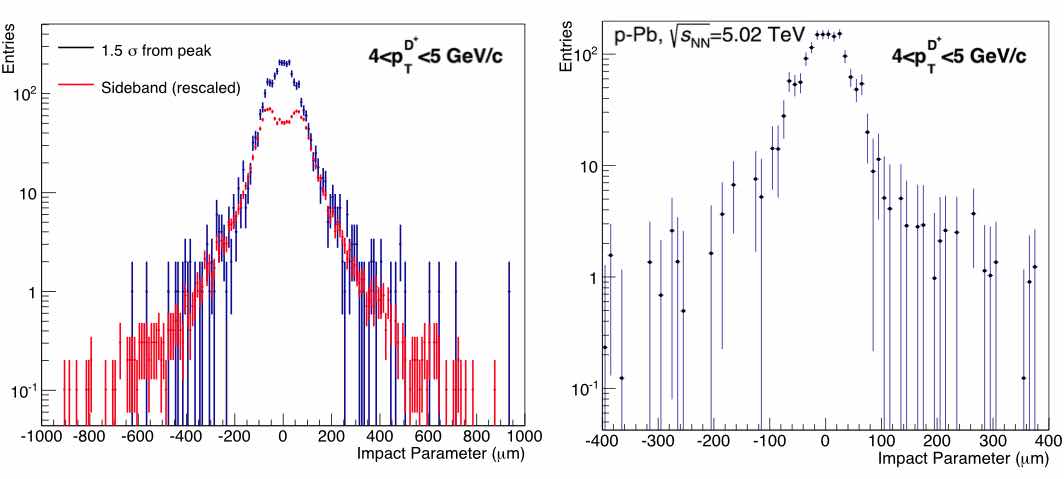} 
\caption{Left:  impact parameter distributions of  candidates in the signal region, i.e. within 1.5 $\sigma^{peak}$ from the  \dplus peak (black) together with the  impact parameter distribution of the candidates in the sideband region, i.e. with 5<$|M-M_{\rm peak}|/\sigma^{\rm peak}$<10 rescaled to the total amount of background present within 1.5 $\sigma^{peak}$ from the invariant mass peak . Left: impact parameter distribution of candidates in the signal region after sideband subtraction.}
\label{SBSub5}
\end{figure}
Figure \ref{SBSub2}  (right) shows the impact parameter distributions of candidates with 4<\pt<5 GeV/c for different invariant mass intervals in the sideband regions. Each interval is characterized by a $n\sigma^{\rm peak}$ value and has a width of 5$\sigma^{\rm peak}$ (50 MeV/c$^{2}$) as indicated in the left panel of Figure \ref{SBSub2}. In Figure \ref{SBSub3} (left) the four impact parameter distributions are normalized to their integral to see if their shape changes significantly depending on the invariant mass interval considered. This is not the case, as clearly visible in  the right panel of Figure \ref{SBSub3} where the ratio of the distributions at 6,8 and 10 $\sigma^{\rm peak}$ to the one obtained at 4$\sigma^{\rm peak}$ is shown. This proves the stability of the impact parameter distribution in different intervals of invariant mass. In the following I will use the impact parameter  distributions of candidates with $M$ in 5<|$M$-$M^{\rm peak}$|/$\sigma^{\rm peak}$<10 to estimate the background under the \dplus peak.\\
\begin{figure}[b]
 \includegraphics[width=0.99\textwidth]{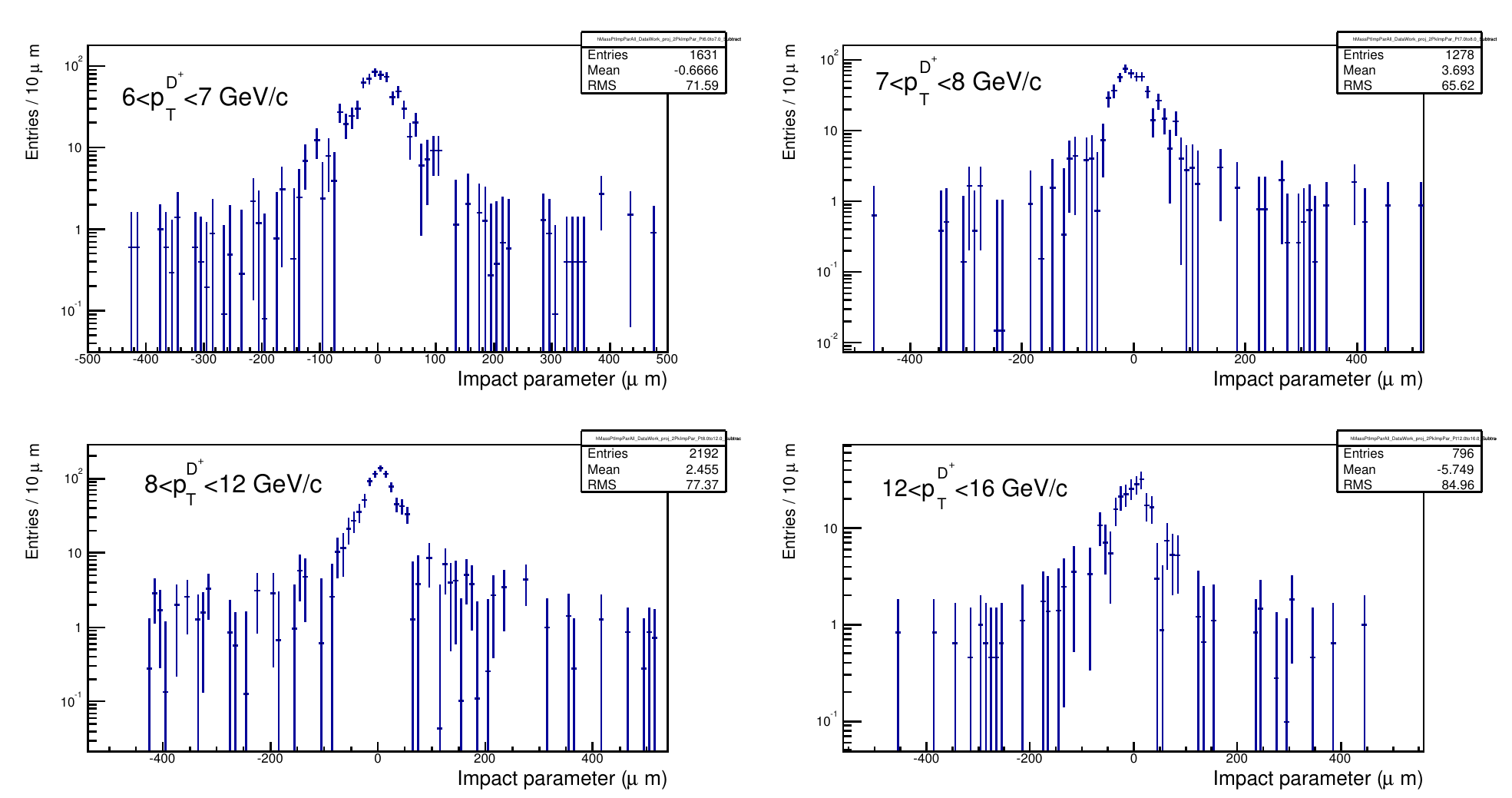} 
 \includegraphics[width=0.99\textwidth]{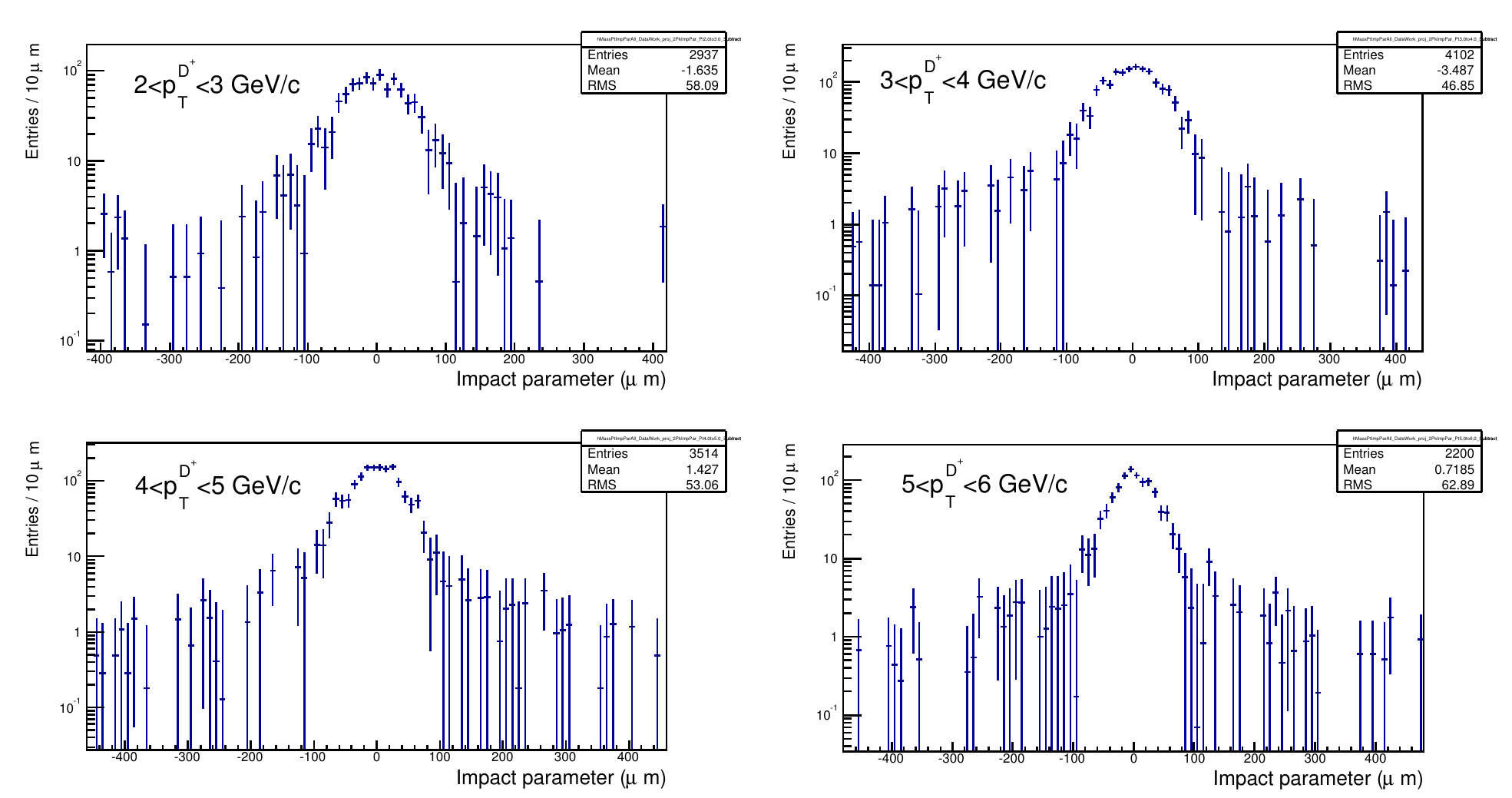} 
\caption{Data impact parameter distributions after sideband subtraction in eight \pt intervals.}
\label{SBSub6}
\end{figure}
 Figure \ref{SBSub4}  (right) shows the impact parameter distribution for candidates in the signal region, defined as    |$M$-$M^{\rm peak}$|<1.5$\sigma^{\rm peak}$. The signal  invariant mass range is the one included in the green area of Figure \ref{SBSub4} (left). Within 1.5 $\sigma^{\rm peak}$ from the mass peak we expect to have 1449 signal candidates and 1582 background candidates.  The  impact parameter distribution of background candidates estimated from the side-bands was normalized  to the  number of background candidate expected in the signal region (red histogram of Figure \ref{SBSub5} (left)) and was  subtracted from the impact parameter distribution of candidates with |$M$-$M^{\rm peak}$|<1.5$\sigma^{\rm peak}$ to obtain the  impact parameter distribution of true \dplus mesons\footnote{One can object that the impact  parameter distribution of  \dplus mesons with reconstructed invariant mass $M$ within 1.5 $\sigma^{\rm peak}$ from the mass peak does not include all the true \dplusm candidates in the \pt bin considered. The variation of $n\sigma^{\rm peak}$ will be considered to evaluate the systematic.}. The result of the sideband subtraction   is shown in Figure \ref{SBSub5} (right) for the \pt bin 4<\pt<5 GeV/c.\\
The result of the sideband subtraction in the eight \pt bins considered in this analysis is shown in Figure \ref{SBSub6}.
\section{Fit method}
\label{ref:fit}
The  method used to fit the impact parameter distribution of \dplus mesons to extract the fraction of prompt yield is composed of two steps: a prefit phase performed on the MC distributions, and fit phase performed on the sideband subtracted impact parameter distributions of data.\\
In the prefit step, the impact parameter distributions of prompt \dplus mesons from MC simulations are fitted with the functional form of Equation \ref{eq:gausexp},  the fit results for \textbf{$F^{\rm prompt}$} are reported in Figure \ref{PlotPromptImpParGE}. This fit yields three parameters: $\lambda^{\rm prompt}$, $\sigma^{\rm prompt}$ and $\mu^{prompt}$, which characterize the slope of the exponential tales  and the  width and mean of the Gaussian function.\\
Then the impact parameter distributions of true feed-down \dplus mesons from MC simulations are fitted with the functional form of Equation \ref{eq:fdfucn}. The fit results are  reported in Figure \ref{PlotFDTrue}. This fit yields three parameters: $\lambda^{\rm FD}_1$, $\lambda^{\rm FD}_2$ and $\mu^{\rm FD}$. Keeping these  parameters fixed,  the function obtained from the fit is convoluted with a gaussian according to Equation \ref{convolution}. The width of the convoluted gaussian is fixed to $\sigma^{\rm prompt}$. The result of the convolution \textbf{$F^{\rm FD}$} are shown in Figure \ref{PlotFDReco}. \\
At this point the final fit on the impact parameter distributions of data  can be performed. The sideband subtracted impact parameter distribution is fitted with the following function
\begin{equation}
F^{D^{+}}(d_0)=A\bigg[(1-f_{\rm prompt})F^{\rm FD}(d_0)+f_{\rm prompt}F^{\rm prompt}(d_0)\bigg]
\end{equation}
where A is the integral of the sideband subtracted impact parameter distribution and \fprompt is the fraction of prompt \dplus mesons. The fit is performed under the following conditions:
\begin{itemize}
\item the parameters  $\lambda^{\rm FD}_1, \lambda^{\rm FD}_2$ and $\lambda^{\rm prompt}$ are fixed to those obtained in the prefit phase
\item the parameters $\mu^{\rm Prompt}$, $\mu^{\rm FD}$ are set to 0
\item the  parameter A is fixed to the integral of the sideband subtracted impact parameter distribution
\item the  parameter $\sigma_{\rm prompt}$ is free to vary within 20\% around the value obtained in the prefit step on  the prompt \dplusm impact parameter distribution from MC simulations
\item the parameter \fprompt is bound between 0 and 1
\end{itemize}
The results of the fit are shown as red curves in Figure \ref{FinalFit} in the eight \pt intervals considered in the analysis. The fit range gradually increases at high \pt to better account for the broadening of the distributions. The fraction of the total fit function corresponding to the feed-down component is represented by the green curves.\\
The values of $\chi^2$/n.d.f. do not exceed 1.2, reaching a minimum of 0.64  in the \pt interval 12<\pt<16 GeV/c where the relative statistical uncertainties of the sideband subtracted distribution become large  due to the limited statistics.
The values of $\sigma_{\rm prompt}$ obtained from the final fit in the different \pt bins are compared to those obtained in the prefit phase in Figure \ref{FitParamFD} (left). None of the final values of $\sigma_{\rm prompt}$ is at limit with respect to the 20\% bound imposed in the fit.\\
\begin{figure}
 \includegraphics[width=0.99\textwidth]{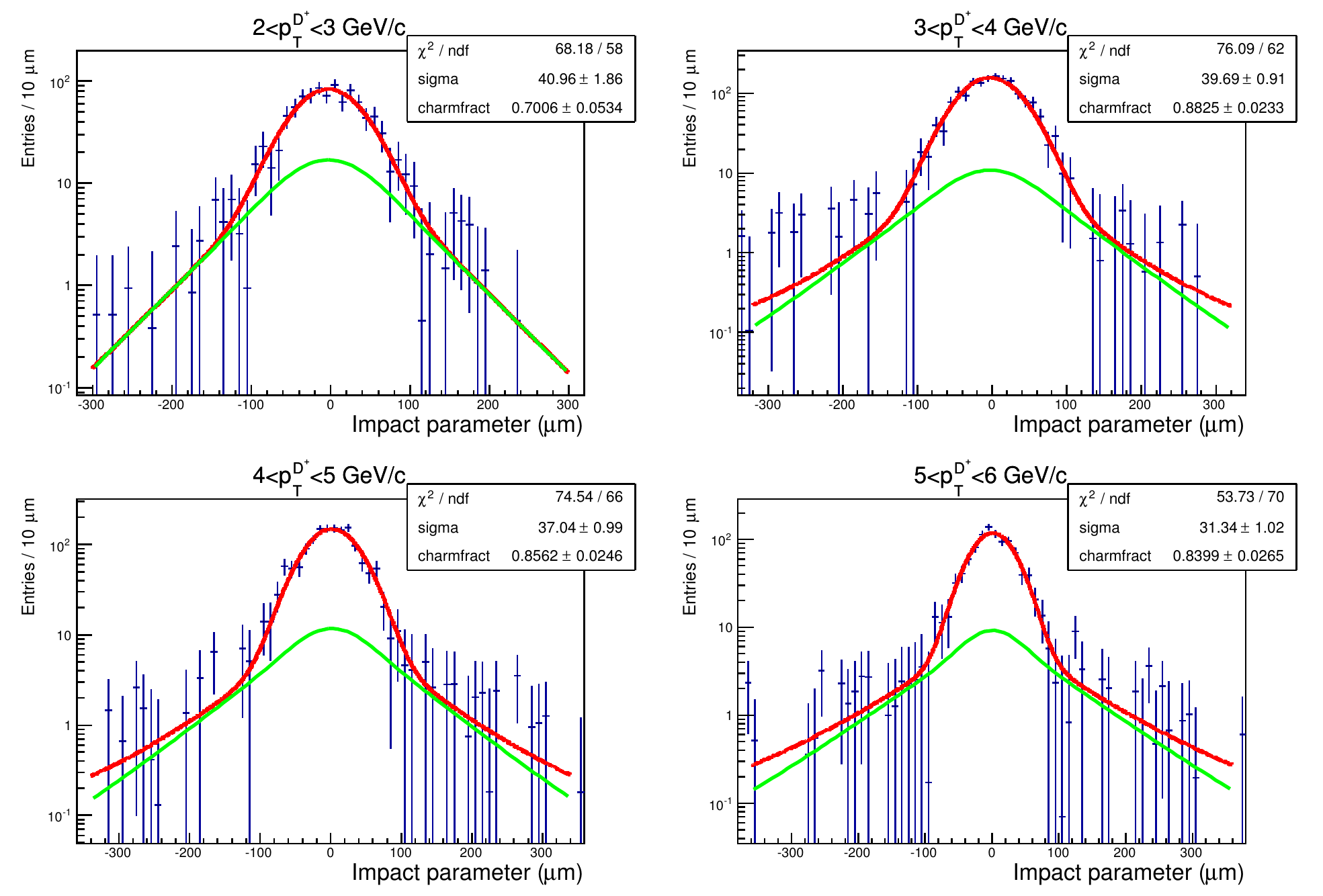} 
 \includegraphics[width=0.99\textwidth]{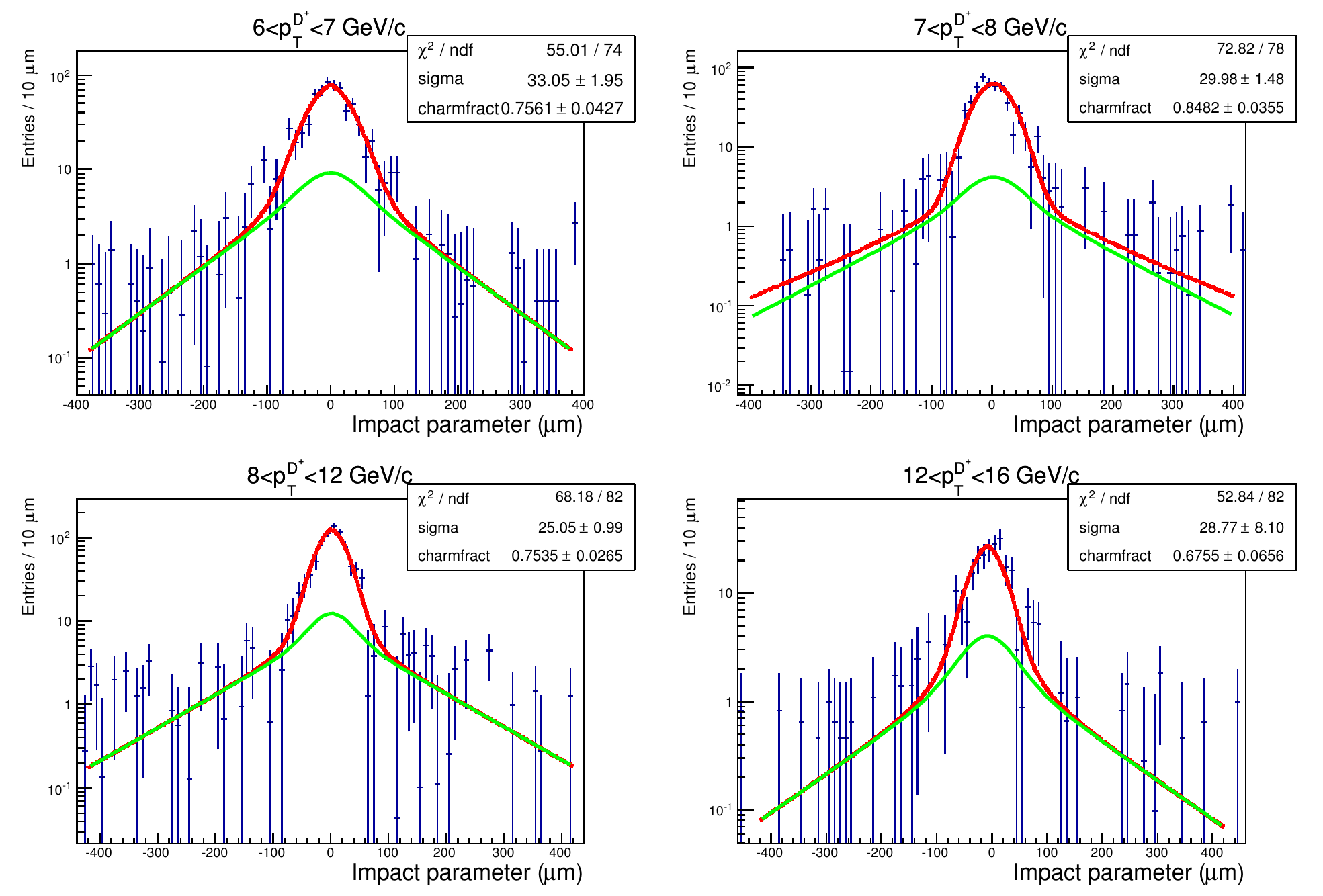} 
\caption{Fit to the sideband subtracted impact parameter  distributions in eight \pt intervals. The red curves are the total fit function, the green ones represent the  feed-down contribution.}
\label{FinalFit}
\end{figure}
\begin{figure}[t]
 \includegraphics[width=0.99\textwidth]{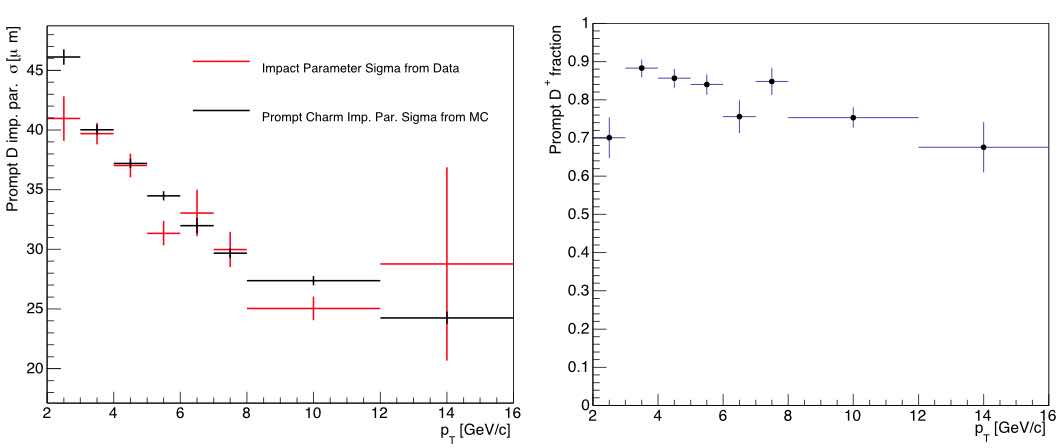} 
\caption{Left: values of $\sigma_{\rm prompt}$ obtained from the final fit to the data in the different \pt intervals  compared to those obtained in the prefit phase. Right: \fprompt obtained from the fit as a function of \pt.}
\label{FitParamFD}
\end{figure}
Finally Figure \ref{FitParamFD} (right) shows the resulting values of \fprompt as a function of \pt. 
\section{Systematic uncertainties}
Three sources of systematic uncertinties are considered in this analysis:
\begin{itemize}
\item systematic uncertainty due to the fit method used
\item systematic uncertainty due to the   sideband subtraction
\item systematic uncertainty due to the \pt shape of generated \dplus mesons in MC
\end{itemize}
\subsection{Systematic uncertainty on fit method}
The systematic uncertainty on the fit method described in  Section \ref{ref:fit} was evaluated in the following way:
\begin{itemize}
\item the impact parameter  fit range was reduced by 100 $\mu$m in each \pt interval
\item the impact parameter fit range was increased by 100 $\mu$m in each \pt interval
\item the bin width used for the impact parameter distribution was reduced from 10 to  5 $\mu$m
\item the  parameter $\sigma_{\rm prompt}$  was fixed to the one obtained in the prefit phase on MC simulations
\item  The function of Equation  \ref{eq:gausexp} was used to describe the feed-down contribution in the fit instead of the convolution of  Equation \ref{convolution}. The function parameters were initialized fitting the impact parameter distribution for reconstructed feed-down \dplusm candidates, as shown in Figure \ref{PlotFDReco2}
\end{itemize}
Figure \ref{SideBandSubSys} (left)   shows the ratio of the \fprompt values obtained with the variations described above and the central values shown in Figure \ref{FitParamFD} (right).  The systematics were assigned by the best estimate of the R.M.S. of the results as 10\% for 3<\pt<12 GeV/c and 15\% for \pt>12 GeV/c and \pt<3 GeV/c.
\subsection{Systematic uncertainty on  sideband subtraction}
\begin{figure}
 \includegraphics[width=0.99\textwidth]{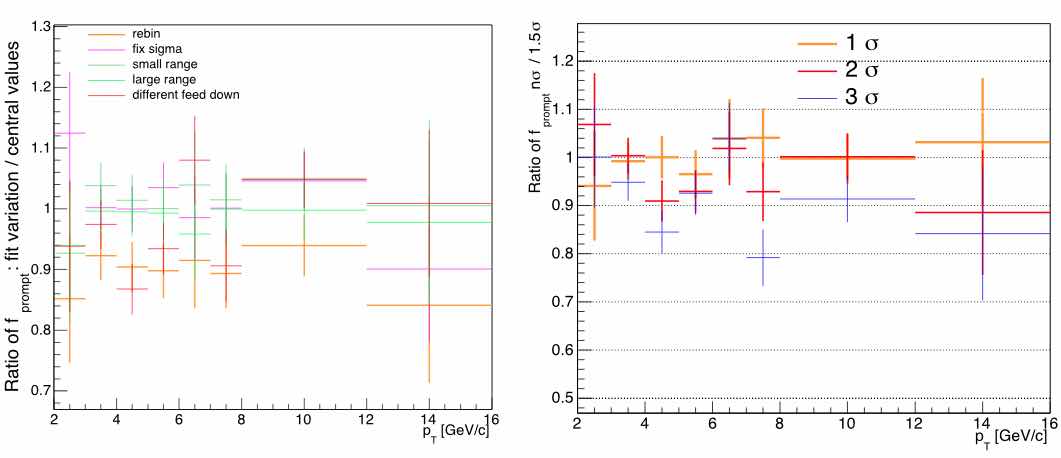} 
\caption{Left: ratio of the \fprompt values obtained varying the fit conditions and the central values. Right: ratio of \fprompt  from the fit performed to \dplus mesons in the invariant mass ranges within 1, 2 and 3 $\sigma^{\rm peak}$ from the \dplus peak and  the \fprompt central values (1.5 $\sigma^{\rm peak}$ from the \dplus peak).}
\label{SideBandSubSys}
\end{figure}
This systematic uncertainty due to the definition of the invariant mass regions from which the signal and background impact parameter are extracted was estimated by repeating the  fit described in Section \ref{ref:fit} changing the number of $\sigma$ from the peak in the invariant mass distribution in which the impact parameter distribution for signal is evaluated. The background distribution was proved not to change significantly in different invariant mass intervals of the sideband region. \\
Figure \ref{SideBandSubSys} shows the ratio of the  \fprompt values extracted   from  fits to impact parameter distributions of \dplus mesons measured within 1, 2 and 3 $\sigma^{\rm peak}$ from the invariant mass peak and the \fprompt value obtained with the standard fit performed in Section \ref{ref:fit} within 1.5 $\sigma^{\rm peak}$. The values of \fprompt obtained using candidates within 1 and 2 $\sigma^{\rm peak}$ seem to fluctuate around unity and do not show significant discrepancies from the central values. The values of \fprompt measured from \dplus candidates in a range within 3 $\sigma^{\rm peak}$  indicate a systematic shift to lower values of \fprompt for \pt > 4 GeV/c. For this reason, no systematic was assigned for \pt<4 GeV/c, while a systematic uncertainty of $_{-10}^{+0}$\% was assigned for \pt>4 GeV/c. 
\subsection{Systematic uncertainty on the p$_{\rm T}$ shape of generated D$^{+}$ mesons in MC}
\begin{figure}
 \includegraphics[width=0.99\textwidth]{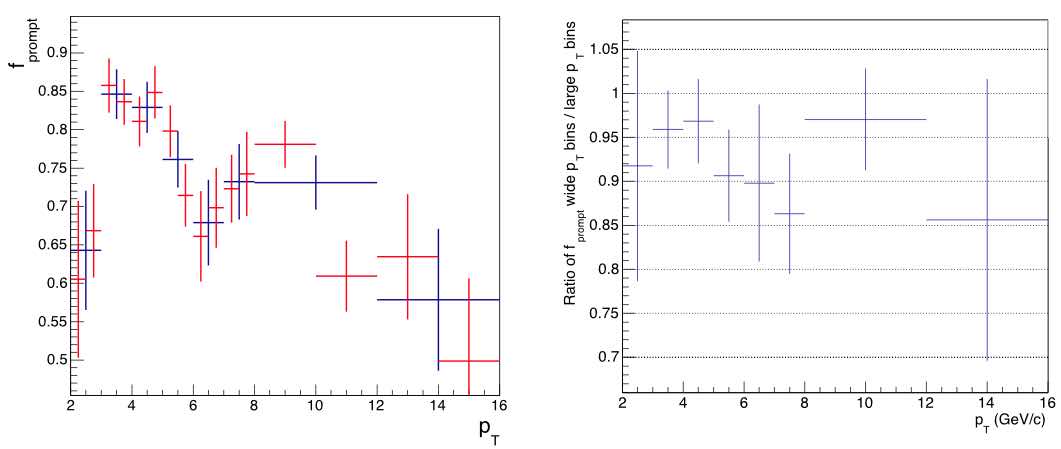} 
\caption{Left: \fprompt values obtained performing the analysis in finer \pt bins (red) and \fprompt values obtained in the standard \pt bins by combining the results in the finer bins (blue). Right: ratio of \fprompt values obtained combining the results in finer \pt bins to the central values}
\label{widefine}
\end{figure}
The fit relies on the MC templates shown in Section \ref{sec:distr}, which are obtained in \pt intervals of finite size. Since the impact parameter distributions vary as a function of \pt, a discrepancy between the \pt shape of real and simulated \dplus mesons would result in a biased shape of the impact parameter distributions of the  MC templates,  especially in large bins like 12<\pt<16 GeV/c.\\
To quantify this effect, the fit described in Section  \ref{ref:fit} was repeated in finer \pt intervals resulting in the \fprompt values shown by the red points in Figure \ref{widefine} (left). The values of \fprompt in the fine \pt intervals were then combined together  to obtain the \fprompt values in the wider \pt bins also shown in  Figure \ref{widefine}  (blue points - left). This was done by making a weighted average of the \fprompt values in the corresponding fine \pt bins, using as weight the amount of signal in each of them. \\
The right plot of Figure  \ref{widefine}  shows the ratio of \fprompt values obtained in this way to those obtained from the standard fit procedure (Figure \ref{FitParamFD}). Despite the fluctuations of this ratio, all values indicate a systematic shift to lower values of \fprompt using finer \pt bins. This shift to lower values of \fprompt is smaller ($^{+0}_{-4}$\%) for \pt<5 GeV/c, and increases  for \pt>5 GeV/c,  where a systematic of $^{+0}_{-10}$\% was assigned. 
\\
\\
The three sources of systematic uncertainties are considered uncorrelated and summed in quadrature, as reported in Table   \ref{SysDataDrive}.
\begin{table}[h]
\centering
\begin{tabular}{|c|c|c|c|c|c|c|c|c|c|}
\hline
 \pt (GeV/c) & [2,3] & [3,4] & [4,5] & [5,6]& [6,7] & [7,8] & [8,12] & [12,16] \\
\hline 
Fit method (\%) &  $^{+15}_{-15}$& $^{+10}_{-10}$&$^{+10}_{-10}$&$^{+10}_{-10}$&$^{+10}_{-10}$&$^{+10}_{-10}$&$^{+10}_{-10}$&$^{+15}_{-15}$\\
\hline
Sideband subtraction (\%) &  $^{+0}_{-0}$& $^{+0}_{-0}$&$^{+0}_{-10}$&$^{+0}_{-10}$&$^{+0}_{-10}$&$^{+0}_{-10}$&$^{+0}_{-10}$&$^{+0}_{-10}$\\
\hline
\pt shape (\%) &  $^{+0}_{-4}$& $^{+0}_{-4}$&$^{+0}_{-4}$&$^{+0}_{-10}$&$^{+0}_{-10}$&$^{+0}_{-10}$&$^{+0}_{-10}$&$^{+0}_{-10}$\\
\hline
Total (\%) &  $^{+15}_{-16}$& $^{+10}_{-11}$&$^{+10}_{-15}$&$^{+10}_{-17}$&$^{+10}_{-17}$&$^{+10}_{-17}$&$^{+10}_{-17}$&$^{+15}_{-21}$\\
\hline
\end{tabular}
\caption{Summary table of the systematic uncertainties  on the data-driven beauty feed-down analysis.}
  \label{SysDataDrive}
\end{table}
\section{Results}
\begin{figure}
\centering
 \includegraphics[width=0.8\textwidth]{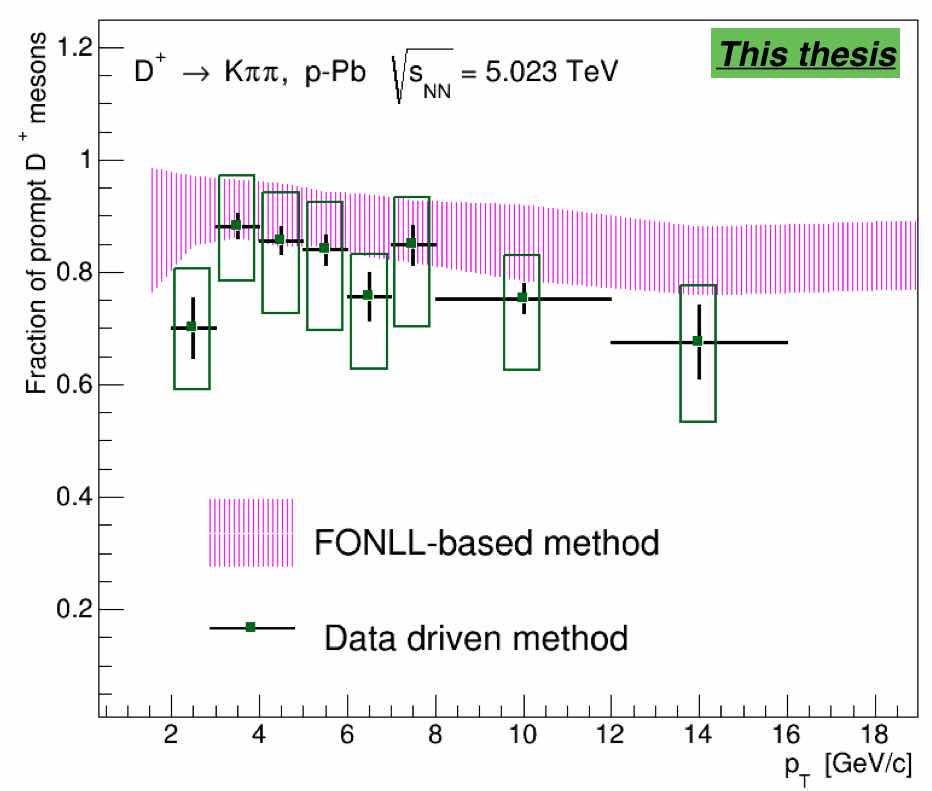} 
\caption{Results of \fprompt obtained from the data-driven approach. The solid bars represent the stastical errors, the green boxes the systematic errors. The FONLL predictions are represented by the pink band. }
\label{FinalPlotFPrompt}
\end{figure}
The results for \fprompt obtained with the data-driven approach described in the previous sections are shown in Figure \ref{FinalPlotFPrompt}. The error bars represent the statistical uncertainties, the green boxes the systematic ones. \\
The \fprompt values obtained from the fit are compared to the FONLL ones, which were used in the previous Chapter to subtract the feed-down contribution and are represented by the magenta band. This magenta band  includes the envelope of the systematic uncertinties on the \fprompt theory-driven calculation (pQCD FONLL parameters and feed-down \dplus energy loss hypothesis).\\
The two methods to determine \fprompt provide results which  are compatible within uncertainties in all  \pt intervals except in 2<\pt<3 GeV/c. \\
The total systematic uncertinties of both methods are compared in Table   \ref{SysDataDrive2}. The theory-driven method shows smaller uncertinties in the whole \pt range considered. \\
\begin{table}[h]
\centering
\begin{tabular}{|c|c|c|c|c|c|c|c|c|c|}
\hline
 \pt (GeV/c) & [2,3] & [3,4] & [4,5] & [5,6]& [6,7] & [7,8] & [8,12] & [12,16] \\
\hline 
Theory-driven         (\%)      &  $_{-11\%}^{+3\%}$& $_{-7\%}^{-3\%}$&$_{-5\%}^{+3\%}$&$_{-4\%}^{+3\%}$&$_{-4\%}^{+3\%}$&$^{+3\%}_{-3\%}$&$_{-3\%}^{+4\%}$&$_{-3\%}^{+5\%} $   \\
 \hline
Data-driven         (\%)  &  $^{+15}_{-16}$& $^{+10}_{-11}$&$^{+10}_{-15}$&$^{+10}_{-17}$&$^{+10}_{-17}$&$^{+10}_{-17}$&$^{+10}_{-17}$&$^{+15}_{-21}$\\
\hline
\end{tabular}
\caption{Summary table of the systematic uncertainties on the data-driven and FONLL driven  beauty feed-down analysis.}
  \label{SysDataDrive2}
\end{table}
\\
However, the higher statistics expected for LHC Run II will help reduce all sources of systematic uncertainties on the data-driven measurement for the following reasons:
\begin{itemize}
\item the analysis could be performed in narrower \pt intervals, expecially at high \pt, thus reducing the influence of the MC \pt shape of generated \dplus mesons on the templates
\item the larger number of \dplus mesons would result in a higher population in the regions of the sideband subtracted histogram at large values of impact parameter, which are crucial in the fit since they correspond to the intervals  dominated by feed-down \dplus mesons. This would give a tighter constraint to the fit and would reduce the systematic on the fit  method
\item a higher signal over background could be obtained since with a higher statistic tighter cuts could be used. This  would reduce the amount of background to be subtracted and would lead to smaller fluctuations in the sideband subtracted impact parameter distribution
\end{itemize}
If we then consider a slightly larger timescale ($\sim$ 2018), the  improvements expected from the upgrade of the ALICE Inner Tracking System will provide a substantial improvement in the tracking and vertexing performance of the ALICE detector, and consequently on the impact parameter which is crucial in the discrimination between the prompt and feed-down \dplus (and also D$^0$, D$^{*+}$...) mesons. This, together with the higher acquisition rate expected after the ALICE TPC upgrades, would reduce the error on the \fprompt measurement and could also allow a measurement of $f_{\rm feed-down}$ with reasonable error, that in turn can be used to compute the cross section of \dplus mesons from beauty decays.\\ 
\chapter{D$^{+}$-meson production as a function of multiplicity}
\lhead{Chapter 7. \emph{D$^{+}$-meson production as a function of multiplicity}} 
As discussed in the introduction of Chapter 5, some of the results obtained in p--Pb collisions at the LHC may indicate the presence of finale state effects: the ridge-like structure observed in the two-particle correlation function at high multiplicities \cite{RidgepPbCMS} may be the consequence of a hydrodynamic evolution after the collision, while the   $\Psi$' nuclear modification factor at forward and backward rapidities \cite{psiprimoLHC} revealed   a larger suppression of this meson species  with respect to  J/$\Psi$ suggesting the presence of a mechanism acting in the final state, after the formation of charmonia. In addition, also some of the initial state effects that are expected to modify charm quark production, such as nuclear shadowing and $k_{T}$ broadening, are predicted to depend on the geometry of the collision, which is usually characterized in terms of centrality or impact parameter. It is thus worth to study the dependence of \dplusm production on multiplicity and event activity (Section \ref{sub:biaspPb}), to verify if $c$-quark production and dynamic in p--Pb collisions is influenced by final state effects, such as  energy loss and hydrodinamic expansion, to assess the dependence of initial state effects on the collision geometry, and to study the role of  multiple hard partonic interactions occuring in a singe collision. \\ 
In the first part of this chapter I will show the results of \dplusm $Q_{\rm pPb}^{\rm mult}$ (as defined in Section \ref{sub:biaspPb}) as a function of event activity measured with the ZNA estimator.  This measurement could reveal a dependence of \dplusm production on the event activity  due to final state effects, like hydrodynamic flow or energy loss of $c$-quarks for high event activity values, in case the conditions to form a QGP are attained. Moreover, as discussed in Section \ref{sub:biaspPb}, several sources of bias arise when trying to define $\langle N_{\rm coll}\rangle$ in event activity classes  in p--Pb collisions  via multiplicity measurements. It is thus also interesting to study these biases using heavy flavour particles production, which is expected to scale with \ncoll. Therefore the \dplusm $Q_{\rm pPb}^{\rm V0A}$ and $Q_{\rm pPb}^{\rm CL1}$ will be reported. \\
In the second part of this chapter, I will present a study of \dplusm production as a function of  charged-particle mutiplicity. Besides the possible presence of hydrodynamic effects at high multiplicities, this measurement is useful to asses several QCD mechanisms that have been observed in pp collisions and that will be now briefly discussed.\\
As discussed in Chapter 2, pQCD models describe the final state particles produced in hadronic collisions  as the products of a hard partonic scattering process with large momentum transfer and an underlying event governed by  energy scales below those at which pQCD is applicable. The measurement of heavy flavour production as a function of multiplicity of charge particles provides insight into  the interplay between soft and hard mechanisms governing particle production in hadronic collisions, which is expected to depend on the centre of mass energy and on the impact parameter of the collision. The generalized Parton Distribution Functions, corresponding to the nucleon transverse partonic structure  measured at HERA via hard exclusive electroproduction of vector mesons and photoproduction of heavy quarkonia, show that gluons with 10$^{-4}$<$x$<10$^{-1}$ are localized at small transverse distances  (0.4-0.5 fm) from the centre of the nucleon while gluons with lower values of Bjorken $x$ occupy more distant regions. As shown in  \cite{gendist}, in pp scattering this  picture implies that hard processes mostly occur in central collisions, i.e. collisions with small impact parameter, where the areas occupied by partons in the relevant Bjorken $x$ range overlap. Peripheral collisions, that constitute the dominant part of the overall inelastic cross section, are characterized by softer scales of momentum transfer. \\
In 1988 the NA27 Collaboration observed that events with open charm production have on average a higher charged particle multiplicity associated to a softening of the momentum spectra of produced hadrons in pp collisions at \sqs= 28 GeV \cite{NA27}. At that  time the results were interpreted as due to the fact that pp collisions in which charm is produced are on average more central than minimum-bias ones.\\
At LHC energies, two additional contributions to the multiplicity dependence of charm production  in pp collisions have to be considered. The first is the larger amount of QCD radiation associated to the hard parton scattering processes leading to charm quark production. The second is the possible presence of Multi-Parton Interactions (MPI), i.e. several hard partonic interactions occurring in a single pp interaction. The presence of MPI constitutes a big difference with respect to the situation described in Figure \ref{CollSketch}, where one parton-parton scattering occurs and all other processes constitute the underlying event governed by soft energy scales.\\
\begin{figure}[t]
\centering
 \includegraphics[width=0.9\textwidth]{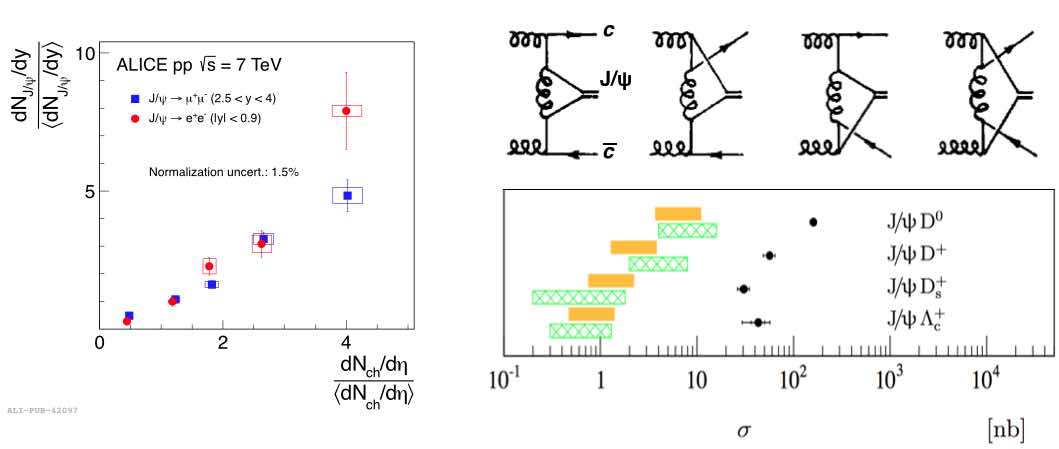}
\caption{Left: J/$\Psi$ yield as a function of charged particles multiplicity in pp collisions at \sqs= 7 TeV \cite{jpsimult}. Right: production cross section of events with  J/$\Psi$ + D mesons measured by the LHCb Collaboration, compared to predictions taking into account gluon fusion processes \cite{LHCbDC}.}
\label{JPMultLHCb}
\end{figure}
CMS measurements of jet and underlying event properties have shown better agreement with models including MPIs \cite{CMSMPI} while ALICE studies on minijets point to an increase of MPIs with increasing charged-particle multiplicity \cite{CMSMPI}. \\
If MPI were mainly affecting processes involving only light quarks and gluons, as implemented e.g. in PYTHIA 6.4, processes like J/$\Psi$ and open heavy flavour production should not be influenced and their rates would be independent of the overall event multiplicity. However  ALICE found an increase with approximately  linear trend of J/$\Psi$ yield as a function of multiplicity in pp collisions at \sqs= 7 TeV (left panel of Figure \ref{JPMultLHCb} \cite{jpsimult}) that is not reproduced by PYTHIA 6.4.  The LHCb Collaboration  measured the   production cross section of J/$\Psi$ mesons accompanied by open charm, and of pairs of open charm hadrons in pp collisions at \sqs= 7 TeV \cite{LHCbDC}. The cross sections are shown in Figure \ref{JPMultLHCb} (right) together with predictions (yellow and green bands) from theoretical calculations taking into account gluon fusion processes ($gg\rightarrow$J/$\Psi$+D) only. Some examples of the gluon fusion processes leading to the production of J/$\Psi$ mesons  and open charm are shown in  Figure \ref{JPMultLHCb}. These predictions clearly underestimate the cross section measurements. On the other hand models taking into account double parton scattering better agree with the data.\\
In pA collisions we can expect a significant enhancement of MPIs. This is due to the fact that the rates of multi-parton interactions depend both on the initial-state partonic multiplicities present in the colliding system and on the density of partons in transverse space, which is higher in nuclei.  However in p--Pb collisions events at high multiplicity can also be due to fluctuations in the number of binary collisions \ncoll. It is also worth  noting that in p--Pb  events at \sqnn= 5.02 TeV the highest multiplicity values obtained are similar to the ones of Pb--Pb peripheral collisions at \sqnn= 2.76 TeV, keeping in mind that, as discussed in Chapter 5, the presence of final state effect such as hydrodynamic expansion is not excluded in p--Pb collisions.\\
It is thus interesting to verify if D$^{+}$ meson production  depends on the overall event multiplicity in p--Pb collisions at \sqnn=5.02 TeV, and to compare these results  to those obtained in pp  collisions.
\section{D$^{+}$-meson $Q_{\rm pPb}$}
In this section I will discuss  the analysis aimed at the measurement of three different observables already shown and discussed in Section \ref{sub:biaspPb} for charged hadrons:
\begin{itemize}
\item  $Q_{\rm pPb}^{V0A}(\pteq)$, obtained in four different intervals of event activity defined as percentiles of the V0A amplitude distribution. In this measurement the $\langle N_{\rm coll}\rangle$ is estimated via a Glauber MC  fit to the V0A amplitude distribution, as discussed in Section \ref{sub:centpPb}
\item $Q_{\rm pPb}^{CL1}(\pteq)$, obtained in four different intervals of event activity defined as percentiles of the  distribution of the number of clusters in the outer layer of the SPD in the range |$\eta$|<1.4 (CL1). In this measurement the $\langle N_{\rm coll}\rangle$ is estimated via a Glauber MC  fit to the CL1 distribution
\item  $Q_{\rm pPb}^{\rm mult}(\pteq)$, obtained in four different intervals of event activity defined as percentiles of the ZNA energy deposition distribution. In this measurement the $\langle N_{\rm coll}\rangle$ is estimated with the hybrid method  discussed in Section \ref{sub:biaspPb}
\end{itemize}
Four event activity classes have been considered here, namely 0-20\%, 20-40\%, 40-60\% and 60-100\%. The first three classes (0-20\%,20-40\%, 40-60\%) contain about 20$\times$10$^{6}$ events each, while the multiplicity class 60-100\% contains about  40$\times$10$^{6}$ events.\\ 
The corrected per-event yields of prompt \dplus mesons in p--Pb collisions were obtained starting
from the raw yields $N^{D^{+/-}}_{\textrm{raw, mult. class}}$(\pt) as follows:
\begin{equation}
\frac{dN^{D^{+}}_{\rm mult. class, est.}}{d\pteq}=\frac{1}{2}\frac{1}{\Delta y\Delta\pteq}\frac{f^{\textrm{prompt}}_{\rm mult. class, est}\cdot N^{D^{+/-}}_{\textrm{raw, mult. class}}|_{|y|<y_{fid}}}{(\textrm{Acc}\times \epsilon)^{\textrm{prompt}}_{\rm mult class, est.}\cdot BR\cdot N_{\textrm{ev}}^{\textrm{mult. class. est}}}
\label{eq:corssp2}
\end{equation}
where  $\Delta y$ and $\Delta\pteq$ are the rapidity and \pt interval width, respectively; \fprompt is the fraction of prompt \dplus mesons, $(\textrm{Acc}\times \epsilon)^{\textrm{prompt}}_{\rm mult class, est.}$ is the acceptance and efficiency correction for prompt \dplus mesons in the respective event activity class, BR is the branching ratio of the D$^{+} \rightarrow K^{-}\pi^{+}\pi^{+}$  decay channel, the factor 1/2 accounts for the fact that we are measuring the raw yields for the sum of  \dplus and D$^{-}$, $N_{\rm ev}^{\textrm{mult. class. est}}$ is the number of analyzed events in the  event activity class estimated with the corresponding estimator, computed as reported in Equation \ref{eq:norm}. The \qppb values are obtained from the corrected per-event yields as:
\begin{equation}
Q_{\textrm{pPb}}^{\textrm{est.}}(\pteq, \textrm{mult. class})=\frac{\frac{dN^{D^{+}_{\rm mult. class, est.}}}{d\pteq}}{\langle T_{\rm pA}^{\rm mult. class, est.}\rangle d\sigma^{\rm pp}/d\pteq}
\label{eq:qppbequartion}
\end{equation}
where $d\sigma^{\rm pp}/d\pteq$ is the pp reference at \sqnn= 5.02 TeV and $\langle T_{\rm pA}^{\rm mult. class, est.}\rangle$ is the nuclear overlap function computed for the corresponding event activity class and estimator. 
\subsection{Signal Extraction}
\begin{table}[b]
\footnotesize
\centering
\begin{tabular}{|c|c|c|c|c|c|c|c|}
\hline
 \pt (GeV/c) & [1,2] & [2,4] & [4,6] & [6,8] & [8,12]& [12,16] & [16,24] \\
\hline 
 $|\Delta M_{D^{+}}|$ (GeV/$c^{2}$) & 0.2& 0.2&0.2&0.2&0.2&0.2 &0.2 \\
\hline
$\sigma_{vertex}$ (cm) & 0.03& 0.034 & 0.034 &0.03 & 0.02 & 0.015 & 0.025 \\
\hline
$p_{\rm T}^K$ (GeV/c) & 0.2 & 0.2 & 0.2 & 0.2 & 0.6 &0.8 &0.8  \\
\hline
$p_{\rm T}^\pi$ (GeV/c) & 0.2& 0.35 & 0.35 & 0.35 & 0.35 & 0.35 &0.35  \\
\hline
Decay Length  (cm) & 0 & 0.04 & 0.04 & 0.05 & 0.08 & 0.1 & 0.1  \\
\hline 
L$_{xy}$ & 9 & 8 & 8 & 8& 3 &3 & 0 \\
\hline 
$\cos(\theta_{pointing})$ & 0.99 & 0.99 & 0.99 & 0.99 & 0.98 & 0.98 & 0.98 \\
\hline 
$\cos(\theta_{pointing, xy})$ & 0.995 & 0.99 & 0.99 & 0.99 & 0.995 & 0.995 & 0.99  \\
\hline
\end{tabular}
\caption{Summary table of the D$^+$  cut values in the event activity classes 0-20\%, 20-40\% and 40-60\%.}
  \label{TableCutsDplus020}
\end{table}
\begin{table}[b]
\footnotesize
\centering
\begin{tabular}{|c|c|c|c|c|c|c|c|}
\hline
\pt (GeV/c) & [1,2] & [2,4] & [4,6] & [6,8] & [8,12]& [12,16] & [16,24] \\
\hline 
 $|\Delta M_{D^{+}}|$ (GeV/$c^{2}$) & 0.2& 0.2&0.2&0.2&0.2&0.2&0.2  \\
\hline
$\sigma_{vertex}$ (cm) & 0.03& 0.034 & 0.034 &0.03 & 0.02 & 0.015 & 0.025 \\
\hline
$p_{\rm T}^K$(GeV/c) & 0.2 & 0.2 & 0.2 & 0.2 & 0.6 &0.8 &0.8  \\
\hline
$p_{\rm T}^\pi$ (GeV/c) & 0.2& 0.35 & 0.35 & 0.35 & 0.35 & 0.35 &0.35  \\
\hline
Decay Length  (cm) & 0 & 0.04 & 0.04 & 0.05 & 0.08 & 0.1 & 0.1  \\
\hline 
L$_{xy}$ & 7 & 6 & 6 & 6& 1 &1 & 0 \\
\hline 
$\cos(\theta_{pointing})$ & 0.99 & 0.99 & 0.99 & 0.99 & 0.98 & 0.98 & 0.98 \\
\hline 
$\cos(\theta_{pointing, xy})$ & 0.99 & 0.985 & 0.985 & 0.985 & 0.99 & 0.99 & 0.988  \\
\hline
\end{tabular}
\caption{Summary table of the D$^+$  cut value in the event activity class 60-100\%.}
  \label{TableCutsDplus60100}
\end{table}
The analysis was performed using the p--Pb data sample collected in 2013 with a minimum-bias trigger (V0AND - Section \ref{sub:Trigger}).  Events were selected as described in Section \ref{sub:EvSel}, rejecting beam-gas interactions and events with pile-up, and keeping only events with |$z_{\rm vert}^{\rm reco}$|<10 cm. At this point events were divided in the four event activity classes according to V0A, ZNA or CL1 estimators.  The secondary vertices of D$^+$ meson candidates are reconstructed using ITS-TPC tracks selected as described in Section \ref{sub:TrackSel} and the same fiducial acceptance   selection on the rapidity of \dplusm candidates  described in Chapter 5 is applied,  ranging from |$y_{\rm lab}$|<0.5 at low \pt (1<\pt<2 GeV/c) to |$y_{\rm lab}$|<0.8 above 4 GeV/c. \\
\begin{figure}[t]
\centering
 \includegraphics[width=1.05\textwidth]{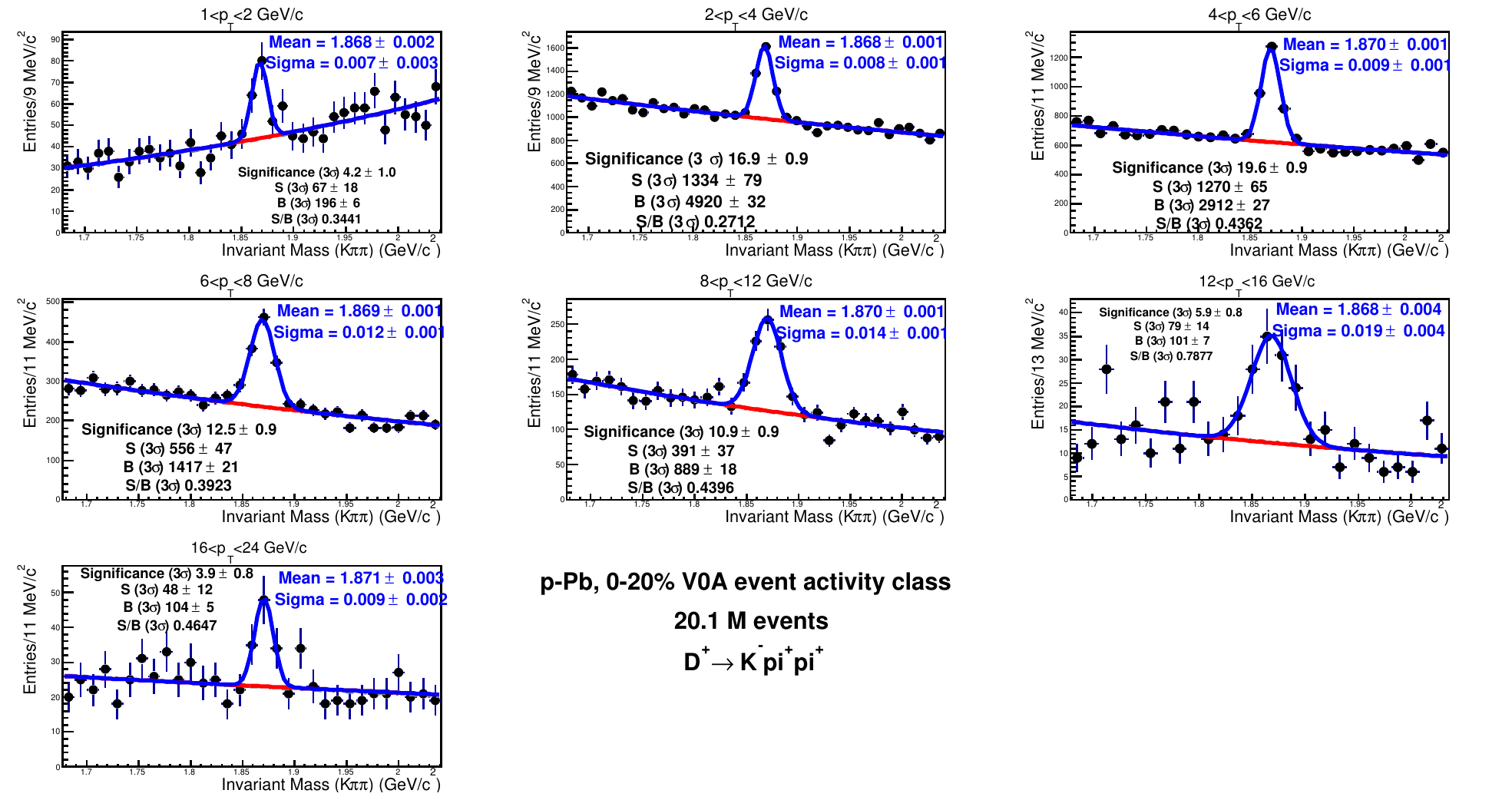}
\caption{\dplus candidates invariant mass distributions from  p--Pb collisions at \sqnn= 5.02 TeV  in seven \pt intervals within the range 1<\pt<24 GeV/c,  for events in the 0-20\% event activity class determined with the V0A estimator.}
\label{V0A020}
\end{figure}
\begin{figure}[b]
\centering
 \includegraphics[width=0.99\textwidth]{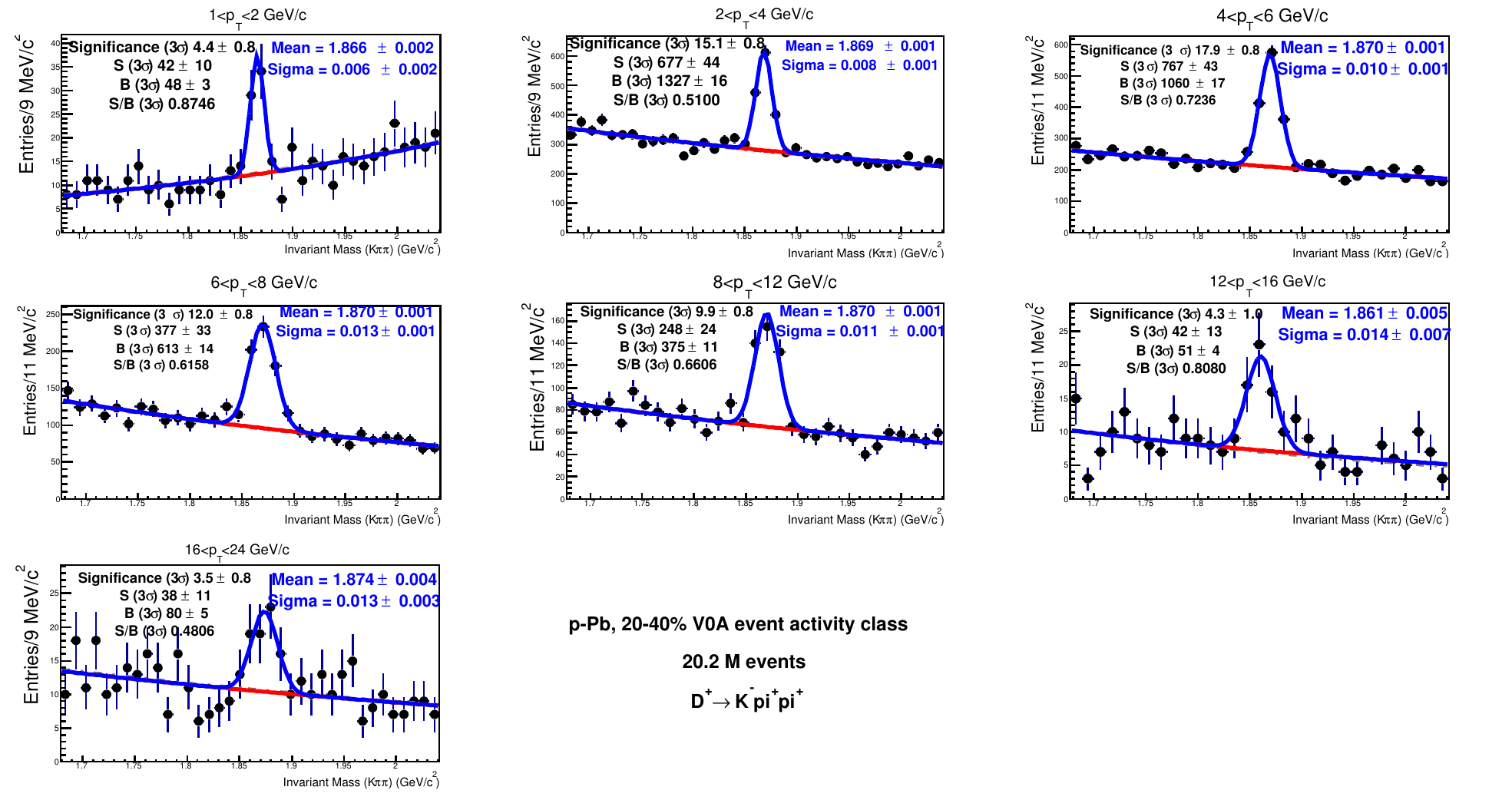}
\caption{\dplus candidates invariant mass distributions from  p--Pb collisions at \sqnn= 5.02 TeV  in seven \pt bins  within the range 1<\pt<24 GeV/c,  for events in the 20-40\% event activity class determined with the V0A estimator.}
\label{V0A2040}
\end{figure}
At this point a cut optimization was performed in seven \pt intervals ranging from 1 to 24 GeV/c, resulting in the cut values of Table   \ref{TableCutsDplus020}, that also reports the \pt intervals in which the analysis is performed. The same cut values were used in the 0-20\%, 20-40\% and 40-60\% event activity classes for all estimators, while in the 60-100\% (Table  \ref{TableCutsDplus60100}) event activity class the cut values on both $L_{\rm xy}$ and $\cos{\theta^{\rm pointing}_{\rm xy}}$ were made looser since this event activity class corresponds on average to a lower event multiplicity and consiquently to a lower combinatorial background. \\
The invariant mass distributions obtained for the V0A estimator in the four event activity classes are shown in Figures \ref{V0A020}, \ref{V0A2040}, \ref{V0A4060} and \ref{V0A60100}, while Table \ref{SignalQpPb} shows the  extracted raw yields for all estimators and event activity classes. Due to the limited statistics the signal could not be extracted in the entire \pt range 1<\pt<24 GeV/c for all event activity classes, in particular at low/high \pt and in the 60-100\% event activity class.
\begin{figure}[t]
\centering
 \includegraphics[width=0.99\textwidth]{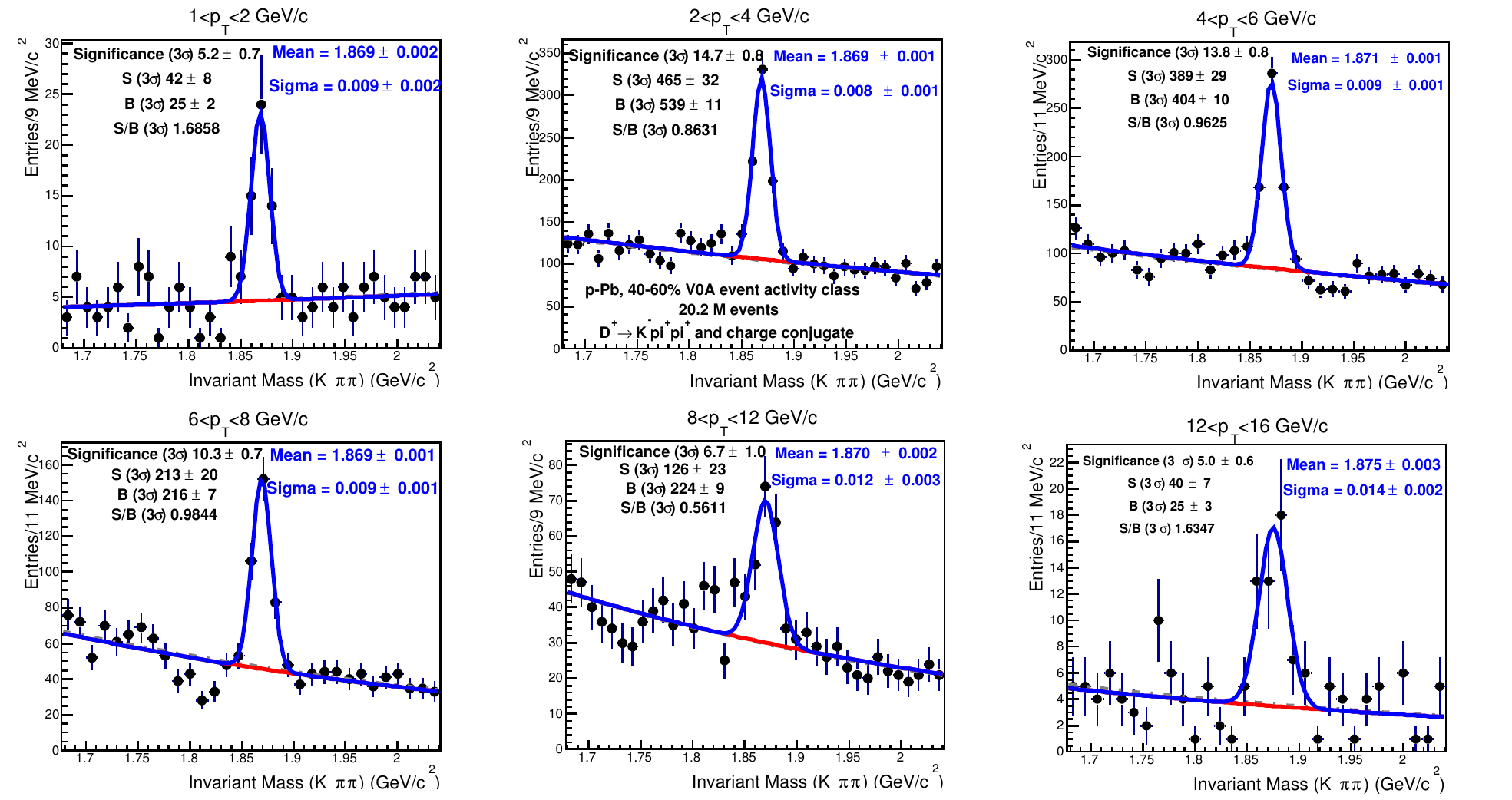}
\caption{\dplus candidates invariant mass distributions from  p--Pb collisions at \sqnn= 5.02 TeV  in six \pt bins  within the range 1<\pt<16 GeV/c,  for events in the 40-60\% event activity class determined with the V0A estimator.}
\label{V0A4060}
\end{figure}
\begin{figure}[b]
\centering
 \includegraphics[width=0.99\textwidth]{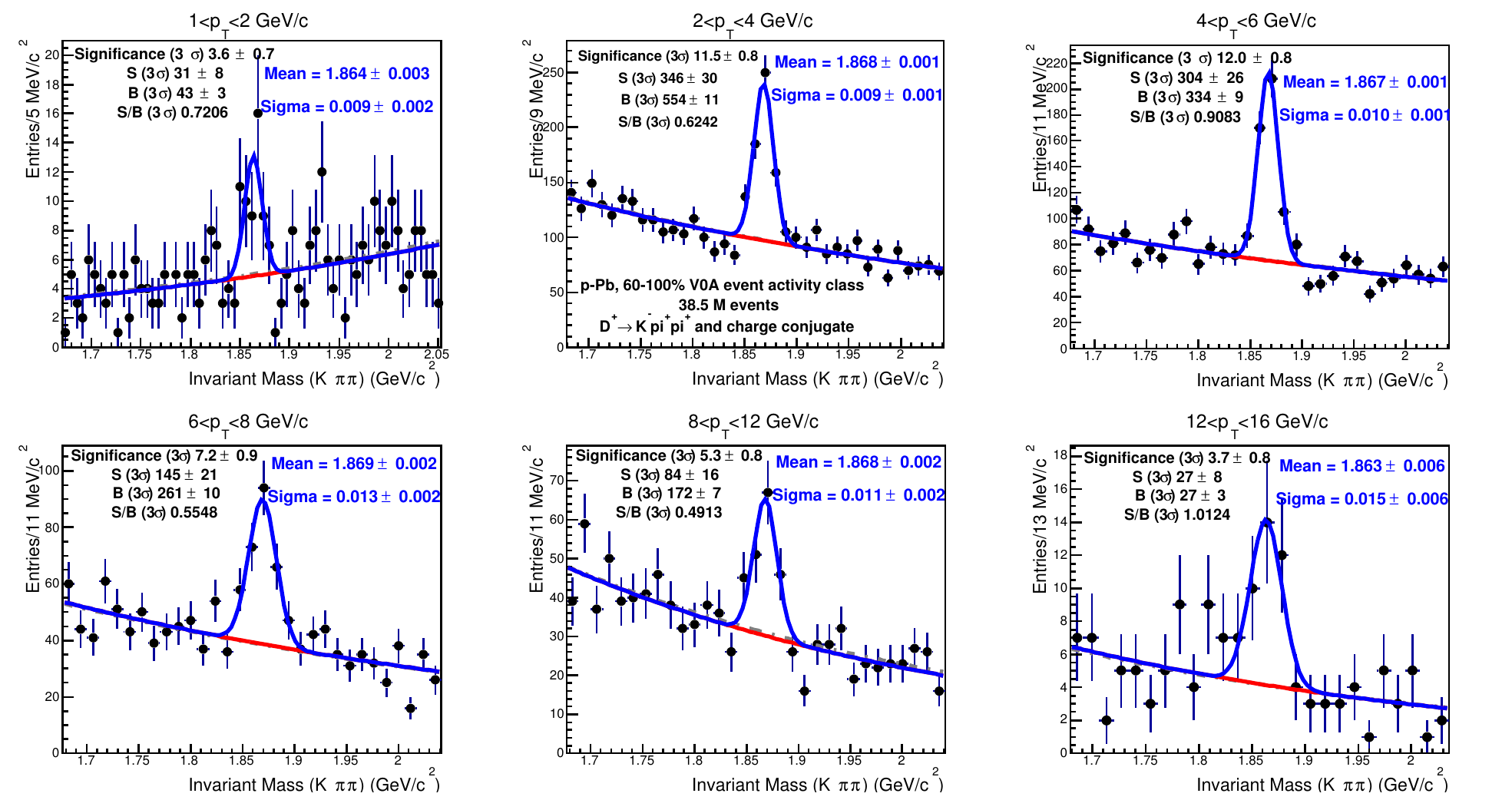}
\caption{\dplus candidates invariant mass distributions from  p--Pb collisions at \sqnn= 5.02 TeV  in six \pt bins  within the range 1<\pt<16 GeV/c,  for events in the 60-100\% event activity class determined with the V0A estimator.}
\label{V0A60100}
\end{figure}
\begin{table}[t]
\small
\begin{tabular}{|c|c|c|c|c|c|c|c|}
\hline
p$_{\rm T}$ (GeV/c) & 1-2       & 2-4         & 4-6         & 6-8        & 8-12       & 12-16     & 16-24     \\ \hline
Event Activity (\%) & \multicolumn{7}{c|}{V0A}                                                                \\ \hline
0-20                & 67$\pm$18 & 1334$\pm$79 & 1270$\pm$65 & 556$\pm$47 & 391$\pm$37 & 79$\pm$14 & 48$\pm$12 \\ \hline
20-40               & 42$\pm$10 & 677$\pm$44  & 767$\pm$43  & 377$\pm$33 & 248$\pm$24 & 42$\pm$13 & 38$\pm$11 \\ \hline
40-60               & 42$\pm$8  & 465$\pm$32  & 389$\pm$29  & 213$\pm$20 & 126$\pm$23 & 40$\pm$7  & X         \\ \hline
60-100              & 31$\pm$8  & 346$\pm$30  & 304$\pm$26  & 145$\pm$21 & 84$\pm$16  & 27$\pm$8  & X         \\ \hline
Event Activity (\%) & \multicolumn{7}{c|}{ZNA}                                                                \\ \hline
0-20                & 46$\pm$12 & 1020$\pm$65 & 969$\pm$56  & 501$\pm$41 & 272$\pm$30 & 68$\pm$14 & X         \\ \hline
20-40               & 58$\pm$12 & 769$\pm$55  & 874$\pm$49  & 336$\pm$33 & 272$\pm$29 & 57$\pm$11 & X         \\ \hline
40-60               & 27$\pm$9  & 557$\pm$42  & 485$\pm$34  & 253$\pm$25 & 165$\pm$23 & 33$\pm$8  & X         \\ \hline
60-100              & X         & 472$\pm$42  & 457$\pm$34  & 243$\pm$27 & 120$\pm$20 & 34$\pm$8  & 30$\pm$9  \\ \hline
Event Activity (\%) & \multicolumn{7}{c|}{CL1}                                                                \\ \hline
0-20                & 90$\pm$21 & 1515$\pm$83 & 1537$\pm$72 & 687$\pm$52 & 494$\pm$41 & 98$\pm$16 & 66$\pm$16 \\ \hline
20-40               & 40$\pm$8  & 724$\pm$44  & 720$\pm$38  & 345$\pm$28 & 218$\pm$24 & 41$\pm$10 & 45$\pm$12 \\ \hline
40-60               & 18$\pm$5  & 389$\pm$25  & 335$\pm$25  & 175$\pm$18 & 96$\pm$15  & 31$\pm$6  & X         \\ \hline
60-100              & 22$\pm$6  & 203$\pm$19  & 196$\pm$16  & 72$\pm$12  & 36$\pm$10  & X         & X         \\ \hline
\end{tabular}
\caption{\dplusm raw yields for all estimators and event activity classes}
\label{SignalQpPb}
\end{table}
\subsection{Acceptance and efficiency corrections}
The raw yields obtained in the previous section are corrected for the reconstruction and selection efficiency according to Equation \ref{eq:corrsp}. The \dplusm efficiencies as a function of \pt and $N_{\rm trk}$ were obtained from the   MC data sample generated using HIJING for simulating the underlying p--Pb event and PYTHIA v6.4.21 to injiect a pp collision with a $c\bar{c}$ pair, as described in Section \ref{sub:MCsample}.  The efficiency correction factor is obtained as the ratio of the \dplus mesons counted at the steps \textit{kStepRecoPID} and  \textit{kStepAcceptance}. The same reweighting on \pt shape of generated \dplus mesons discussed in Section \ref{AccEffRppB} were applied.  However in the Monte Carlo it was not possible to select events according to the  event activity classes defined for  V0A and ZNA since detectors at forward rapidity were not included in the simulation.
\begin{figure}[b]
\centering
 \includegraphics[width=0.6\textwidth]{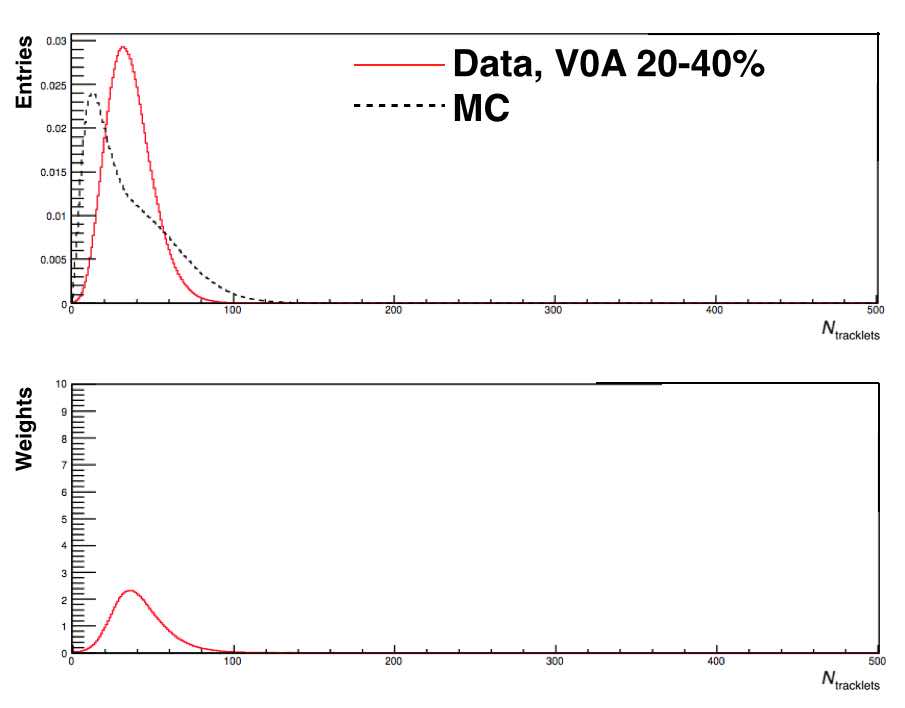}
\caption{Top: distribution of number of tracklets in |$\eta$| < 1 from MC (multiplicity integrated - black) and from data (red) for events in the V0A 20-40\% event activity class. Bottom:  ratio  of the black and red histograms shown in the top panel, which is used to reweight the efficiencies in the MC.}
\label{weightqppb}
\end{figure}
\begin{figure}[t]
\centering
 \includegraphics[width=0.99\textwidth]{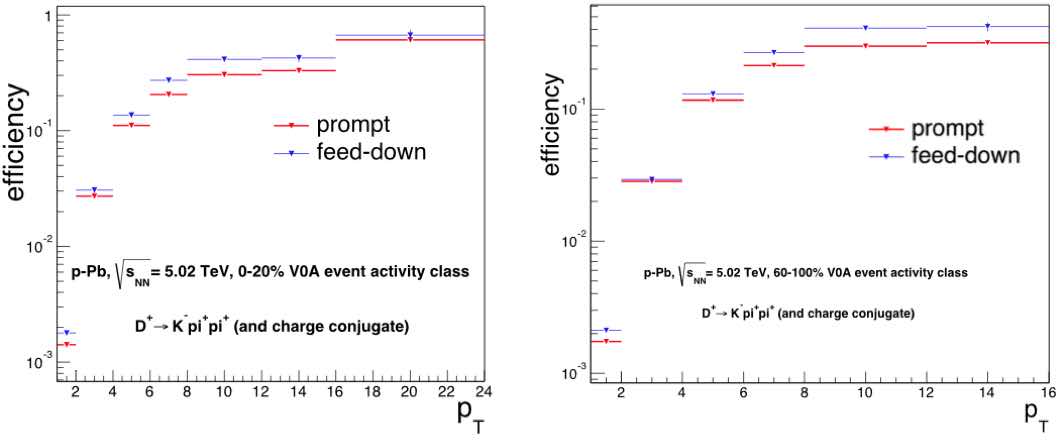}
\caption{\dplus reconstruction and  selection efficiencies  in  p--Pb collisions at \sqnn= 5.02 TeV  in seven \pt bins in the range  1<\pt<16 GeV/c,  for prompt and feed-down \dplus mesons  for events in the 0-20\% (left) and 60-100\% (right) event activity classes determined with the V0A estimator.}
\label{EffV0A}
\end{figure}
Therefore, to reproduce in the Monte Carlo the same multiplicity observed in data in each event activity class a weighting procedure was used. The distribution of the number of tracklets ($N_{\rm trk}$) measured in the data in the pseudorapidity range |$\eta$| < 1 in each event activity class (estimated with  V0A, ZNA and CL1) were used to calculate the weights. These distributions were divided by the  distribution of the number of tracklets in |$\eta$| < 1 extracted from the Monte Carlo to get the weights used to reweight the multiplicity value of each event. As an example, the top panel of Figure \ref{weightqppb} shows the $N_{\rm trk}$ distribution in |$\eta$| < 1 from MC (multiplicity integrated, black) and from data for events in the V0A 20-40\% event activity class (red). Their ratio, which is used to reweight the efficiencies for the 20-40\% V0A event class, is shown in the bottom panel. \\
The values of the efficiency for both prompt and feed-down \dplus mesons in the 0-20\% and 60-100\% V0A event activity classes obtained with this procedure are shown in Figure \ref{EffV0A} as a function of \pt. 
\subsection{Systematic Uncertainties}
\begin{table}[t]
\begin{center}ls
\begin{tabular}{|c|c|c|c|c|c|c|c|c|}
\hline
estimator & class & 1-2 & 2-4& 4-6 & 6-8& 8-12 & 12-16 & 16-24  \\
\hline
V0A & 0-20\%   &  8\%& 3\% & 3\% & 6\% & 6\% & 8\% & 8\% \\
V0A & 20-40\%   & 8\%& 3\% & 3\% & 6\% &6\% & 8\% & 8\% \\
V0A & 40-60\%   & 8\% & 3\% & 3\% & 6\% & 6\% & 8\% & X \\
V0A & 60-100\%   & 8\% & 3\% & 3\% & 6\% &  6\% & 8\% & X \\
\hline
CL1 & 0-20\%     &  8\%   & 3\% & 3\% & 6\% & 6\% & 8\% & 8\%\\
CL1 & 20-40\%   & 8\%  & 3\% & 3\% & 6\% & 6\% & 8\% & 8\%\\
CL1 & 40-60\%   &  8\%   & 3\% & 3\% & 6\% & 6\% & 8\% & X\\
CL1 & 60-100\% &  8\%   & 3\% & 3\% & 6\% & 6\% & X & X\\
\hline
ZNA & 0-20\%     &  8\%   & 3\% & 3\% & 6\% & 6\% & 8\% & X\\
ZNA & 20-40\%   & 8\%  & 3\% & 3\% & 6\% & 6\% & 8\% & X\\
ZNA & 40-60\%   &  8\%   & 3\% & 3\% & 6\% & 6\% & 8\% & X\\
ZNA & 60-100\% &  X  & 3\% & 3\% & 6\% & 6\% & 8\% & 8\%\\
\hline
\end{tabular}
\caption{Summary table of the \dplus  systematics from yield
  extraction for the different centrality classes defined with V0A, ZNA and CL1.}
\label{tab:Dplussyst}
\end{center}
\end{table}%
\begin{table}[b]
\begin{center}
\begin{tabular}{|c|c|c|c|c|c|c|c|c|}
\hline
estimator & class & 1-2 & 2-4& 4-6 & 6-8& 8-12 & 12-16 & 16-24  \\
\hline
V0A & 0-20\%   &  10\%& 5\% & 5\% & 5\% & 5\% & 5\% & 5\% \\
V0A & 20-40\%   & 10\%& 5\% & 5\% & 5\% &5\% & 5\%  & 5\%\\
V0A & 40-60\%   &  10\% & 5\% & 5\% & 5\% & 5\% & 5\% & X \\
V0A & 60-100\%   &  10\% & 5\% & 5\% & 5\% &  5\% & 5\% & X\\
\hline
ZNA & 0-20\%     &  10\% & 5\%   & 5\%   & 5\%   & 5\%   & 5\%   & 5\% \\
ZNA & 20-40\%   &  10\% & 5\%   & 5\%   & 5\%   & 5\%   & 5\%   & 5\% \\
ZNA & 40-60\%   &  10\% & 5\% & 5\% & 5\%& 5\% & 5\%   & X \\
ZNA & 60-100\% &  10\% & 5\% & 5\% & 5\% & 5\% & 5\% & X\\
\hline
CL1 & 0-20\%     &  10\% & 5\%   & 5\%   & 5\%   & 5\%   & 5\%   & X \\
CL1 & 20-40\%   &  10\% & 5\%   & 5\%   & 5\%   & 5\%   & 5\%   & X\\
CL1 & 40-60\%   &  10\% & 5\% & 5\% & 5\%& 5\% & 5\%   & X\\
CL1 & 60-100\% &  X & 5\% & 5\% & 5\% & 5\% & 5\% & 5\%\\
\hline
\end{tabular}
\caption{Summary table of the \dplus  systematics from cut
  efficiency for the different centrality classes defined with V0A, ZNA and CL1.}
\label{tab:Dplussyst_cutvariation}
\end{center}
\end{table}%
The sources of  systematic uncertainties are the same  discussed in Section \ref{sec:sysrppb}. \\
The systematic uncertainty on the yield extraction was evaluated using the same strategy described in Section  \ref{sub:systYield} in each of the \pt and event activity intervals, and its values are listed in Table \ref{tab:Dplussyst}. \\The systematic uncertainty on topological cut efficiency was evaluated following the same strategy described in Section \ref{sub:systCut}, i.e. comparing the corrected yields with those obtained using  two looser sets of cuts and two tighter sets of cuts. The values of the estimated uncertainties are reported in Table \ref{tab:Dplussyst_cutvariation} and are found to be in agreement with those assigned in the different \pt and event activity intervals to D$^0$ and D$^{*+}$ mesons. The uncertainty is larger in the \pt bin 1<\pt<2 GeV/c where tighter cuts have been used. \\ For what concerns PID, the systematic uncertainty was evaluated comparing the corrected yields obtained in the different event activity classes with and without PID selections. The situation is similar to the one obtained in the minimum bias analysis, and no systematic uncertainty is assigned for \pt>2 GeV/c. However, in the \pt bin 1<\pt<2 GeV/c it is impossible to extract \dplus signal without applying PID selections in the four event activity intervals for all estimators. The same systematic uncertainty from PID estimated in the minimum bias analysis (10\%) was assigned in each event activity class in the \pt bin 1<\pt<2 GeV/c.\\
 The systematic uncertainty due to  the \pt shape of generated \dplus mesons was evaluated following the same strategy described in Section \ref{sub:systMCPtShape}. The  assigned uncertainties   shown in Table \ref{sysptqppb} and are independent of the event activity estimator.  The uncertainties are  larger in the 60-100\% event activity class, since, despite the looser topological cut values,  the lower multiplicity implies worse resolution on cut variables such as decay length, and makes the efficiency values more sensitive to the \pt shape used to generate \dplus mesons. \\
\begin{table}[t]
\centering
\begin{tabular}{|c|c|c|c|c|c|c|c|}
\hline
p$_{\rm T}$ (GeV/c) & 1-2 & 2-4 & 4-6 & 6-8 & 8-12 & 12-16 & 16-24 \\ \hline
0-60\% sys. (\%)    & 3   & 2   & 1   & 1   & 1    & 1     & 1     \\ \hline
60-100\% sys. (\%)  & 10  & 6   & 3   & 2   & 2    & 2     & 2     \\ \hline
\end{tabular}
\caption{Systematic uncertainty on the \pt shape of generated \dplus mesons}
\label{sysptqppb}
\end{table}
The systematic uncertainty assigned due to the tracking efficiency is the same as in the minimum bias analysis, i.e. 9\% independently on \pt and avent acitivity class. \\
The uncertainties on the beauty feed-down subtraction and the nuclear modification of the feed-down \dplusm (\rppb of \dplus from B), evaluated as discussed in Section \ref{sub:fds}, are reported in Table \ref{sysfdqppb}. \\
The  pp reference used as the denominator in the \qppb formulas is the same that has been discussed in Section \ref{sub:sc502}, and is affected by the same systematic uncertainties reported in Table   \ref{SysPP}.\\ Since the \qppb is computed from Equations \ref{eq:corssp2} and \ref{eq:qppbequartion},  the systematic on the integrated  luminosity determination does not  affect the measurement. The systematic on the branching ratio is 2.1\%, while the systematic uncertainties on $\langle T_{\rm pPb}^{\rm est}\rangle$ depend on the event activity estimator and on the event activity class, and are reported in Tables \ref{systppb} and \ref{systppb2} together with the  $\langle T_{\rm pPb}^{\rm est.}\rangle$ values used in the next section to compute the \qppb values. 
\begin{table}[t]
\centering
\begin{tabular}{|c|c|c|c|c|c|c|c|}
\hline
p$_{\rm T}$ (GeV/c) & 1-2                             & 2-4                             & 4-6                            & 6-8                            & 8-12                           & 12-16                          & 16-24                          \\ \hline
Feed-down sys. (\%) &  $^{+1}_{-30}$ &  $^{+3}_{-10}$ &  $^{+3}_{-6}$ &  $^{+3}_{-4}$ &  $^{+3}_{-4}$ &  $^{+5}_{-3}$ &  $^{+4}_{-3}$ \\ \hline
\end{tabular}
\caption{Systematic uncertainty on beauty feed-down subtraction.}
\label{sysfdqppb}
\end{table}
\begin{table}[h]
\footnotesize
\centering
\begin{tabular}{|c|c|c|c|}
\hline
Event Activity (\%) & $\langle T_{\rm pPb}^{\rm V0A}\rangle$ (mb$^{-1}$) & $\langle T_{\rm pPb}^{\rm CL1}\rangle$ (mb$^{-1}$) & Unc.  V0A, CL1 (\%) \\ \hline
0-20                & 0.183                                              & 0.190                                              & 4                                           \\ \hline
20-40               & 0.134                                              & 0.136                                              & 3.7                                      \\ \hline
40-60               & 0.092                                              & 0.088                                              & 5.6                                       \\ \hline
60-100              & 0.041                                              & 0.037                                              & 22.95                                           \\ \hline
\end{tabular}
\caption{$\langle T_{\rm pPb}^{\rm V0A}\rangle$ and $\langle T_{\rm pPb}^{\rm CL1}\rangle$ values used for \qppb calculation and relative uncertainties.}
 \label{systppb}
\end{table}
\begin{table}[h]
\footnotesize
\centering
\begin{tabular}{|c|c|c|}
\hline
Event Activity (\%)& $\langle T_{\rm pPb}^{\rm mult}\rangle$ (mb$^{-1}$) & Unc  hybrid (\%) \\ \hline
0-20                       & 0.164                                               & 6.6                      \\ \hline
20-40                 & 0.137                                               & 3.9                      \\ \hline
40-60               &  0.101                                               & 5.9                      \\ \hline
60-100              &  0.046                                               & 6.34                     \\ \hline
\end{tabular}
\caption{$\langle T_{\rm pPb}^{\rm mult}\rangle$ values used for \qppb calculation and relative uncertainties.}
 \label{systppb2}
\end{table}
\subsection{Results}
\begin{figure}[b]
\centering
 \includegraphics[width=0.99\textwidth]{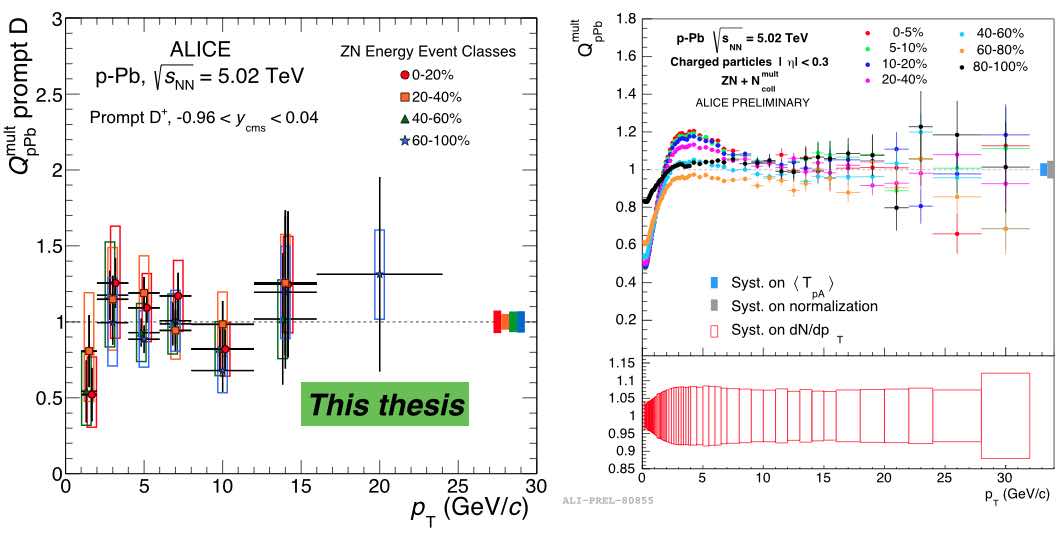}
\caption{Left: $Q_{\rm pPb}^{\rm mult}(\pteq)$ of \dplusm as a function of \pt in four ZNA event activity classes. Right:  $Q_{\rm pPb}^{\rm mult}(\pteq)$ of charged hadrons as a function of \pt in seven ZNA event activity classes}
\label{QpPbResZNA}
\end{figure}
The results of \dplusm   $Q_{\rm pPb}^{\rm mult}(\pteq)$, $Q_{\rm pPb}^{\rm V0A}(\pteq)$ and $Q_{\rm pPb}^{\rm CL1}(\pteq)$  in p--Pb collisions at \sqnn= 5.02 TeV are shown in the left panels of Figures \ref{QpPbResZNA},  \ref{QpPbResV0A} and  \ref{QpPbResCL1} as a function of \pt in the four event activity classes considered in this analysis. The vertical bars represent the statstical uncertainties, the colored boxes the systematic uncertainties. The uncertainty on $\langle T_{\rm pPb}\rangle$ is represented by the filled boxes at \qppb = 1 on the right of each figure. \\
The right panels of Figures \ref{QpPbResZNA},  \ref{QpPbResV0A} and  \ref{QpPbResCL1} report the \qppb values of charged hadrons measured with the same estimators used in the \dplusm analysis already discussed in Section \ref{sub:biaspPb}. The comparison of the \qppb results of \dplus mesons and charged hadrons lead to the following conclusions:
\begin{itemize}
\item for what concerns $Q_{\rm pPb}^{\rm mult}(\pteq)$, the \dplusm results are compatible with unity for  \pt>2 GeV/c. This observation is in agreement with what observed for charged hadrons for \pt>7$\div$8 GeV/c. In the intermediate \pt region (3<\pt<7 GeV/c) charged hadrons show a Cronin enhancement that increases with event activity, and might be due to $k_{\rm T}$ broadenind and/or radial flow effects. The current systematic uncertainties on the \dplusm measurement do not allow to conclude on the possible presence of these effects. It is worth noting that for the \dplus measurement the  \pt bin 1<\pt<2 GeV/c, where for the event activity classes 0-20\%, 40-60\% and 60-100\% the $Q_{\rm pPb}^{\rm mult}(\pteq)$ values are systematically lower than one, at mid-rapidity corresponds to a \pt region  in which Bjorken $x$ values of the order of 10$^{-3}$ are explored. 
\item for what concerns $Q_{\rm pPb}^{\rm V0A}(\pteq)$, the \dplusm results in the \pt range 2<\pt<12 GeV/c follow the same hierarchy with respect to the event activity class observed for charged hadrons, i.e. \qppb values from larger   event activity classes are higher than those from lower event activity class. The values of \dplusm \qppb for different event activity classes reduce their discrepancies for \pt> 12 GeV/c, as observed for charged hadrons.
\begin{figure}[t]
\centering
 \includegraphics[width=0.99\textwidth]{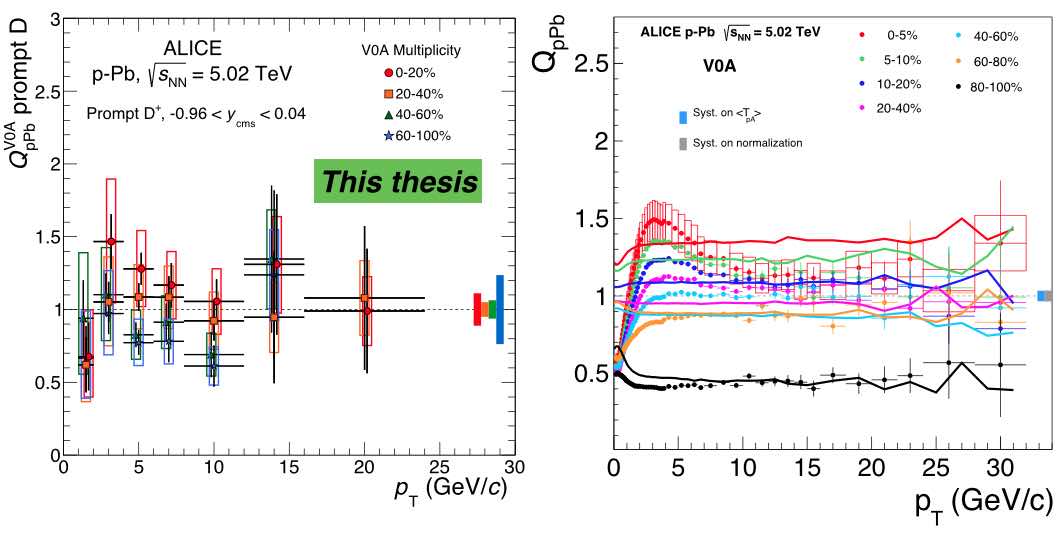}
\caption{Left: $Q_{\rm pPb}^{\rm V0A}(\pteq)$ of \dplusm as a function of \pt in four V0A event activity classes. Right:  $Q_{\rm pPb}^{\rm V0A}(\pteq)$ of charged hadrons as a function of \pt in seven V0A event activity classes}
\label{QpPbResV0A}
\end{figure}
\begin{figure}[t]
\centering
 \includegraphics[width=0.99\textwidth]{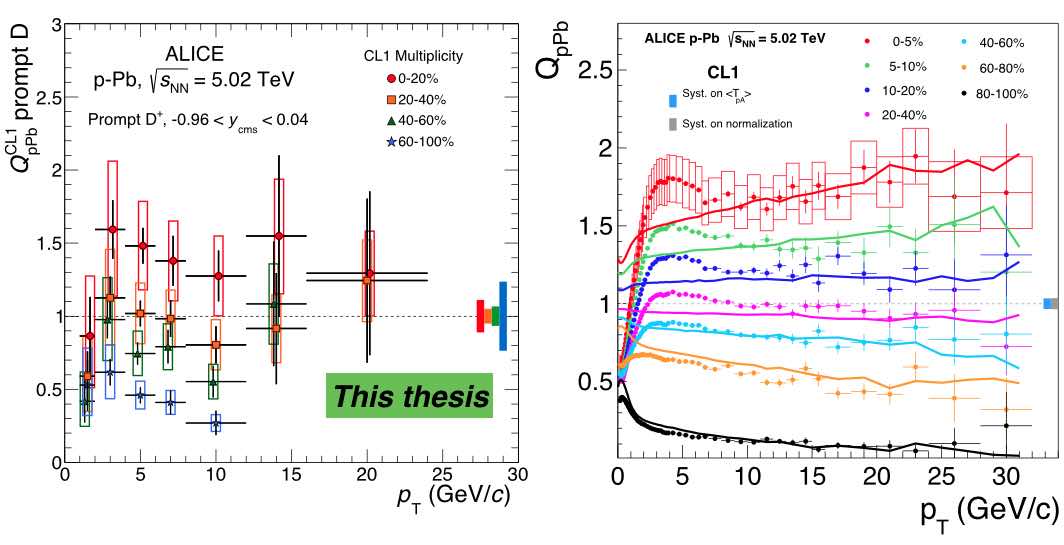}
\caption{Left: $Q_{\rm pPb}^{\rm CL1}(\pteq)$ of \dplusm as a function of \pt in four CL1 event activity classes. Right:  $Q_{\rm pPb}^{\rm CL1}(\pteq)$ of charged hadrons as a function of \pt in seven CL1 event activity classes}
\label{QpPbResCL1}
\end{figure}
\item for what concerns $Q_{\rm pPb}^{\rm CL1}(\pteq)$, the \dplusm results for \pt>2 GeV/c  follow the same hierarchy with respect to event activity class observed for  charged hadrons, i.e. \qppb values from higher event activity class are higher than those from lower event activity class. Focusing on the 0-20\% and 60-100\% event activity classes, in the first \dplusm \qppb is systematically higher than unity for \pt>2 GeV/c, in the latter it is systematically below unity for \pt>2 GeV/c
\end{itemize}
The spread among the \dplusm \qppb values  between high and low event activity classes is reduced when going from the mid-rapidity event activity estimator (CL1) to an event activity estimator located at higher rapidity (V0A), and it is further reduced using ZNA energy to estimate event activity and  the hybrid method for determining $\langle T_{\rm pPb}\rangle$ . We can conclude that the same  biases in the determination of  $\langle N_{\rm coll}\rangle$ and $\langle T_{\rm pPb}\rangle$ observed in the measurement of charged hadrons \qppb is present in measurements related to \dplus mesons. These conclusions are confirmed by the D$^{0}$-meson $Q_{\rm pPb}^{\rm mult}(\pteq)$, $Q_{\rm pPb}^{\rm V0A}(\pteq)$ and $Q_{\rm pPb}^{\rm CL1}(\pteq)$ measurements, shown in Figures \ref{QpPbD0} and \ref{QpPbResjpsi}.\\
For the least biased estimator, ZNA energy, the  $Q_{\rm pPb}^{\rm mult}(\pteq)$ values obtained calculating $\langle T_{\rm pPb}\rangle$ with the hybrid method are compatible with unity within the current uncertainties and no nuclear effects are observed.\\
\begin{figure}[t]
\centering
 \includegraphics[width=0.99\textwidth]{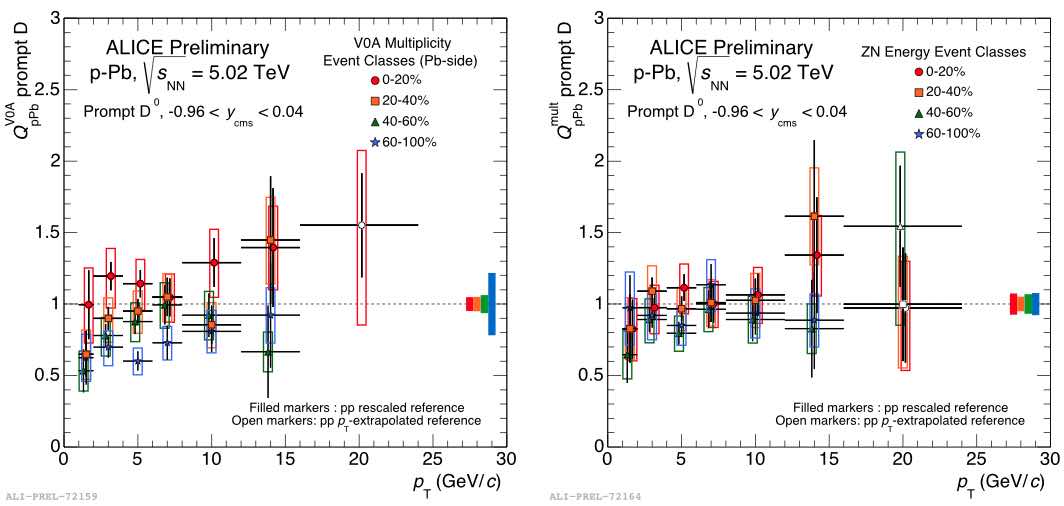}
\caption{Left: $Q_{\rm pPb}^{\rm V0A}(\pteq)$ of D$^0$ as a function of \pt in four V0A event activity classes. Right:  $Q_{\rm pPb}^{\rm ZNA}(\pteq)$ of D$^0$ as a function of \pt in four ZNA event activity classes.}
\label{QpPbD0}
\end{figure}
\begin{figure}[b]
\centering
 \includegraphics[width=0.99\textwidth]{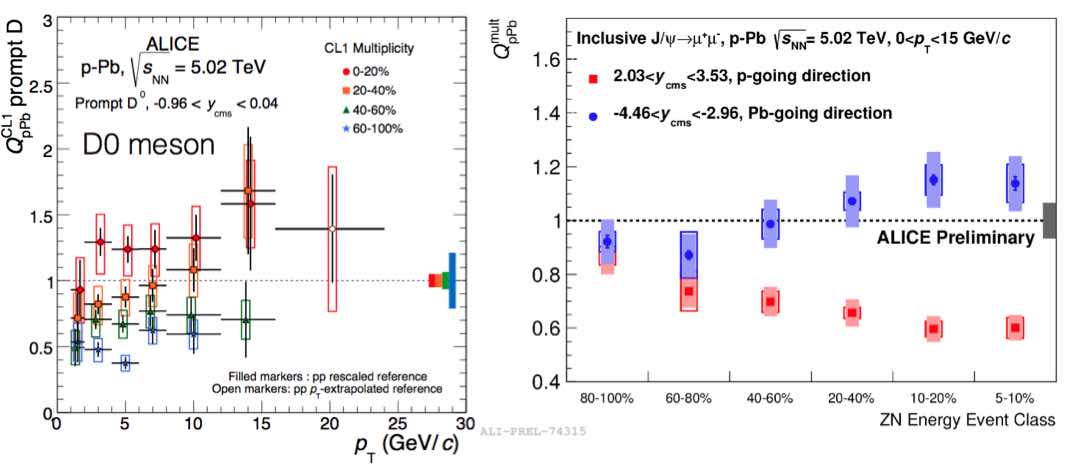}
\caption{Left: $Q_{\rm pPb}^{\rm CL1}(\pteq)$ of D$^0$ as a function of \pt in four CL1 event activity classes. Right: $Q_{\rm pPb}^{\rm mult}(\pteq)$ of J/$\Psi$ as a function of  ZNA event activity classes in the \pt range 0<\pt<15 GeV/c at forward (2.03<$y$<3.53) and backward (-4.46<$y$<-2.96) rapidity.}
\label{QpPbResjpsi}
\end{figure}
Further studies as a function of event activity  in the heavy-flavour sector are currently ongoing in the ALICE Collaboration, but their results are not public yet except for those relative to J/$\Psi$ production at forward and backward rapidity. For completeness these results will be now discussed, although they can not be directly compared to those relative to $D$ mesons because of the different rapidity coverage of the measurements. \\ 
I recall that the minimum bias nuclear modification factor of inclusive  $J/\Psi$ in p--Pb collisions at \sqnn = 5.02 TeV was measured by the ALICE collaboration (\cite{jpsipPb} - Section \ref{sub:pAresults})  at  forward rapidity (2.03<$y_{\rm cms}$< 3.53), corresponding to the p-going direction,  and negative rapidity (-4.46<$y_{\rm cms}$< -2.96) corresponding to the Pb-going direction. The results of this measurement showed an enhancement above unity of the nuclear modification factor of $J/\Psi$ at backward rapidity and a suppression below unity at forward rapidity, respectively, in agreement with models taking into account nuclear shadowing of PDFs and cold nuclear matter energy loss. \\
Figure \ref{QpPbResjpsi} shows the $Q_{\rm pPb}^{\rm mult}(\pteq)$ values of  inclusive  $J/\Psi$ at forward and backward rapidity as a function of the event activity class. At backward rapidity the  inclusive  $J/\Psi$ $Q_{\rm pPb}^{\rm mult}(\pteq)$ shows a  feeble dependence on event activity, with the higher  $Q_{\rm pPb}^{\rm mult}(\pteq)$ values in the 5-10\% event activity class. The $Q_{\rm pPb}^{\rm mult}(\pteq)$ of inclusive  $J/\Psi$  at forward rapidity shows a decreasing trend at backward rapidity, with all measurements below one. These results show  a stronger modification (with respect to pp collisions)  of $J/\Psi$ production at forward rapidity in the highest event activity classes, that might be due to a stronger influence of effects like nuclear shadowing of PDFs and cold nuclear matter energy loss.
\section{\dplusm  production as a function of charged particle multiplicity}
The  results will be presented as a function of relative primary charged particle multiplicity at central rapidity (|$\eta$|<1) defined as d$N_{\rm ch}/\textrm{d}\eta^{j}$/$\langle\textrm{d}N_{\rm ch}/\textrm{d}\eta\rangle$,  where $j$ is a given primary charged particle multiplicity interval, measured in |$\eta$|<1.0, and $\langle\textrm{d}N_{\rm ch}/\textrm{d}\eta\rangle$ is the average primary charged particle multiplicity of the  minimum bias data sample. Primary charged particles are defined as prompt particles produced in the collisions, including their decay products, except those from weak decays of strange particles.\\
The results are presented in form of the \dplusm self-normalized yield  in  inelastic p--Pb collisions, defined as
\vspace{-0.2cm}
\begin{equation}
\frac{(\rm d^2\textit{N}^{\rm D^{+}}/d\textit{y}d\textit{p}_{\rm T})^{j}}{\langle \rm d^2\textit{N}^{\rm D^{+}}/d\textit{y}d\textit{p}_{\rm T} \rangle}=\frac{Y^{j}/(\epsilon^{j})f^{\rm prompt}_{j}}{Y^{\rm tot}/(\epsilon^{\rm tot}\times \epsilon^{\rm trigger})f^{\rm prompt}_{\rm mult \ int}}
\label{eq:sfy}
\end{equation}
where $Y^{j}$ is the \dplusm  yield in the considered multiplicity interval, $Y^{\rm tot}$ the multiplicity integrated \dplus  yield, $\epsilon^{j}$ and $\epsilon^{\rm tot}$ are the corresponding reconstruction and selection efficiencies,  $f^{\rm prompt}_{j}$ ($f^{\rm prompt}_{\rm mult \ int}$) is the fraction of prompt \dplusm yield in the multiplicity interval considered (multiplicity integrated data sample) and $\epsilon^{\rm trigger}$ is the V0AND trigger efficiency for non-single diffractive events in p--Pb collisions at \sqnn= 5.02 TeV, which was measured to be 96.4\%, with a systematic uncertainty of 3.1\% \cite{triggermult}. 
\subsection{Charged particle multiplicity determination and correction}
\begin{table}[b]
\centering
\begin{tabular}{|c|c|c|}
\hline 
Period & $\langle N_\mathrm{trk}\rangle$ before correction& $\langle N_\mathrm{trk}\rangle$ after correction \\ [0.5 ex]
\hline
LHC13b &  28.29 & 27.87 \\
\hline
LHC13c & 27.87 & 27.87 \\ [1ex]
\hline
\end{tabular}
\caption{Mean multiplicity $\langle N_{\rm trk}\rangle$ as a function of $Z_\mathrm{vtx}$ before and after correction.}
\label{tab:multavg}
\end{table}
This analysis is performed on the p--Pb data sample at \sqnn= 5.02 TeV collected with the V0AND trigger, after applying the physics selection and pile-up rejection criteria described in Section \ref{sub:EvSel}. The whole p--Pb data sample is divided in two sub-samples: the first, LHC13b, collected between 16 and 22 January 2013, the second, LHC13c, collected between 22 and 25 January 2013.	The detector conditions were similar in this two periods.\\
\begin{figure}[t]
\centering
\includegraphics[width=0.7\textwidth]{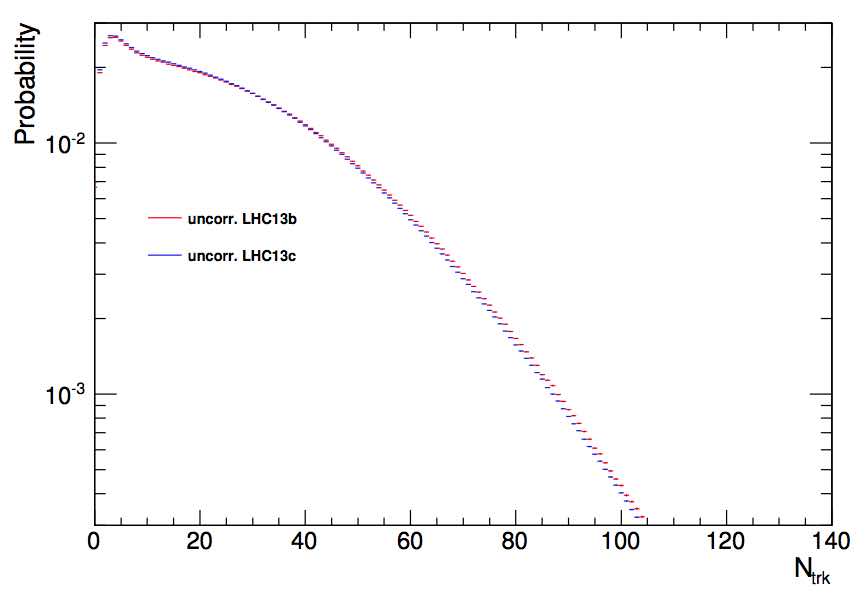}
\caption{Distribution of uncorrected number of tracklets  for LHC13b and LHC13c, normalised to their respective integrals.}
\label{fig:rawmult}
\end{figure}
\begin{figure}[b]
\centering
\includegraphics[width=0.8\columnwidth ]{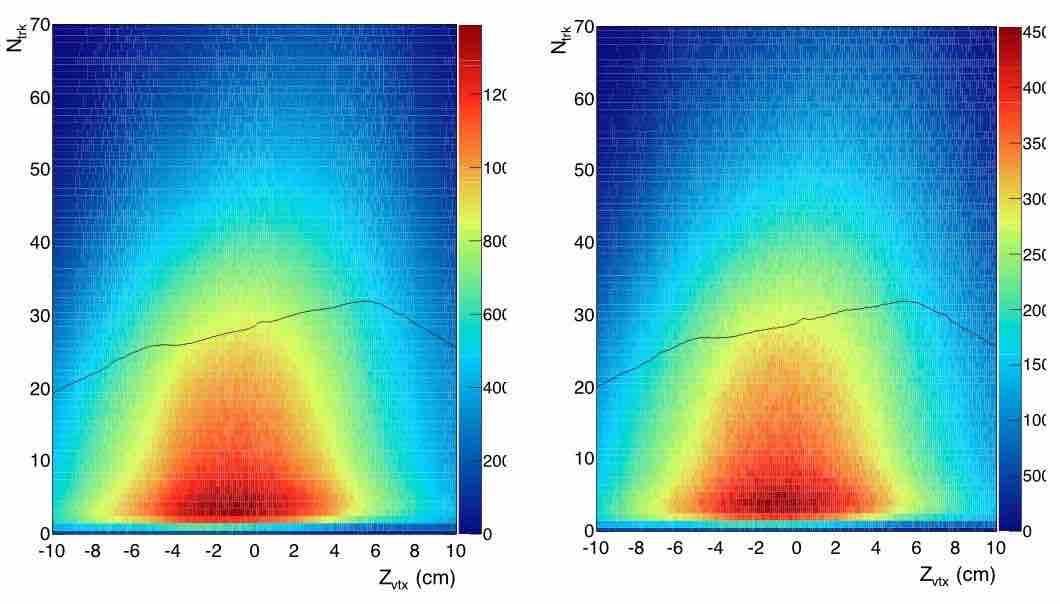}
\caption{ Distribution of number of events as a function of  $z_\mathrm{vtx}$ and of the  uncorrected number of tracklets $N_\mathrm{trk}$, for  LHC13b (left) and  LHC13c (right).}
\label{fig:rawmult2d}
\end{figure}
Only events with  vertex reconstructed with ITS-TPC tracks with $z_{\rm vtx}^{\rm Reco}$<10 cm are selected for the analysis. Expecially at low multiplicity, the inclusion of the \dplusm decay tracks introduces a bias in the calculation of the primary vertex position and covariance matrix. For this reason, in this analysis for each \dplusm candidate the primary vertex was recomputed excluding its decay tracks.\\
The multiplicity estimator used for this analysis is based on the number of tracklets reconstructed in the Silicon Pixel Detector (SPD) within a pseudorapidity range of $|\eta|$<1, $N_\mathrm{trk}$|$_{|\eta|<1}$ (in the following $N_\mathrm{trk}$ for brevity). An SPD tracklet is obtained by joining hits in the two SPD layers  aligned with the reconstructed primary vertex.
The measured  distributions of the number of tracklets ($N_{\rm trk}$) for each of the periods LHC13b and LHC13c are shown in Figure \ref{fig:rawmult}.  As can be seen in Table \ref{tab:multavg} the difference between the average multiplicities of the two periods before corrections is very small.\\
\begin{figure}[t]
\centering
\includegraphics[width=0.8\columnwidth ]{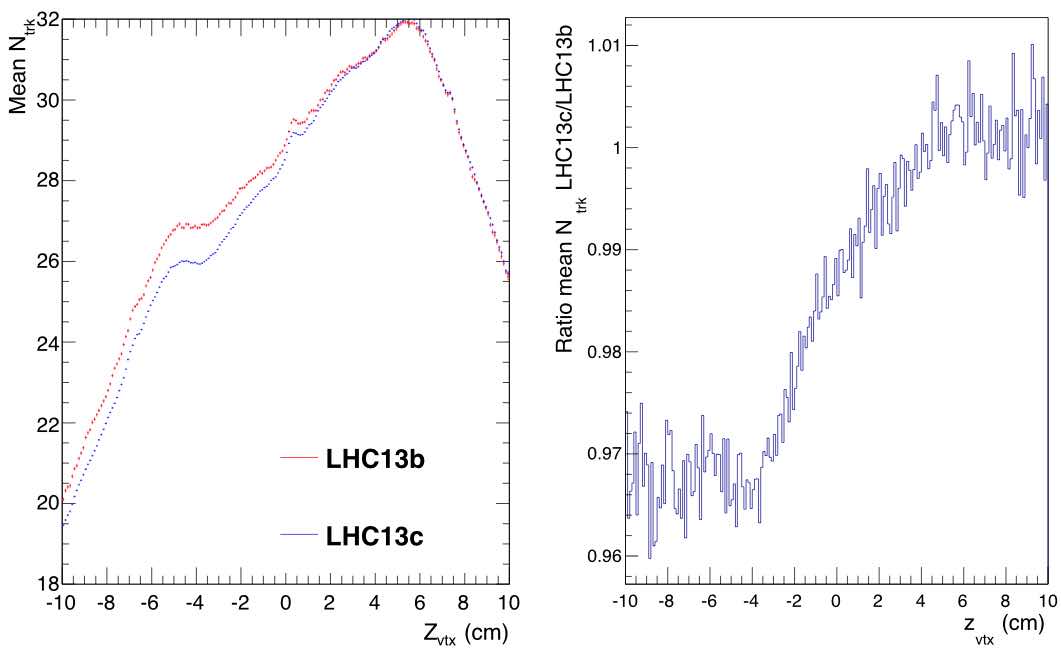}
\caption{Left: average uncorrected number of tracklets per event as a function of $z_{\rm vtx}$ (left). Right:  ratio of  the LHC13b profile over LHC13c profile.}
\label{fig:rawmultprofiles}
\end{figure}
\begin{figure}[b]
\centering
\includegraphics[width=0.8\columnwidth ]{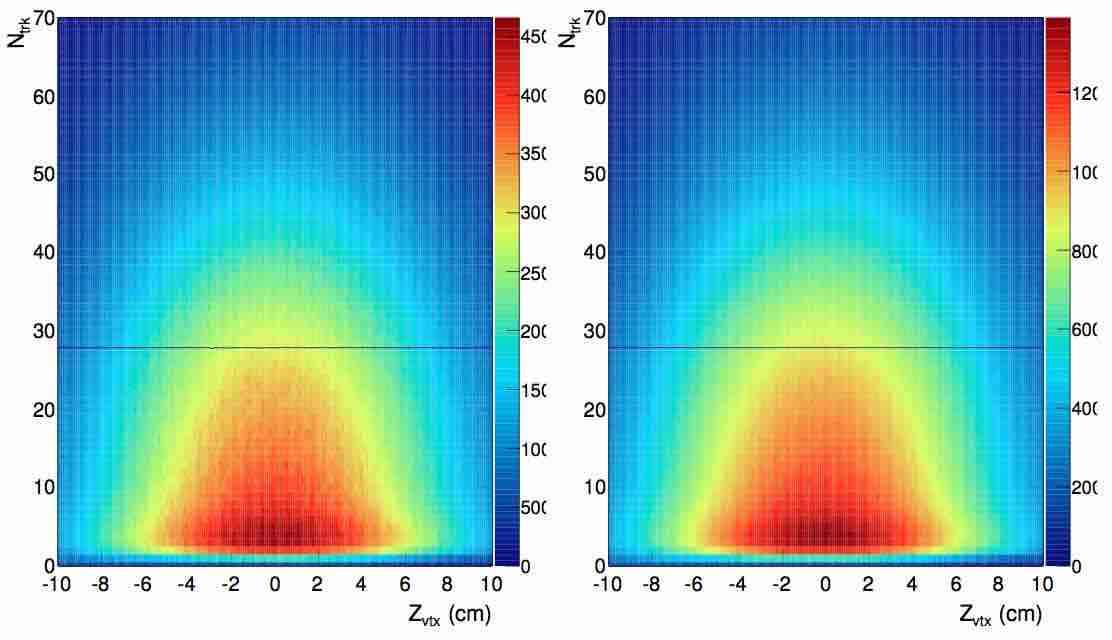}
\caption{Distribution of corrected $N_\mathrm{trk}$ as a function of $z_\mathrm{vtx}$, for LHC13b (left)  and LHC13c (right) periods.}
\label{fig:multcorr2d}
\end{figure}
In order to extract a correction for the number of tracklets that account for the $z_\mathrm{vtx}$ dependence of the SPD acceptance, the uncorrected number of tracklets was studied as a function of $z_\mathrm{vtx}$. This is shown in Figure \ref{fig:rawmult2d} for LHC13b (left) and LHC13c (right). In addition, the profile of these two-dimensional histograms, defined as the mean of $N_{\rm trk}$ as a function of $z_{\rm vtx}^{\rm Reco}$, is drawn in Figure \ref{fig:rawmult2d}.\\
\begin{figure}[t]
\centering
\includegraphics[width=0.6\columnwidth ]{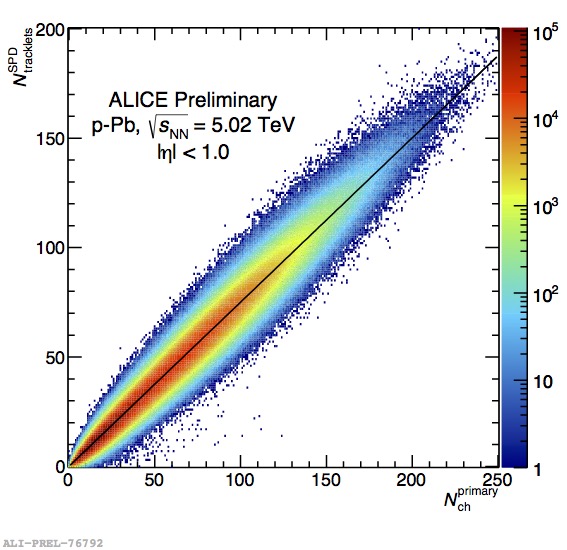}
\caption{Scatter plot of   number of tracklets $N_\mathrm{trk}$|$_{|\eta|<1}$, as a function of the generated primary charged particle multiplicity at mid-rapidity $N_\mathrm{ch}^\mathrm{primary}$|$_{|\eta|<1}$,  from MC simulations.}
\label{MCMultNch}
\end{figure}
\begin{table}[b]
\begin{center}
\makebox[\textwidth][c]{\begin{tabular}{|c|c|c|c|c|c|c|c|}
\hline
$N_{\rm trk}$ & $\langle{}N_\mathrm{trk}\rangle$ & ${\rm d}N_{\rm ch}/{\rm d}\eta$ & $ \langle {\rm d}N_{\rm ch}/{\rm d}\eta \rangle$ & 
	$\langle{\rm d}N_{\rm ch}/{\rm d}\eta\rangle \big/ \langle {\rm d}N_{\rm ch}/{\rm d}\eta \rangle^{mult}_{int}$ 
	& $N_{events} \cdot 10^6$  \\ \hline
$[1,24]$  		&12.11	& $0.7-16.4$	& 8.09	&  0.46		& 50.92		\\ \hline
$[25,44]$		&33.58	& $16.4-29.7$	& 22.43	&  1.27	 	& 27.23 		\\ \hline
$[45,59]$		&51.17	& $29.7-39.8$	& 34.19	&  1.94 	& 10.83 	\\ \hline
$[60,74]$ 	&65.95	& $39.8-49.8$	& 44.07	&  2.50	 	& 5.17 		\\ \hline
$[75,99]$ 	&83.54	& $49.8-66.5$	& 55.82	&  3.16	& 2.74 	 	\\  \hline
$[100,199]$	&110.75	& $66.5-133.0$ & 74.00	&  4.20		& 0.53		\\  \hline
Integrated 	&		&    			& 17.64 	& 					& 99.26 	 	\\ \hline
\end{tabular}}
\caption{Summary table of the intervals of   number of tracklets reconstructed in the SPD used in the analysis. The values of average number tracklets, $\langle N_\mathrm{trk}\rangle$, of ${\rm d}N_{\rm ch}/{\rm d}\eta$ and of  	${\rm d}N_{\rm ch}/{\rm d}\eta \big/ \langle {\rm d}N_{\rm ch}/{\rm d}\eta \rangle^{mult}_{int}$ for $|\eta_\mathrm{lab}|<1.0$ are reported for each interval.}
\label{tab:NchCorr}
	\end{center}
\end{table}
This dependence of the $N_{\rm trk}$ distributions on  $z_{\rm vtx}^{\rm Reco}$ and data taking period needs to be corrected in order to define consistently the $N_{\rm trk}$ intervals in which the analysis is performed, otherwise a given $N_{\rm trk}$ interval would correspond to a different real charged particle multiplicity range   depending on $z_{\rm vtx}^{\rm Reco}$ or data taking period.
A $z_{\rm vtx}^{\rm Reco}$-dependent correction factor was applied to make the $N_{\rm trk}$ distributions uniform in $z_{\rm vtx}^{\rm Reco}$.  The  reference multiplicity was set so that  the  mean of  the $N_{\rm trk}$ distribution is 27.87 (which is the  $z_{\rm vtx}^{\rm Reco}$-integrated   $N_{\rm trk}$ mean for period LHC13c, see Table \ref{tab:multavg}) for both data taking periods and for all $z_{\rm vtx}^{\rm Reco}$ values. Fig.~\ref{fig:multcorr2d} shows the corrected $z_\mathrm{vtx}$ distribution of $N_\mathrm{trk}$ for LHC13b (left) and LHC13c (right) periods.\\
Once this correction is done, the events of the p--Pb minimum bias data sample are divided in six $N_{\rm trk}$ intervals. The intervals were chosen in order to have  sufficient statistics for the \dplusm yield extraction, and are listed in the first column of Table \ref{tab:NchCorr}, while the second column reports the average $\langle N_\mathrm{trk}\rangle$ inside each interval.  \\
Figure \ref{MCMultNch} shows the scatter plot of   $N_\mathrm{trk}$|$_{|\eta|<1}$ as a function of the generated primary charged particle multiplicity at mid-rapidity $N_\mathrm{ch}^\mathrm{primary}$|$_{|\eta|<1}$ obtained from the MC data sample described in Section  \ref{sub:MCsample}.  A linear trend is observed between $N_\mathrm{trk}$|$_{|\eta|<1}$ and $N_\mathrm{ch}^\mathrm{primary}$|$_{|\eta|<1}$, with a factor of proportionality ($N_\mathrm{trk}/N_\mathrm{ch}^{\rm primary}$)|$_{|\eta|<1}$ of 0.75 obtained from a linear fit  to the average $N_\mathrm{trk}$ values as a function of $N_\mathrm{ch}^\mathrm{primary}$  represented by the black line. The $\langle N_\mathrm{trk}\rangle$ values were divided by this value to convert them into the number of generated primary charged particles $N_\mathrm{ch}$, and then by a  factor of two (the width of the pseudorapidity range studied) to give d$N_\mathrm{ch}$/d$\eta$. The third column of Table  \ref{tab:NchCorr}  shows the limits of the multiplicity intervals in terms of d$N_\mathrm{ch}$/d$\eta$, the fourth column the average d$N_\mathrm{ch}$/d$\eta$ within the respective multiplicity interval ($\langle$d$N_\mathrm{ch}$/d$\eta\rangle$) and the fifth column reports the ratio of $\langle$d$N_\mathrm{ch}$/d$\eta\rangle$ to the  average d$N_\mathrm{ch}$/d$\eta$  of the minimum bias data sample, $\langle\textrm{d}N_{\rm ch}/\textrm{d}\eta\rangle^{mult}_{int}$. The last column reports the number of events in each multiplicity class. 
\begin{table}[h]
\footnotesize
\centering
\begin{tabular}{|c|c|c|c|c|c|c|c|}
\hline
\pt (GeV/c) & [1,2] & [2,4] & [4,8] & [8,12] & [12,24]\\
\hline 
 $|\Delta M_{D^{+}}|$ (GeV/$c^{2}$) & 0.2& 0.2&0.2&0.2&0.2 \\
\hline
$\sigma_{vertex}$ (cm) & 0.03& 0.03 & 0.03 &0.03 & 0.03 \\
\hline
$p_{\rm T}^K$ (GeV/c) & 0.2 & 0.2 & 0.2 & 0.2 & 0.2   \\
\hline
$p_{\rm T}^\pi$ (GeV/c) & 0.2& 0.35 & 0.35 & 0.35 & 0.35 \\
\hline
Decay Length  (cm) & 0 & 0.0 & 0.0 & 0.0 & 0.0 \\
\hline 
L$_{xy}$ & 9 & 9 & 9 & 9& 9 \\
\hline 
$\cos(\theta_{pointing})$ & 0.99 & 0.99 & 0.99 & 0.99 & 0.99  \\
\hline 
$\cos(\theta_{pointing, xy})$ & 0.995 & 0.995 & 0.995 & 0.995 & 0.995  \\
\hline
\end{tabular}
\caption{Summary table of the cut values in D$^+$ vs multiplicity analysis.}
  \label{TableCutsDplusMult}
\end{table}
\begin{figure}[b]
\centering
 \includegraphics[width=0.8\textwidth]{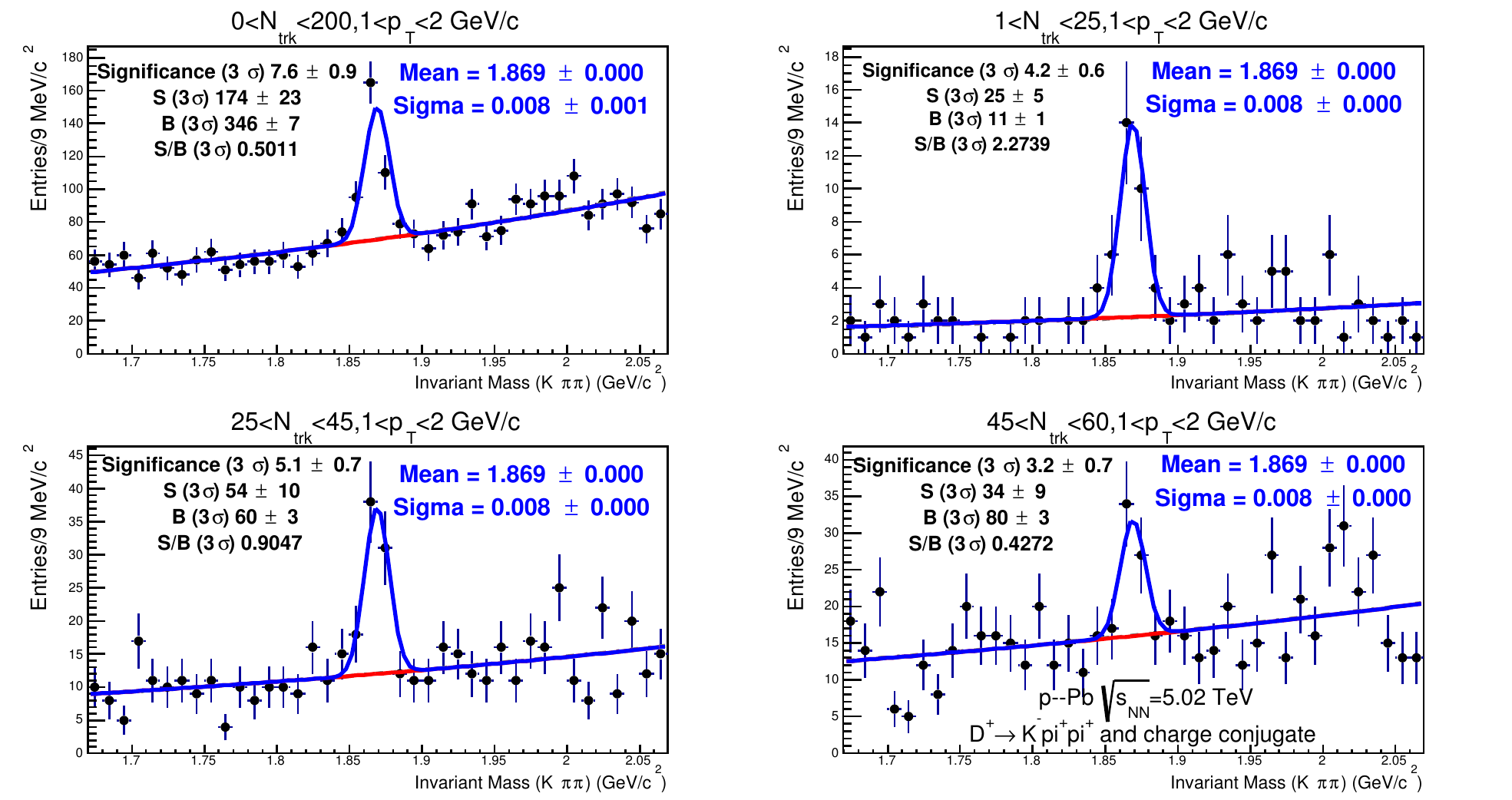}
\caption{\dplus candidate invariant mass distributions from  p--Pb collisions at \sqnn= 5.02 TeV in the \pt bin 1<\pt<2 GeV/c  integrated over multiplicity and in three multiplicity intervals ranging form 1 to 60 $N_{\rm trk}$.}
\label{12Tesi}
\end{figure}
\subsection{Signal Extraction}
 The decay vertices of D$^+$ meson candidates are reconstructed using ITS-TPC tracks selected as described in Section \ref{sub:TrackSel}. The same  selection on \dplusm candidates rapidity,  ranging from |$y_{\rm lab}$|<0.5 at low \pt (1<\pt<2) to |$y_{\rm lab}$|<0.8 above 4 GeV/c, is applied.\\
\begin{figure}[t]
\centering
 \includegraphics[width=0.99\textwidth]{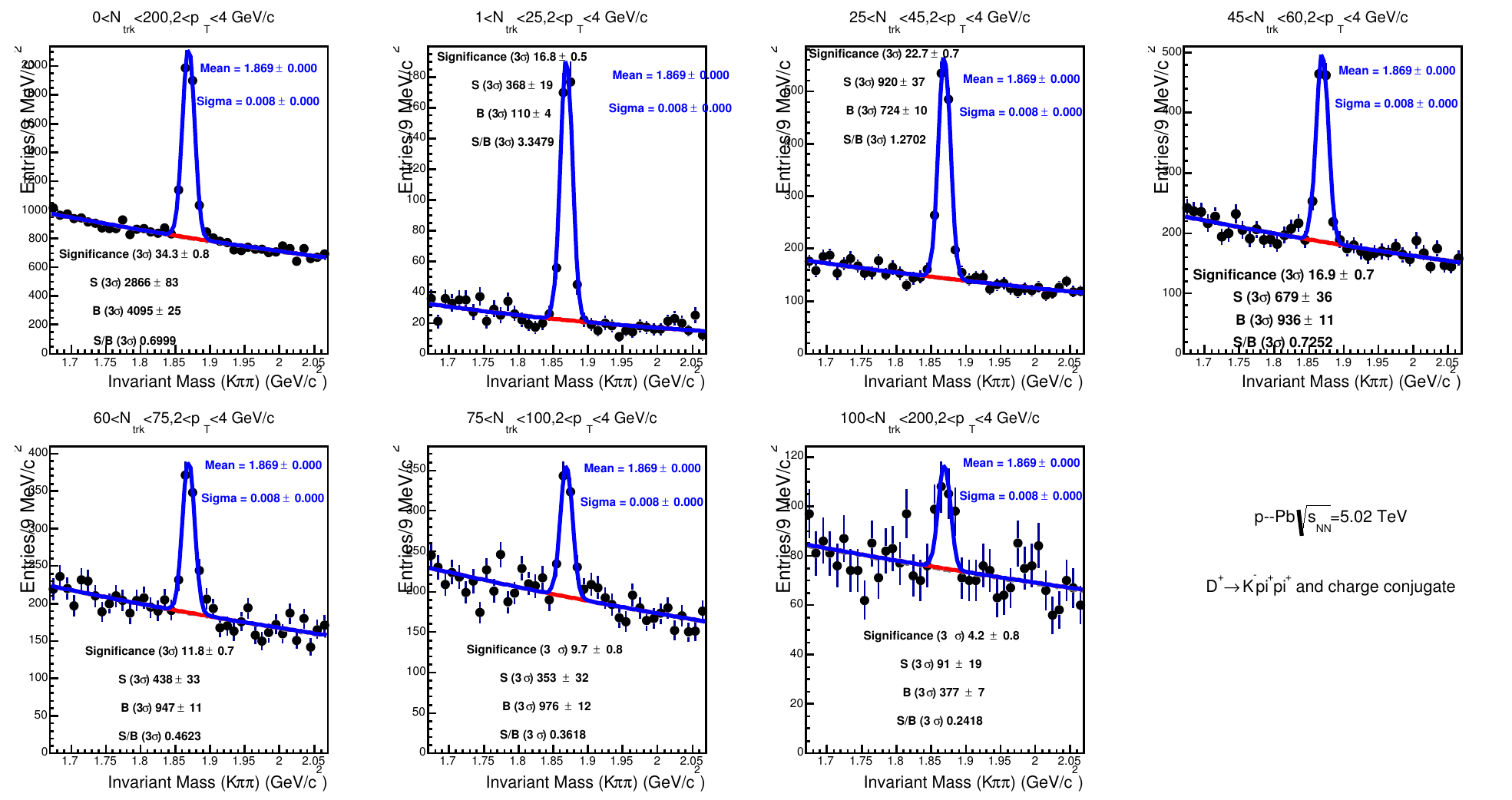}
\caption{\dplus candidate invariant mass distributions from  p--Pb collisions at \sqnn= 5.02 TeV in the \pt bin 2<\pt<4 GeV/c  integrated over multiplicity and in six multiplicity intervals ranging form 1 to 200 $N_{\rm trk}$.}
\label{24Tesi}
\end{figure}
\begin{figure}[b]
\centering
 \includegraphics[width=0.99\textwidth]{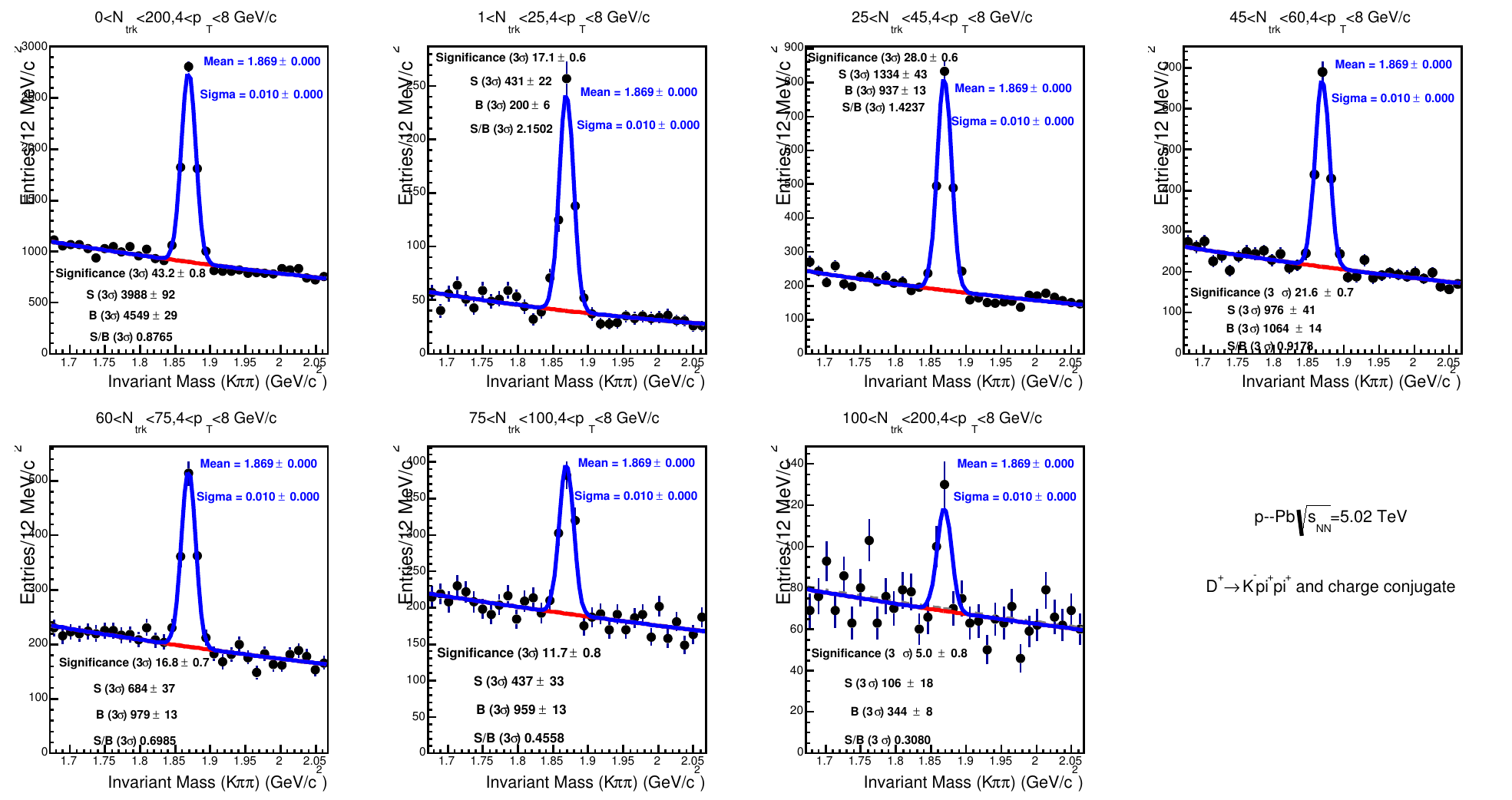}
\caption{\dplus candidate invariant mass distributions from  p--Pb collisions at \sqnn= 5.02 TeV in the \pt bin 4<\pt<8 GeV/c  integrated over multiplicity and in six multiplicity intervals ranging form 1 to 200 $N_{\rm trk}$.}
\label{48Tesi}
\end{figure}
The values of the topological cuts were optimized in six \pt intervals ranging from 1 to 24 GeV/c as described in Section  \ref{sub:TopSel}. The same selections, reported in Table  \ref{TableCutsDplusMult} were used in all multiplicity intervals in order to minimize the effect of efficiency corrections in the ratio of the yields of Equation \ref{eq:sfy}.  The Particle Identification strategy discussed in Section \ref{sub:pid} was applied to further reduce the background: the STANDARD PID selection was used in the \pt range  2<\pt<24 GeV/c, the STRONG PID  selection was used in the \pt range 1<\pt<2 GeV/c.\\
Figures \ref{12Tesi}, \ref{24Tesi}, \ref{48Tesi}, \ref{812Tesi} and \ref{1224Tesi} show the fits to the invariant mass distributions of \dplus meson candidates (and their charge conjugates) obtained after applying the selections described above in the five \pt intervals used in the analysis. The fitting function is composed of an exponential function for the background and a Gaussian function for the signal.  In order to reduce the influence of statistical fluctuations, the raw yields were determined by fixing in the fit the centroid of the signal Gaussians to the world-average \dplus mesons mass from PDG, and its width to the value obtained from a fit to the invariant mass distribution in minimum-bias events, where the signal statistical significance is larger.\\
\begin{figure}[h]
\centering
 \includegraphics[width=0.99\textwidth]{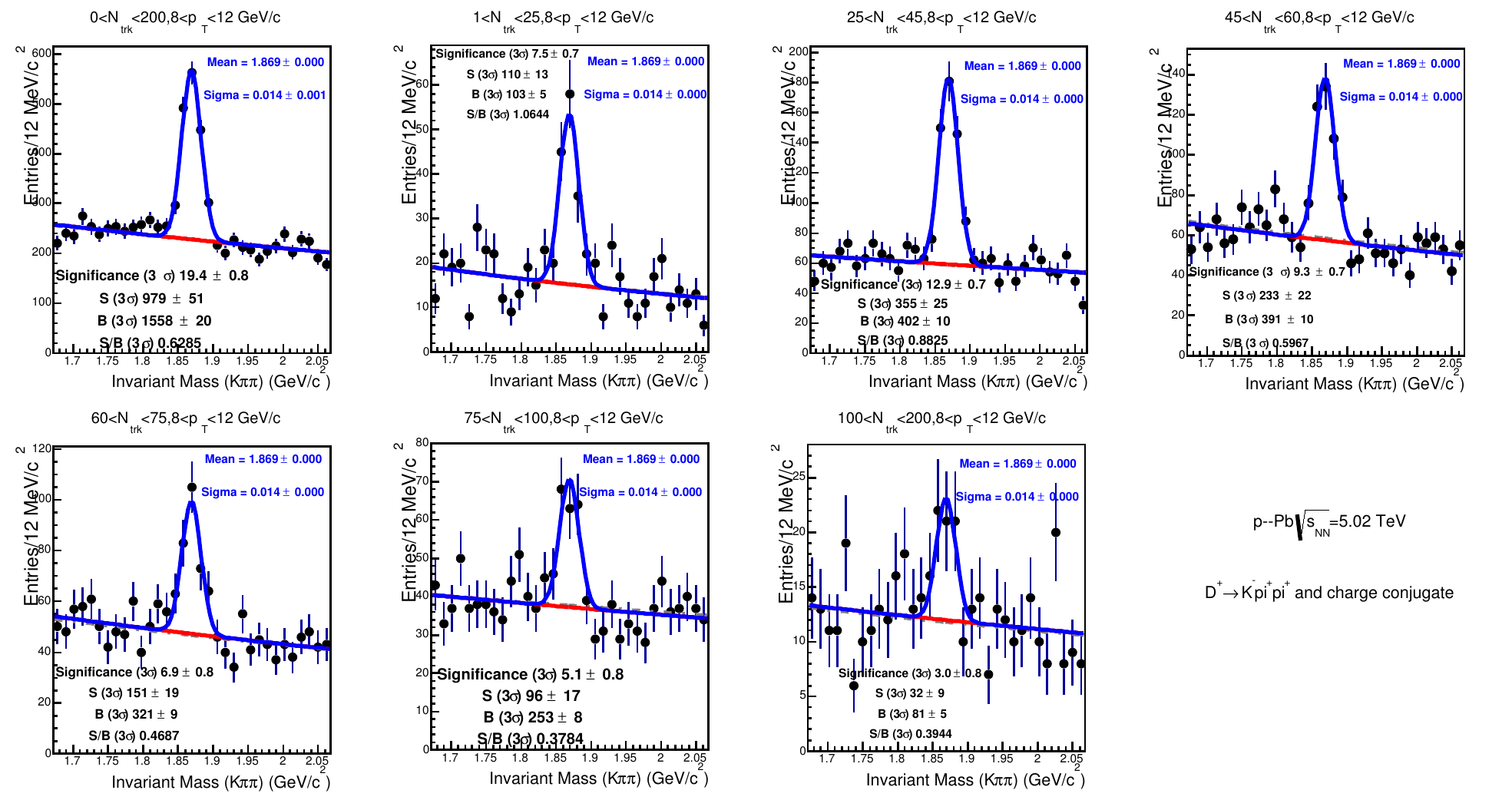}
\caption{\dplus candidate invariant mass distributions from  p--Pb collisions at \sqnn= 5.02 TeV in the \pt bin 8<\pt<12 GeV/c  integrated over multiplicity and in six multiplicity intervals ranging form 1 to 200 $N_{\rm trk}$.}
\label{812Tesi}
\end{figure}
\begin{figure}[h]
\centering
 \includegraphics[width=0.99\textwidth]{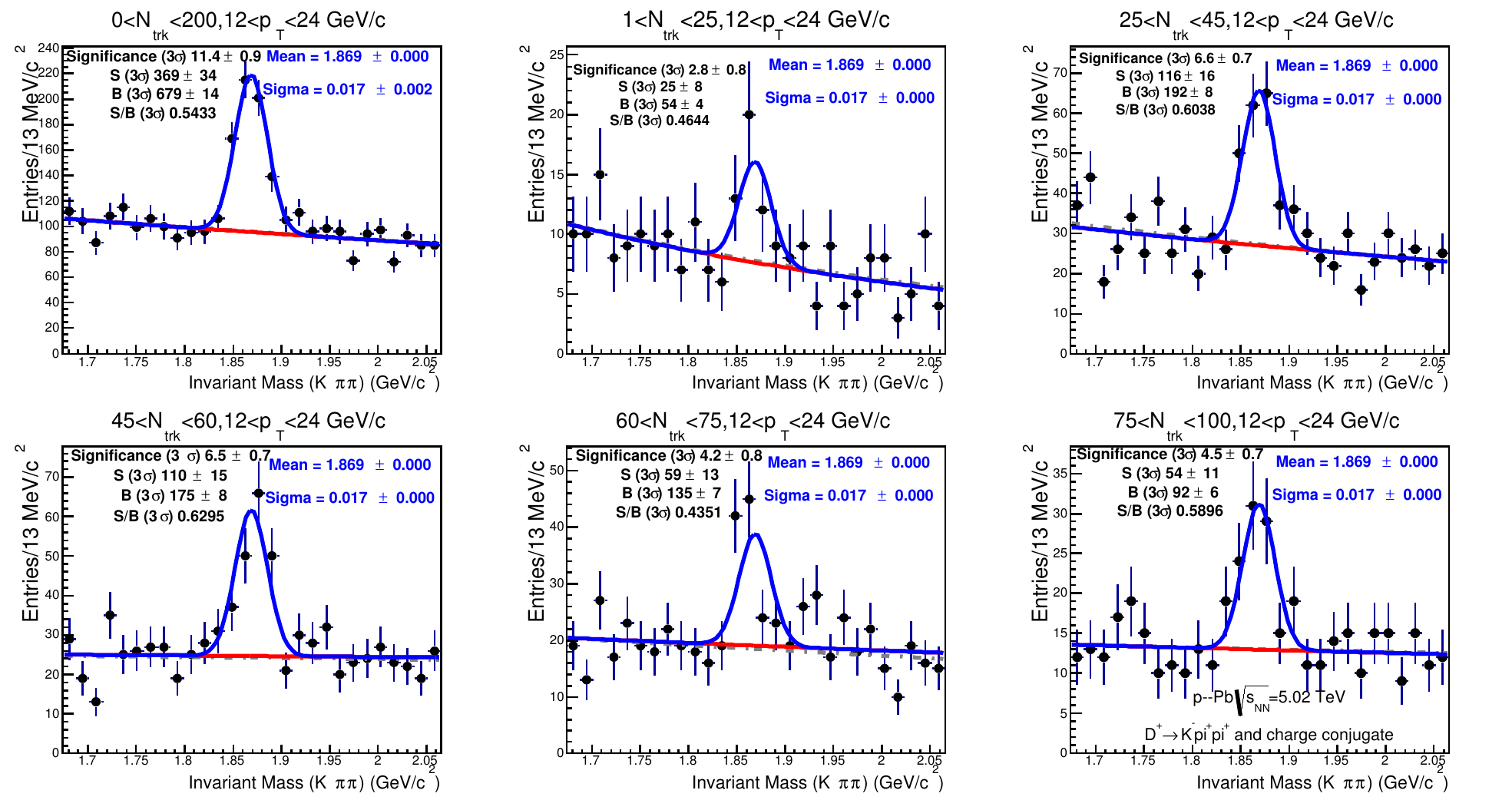}
\caption{\dplus candidate invariant mass distributions from  p--Pb collisions at \sqnn= 5.02 TeV in the \pt bin 12<\pt<24 GeV/c  integrated over multiplicity and in six multiplicity intervals ranging form 1 to 200 $N_{\rm trk}$.}
\label{1224Tesi}
\end{figure}
\subsection{Selection efficiency correction and beauty feed-down subtraction}
The raw yields obtained in the previous section are corrected for the reconstruction and selection efficiency according to Equation \ref{eq:sfy}. The \dplusm efficiencies as a function of \pt and $N_{\rm trk}$ were obtained from the   MC data sample described in Section \ref{sub:MCsample}.  The efficiency correction factor is obtained as the ratio of the \textit{kStepRecoPID} and  \textit{kStepAcceptance} containers. In Figure \ref{effvsmult2} the efficiency is shown as a function of  $N_{\rm trk}$ in the five \pt intervals considered in the analysis. These efficiencies are obtained applying the same reweighting on \pt shape of generated \dplus mesons and $N_{\rm trk}$ distributions discussed in Section \ref{AccEffRppB}.  At low \pt the selection efficiency increases with increasing $N_{\rm trk}$. This is due to the fact that, as shown in Figure \ref{seco}, the resolution on the position of the primary vertex increases with multiplicity.\\
 The  acceptance correction (\textit{kStepAcceptance} over \textit{kStepGeneratedLimAcc}) does not depend on multiplicity\footnote{This is strictly true only if the \pt distribution of \dplusm does not depend on multiplicity. The $Q_{\rm pPb}^{\rm mult}$ values shown previously indicate that the \pt shape of \dplusm does not depend on multiplicity, within the uncertainty of the measurement.} and simplifies in the ratio of Equation \ref{eq:sfy}.\\
\begin{figure}[t]
\centering
 \includegraphics[width=0.7\textwidth]{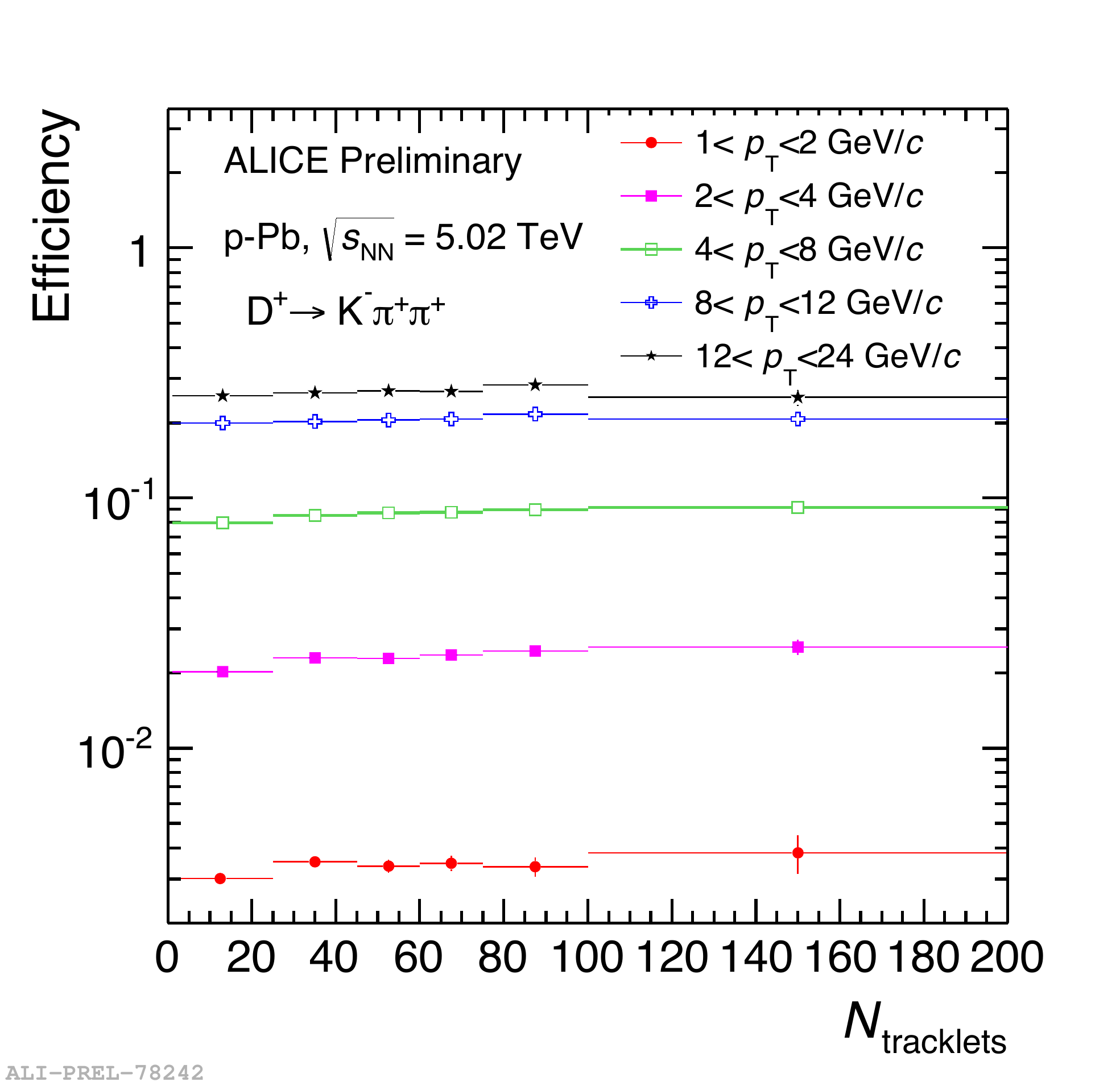}
\caption{\dplus mesons selection efficiencies as a function of $N_{\rm trk}$ in five \pt intervals ranging from 1 to 24 GeV/c.}
\label{effvsmult2}
\end{figure}
For the computation of the central values of  \dplusm self-normalized yield, the assumption that \fprompt does not depend on multiplicity is used, so that \fprompt cancels out in the numerator and denominator of Equation  \ref{eq:sfy}. However, the production of  B hadrons and of prompt \dplus mesons can have a different trend  with  charged particles multiplicity. A systematic uncertainty  due to this possible effect has been estimated and assigned, as discussed in next section.
\subsection{Systematic Uncertainties}
\subsubsection{Systematic uncertainty on the yield extraction}
The yield extraction systematic uncertainties were evaluated in the various multiplicity and \pt intervals with the same strategy described in Section \ref{sub:systYield}  with two differences:
\begin{itemize}
\item since here the raw yields in the multiplicity intervals are evaluated fixing the sigma of the invariant mass peak to the value obtained in the multiplicity integrated case, the systematic uncertainty is evaluated letting the sigma free to vary in the fit 
\item the effect of the variation of the fit configuration was studied on the quantity
\begin{equation}
R^{\rm Yield extraction} = \frac{Y^{\rm mult \ bin}_{\rm var}/Y^{\rm mult \ int}_{\rm var}}{Y^{\rm mult \ bin}_{\rm central}/Y^{\rm mult \ int}_{\rm central}}
\end{equation}
where $Y^{\rm mult \ bin}_{var}$ is the raw yield in a given multiplicity interval evaluated with the different fit strategies described in Section \ref{sub:systYield},  $Y^{\rm mult \ int}_{var}$ is the \dplus raw yield in the multiplicity integrated sample measured with the same variation of the fit configuration and $Y^{\rm mult \ bin \ (int)}_{central}$ are the central values in the multiplicity intervals (multiplicity integrated)
\end{itemize}
Figure \ref{YieldExMult} shows the relative variations of $R^{\rm Yield \ extraction}$ as a function of \pt in the multiplicity intervals 1<$N_{\rm trk}$<25 (left) and 45<$N_{\rm trk}$<60 (right).  The systematic uncertainties  assigned to each multiplicity and \pt interval are reported in Table \ref{tab:YieldExtrSyst}.\\
\begin{table}[b]
\begin{center}
\begin{tabular}{lc|c|c|c|c|c|}
\hline
\pt (GeV/c) & \multicolumn{5}{c}{Multiplicity bin} \\ \hline
&   1-24 & 25-44 & 45-60 & 60-75& 75-99 & 100-200 \\ \hline
1-2   & 8 \% & 8\% & 15\% & - & - & -\\ \hline
2-4   &6 \%&4\% &4\% &6\% &8\% & 10\% \\ \hline
4-8   &6 \%&4\% &4\% &6\% &8\% & 10\% \\ \hline
8-12   &8 \%&6\% &6\% &6\% &8\% & 10\% \\ \hline
12-24   &10 \%&10\% & 10\%&15\% & 15\% & -  \\ \hline
\end{tabular}
\caption{Summary table of the yield extraction systematic.
	\label{tab:YieldExtrSyst}
	}
\end{center}
\label{default}
\end{table}
\begin{figure}[t]
\centering
 \includegraphics[width=0.9\textwidth]{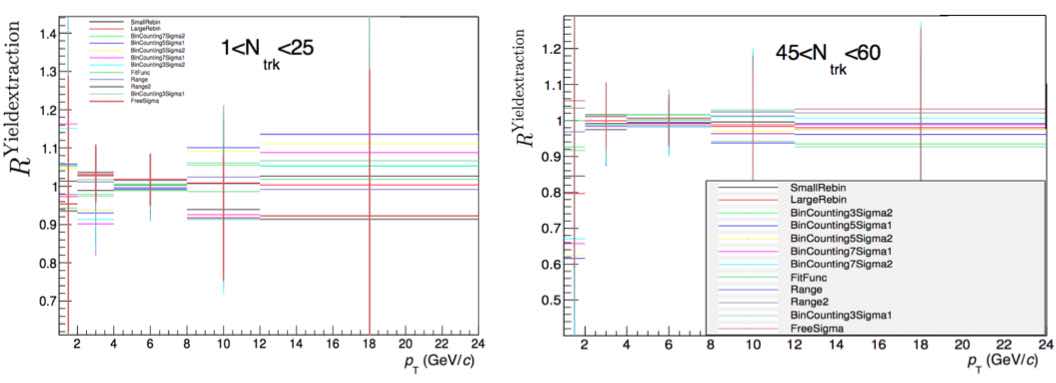}
\caption{ $R^{\rm Yield \ extraction}$  as a function of \pt in the multiplicity intervals 1<$N_{\rm trk}$<25 and 45<$N_{\rm trk}$<60.}
\label{YieldExMult}
\end{figure}
\subsubsection{Systematic uncertainty on the topological cut efficiency}
Since in this analysis the same cuts are used in each multiplicity interval and in the multiplicity integrated sample, it is expected that the cut efficiency systematic errors will cancel out in the ratio of Equation  \ref{eq:sfy} and no systematic uncertainty due to the description of the cut variables in the MC is present. As a check,  the corrected yields were calculated, both in multiplicity bins and
 integrated over multiplicity, using different sets of cuts (tighter/looser than the central ones). To do this the raw yields and the efficiencies  and the raw yields  were combined according to the following formula
\begin{equation}
R^{\rm cut var}=\frac{Y^{\rm mult}_{loose/tight}/(\epsilon^{\rm mult}_{loose/tight})/Y^{\rm tot}_{loose/tight}/(\epsilon^{\rm tot}_{loose/tight})}{Y^{\rm mult}_{central}/(\epsilon^{\rm mult}_{central})/Y^{\rm tot}_{central}/(\epsilon^{\rm tot}_{central})}
\label{eq:cvmult}
\end{equation}
The values of $R^{\rm cur var}$ are shown in Figure \ref{CVMult} for the six multiplicity intervals as a function of \pt. In the higher multiplicity bin the \dplus signal could not be extracted with the tight set of topological cuts. All values  are compatible with unity and no systematic uncertainty is therefore assigned. \\   
\begin{figure}[t]
\centering
 \includegraphics[width=0.9\textwidth]{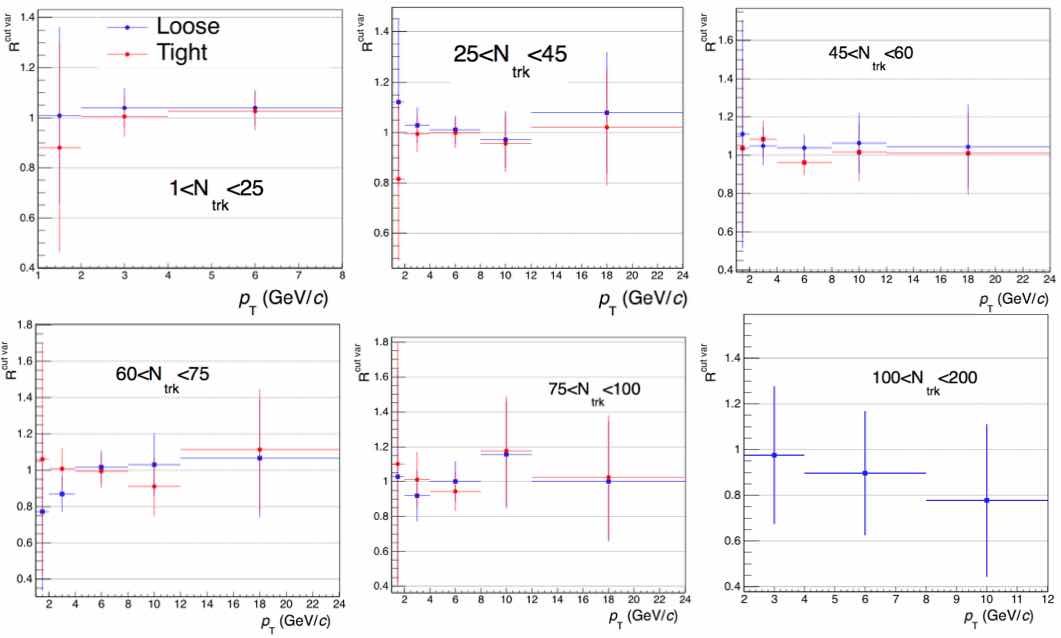}
\caption{$R^{\rm cur var}$  (i.e. variation of the self normalized yieds when varying the topological cuts)  for the six multiplicity bins as a function of \pt.}
\label{CVMult}
\end{figure}
\subsubsection{Systematic uncertainty on PID}
The systematic due to PID is also expected to cancel out in the ratio of Equation   \ref{eq:sfy}. In fact,  the TPC response is stable up to multiplicities of at least 400 tracks, so  no variation  of its response in the range considered in this analysis is expected. For what concerns the TOF information, an  effect on the resolution of the  start of time information $t_0$ (Section \ref{sub:TOF}) could be
 expected, with better resolution at high multiplicities. However this effect has a small influence on the PID strategy discussed in Section  \ref{sub:pid}.  As a cross chech, the same ratio of Equation  \ref{eq:cvmult} (here called $R^{PID}$) was evaluated using \dplusm the yields obtained with and without PID in all the \pt and multiplicity intervals. The result is shown in Figure \ref{SisPidMult}. The \pt bin 1<\pt<2 GeV/c is not shown since here the \dplus signal could not be extracted without PID. All ratios $R^{\rm PID}$ are compatible with unity (the big discrepancy in 4<\pt<8 GeV/c for the interval 100<$N_{\rm trk}$<200 is due to the fact the fit on the invariant mass distribution obtained without PID is not stable) and no PID uncertainty is assigned.\\
\begin{figure}[t]
\centering
 \includegraphics[width=0.65\textwidth]{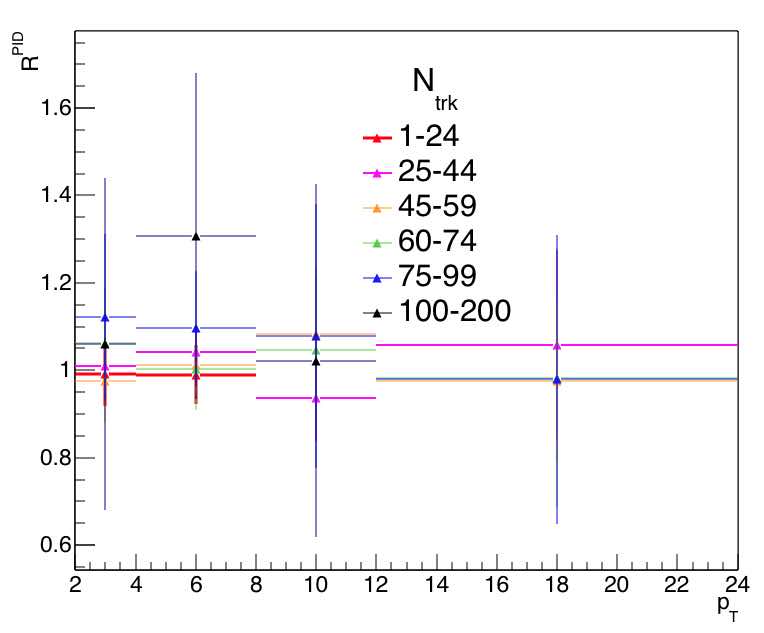}
\caption{Relative variation of $R^{\rm PID}$  for the six multiplicity bins as a function of \pt.}
\label{SisPidMult}
\end{figure}
\subsubsection{Systematic uncertainty on beauty feed-down subtraction}
The \dplusm self-normalized yields are calculated  assuming that the relative contribution of B decays to the \dplusm yields in the different multiplicity bins is constant and equal to that in the multiplicity integrated sample.
However, the dependence of B hadron and prompt \dplusm production on the charged particles density could differ. Simulations performed with PYTHIA 6 in the extreme scenario of the  hard-QCD tune \cite{PYTHIA} show that  the ratio of the B to \dplusm relative yield evolution with multiplicity presents a linear increase with increasing multiplicity. \\
Therefore, to evaluate the systematic the relative fraction of \dplus from B hadrons decays was varied with  charged particle multiplicity by a factor of 0.5 (at low multiplicity) to 2 (at high multiplicity)  as shown in Figure \ref{flastsist} (left). 
The relative fraction of prompt \dplus yield, $f_{\rm prompt}$, in the multiplicity integrated sample was computed with both $N_b$ and $f_c$  methods as described inSection \ref{sec:bfd}.
Starting from these multiplicity integrated \fprompt values, the \fprompt values in the different multiplicity intervals were obtained according  to the assumptions of  Figure \ref{flastsist} (left), 
 obtaining  a range of \fprompt values that is  converted into a relative uncertainty on the  \dplusm self-normalized yields.\\
\begin{figure}[t]
\centering
 \includegraphics[width=0.9\textwidth]{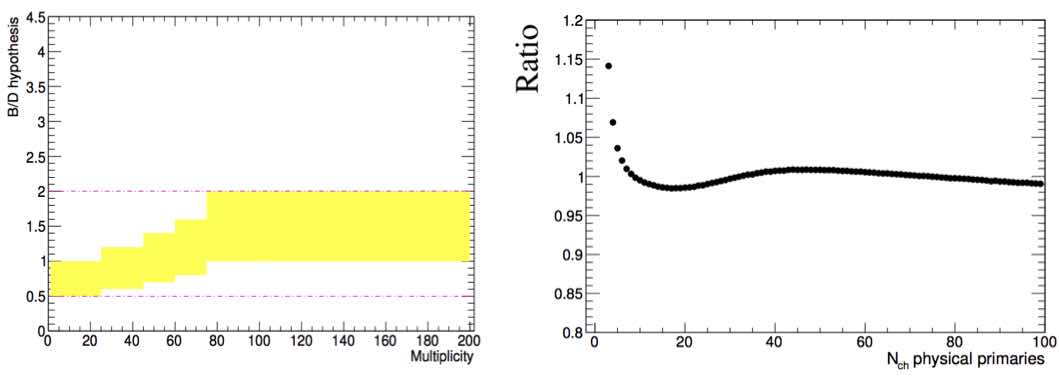}
\caption{Left: hypothesis on   the relative fraction of \dplus from B hadrons as a function of multiplicity used to evaluate the \fprompt systematic uncertainty. Right:   deviations from the linear dependence of $N_\mathrm{trk}$ on $N_\mathrm{ch}$.}
\label{flastsist}
\end{figure}
The calculation was done independently for:
\begin{itemize}
\item the upper and lower values of the uncertainties on \fprompt,  which originate from the variation of the pQCD parameters in the FONLL  calculation discussed in Section \ref{sub:fds}
\item the two methods $N_b$ and $f_c$ 
\end{itemize}
The final relative uncertainties on the  \dplusm self-normalized yields  are assigned as the envelope of these cases.\\
\subsubsection{Systematic uncertainty on the determination of $N_\mathrm{ch}$}
The systematic uncertainty on the variable used to quantify the multiplicity of
the collision ${\rm d}N_{\rm ch}/{\rm d}\eta \big/ \langle {\rm d}N_{\rm ch}/{\rm d}\eta \rangle^{mult}_{int}$  was assigned evaluating   the deviations from the linear dependence of $N_\mathrm{trk}$ on $N_\mathrm{ch}$ shown in Figure \ref{MCMultNch}. This was checked by dividing the actual mean of $N_\mathrm{trk}$ at each value of $N_\mathrm{ch}$ by the value of the linear fit in Figure \ref{MCMultNch}. The resulting ratio  is shown in Figure \ref{flastsist} (right). A systematic uncertainty of 3\% was assigned to each multiplicity interval.
\subsection{Results}
The results of the \dplusm self normalized  yields for the five \pt intervals are presented in Figure \ref{ResMult1}  as a function of the relative primary charged particle multiplicity (d$N_{\rm ch}/\textrm{d}\eta)$/$\langle\textrm{d}N_{\rm ch}/\textrm{d}\eta\rangle$.  The self normalized yields are presented in the top panels with their statistical (vertical bars) and systematic (boxes) uncertaintites except  for the uncertainty on the feed-down fraction, which is drawn separately in the bottom panels in the form of relative uncertainties. The points are located on the $x$-axis at the average value of the relative charged particle multiplicity,  (d$N_{\rm ch}/\textrm{d}\eta)/\langle\textrm{d}_{\rm ch}/\textrm{d}\eta\rangle$ for every $N_\mathrm{trk}$ interval as shown in Table \ref{tab:NchCorr}.  The \dplusm self normalized  yields  in different \pt  intervals are in agreement within uncertaintes for all multiplicity intervals.\\
\begin{figure}[t]
\centering
 \includegraphics[width=0.65\textwidth]{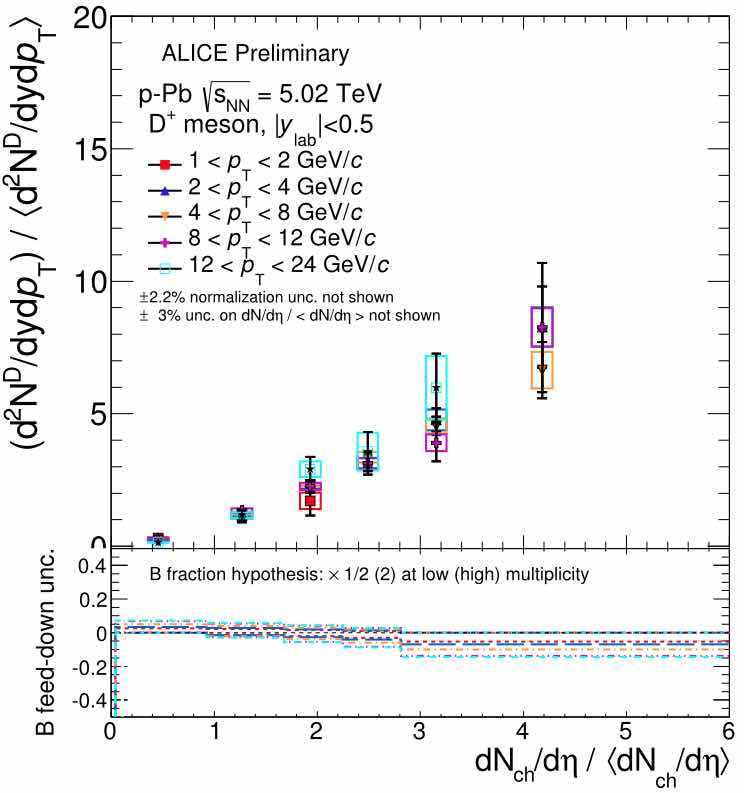}
\caption{\dplusm self normalized  yields in five \pt intervals  as a function of the relative primary charged particle multiplicity in p--Pb collisions at \sqnn=5.02 TeV. Right: average \dplus,  D$^0$ and D$^{*+}$ mesons self normalized  yields in five \pt intervals  as a function of the relative primary charged particle multiplicity in p--Pb collisions at \sqnn=5.02 TeV.}
\label{ResMult1}
\end{figure}
The  self normalized  yields were also computed for D$^0$ and D$^{*+}$ mesons. The average \dplus,  D$^0$ and D$^{*+}$ mesons self normalized yields were computed for each \pt interval using as weights the inverse square of their relative uncertainties and is shown in Figure \ref{ResMult2} (left). The average D meson self normalized yields increase with the primary charged particles multiplicity by about a factor of eight in the range  0.5 $\div$ 4.5 $\times$ (d$N_{\rm ch}/\textrm{d}\eta)/\langle\textrm{d}N_{\rm ch}/\textrm{d}\eta\rangle$. \\  In order to compare the trend with multiplicity in the different \pt intervals, the average $D$-meson self normalized yields in the different \pt intervals were divided by the one in 2<\pt<4 GeV/c. The result is shown in Figure  \ref{ResMult2} (right). Within current uncertainties no \pt trend of the self normalized yields is visible.\\
It is interesting to compare the trend of the self normalized yields as  a function  of the relative primary charged particle multiplicity in pp and p--Pb collisions. Figure \ref{ResMult3} (left) shows \dplusm self normalized  yields as a function of multiplicity in pp collisions at \sqs= 7 TeV, measured in three \pt bins from 2 to 12 GeV/c. The \dplusm self normalized yields as  a function of  the relative primary charged particle multiplicity show a trend similar to the one observed in p--Pb collisions. This observation is confirmed in Figure \ref{ResMult3} (right) where the D$^0$ self normalized  yields are shown together for pp and p--Pb collisions in the \pt interval 2<\pt<4 GeV/c.  The increasing trend present in pp collisions is interpreted as due to MPIs and/or to a larger amount of gluon radiation in collisions where heavy quarks are produced, while in p--Pb collisions it is also  due to a higher number of binary collisions.  \\
\begin{figure}[b]
\centering
 \includegraphics[width=0.9\textwidth]{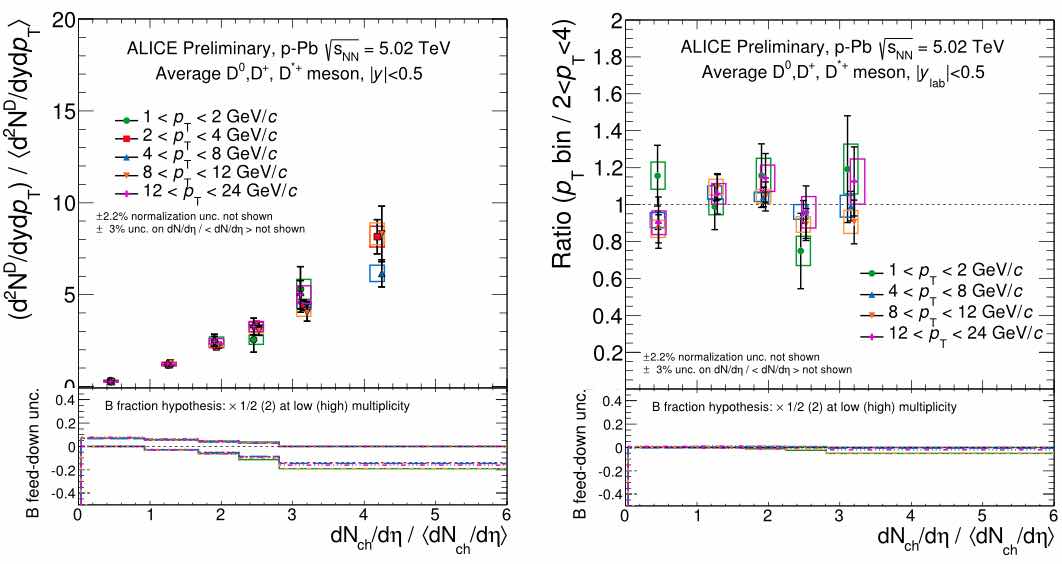}
\caption{Left: average \dplus,  D$^0$ and D$^{*+}$ mesons self normalized  yields in five \pt intervals  as a function of the relative primary charged particle multiplicity in p--Pb collisions at \sqnn=5.02 TeV. Right: average \dplus,  D$^0$ and D$^{*+}$ mesons self normalized yields in the different \pt intervals  divided by the average  self normalized yield in 2<\pt<4GeV/c in p--Pb collisions at \sqnn=5.02 TeV. }
\label{ResMult2}
\end{figure}
To conclude we will compare the results relative to open charm production as a function of multiplicity to those relative to hidden charm (J/$\Psi$). Figure  \ref{ResMult4} (left) shows the average D meson and inclusive (prompt + feed-down) J/$\Psi$ self normalized yields as a function of relative charged particle multiplicity, for D mesons in the range 2<pt<4 GeV/c  and inclusive J/$\Psi$ for \pt>0 GeV/c. in two rapidity intervals, namely 2.5<$y_{\rm lab}$<4 and -4<$y_{\rm lab}$<-2.5. The inclusive J/$\Psi$ self normalized yields also show an increase as a function of relative charged particle multiplicity, with higher values of the self normalized yields at backward rapidity, i.e. in the Pb-going direction. The different increase of  inclusive J/$\Psi$ self normalized yields at forward and backward rapidities is coherent with the  $Q_{\rm pPb}^{\rm mult}(\pteq)$ results shown in Figure \ref{QpPbResjpsi} and it is due to the different cold nuclear matter effects in the two rapidity regions. The situation is different in pp collisions at \sqs= 7 TeV: as shown in  Figure  \ref{ResMult4} (right), a similar increase of the relative yield with the charged particle multiplicity is observed for open (D$^{0}$-meson) and hidden (inclusive J/$\Psi$) charm production, the latter measured  both at central (|$y_{\rm lab}$|=|$y_{\rm cms}$|<0.9) and forward (2.5<|$y_{\rm lab}$|<4) rapidities.

\begin{figure}[b]
\centering
 \includegraphics[width=0.9\textwidth]{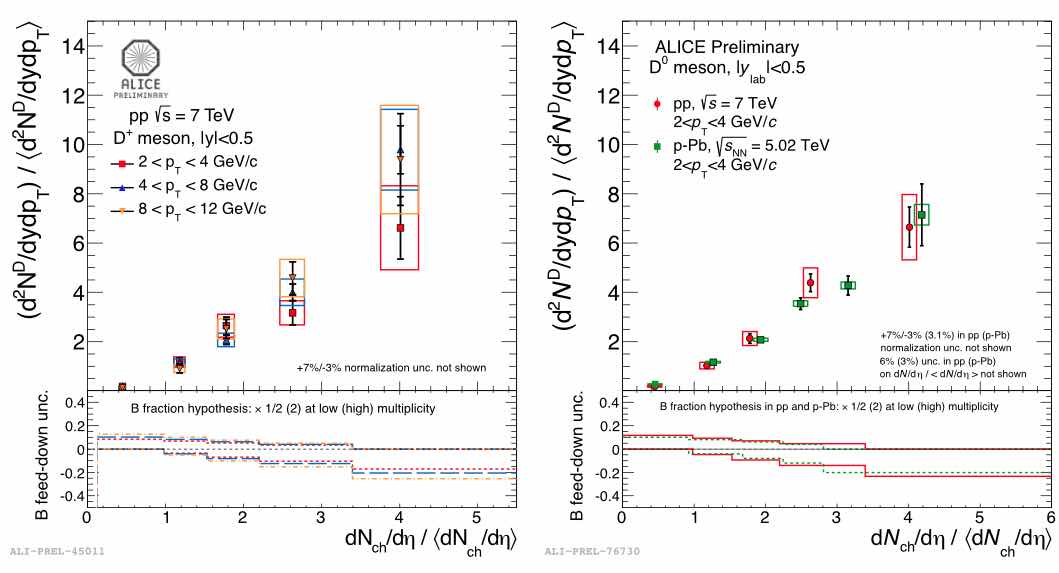}
\caption{Left: \dplusm self normalized  yields in three \pt intervals  as a function of the relative primary charged particle multiplicity in pp collisions at \sqnn=7 TeV. Right: D$^0$ self normalized  yields  in pp collisions at \sqs= 7 TeV and p--Pb collisions at \sqnn= 5.02 TeV in the \pt interval 2<\pt<4 GeV/c. }
\label{ResMult3}
\end{figure}
\begin{figure}[b]
\centering
 \includegraphics[width=0.9\textwidth]{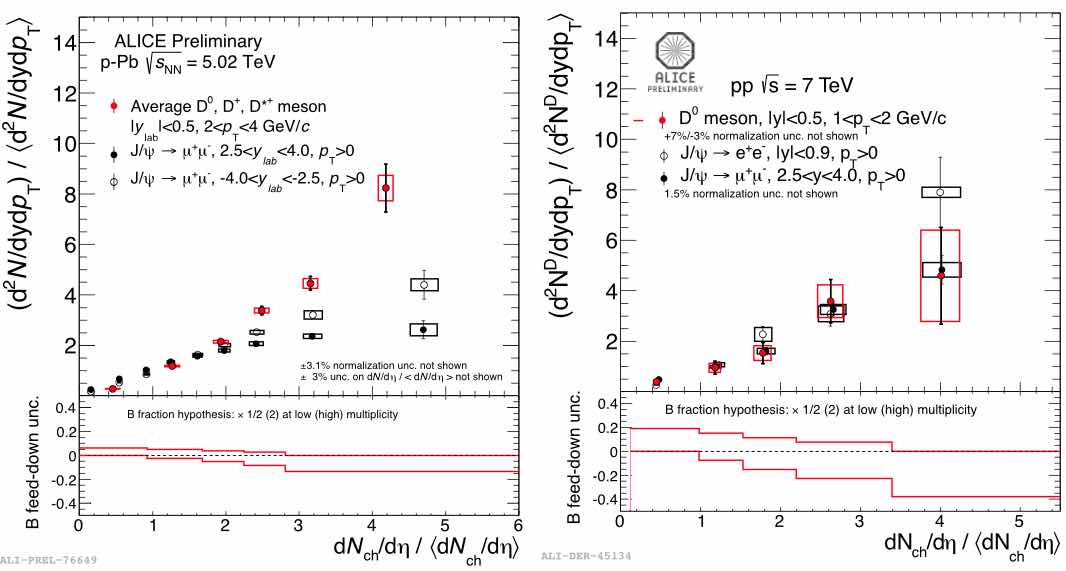}
\caption{Left: average D meson and inclusive J/$\Psi$ self normalized yields as a function of relative charged particle multiplicity in p--Pb collisions at \sqnn= 5.02 TeV. Right:  D$^{0}$ meson and inclusive J/$\Psi$ self normalized yields as a function of relative charged particle multiplicity in pp collisions at \sqs= 7 TeV.}
\label{ResMult4}
\end{figure}

 \chapter{Conclusions}
In this thesis several results were obtained by reconstructing \dplus mesons (and their antiparticles) in their $K^{-}\pi^{+}\pi^{+}$ hadronic decay channel at mid-rapidity in p--Pb collisions at \sqnn= 5.02 TeV. The vertex reconstruction and particle identification capabilities of the ALICE central barrel were exploited in order to reduce the combinatorial background, so as to be able to extract \dplusm signal with significance larger than three (up to 24) in different transverse momentum, rapidity and multiplicity intervals. \\
\vskip 1 em
The \dplusm production cross section was measured in five rapidity intervals (measured in the centre of mass  of the p--Pb system) for three wide \pt intervals showing no significant rapidity dependence  in  the range -1.265<$y_{\rm cms}$<0.335.\\
\vskip 1 em
The \dplusm production cross section was also measured in the rapidity range \\ -0.96<$y_{\rm cms}$<0.04 in ten transverse momentum intervals in the range 1<\pt<24 GeV/c, and compared to the pp cross-section at \sqnn= 5.02 TeV/c scaled by mass number A of the Pb nucleus. The resulting nuclear modification factor of \dplusm is compatible with unity within uncertainties, showing that \dplusm production in p--Pb collisions at \sqnn= 5.02 TeV scales with  the number of binary nucleon-nucleon collisions  with respect to pp collisions at the same energy. The nuclear modification factor of \dplusm is  described by models taking into account nuclear shadowing of PDFs and saturation models, $k_{\rm T}$ broadening and  cold nuclear matter energy loss as well as by calculations based on gluon saturation in the initial state. This measurement allows us to conclude that the suppressions of $D$-meson yield at high \pt in central Pb--Pb collisions can not be due to initial state effects only, implying that hot nuclear matter effects must be present. \\
\vskip 1 em
A data-driven method for subtracting the fraction of the \dplus mesons yield coming from B-hadron decays has been developed. The results  are compatible with those obtained with a theory-driven method based on pQCD, although with larger uncertainties. However the uncertainty might be significantly reduced if this method is applied to a larger data sample, like the one expected from the LHC Run II.\\
\vskip 1 em
After observing that centrality determination in p--Pb  collisions is affected by several sources of bias when using centrality estimators based on multiplicity, an almost bias-free method for determining centrality has been presented, the hybrid method. The results of $Q_{\rm pPb}^{\rm mult}$ obtained with the hybrid method are compatible with unity within uncertainties in the four event activity classes studied, showing no evidence of strong cold nuclear matter effects in the highest event activity classes for p--Pb collisions. This result is in agreement with the $Q_{\rm pPb}^{\rm mult}$ measurement for charged particles. \\
\vskip 1 em
Finally, the \dplusm yield has been extracted as a function of the multiplicity of charged particles produced in the collisions. The results show an increase of \dplusm production rate with increasing charged particle multiplicities, that might be due to the higher number of nucleon-nucleon collisions and/or to a higher number of multi-parton interactions (MPIs)  for high multiplicity events. By comparing these results with those obtained in pp collisions and with model calculations, valuable information can be obtained on the role of MPIs in $c\bar{c}$ pair production and on the interplay of hard and soft processes in hadronic collisions.

%
%
\appendix 
%
%

\chapter{$R_{\rm pPb}$ equations}
\label{AppendixA} 
I would like to demonstrate here the second equality of Equation \ref{Eq:rppb}:
\begin{equation}
R_{\rm pA}(\pteq) = \frac{dN^{\rm pA}/d\pteq}{\langle N_{\rm coll}\rangle dN^{\rm pp}/d\pteq} =  \frac{d\sigma^{\rm pA}/d\pteq}{{\rm A} \ d\sigma^{\rm pp}/d\pteq}
\end{equation}
I recall the equations for the corrected per-event yield for prompt \dplus mesons:
\begin{equation}
\frac{dN^{D^{+}}}{d\pteq}=\frac{1}{2}\frac{1}{\Delta y\Delta\pteq}\frac{f_{\textrm{prompt}}\cdot N^{D^{+/-}}_{\textrm{raw}}|_{|y|<y_{fid}}}{(\textrm{Acc}\times \epsilon)^{\textrm{prompt}}\cdot BR\cdot N_{\textrm{ev}}}
\label{eq:corrspapp}
\end{equation}
and for the corrected invariant cross section as:
\begin{equation}
\frac{d\sigma^{D^{+}}}{d\pteq}=\frac{1}{2}\frac{1}{\Delta y\Delta\pteq}\frac{f_{\textrm{prompt}}\cdot N^{D^{+/-}}_{\textrm{raw}}|_{|y|<y_{fid}}}{(\textrm{Acc}\times \epsilon)^{\textrm{prompt}}\cdot BR\cdot L_{\textrm{int}}}
\end{equation}
Then 
\begin{equation}
\frac{\frac{d\sigma^{D^{+}}}{d\pteq}}{\frac{dN^{D^{+}}}{d\pteq}}=\frac{N_{\rm ev}}{L_{\rm int}}
\end{equation}
The last equation can be used to express  $R_{\rm pA}(\pteq)$ in terms of  $d\sigma^{\rm pA}/d\pteq$
\begin{equation}
R_{\rm pA}(\pteq) = \frac{dN^{\rm pA}/d\pteq}{\langle N_{\rm coll}\rangle dN^{\rm pp}/d\pteq} = \frac{L_{\rm int}^{\rm pA}}{N_{\rm ev}^{\rm pA}} \frac{d\sigma^{\rm pA}/d\pteq}{\langle N_{\rm coll}\rangle dN^{\rm pp}/d\pteq}=
\frac{1}{\sigma_{\rm tot}^{\rm pA}}\frac{d\sigma^{\rm pA}/d\pteq}{\langle N_{\rm coll}\rangle dN^{\rm pp}/d\pteq}
\end{equation}
Doing the same steps for $dN^{\rm pp}/d\pteq$ in the denominator, one obtains
\begin{equation}
R_{\rm pA}(\pteq) = \frac{\sigma_{\rm tot}^{\rm pp}}{\sigma_{\rm tot}^{\rm pA}}\frac{d\sigma^{\rm pA}/d\pteq}{\langle N_{\rm coll}\rangle d\sigma^{\rm pp}/d\pteq}
\end{equation}
Now, observing that for minimum-bias p--Pb collisions $\langle N_{\rm coll}\rangle$ = A $\frac{\sigma_{\rm tot}^{\rm NN}}{\sigma_{\rm tot}^{pA}}$ and assuming that at this energies $\sigma_{\rm tot}^{\rm NN}=\sigma_{\rm tot}^{\rm pp}$, we finally get
\begin{equation}
R_{\rm pA}(\pteq) = \frac{d\sigma^{\rm pA}/d\pteq}{{\rm A} \ d\sigma^{\rm pp}/d\pteq}
\end{equation}
\lhead{Appendix A. \emph{Appendix Title Here}} 


\chapter{ALICE Coordinate System}
\label{AppendixB}
The ALICE coordinate system is a right-handed orthogonal Cartesian system with point of origin $x$, $y$, $z$ = 0 at the beams interaction point (IP). The axes, azimuthal angle $\varphi$ and polar angle $\theta$ are defined as follows: 
\begin{itemize}
\item \textbf{$x$ axis}: perpendicular to the mean beam direction, aligned with the local horizontal and pointing to the accelerator centre. Positive $x$ is from the point of origin toward the accelerator centre, negative $x$ is from the point of origin outward
\item \textbf{$y$ axis}: perpendicular to the $x$ axis and the mean local beam direction, pointing upward. Positive $y$ is from the point of origin upward, negative $y$ is from the point of origin downward
\item  \textbf{$z$ axis}:  parallel to the mean beam direction. Negative $z$ is from the point of origin  toward  the muon arm 
\item  \textbf{azimuthal angle $\varphi$}: increases counter-clockwise from $x$ ($\varphi$=0) to $y$ ($\varphi$=$\pi$/2) with the observer standing at positive $z$  and looking in direction of the muon arm
\item  \textbf{polar angle $\theta$}:  increases from $z$ ($\theta$=0), to $xy$ plane ($\theta$=$\pi$/2) to -$z$ ($\theta$=$\pi$)
\end{itemize}
The conversion from spherical to Cartesian coordinates is done through:
\begin{equation}
\begin{aligned}
x = r \sin{\theta}\cos{\varphi}\\
y = r \sin{\theta}\sin{\varphi}\\
z = r \cos{\theta}\\
\end{aligned}
\end{equation}
The inverse conversion from Cartesian to spherical coordinates is:
\begin{equation}
\begin{aligned}
r =x^2+y^2+z^2\\
\theta=\arccos{z/r}\\
\varphi=\arctan{y/x}\\
\end{aligned}
\end{equation}
\begin{figure}
\centering
 \includegraphics[width=0.6\textwidth]{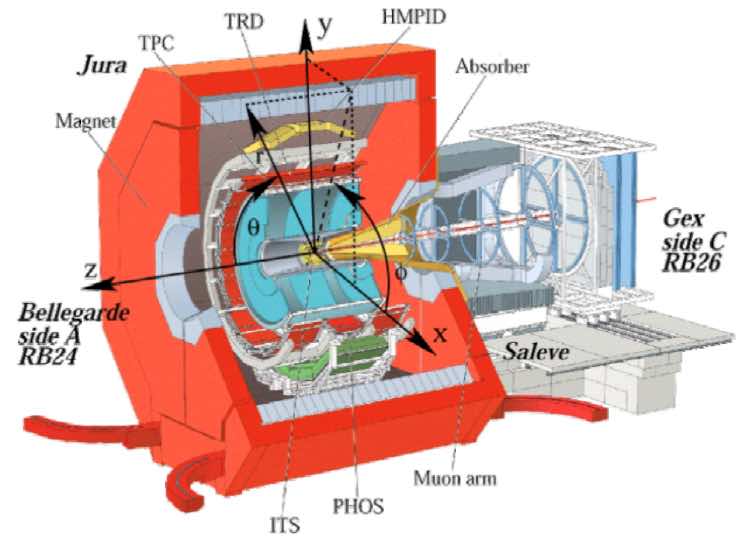} 
\caption{ALICE Coordinate System (in this figure $\varphi$ is tagged as $\phi$).}
\label{ALICECOOR}
\end{figure}
\end{spacing}{1.0}
%
%
%

\label{Bibliography}

\lhead{\emph{Bibliography}} 

  \bibliographystyle{unsrtnat} 

\bibliography{Bibliography} 

\begin{thebibliography}{149}
\providecommand{\natexlab}[1]{#1}
\providecommand{\url}[1]{\texttt{#1}}
\expandafter\ifx\csname urlstyle\endcsname\relax
  \providecommand{\doi}[1]{doi: #1}\else
  \providecommand{\doi}{doi: \begingroup \urlstyle{rm}\Url}\fi

\bibitem[Collaboration(2014{\natexlab{a}})]{PaperpPb}
ALICE Collaboration.
\newblock Measurement of prompt {D}-meson production in p--{Pb} collisions at
  $\sqrt{s_{NN}}$ = 5.02 {TeV}.
\newblock \emph{PRL 113, 232301}, 2014{\natexlab{a}}.

\bibitem[Mangano(2006)]{Mangano}
M.L. Mangano.
\newblock Introduction to {QCD} in hadronic collisions.
\newblock 2006.

\bibitem[F.~Halzen(1984)]{hm}
A.~Martin F.~Halzen.
\newblock Quarks and leptons.
\newblock \emph{John Wiley {\&} Sons}, 1984.

\bibitem[et~al.(1985)]{dis}
E.D.~Bloom et~al.
\newblock High energy inelastic e-p scattering at 6$\,^{\circ}$ and
  10$\,^{\circ}$.
\newblock \emph{Phys. Rev. Lett.}, 1985.

\bibitem[Muller(1985)]{muller}
B.~Muller.
\newblock The physics of {Quark} {Gluon} {Plasma}.
\newblock \emph{Springer Verlag}, 1985.

\bibitem[Griffiths(1987)]{gr}
D.~Griffiths.
\newblock Introduction to elementary particles.
\newblock \emph{John Wiley {\&} Sons}, 1987.

\bibitem[D.~Gross(1973)]{Asimp}
F.~Wilczek D.~Gross.
\newblock Asymptotically free gauge theories.
\newblock \emph{Phys. Rev. D}, 8:\penalty0 3633--3652, Nov 1973.

\bibitem[Bartke(2009)]{ba}
J.~Bartke.
\newblock Introduction to relativistic heavy ion physics.
\newblock \emph{World Scientific}, 2009.

\bibitem[Hagedorn(1965)]{hagedorn}
Rolf Hagedorn.
\newblock \emph{Suppl. Nuovo Cim. 3}, 1965.

\bibitem[Cabibbo and Parisi(1975)]{Cabibbo}
N.~Cabibbo and G.~Parisi.
\newblock Exponential hadronic spectrum and quark liberation.
\newblock \emph{Physics Letters B}, 59\penalty0 (1):\penalty0 67 -- 69, 1975.

\bibitem[A.~Chodos and Thorn(1974)]{MITBag}
K.~Johnson A.~Chodos, R. L.~Jaffe and C.~B. Thorn.
\newblock Baryon structure in the bag theory.
\newblock \emph{Phys. Rev. D 12 (1974) 2599}, 1974.

\bibitem[Hands(2001)]{sh}
S.~Hands.
\newblock The phase diagram of {QCD}.
\newblock \emph{Contemp.Phys.42:209-225}, 2001.

\bibitem[Yagy(2005)]{ya}
K.~Yagy.
\newblock {Quark-Gluon} {Plasma}.
\newblock \emph{Cambridge University Press}, 2005.

\bibitem[Wong(1995)]{wo}
C.~Wong.
\newblock Introduction to high-energy heavy-ion collisions.
\newblock \emph{World Scientific}, 1995.

\bibitem[et~al.(2008)]{csc}
Mark G.~Alford et~al.
\newblock Color superconductivity in dense quark matter.
\newblock \emph{Rev.Mod.Phys. 80}, 2008.

\bibitem[Vogt(2007)]{vo}
R.~Vogt.
\newblock Ultrarelativistic heavy-ion collisions.
\newblock \emph{Elsevier}, 2007.

\bibitem[collaboration(2004)]{ppr}
ALICE collaboration.
\newblock Physics performance report.
\newblock \emph{Journal of Physics G}, 2004.

\bibitem[Sourav~Sarkar(2010)]{Chiral}
Bikash~Sinha Sourav~Sarkar, Helmut~Satz.
\newblock The physics of the {Quark-Gluon} {Plasma}.
\newblock \emph{Springer Volume 785}, 2010.

\bibitem[{B. Gossiaux, R. Bierkandt, J. Aichelin}(2009)]{adscft}
{B. Gossiaux, R. Bierkandt, J. Aichelin}.
\newblock Tomography of quark gluon plasma at energies available at the {BNL}
  {Relativistic Heavy Ion Collider (RHIC)} and the {CERN Large Hadron Collider
  (LHC)}.
\newblock \emph{PHYSICAL REVIEW C 79, 044906 (2009)}, 2009.

\bibitem[{M. Miller, K. Reygers, S. Sanders, P. Steinberg}(2007)]{Glauber}
{M. Miller, K. Reygers, S. Sanders, P. Steinberg}.
\newblock {Glauber Modeling in High Energy Nuclear Collisions}.
\newblock \emph{Ann.Rev.Nucl.Part.Sci.57:205-243}, 2007.

\bibitem[for~the ALICE~Collaboration(2013)]{thermalphotontheory}
Nicolas~Arbor for~the ALICE~Collaboration.
\newblock {Recent photon physics results from the ALICE experiment at the LHC}.
\newblock \emph{EPJ Web of Conferences Volume 60}, 2013.

\bibitem[(for~the PHENIX~collaboration)(2014)]{photonsPHENIX}
Benjamin~Bannier (for~the PHENIX~collaboration).
\newblock Measurements of direct photons in {Au + Au} collisions with {PHENIX}.
\newblock \emph{Nuclear Physics A 00 1-4}, 2014.

\bibitem[University(2013)]{PhotonsALICE}
Munster University.
\newblock Measurement of direct photons in pp and {Pb-Pb Collisions with
  ALICE}.
\newblock \emph{Nucl.Phys. A904-905 573c-576c}, 2013.

\bibitem[Becattini(2009)]{Becattini}
F.~Becattini.
\newblock An introduction to the statistical hadronization model.
\newblock \emph{arXiv:0901.3643}, 2009.

\bibitem[{A. Andronic, P. Braun-Munzinger, K. Redlich, J.
  Stachel}(2013)]{ThermalFits}
{A. Andronic, P. Braun-Munzinger, K. Redlich, J. Stachel}.
\newblock The statistical model in {Pb-Pb collisions at the LHC}.
\newblock \emph{Nucl.Phys. A904-905}, 2013.

\bibitem[Floris(2014)]{Floris}
Michele Floris.
\newblock Hadron yields and the phase diagram of strongly interacting matter.
\newblock \emph{Proceedings of the XXIV International Conference On
  Ultrarelativistic Nucleus-Nucleus Collisions (Quark Matter 2014)}, 2014.

\bibitem[Collaboration(2012{\natexlab{a}})]{ALICESpectra}
ALICE Collaboration.
\newblock {Pion, Kaon and Proton Production in Central Pb--Pb Collisions at
  $\sqrt{s_{\rm NN}}$ = 2.76 TeV}.
\newblock \emph{PRL 109, 252301}, 2012{\natexlab{a}}.

\bibitem[{Ekkard Schnedermann, Josef Sollfrank, and Ulrich
  Heinz}(1993)]{BlastWave}
{Ekkard Schnedermann, Josef Sollfrank, and Ulrich Heinz}.
\newblock {Thermal phenomenology of hadrons from 200A GeV S+S collisions}.
\newblock \emph{Phys. Rev. C 48, 2462}, 1993.

\bibitem[{R. J. Fries, B. Muller, and C. Nonaka}(2003)]{recombination1}
{R. J. Fries, B. Muller, and C. Nonaka}.
\newblock Hadron production in heavy ion collisions: Fragmentation and
  recombination from a dense parton phase.
\newblock \emph{Phys.Rev.C68}, 2003.

\bibitem[Collaboration(2013{\natexlab{a}})]{poverpPHENIX}
PHENIX Collaboration.
\newblock {Spectra and ratios of identified particles in Au+Au and d+Au
  collisions at $\sqrt{s_{\rm NN}}$ = 200 GeV}.
\newblock \emph{Phys. Rev. C 88, 024906 (2013)}, 2013{\natexlab{a}}.

\bibitem[Collaboration(2014{\natexlab{b}})]{poverpALICE}
ALICE Collaboration.
\newblock {Production of charged pions, kaons and protons at large transverse
  momenta in pp and Pb--Pb collisions at $\sqrt{s_{\rm NN}}$ = 2.76 TeV}.
\newblock \emph{Physics Letters B 736 196-207}, 2014{\natexlab{b}}.

\bibitem[{V. Greco, C.M. Ko, P. Levai}(2003)]{recombination2}
{V. Greco, C.M. Ko, P. Levai}.
\newblock {Parton Coalescence at RHIC}.
\newblock \emph{Phys.Rev. C68 (2003) 034904}, 2003.

\bibitem[{Piotr Bozek, Iwona Wyskiel-Piekarska}(2012)]{pioverpHydro}
{Piotr Bozek, Iwona Wyskiel-Piekarska}.
\newblock {Particle spectra in Pb-Pb collisions at 2.76 TeV}.
\newblock \emph{Phys.Rev.C85:064915}, 2012.

\bibitem[Collaboration(2010{\natexlab{a}})]{FlowALICE}
ALICE Collaboration.
\newblock {Elliptic flow of charged particles in Pb-Pb collisions at 2.76 TeV}.
\newblock \emph{Phys.Rev.Lett.105}, 2010{\natexlab{a}}.

\bibitem[Snellings(2007)]{snellings}
Raimond Snellings.
\newblock {Anisotropic Flow from RHIC to the LHC}.
\newblock \emph{Eur.Phys.J.C49:87-90}, 2007.

\bibitem[Collaboration(2013{\natexlab{b}})]{IDFlow}
ALICE Collaboration.
\newblock {Anisotropic flow of charged hadrons, pions and (anti-)protons
  measured at high transverse momentum in Pb-Pb collisions at $\sqrt{s_{\rm
  NN}}$=2.76 TeV}.
\newblock \emph{Phys.Lett. B719 (2013) 18-28}, 2013{\natexlab{b}}.

\bibitem[{D. Molnar, S.A. Voloshin}(2003)]{flowhierachy}
{D. Molnar, S.A. Voloshin}.
\newblock {Elliptic Flow at Large Transverse Momenta from Quark Coalescence}.
\newblock \emph{Phys. Rev. Lett. 91}, 2003.

\bibitem[{K. Kovtun, D.T. Son, and A.O. Starinets}(2005)]{etas}
{K. Kovtun, D.T. Son, and A.O. Starinets}.
\newblock {Viscosity in Strongly Interacting Quantum Field Theories from Black
  Hole Physics}.
\newblock \emph{Phys. Rev. Lett. 94, 111601}, 2005.

\bibitem[Luzum and Romatschke(2009)]{luzum}
Matthew Luzum and Paul Romatschke.
\newblock {Viscous Hydrodynamic Predictions for Nuclear Collisions at the LHC}.
\newblock \emph{Phys. Rev. Lett. 103, 262302}, 2009.

\bibitem[{C. Shen, U. Heinz, P Huovinen, H. Song}(2005)]{Pasi}
{C. Shen, U. Heinz, P Huovinen, H. Song}.
\newblock {Radial and elliptic flow in Pb+Pb collisions at the Large Hadron
  Collider from viscous hydrodynamics}.
\newblock \emph{Phys.Rev. C84, 044903}, 2005.

\bibitem[{Burak Han Alver, Cl{\'e}ment Gombeaud, Matthew Luzum, and Jean-Yves
  Ollitrault}(2010)]{Triangular}
{Burak Han Alver, Cl{\'e}ment Gombeaud, Matthew Luzum, and Jean-Yves
  Ollitrault}.
\newblock Triangular flow in hydrodynamics and transport theory.
\newblock \emph{Phys. Rev. C 82, 034913}, 2010.

\bibitem[Collaboration(2011{\natexlab{a}})]{ALICEHarmonics}
ALICE Collaboration.
\newblock {Higher harmonic anisotropic flow measurements of charged particles
  in Pb-Pb collisions at $\sqrt{s_{\rm NN}}$=2.76 TeV}.
\newblock \emph{Phys.Rev.Lett. 107:032301}, 2011{\natexlab{a}}.

\bibitem[{Bjoern Schenke, Sangyong Jeon, Charles Gale}(2011)]{PredHarmonics}
{Bjoern Schenke, Sangyong Jeon, Charles Gale}.
\newblock {Anisotropic flow in $\sqrt{s_{\rm NN}}$=2.76 TeV Pb+Pb collisions at
  the LHC}.
\newblock \emph{Phys.Lett. B702 59-63}, 2011.

\bibitem[Rafelski(2012)]{strangeQGP}
Jan Rafelski.
\newblock Strangeness and quark--gluon plasma.
\newblock \emph{Acta Physica Polonica B vol. 43}, 2012.

\bibitem[Collaboration((2014))]{MultiStrange}
ALICE Collaboration.
\newblock {Multi-strange baryon production at mid-rapidity in Pb-Pb collisions
  at $\sqrt{s_{\rm NN}}$ = 2.76 TeV}.
\newblock \emph{Phys. Lett. B 728 216--227}, (2014).

\bibitem[Collaboration(1996)]{RhoNA60}
NA60 Collaboration.
\newblock {First Measurement of the $\rho$ Spectral Function in High-Energy
  Nuclear Collisions}.
\newblock \emph{Phys. Rev. Lett. 96, 162302}, 1996.

\bibitem[Rapp(2013)]{RappRho}
Ralf Rapp.
\newblock Update on chiral symmetry restoration in the context of dilepton
  data.
\newblock \emph{J.Phys.Conf.Ser. 420 (2013) 012017}, 2013.

\bibitem[Collaboration(2007{\natexlab{a}})]{jpsiPHENIX}
PHENIX Collaboration.
\newblock {J/$\Psi$ Production vs Centrality, Transverse Momentum, and Rapidity
  in Au+Au Collisions at $\sqrt{s_{\rm NN}}$ = 200 GeV}.
\newblock \emph{Phys.Rev.Lett.98:232301}, 2007{\natexlab{a}}.

\bibitem[Collaboration(2012{\natexlab{b}})]{upsilonCMS}
CMS Collaboration.
\newblock {Observation of sequential $\Upsilon$ suppression in Pb--Pb
  collisions}.
\newblock \emph{Phys. Rev. Lett. 109 (2012)}, 2012{\natexlab{b}}.

\bibitem[Collaboration(2014{\natexlab{a}})]{jpsiALICE}
ALICE Collaboration.
\newblock {Centrality, rapidity and transverse momentum dependence of J/$\Psi$
  suppression in Pb--Pb collisions at $\sqrt{s_{\rm NN}}$=2.76 TeV}.
\newblock \emph{Phys. Lett. B 734 (2014) 314-327}, 2014{\natexlab{a}}.

\bibitem[Grigorescu(2008)]{tesirum}
Tiberiu~Gabriel Grigorescu.
\newblock {Measurement of Charm Production in Deep Inelastic Scattering at HERA
  II}.
\newblock \emph{Ph.D. Thesis, Universiteit van Amsterdam}, 2008.

\bibitem[Mangano(1997)]{Mangano2}
M.L. Mangano.
\newblock Two lectures on heavy quark production in hadronic collisions.
\newblock \emph{hep-ph/9711337}, 1997.

\bibitem[{J. Pumplin, D.R. Stump, J. Huston, H.L. Lai, P. Nadolsky, W.K.
  Tung}(2002)]{cteqm6}
{J. Pumplin, D.R. Stump, J. Huston, H.L. Lai, P. Nadolsky, W.K. Tung}.
\newblock {New Generation of Parton Distributions with Uncertainties from
  Global QCD Analysis}.
\newblock \emph{JHEP 0207:012}, 2002.

\bibitem[{Matteo Cacciari, Paolo Nason, and Ramona Vogt}(2005)]{FONLL}
{Matteo Cacciari, Paolo Nason, and Ramona Vogt}.
\newblock {QCD Predictions for Charm and Bottom Quark Production at RHIC}.
\newblock \emph{Phys. Rev. Lett. 95, 122001}, 2005.

\bibitem[{B. A. Kniehl, G. Kramer, I. Schienbein, and H.
  Spiesberger}(2006)]{GM-VFNS}
{B. A. Kniehl, G. Kramer, I. Schienbein, and H. Spiesberger}.
\newblock {Reconciling Open-Charm Production at the Fermilab Tevatron with
  QCD}.
\newblock \emph{Phys. Rev. Lett. 96, 012001}, 2006.

\bibitem[Vogt(2014)]{VogtFact}
R.~Vogt.
\newblock {Open and Hidden Heavy Flavor Production in pp, pA and AA
  Collisions}.
\newblock \emph{Journal of Physics: Conference Series 509 (2014) 012007}, 2014.

\bibitem[Collaboration(2006)]{BelleFrag}
Belle Collaboration.
\newblock {Charm Hadrons from charm fragmentation and beauty decays in
  $e^{+}e^{-}$ annihilation at $\sqrt{s}$=10.6 GeV}.
\newblock \emph{Phys.Rev.D73:032002}, 2006.

\bibitem[{T. Kneesch, B.A. Kniehl, G. Kramer, I.
  Schienbein}(2008{\natexlab{a}})]{OPAL}
{T. Kneesch, B.A. Kniehl, G. Kramer, I. Schienbein}.
\newblock {Charmed-Meson Fragmentation Functions with Finite-Mass Corrections}.
\newblock \emph{Nucl.Phys.B799:34-59}, 2008{\natexlab{a}}.

\bibitem[Collaboration(2000)]{ALEPHFrag}
ALEPH Collaboration.
\newblock {Study of charm production in Z decays}.
\newblock \emph{Eur.Phys.J.C16:597-611}, 2000.

\bibitem[{M.Cacciari, P. Nason}(2003)]{FragCacciari}
{M.Cacciari, P. Nason}.
\newblock {Charm Cross Sections for the Tevatron Run II}.
\newblock \emph{JHEP 0309 (2003) 006}, 2003.

\bibitem[{E. Braaten, K. Cheung, S. Fleming, T.C. Yuan}(1995)]{formulaccia}
{E. Braaten, K. Cheung, S. Fleming, T.C. Yuan}.
\newblock {Perturbative QCD Fragmentation Functions as a Model for Heavy-Quark
  Fragmentation}.
\newblock \emph{Phys.Rev.D51:4819-4829}, 1995.

\bibitem[{T. Kneesch, B.A. Kniehl, G. Kramer, I.
  Schienbein}(2008{\natexlab{b}})]{FragGM}
{T. Kneesch, B.A. Kniehl, G. Kramer, I. Schienbein}.
\newblock {Charmed-Meson Fragmentation Functions with Finite-Mass Corrections}.
\newblock \emph{Nucl.Phys.B799:34-59}, 2008{\natexlab{b}}.

\bibitem[Collaboration(2003{\natexlab{a}})]{OpalBeauty}
Opal Collaboration.
\newblock {Inclusive Analysis of the b Quark Fragmentation Function in Z Decays
  at LEP}.
\newblock \emph{Eur.Phys.J.C29:463-478}, 2003{\natexlab{a}}.

\bibitem[{M. Soleymaninia, A. N. Khorramian, S. M. Moosavi Nejad, F.
  Arbabifar}(2013)]{pionfrag}
{M. Soleymaninia, A. N. Khorramian, S. M. Moosavi Nejad, F. Arbabifar}.
\newblock {Determination of pion and kaon fragmentation functions including
  spin asymmetries data in a global analysis}.
\newblock \emph{Phys.Rev. D88 5}, 2013.

\bibitem[Bowler(1981)]{BowlerFrag}
M.G. Bowler.
\newblock {e$^{+}$ e$^{-}$ Production of Heavy Quarks in the String Model}.
\newblock \emph{Z. Phys. C11 169.}, 1981.

\bibitem[Collaboration(2003{\natexlab{b}})]{CDFD}
CDF Collaboration.
\newblock {Measurement of Prompt Charm Meson Production Cross Sections in pp
  Collisions at $\sqrt{s}$=1.96 TeV}.
\newblock \emph{Phys. Rev. Lett. 91, 241804}, 2003{\natexlab{b}}.

\bibitem[Collaboration(2010{\natexlab{b}})]{STARD}
STAR Collaboration.
\newblock {Measurement of open heavy flavour production in the STAR experiment
  at RHIC}.
\newblock \emph{Pos (DIS 2010)182}, 2010{\natexlab{b}}.

\bibitem[Collaboration(2012{\natexlab{c}})]{D7TeV}
ALICE Collaboration.
\newblock {Measurement of Charm Production at central rapidity in proton-proton
  collisions at $\sqrt{s}$ = 7 TeV}.
\newblock \emph{JHEP01 128}, 2012{\natexlab{c}}.

\bibitem[Rafal~Maciula(2013)]{ALICEkt}
Antoni~Szczurek Rafal~Maciula.
\newblock {Open charm production at the LHC - k$_{\rm T}$-factorization
  approach}.
\newblock \emph{Phys.Rev. D87 9}, 2013.

\bibitem[Collaboration(2013{\natexlab{c}})]{LHCb}
LHCb Collaboration.
\newblock {Prompt charm production in pp collisions at $\sqrt{s}$=7 TeV}.
\newblock \emph{Nuclear Physics, Section B 871}, 2013{\natexlab{c}}.

\bibitem[L{\'e}vai and Vogt(1997)]{cctherm}
P{\'e}ter L{\'e}vai and Ramona Vogt.
\newblock Thermal charm production by massive gluons and quarks.
\newblock \emph{Phys. Rev. C 56, 2707}, 1997.

\bibitem[{Ben-Wei Zhang, Che Ming Ko, and Wei Liu}(2008)]{cctherm2}
{Ben-Wei Zhang, Che Ming Ko, and Wei Liu}.
\newblock {Thermal charm production in a quark-gluon plasma in Pb-Pb collisions
  at $\sqrt{s_{\rm NN}}$=5.5 TeV}.
\newblock \emph{Phys. Rev. C 77, 024901}, 2008.

\bibitem[{Alberto Accardi et. al. }(2012)]{StrucFun}
{Alberto Accardi et. al. }.
\newblock {Impact of nuclear dependence of R=$\sigma$L/$\sigma$T on
  anti-shadowing in nuclear structure functions}.
\newblock \emph{Phys.Rev. C86 (2012) 045201}, 2012.

\bibitem[Rith(2014)]{EMCStatus}
Klaus Rith.
\newblock {Present Status of the EMC effect}.
\newblock \emph{arXiv:1402.5000}, 2014.

\bibitem[Gelis(2012)]{CGC}
F.~Gelis.
\newblock {Color Glass Condensate and Glasma}.
\newblock \emph{arXiv:1211.3327}, 2012.

\bibitem[{}K. J.~Eskola(2009)]{EPS09}
C.~A.~Salgado {}K. J.~Eskola, H.~Paukkunen.
\newblock {EPS09 - a New Generation of NLO and LO Nuclear Parton Distribution
  Functions}.
\newblock \emph{HEP 0904:065}, 2009.

\bibitem[{M. Hirai, S. Kumano, T.-H. Nagai}(2007)]{HKN07}
{M. Hirai, S. Kumano, T.-H. Nagai}.
\newblock Determination of nuclear parton distribution functions and their
  uncertainties at next-to-leading order.
\newblock \emph{Phys.Rev.C76:065207}, 2007.

\bibitem[Dainese(2003)]{TesiDainese}
A.~Dainese.
\newblock {Charm production and in-medium QCD energy loss in nucleus-nucleus
  collisions with ALICE. A performance study}.
\newblock \emph{arXiv:nucl-ex/0311004}, 2003.

\bibitem[Dainese(2005)]{ProcDainese}
A.~Dainese.
\newblock {Charm and beauty of the Large Hadron Collider}.
\newblock \emph{J.Phys. G31}, 2005.

\bibitem[{J. W. Cronin et $al$.}(1975)]{CroninEffect}
{J. W. Cronin et $al$.}
\newblock {Production of hadrons at large transverse momentum at 200, 300, and
  400 Gev}.
\newblock \emph{Phys. Rev. D 11}, 1975.

\bibitem[Lev and Petersson(1983)]{CroninTheory}
M.~Lev and B.~Petersson.
\newblock {Nuclear Effects at Large Transverse Momentum in a QCD Parton Model}.
\newblock \emph{Z. Phys. C Particles and Fields 21, 155-161}, 1983.

\bibitem[{R. Baier, Yu. L. Dokshitzer, A.H. Mueller, S. Peign, D.
  Schiff}(1997)]{CNMEL}
{R. Baier, Yu. L. Dokshitzer, A.H. Mueller, S. Peign, D. Schiff}.
\newblock {Radiative energy loss and p$_{\rm T}$-broadening of high energy
  partons in nuclei}.
\newblock \emph{Nucl.Phys.B484:265-282}, 1997.

\bibitem[Bjorken(1982)]{BjorkenColl}
J.~D. Bjorken.
\newblock {Energy Loss of Energetic Partons in Quark-Gluon Plasma: Possible
  Extinction of High p$_{\rm T}$ Jets in Hadron-Hadron Collisions}.
\newblock \emph{FERMILAB-Pub-82/59-THY}, 1982.

\bibitem[{Stephane Peigne, Andre Peshier}(2008)]{PeigneEnLoss}
{Stephane Peigne, Andre Peshier}.
\newblock {Collisional energy loss of a fast heavy quark in a quark-gluon
  plasma}.
\newblock \emph{Phys.Rev.D77:114017}, 2008.

\bibitem[{Y.L. Dokshitzer, D.E. Kharzeev}(2001)]{DokKha}
{Y.L. Dokshitzer, D.E. Kharzeev}.
\newblock {Heavy quark colorimetry of QCD matter}.
\newblock \emph{Phys.Lett.B519}, 2001.

\bibitem[{R.Baier, Yu.L.Dokshitzer, A.H.Mueller, S.Peign{\'e},
  D.Schiff}(1997)]{BaierEnLoss}
{R.Baier, Yu.L.Dokshitzer, A.H.Mueller, S.Peign{\'e}, D.Schiff}.
\newblock {Radiative energy loss of high energy quarks and gluons in a finite
  volume quark-gluon plasma}.
\newblock \emph{Nucl.Phys. B483 291-320}, 1997.

\bibitem[{N{\'e}stor Armesto, Andrea Dainese, Carlos A. Salgado, Urs Achim
  Wiedemann}(2005)]{BDMPS}
{N{\'e}stor Armesto, Andrea Dainese, Carlos A. Salgado, Urs Achim Wiedemann}.
\newblock {Testing the Color Charge and Mass Dependence of Parton Energy Loss
  with Heavy-to-light Ratios at RHIC and LHC}.
\newblock \emph{Phys.Rev.D71:054027}, 2005.

\bibitem[{Xiang Wen-Changet al.}(2005)]{Cinesi}
{Xiang Wen-Changet al.}
\newblock {Radiative Energy Loss of Heavy Quark and Dead Cone Effect in
  Ultra-relativistic Heavy Ion Collisions}.
\newblock \emph{Chin. Phys. Lett, Vol 22, N.1 (2005) 72}, 2005.

\bibitem[{Cao, Qin, Bass}(2013)]{Cao}
{Cao, Qin, Bass}.
\newblock {Heavy-quark dynamics and hadronization in ultrarelativistic
  heavy-ion collisions: Collisional versus radiative energy loss}.
\newblock \emph{Phys.Rev. C88 (2013) 044907}, 2013.

\bibitem[{Cao, Bass}(2011)]{CaoTherm}
{Cao, Bass}.
\newblock {Thermalization of charm quarks in infinite and finite QGP matter}.
\newblock \emph{Phys. Rev. C 84, 064902}, 2011.

\bibitem[{Greco, Ko, Rapp}(2004)]{GKR}
{Greco, Ko, Rapp}.
\newblock {Quark coalescence for charmed mesons in ultrarelativistic heavy-ion
  collisions}.
\newblock \emph{Phys. Lett. B 595 202-208}, 2004.

\bibitem[Collaboration(2003{\natexlab{c}})]{PHENIXRdAu}
PHENIX Collaboration.
\newblock {Absence of Suppression in Particle Production at Large Transverse
  Momentum in $\sqrt{s_{\rm NN}}$ = 200 GeV d+Au Collisions}.
\newblock \emph{Phys.Rev.Lett.91}, 2003{\natexlab{c}}.

\bibitem[Collaboration(2013{\natexlab{d}})]{ALICERpPbLight}
ALICE Collaboration.
\newblock {Transverse Momentum Distribution and Nuclear Modification Factor of
  Charged Particles in p--Pb Collisions at $\sqrt{s_{\rm NN}}$= 5.02 TeV}.
\newblock \emph{Phys.Rev.Lett. 110 (2013) 082302}, 2013{\natexlab{d}}.

\bibitem[Collaboration(2012{\natexlab{d}})]{PHENIXEl}
PHENIX Collaboration.
\newblock {Cold-nuclear-matter effcts on heavy-quark production in d+Au
  collisions at $\sqrt{s_{\rm NN}}$ = 200 GeV}.
\newblock \emph{Phys.Rev.Lett. 109 (2012) 24}, 2012{\natexlab{d}}.

\bibitem[Collaboration(2014{\natexlab{b}})]{jpsipPb}
ALICE Collaboration.
\newblock {J/$\Psi$ production and nuclear effects in p-Pb collisions at
  $\sqrt{s_{\rm NN}}$=5.02 TeV}.
\newblock \emph{JHEP 02 (2014) 073}, 2014{\natexlab{b}}.

\bibitem[Collaboration(2011{\natexlab{b}})]{ALICEChargedRaa}
ALICE Collaboration.
\newblock {Suppression of charged particle production at large transverse
  momentum in central Pb--Pb collisions at $\sqrt{s_{\rm NN}}$=2.76 TeV}.
\newblock \emph{Physics Letters B 696 30--39}, 2011{\natexlab{b}}.

\bibitem[Collaboration(2003{\natexlab{d}})]{STARCharged}
STAR Collaboration.
\newblock {Transverse momentum and collision energy dependence of high p$_{\rm
  T}$ hadron suppression in Au+Au collisions at ultrarelativistic energies}.
\newblock \emph{Phys.Rev.Lett.91:172302}, 2003{\natexlab{d}}.

\bibitem[Collaboration(2012{\natexlab{e}})]{CMSCharged}
CMS Collaboration.
\newblock {Study of high-p$_{\rm T}$ charged particle suppression in Pb--Pb
  compared to pp collisions at $\sqrt{s_{\rm NN}}$=2.76TeV}.
\newblock \emph{Eur. Phys. J. C 72:1945}, 2012{\natexlab{e}}.

\bibitem[Collaboration(2007{\natexlab{b}})]{PHENIXAuAu}
PHENIX Collaboration.
\newblock {Energy Loss and Flow of Heavy Quarks in Au+Au Collisions at
  $\sqrt{s_{\rm NN}}$ = 200 GeV}.
\newblock \emph{Phys. Rev. Lett. 98, 172301}, 2007{\natexlab{b}}.

\bibitem[Collaboration(2014{\natexlab{c}})]{STARD0}
STAR Collaboration.
\newblock {Observation of D$^{0}$ meson nuclear modifications in Au+Au
  collisions at $\sqrt{s_{\rm NN}}$ = 200 GeV}.
\newblock \emph{Phys.Rev.Lett. 113 (2014) 142301}, 2014{\natexlab{c}}.

\bibitem[{Min He, Rainer J. Fries, Ralf Rapp}(2012{\natexlab{a}})]{TAMU}
{Min He, Rainer J. Fries, Ralf Rapp}.
\newblock {Heavy Flavor at the Large Hadron Collider in a Strong Coupling
  Approach}.
\newblock \emph{arXiv:1401.3817}, 2012{\natexlab{a}}.

\bibitem[{Min He, Rainer J. Fries, Ralf Rapp}(2012{\natexlab{b}})]{TAMU2}
{Min He, Rainer J. Fries, Ralf Rapp}.
\newblock {Heavy-Quark Diffusion and Hadronization in Quark-Gluon Plasma}.
\newblock \emph{Phys.Rev. C86 014903}, 2012{\natexlab{b}}.

\bibitem[{M. Djordjevic, M. Djordjevic}(2014)]{Djordjevic}
{M. Djordjevic, M. Djordjevic}.
\newblock {LHC jet suppression of light and heavy flavor observables}.
\newblock \emph{Physics Letters B 734 (2014) 286--289}, 2014.

\bibitem[{Horowitz, W.A. and Gyulassy, Miklos}(2011)]{WHDG}
{Horowitz, W.A. and Gyulassy, Miklos}.
\newblock {The Surprising Transparency of the sQGP at LHC}.
\newblock \emph{Nucl.Phys. A872 (2011) 265-285}, 2011.

\bibitem[{Simon Wicks, William Horowitz, Magdalena Djordjevic, Miklos
  Gyulassy}(2007)]{DGLV}
{Simon Wicks, William Horowitz, Magdalena Djordjevic, Miklos Gyulassy}.
\newblock {Elastic, Inelastic, and Path Length Fluctuations in Jet Tomography}.
\newblock \emph{Nucl.Phys.A784:426-442}, 2007.

\bibitem[{Marlene Nahrgang, Joerg Aichelin, Pol Bernard Gossiaux, Klaus
  Werner}(2014)]{MCat}
{Marlene Nahrgang, Joerg Aichelin, Pol Bernard Gossiaux, Klaus Werner}.
\newblock {Influence of hadronic bound states above T$_c$ on heavy-quark
  observables in Pb--Pb collisions at at the CERN Large Hadron Collider}.
\newblock \emph{Phys. Rev. C 89, 014905}, 2014.

\bibitem[{K. Werner, Iu. Karpenko, T. Pierog, M. Bleicher, and K.
  Mikhailov}(2010)]{FluxTube}
{K. Werner, Iu. Karpenko, T. Pierog, M. Bleicher, and K. Mikhailov}.
\newblock {Event-by-event simulation of the three-dimensional hydrodynamic
  evolution from flux tube initial conditions in ultrarelativistic heavy ion
  collisions}.
\newblock \emph{PHYSICAL REVIEW C 82 , 044904}, 2010.

\bibitem[Vitev(2006)]{Vitev}
I.~Vitev.
\newblock {Testing the theory of QGP-induced energy loss at RHIC and the LHC}.
\newblock \emph{Phys.Lett.B639:38-45}, 2006.

\bibitem[et~al.(2011)]{POWLANG}
W.M.~Alberico et~al.
\newblock {Heavy-flavor dynamics in nucleus-nucleus collisions: from RHIC to
  LHC}.
\newblock \emph{J.Phys. G38 (2011) 124144}, 2011.

\bibitem[{Stefano Frixione, Paolo Nason, Giovanni Ridolfi}(2007)]{POWHEG}
{Stefano Frixione, Paolo Nason, Giovanni Ridolfi}.
\newblock A positive-weight next-to-leading-order {Monte} {Carlo} for heavy
  flavour hadroproduction.
\newblock \emph{JHEP 0709:126}, 2007.

\bibitem[{Jan Uphoff, Oliver Fochler, Zhe Xu, Carsten Greiner}(2012)]{BAMPS}
{Jan Uphoff, Oliver Fochler, Zhe Xu, Carsten Greiner}.
\newblock {Open heavy flavor in Pb--Pb collisions at $\sqrt{s_{\rm NN}}$=2.76
  TeV within a transport model}.
\newblock \emph{Physics Letters B 717}, 2012.

\bibitem[Collaboration(2012{\natexlab{f}})]{nonpromptCMS}
CMS Collaboration.
\newblock {Suppression of non-prompt J/$\Psi$, prompt J/$\Psi$, and
  $\Upsilon$(1S) in Pb--Pb collisions at $\sqrt{s_{\rm NN}}$ = 2.76 TeV}.
\newblock \emph{JHEP 05 (2012) 063}, 2012{\natexlab{f}}.

\bibitem[Collaboration(2014{\natexlab{d}})]{ALICEv2}
ALICE Collaboration.
\newblock {Azimuthal anisotropy of D meson production in Pb--Pb collisions at
  $\sqrt{s_{\rm NN}}$=2.76 TeV}.
\newblock \emph{Phys. Rev. C 90 034904}, 2014{\natexlab{d}}.

\bibitem[Collaboration(2010{\natexlab{c}})]{mult}
ALICE Collaboration.
\newblock {Charged Particle Multiplicity Density at Midrapidity in Central
  Pb--Pb Collisions at $\sqrt{s_{\rm NN}}$=2.76 TeV}.
\newblock \emph{Phys. Rev. Lett. 105}, 2010{\natexlab{c}}.

\bibitem[Collaboration(2014{\natexlab{e}})]{ALICEPerfPaper}
ALICE Collaboration.
\newblock {Performance of the ALICE Experiment at the CERN LHC}.
\newblock \emph{Int. J. Mod. Phys. A 29}, 2014{\natexlab{e}}.

\bibitem[Collaboration(2008)]{ALICEJinst}
ALICE Collaboration.
\newblock {The ALICE experiment at the CERN LHC}.
\newblock \emph{JINST3 S08002}, 2008.

\bibitem[et~al.(2013)]{TOFBologna}
A.~Akindinov et~al.
\newblock {Performance of the ALICE Time-Of-Flight detector at the LHC}.
\newblock \emph{Eur. Phys. J. Plus 128: 44}, 2013.

\bibitem[Fruhwirth(1987)]{Kalman}
R.~Fruhwirth.
\newblock {Application of Kalman filtering to track and vertex fitting}.
\newblock \emph{Nucl. Instr. and Meth. A 262, 444}, 1987.

\bibitem[Collaboration(2013{\natexlab{e}})]{CentPbPb}
ALICE Collaboration.
\newblock {Centrality determination of Pb--Pb collisions at $\sqrt{s_{\rm
  NN}}$= 2.76 TeV with ALICE}.
\newblock \emph{Phys. Rev. C 88}, 2013{\natexlab{e}}.

\bibitem[{M. Alvioli, M. Strikman}(2013)]{FlucGlauber}
{M. Alvioli, M. Strikman}.
\newblock {Color fluctuation effects in proton-nucleus collisions}.
\newblock \emph{Phys. Lett. B 722}, 2013.

\bibitem[{Ferenc Sikler}(2003)]{SlowNucleons}
{Ferenc Sikler}.
\newblock {Centrality control of hadron-nucleus interactions by detection of
  slow nucleons}.
\newblock \emph{arXiv:hep-ph/0304065}, 2003.

\bibitem[Collaboration(2014{\natexlab{f}})]{CentpPb}
ALICE Collaboration.
\newblock {Centrality dependence of particle production in p--Pb collisions at
  $\sqrt{s_{\rm NN} }$= 5.02 TeV}.
\newblock \emph{arXiv:1412.6828v1}, 2014{\natexlab{f}}.

\bibitem[{Particle Data Group}(2014)]{PDG}
{Particle Data Group}.
\newblock {Review of Particle Physics}.
\newblock \emph{Chinese Physics C, Volume 38, Number 9}, 2014.

\bibitem[{E.Bruna, A.Dainese, M.Masera, F.Prino}(2009)]{secver}
{E.Bruna, A.Dainese, M.Masera, F.Prino}.
\newblock {Vertex reconstruction for proton-proton collisions in ALICE}.
\newblock \emph{ALICE-INT-2009-018}, 2009.

\bibitem[{Xin-Nian Wang and Miklos Gyulassy}(1991)]{HIJING}
{Xin-Nian Wang and Miklos Gyulassy}.
\newblock {HIJING: A Monte Carlo model for multiple jet production in p p, p A
  and A A collisions}.
\newblock \emph{Phys.Rev.D 44, 3501}, 1991.

\bibitem[Collaboration(2014{\natexlab{g}})]{psiprimoLHC}
ALICE Collaboration.
\newblock {Suppression of $\Psi$(2S) production in p-Pb collisions at
  $\sqrt{s_{NN}}$ = 5.02 TeV}.
\newblock \emph{JHEP 1412 073}, 2014{\natexlab{g}}.

\bibitem[Collaboration(2013{\natexlab{f}})]{PsiPrimoRHIC}
PHENIX Collaboration.
\newblock {Nuclear Modification of $\Psi$', $\chi_c$, and J/$\Psi$ Production
  in d+Au Collisions at $\sqrt{s_{\rm NN}}$ = 200 GeV}.
\newblock \emph{PRL 111, 202301}, 2013{\natexlab{f}}.

\bibitem[Collaboration(2013{\natexlab{g}})]{RidgepPbCMS}
CMS Collaboration.
\newblock {Observation of long-range, near-side angular correlations in p--Pb
  collisions at the LHC}.
\newblock \emph{Volume 718, Issue 3, 8 January 2013, Pages 795--814},
  2013{\natexlab{g}}.

\bibitem[Sickles(2014)]{Sickles}
A.M. Sickles.
\newblock {Possible evidence of radial flow of heavy mesons in d+Au
  collisions}.
\newblock \emph{Phys. Lett. B 731 51-56}, 2014.

\bibitem[Collaboration(2010{\natexlab{d}})]{RidgeppCMS}
CMS Collaboration.
\newblock {Observation of long-range, near-side angular correlations in
  proton-proton collisions at the LHC}.
\newblock \emph{JHEP 1009 (2010) 091}, 2010{\natexlab{d}}.

\bibitem[Collaboration(2012{\natexlab{g}})]{RidgePbPbCMS}
CMS Collaboration.
\newblock {Centrality dependence of dihadron correlations and azimuthal
  anisotropy harmonics in PbPb collisions at $\sqrt{s_{\rm NN}}$=2.76 TeV}.
\newblock \emph{Eur. Phys. J. C (2012) 72:2012}, 2012{\natexlab{g}}.

\bibitem[Collaboration(2013{\natexlab{h}})]{RidgeALICE}
ALICE Collaboration.
\newblock Long-range angular correlations on the near and away side in p--pb
  collisions at view the mathml source.
\newblock \emph{Physics Letters B 719 (2013) 29--41}, 2013{\natexlab{h}}.

\bibitem[Collaboration(2013{\natexlab{i}})]{RidgeATLAS}
ATLAS Collaboration.
\newblock {Observation of Associated Near-Side and Away-Side Long-Range
  Correlations in $\sqrt{s_{\rm NN}}$= 5.02 TeV Proton-Lead Collisions with the
  ATLAS Detector}.
\newblock \emph{Phys. Rev. Lett. 110, 182302}, 2013{\natexlab{i}}.

\bibitem[{S. Voloshin, Y. Zhang}(1996)]{modella}
{S. Voloshin, Y. Zhang}.
\newblock {Flow study in relativistic nuclear collisions by Fourier expansion
  of azimuthal particle distributions}.
\newblock \emph{Zeitschrift f{\"u}r Physik C Particles and Fields December
  1996, Volume 70, Issue 4, pp 665-671}, 1996.

\bibitem[{D.J. Lange}(2011)]{evtgen}
{D.J. Lange}.
\newblock {The EvtGen particle decay simulation package}.
\newblock \emph{Nucl. Instrum. Methods A462 152}, 2011.

\bibitem[{R. Averbeck et al.}(2011)]{ppref}
{R. Averbeck et al.}
\newblock {Reference heavy flavour cross sections in pp collisions at
  $\sqrt{s}$ = 2.76 TeV, using a pQCD-driven $\sqrt{s}$-scaling of ALICE
  measurements at $\sqrt{s}$ = 7 TeV}.
\newblock \emph{arXiv:1107.3243}, 2011.

\bibitem[Collaboration(2014{\natexlab{h}})]{V0AND}
ALICE Collaboration.
\newblock {Measurement of visible cross sections in proton-lead collisions at
  $\sqrt{s_{\rm NN}}$= 5.02 TeV in van der Meer scans with the ALICE detector}.
\newblock \emph{JINST 9 (2014) 11, P11003}, 2014{\natexlab{h}}.

\bibitem[Collaboration(2012{\natexlab{h}})]{citare}
ALICE Collaboration.
\newblock {$D_{s}^+$ meson production at central rapidity in proton--proton
  collisions at $\sqrt{s}=7$ TeV}.
\newblock \emph{Physics Letters B}, 718\penalty0 (2):\penalty0 279 -- 294,
  2012{\natexlab{h}}.

\bibitem[{H. Fujii, K Watanabe}(2013)]{CGCpPn}
{H. Fujii, K Watanabe}.
\newblock {Heavy quark pair production in high energy pA collisions: Open heavy
  flavors}.
\newblock \emph{Nuclear Physics A 920, 78-93}, 2013.

\bibitem[{R. Sharma, I. Vitev, B. Zhang}(2009)]{VitevpPb}
{R. Sharma, I. Vitev, B. Zhang}.
\newblock {Light-cone wave function approach to open heavy flavor dynamics in
  QCD matter}.
\newblock \emph{PHYSICAL REVIEW C 80 , 054902}, 2009.

\bibitem[{Gian Michele Innocenti for the CMS Collaboration}(2014)]{gian}
{Gian Michele Innocenti for the CMS Collaboration}.
\newblock {B-meson reconstruction performance and spectra in pp and pPb
  collisions in CMS}.
\newblock \emph{Nuclear Physics A Volume 931, November 2014, Pages 1184--1188},
  2014.

\bibitem[Collaboration(2012{\natexlab{i}})]{RAAvecchio}
ALICE Collaboration.
\newblock {Suppression of high transverse momentum D mesons in central Pb--Pb
  collisions at $\sqrt{s_{\rm NN}}$=2.76 TeV}.
\newblock \emph{JHEP 09 (2012) 112}, 2012{\natexlab{i}}.

\bibitem[{L. Frankfurt, M. Strikman, and C. Weiss}(2011)]{gendist}
{L. Frankfurt, M. Strikman, and C. Weiss}.
\newblock {Transverse nucleon structure and diagnostics of hard parton--parton
  processes at LHC}.
\newblock \emph{Phys.Rev. D83 054012}, 2011.

\bibitem[Collaboration(1988)]{NA27}
NA27 Collaboration.
\newblock {Comparative properties of 400 GeV/c proton-proton interactions with
  and without charm production}.
\newblock \emph{Z. Phys. C Particles and Fields 41,191-196}, 1988.

\bibitem[Collaboration(2012{\natexlab{j}})]{jpsimult}
ALICE Collaboration.
\newblock {J/$\Psi$ production as a function of charged particle multiplicity
  in pp collisions at $\sqrt{s}$=7 TeV}.
\newblock \emph{Physics Letters B Volume 712, Issue 3, 6 June 2012, Pages
  165--175}, 2012{\natexlab{j}}.

\bibitem[Collaboration(2012{\natexlab{k}})]{LHCbDC}
LHCb Collaboration.
\newblock {Observation of double charm production involving open charm in pp
  collisions at $\sqrt{s}$ = 7 TeV}.
\newblock \emph{JHEP 1206 (2012) 141}, 2012{\natexlab{k}}.

\bibitem[Collaboration(2013{\natexlab{j}})]{CMSMPI}
CMS Collaboration.
\newblock {Jet and underlying event properties as a function of
  charged-particle multiplicity in proton--proton collisions at $\sqrt{s}$= 7
  TeV}.
\newblock \emph{Eur.Phys.J. C73 2674}, 2013{\natexlab{j}}.

\bibitem[Collaboration(2013{\natexlab{k}})]{triggermult}
ALICE Collaboration.
\newblock {Pseudorapidity density of charged particles in p--Pb collisions at
  $\sqrt{s_{NN}}=5.02$ TeV}.
\newblock \emph{Phys.Rev.Lett. 110 (2013) 3, 032301}, 2013{\natexlab{k}}.

\bibitem[{T. Sjostrand, S. Mrenna, P. Skands}(2006)]{PYTHIA}
{T. Sjostrand, S. Mrenna, P. Skands}.
\newblock {PYTHIA 6.4 physics and manual }.
\newblock \emph{JHEP 05 026}, 2006.

\end{thebibliography}

\end{document}